\documentclass[11pt,twoside,openright]{book}
\usepackage[english]{babel}
\usepackage[utf8x]{inputenc}

\usepackage{breakurl}
\usepackage{bold-extra}
\usepackage{float}
\usepackage{graphicx}

\usepackage{longtable}
\usepackage{lscape}
\usepackage{amssymb}
\usepackage[nointegrals]{wasysym} 
\usepackage{gensymb}
\usepackage{hyperref}
\hypersetup{%
  colorlinks = true,
  linkcolor  = blue
}
\hypersetup{colorlinks,citecolor=blue}
\usepackage{natbib}
\bibpunct{(}{)}{;}{a}{}{,}
\usepackage{fancyhdr}
\usepackage{framed}
\usepackage{subcaption}
\usepackage{minibox}
\usepackage{textcomp}
\usepackage{makeidx}
\usepackage{longtable}
\usepackage{wrapfig}
\usepackage{booktabs}
\usepackage{geometry}
\usepackage{gensymb}
\usepackage[titletoc,title]{appendix}
\usepackage{color}
\usepackage{url}
\usepackage{fourier-orns}
\usepackage{amsmath}
\usepackage{alltt, xcolor}
\usepackage{scrextend}
\usepackage{blindtext}
\usepackage{enumerate} 
\usepackage{tablefootnote}
\usepackage{txfonts}
\usepackage{chemformula} 
\usepackage[version=4]{mhchem} 
\usepackage{pdfpages}
\usepackage{lettrine} 

\usepackage{ulem}

\usepackage{multirow}

\usepackage{titlesec}
\usepackage{sectsty} 

\usepackage{multicol} 

\usepackage{booktabs}
\usepackage[referable]{threeparttablex}
\usepackage{tablefootnote}
\usepackage{tabularx}
\usepackage{wasysym} 
\usepackage{hyperref} 

\usepackage{pdfpages}
\usepackage{multicol}
\usepackage{ragged2e}
\usepackage{caption}
\usepackage{xurl}    
\usepackage{marvosym}

\counterwithout*{footnote}{chapter} 

\usepackage{footmisc}
\usepackage{multicol}
\usepackage[printonlyused]{acronym}
\makeindex
\usepackage{cleveref}
\crefname{section}{§}{§§}
\Crefname{section}{§}{§§}
\usepackage{color}

\usepackage{courier}
\usepackage{setspace}
\usepackage{listings}
\usepackage{subcaption}

\usepackage{afterpage}

\usepackage{changepage}

\usepackage{yfonts}

\usepackage{appendix}

\usepackage{imakeidx}
\makeindex[columns=3, title=Alphabetical Index, intoc]

\makeindex

\begin{document}
\bibliographystyle{My_apj} 

\begin{titlepage}
\renewcommand{\thefootnote}{\Roman{footnote}} 

{\fontfamily{ptm}\selectfont

\begin{center}
\begin{LARGE}
\textsc{\textbf{Complutense University of Madrid
}}\\\vspace{.4cm}
\end{LARGE}

\begin{Large}
\textbf{Faculty of Physical Sciences}\\\vspace{.4cm}
\end{Large}

\begin{figure} 
\vspace{0.8cm}
  \centering
    \includegraphics[width=.3 \textwidth]{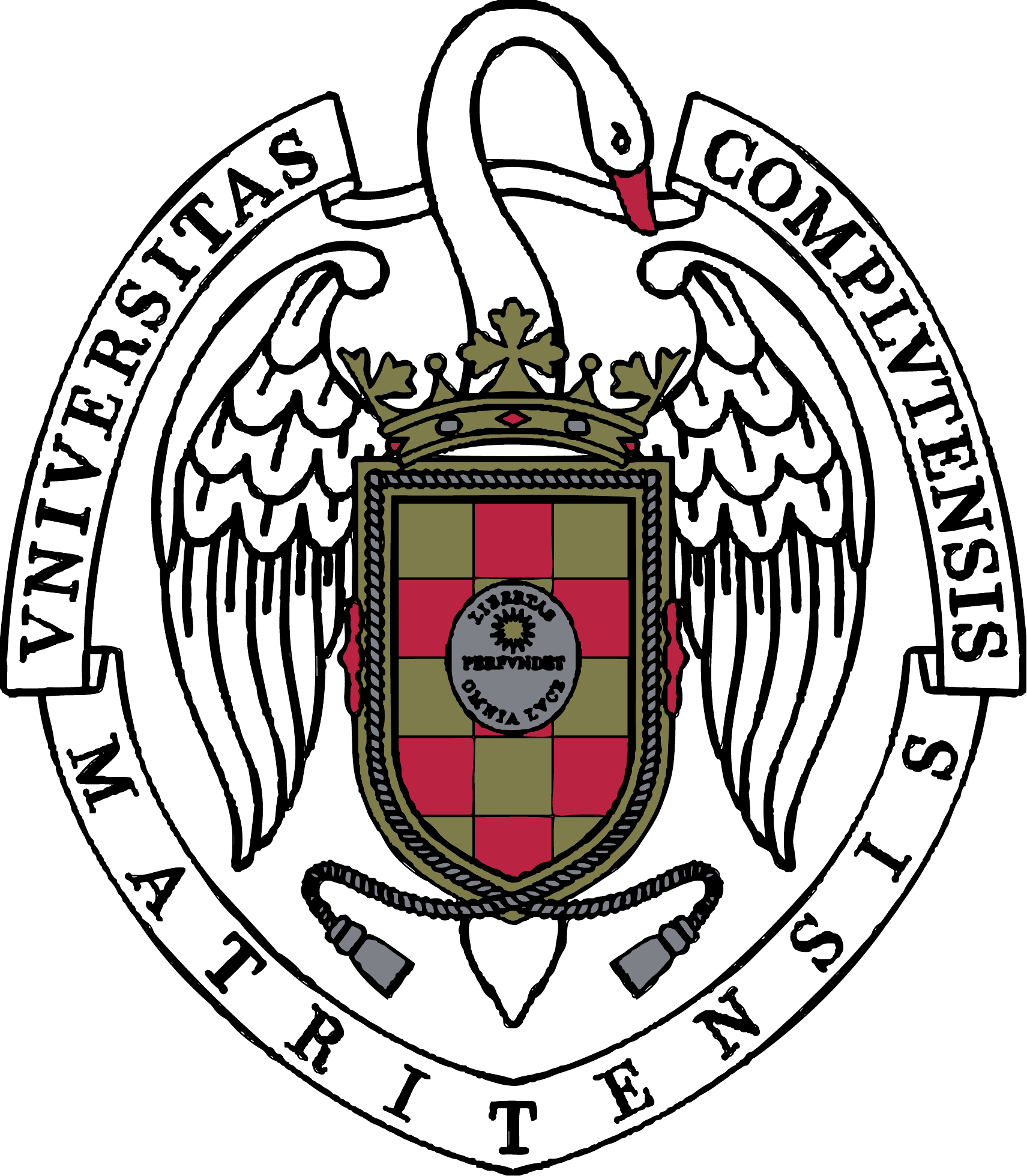}
\end{figure}

\begin{Large}
\vspace{2.7cm}
\textsc{\textbf{Doctoral thesis}}\\\vspace{.4cm}
\end{Large}

\begin{Huge}
\textit{\textbf{Exoplanet exploration \\ in the context of the PLATO mission: \\ From detection to population studies}}\\
\end{Huge}

\vspace{1 cm}
\vspace{5.2 cm}
\begin{large}
\end{large}
\begin{huge}
\textbf{Amadeo Castro González}\\
\end{huge}

\vspace{0.6 cm}

\begin{large}

\textbf{Madrid, 2025}\\[1.5 cm]
\end{large}

\end{center}\vfill 

}

\end{titlepage}


\clearpage\hbox{}\thispagestyle{empty}\newpage 

{\fontfamily{ptm}\selectfont

\begin{center}
\begin{LARGE}
\textsc{\textbf{Universidad Complutense de Madrid}}\\\vspace{.4cm}
\end{LARGE}

\begin{Large}
\textbf{Facultad de Ciencias Físicas}\\\vspace{.4cm}
\end{Large}

\begin{figure} 
\vspace{0.8cm}
  \centering
    \includegraphics[width=.3 \textwidth]{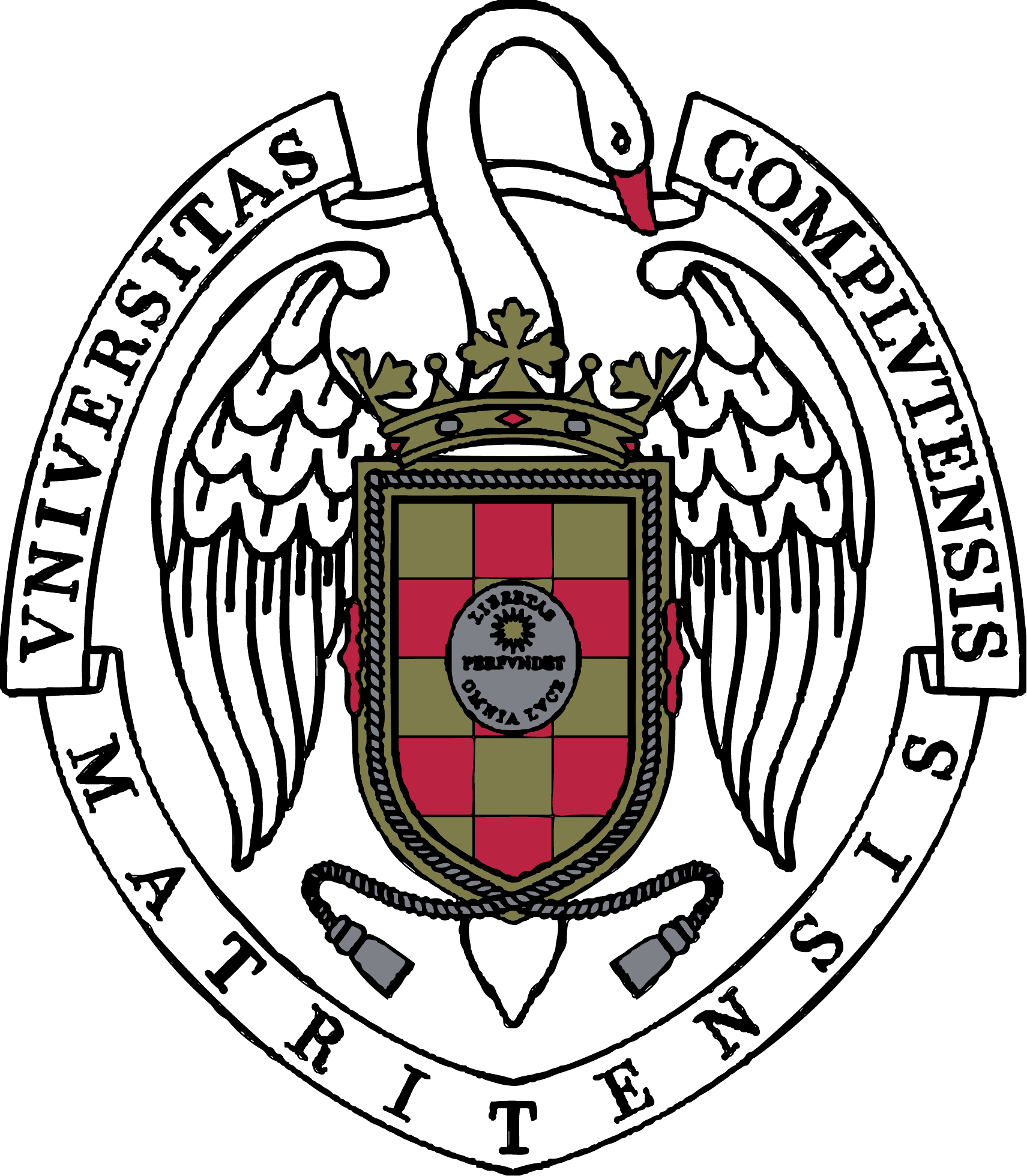}
\end{figure}

\begin{huge}
\textsc{\textbf{doctorado en astrofísica}}\\\vspace{.4cm}
\end{huge}

\begin{LARGE}
\textit{\textbf{Exploración de exoplanetas en el contexto de la misión PLATO: De la detección al estudio de poblaciones}}\\
\vspace{.6cm}
\textit{\textbf{Exoplanet exploration in the context of the PLATO mission: From detection to population studies}}\\
\end{LARGE}

\vspace{1 cm}
\begin{large}
Memoria para optar al grado de doctor presentada por:\\\vspace{0.5 cm}
\end{large}
\begin{huge}
\textbf{Amadeo Castro González}\\
\end{huge}

\vspace{0.8 cm}

\begin{large}
\textbf{Directores:} \\\vspace{0.4 cm}
Dr. Jorge Lillo Box \\
Dr. David Barrado Navascués\\\vspace{0.4 cm}

\textbf{Madrid, 2025}\\[0.4cm]
\end{large}

\thispagestyle{empty} 

\end{center}\vfill 

}

\clearpage\hbox{}\thispagestyle{empty}\newpage 

\thispagestyle{empty} 
\includegraphics[width=\textwidth]{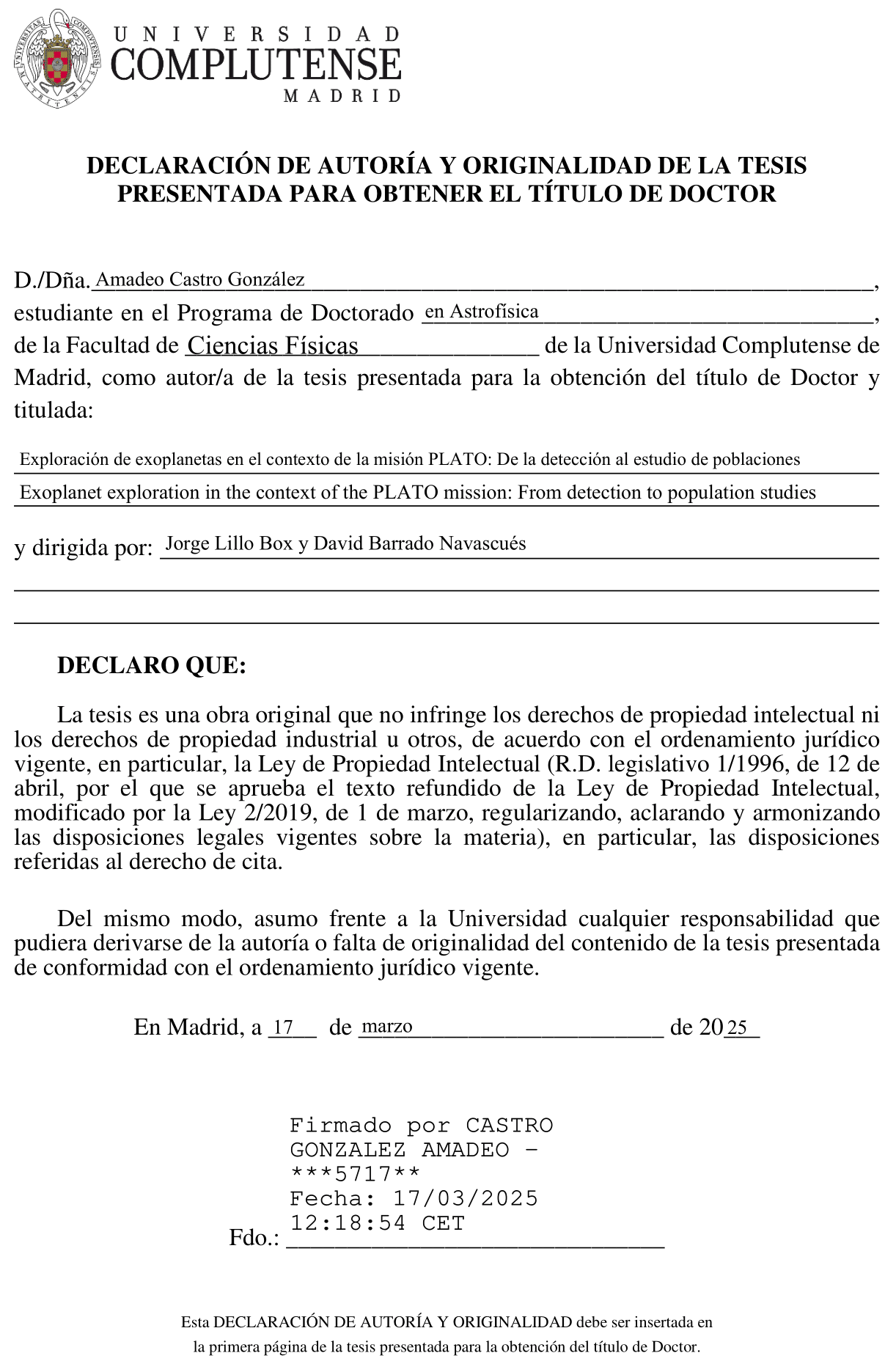}

\clearpage\hbox{}\thispagestyle{empty}\newpage 

\setcounter{footnote}{0}

\thispagestyle{empty}
\vspace*{0.2cm}


\begin{center}
\begin{large}
    \textit{Unos buenos padres valen por cien maestros} \\ \vspace{2mm}
\end{large}
    Jean-Jacques Rousseau
\end{center}

\begin{figure*}[h]
    \centering
    \includegraphics[width=0.85\textwidth, angle = 22]{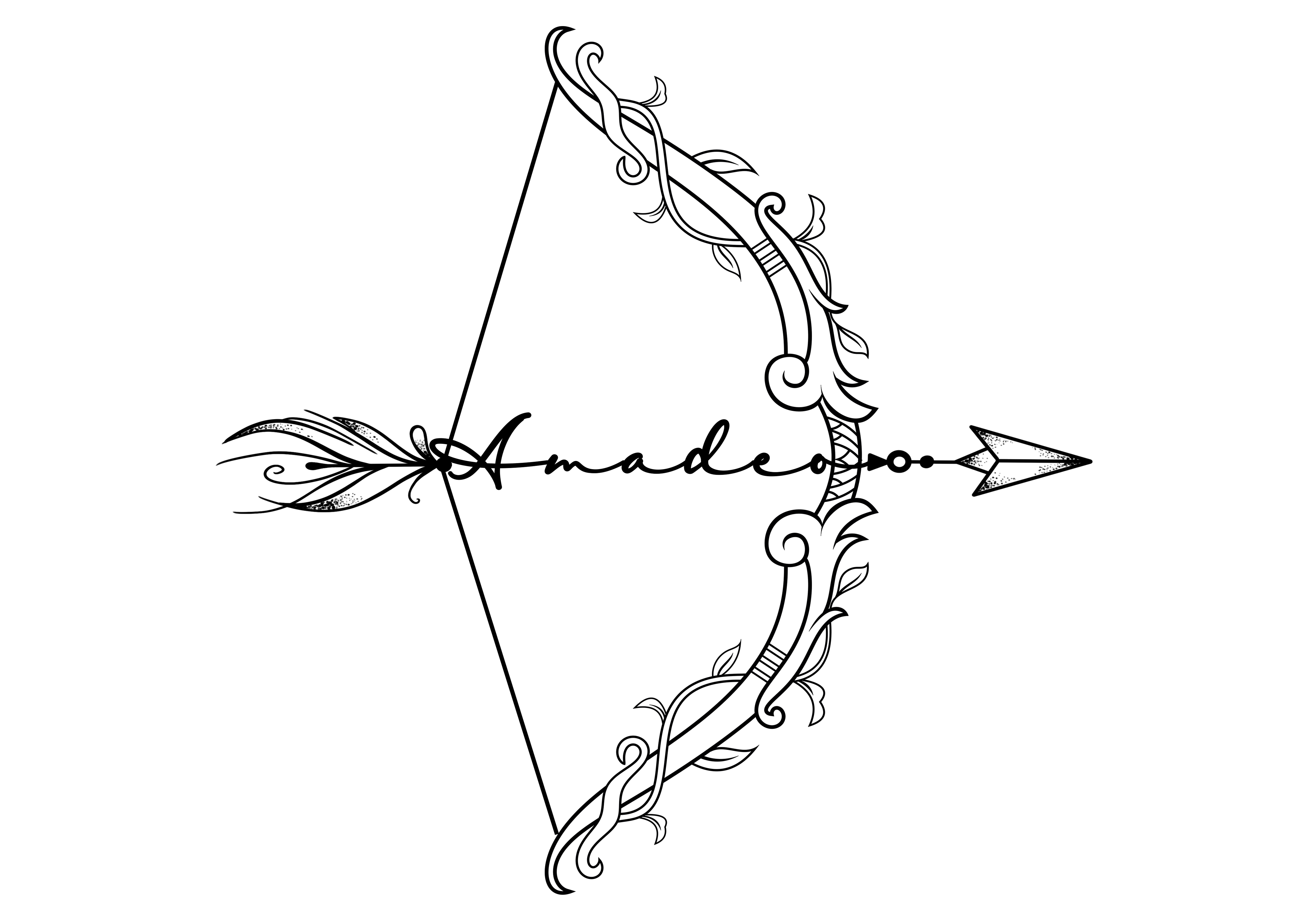}
\end{figure*}

\begin{center}
\begin{large}
    \textit{La misión del arco es impulsar la flecha, darle fuerza y velocidad; también orientarla bien en su ruta hacia el blanco. La de la flecha es seguir su propio camino sin apegarse al arco. De no hacerlo así, ni el arco ni la flecha habrían cumplido su misión} 
    \end{large}
    \\ \vspace{3mm}
    Jaume Soler $\&$ M. Mercé Conangla

\end{center}

\vspace{7mm}

\begin{center}
\begin{Large}
    Esta tesis se la dedico a mis padres, por el impulso, la orientación, la dedicación, la ayuda y el amor
\end{Large}
\end{center}

\chapter*{Agradecimientos}
\vspace{2.5cm}
\pagestyle{fancy}
\fancyhf{}

Agradezco, en primer lugar, a mis padres el haberme inculcado la importancia de vivir de manera consciente con sentido de gratitud. Gracias a mis tíos José Antonio, María Jesús, Manuel Ángel e Isabel, y a mis primas Diana y Begoña por su apoyo. Gracias a mis abuelos por su cariño.

Aprovecho la oportunidad que me brinda esta tesis para expresar mi agradecimiento hacia personas, lugares y experiencias que he vivido durante la etapa de doctorado.

Estoy muy agradecido por las oportunidades científicas recibidas y por haber tenido la ocasión de conocer y ser acompañado por personas que me han aportado valiosos tesoros inmateriales, en forma de modelos de referencia ética a seguir, tanto científica como personal. Agradezco de igual modo todas las experiencias vividas durante esta etapa, tanto las buenas como aquellas otras más oscuras y dolorosas de las que he sacado valiosos aprendizajes. En algunas batallas, lo importante no es la espada, sino la mano que la dirige y el corazón que la sostiene. Gracias, David Barrado Navascués.

Agradezco a Jorge Lillo Box, David Barrado Navascués y Miguel Mas Hesse por abrirme las puertas del Centro de Astrobiología (CAB) con un contrato FPI, por ofrecerme la oportunidad de desarrollar mi doctorado en diversos consorcios y por facilitarme la asistencia a cursos, congresos, seminarios y estancias en el extranjero. Sin duda, el valor de las estancias tanto en el CAUP (Oporto) como en el Observatorio de Ginebra ha sido extraordinario. Estas visitas han influido mucho en mi trayectoria profesional, por los conocimientos adquiridos, por las colaboraciones actuales y futuras, por abrir mi mente a nuevas formas de trabajar y hacer ciencia, y sobre todo por la acogida que recibí. ¡Cómo olvidarme de esas tardes de fútbol en el Estádio Universitário y de los viernes de cervezas y pinchos en Casa Agrícola, y cómo olvidarme de esos paseos en bicicleta y comidas en Manor Chavannes, las fondues y las tardes de cervezas en Versoix! Muchas gracias, Nuno, Vardan, Sérgio, Olivier, Elisa, Susana, Elsa, André, Angela, Ana Rita, Bárbara, Pedro, João, Vincent, Monika, François, Jean Baptiste, Xavier, Adrien Leleu, Adrien Deline, Nicolas, Christoph, Baptiste, Babatunde, Léna, Melissa, Julia, José, David, Stéphane y Francesco. 

Quiero destacar a Pedro Figueira, João Faria y André Silva. En nuestro último reencuentro en Ginebra nos convertimos en los \textit{four horsemen of the beer-ocalipse} y forjamos algo más que una colaboración. 

Quiero también mencionar especialmente al Profesor Vincent Bourrier por ser un referente, un modelo de hacer buena ciencia. Ahora cala en mí, más profundamente, la idea de Benjamín Disraeli de que ‘lo mejor que podemos hacer por el otro no es solo compartir nuestras riquezas, sino mostrarle las suyas'. Muchas gracias por todo lo que me has enseñado, y especialmente por la confianza depositada en mí. En relación a la enseñanza, siempre me ha gustado la metáfora de que ‘la mente no es un recipiente para llenar, sino un fuego por encender'. Y en este sentido, Profesor Bourrier, tú has sido un auténtico ‘avivador de brasas'.

Estoy muy agradecido a todos los colaboradores que han contribuido a esta tesis, en forma de análisis y múltiples discusiones. Gracias a Vincent Bourrier, Alexandre Correia, Lorena Acuña, Artem Aguichine, Jean Baptiste Delisle, Vardan Adibekyan, Sérgio Sousa, Hugo Tabernero, María Rosa Zapatero Osorio, Alexandros Antoniadis-Karnavas, Elisa Delgado-Mena, Andy Moya, Olivier Demangeon, Siddharth Gandhi, Evgenya Shkolnik, Magali Deleuil, Jeanne Davoult, Yann Alibert, Sara Seager, James Jenkins, John Livingston, Jerome de Leon, Judith Korth, Enrique Díez Alonso, Juan Menéndez Blanco, Susana Fernández Menéndez y Manel Recio. 

Quiero hacer una mención especial a David Armstrong. Gracias por la confianza y por transmitirme la pasión por los exo-Neptunos.

Agradezco también la contribución desinteresada de astrónomos amateurs de la Sociedad Astronómica Asturiana Omega, Vidal, Saray, Ramón, Javier, Faustino, Alfonso, Raúl, Pilar, Nabila y David, quienes
detectaron las señales de cuatro nuevos planetas presentados en esta tesis. Esto ha sido posible gracias a la colaboración ProAm con el Instituto de Ciencias y Tecnologías Espaciales de Asturias (ICTEA) facilitada por su director, Javier de Cos, donde di mis primeros pasos en el mundo de la investigación cuando aún era un estudiante de Física en la Universidad de Oviedo.

Gracias también a Nuno Santos, Ilaria Carleo, Mario Damaso, Miriam Cortes y David Montes por vuestra generosidad al haber accedido a formar parte de mi tribunal. 

En estos momentos finales del doctorado, también me siento feliz por el acompañamiento de personas importantes para mí. Me estoy refiriendo a esa red de apoyo de las personas de mi círculo más cercano que ‘siempre están ahí'. Gracias, Raquel, por apoyarme y por estar siempre a mi lado, en los buenos momentos y los no tan buenos. Gracias a mis amigos de toda la vida, Andrea, Nacho y Toni. Gracias, Gito, Jorge, José, Juan, Carmen, Coral y Javier. Gracias, Miguel, sin duda el recuerdo que tengo de mi llegada a San Francisco habría sido muy diferente sin tu ayuda. 

Quiero agradecer también a las personas que nos hacen fáciles y agradables las tareas burocráticas y que siempre están dispuestas a ayudar, como la coordinadora de este doctorado, Mariángeles, mi tutor, Nicolás, y Margie en el CAB. Margie, he tenido siempre tu ayuda, antes, durante y después de todos mis viajes. Mariángeles, tu apoyo, asesoramiento y ánimos han sido inestimables para sobrellevar mejor la última etapa de mi doctorado. Gracias a los tres.

Villafranca ha sido mi casa durante los últimos cuatro años, una etapa de mi vida que sin duda recordaré con mucho cariño. Esto es, en gran medida, gracias a mis compañeros del CAB, con quienes he podido compartir gratos momentos. Gracias, Almudena, por tus buenos consejos y reflexiones profundas. Gracias, María Rosa, por compartir esos descansos durante las largas jornadas de trabajo que se acababan convirtiendo en conversaciones muy valiosas y que se alargaban hasta altas horas de la noche en el CAB. Gracias, Nuria. Eres para mí un ejemplo a seguir, actuando siempre con criterio, espíritu crítico y valentía. Gracias, Sergio y Antonio, por el inestimable soporte informático; siempre dispuestos a ayudar, a cualquier hora y cualquier día (¡incluso fines de semana!). Gracias, Jaime. Recuerdo al detalle una de nuestras últimas conversaciones en UniOvi, a la espera de entrar a la revisión de aquel famoso examen. El futuro era incierto, pero ambos ya soñábamos con trabajar de astrofísicos. ¡Parece que lo hemos conseguido (al menos durante cuatro años)! Espero que nuestros caminos se vuelvan a cruzar, por lo menos, y como tantas otras veces, en la calle Mon. Gracias, Pedro, gracias, Alberto, por compartir el fútbol y la música, y por todas las tardes de entrenamiento. Gracias, Javier, gracias, Carlos, por, entre muchas otras cosas, esta plantilla \LaTeX\ en la que hoy redacto esta tesis. Gracias, José (a.k.a xo-sé ka-ba-jé-ro). Entre datos y análisis, tus ocurrencias creativas siempre han sido un soplo de aire fresco y un generador de sonrisas. Gracias también a mis compañeros de doctorado, postdocs e investigadores seniors con los que he coincidido estos años en el CAB. Gracias a todos.

También quiero reflejar mi gratitud hacia un lugar especial que es la pequeña isla de Ibiza donde nací (‘Roqueta, sa meua roca'), donde vuelvo siempre que necesito recargar pilas y donde se gestaron, al vaivén de las olas, algunas de las ideas que se plasman en esta tesis.

Al recordar la imagen de aquel chico que se incorporó a su doctorado hace cuatro años con miedos e ilusiones y con un alto grado de síndrome del impostor, incluso antes de comenzar, me viene a la memoria aquella frase de Víctor Hugo que decía ‘en los ojos del joven arde la llama y en los del viejo brilla la luz'. Observo la evolución, y ahora que se acerca el final de este periodo, siento la satisfacción de mantener viva la llama en mis ojos y la pasión por la investigación en mi corazón, y en gran parte es gracias a todos vosotros. Sin duda, os llevaré en mi recuerdo, pues quien escribe en el alma, escribe para siempre.

Gracias.

\clearpage\hbox{}\thispagestyle{empty}\newpage 

\setcounter{footnote}{0}

\thispagestyle{empty}
\vspace*{7cm}
\begin{flushright}
\textit{Antes de la iluminación, corto leña y acarreo agua. \\
Después de la iluminación, corto leña y acarreo agua. } \\
Proverbio Zen
\\

\textit{---}
\end{flushright}

\thispagestyle{empty}

\newpage

\setlength{\parskip}{2mm} 
 
 \makeatletter
\newcommand\Desmesurado{\@setfontsize\Desmesurado{27}{42}}
 \makeatother
 \makeatletter
 \newcommand\MasDesmesurado{\@setfontsize\MasDesmesurado{50}{42}}
 \makeatother

\renewcommand{\chaptername}{}
\titleformat{\chapter}[display]
  {\normalfont\MasDesmesurado\centering\color{black}}{\centering\chaptertitlename\ \thechapter}{80pt}{\bfseries\Desmesurado\color{black}}
\titlespacing*{\chapter} 
  {0pt}{50pt}{40pt}

\hypersetup{linkcolor=black} 

\pagestyle{empty}
{
  \renewcommand{\thispagestyle}[1]{}
  \tableofcontents
}
\clearpage
\pagestyle{plain}


\normalsize
\hypersetup{linkcolor=black} 

\thispagestyle{empty}

\clearpage\hbox{}\thispagestyle{empty}\newpage 
\vspace{9cm}
\phantomsection
\addcontentsline{toc}{chapter}{List of Figures}
\listoffigures

\thispagestyle{plain}
\vspace{9cm}
\phantomsection
\addcontentsline{toc}{chapter}{List of Tables}
\listoftables

\newpage
\thispagestyle{empty}

\printindex






\large
\setcounter{footnote}{0}
\thispagestyle{empty}
\begin{center}
\textbf{\Large{Resumen}}\\
\addcontentsline{toc}{chapter}{Resumen}
\end{center}
\normalsize

\vspace{0.5cm}

Uno de los principales objetivos de la investigación exoplanetaria consiste en comprender cómo se forman y evolucionan los sistemas planetarios. El descubrimiento de 51 Peg b en 1995, un planeta similar a Júpiter pero con una órbita siete veces más pequeña que la de Mercurio, cambió el paradigma histórico, hasta entonces fundamentado en el Sistema Solar. Las teorías de formación planetaria indican que los planetas gigantes como 51 Peg b solo pueden formarse en las regiones más externas de los sistemas planetarios. Sin embargo, todavía quedan muchas dudas sobre cómo algunos de estos planetas migraron hasta sus órbitas cercanas actuales. Los planetas pequeños, por el contrario, pudieron haberse formado más cerca de sus estrellas, pero estos también plantean un desafío importante: la existencia de una gran incertidumbre en sus composiciones y estructuras internas. De este modo, el camino hacia una comprensión profunda de cómo se forman y evolucionan los planetas pasa inequívocamente por el estudio de los mecanismos de migración de los planetas gigantes y de las estructuras internas de los planetas pequeños, dos cuestiones relevantes para las que aún carecemos de una respuesta definitiva.

Esta tesis constituye un esfuerzo observacional para avanzar en nuestra comprensión de la formación y evolución planetaria en diferentes regímenes. Para ello, explotamos los datos de Kepler/K2 y TESS en combinación con observaciones desde tierra, sirviendo como preparación para la misión PLATO. Este trabajo se desarrolló en el contexto del proyecto K2-OjOS, una colaboración pro-am en la que 10 astrónomos amateurs inspeccionaron 20\,427 curvas de luz de la campaña 18 de K2 (C18) para buscar señales de tipo tránsito. Realizamos un análisis completo de validación estadística para 42 de las señales encontradas, lo cual implicó diferentes tareas tales como caracterizar las estrellas anfitrionas a través de datos fotométricos y espectroscópicos de archivo, modelar los tránsitos de K2 a través de toda la fotometría disponible, estudiar la presencia de estrellas contaminantes y, en última instancia, estimar las probabilidades de que los candidatos sean falsos positivos. Esta tesis también se llevó a cabo en el contexto de las colaboraciones HARPS-NOMADS y ESPRESSO, las cuales se centran en la exploración del ‘desierto' neptuniano $-$las órbitas cercanas donde apenas se encuentran planetas gigantes con tamaños entre Neptuno y Júpiter$-$ y de la población de tierras/supertierras, respectivamente. El trabajo en estas colaboraciones implicó diferentes tareas como caracterizar las propiedades de las estrellas anfitrionas y las señales de actividad inducidas por la rotación, realizar análisis comparativos, y usar los mejores modelos para inferir las propiedades de los sistemas. Los resultados obtenidos nos condujeron a realizar estudios poblacionales que nos permitieron contextualizar mejor nuestros hallazgos y obtener más información sobre los posibles mecanismos de formación y evolución que dieron lugar a los sistemas estudiados.

Este trabajo desembocó en el descubrimiento y validación estadística de cuatro planetas (K2-355~b, K2-356~b, K2-357~b y K2-358~b), la confirmación y caracterización precisa de otros dos planetas (TOI-244~b y TOI-5005~b), la detección de 14 nuevos candidatos a planetas y la revisión de las propiedades (incluyendo mejoras en las efemérides de los tránsitos) de 25 planetas previamente detectados. Entre ellos, determinamos que TOI-244~b es una supertierra con una densidad de $\rho$ = 4.2 $\pm$ 1.1 $\rm g \cdot cm^{-3}$. Este valor está por debajo del que se esperaría para una composición similar a la de la Tierra para un planeta con su masa, por lo que nos referimos a TOI-244~b como una supertierra de baja densidad (LDSE, por sus siglas en inglés). También medimos la masa y confirmamos el tránsito del superneptuno TOI-5005~b. En este sistema, encontramos variabilidad fotométrica que coincide con el período orbital planetario, lo que sugiere la existencia de interacciones magnéticas estrella-planeta (MSPI, por sus siglas en inglés). También detectamos señales sincronizadas similares en el sistema excéntrico HD 118203, las cuales aparecen y desaparecen de una órbita a otra, en concordancia con la conocida naturaleza ‘on/off' de las MSPIs. También encontramos variaciones significativas en el tiempo de tránsito en la estrella brillante K2-184 y detectamos un nuevo tránsito solitario en K2-274, lo que sugiere la presencia de compañeros de largo período. Además, desvalidamos el planeta K2-120~b, ya que no es posible identificar el origen de la señal detectada. A nivel poblacional, encontramos que las LDSE tienden a orbitar estrellas con metalicidades subsolares (aunque esta tendencia debe tomarse con cuidado ya que la mayoría de estrellas en nuestra muestra son enanas M) y que las LDSE de menor densidad tienden a recibir flujos de insolación bajos (típicamente $S$ $<$ 10 $\rm S_{\oplus}$), lo que proporciona indicios sobre sus composiciones. También identificamos una sobredensidad de planetas neptunianos en órbitas correspondientes a períodos entre $\simeq$3.2 y $\simeq$5.7 días, a la que nos referimos como la ‘cordillera' neptuniana. Además, encontramos que los planetas en esta ‘cordillera' tienden a tener densidades mayores que los planetas a distancias orbitales mayores $-$dentro de la llamada ‘sabana' neptuniana$-$, lo que sugiere la existencia de procesos evolutivos particulares poblando estas regiones.

Los resultados obtenidos en esta tesis nos han permitido aumentar y entender mejor la demografía de exoplanetas. Hemos presentado nuevos planetas y candidatos de interés para observaciones futuras, así como caracterizaciones en profundidad de sistemas en nichos poco explorados. Además, el estudio de las LDSEs y Neptunos calientes a nivel poblacional nos ha permitido llegar a diferentes conclusiones sobre su naturaleza y evolución. Por un lado, basándonos en las correlaciones encontradas y en el hecho de que los discos protoplanetarios pobres en hierro tienden a ser ricos en agua, proponemos que las LDSEs podrían explicarse a través de la existencia de grandes reservas de agua preservadas en las estructuras planetarias (presumiblemente en las atmósferas) gracias a las condiciones favorables de baja irradiación. Por otra parte, basándonos en una evidencia tentativa de que los planetas en la ‘cordillera' neptuniana son más excéntricos y tienen órbitas más desalineadas que los planetas en la ‘sabana', proponemos que ciertos procesos de migración de alta excentricidad podrían estar poblando preferentemente la ‘cordillera'. Independientemente de estas constricciones, la ‘cordillera', coincidiendo con el rango orbital de la acumulación de Júpiters calientes (una sobredensidad similar en la población de planetas Jovianos), sugiere un camino común en la evolución de los planetas gigantes más cercanos, desde tamaños de Neptuno a Júpiter. Por su parte, la correlación con el flujo de insolación de las LDSE sugiere que la gran mayoría de elementos volátiles en estos planetas está directamente expuesta a (o al menos, influenciada por) la irradiación recibida de sus estrellas anfitrionas. Se necesitan más observaciones para averiguar qué mecanismos llevan a los planetas neptunianos preferentemente a la ‘cordillera' y a otras órbitas cercanas, así como cuáles son los principales elementos volátiles que disminuyen las densidades de las LDSE. Resolver estas cuestiones contribuirá en gran medida a completar el puzle de la evolución de los sistemas planetarios. La próxima misión espacial de fotometría, PLATO, con su amplio campo de visión y su enfoque en planetas pequeños y de largo período, será clave para responder estas preguntas.

\vfill
\noindent\textbf{Palabras clave:} planetas y satélites: detección $-$ planetas y satélites: composición $-$ planetas y satélites: formación $-$ planetas y satélites: evolución dinámica y estabilidad $-$ planetas y satélites: planetas gaseosos $-$ planetas y satélites: campos magnéticos $-$ planetas y satélites: interacciones planeta-estrella $-$ planetas y satélites: evolución física $-$ técnicas: velocidades radiales $-$ técnicas: fotometría

\newpage


\large
\thispagestyle{empty}
\begin{center}
\renewcommand{\thefootnote}{\arabic{footnote}}
\textbf{\Large{Abstract}}\\
\addcontentsline{toc}{chapter}{Abstract}
\end{center}
\normalsize

\vspace{0.5cm}

Understanding how planetary systems form and evolve is one of the main goals of exoplanet research. The discovery of 51 Peg b in 1995, a Jupiter-mass planet with an orbit seven times shorter than Mercury's, changed our long-standing paradigm based on the Solar System. Formation theories show that giant planets such as 51 Peg b can only form in the outer regions of planetary systems. However, many doubts remain about how a fraction of these planets migrated towards their present-day close-in orbits. Small planets, in contrast, could have been formed closer to their stars, but still pose an important challenge: the existence of large degeneracies in their possible internal structures. Therefore, the path towards a deep understanding of how planets form and evolve unequivocally passes through the study of the migration mechanisms of giant planets and the internal structures of small planets, two relevant topics for which we still lack a definitive picture.

This thesis constitutes an observational effort to advance our understanding of planet formation and evolution across different regimes. To do so, we exploited Kepler/K2 and TESS data in combination with ground-based follow-up observations, serving as a preparation for the PLATO mission. This work was carried out in the context of the K2-OjOS project, a pro-am collaboration where 10 amateur astronomers inspected 20\,427 light curves of K2 campaign 18 (C18) to search for transit-like signals. We here carried out a complete statistical validation analysis for 42 signals found, which involved different tasks such as characterising the planet candidate hosts through archival photometric and spectroscopic data, modelling the K2 transits through all available photometry, studying the presence of contaminant stars, and ultimately obtaining reliable false positive probabilities. This work was also carried out in the context of the HARPS-NOMADS and ESPRESSO collaborations, which focus on exploring the Neptunian ‘desert' $-$the closer-in orbits where giant planets with sizes between Neptune and Jupiter are rarely found$-$ and the Earth/super-Earth population, respectively. The work in these collaborations involved different tasks such as characterising the host star’s properties and rotation-induced activity signals, performing model comparison analyses, and using the best models to infer the properties of the systems. The results led us to embark on different population studies that allowed us to contextualise our findings better and gain further insight into the possible formation and evolution mechanisms behind our studied systems. 

This endeavour led to the discovery and statistical validation of four planets (K2-355 b, K2-356 b, K2-357~b, and K2-358 b), the confirmation and precise characterization of two other planets (TOI-244 b and TOI-5005~b), the detection of 14 new planet candidates, and the revision of the properties (including ephemeris improvements) of 25 previously reported planets. Among them, we find TOI-244~b to be a super-Earth with a density of $\rho$ = 4.2 $\pm$ 1.1 $\rm g \cdot cm^{-3}$. This value is below what would be expected for an Earth-like composition for a planet of its mass, and we thus refer to TOI-244~b as a low-density super-Earth (LDSE). We also measure the mass and confirm the transiting super-Neptune TOI-5005~b. In this system, we find photometric variability that matches the planetary orbital period, suggesting the existence of magnetic star-planet interactions (MSPIs). We also detect similar synchronised signals in the eccentric HD 118203 system, which appear and disappear across different orbits, in agreement with the well-known ‘on/off' nature of MSPIs. We find significant transit timing variations in the bright star K2-184 and detect a new single transit in K2-274, suggesting the presence of long-period companions. We also devalidate the planet K2-120~b, as the signal's origin cannot be identified. At a population level, we find that LDSEs tend to be hosted by stars with subsolar metallicities (although this trend has to be taken with care since most hosts in our sample are M-dwarfs) and that the lowest dense LDSEs tend to receive low insolation fluxes (i.e. $S$ $<$ 10 $\rm S_{\oplus}$), which may hint at their composition. We also identify an over-density of Neptunian planets in orbits corresponding to periods between $\simeq$3.2 and $\simeq$5.7 days, which we refer to as the Neptunian ‘ridge'. Additionally, we find that planets in this ‘ridge' tend to have larger densities than planets at larger orbital distances $-$within the so-called Neptunian ‘savanna'$-$ suggesting the existence of particular processes populating these regions.  

The results obtained in this thesis allowed us to increase and better understand the exoplanet demographics. We presented new planets and candidates of interest for follow-up observations, as well as in-depth characterisations of systems in poorly-explored niches. In addition, the study of LDSEs and hot Neptunes at a population level allowed us to reach different conclusions on their nature and evolution. On the one hand, based on the fact that iron-poor protoplanetary disks tend to be water-rich, we propose that LDSEs could be explained through the existence of large reservoirs of water preserved in the planetary structures (presumably in the atmospheres), thanks to the favourable low-irradiation conditions. On the other hand, based on tentative evidence that planets in the ‘ridge' are more eccentric and have more misaligned orbits with the orbital plane than planets in the ‘savanna', we propose that high-eccentricity tidal migration processes could be preferentially populating the ‘ridge'. Regardless of these additional constraints, the ‘ridge', coinciding with the orbital range of the hot Jupiter pileup (a similar over-density within the hot Jupiter population), suggests a common path in the evolution of the closest giant planets, from Neptune to Jupiter sizes. On its side, the low-insolation trend of LDSEs suggests that the largest fraction of volatile elements in these planets is directly exposed to (or at least, influenced by) the received irradiation from their host stars. More observations are needed to unveil which mechanisms bring Neptunian planets preferentially to the ‘ridge' and other close-in orbits, and which are the main volatiles decreasing the densities of LDSEs. Solving these two problems will greatly contribute to completing the full picture of the evolution of planetary systems. The upcoming PLATO mission, with its large field of view and focus on small and long-period planets, will be key to answering these questions.

\vfill
\noindent\textbf{Keywords:} planets and satellites: detection $-$ planets and satellites: composition $-$ planets and satellites: formation $-$ planets and satellites: dynamical evolution and stability $-$ planets and satellites: gaseous planets $-$ planets and satellites: magnetic fields $-$ planets and satellites: planet-star interactions $-$ planets and satellites: physical evolution $-$ techniques: radial velocities $-$ techniques: photometric  \\

\chapter{Introduction} 
\label{ch:introduction}
\vspace{2cm}
\pagestyle{fancy}
\fancyhf{}
\lhead[\small{\textbf{\thepage}}]{\small{\textbf{\nouppercase{\leftmark}}}}
\rhead[\small{\textbf{\nouppercase{\rightmark}}}]{\small{\textbf{\thepage}}}

\bigskip

\lettrine[lines=3, lraise=0, nindent=0.1em, slope=0em]{T}{he study of stars} has been present in human history for millennia. However, detecting planets beyond the Solar System (also known as exoplanets) has always been a difficult task. This is because of their intrinsic faintness and the large brightness contrast
with their host stars. Today, even with the most powerful telescopes, obtaining an image of an exoplanet separated from its star is a big challenge. Instead of directly visualising the planet, its existence is typically deduced through indirect methods, that is, by observing the changes induced in our view of its star. In addition to the planet detections themselves, indirect methods allow us to infer different orbital, physical, and recently physico-chemical properties of the planetary systems, which has led to the identification of different planet populations and the discovery of unusual systems with unique characteristics. In this regard, understanding the rules and the exceptions is critical to bringing knowledge on one of the main questions of exoplanet exploration: how planets form and evolve. 

The era of exoplanet exploration started three decades ago. In the late 1980s, \citet{Campbell1988} presented the first exoplanet candidate that was decades later confirmed to be a real planet: $\gamma$~Cephei~A~b \citep{Hatzes2003}. In 1992, \citet{WolszczanFrail1992} confirmed the first exoplanetary system around the millisecond pulsar PSR1257 + 12, and three years later \citet{1995Natur.378..355M} discovered the first exoplanet orbiting a main-sequence star, 51 Peg b. Having an orbital period of 4.2 days and a mass of at least half that of Jupiter, 51 Peg b changed the paradigm based on our interpretation of the Solar System. The basic dichotomy between small/rocky planets in shorter orbits and giant/gaseous planets in more distant orbits became obsolete, making it necessary to build new formalisms and reconsider old theories of planet formation and migration to explain this ground-breaking discovery. Gas giant planets orbiting close to their host stars (a $\lesssim$ 0.1 au) like 51~Peg~b are commonly known as hot Jupiters and dominated the exoplanet demographics in the
early years of exoplanet exploration since they were the easiest to detect \citep[e.g.][]{1997ApJ...474L.115B,2000ApJ...529L..41H,2000ApJ...529L..45C,2000A&A...361..265S}. As long as the detection methods were refined and the instrumental precision improved, exoplanet detection quickly expanded towards smaller sizes and longer orbital distances, allowing us to unveil new types of planets, which frequently had no analogue in the Solar System. Population studies and theoretical works have allowed us to gain important knowledge on the plethora of planets known to date, but many questions about their specific origins and evolution remain open.

\section{Detection methods}

\begin{figure}

\includegraphics[width=0.47\textwidth]{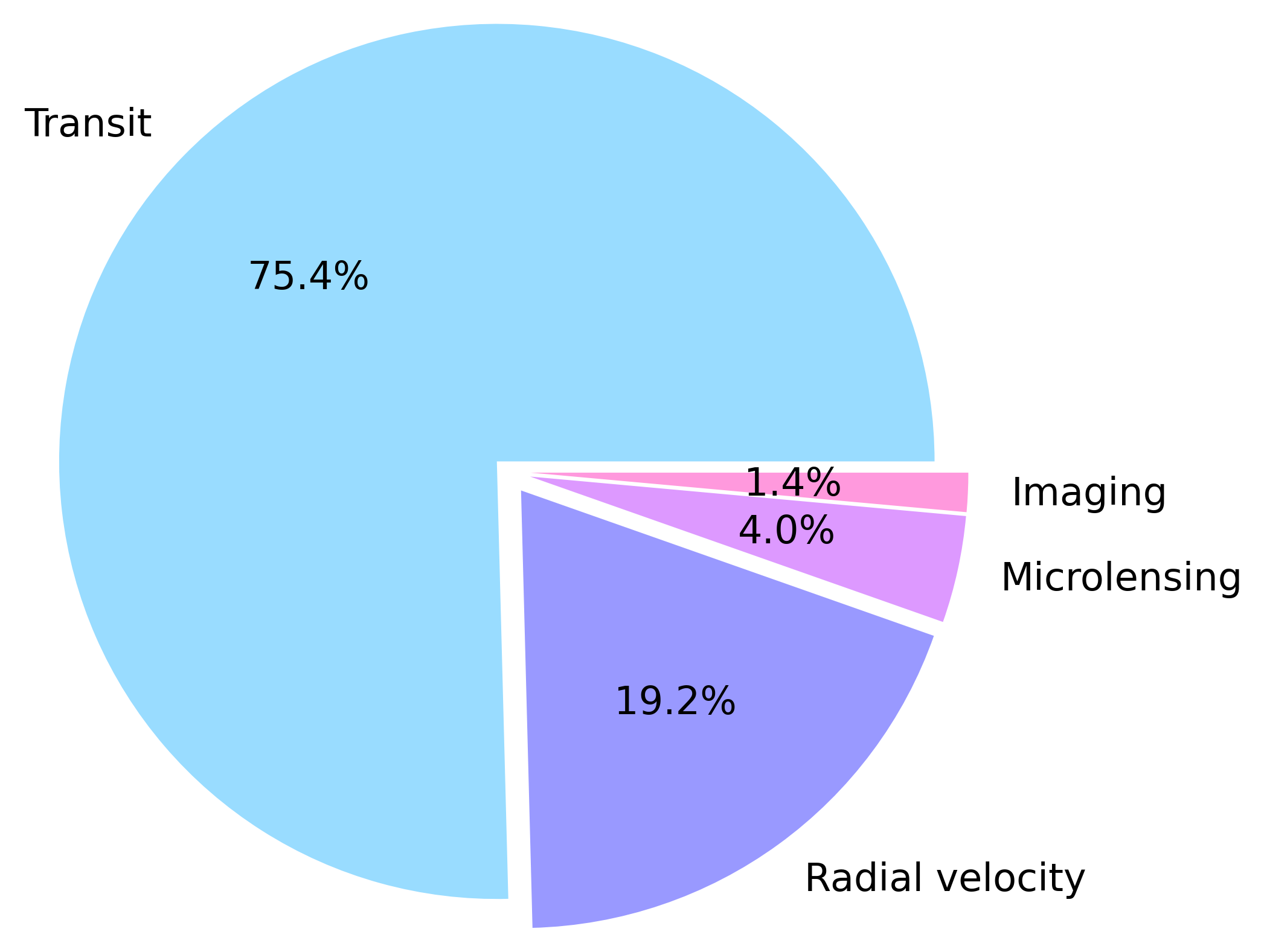}
    \includegraphics[width=0.50\textwidth]{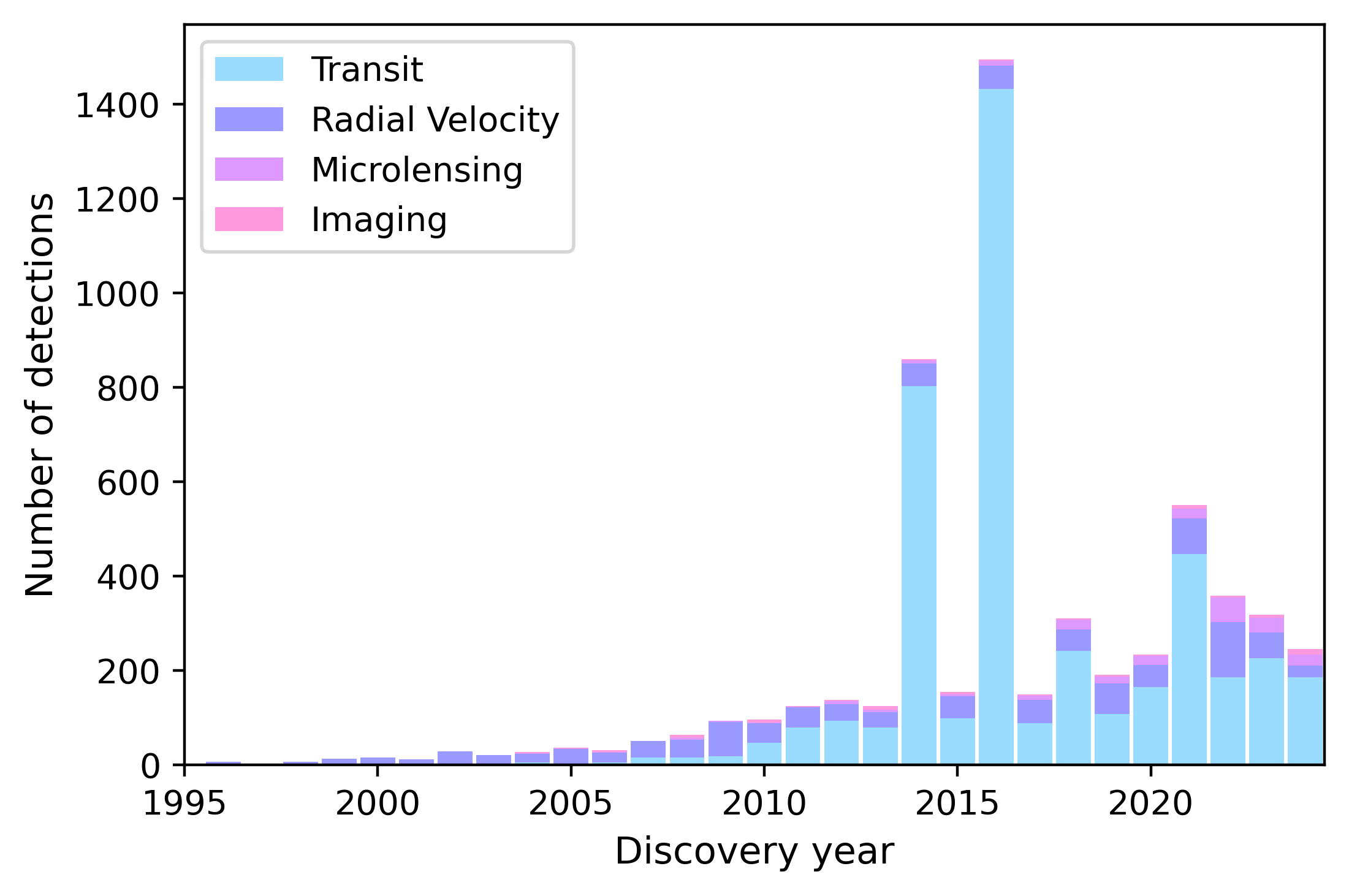}
    
    \caption[Exoplanets discovered by the most prolific techniques.]{Total number of confirmed and statistically validated exoplanets discovered by the most prolific techniques. The data were acquired from the NASA Exoplanet Archive. The detection peaks in 2014 and 2016 are largely due to the large-scale planet validation works by \citet{2014ApJ...784...45R} and \citet{2016ApJ...822...86M}, based on data from the \textit{Kepler} satellite (see Sect.~\ref{sec:intro_kepler}).}
    \label{fig:discovery_methods}
\end{figure}

Today, there are more than 5800 planets collected in the NASA Exoplanet Archive\footnote{\url{https://exoplanetarchive.ipac.caltech.edu/}} \citep{2013PASP..125..989A,Christiansen2022,2025arXiv250603299C}, which have been reported as “confirmed" or “statistically validated" planets in peer-reviewed journals. Only 1.4$\%$ of these planets have been detected through a direct method, the so-called direct imaging technique, while the existence of the remaining 98.6$\%$ has been inferred indirectly. In Fig.~\ref{fig:discovery_methods} (left plot), we show a pie chart with the fraction of planets discovered by the most prolific techniques to date (i.e. those with contributions $>$ 1 $\%$ to the total population). In the same figure (right plot), we represent a histogram with the year-by-year discoveries. The so-called radial velocity method, which was used to detect 51 Peg b, dominated the planet discoveries during the first decade of exoplanet exploration. In 2000, the so-called transit method was used for the first time to detect an exoplanet \citep{2000ApJ...529L..45C,2000ApJ...529L..41H}, and this technique was definitely exploited with the launch of space-based observatories which monitor thousands of stars simultaneously, quickly becoming the most prolific method with more than 75$\%$ of the total discoveries. 

In the following, we introduce the transit and radial velocity techniques, on which this thesis is based. For a description of other techniques used to detect exoplanets, we refer to \citet{2011exha.book.....P,2014exha.book.....P,2018exha.book.....P}.

\subsection{The transit method}
\label{subsec:transit_method}

The transit method consists of measuring the stellar flux over time. The objective is to detect periodic flux decreases, which, after discarding other possible origins, can be interpreted as the existence of a planet blocking part of the starlight.

\subsubsection{Transit condition and probability}

\begin{figure}
    \centering
    \includegraphics[width=0.9\textwidth]{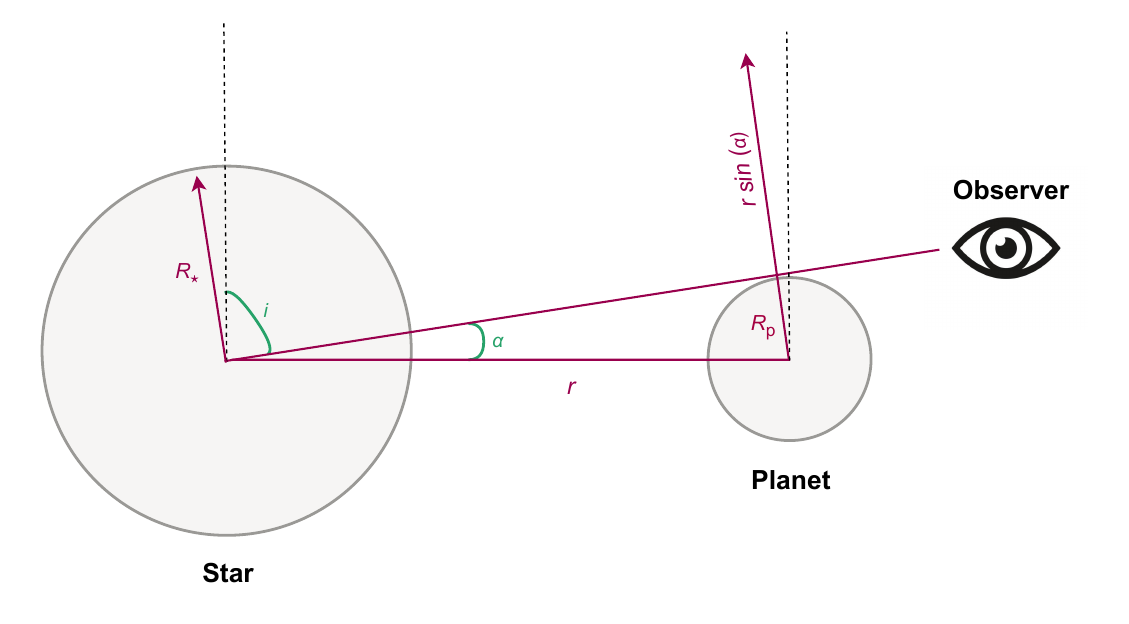}
    \caption[Geometric condition of a transit.]{Scheme of a three-body system composed of a star, an orbiting planet, and an observer, showing the necessary geometric condition to observe a transit. Distances are coloured in garnet and angles in green.}
    \label{fig:transit_condition}
\end{figure}

The transit method can only be successful in the particular case where there is a chance alignment among the host star, its orbiting planet, and the observer (i.e. the Earth). In Fig.~\ref{fig:transit_condition}, we represent a scheme with these three bodies and the different distances and angles between them. A simple inspection of the figure reveals that the necessary geometric condition for a planet to transit its star is that the star-planet separation projected in the observer's plane of vision must be less than the sum of the radii of the planet ($R_{\rm p}$) and the star ($R_{\rm \star}$). This projected separation can be written as a function of the star-planet distance ($r$) and the angle $\alpha$ as $r \, sin(\alpha)$. We can also link $\alpha$ to the angle $i$ (which represents the inclination between the observation line and the perpendicular line to the orbital plane) as $sin(\alpha) = cos(i)$. Therefore, we can write the transit condition by explicitly using the star-planet separation in the following compact form: 

\begin{equation}
    r \, cos(i) \leq R_{\rm \star} + R_{\rm p}.
    \label{ec:transit_condition}
\end{equation}

We can now try to estimate the probability that the transit condition is satisfied, that is, how likely it is that cos($i$) is lower than or equal to the sum of the radii of the planet and star divided by the distance between them. Assuming random orbital orientations, for an unknown angle $i$, the probability that equation \ref{ec:transit_condition} is met (i.e. the transit probability; $P_{\rm transit}$) can be estimated as:  

\begin{equation}
    P_{\rm transit}  \simeq \frac{R_{\rm \star}}{r}
    \label{eq:transit_probability}.
\end{equation}

Therefore, the larger the stellar radius compared to the star-planet distance, the larger the probability of observing a planetary transit. If we evaluate equation \ref{eq:transit_probability} with the distances corresponding to the Earth-Sun system, we obtain $P_{\rm transit}^{\rm Earth}$ $\simeq$ 0.005, that is, only a 0.5$\%$ probability of seeing the Earth transiting the Sun. As mentioned before, despite the typically low transit probabilities, searching for planetary transits has been the most prolific detection technique to date, with three-quarters of all detections. This is, to a greater extent, due to our capabilities and funding efforts of the space agencies to send high-precision photometers to outer space, and their ability to observe a large number of stars simultaneously. 

\subsubsection{Photometric variations due to transits}

\begin{figure}
    \centering
    \includegraphics[width=\textwidth]{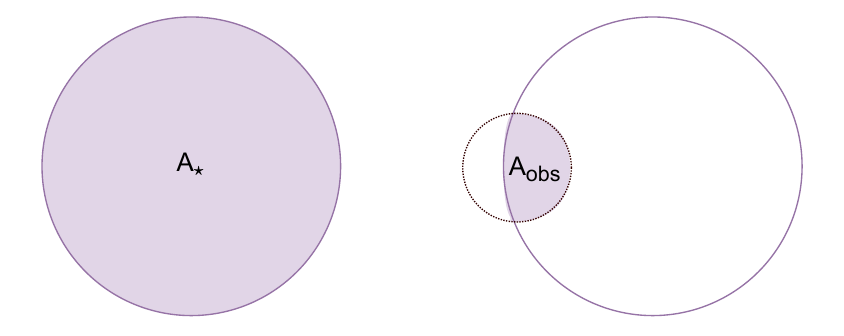}
    \caption[Area of a bi-dimensional star $A_{\star}$ and the area that a planet obstructs the star $A_{\rm obs}$.]{Graphic representation of the area of a bi-dimensional star $A_{\star}$ and the area that the planet obstructs the star $A_{\rm obs}$.}
    \label{fig:areas}
\end{figure}

The observed stellar flux during a transit depends on different phenomena. In a first approximation, we can consider the star as a source of light with uniform brightness and a circular geometry. The luminous flux emitted by a bi-dimensional source is proportional to its area. We here call $A_{\star}$ to the area of the star, and $A_{\rm obs}$ to the area that the planet obstructs the star (see Fig.~\ref{fig:areas}).

The value of $A_{\rm obs}$ changes as the planet transits its star, so that $A_{\rm obs}$ $\equiv$ $A_{\rm obs} (t)$. Hence, the fluxes emitted by the star without obstruction $F_{\star}$ and the star when is obstructed $F(t)$:

\begin{equation}
    F_{\star} \propto A_{\star} \quad \textrm{and}  \quad F(t) \propto  A_{\star} -  A_{\rm obs}(t).  
\end{equation}

Dividing both fluxes, we obtain:

\begin{equation}
    \frac{F(t)}{F_{\star}} = 1 - \frac{A_{\rm obs}(t)}{A_{\star}}  \rightarrow F(t) = F_{\star} \left(1 - \frac{A_{\rm obs}(t)}{A_{\star}}  \right).
\end{equation}

The fraction $\frac{A_{\rm obs}(t)}{A_{\star}}$ is typically denoted in the literature as $\Delta F (t)$, so that for an emitting source with uniform brightness its emitted flux is commonly written as:

\begin{equation}
    F(t) = F_{\star} \left(1 -  \Delta F (t) \right).
    \label{ec:stellar_flux}
\end{equation}

From equation \ref{ec:stellar_flux} we can infer that the observed stellar flux decreases as $\Delta F (t)$ increases, so that the smaller planets respect to their stars ($A_{\rm obs}$ $\ll$ $A_{\star}$) are more difficult to detect with this method. 

\begin{figure}
    \centering
    \includegraphics[width=0.45\textwidth]{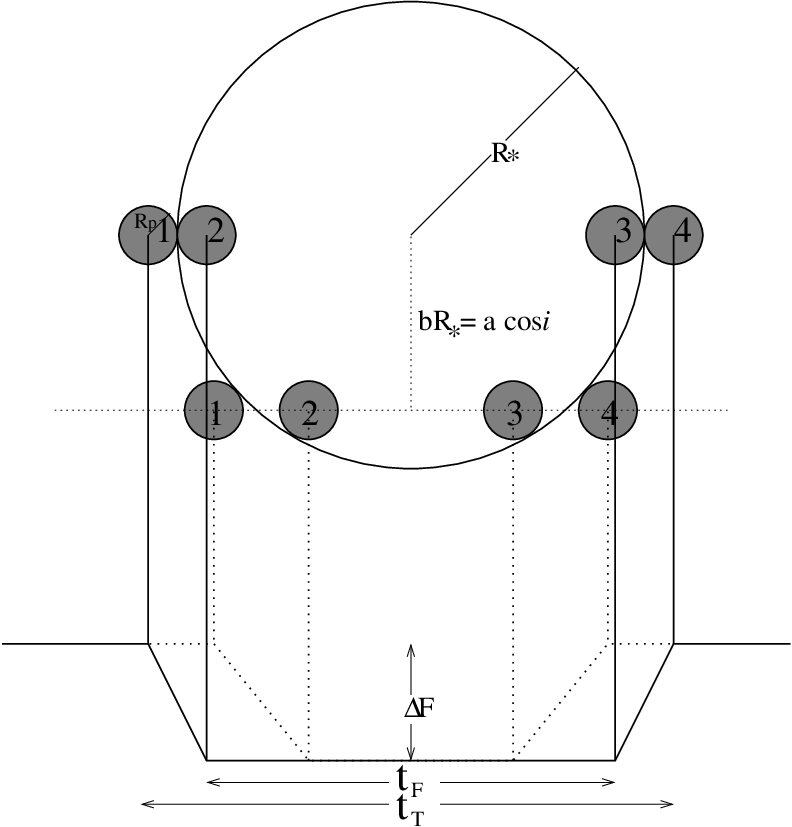}
    \caption[Representation of the observables $t_{T}$, $t_{F}$, and $\Delta F$.]{Representation of the observables $t_{T}$, $t_{F}$, and $\Delta F$. The upper part shows two different situations where a planet transits near the disk equator and where a planet transits closer to the edge. The lower part shows the corresponding stellar flux in both situations. Source: \citet{SeagerMallen2003}.}
    \label{fig:seaquer_scheme}
\end{figure}

The analytic form of $\Delta F (t)$ can be trivially inferred in two different star-planet-observer configurations. When the planet is not blocking any stellar light (i.e. is not transiting), $A_{\rm obs}(t)$ is 0, then $\Delta F (t)$ is also~0. In this case, we would simply observe the emitted stellar flux $F(t)$ = $F_{\star}$. When the whole planet blocks the star, $A_{\rm obs}(t)$ is equal to the area of the planet's disk, that is, $ \pi R_{\rm p}^{2}$, and hence $\Delta F (t)$ = $ R_{p}^{2} R_{\star}^{-2}$. Therefore, by knowing the emitted stellar flux when the planet is not transiting, the stellar flux when the entire planet blocks the star, and the stellar radius, it can be easily estimated the radius of the planet. The analytic form of $\Delta F (t)$ when only a fraction of the planet is obscuring the stellar disk requires some geometrical deductions, and can be written as \citep[e.g.][]{2002ApJ...580L.171M}:

\begin{equation}
   \Delta F = \frac{1}{\pi} \left[  p^2 k_{0} + k_{1} - \sqrt{\frac{4 z^2- (1+z^2-p^2)^2}{4}}\right],
\end{equation}
being 
\begin{equation}
    k_{0} = cos^{-1} \left(  \frac{p^2+z^2-1 }{2pz} \right) \quad \textrm{and} \quad k_{1} = cos^{-1}\left( \frac{1-p+z^2}{2z} \right),
\end{equation}

with $p$ defined as the planet-star radius ratio ($p$ = $\frac{R_{\rm p}}{R_{\star}}$) and $z$ as the ratio between the planet-star projected distance and the radius of the star ($z$ = $\frac{d}{R_{\star}}$). Studying the planet's transit across the edges of the stellar disk is especially relevant to estimating other physical parameters. For example, by defining $t_{T}$ (also known as $t_{14}$) as the total transit duration and $t_{F}$ (also known as $t_{23}$) as the duration of the transit completely inside ingress and
egress (see Fig.~\ref{fig:seaquer_scheme}), and under the assumption of a circular orbit and $\frac{t_{T} \pi}{P_{\rm orb}} \ll 1$, \citet{SeagerMallen2003} showed that the star-planet distance can be written as:

\begin{equation}
    a = \frac{2P_{\rm orb}}{\pi} \frac{\Delta F^{1/4}}{\left( t_{T}^{2} - t_{F}^{2}  \right)} R_{\star}.
\end{equation}

We note that the geometric deductions described above can be used to obtain first-order estimations of the planetary parameters. However, the reality is more complex. This is because the stellar flux is not constant, and the planet can also emit and reflect light, contributing to the total observed flux. 

The intensity of the stellar irradiation is not uniform across the stellar disk. Instead, it decreases towards the edges (typically referred to as the limbs). This phenomenon, known as limb darkening, is related to the higher depth of a sphere at its centre than at its borders and the lower density and temperature of a star as its radius increases. To account for this effect, the stellar disk intensity $I$ is typically parametrised through laws that depend on $\mu = cos \theta$, where $\theta$ is the angle between the line of sight and the perpendicular line to a certain point of the stellar surface. The limb darkening laws have been parametrised through linear, quadratic, logarithmic, and exponential laws \citep[e.g.][]{Claret2000}. In the exoplanet field, the most commonly used law is the quadratic expression proposed by \citet{2002ApJ...580L.171M}:

\begin{equation}
    I = 1 - u_{1} \left(1-\mu \right) - u_{2} \left(1- \mu \right)^{2},
    \label{ec:LD_law}
\end{equation}

where $u_{1}$ and $u_{2}$ are the limb-darkening coefficients, which depend on the stellar spectral type and the wavelength of the radiation. Hence, the analytical solution of $F(t)$ described in equation \ref{ec:stellar_flux} has to integrate the intensity expression described in equation \ref{ec:LD_law}, so that the limb-darkened transit model can be written as \citep[e.g.][]{2002ApJ...580L.171M}:

\begin{equation}
\label{ec:LD}
F(t)=F_{*} \left( 1- \frac{(1-u_{1}-2u_{2}) \Delta F +(u_{1}+2u_{2})\left[ \lambda^{d}+\frac{2}{3}\Theta(p-z)\right] -u_{2}\eta^{d}}{1-u_{1}/3-u_{2}/6} \right), 
\end{equation}

where $\Theta$ is a step function that takes the value 1 when $p$ $>$ $z$ and 0 in any other case, and $\lambda^{d}$ and $\eta^{d}$ are expressions that depend on complete elliptic integrals of the third kind. Both $\lambda^{d}$ y $\eta^{d}$ are tabulated in Table 1 of \citet{2002ApJ...580L.171M} for any possible star-planet-observer geometry. 

\begin{figure}
    \centering
    \includegraphics[width=0.98\textwidth]{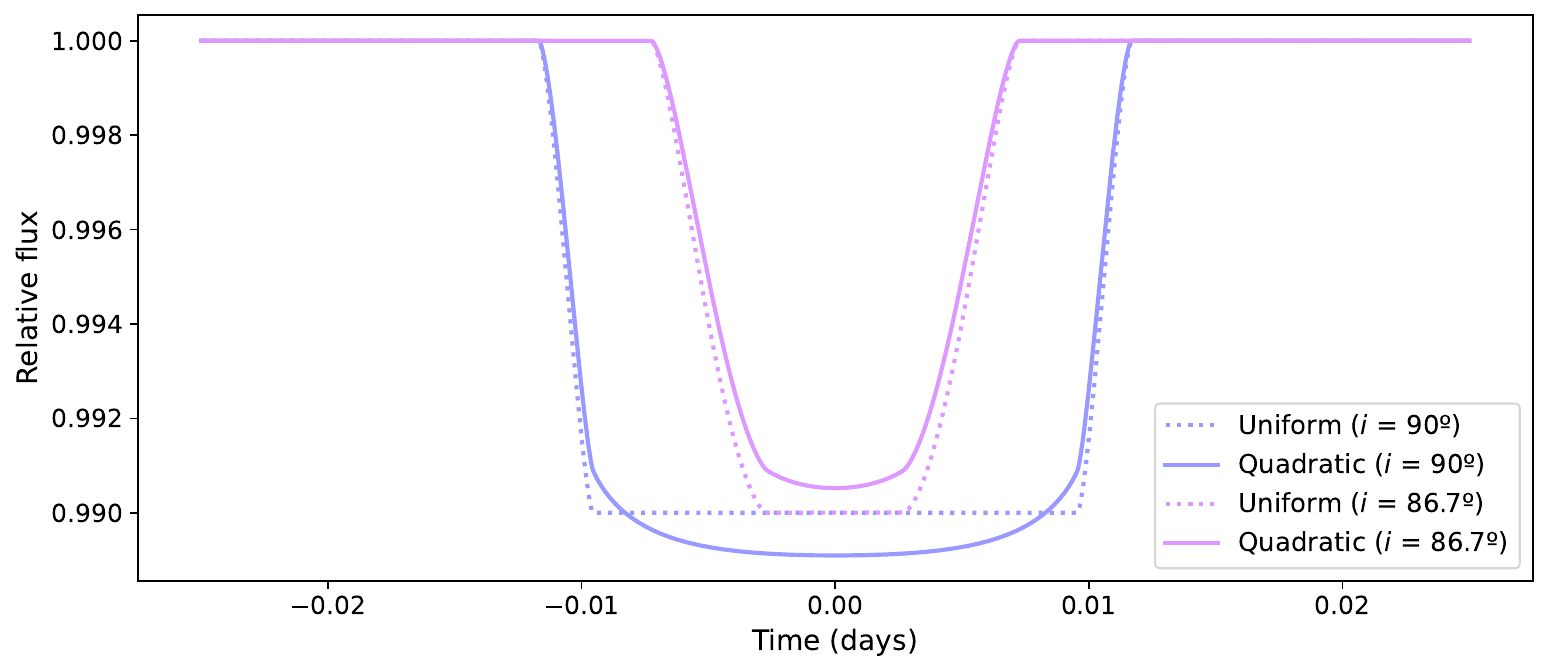}
    \caption[Transits with uniform and limb-darkened sources, and with different inclination angles.]{Simulated planetary transits based on \citet{2002ApJ...580L.171M}'s model showing the different transit shapes generated by uniform and limb-darkened sources, and the different transit depths generated by different inclination angles $i$ in limb-darkened sources. The models were implemented through \texttt{batman} \citep{2015PASP..127.1161K}.}
    \label{fig:LD_example}
\end{figure}

The limb-darkening effect modifies the shape of the transit. Instead of being flat when the entire planet blocks the star (i.e. during $t_{\rm 23}$), the transit signal becomes round in this region and reaches its minimum at mid-transit (Fig.~\ref{fig:LD_example}). Interestingly, not only the transit shape but also the absolute transit depth can change depending on the inclination angle $i$ (i.e. how far from the centre of the stellar disk the planet transits the star). This is because the observed intensity emitted by a limb-darkened star is the average of the intensities emitted in different regions of the stellar disk. If the planet transits its star occulting the central region of the disk (i.e. with $i \simeq 90^{\circ}$), where the intensity is maximum, the total flux decrease will be larger than if it transits closer to the stellar limb (i.e. with $a/R_{\star}$ cos($i$) $\simeq$ 1), where the intensity is lower. We illustrate this phenomenon in Fig.~\ref{fig:LD_example}. As we can see, there can be transit depths larger and shallower than that corresponding to a uniform stellar emission. This depends on whether the planet mostly occults a region of the stellar disk with an intensity higher or lower than the average. 

\begin{figure}
    \centering
    \includegraphics[width=0.38\textwidth]{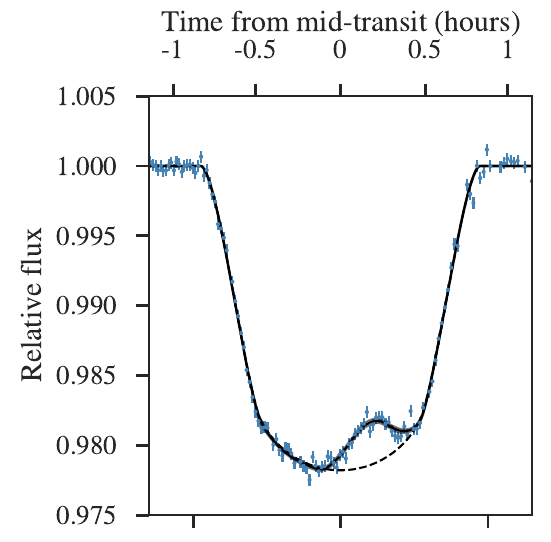}
    \includegraphics[width=0.37\textwidth]{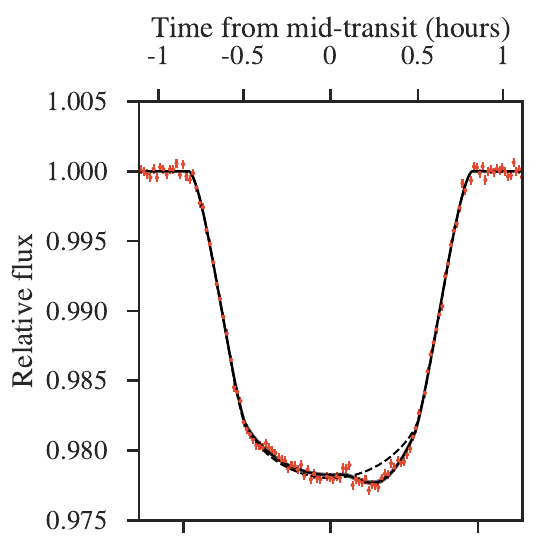}
    \includegraphics[width=0.38\textwidth]{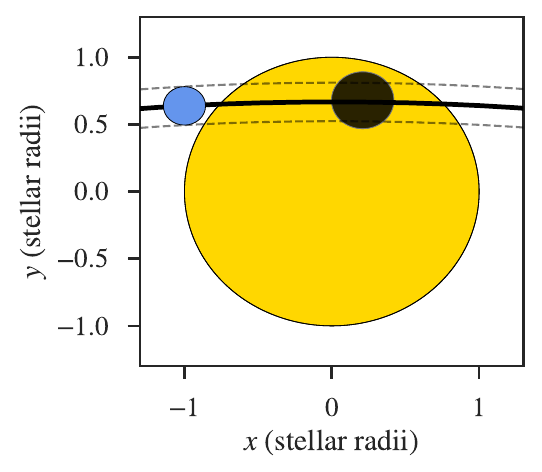}
    \hspace{0.5pt}
    \includegraphics[width=0.38\textwidth]{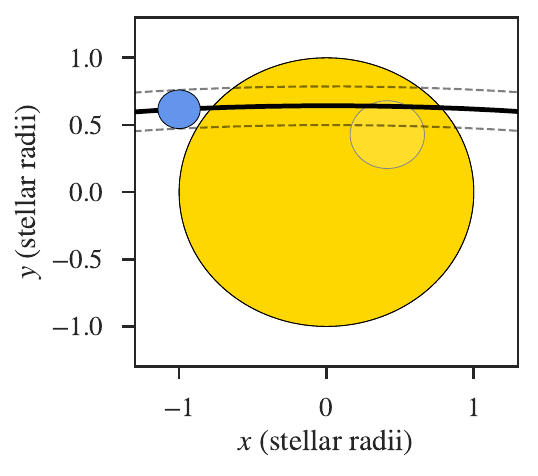}
    \caption[Spot-crossing event detected in WASP-19 b.]{Spot-crossing event detected in WASP-19 b. The top panels show the light curve distortions caused by a dark spot (left) and bright plage (right) in the stellar surface. Image from \citet{Espinoza2019}.}
    \label{fig:spot-crossing}
\end{figure}

In addition to limb darkening, the stellar surface typically presents localized brightness inhomogeneities, which can also alter the shape of a transit \citep[e.g.][]{2003ApJ...585L.147S,2011ApJS..197...14D,Silva-ValioLanza2011,2011MNRAS.416.1443S,Oshagh2013,Espinoza2019}. These features can range from few kilometres up to about 30$\%$ of the stellar surface, and their evolution across the stellar surface is driven by the magnetic field of the star. These regions are often called spots and plages depending on whether they have a dark or bright contribution, respectively. If, for example, a planet transits a star with spots and there is a chance-alignment with one of them, the observed stellar flux would exhibit a small increase since the planet would go from blocking a brighter area of the star to a darker one (i.e.~the spot), a phenomenon often called spot-crossing event (see Fig.~\ref{fig:spot-crossing}).

While there are several cases of relatively high-amplitude spot-crossing events, this phenomenon is negligible in most exoplanet detections since the photometric precisions of state-of-the-art facilities are only capable of detecting them when the spot coverage is large (e.g.~Fig.~\ref{fig:spot-crossing}). In addition, the effect of those events happening in specific transits can be minimised through averaging several transit events. We note, however, that while the effect of spot-crossing events in the planet radius is typically within the uncertainty \citep[i.e. $\sim 4 \, \%$;][]{Oshagh2013}, these events can mimic the presence of moons or other planets in resonance, and produce considerable distortions in the apparent time of transits, potentially leading to an erroneous interpretation of the data (see Sect.~\ref{TTVs}). 

\subsubsection{Photometric variations due to other phenomena}

The observed stellar flux can also vary due to other phenomena. Understanding these non-transit-related variations is of crucial importance to disentangle them from planetary transits, and, in case they appear together with transit signals, to properly characterise their contributions to the in-transit and out-of-transit regions. These additional variations can have a stellar, planetary, or instrumental origin. 

Star spots and plages are not necessarily evenly distributed across the stellar surface and can therefore generate brightness imbalances between different regions. These regions are not static in the stellar disk. Instead, they are displaced on time scales of days driven by the rotation of the star and on time scales of years driven by the stellar magnetic cycle. Their latitudes can also vary on time scales of months. Characterising the flux variations caused by stellar activity is not an easy task. Several formalisms have been designed for this purpose (e.g. \texttt{SOAP} from \citealt{2012A&A...545A.109B}, \texttt{SOAP2} from \citealt{2014ApJ...796..132D}, \texttt{starry} from \citet{2021AJ....162..123L}, and \texttt{FENRIR} from \citealt{2023arXiv230408489H}), which have been able to successfully infer different statistical properties of the surface of particular stars. However, many degeneracies still prevent us from obtaining precise characterisations. 

Fortunately, activity-related flux variations typically have much lower frequencies than transit signals. Therefore, it is common practice to use simple $-$and not necessarily physically motivated$-$ models to detrend the flux observed from activity signals before analysing the planetary transits. Median filters, spline-based methods \citep[e.g.][]{2019AJ....158..143H},  M-estimators \citep[e.g.][]{doi:10.1080/01621459.1974.10482962,mosteller1977data,doi:10.1080/03610917808812083,huber1981robust}, and Gaussian Process regressions \citep[GPs; e.g.][]{2006gpml.book.....R,2012RSPTA.37110550R} are among the most used de-trending techniques. We note that it is increasingly common to not detrend the activity-related flux but instead model it $-$typically through GPs$-$ jointly with the transits. This has been proven to be very successful in preserving the transit shapes \citep[e.g.][]{2021A&A...649A.144S,2023A&A...669A.109L}, especially when the transit signals have low S/N \citep[e.g.][]{2022MNRAS.515.1328D,2022A&A...668A..85C}. In addition, modelling these two components together ensures a correct propagation of the uncertainties of the model parameters \citep[e.g.][]{2021A&A...649A..26L}.

Instrumental systematic errors can also induce flux variations, potentially affecting the detection and characterisation of planetary transits. In ground-based instrumentation, Earth's atmosphere is one of the main factors causing these variations. This is because there are different environmental factors, such as humidity, temperature, wind, and atmospheric turbulence, which can alter the measured stellar flux. These atmospheric conditions can evolve on short time scales, provoking considerable systematics during the observing night. To correct for these effects, researchers using ground-based facilities usually perform differential photometry, which consists of dividing the flux of the target star by the flux of one or more nearby stars (commonly called reference stars) known to have a stable emission. Atmospheric conditions affect the flux of nearby stars similarly, so this methodology has been proven successful in neutralising the main systematics affecting both the target and reference stars. We note that this is a first-order correction, and it is frequent that residual systematics still hold. Residual systematics have been frequently found to correlate with the seeing, airmass, and background flux of the observations. Therefore, ground-based surveys often collect these ancillary data to detrend (or jointly model) the target's photometry. 

In space-based instrumentation, where Earth's atmosphere does not have an impact, instrumental systematics mainly come from the spacecraft's pointing. Depending on the orbit, scattered light from Earth and the Moon can also induce systematic effects. Different data-driven techniques have been developed to mitigate spacecraft systematics. In Sect.~\ref{sec:instrumentation}, we introduce different approaches developed to correct specific features in the space-based instrumentation used in this dissertation. We note that, as for the ground-based facilities, flux corrections in space-based facilities are not perfect, and some instrumental trends can still hold. It is hence fairly common to model these data through GPs with flexible kernels (e.g. Matérn-3/2; \citealt{2017AJ....154..220F}) able to describe an unknown mixture of stellar variability and residual instrumental systematics.

Planets themselves may also induce flux variations that can appear together with their transit signatures. For example, planets accumulate heat and therefore produce a certain amount of blackbody (or thermal) emission. They can also reflect a fraction of the received stellar light with an efficiency determined by their albedos \citep[e.g.][]{2012ApJ...754...22K}. Massive close-in planets can also induce tidal effects in their host stars, deforming their shapes along their orbits and periodically changing the effective area of the stellar disk from a circle to an ellipsoid. This phenomenon produces periodic changes in the stellar flux, known as ellipsoidal variations \citep[see][and references therein]{2008ApJ...679..783P}. The reflex motion of the star when it is orbited by a close-in massive planet can also provoke changes in the flux observed in a particular photometric filter (i.e. wavelength range), an effect known as Doppler beaming \citep[e.g.][]{2003ApJ...588L.117L}. These planet-induced effects require of particular planet-star properties and hence have been only detected in a limited number of planetary systems \citep[e.g.][]{2007Natur.447..183K,2010ApJ...713L.145W,2012ApJ...761...53B,2013ApJ...772...51E,2015ApJ...804..150E,2014A&A...562A.109L,2017AJ....154...83M,2021A&A...648A..71V,2024MNRAS.531.1133C}. In addition, if a massive close-in planet (i.e. inside the Alfvén radius) is highly magnetized, it can induce co-rotating stellar spots in the planetary surface through a phenomenon of magnetic energy channelling \citep[e.g.][]{2018haex.bookE..25S}, thus causing a flux variation modulated with the planetary orbit. This phenomenon is known as magnetic star-planet interaction (MSPI) and has been detected photometrically in several systems with close-in giant planets \citep[e.g.][]{2008A&A...482..691W,2009EM&P..105..373P,2015ApJ...811L...2M}. Notably, while the theoretical formalisms describing ellipsoidal variations, Doppler beaming, and phase curves are known to describe the observations fairly well \citep[e.g.][]{2011MNRAS.415.3921F,2013ApJ...767..137Q,2014A&A...562A.109L}, describing MSPIs poses great challenges yet to be solved \citep[e.g.][]{2009A&A...505..339L,2012A&A...544A..23L,2015ApJ...815..111S,2018haex.bookE..25S}.

\subsection{The radial velocity method}
\label{subsec:rv_method}

\begin{figure}
    \centering
    \includegraphics[width=1\textwidth]{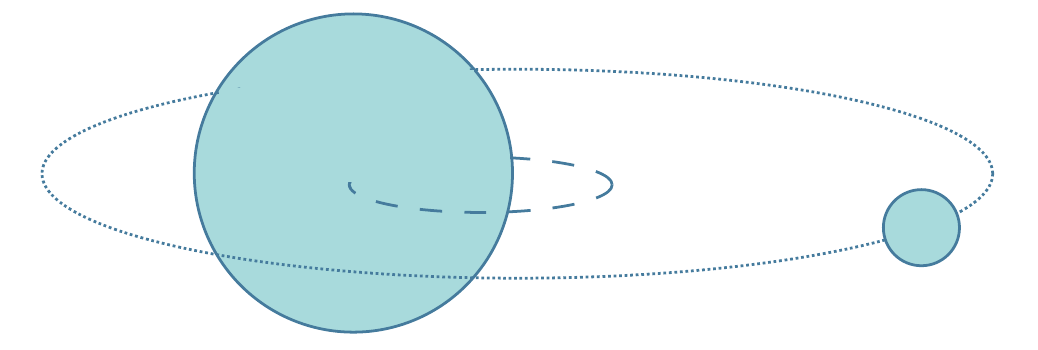}
    \caption[Orbital motion of a planet and its host stars around the centre of mass.]{Graphic representation of the orbital motion of a planet and its host stars around the centre of mass.}
    \label{fig:rv_scheme}
\end{figure}

The radial velocity (RV) method consists of measuring the radial component of the velocity of a star in the search for periodic variations, which, after discarding other possible origins, can be interpreted through a gravitational disturbance caused by a planet orbiting it. Newton's Third Law implies that the orbital motion of a planet makes its host star wobble around the centre of mass of the planetary system with the same periodicity as the planet's orbit (see Fig~\ref{fig:rv_scheme}). Contrary to the transit method, there is no restricted number of star-planet-observer configurations where this method can succeed. Indeed, all stars with planets have a radial velocity signature, except in the particular case where the planet is face-on to the observer's line-of-sight, being this the only case where the RV is zero.  

\subsubsection{Measuring the radial velocities}

\begin{figure}
    \centering
    \includegraphics[width=1\textwidth]{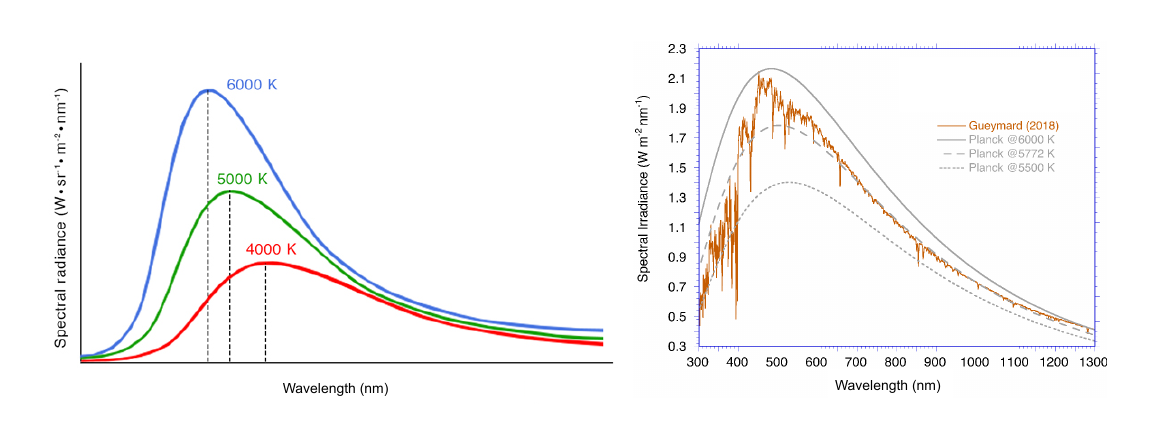}
    \caption[Blackbody emission at different temperatures and a solar spectrum.]{Left: Blackbody emission at different temperatures. Source: \url{https://www.sciencefacts.net/blackbody-radiation.html}. Right: Comparison between blackbody emission and a solar spectrum. Source: \citet{2018SoEn..169..434G}.}
    \label{fig:spectrum_example}
\end{figure}

The stellar RVs encoding the orbital motion of a planet are measured through spectra obtained in different epochs. All stars, in blackbody approximation, emit light according to Planck's function:

\begin{equation}
    B_{\nu} (T) = \frac{2h\nu^{3}}{c^{2}} \frac{1}{e^{\frac{h \nu}{k_{B}T}}-1}, 
    \label{eq:plank}
\end{equation} 

where $B_{\nu} (T)$ is the amount of energy per unit of time and area (i.e. flux) emitted by a blackbody at a temperature $T$, and is commonly referred to as spectral radiance. For a given temperature, $B_{\nu} (T)$ is distributed according to the radiation frequency $\nu$ (see Fig.~\ref{fig:spectrum_example}, left), and the maximum of this distribution is described by Wien's law. In addition, stellar spectra have different atomic lines and molecular bands. This is because the atoms and molecules that make up the star absorb photons with specific wavelengths while yielding excited electronic states. Therefore, for specific wavelengths corresponding to atomic and molecular transitions, the stellar flux decreases, creating particular features (see Fig.~\ref{fig:spectrum_example}, right).

\begin{figure}
    \centering
    \includegraphics[width=0.45\textwidth]{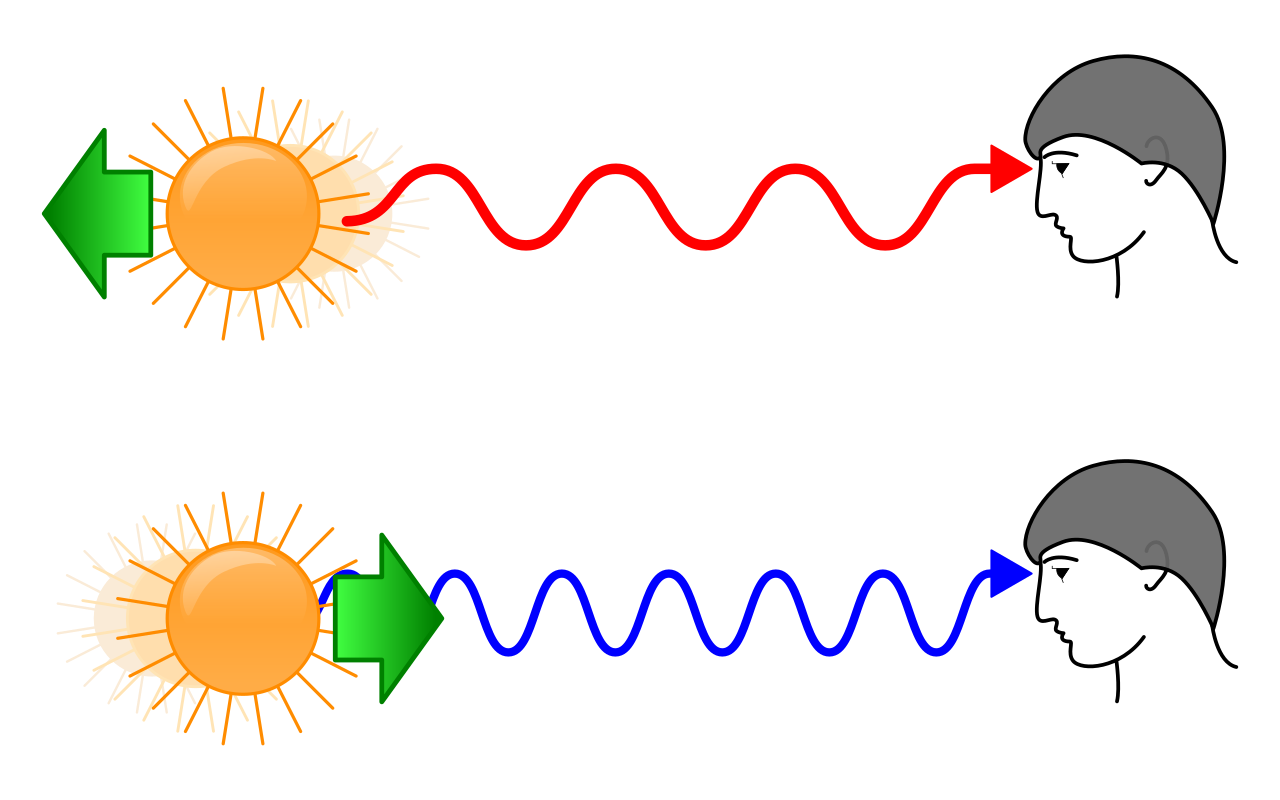}
    \includegraphics[width=0.47\textwidth]{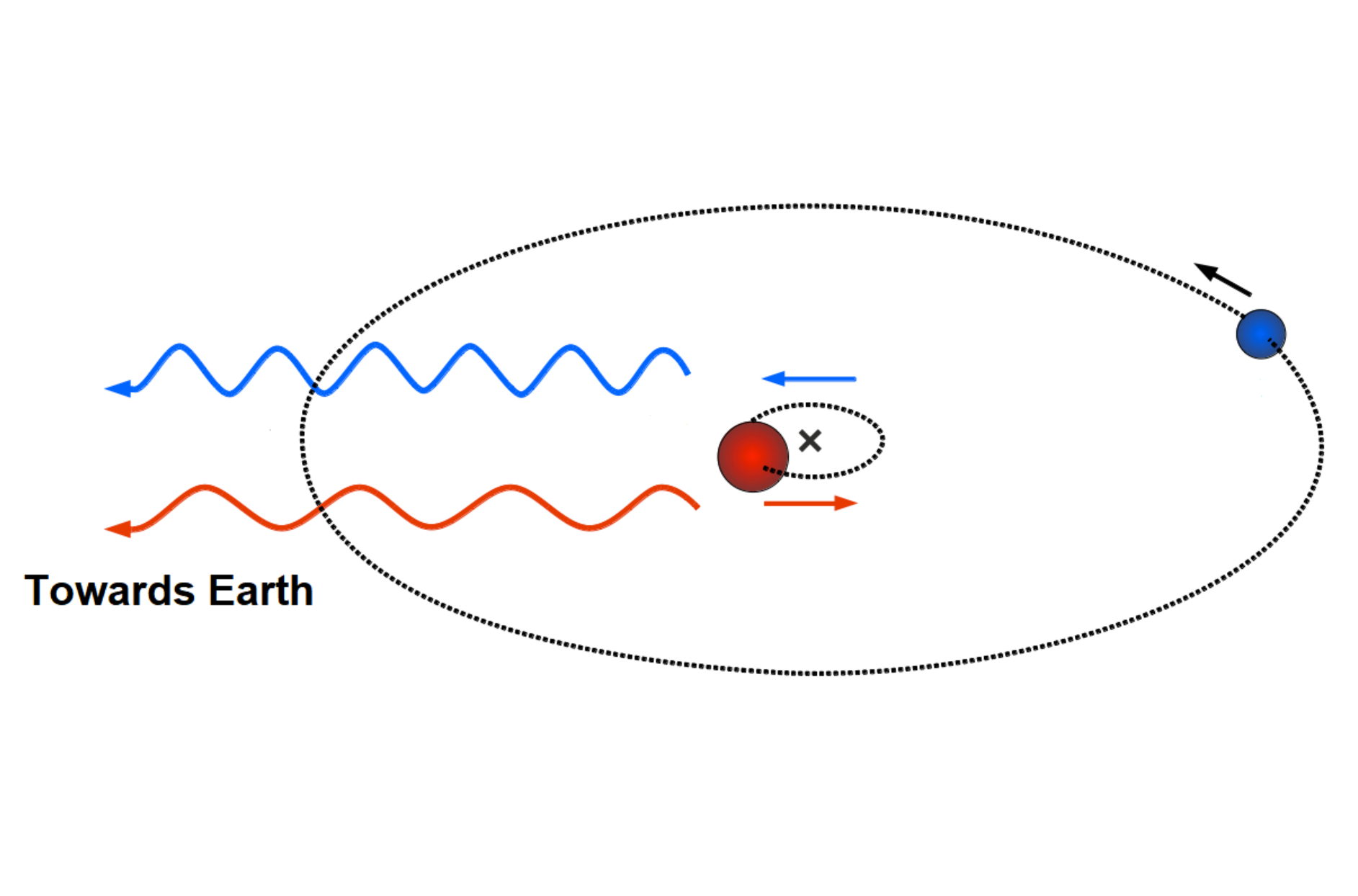}
    \caption[Representation of the Doppler effect.]{Left: Graphic representation of the Doppler effect when the emitting source moves away and approaches the observer. Right: Doppler effect when the source is a star and the movements of approach and distancing are caused by an orbiting planet. Credit: Ales Tosovsky (left) and Hans Deeg (right).}
    \label{fig:doppler}
\end{figure}

An observed stellar spectrum is shifted in wavelength (frequency) relative to the rest frame due to the Doppler effect. The magnitude of the shift depends on the velocity with which the star is approaching or moving away from the observer. This is explained through the wave nature of light. If the emitting source is moving towards the observer, the maxima of the emitted wave are observed closer to each other, so the wavelength is shorter (i.e. the light turns blueish). The opposite occurs if the source is moving away from the observer (i.e. the light turns reddish), see Fig.~\ref{fig:doppler}.

To quantify the displacement between the measured and emitted spectrum, we can take as reference different absorption lines located at well-known wavelengths ($\lambda_{\rm em}$) and compare them with the wavelengths of the observed lines ($\lambda_{\rm obs}$). Thus, if the observed star is approaching or moving away from the observer, the spectral lines of the spectrum will be shifted according to the relative velocity between both bodies. The non-relativistic relation between the relative observer-star radial velocity $v_{\rm r}$ and the observed wavelength $\lambda_{\rm obs}$ corresponding to an emitted wavelength $\lambda_{\rm em}$ is given by:

\begin{equation}
    \frac{v_{\rm r}}{c} = \frac{\lambda_{\rm obs}-\lambda_{\rm em}}{\lambda_{\rm em}}.
    \label{ec:doppler}
\end{equation}

Hence, by measuring the spectra of a star in different epochs, and using Eq.~\ref{ec:doppler}, we can measure $v_{\rm r}$ over time. To measure the actual RV of the observed star, we need to correct $v_{\rm r}$ for the observer's motion. Stellar spectra are taken from Earth, which orbits the Sun, rotates, and is orbited by the Moon. Notably, Earth's motion around the Sun can add up to $\pm$30 $\rm km\,s^{-1}$, while Earth's rotation, depending on the latitude, can add about 0.5 $\rm km\,s^{-1}$. These velocities are significantly larger than those typically induced by planets on their stars. Therefore, the observed wavelengths need to be converted from the observer's reference frame (i.e. the Earth) to an inertial frame, which is chosen to be the centre of mass of the Solar System. This correction is commonly known as the Barycentric Earth Radial Velocity (BERV) correction.

As described above, the process to extract RVs is conceptually simple, since it is only necessary to measure distances (i.e. wavelength shifts) in the spectra and convert them into velocities through Eq.~\ref{ec:doppler}. However, these shifts are often very small $-$e.g. several times smaller than the width of the spectral lines$-$ and thus require advanced statistical techniques to obtain accurate values. Different approaches have been proposed to date, with the so-called cross-correlation function (CCF) and the Template Matching (TM) techniques being the most commonly used ones.

Originally, the CCF method used a binary mask that associated a weight of one to the expected wavelength locations in a rest frame of a set of absorption lines and a weight of zero to the rest of the spectrum \citep{1996A&AS..119..373B}. This mask was then cross-correlated with the observed spectra, allowing the generation of a CCF profile that was fit to Gaussian functions to extract the RVs (i.e. the centre of the Gaussian). Today, spectral masks are not binary anymore and instead are weighed according to the RV information of each line \citep[i.e. deep and sharp lines constrain the wavelength shifts better than broad and shallow lines;][]{2001A&A...374..733B,2002A&A...388..632P}. 

The TM method is based on the construction of a high-S/N stellar template built from the observations of the star. This template is then compared to the individual spectra, and the RVs are extracted through minimisation processes \citep{2012ApJS..200...15A,2018A&A...609A..12Z,2022A&A...663A.143S}. The CCF approach performs very well in Solar-type stars, where the spectral lines are well-defined and generally well-spaced. However, cool stars exhibit a heavy line blending and a large number of molecular bands, making it difficult to construct weighted masks. For these stars, the TM approach $-$which automatically uses the whole spectrum$-$ has the potential of reaching a better performance. We note that apart from the CCF and TM techniques, novel approaches have been proposed recently, such as line-by-line RV extraction \citep{2018A&A...620A..47D} and GP-based algorithms \citep{2020MNRAS.492.3960R}.

\subsubsection{Radial velocity variations due to an orbiting planet}

In the following, we describe how the stellar RVs change due to an orbiting planet, and which planetary parameters we can infer from their study. We first start with the case of circular orbits, which, given its simplicity, is very suitable to introduce the main concepts. We then present the two main equations of the most general case, where planets can orbit with non-circular trajectories. 

\subsubsection*{Circular orbits}
\label{sec:circular_orbits}

\begin{figure}
    \centering
    \includegraphics[width=\textwidth]{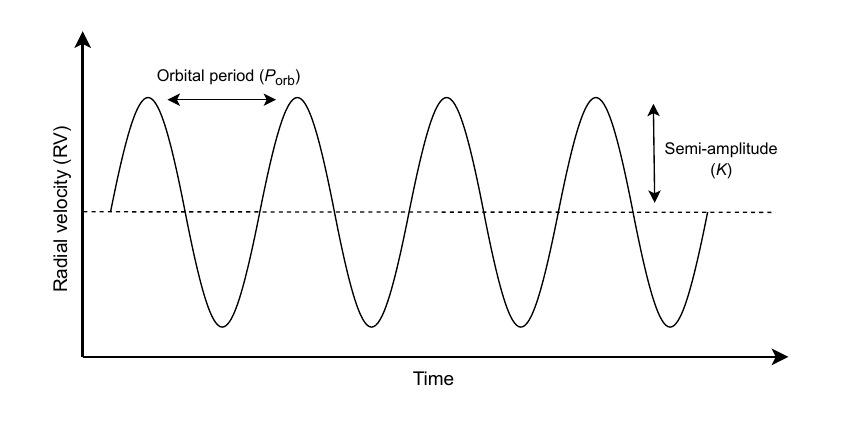}
    \caption[Radial velocity of a star induced by an orbiting planet in a circular orbit.]{Graphic representation of the radial velocity of a star induced by an orbiting planet in a circular orbit.}
    \label{fig:rv_curve}
\end{figure}

If we consider a planet of mass $M_{\rm p}$ orbiting a star of mass $M_{\rm \star}$ in a circular orbit of semi-major axis (or star-planet distance) $a$, the orbital velocity $v_{\rm p}$ of the planet can be obtained by simply equating centripetal force to gravitational force. In the case of $M_{\rm p}$ $\ll$ $M_{\rm \star}$, the centre of mass is near the centre of gravity of the star, and, using Kepler's third law, the orbital velocity can be written as: 
\begin{equation}
    v_{\rm p} = \sqrt{\frac{GM_{\star}}{a}}.
\end{equation}
Conservation of linear momentum implies that the orbital velocity of the star around the centre of mass ($v_{\star}$) is described by 
\begin{equation}
    M_{\rm \star} v_{\rm \star} = M_{\rm p} v_{\rm p}.
\end{equation}
If the plane of the sky and the orbital plane form an angle $i$ with the line of sight, the radial velocity measured by the observer would vary following a sinusoid with a semi-amplitude $K$ = $v_{\star}$ sin$\,i$, that is,

\begin{equation}
    K = \left( \frac{M_{\rm p}}{M_{\star}}  \right) \sqrt{\frac{GM_{\star}}{a}} \textrm{sin} \, i , 
    \label{eq:semi-amplitude}
\end{equation}

and a periodicity determined by the planetary orbit (see Fig.~\ref{fig:rv_curve}), which, following Kepler's Third Law, can be written as

\begin{equation}
    P_{\rm orb} = 2 \pi \sqrt{\frac{a^{3}}{GM_{\star}}}. 
    \label{eq:porb}
\end{equation}

By measuring an RV curve and managing to infer the semi-amplitude of the variations ($K$) and its periodicity ($P_{\rm orb}$), in case the mass of the star can be estimated independently, we have two equations (Eqs. \ref{eq:semi-amplitude} and \ref{eq:porb}) and three unknowns ($a$, $M_{\rm p}$, and sin $i$). The best that we can do is to substitute $a$ from Eq.~\ref{eq:porb} in Eq.~\ref{eq:semi-amplitude} and obtain a lower limit for the planet mass through the product $M_{\rm p} \, \textrm{sin} \, i$:

\begin{equation}
    M_{\rm p} \, \textrm{sin} \, i = \left( \frac{M_{\star}^{2} P}{2 \pi G} \right)^{\frac{1}{3}} K.
\end{equation}

The average value of $\textrm{sin} \, i$ for randomly inclined orbits is $\pi / 4$, so, statistically speaking, the difference between the minimum and true masses is not large. We note, however, that orbiting bodies with masses beyond the planetary domain ($M_{\rm p}$ $>$ 13$\rm M_{\rm J}$) and considerably inclined orbits could be wrongly interpreted as planets through this method, especially near the high-mass end.

Replacing the masses of the Sun and the Solar System planets in Eq.~\ref{eq:semi-amplitude} can give us an idea of the typical amplitude of the RV signal. For Jupiter, $v_{\star}$ = 12.5 $\rm m\,s^{-1}$, and for Earth, $v_{\star}$ = 0.09 $\rm m\,s^{-1}$. This large difference factor, of two orders of magnitude, is caused by the mass difference between the two bodies. As we can see in Eq.~\ref{eq:semi-amplitude}, the induced velocities scale inversely with the square root of the star-planet distance and scale linearly with the planet's mass. Therefore, small planets in large orbits induce the smallest RVs and thus are the most difficult to detect through this technique. 

\subsubsection*{Eccentric orbits}
\label{sec:eccentric_orbits}

The complete derivation of the RV expression for eccentric orbits is lengthy and requires introducing non-intuitive celestial mechanics concepts that are not explicitly used in this dissertation. Its derivation involves solving the well-known two-body problem, and hence we here only present the final results. For the complete deduction, we refer to \citet{1999ssd..book.....M} and \citet{2010exop.book...15M}.

\begin{figure}
    \centering
    \includegraphics[width=0.45\linewidth]{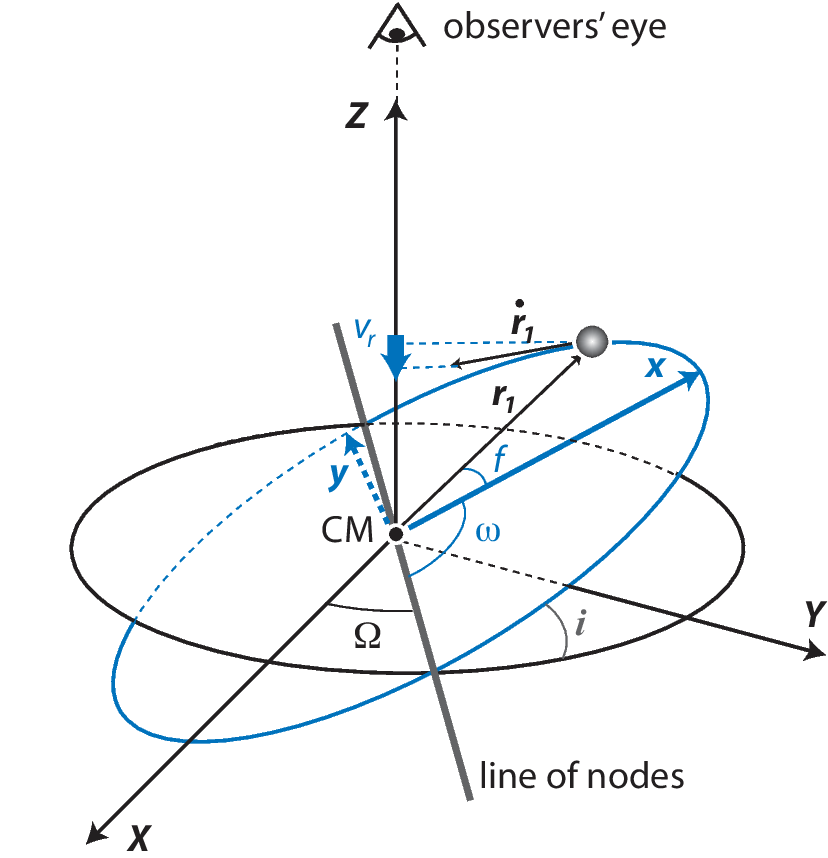}
    \caption[Orbital elements describing an eccentric orbit.]{Representation of different orbital elements describing an eccentric orbit, including the longitude of the periastron ($\omega$) and the true anomaly ($f$). Source: \citet{2010exop.book...15M}.}
    \label{fig:orbital_mechanics}
\end{figure}

The radial velocity of a star with an orbiting planet in an eccentric orbit can be written as: 

\begin{equation}
    v_{\star} (t) = K \left[ \textrm{cos}\left( f+\omega \right) + e \, \textrm{cos} \, \omega \right], 
\end{equation}

where $e$ is the eccentricity of the planet, $\omega$ is the longitude of the periastron, and $f$ is the true anomaly. The planet eccentricity $e$ is defined so that the star-planet distance in an eccentric orbit varies between $a(1+e)$ (i.e. furthest point between the planet and the star, commonly known as apoapsis) and  $a(1-e)$ (i.e. the closest point between the planet and the star, commonly known as periapsis). The longitude of the periastron $\omega$ is defined as the angle between the line of nodes and the periastron (see Fig.~\ref{fig:orbital_mechanics}), and the true anomaly $f$ is defined as the angle between the line joining the star and the planet and the direction of the periapsis (see Fig.~\ref{fig:orbital_mechanics}). 

The generalisation of the semi-amplitude for an eccentric orbit can be written as: 

\begin{equation}
    K = \frac{1}{\sqrt{1-e^{2}}}  \left( \frac{M_{\rm p}}{M_{\star}}  \right) \sqrt{\frac{GM_{\star}}{a}} \textrm{sin} \, i .
    \label{eq:semi-amplitude-general}
\end{equation}

Therefore, compared to a circular orbit with an equal period, a planet on an eccentric orbit produces RV signals of larger amplitude due to the rapid planetary motion near the periapsis. 

\subsubsection{Radial velocity variations due to other phenomena}

The shape and location of the measured spectral lines can also be altered by other phenomena not related to the gravitational pull of an orbiting body, thus generating non-planetary RV variations. Understanding these variations is crucial since they can complicate the characterisation of planets and even mimic their RV signatures, potentially leading to false detections. 

The main source of non-planetary RVs is stellar activity, which can appear in multiple manifestations, such as spots and plages, granulation, supergranulation, pulsations, and magnetic cycles. A detailed description of these effects is outside the scope of this work. However, we introduce the main concepts of spot- and plage-induced stellar activity since their effects are typically the most prominent in RV studies and affect the analysis presented in this dissertation.

When a star rotates, half the star approaches the observer, creating a blueshift effect, while the other half moves away, creating a redshift effect. This phenomenon causes the spectral lines to broaden. This rotational broadening generates symmetric line profiles if there is a symmetrical emission between the approaching and receding regions. However, the stellar flux is not necessarily symmetrical across the stellar disk. For example, stellar spots have lower temperatures than the effective temperature of the star. In the Sun, for example, the temperature difference is about 700 K \citep{2010A&A...512A..39M}. According to Plank's function (Eq.~\ref{eq:plank}), and as we can see in Fig.~\ref{fig:spectrum_example}, lower temperatures imply lower thermal radiation. Therefore, as the star rotates, a stellar spot can break the flux balance between the blue-shifted approaching limb and the redshifted receding limb, thus inducing an RV variation \citep[e.g.][]{1997ApJ...485..319S,2002AN....323..392H,2007A&A...473..983D,2012A&A...545A.109B,2014ApJ...796..132D}. 

\begin{figure}
    \centering
    \includegraphics[width=0.32\textwidth]{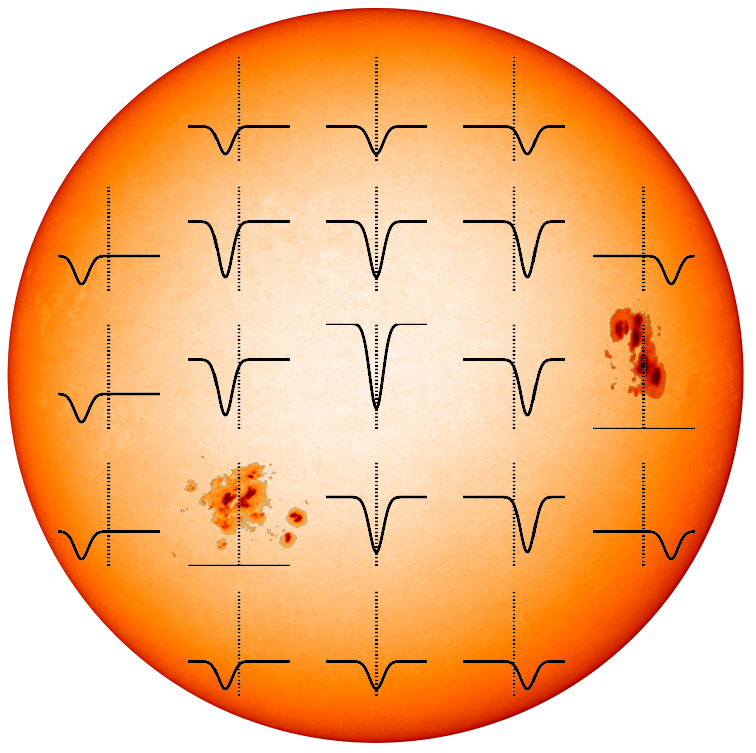}
    \hspace{1.5cm}
    \includegraphics[width=0.45\textwidth]{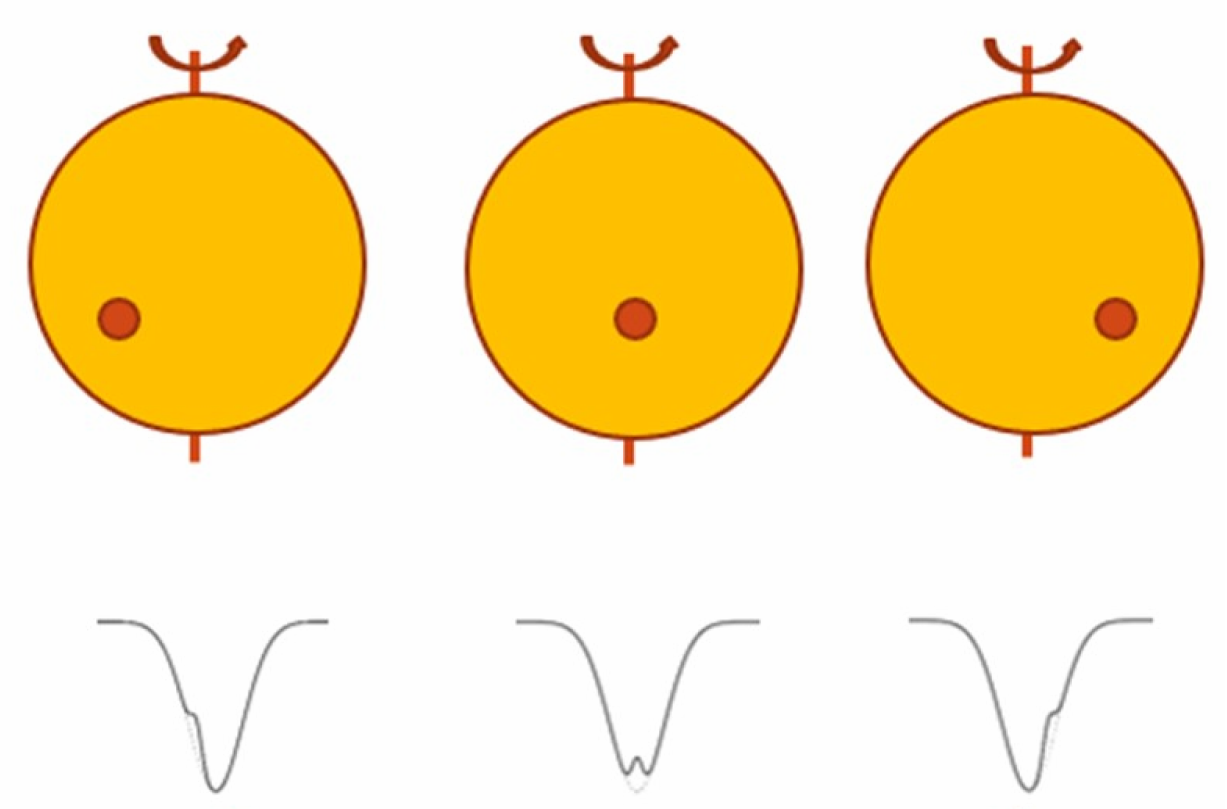}
    \caption[Effect of stellar spots in the measured spectral line profile.]{Effect of stellar spots in the measured line profiles. The left plot represents local CCF contributions from different stellar locations \citep[Source:][]{2014ApJ...796..132D}. In the regions with spots,  the local CCFs are suppressed, thus creating an asymmetry in the overall CCF, as shown in the right plot \citep[Source:][]{2023spi..conf...22M}.  }
    \label{fig:stellar_spots}
\end{figure}

A similar phenomenon occurs with stellar plages, which have hotter temperatures than the average effective temperature. We note, however, that the absolute temperature imbalance is much greater for spots than plages, so, despite they can be an order of magnitude larger \citep[e.g.][]{2001ApJ...555..462C}, plages typically induce smaller flux imbalances than spots. In addition to this effect, stellar spots can also induce RV variations due to the inhibition of stellar convection in the spot region due to strong local magnetic fields \citep[e.g.][]{2010A&A...520A..53L,2012MNRAS.419.3147A,2014ApJ...796..132D}. In Fig.~\ref{fig:stellar_spots}, we show how stellar spots alter the line profiles (or CCFs). 

The spot-induced asymmetries in the absorption lines can be translated into an RV signal with the periodicity of the stellar rotation. Even though some specific spot/plage configurations can lead to specific signal shapes, RV rotation modulations typically show a sinusoidal behaviour, making them difficult to differentiate from planetary signals. If the observing baseline is large enough, we would observe quasi-periodic variations since the spots and plages evolve $-$i.e. they
move on the stellar surface and appear and disappear$-$ throughout the magnetic cycle time scale. To differentiate between planet-induced and activity-induced RV signals, it is common to analyse different proxies for stellar activity, commonly known as activity indicators. These indicators typically measure changes in the shape of the mean spectral profile $-$i.e. the CCF$-$ or measure the flux in specific lines particularly sensitive to activity. As we can see in Fig.~\ref{fig:stellar_spots}, the width of the CCF can undergo periodic variations due to stellar spots. Hence, the full width at half maximum (FWHM) of the CCF is commonly used as an indicator. Another widely used indicator is the bisector (BIS), which was designed to characterise the CCF asymmetry by measuring the location of the intermediate points between its wings. The amplitude of the CCF is also used as a proxy for stellar activity and is commonly referred to as the CCF contrast. Regarding flux emissions of lines particularly sensitive to activity, the $\rm H_{\alpha}$ line (656.5 nm) and the Ca II H$\&$K (393.3 and 396.9 nm, respectively) doublet are among the most commonly used ones.

Finding correlations or coincident periodicities between activity indicators and RVs typically reflects a non-planetary RV origin \citep[e.g.][]{2001A&A...379..279Q,2008A&A...489L...9H,2009A&A...493..645F,2014A&A...566A..35S}. It is also common to find RV planetary signals $-$whose existence is corroborated by other methods (e.g. transit)$-$ in combination with activity-related RVs due to spots and plages. These RVs are often comparable to the planetary RVs $-$especially in cool stars$-$ thus requiring them to be treated as signals instead of white noise. In general, analysing stellar signals embedded in RV data is more challenging than analysing stellar signals in transit observations. This is, to a greater extent, due to the different time scales between the planetary and stellar signals and the lower cadence in which RV observations are usually taken. Stellar signals typically vary on time scales of several days $-$although young stars can rotate faster, with periodicities below 1 day$-$, while transit signatures only span a few hours. In contrast, RV signals do not occur in particular discrete time scales but continuously vary as the planet orbits its star in typical time scales of several days, comparable to that of the stellar rotation, making planetary and stellar signals more difficult to decode in RV data \citep[e.g.][]{2016A&A...588A..31F,2020A&A...635A..13F}. 

RV activity signals are typically modelled through GPs. In particular, over the last few years, a GP with a quasi-periodic autocorrelation function or kernel \citep{2015ITPAM..38..252A} has become the most popular choice to analyse RV time series affected by stellar activity. This kernel is physically motivated, and hence its parameters can be directly interpreted in terms of different stellar properties such as the rotation period of the star and the time scale of growth and decay of the active regions $-$i.e. a proxy for the magnetic cycle time scale \citep[e.g.][]{2020A&A...635A..13F,2024A&A...685A..56S}.

Instrumental systematic errors can also induce non-planetary RV variations, which often come from inaccurate wavelength calibrations. High-resolution spectrographs are highly sensitive to pressure and temperature changes, so they are typically isolated in vacuum chambers. In addition, nightly or simultaneous wavelength calibrations are frequently carried out $-$typically through Th-Ar hollow cathode lamps or Fabry-Pérot interferometers$-$ to minimise instrumental effects. In some cases, wavelength calibrations are not enough, and still, instrumental RVs can show large variations. These variations are not yet well understood and are typically corrected by observing different standard stars (i.e. quiet stars with little activity) every observing night and subtracting their RVs from the targets observed that night \citep[e.g.][]{2018A&A...609A.117T,2024A&A...687A.148G}. We note, however, that the instrumentation used in this work is not subjected to these nightly instrumental zero points.

\section{Instrumentation and collaborations}
\label{sec:instrumentation}

The great success achieved by the transit and radial velocity techniques during the last years has been largely due to an impressive instrumental advance. Exoplanet detection techniques were developed long before the first planet discoveries, and were extensively used to search for and characterise stellar binary systems. However, the precision of the early astronomical instrumentation did not allow us to reach the planetary domain. Another important factor that allowed exoplanet exploration to thrive is large-scale collaborations, which often provided the funds necessary to build state-of-the-art instrumentation and facilitated resource sharing to maximise the amount and quality of the scientific output.

In this thesis, we started the K2-OjOS collaboration, which is composed of amateur and professional astronomers and is aimed at exploiting public space-based high-precision photometry from \textit{Kepler}/K2. This thesis has been also carried out in the context of the ESPRESSO Consortium and the HARPS-NOMADS collaboration, which, among other objectives, used $-$and in the case of ESPRESSO, built$-$ state-of-the-art high-resolution spectrographs to characterize exoplanets in two key and poorly explored regions of the parameter space that we will describe later. In this section, we introduce the main instruments that helped us to achieve the objectives of this work and describe the collaborations that made it possible. 

\subsection{Space-based high-precision photometers}
\label{sec:space_based_photometers}

Space-based photometers rely on removing the atmospheric contribution that ground-based observations have to deal with, which is relevant to detect small planets through the transit technique \citep[e.g.][]{1971Icar...14...71R,1984Icar...58..121B}. Pioneer missions such as Spitzer \citep{2004ApJS..154....1W}, launched in 2003, and CoRoT \citep{2009A&A...506..411A}, launched in 2006, detected 36 transiting planets $-$one by Spitzer\footnote{Detecting planets was not Spitzer's goal, but it turned out to be an excellent instrument to characterize known systems.} and 35 by CoRoT, including groundbreaking discoveries such as the first transiting super-Earth CoRoT-7~b \citep{2009A&A...506..287L}$-$ and caved the path towards new-generation large-scale space-based photometers that revolutionised the exoplanet field. 

\subsubsection{The \textit{Kepler} spacecraft and the K2-OjOS collaboration}
\label{sec:intro_kepler}

The \textit{Kepler} spacecraft \citep{2010Sci...327..977B} was launched in March 2009 into a 372.5-day heliocentric Earth-trailing orbit. It quickly became the most prolific planet-hunting facility thanks to its unprecedented photometric precision, able to achieve the detection of the 80 parts per million (ppm) signal from an Earth-Sun equivalent transit. Its primary mission lasted until August 2013, when two of its four reaction wheels failed. This issue prevented the necessary pointing precision and threatened the continuation of the mission. Fortunately, thanks to the engineering team of the mission \citep{2014AAS...22322801H}, between 2014 and 2018, the satellite kept operating on its second $-$or extended$-$ mission, K2, taking advantage of solar photon pressure to compensate for the lack of reaction wheels \citep{2014PASP..126..398H}. \textit{Kepler}/K2 detected a total of 6282 planet candidates, of which 3326 have been validated or confirmed. These planets, almost seven years after its decommissioning, still make up about 60$\%$ of all known planets. 

\begin{figure}
    \centering
    \includegraphics[width=0.7\textwidth]{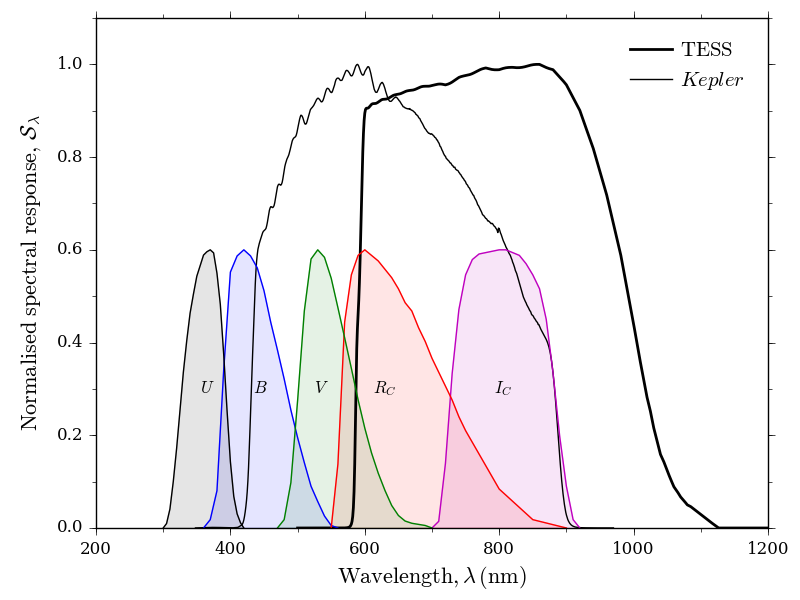}
    \caption[Spectral response functions for \textit{Kepler} and TESS.]{Spectral response functions for \textit{Kepler} and TESS normalised to 1, compared to the standard Johnson-Cousins $UBVR_{C}I_{C}$ photometric systems \citep{1990PASP..102.1181B} normalised to 0.6. Source: \citet{2017EPJWC.16001005L}.}
    \label{fig:kepler_tess_filters}
\end{figure}

\textit{Kepler} was a 0.95m Schmidt telescope equipped with a photometer made of 42 charge-coupled devices (CCDs) of 2200 $\times$ 1024 pixels. It had a pixel scale of 3.98 arcsecond $\rm pixel^{-1}$, which corresponds to a total field of view (FoV) of 105 $\rm deg^{2}$. It used one broad bandpass in the visible, which ranged from 420 to 900 nm (see Fig.~\ref{fig:kepler_tess_filters}). The main scientific goal of \textit{Kepler} was to detect Earth-like planets around Sun-like stars and measure their occurrence rate in the Galaxy. To that aim, it was necessary to observe a particular region of the sky for more than two years, since this is the minimum temporal baseline needed to measure three consecutive transits that would confirm the periodicity of a planet candidate. Therefore, the \textit{Kepler} observing strategy consisted of observing a fixed sky region during the entire primary mission, which was chosen to be in the northern sky centred at RA = 19h 22m 40s and Dec = +44º 30' 00" (J2000). 

The spacecraft rolled by 90 degrees every three months to optimise solar panel efficiency, dividing the data into 18 Quarters (Q0-Q17) separated by up to about 1-day gaps. The exposure time of the observations was 6.02 seconds with a 0.52-second readout. However, this short-cadence data was not released due to downlink restrictions, and instead \textit{Kepler} data products consisted of images and photometry stacked in intervals of 30 minutes (commonly referred to as long cadence data). Only a small set of 512 targets was chosen to be stacked with a 1-minute cadence. The obtained images were processed in the \textit{Kepler} Science Operations Center (SOC) at NASA Ames Research Center by the Kepler Science Processing Pipeline \citep{2010ApJ...713L..87J}, which performed pixel-level calibration,  aperture photometry extraction and systematic error correction \citep{2012PASP..124.1000S,2012PASP..124..985S}, and searches for transit signatures through the Transit Planet Search (TPS) module \citep{2010SPIE.7740E..0DJ}. This module searched for planetary transits using the Multiple Event Statistic (MES) metric, which is a measure of the S/N of a set of transits. Transit-like signals with an MES above 7.1 were selected as Threshold Crossing Events (TCEs) and subjected to different diagnostic tests to reject false positives caused by background eclipsing binaries \citep{2010SPIE.7740E..23T}. If passed, TCEs became Kepler Objects of Interest (KOIs).

KOIs are not considered planets, but instead planet candidates. This is mainly because of the large \textit{Kepler} pixel sizes, which can potentially collect the flux emitted by nearby stars inside and surrounding the photometric aperture, and be the actual source of the observed transit signals. Follow-up ground-based RV observations can collect spectra through tiny fibres with $<$ 1"  diameters and hence confirm the planetary nature of the \textit{Kepler} transiting signals\footnote{We note that before investing time in RV confirmation, high-resolution images were typically taken as a first screening.} \citep[e.g.][]{2010ApJ...713L.126B,2011ApJS..197...14D,2011ApJS..197...13E,2014ApJS..210...20M,2014A&A...571A..37S,2015A&A...577A.105L,2019MNRAS.484.3233B,2023A&A...677A..33B}. However, most \textit{Kepler} candidate-hosting stars are too faint to confirm their signals through RVs. Instead, \textit{Kepler} candidates were subjected to statistical validation studies, which allowed to estimate the probability that the candidates had a planetary origin based on a series of assumptions and modest follow-up observations \citep[e.g.][]{2011ApJ...727...24T,2014MNRAS.441..983D,2014ApJ...784...45R,2015AJ....149...55E,2016ApJ...822...86M,2021MNRAS.504.5327A,2022ApJ...926..120V}. 

Among the \textit{Kepler} detections, there were some cases of small planets in warm regions of their host stars \citep{2013ApJ...768..101B,2013Sci...340..587B,2015AJ....150...56J,2017AJ....153..180S}. However, no Earth analogue was detected. Still, the \textit{Kepler} mission provided invaluable insight into planet demographics. Its fixed long-term focus on a particular region of the sky, together with its outstanding precision and homogeneous and well-characterized pipeline performance made it an excellent opportunity to perform planet occurrence studies ranging from sub-Earths to super-Jupiters \citep[e.g.][]{2012ApJS..201...15H,2012ApJ...753...90M,2013ApJS..204...24B,2013ApJ...766...81F,2013ApJ...767L...8K,2013PNAS..11019273P,2014ApJ...795...64F,2014ApJ...791...10M,2015ApJ...810...95C,2015ApJ...807...45D,2015ApJ...798..112M,2018AJ....155..205H,2020AJ....159..248K,2021AJ....161...36B,2023AJ....166..122D}, which provided important constraints to planetary evolution and formation theories. 

The loss of two of the four reaction wheels and the subsequent transition from the \textit{Kepler} primary mission to the K2 extended mission brought different drawbacks, but also new opportunities. The spacecraft underwent continuous drifts that translated into larger systematic errors than those observed in the primary mission. These systematics were not corrected by the \textit{Kepler} pipeline. Instead, the community developed new techniques $-$e.g. the self-flat fielding technique \citep[SFF;][]{2014PASP..126..948V}$-$ and used previously developed methods $-$e.g. the pixel-level decorrelation \citep[PLD;][]{2015ApJ...805..132D,2016AJ....152..100L,2018AJ....156...99L} or Gaussian Process Regression methods \citep{2016MNRAS.459.2408A}$-$ to obtain systematics-corrected photometry. This dissociation between the basic calibration performed by the \textit{Kepler} pipeline and the drift-related correction performed by the community provoked the \textit{Kepler} science team at SOC to stop releasing KOIs. Therefore, the community was in charge of both correcting the data from drift-related systematics and searching for planetary signals. 

\begin{figure}
    \centering
    \includegraphics[width=\textwidth]{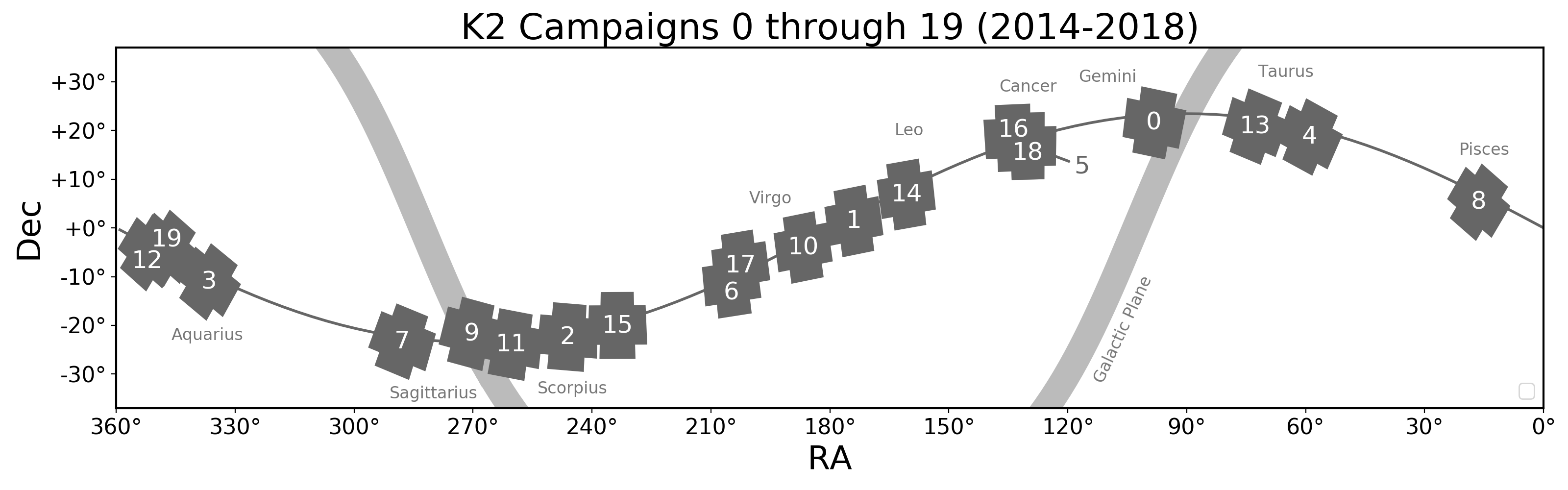}
    \caption[Right ascension and declination of the K2 campaign fields over the ecliptic plane.]{Right ascension (RA) and declination (Dec) of the K2 campaign fields over the ecliptic plane (thin line). The galactic plane is also shown. Source: \url{https://keplergo.github.io/KeplerScienceWebsite/k2-fields}.}
    \label{fig:k2_campaigns}
\end{figure}

The K2 observing strategy also changed relative to that of the primary mission. Instead of observing a fixed region of the sky with a long temporal baseline, K2 observed different fields across the ecliptic plane during 20 observing campaigns (C) of $\sim$80 days each. In Fig.~\ref{fig:k2_campaigns}, we show the coordinates of the K2 campaign fields across the ecliptic. Most of the fields were observed once, which considerably limited the detectable orbital periods of the planet candidates, but in some cases, the same field was re-visited once or twice (e.g. C5, C16, and C18), allowing us to identify longer period candidates and obtain more accurate physical and orbital parameters. The particular K2 properties described above considerably changed the main scientific output of the spacecraft. The reduced observing baseline, together with the varying fields of view and the lack of homogeneous large-scale transit searches, prevented the study of planet occurrences. In contrast, the larger monitored sky area allowed us to detect a larger number of bright planet-hosting stars amenable for follow-up RV studies and mass measurements \citep[e.g.][]{2017A&A...608A..25B,2018MNRAS.476L..50D,2018MNRAS.480L...1D,2019A&A...623A..41P,2019A&A...623A.114L,2020A&A...640A..48L}.

K2 transit-like signals were typically searched through specific algorithms (e.g. the Box Least Squares, BLS, \citealt{2002A&A...391..369K}, and the Transit Least Squares, TLS, \citealt{2019A&A...623A..39H}) that were systematically applied to a large number of light curves \citep[e.g.][]{2016A&A...594A.100B,2016MNRAS.461.3399P,2018AJ....155..136M}. In addition, the lack of official K2 planet candidates triggered a large interest in the astronomical community to search for planetary signals manually, that is, through a direct visual inspection of the photometric time series. 

The process of scrutinising satellite data in the hunt for new worlds is a fascinating endeavour for astronomy enthusiasts, but the huge amounts of K2 data pose a big challenge. In this regard, amateur astronomers and citizen scientists played a crucial role. These communities largely outnumber professional astronomers, which allows for distributing and reducing the workload of the planet hunters through professional-amateur (pro-am) collaborations. Pioneer large-scale collaborations such as Planet
Hunters \citep[e.g.][]{2012MNRAS.419.2900F,2012ApJ...754..129S,2014AJ....148...28S,2014ApJ...795..167S,2020MNRAS.494..750E,2021MNRAS.501.4669E,2021MNRAS.505.1827E,2022MNRAS.511.4710E,2024AJ....167..241E,2024AJ....167..238O} and Exoplanet Explorers \citep[e.g.][]{2018AJ....155...57C,2019AJ....157...40F,2019RNAAS...3...43Z,2021AJ....161..219H} resulted in an impressive scientific output. These collaborations had a worldwide reach thanks to the implementation of the necessary tools and graphic interfaces in online public platforms specialised in pro-am collaborations.

In this work, we initiated the K2 OjímetrO Survey \citep[K2-OjOS\footnote{K2-OjOS website: \url{https://sites.google.com/view/k2-ojos/english}};][]{2020sea..confE..97C}, a pro-am collaboration between the University of Oviedo and the Sociedad Astronómica Asturiana Omega with the ultimate goal of identifying variable stars and planets in K2 data. The K2-OjOS project represents a \textit{rara avis} pro-am collaboration in the exoplanet field. Unlike the worldwide projects mentioned above, which have thousands of volunteer citizen scientists, K2-OjOS was composed of just 10 amateur astronomers. This naturally limits our planet hunting capacity compared to the global collaborations, but allowed the amateur volunteers to go a step further and participate in more specialised tasks typically reserved for professional astronomers. Once a transit-like signal is identified, different diagnostic tests (or vetting) are typically performed before subjecting the data to a statistical validation analysis. The K2-OjOS members were thus trained to participate in the vetting process with the goals of making a first screening of the signals found, identifying their possible causes, and discarding false positives. In Sect.~\ref{project}, we describe in more detail this collaboration. 

\subsubsection{The TESS mission}
\label{sec:intro_tess}

The Transiting Exoplanet Survey Satellite \citep[TESS;][]{2014SPIE.9143E..20R} is the successor of \textit{Kepler}. It was launched in 2018 into a highly elliptical 13.7-day orbit around the Earth, and currently continues operations in its extended mission. TESS and \textit{Kepler} have different scientific goals, which translate into different observing strategies. 

\begin{figure}
    \centering
    \includegraphics[width=\textwidth]{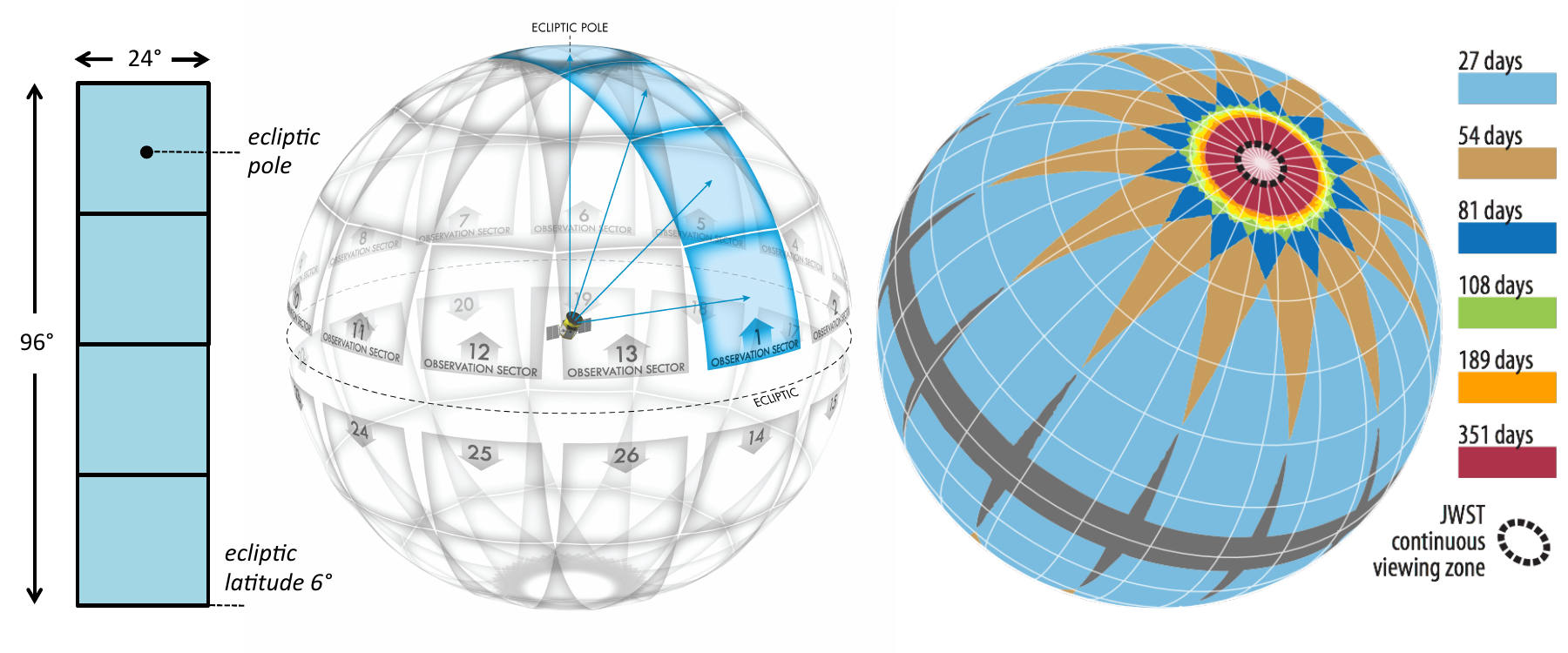}
    \caption[Field of view of the four TESS cameras.]{Left: Instantaneous combined field of view of the four TESS cameras. Middle: Division of the celestial sphere into 26 observation sectors (13 per hemisphere). Right: Duration of observations on the celestial sphere, taking into account the overlap between sectors. Source: \citet{2014SPIE.9143E..20R}.}
    \label{fig:tess_fiels}
\end{figure}

One of the main results from \textit{Kepler} was to find that small planets ($R_{\rm p}$ $<$ 4$\rm R_{\oplus}$) are abundant in our Galaxy \citep[e.g.][]{2012ApJS..201...15H,2013ApJS..204...24B,2013PNAS..11019273P,2015ApJ...809....8B}, confirming previous hints from early RV surveys \citep{2011arXiv1109.2497M}. However, very little was known about the properties of these planets, given that the typically faint \textit{Kepler} hosts prevented us from conducting systematic follow-up studies. Planets in this radius range are commonly referred to as super-Earths (1$\rm R_{\oplus}$ $<$ $R_{\rm p}$ $<$ 2$\rm R_{\oplus}$) and mini-Neptunes (2$\rm R_{\oplus}$ $<$ $R_{\rm p}$ $<$ 4$\rm R_{\oplus}$), and have no counterpart in the Solar System. The primary objective of TESS was to discover 50 transiting planets smaller than Neptune with host stars bright enough for follow-up spectroscopy to measure planetary masses and atmospheric compositions. To achieve this goal, TESS was designed as an all-sky survey able to acquire photometry for hundreds of thousands of bright stars. In particular, TESS observes a certain region of the sky (i.e. Sector) every about 27 days, managing to cover one hemisphere (i.e. Cycle) every year (see Fig.~\ref{fig:tess_fiels}). Contrary to \textit{Kepler}, the main TESS value is not statistical completeness, but rather the relative ease of following up on discoveries with current and forthcoming instruments.

TESS scans the sky through four cameras of 10 cm diameter, consisting of a lens assembly with seven optical elements and a detector assembly with four CCDs. Each lens forms a 24 $\times$ 24 deg image on the four-CCD mosaic in its focal plane, making up a combined FoV of 24 $\times$ 96 deg = 2300 $\rm deg^{2}$, which is about 22 times larger than \textit{Kepler}'s FoV (see Fig.~\ref{fig:tess_fiels}). Before reaching the CCDs, the starlight goes through a filter with a broad bandpass, redder than \textit{Kepler}'s, ranging from 600 to 1000 nm (see Fig.~\ref{fig:kepler_tess_filters}). In all, TESS can reach a 200 ppm photometric precision in 1 hour in a star with $I_{C}$ = 10 mag, with systematic noise sources below 60 ppm per hour. This implies a worse photometric performance than \textit{Kepler}, but it is still good enough to detect hundreds of super-Earths and larger planets in the brightest stars of our neighbourhood \citep{2021ApJS..254...39G}.

TESS data is downloaded every 13.7 days, when the satellite is at the perigee of its orbit.  This coincides with the change of sector and with the mid sectors, generating data gaps of about 1-3 days. TESS images are processed at the Science Processing
Operation Center \citep[SPOC;][]{2016SPIE.9913E..3EJ} at NASA Ames Research Center by closely following the data reduction and analysis developed for the \textit{Kepler} mission \citep{10.1117/12.857625,2010ApJ...713L..87J,2010SPIE.7740E..0DJ,2010SPIE.7740E..23T,2012PASP..124.1000S,2012PASP..124..985S,Stumpe2014,2020RNAAS...4..201C,morris:PA2020KDPH}. TESS data products include aperture photometry and systematics-corrected photometry for a selected number of targets, and full-frame images (FFIs) of the entire sky. The TESS observing cadence is 2 sec, but the corrected photometry is released with cadences of 2~min or 20~sec, depending on the target, and the uncalibrated FFIs are provided with cadences of 30~min, 10~min, or 200~sec, depending on the Cycle. Based on the reduced data, the TESS pipeline systematically searches for transit-like events, which, once they pass the diagnostic tests, receive a TESS Object of Interest (TOI) identifier.

\begin{figure}
    \centering
    \includegraphics[width=0.48\textwidth]{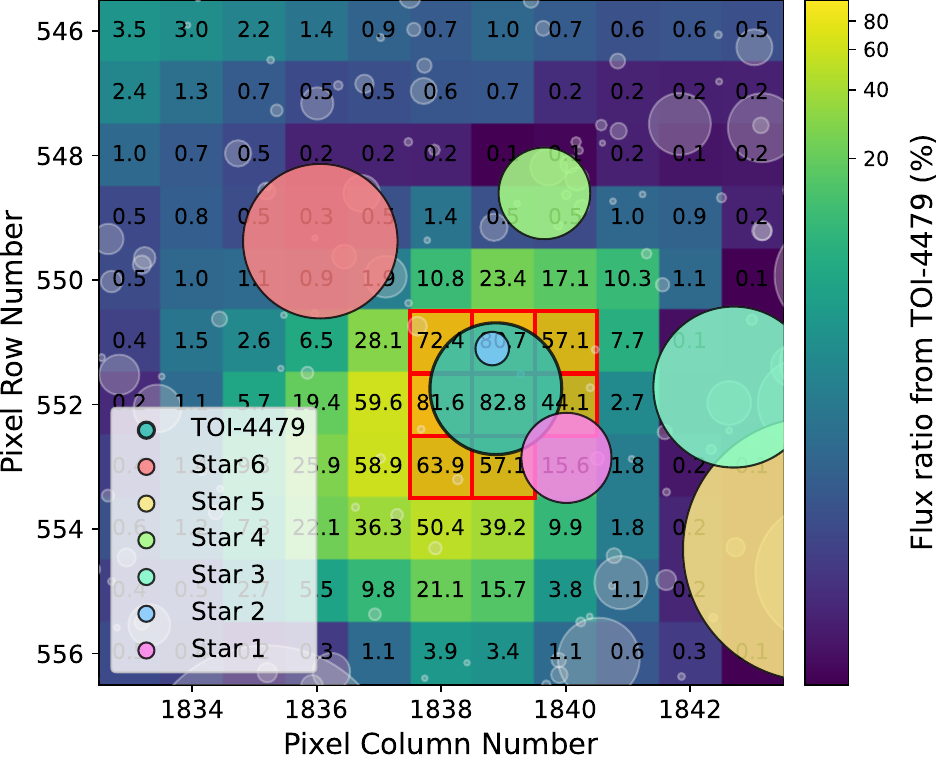}
    \includegraphics[width=0.49\textwidth]{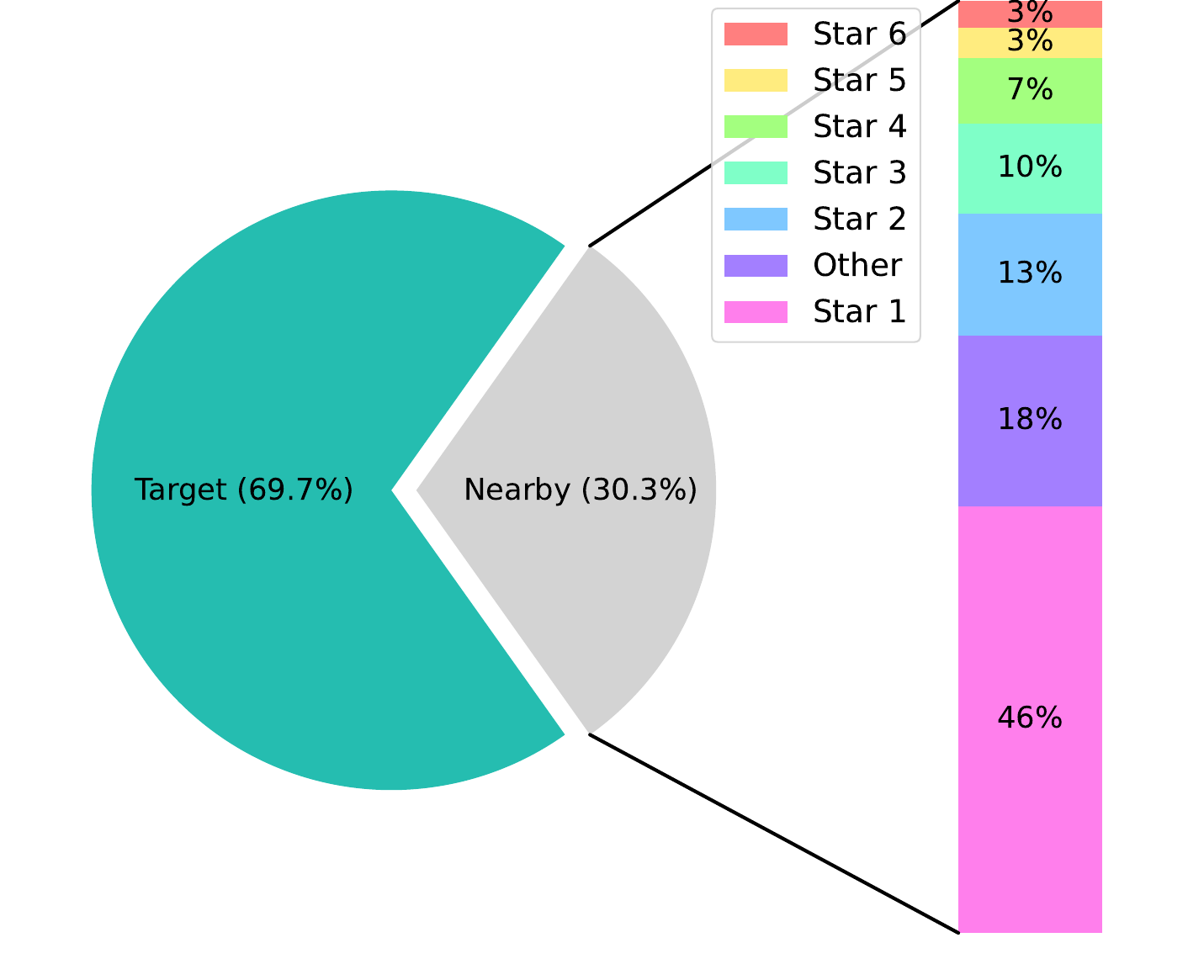}
    \caption[Flux contamination in the photometry of a TESS Object of Interest.]{Flux contamination from nearby sources in the photometry of a TESS Object of Interest, TOI-4479. Left: heat map with the pixel-by-pixel flux fraction coming from TOI-4479 in sector 41. The red grid is the SPOC aperture. The white disks represent all the nearby sources, and the six sources that most contribute to the aperture flux are highlighted in different colours. Right: Flux contributions to the SPOC aperture from the target and most contaminant stars. This plot was created through \texttt{TESS-cont} (\url{https://github.com/castro-gzlz/TESS-cont}).}
    \label{fig:TESS-cont_example}
\end{figure}

Similar to the \textit{Kepler} KOIs, the TESS TOIs are considered as planet candidates given the impossibility to assess the true origin of the transits found based on satellite data alone. Indeed, TESS has a pixel scale of 21 arcsecond $\rm pixel^{-1}$, which is more than five times larger than \textit{Kepler}'s (3.98 arcsecond $\rm pixel^{-1}$), and thus increases the flux contamination issue (see Fig.~\ref{fig:TESS-cont_example}). Therefore, given the greater need for high-spatial resolution observations to confirm TESS planets, the excellent follow-up opportunities that its bright planet-hosting stars offer, and the quick development of precise ground-based instrumentation during the last few years, TESS has completely changed the course of exoplanet detection from large-scale statistical validations to more dedicated characterisations of targets of particular interest. Overall, TESS results, together with ground-based follow-up efforts, are allowing us to expand the dimensionality of the exoplanet parameter space beyond the orbital period and radius spaces previously exploited by \textit{Kepler}. This expansion is allowing us to detect unusual systems and new populations in previously unexplored niches, which are bringing insightful information on the formation and evolution of exoplanets. 

In 2026, the European PLATO mission \citep{2014ExA....38..249R} will take over from \textit{Kepler} and TESS to become the flagship mission of exoplanet exploration. With a similar photometric precision and strategy to \textit{Kepler}, but focusing on brighter stars, PLATO will be crucial to explore the population of small planets in large orbits, a key niche for planet formation and habitability studies. This thesis is contextualised within the preparatory work for PLATO. In Sect.~\ref{sec:PLATO}, we introduce the most relevant technical aspects of the instrument and discuss the synergies between our work and the mission objectives.

\subsection{Ground-based high-resolution spectrographs}
\label{sec:ground_based_spectrographs}

\subsubsection{The ESPRESSO Consortium}
\label{espresso-consortium}

The Echelle SPectrograph for Rocky Exoplanets and Stable Spectroscopic Observations \citep[ESPRESSO;][]{2021A&A...645A..96P} is an European endeavour with contributions from Italy, Portugal, Spain, and Switzerland, which make up the ESPRESSO Consortium. Similar to \textit{Kepler}, it was conceived with the scientific goal of finding and characterising terrestrial planets in warm regions of their stars.

Building ESPRESSO, a HARPS-like instrument on a 8.2 m telescope, was the natural step forward in the exoplanet research by that time, since it allowed bridging the large gap between the small and medium-size telescopes ($<$ 4 m diameter) and the upcoming extremely large telescopes, where the inclusion of high-resolution spectrographs for RV and other studies had already begun to be discussed \citep[e.g.][]{2005Msngr.122...10P,2009ASSP....9..395P,2006IAUS..232..198M,2009ASSP....9..389M}. ESPRESSO was also built in response to the exponential growth of exoplanet candidates provided by space-based transit missions. In just a few years, thousands of exoplanets were discovered, but most of them lacked a mass measurement, which is a fundamental property to understand planet formation, evolution, interiors, and atmospheric compositions. This situation emphasised the synergy between RVs and transit observations, and thus the need for ultra-precise ground-based RV follow-up that would allow us to understand the wealth and diversity of small planets detected by space-based transit surveys. 

ESPRESSO is installed in the Very Large Telescope (VLT) in ESO's Paranal Observatory, in Chile. It is enclosed in a vacuum vessel located inside a thermally controlled room to minimise the impact of temperature and pressure variations. ESPRESSO's first light took place on 27 November 2017, when its commissioning phase began, and started scientific operations on 1 October 2018 $-$although improvements to the instrument and recommissioning runs were conducted until July 2019. ESPRESSO can operate with any of the 8.2 m VLT units (1-UT configuration), or with the four units simultaneously  (4-UT configuration), achieving a larger collecting area equivalent to a 16 m diameter telescope.

ESPRESSO is illuminated by two fibres. One carries the light from the science target, and the other carries either the light from the sky background or from a reference source for simultaneous RV drift measurements. The reference source light can be a combination of spectra from a Th-Ar lamp and a white-light illuminated Fabry-Pérot. The former provides accuracy, and the latter ensures local wavelength precision. A laser frequency comb (LFC) is also available to carry out the wavelength calibration. However, due to a lack of reliability and limited spectral coverage, it is currently not integrated into the ESPRESSO operation. We note that in practice, the simultaneous reference source is only needed when the RV error on the science target is expected to be comparable to or lower than the instrumental drifts $-$on the order of a few $\rm m \, s^{-1}$. Otherwise, if no high RV precision is required, the second fibre is fed with sky light so that the sky background can simply be
measured and subtracted. In all, ESPRESSO has been proven to have an on-sky precision better than 25 $\rm cm \, s^{-1}$ on a single night \citep{2021A&A...645A..96P} and better than 45 $\rm cm \, s^{-1}$ in the long term (Figueira et al. 2025). For a detailed description of the instrument, its systems and software, we refer to \citet{2014AN....335....8P,2014SPIE.9147E..1HM,2016SPIE.9913E..2KC,2018SPIE10707E..2GC,2016SPIE.9913E..2HB,2018haex.bookE.157G,2021A&A...645A..96P}.

The ESPRESSO Consortium had access to Guaranteed Time Observation (GTO), a total of 273 observing nights that ESO provided in exchange for funding and building the instrument. The Consortium proposed a program where 80$\%$ of the time would be dedicated to search for and characterise exoplanets, 10$\%$ to study the possible variability of fundamental physical constants, and the remaining 10$\%$ for other science cases of interest. The GTO program was organised into four working groups (WGs) dedicated to different scientific cases. The WG-1 focused on making a blind RV search $-$i.e. a planet search with no prior evidence nor hints on the existence of planetary signals$-$ in bright, nearby, and quiet stars in our Galaxy. The WG-2 was dedicated to the characterisation of planetary atmospheres through the transmission spectroscopy technique. The WG-3 focused on making an RV follow-up of K2 and TESS transiting planets, and the WG-4 was dedicated to studying fundamental constants. 

ESPRESSO's WG-3 took the opportunity that the bright K2 host stars and the new TESS targets provided in 2018-2019, just as ESPRESSO science operations began. A relevant niche of planetary science still not well understood is the rocky planet population, typically defined as those planets with $R_{\rm p}$ $<$~2~$\rm R_{\oplus}$, which receive this name because their typical masses and radii are compatible with rocky internal compositions (see more details in Sect.~\ref{sec:small}). Given their intrinsic scientific interest and the excellent opportunity to showcase the technical capabilities of ESPRESSO, rocky planets were selected to be exploited in this WG. In particular, the original WG-3 planet sample was defined based on the following criteria: a) confirmed or statistically validated planets, b) estimated planet radius $<$ 2 $\rm R_{\oplus}$, c) host star brightness $V$ $<$ 14.5 mag, and d) no precise mass measurement available. We note that the first criterion was relaxed throughout the GTO to also consider TESS's TOIs, given the low TESS false positive rate and the lack of large-scale TESS statistical validation works. In addition, a set of sub-Neptunes (2 $\rm R_{\oplus}$ $<$ $R_{\rm p}$ $<$ 4 $\rm R_{\oplus}$) were added to the original target list to study the rocky-to-gaseous transition. Studying this transition at different irradiation levels would allow us to constrain the evaporation processes that presumably transformed sub-Neptunes into naked rocky cores under the influence of extreme UV (XUV) irradiation from the host star, or, alternatively, support the hypothesis that these two populations are driven by formation instead of evolution processes (see more details in Sect.~\ref{sec:small}).

\subsubsection{The HARPS-NOMADS collaboration}
\label{harps-nomads}

The High Accuracy Radial velocity Planet Searcher spectrograph \citep[HARPS;][]{2003Msngr.114...20M} is one of the first large-scale planet searchers based on the RV technique. It was designed by a European consortium of four French and Swiss institutions in collaboration with ESO. HARPS was built in just three and a half years, and is often referred to as a second-generation instrument since it benefited from the previous experience with the early ELODIE and CORALIE spectrographs. With 222 planet discoveries, HARPS is the most prolific RV planet searcher to date.

HARPS was installed on ESO’s 3.6m Telescope at La Silla Observatory (Chile) in January 2003, and its first light took place one month later. HARPS revolutionised the exoplanet field by achieving an unprecedented RV precision of 1 $\rm m \, s^{-1}$. The major instrumental breakthrough to achieve this precision was the instrument's stability. A lot of effort was put into making the spectrograph intrinsically and extrinsically stable. HARPS was a pioneer in being placed in a pressure- and temperature-controlled room, as well as in using the Th-Ar reference technique, consisting of performing simultaneous wavelength calibration with a Th-Ar lamp (see Sect.~\ref{espresso-consortium}).

The HARPS-NOMADS collaboration is an international effort led by David J. Armstrong from the University of Warwick (United Kingdom). This collaboration uses the HARPS spectrograph to confirm and characterise TESS planet candidates inside and surrounding the Neptunian desert, a region of the period-radius parameter space with a scarcity of intermediate-size planets (see Sect.~\ref{sec:intermediate} for more details). TESS, with its all-sky coverage and a photometric precision sufficient to detect Neptunian planets, is providing a handful of planet candidates inside the Neptunian desert, where it used to be practically devoid of planets, given the much-limited \textit{Kepler}'s sky coverage. This is allowing us to study particular systems in a previously unexplored region of the parameter space \citep[e.g.][]{2020Natur.583...39A,2021A&A...653A..60M,2023Natur.622..255N,2023A&A...669A.109L,2024MNRAS.532.1612H}. Notably, HARPS' precision is ideal to monitor Neptunian planets and get precise masses and orbital parameters with a modest number of observations. Overall, the short-term goal of the HARPS-NOMADS program is to get a sample of confirmed planets with accurate masses that allow for follow-up atmospheric studies in an unexplored region of the parameter space. The long-term goal is to make statistical studies of the planet's and host star's properties inside and surrounding the Neptunian desert, with the ultimate objective of unveiling its origin and hence shedding some light on the formation and evolution of close-in Neptunian planets.

This thesis has also been developed in the context of other international collaborations, such as CARM-TESS \citep[see e.g.][]{2023A&A...677A.182M}, a legacy program with the CARMENES spectrograph \citep{2016SPIE.9908E..12Q} to confirm and validate TESS Objects of Interest, the KOBE experiment \citep{2022A&A...667A.102L}, another CARMENES legacy program focused on finding planets in the warm orbital regions of late-K dwarfs, the KESPRINT collaboration \citep[see e.g.][]{2021MNRAS.508..195D}, a consortium devoted to the detection and characterization of transiting exoplanets found by space-based missions, the ATREIDES collaboration (Bourrier et al., in prep), an ESO large program with ESPRESSO to obtain a census of obliquities of close-in Neptunian planets, the TROY project \citep{2018A&A...609A..96L}, a global effort to search for the first co-orbital planets (i.e. planets sharing the same orbit), and the NCORES collaboration \citep{2025MNRAS.tmp..168A}, an ESO large program with HARPS to explore the naked cores of Neptunes inside the desert.

\section{Exoplanet populations}
\label{exoplanet_populations}

The orbital and physical properties of the 5800 confirmed and validated exoplanets known to date are not typically homogeneously distributed across the parameter space. Instead, they often show particularities, and in some cases are correlated with other planetary or stellar properties. The observed distributions can be significantly altered by observational biases. Once accounted for, unveiling the inherent features, patterns, and correlations of the planets' and their host stars' properties can provide relevant clues on the formation and evolution histories of the planetary systems.

Planets can be classified into three main groups according to their sizes: gas giant planets\footnote{The word `gas' in a size-based classification aims at preserving the historical reference of `giant planet' to the Solar System planets larger than Neptune, which includes both the gas (Jupiter and Saturn) and icy (Neptune and Uranus) giants.} ($R_{\rm p}$ $>$ 10~$\rm R_{\oplus}$), intermediate-size planets\footnote{These planets are also referred to as transitional planets. We decided to avoid this nomenclature throughout this work to prevent the possible introduction of connotations related to their evolution.} (4 $\rm R_{\oplus}$ $<$ $R_{\rm p}$ $<$ 10 $\rm R_{\oplus}$), and small planets\footnote{These planets are also referred to as $\textit{Kepler}$ planets, given the relevance of the survey in their discovery. We also avoid this nomenclature to preserve instrument-independent terms.  } ($R_{\rm p}$ $<$ 4 $\rm R_{\oplus}$), which are also referred to as Jupiters and super-Jupiters, super-Neptunes and sub-Saturns/sub-Jovians, and super-Earths and sub-Neptunes, respectively. These divisions are highly inspired by the sizes of the Solar System's planets, but are also physically motivated, since there is evidence that these planets formed and/or evolved differently. In the following sections, we introduce these three populations by linking the main theories of planet formation and evolution to the observed planetary and stellar properties in different parameter spaces.

\subsection{Gas giant planets}
\label{sec:giant}

Gas giant planets ($R_{\rm p}$ $>$ 10 $\rm R_{\oplus}$) are known to be primarily composed of hydrogen ($\rm H_{2}$) and helium (He), similar to the widely studied gas giant planets in our Solar System, Jupiter and Saturn. Their formation and evolution are, however, an open question in exoplanet science.

\subsubsection{Formation mechanisms}
\label{sec:formation}

The two most accepted theories for giant planet formation are the core accretion and gravitational instability theories. In brief, the core accretion theory states that proto-planetary disks first form rocky proto-planets, which, if they reach a certain mass threshold before the disk gas has dissipated, their gravitational attraction would produce a runaway accretion of $\rm H_{2}$ and He, significantly increasing the size and mass of the proto-planet in a few million years \citep[e.g.][]{1974Icar...22..416P,1996Icar..124...62P,2006ApJ...648..666R,2014prpl.conf..619C,2014prpl.conf..643H,2015ApJ...811...41L,2019ApJ...878...36L}. In contrast, the gravitational instability theory states that a fraction of the proto-planetary disk can be fragmented into clumps that directly lead to giant planets in a similar way that stars and brown dwarfs form \citep{1997Sci...276.1836B,2007prpl.conf..607D,2016ARA&A..54..271K}. From a theoretical perspective, both theories have been shown to have different issues. For example, on the one hand, the core accretion theory struggles to explain how meter-size bodies can coagulate into larger kilometre-size bodies since the former are expected to reach high velocities and thus collide and destroy each other \citep[e.g.][]{2007A&A...466..413O,2008ARA&A..46...21B,2016SSRv..205...41B,2018MNRAS.479.5272G,2019ApJ...874...60S}, or drift into the star \citep[e.g.][]{1972fpp..conf..211W,1976PThPh..56.1756A,1977MNRAS.180...57W,1986Icar...67..375N}. On the other hand, the disk instability theory requires cold and massive proto-planetary disks, on the order of 10$\%$ of the stellar mass, which is not expected for most disks \citep{2005ApJ...621L..69R,2007ApJ...659..705A}. 

\subsubsection*{Constraints from population studies}

\begin{figure}
    \centering
    \includegraphics[width=0.9\textwidth]{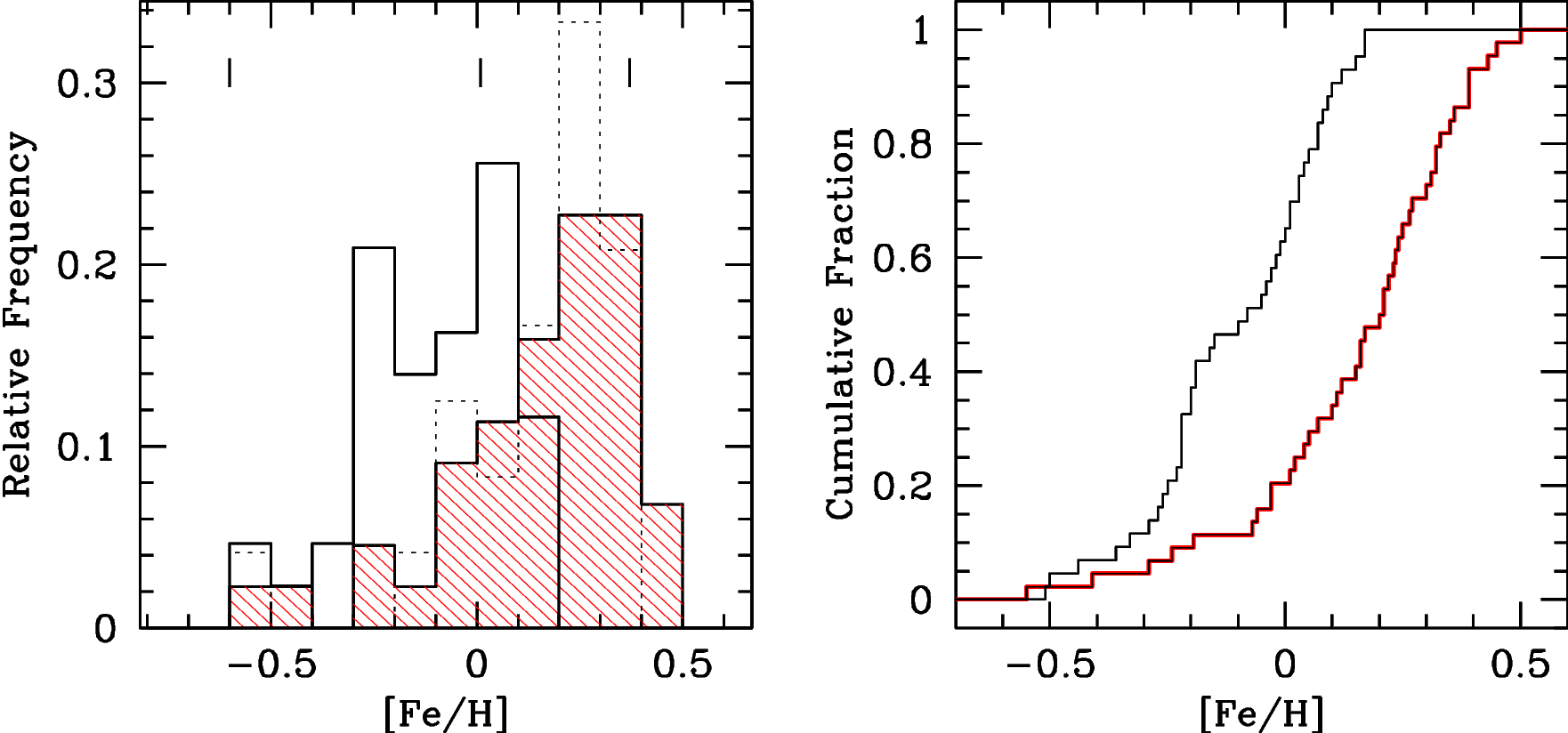}
    \caption[Metallicity distribution of field stars and stars hosting gas giant planets.]{Left: Metallicity distribution of stars hosting gas giant planets (shaded histogram) compared with the distribution of field stars. Right: Cumulative frequencies of both samples. Source: \citet{2001A&A...373.1019S}.}
    \label{fig:feh_ocurrence_santos}
\end{figure}

The properties of the observed gas giant population are, in general, considered to favour the core accretion theory. One of the main observational results supporting this theory is the positive correlation between the giant planet occurrence rate and the metallicity of their host stars first suggested by \citet{1997MNRAS.285..403G} and robustly confirmed by \citealp{2001A&A...373.1019S,2005ApJ...622.1102F} (see Fig.~\ref{fig:feh_ocurrence_santos}). The chemical compositions of the stars are considered to be a good proxy of the original chemistry of the protoplanetary disks from which they were formed, and the presence of a larger amount of solids is known to facilitate the assembling of massive planetary cores able to trigger the runaway gas accretion. Gas giant planets are also scarce around low-mass stars, which is attributed to the same phenomenon. However, the increasing number of giant planets found around M-dwarfs \citep[e.g.][]{1998ApJ...505L.147M,2019Sci...365.1441M,2021A&A...645A..16P,2022MNRAS.511...83G} suggest that other complementary scenarios can also be playing a role. 

Both the core accretion and disk instability theories suggest that giant planets can only form at large orbital distances, beyond the ice line, where, in the case of core accretion there is enough material in the protoplanetary disk to quickly form massive cores, and, in the case of gravitational instability, where the gas disk is cool enough to be fragmented into clumps. For example, \citet{2005ApJ...621L..69R} showed that close-in giant planet formation via gravitational instability is not possible, and, although core accretion could operate close to their stars \cite[e.g.][]{2014ApJ...797...95L,2016ApJ...829..114B}, different authors showed that this process would be very inefficient and unlikely \citep{2014ApJ...795L..15S,2016ApJ...817...90L,2017AREPS..45..359J}. This well-established theoretical prediction strongly contrasts with the observed distribution of gas giants. Exoplanetary gas giants have been detected in a wide range of orbital distances, showing orbital periods from a few days to thousands of days, and thus require additional mechanisms to explain how a fraction of the giant planet population arrived at their close-in locations.

\subsubsection{Orbital migration mechanisms}
\label{sec:migration}

Giant planets are thought to be able to migrate towards close-in orbits through two main mechanisms: disk-driven migration and high-eccentricity tidal migration. 

\begin{figure}
    \centering
    \includegraphics[width=0.38\textwidth]{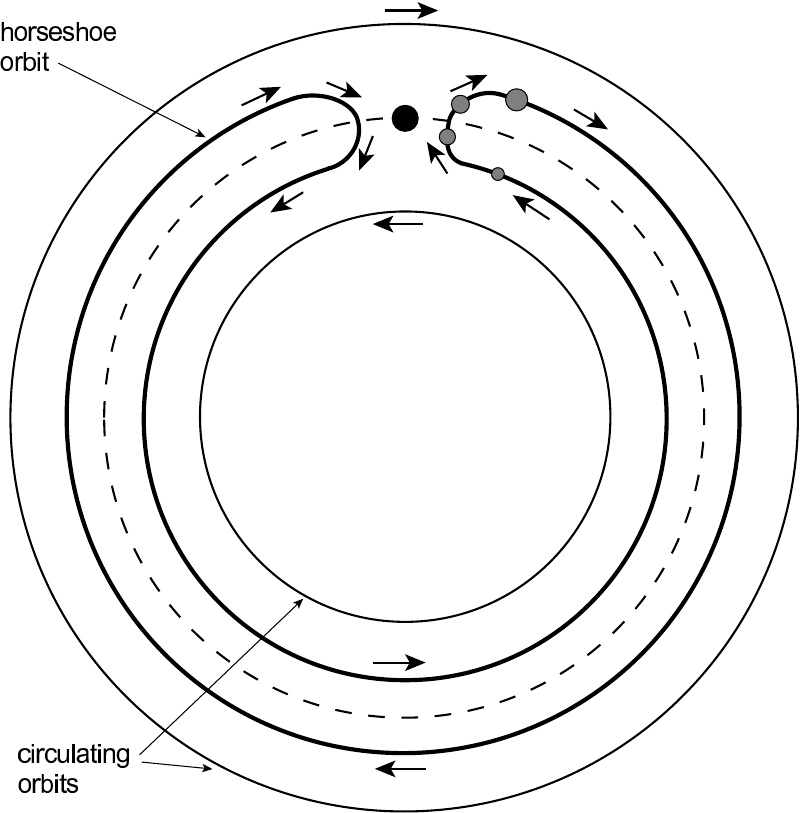}
    \includegraphics[width=0.48\textwidth]{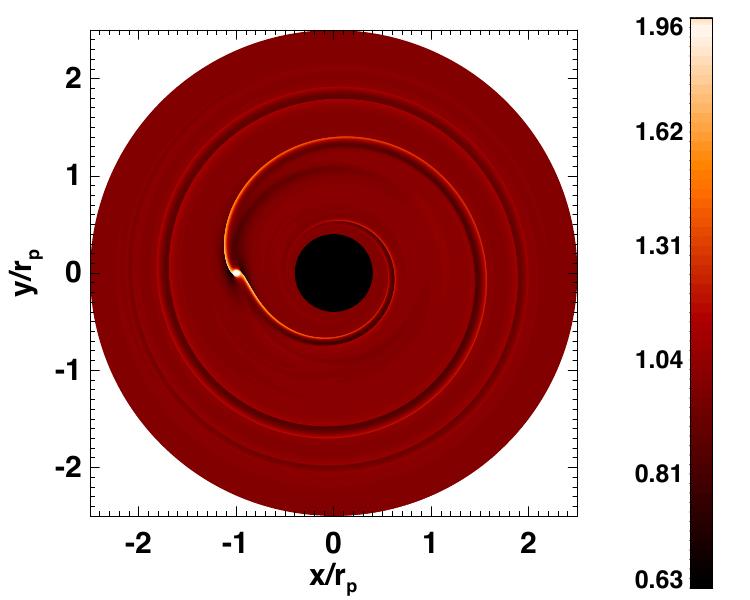}
    \caption[Horseshoe orbit and simulation of disk-driven migration.]{Left: Graphic representation showing a horseshoe orbit relative to a planet. Source: \citet{2007astro.ph..1485A}. Right: Disk-driven migration simulation of a planet embedded in a viscous disk. Source: \citet{2018haex.bookE.139N}.}
    \label{fig:disk-driven}
\end{figure}

In disk-driven migration, soon after the formation of the giant planet and before the dissipation of the gaseous proto-planetary disk, the planet-disk interactions can shrink the semimajor axis of a planet \citep{1979ApJ...233..857G,1986ApJ...309..846L,1996Natur.380..606L,2008ApJ...673..487I}. This is because the planet can exchange angular momentum with the disk through co-rotation torques, perturbing nearby gas onto horseshoe orbits (see Fig.~\ref{fig:disk-driven}, left) and deflecting more distant gas through Lindblad torques \citep[see][for comprehensive reviews]{2010apf..book.....A,2013apf..book.....A,2014prpl.conf..667B}, generating a planet migration as shown in Fig.~\ref{fig:disk-driven} (right). We note that the strength and sign of the co-rotation torques depend on the disk's turbulent viscosity, opacity, and radial entropy profile, so that they can drive the planet into an inward or outward migration \citep[e.g.][]{2006A&A...459L..17P,2015ApJ...812...94D}. 

There are important doubts on the impact of disk-driven migration on the origin of hot Jupiters because the magnitude and sign of the migration rate are highly sensitive to the conditions of the disk \citep[see][for a review]{2018ARA&A..56..175D}. In the case that the migration process is fast compared with the lifetime of the disk, giant planets could be subjected to tidal disruption or engulfment by their stars. In the literature, different mechanisms that could prevent the disappearance of giant planets and allow them to preserve close-in orbits in the long term have been proposed. Soon after the discovery of 51 Peg b, \citet{1996Natur.380..606L} suggested that an angular momentum transfer from the host star and the presence of a magneto-cavity in the inner disk could have halted its migration. In the former mechanism, a gas giant migrating inwards could preserve a close-in orbit and avoid infalling by extracting angular momentum from its host star through tides \citep{1998ApJ...500..428T}. In addition, the giant planet could lose some mass due to Roche lobe overflow, which can also contribute to preventing its orbital decay \citep[e.g.][]{2015ApJ...813..101V}. In the magneto-cavity mechanism, the magnetic field of the star could create a cavity in the innermost disk that would stop the planet when it reaches a 2:1 resonance with the edge of the inner disk \citep[e.g.][]{2008MNRAS.384.1242R,2010ApJ...708.1692C}. \citet{2002ApJ...574L..87K} also proposed that gas giants could stop at the gas sublimation radius, although \citet{2005ApJ...623..952E} found that this radius is too large when compared to the observed distribution. Overall, in case disk-driven migration plays a leading role in the hot Jupiter population, the magnetospheric cavity, mass loss, and tidal interactions are thought to be the main mechanisms in determining their final orbits \citep[e.g.][]{2010ApJ...708.1692C}.

\begin{figure}
    \centering
    \includegraphics[width=\textwidth]{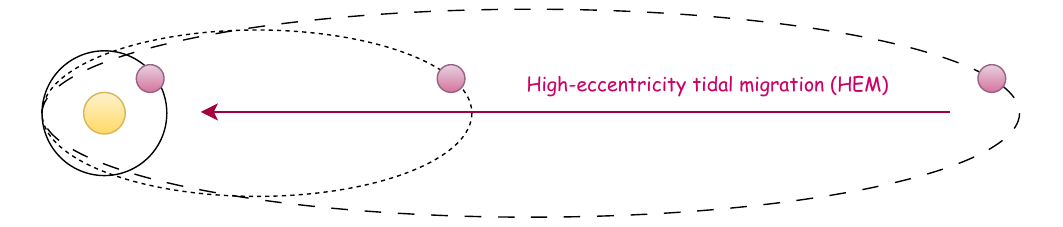}
    \caption[Energy dissipation process of a gas giant planet due to the tidal stellar forces.]{Graphic representation of the energy dissipation process of a gas giant planet due to the tidal forces exerted by the host star during the close encounters in the periapsis of the orbit. }
    \label{fig:HEM}
\end{figure}

High-eccentricity tidal migration (HEM) has also been proposed to explain the orbits of close-in giant planets. Contrary to the disk-driven migration scenario, where the planet-disk interactions generate an angular momentum and energy decay simultaneously, the HEM scenario provokes these decays in a two-step process, first reducing the orbital momentum of the planet and then decreasing its energy. In the first step, an external perturber extracts angular momentum from the planet, turning its orbital path into a highly elliptical orbit. In the second step, the planet progressively dissipates its energy through tidal interactions with its host star. In the periapsis of the orbit, the planet experiences close encounters with the star, which raises tides on the planet. These tides cause the gaseous planet to stretch, dissipating energy and changing the shape of the planet to adapt to the rapidly changing tidal potential. During this process, the orbit of the planet undergoes a circularisation process, reducing its semi-major axis to 
\begin{equation}
    a_{\rm final} = a_{\rm initial} \left(1-e^{2}\right),
    \label{eq:HEM}
\end{equation}
as schematized in Fig.~\ref{fig:HEM}. This means that if a planet initiates the energy dissipation process at 1 AU (5 AU) and circularises to $a = 0.05 \rm AU$, the original perturber must have raised the eccentricity of the gas giant to $e = 0.975 \,(0.995)$. In the literature, two main mechanisms have been proposed to be able to excite the eccentricity of giant planets to those large values: planet-planet scattering and secular interactions. 

The planet-planet scattering mechanism can occur when two or more planets in different orbits get close to each other. The differences between their angular velocities can modify their angular momentum, and “scatter” each other into eccentric orbits \citep[e.g.][]{1996Sci...274..954R,1996Natur.384..619W,2006ApJ...638L..45F,2008ApJ...686..580C,2008ApJ...678..498N,2011ApJ...742...72N,2012ApJ...751..119B,2017MNRAS.464.1709G}. This process can take place in tightly packed systems \citep[e.g.][]{2008ApJ...686..603J} or by close encounters with other stars \citep[i.e. stellar fly-bys;][]{2016ApJ...816...59S}. Planet-planet scattering by itself can alter the semi-major axes of the interacting planets. Even though it is highly unlikely that a close-in giant reaches its present-day orbit solely by scattering \citep[e.g. since the total energy of the system is conserved, a Jupiter-size planet would need to eject 100 planets of its mass at or beyond its original semi-major axis to reduce its semi-major axis by 100; see ][]{2018ARA&A..56..175D}, this process can generate a significant eccentricity increase throughout a series of multiple close encounters, potentially leading to a tidal energy dissipation, orbital circularization and contraction as described before.  We note that the eccentricity increase through this mechanism is limited to an epicyclic velocity that corresponds to the escape velocity from the planetary surface \citep[e.g.][]{2004ARA&A..42..549G,2013ApJ...775...42I,2014ApJ...786..101P}: $e_{\rm scatter}$ $\lesssim$ $\sqrt{2 G M_{\rm p} / R_{\rm p}}$ $(2 \pi a)^{-1}P_{\rm orb}$. If the planetary eccentricity reaches this value, the close planet-planet encounters would preferentially merge them rather than scatter them. However, for orbital separations of the order of one astronomical unit, the limiting eccentricity value exceeds 1, so that Jupiter planets can be easily scattered onto highly elliptical orbits. 

Secular interactions, contrary to the closely interacting bodies in the planet-planet scattering mechanism, consist of slow exchanges of angular momentum between widely separated planets. Depending on the mass and separations between the planets, this process can take over thousands or even millions of years. The angular momentum exchange has been proposed to occur either periodically \citep[e.g.][]{2015ApJ...805...75P} or chaotically \citep[e.g.][]{2011ApJ...735..109W,2017MNRAS.464..688H}, and both processes have been proven to be able to bring the eccentricity of the planet towards highly elliptical orbits. The latter mechanism is called secular chaos, and requires either three interacting planets or hierarchical two-planet systems with large eccentricities and inclinations \citep[see][]{2011ApJ...735..109W}. Periodic angular momentum exchange can occur via Kozai-Lidov cycles \citep{1962AJ.....67..591K,1962P&SS....9..719L,2016ARA&A..54..441N} triggered by a star \citep{2003ApJ...589..605W,2007ApJ...669.1298F,2011PhRvL.107r1101K,2012ApJ...754L..36N} or a planetary companion \citep{2011Natur.473..187N,2013ApJ...779..166T}. This mechanism has been extensively proposed to provoke the high eccentricity increase of a fraction of gas giants that would subsequently lead them to their present-day close-in orbits through tidal interactions, as discussed before.  

\begin{figure}
    \centering
    \includegraphics[width=\textwidth]{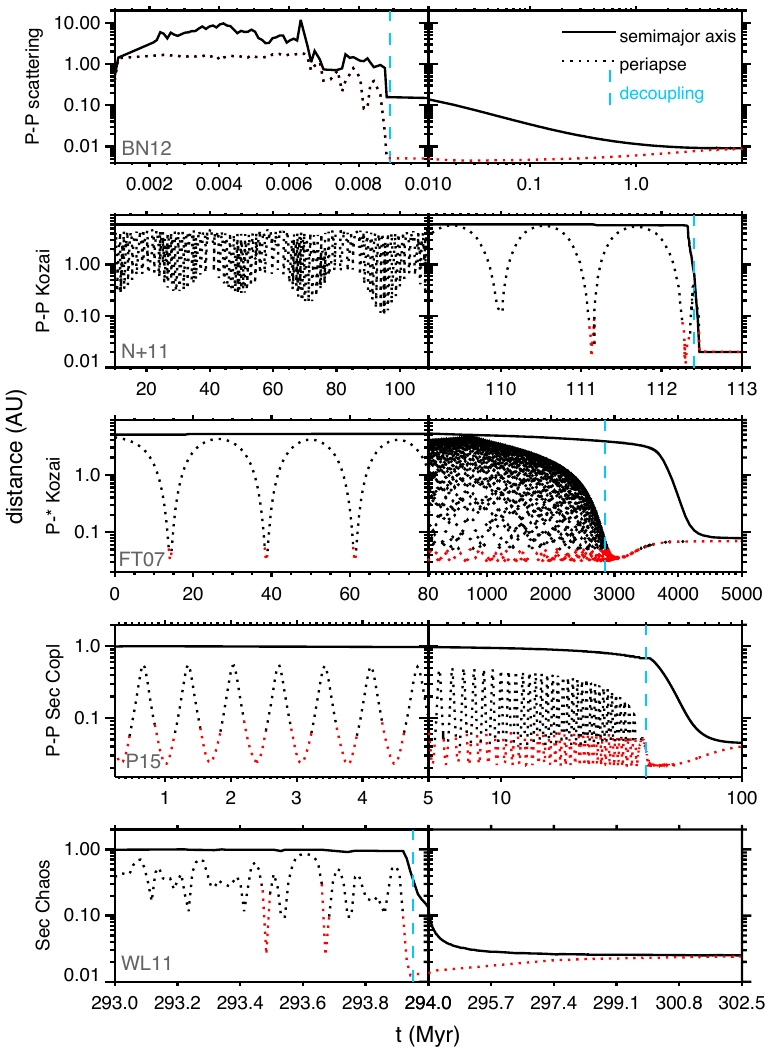}
    \caption[Dynamical histories of hot Jupiters arrived through different HEM processes.]{Simulated dynamical histories of hot Jupiters arrived at their close-in orbits through different HEM processes. First panel: planet-planet scattering (simulation data from \citealt{2012ApJ...751..119B}). Second panel: Kozai-Lidov cycles triggered by a planet (simulation data from \citealt{2011Natur.473..187N}). Third panel: Kozai-Lidov cycles triggered by a star (simulation data by \citealt{2007ApJ...669.1298F}). Fourth panel: planet-planet coplanar secular interactions (simulation data from \citealt{2015ApJ...805...75P}). Fifth panel: secular chaos (simulated data from \citealt{2011ApJ...735..109W}). The time scales depend on the particular mechanisms, the perturber properties, and the angular momentum of the migrated gas giant planet. Source: \citet{2018ARA&A..56..175D}.}
    \label{fig:HEM_processes}
\end{figure}

At the end of the HEM processes, the star-planet tidal interactions end up decoupling the migrated planet from the original perturbations that raised its eccentricity. In the case of planet-planet scattering, decoupling occurs when the semi-major axis has decreased enough to avoid subsequent close encounters with its perturber. In the case of secular interactions, decoupling occurs when the precession from general relativity or tides exceeds the precession caused by the secular perturber. In Fig.~\ref{fig:HEM_processes}, we show the simulated time evolution of the semi-major axis and periapsis of a planet subjected to planet-planet scattering (first panel) and secular interactions (panels 2-5). In all scenarios, the planet loses angular momentum and is perturbed to a small periapsis (dotted line), whose difference with the semi-major axis (solid line) reflects the induced orbital eccentricity. When the planet gets close enough to the star (red dotted line), it loses orbital energy, circularising its trajectory through tidal interactions and decreasing its semi-major axis. The blue vertical dashed line indicates the moment when the planet is decoupled from its perturber.

\subsubsection*{Constraints from population studies}

The migration mechanisms introduced above make different predictions on the distribution of the orbital properties of the migrated gas giant planets. In the following, we introduce these predictions and compare them with the current observational constraints. 

\begin{figure}
    \centering
    \includegraphics[width=0.89\textwidth]{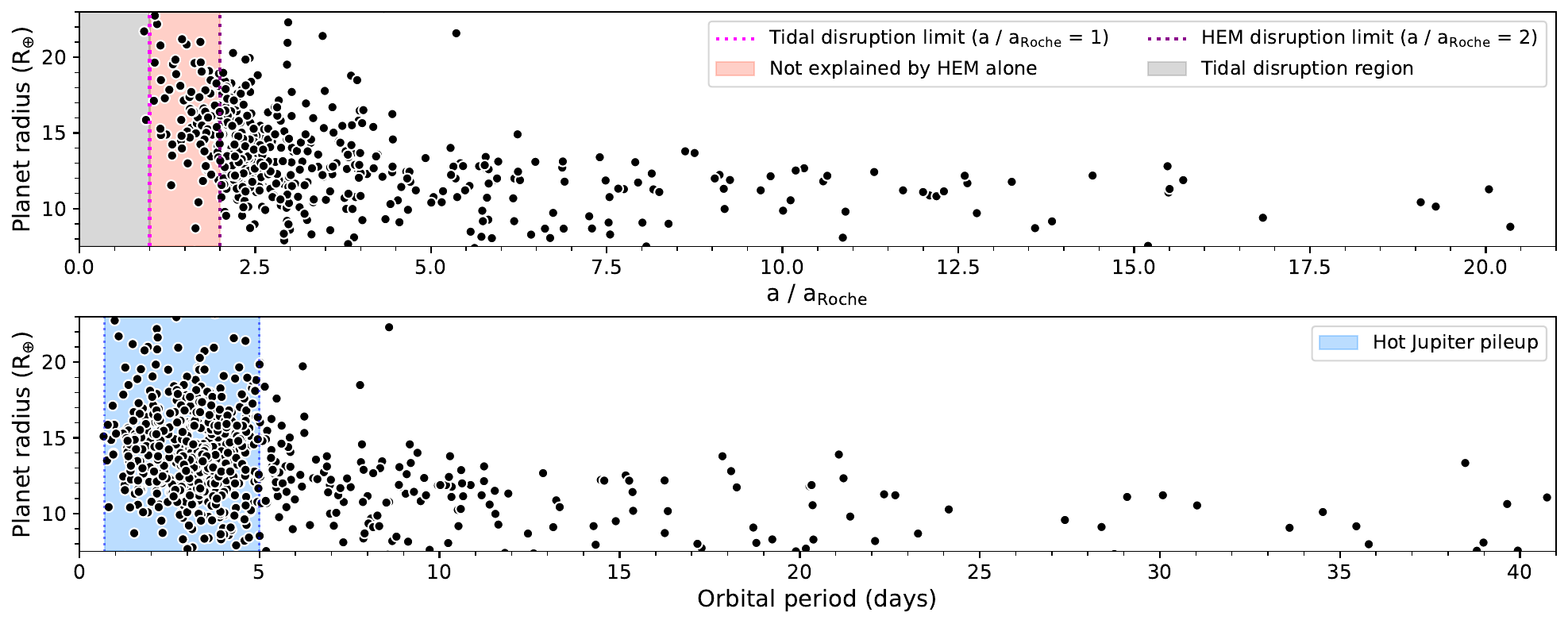}
    \caption[Orbital distribution of gas giant planets.]{Orbital distribution of gas giant planets. Upper panel: Planet radius as a function of semi-major axis ($a$) over $a_{\rm Roche}$. Lower panel: Planet radius as a function of orbital period. Planet and stellar data were acquired from the NEA.}
    \label{fig:a_dist_giant}
\end{figure}

\begin{figure}
    \centering
    \includegraphics[width=0.89\textwidth]{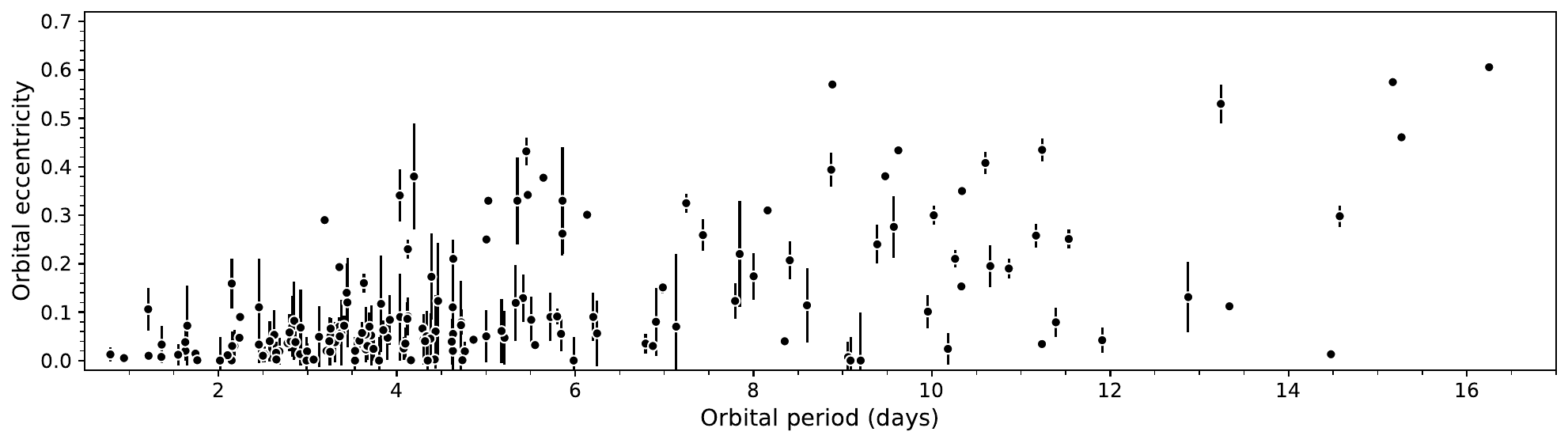}
    \caption[Eccentricity distribution of gas giant planets.]{Orbital eccentricity distribution of gas giant planets ($R_{\rm p}$ $>$ 10 $\rm R_{\oplus}$). Planet data was extracted from the NASA Exoplanet Archive \citep{2013PASP..125..989A}.}
    \label{fig:e_dist_giant}
\end{figure}

The inner edge of the distribution of semi-major axes of migrated gas giant planets is expected to be truncated by the tidal disruption limit (also known as the Roche limit). This limit corresponds to the closer-in orbits where a planet can be destroyed because the tidal forces are higher than its self-binding force. The Roche limit for a hot Jupiter can be written as $a_{\rm Roche} \simeq f_{\rm p} R_{\rm p} M_{\star}^{1/3} M_{\rm p}^{-1/3}$, where $f_{\rm p}$ is a dimensionless scale factor that depends on material properties of the body \citep[e.g.][]{2005Icar..175..248F}. Based on hydrodynamical simulations, \citet{2011ApJ...732...74G} found that $f_{\rm p} \geq 2.7$ for gas giant planets\footnote{Planets at $f_{\rm p}$ $\simeq$ 2.7 were not directly disrupted but destroyed after mass loss and re-accretion after several close encounters, since the re-accretion makes the planet puffier and hence easier to disrupt.}. In Fig.~\ref{fig:a_dist_giant} (upper panel), we show the semi-major axis distribution of currently known gas giants scaled to $a_{\rm Roche}$ assuming $f_{\rm p}$ = 2.7. In agreement with the theoretical prediction, the $a / a_{\rm Roche}$ $<$ 1 region is devoid of planets, with only two exceptions lying right next to the theoretical disruption threshold.

In Fig.~\ref{fig:a_dist_giant} (lower panel), we also show the orbital period distribution of gas giant planets, which shows a clear over-density of planets with orbital periods around 2-4 days, which was identified soon after the first discoveries and designated as the hot Jupiter pileup \citep{1999ApJ...526..890C,2003A&A...407..369U,2005ApJ...623..472G,2006ApJ...646..505B,2007ARA&A..45..397U,2008PASP..120..531C,2011ApJ...732L..24L,2016A&A...587A..64S}. Even though closer-in planets are easier to detect through the transit and RV techniques (see Sect.~\ref{fig:discovery_methods}), this pileup is well known to reflect a true feature of the exoplanet distribution, since it is detected after correcting for observational biases \citep[e.g.][]{2012ApJS..201...15H}. This pileup might reflect the outcome of a specific migration mechanism bringing hot Jupiters to this particular orbital range. However, neither disk-driven migration nor HEM processes are known to necessarily predict an over-density of planets in this region. On the one hand, HEM processes predict planets to migrate at or beyond 2$a_{\rm Roche}$ \citep[e.g.][]{1996Sci...274..954R,2010ApJ...725.1995M}. This is related to the tidal disruption limit. According to Eq.~\ref{eq:HEM}, when planets start to migrate at high eccentricities, their initial periapses are approximately half their final semimajor axes. Therefore, planets migrated through HEM at 2$a_{\rm Roche}$ must have reached the tidal disruption limit $a_{\rm Roche}$ during the migration process. On the other hand, the expected final semi-major axis of hot Jupiters migrated through the disk depends on the considered halting mechanism. For example, the magneto-cavity mechanism would drive the planet towards a 2:1 resonance with the edge of the inner disk, as discussed above. This implies orbital periods of $\sim$5 days, which is more consistent with the observed distribution \citep[e.g.][]{2021JGRE..12606629F}. We warn, however, that even though our current theoretical understanding of the migration mechanisms indicates that the present-day orbital distribution on HJs is better described by disk-driven migration, such distribution could be driven by subsequent tidal evolution, which could erase any semi-major imprint of its migration. For example, \citet{2014ApJ...793L...3V} find that, by considering certain tidal models and parameters, this evolution could subsequently shrink HJ's semi-major axes towards 1$-$2 $a_{\rm Roche}$. Therefore, our current knowledge does not allow us to discriminate between the two main migration mechanisms based on the semi-major axis distribution, and the reason why HJs tend to cluster at the pileup remains an open question. 

\begin{figure}
    \centering
    \includegraphics[width=\textwidth]{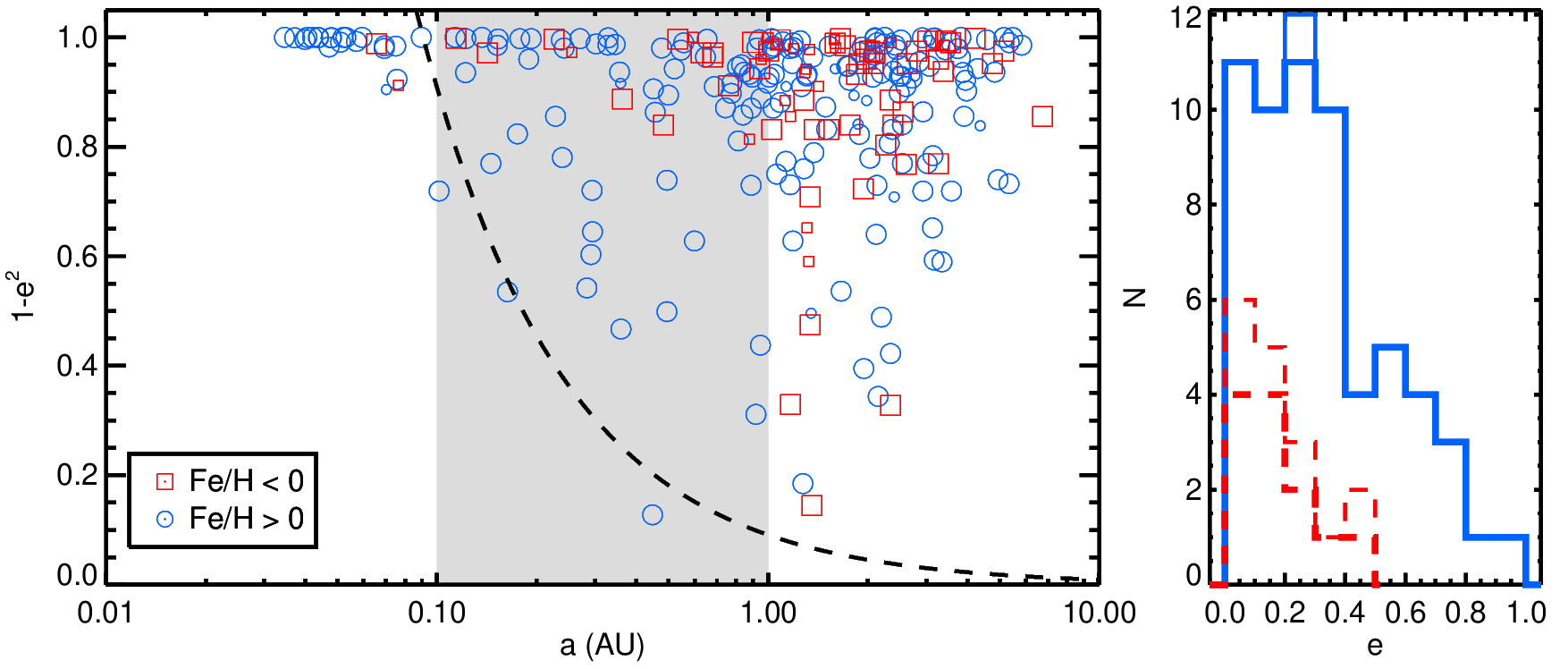}
    \caption[Eccentricities and semi-major axes of giant planets in stars with different metallicities.]{Left: Eccentricities and semi-major axes of giant planets colour coded according to the metallicities of the stellar hosts. In the close-in region where planets start to show elliptical orbits (grey shade), these moderately high eccentricities only appear around metal-rich stars. Conversely, giant planets with circular orbits are found around both metal-rich and metal-poor stars. Right: Eccentricity distribution of the grey-shaded planets orbiting metal-rich (blue solid) and metal-poor (red dashed) stars. Source: \citet{2013ApJ...767L..24D}.}
    \label{fig:ecc-met-trend}
\end{figure}

The present-day orbital eccentricities are considered to be excellent tracers of the origins of HJs, since the migration mechanisms introduced above are expected to act differently in this parameter space. In Fig.~\ref{fig:e_dist_giant}, we plot the eccentricity-period distribution of close-in gas giant planets. In the shortest-period orbits, most HJs have circular orbits. This is commonly considered to be an imprint of the tidal circularization process, which occurs faster at at shorter orbital distances \citep[e.g.][]{1980A&A....92..167H,1998ApJ...499..853E,2011CeMDA.111..105C}, and so it is highly unlikely to observe these planets in elliptical orbits. Gas giants in more distant regions show, however, moderately elliptical orbits (0.2 $<$ $e$ $<$ 0.6). In the disk-driven migration scenario, the interactions between the planet and protoplanetary disk are predicted not to significantly excite the planetary eccentricities \citep[e.g.][]{2013MNRAS.428.3072D,2015ApJ...812...94D}. Also, while these excitations could be produced by an outer planet through secular interactions, the perturber would need to be massive and/or nearby to overcome precession from general relativity and tides \citep[e.g.][]{2010exop.book..217F}. From an observational perspective, such companions have been ruled out for many moderately eccentric HJs \citep[see the discussion in][]{2018ARA&A..56..175D}. In contrast, HEM processes provide a natural explanation for HJs in moderately eccentric orbits, which would still be in the process of tidal circularisation. This scenario is reinforced by the results in \citet{2017A&A...602A.107B}, who found that less massive planets are circularised at larger distances, as predicted by the larger tidal circularisation time scale \citep[e.g.][]{2011MNRAS.414.1278P}. The reason why we find both eccentric and circular orbits at larger distances can be explained in two different ways. On the one hand, HEM could be the main mechanism populating this region of the parameter space, and the eccentric planets could have arrived more recently, so that they did not have time to completely circularise. On the other hand, both HEM and disk-driven migration could be playing different roles in populating this region. Also in favour of this scenario, eccentric HJs orbit metal-rich stars while circular HJs orbit both metal-rich and metal-poor stars \citep[][see Fig.~\ref{fig:ecc-met-trend}]{2013ApJ...767L..24D}, suggesting that HEM was triggered by planet-planet gravitational interactions in a fraction of systems able to form two massive bodies, as their occurrence is correlated with the host star metallically (see Sect.~\ref{sec:formation} and Fig.~\ref{fig:feh_ocurrence_santos}). Overall, the existence of moderately eccentric HJs provides evidence that at least a fraction of these planets arrived at their present-day locations through HEM processes.

The present-day angle between the stellar spin-axis and the normal to the orbital plane (commonly known as the stellar obliquity)\footnote{The stellar obliquity is typically measured through the Rossiter-McLaughlin effect \citep[e.g.][]{1924ApJ....60...15R,1924ApJ....60...22M,2018haex.bookE...2T,2021A&A...654A.152B}, even though different other techniques have been also proposed and successfully implemented \citep[e.g.][]{2009ApJ...705..683B,2010MNRAS.407..507C,2010ApJ...719..602S,2011ApJ...743...61S,2011ApJ...733..127S,2013Sci...342..331H,2015ApJ...801....3M,2016ApJ...818....5L}.} has been long thought to be the best tracer of the migration histories of close-in gas giants. Planets that underwent an early migration through the disk are expected to maintain aligned orbits \citep[e.g.][]{2013A&A...555A.124B,2020A&A...643A..25P,2020ApJ...892L..21Z,2020AJ....160..179M} or moderately misaligned orbits \citep[e.g.][]{2010MNRAS.401.1505B,2011MNRAS.412.2790L,2011MNRAS.417.1817T,2015MNRAS.450.3306F,2012Natur.491..418B,2012ApJ...758L...6R}, while planets that underwent HEM processes would lead to different obliquity distributions depending on the specific process \citep[e.g.][]{2009ApJ...696.1230F,2011ApJ...729..138M,2012ApJ...754L..36N,2012ApJ...757...18A,2017AJ....154..106N}. In theory, contrary to the orbital eccentricities discussed above, stellar obliquities are insensitive to tidal evolution, but observational data shows evidence of correlations with star's tidal properties, suggesting a possible tidal realignment \citep[e.g.][]{2010ApJ...719..602S,2010ApJ...718L.145W,2012ApJ...757...18A} that leaves open an important theoretical problem \citep[e.g.][]{2012MNRAS.423..486L,2015A&A...574A..39D,2017MNRAS.468.1387L}. To date, many HJs are well aligned with their host stars' spin, while others show highly misaligned orbits \citep[e.g.][]{2012ApJ...757...18A}, which, similarly to the observed eccentric HJs, is interpreted as HEM processes having a relevant role in the migration of at least a fraction of these planets. 

Additional non-orbital properties, such as the ages of the planetary systems, have also been tested as tracers of the migration histories of HJs. In principle,  HJs that arrived through the disk could be observed in younger stars than those that arrived through HEM, since the latter processes can occur at any time of a planet's lifetime. However, HEM can also occur soon after the planet's formation, and stellar ages are typically difficult to measure. Despite the difficulties, the recent discoveries of HJs around young T Tauri stars, which still have their gas disk, support that these planets could have arrived recently through the disk \citep[e.g.][]{2016Natur.534..662D,2017MNRAS.465.3343D,2016ApJ...826..206J,2017MNRAS.467.1342Y}. Another promising niche to explore is the different HJ occurrence between young clusters and field stars. Still, current studies did not reach conclusive evidence because of the intrinsic difficulties of the age measurements \citep[see][]{2012MNRAS.427.1587C,2012ApJ...756L..33Q,2013MNRAS.433..867H,2016ApJ...816...59S}.

The presence or absence of planetary and stellar companions in planetary systems with HJs can also provide relevant clues on their migration channels. Disk-driven migration is expected to deliver HJs with resonant companions \citep[e.g.][]{1993Natur.365..819M,2002ApJ...567..596L,2006Sci...313.1413R}. Pairs of small planets have the potential of escaping from the resonance, but giant planets are most likely to preserve it \citep{2014AJ....147...32G}. On the contrary, HEM is expected to eject any small planet situated within the orbit of the migrating gas giant \citep{2015ApJ...808...14M} but requires an outer stellar or giant planet companion to initiate the HEM process, as discussed above. Interestingly, outer massive planets in systems hosting HJs are abundant. \citet{2014ApJ...785..126K} and \citet{2016ApJ...821...89B} found that $\sim$70$\%$ of HJs have an outer companion between 1 and 20 AU and 1-13 $\rm M_{\rm J}$, which are known to be capable of having raised the eccentricities of the present-day HJs presumably formed farther out, and are also known to be compatible with the observed population of eccentric HJs introduced above. Planetary systems with HJs have been also found in binary systems. However, the properties of the stellar companions are typically incompatible with HEM processes. \citet{2016ApJ...827....8N} found that less than $\sim$15$\%$ of HJs with stellar companions would be able to trigger HEM through Kozai-Lidov cycles since most stellar companions are not massive or nearby enough to overcome relativistic and tidal precession. Overall, the abundance of outer massive planetary companions to HJs supports the HEM hypothesis. In addition, HJs have been found to not typically have nearby planetary companions \citep[e.g.][]{2011ApJ...732L..24L,2012PNAS..109.7982S,2016ApJ...825...98H}, which, as mentioned before, is also considered to support the HEM hypothesis, even though there have been found a few examples of nearby companions \citep[e.g.][]{2015ApJ...812L..18B,2016ApJ...823L...7M}, and it has also been shown that this absence do not necessarily rule out disk-driven migration \citep[e.g. see][]{2013ApJ...778L...9O,2014ApJ...787..172O,2016ApJ...825...62S}.

In summary, despite the large number of theoretical and observational efforts devoted to understanding the migration processes that bring HJs to their close-in locations, there is no absolute evidence on the predominant migration mechanism. Different observational constraints, such as the existence of a large fraction of eccentric and misaligned HJs, the absence of nearby planetary companions, and the prevalence of outer massive companions, suggest that HEM processes have a relevant role in the observed population. In contrast, other constraints such as the detection of HJs around T Tauri stars, the rare cases of HJs with nearby companions, or the fraction of HJs observed within 2$a_{\rm Roche}$ are not easy to account for based on HEM, favouring the disk-driven hypothesis instead. Therefore, regardless of whether a predominant migration mechanism could exist, the current consensus is that both HEM and disk-driven migration play non-negligible roles in the observed HJ distribution, whose relative relevance needs to be further studied through additional theoretical and observational efforts. 

\subsubsection{The new challenges of migrated hot Jupiters}
\label{sec:inflation_mspi}

In the previous section, we described how HJs, by simply residing in very short orbits, drastically changed our paradigm about planetary systems. The paradigm shift, however, did not stop there, and had to quickly evolve to explain new phenomena not observed in the Solar System. Some of these phenomena are now relatively well understood, but others require significant theoretical and observational efforts to reach a comprehensive understanding. 

\begin{figure}
    \centering
    \includegraphics[width=0.9\textwidth]{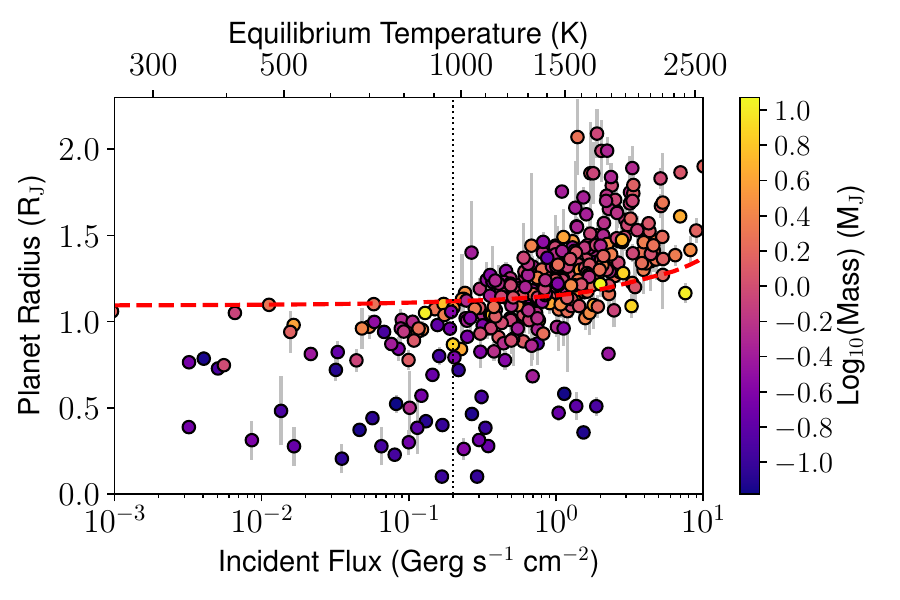}
    \caption[Radius as a function of the incident stellar flux of gas giant planets.]{Radius as a function of the incident stellar flux of gas giant planets. The dotted vertical line is the empirical flux cut-off for inflation. The red dashed line shows an evolutionary model for a Jupiter-mass planet without additional inflation effects. Source: \citet{2018AJ....155..214T}.}
    \label{fig:radius_inflation}
\end{figure}

Several HJs have been found with anomalously large radii that internal structure models cannot explain. These radii require additional heat sources such as tidal heating during HEM \citep[e.g.][]{2001ApJ...548..466B,2003ApJ...588..509G,2009ApJ...700.1921I,2008ApJ...678.1396J,2008ApJ...681.1631J,2010A&A...516A..64L}, thermal tides caused by stellar irradiation \citep[e.g.][]{2010ApJ...714....1A,2013arXiv1304.4121S}, or deposition of stellar energy into the interior \citep[e.g.][]{2002A&A...385..156G,2010ApJ...714L.238B,2010ApJ...721.1113Y,2017ApJ...841...30T}. Interestingly, radius inflation due to intense stellar irradiation was originally predicted by \citet{1996ApJ...459L..35G}, and to date is the leading theory to explain this phenomenon. This is because Jupiter-mass planets with equilibrium temperatures below 1000 K (commonly known as warm Jupiters) are not observed to be inflated \citep[e.g.][]{2011ApJS..197...12D,2011ApJ...729L...7L,2011ApJ...736L..29M}, while the radii of hotter gas giants seem to scale with their equilibrium temperatures \citep[e.g.][]{2013ApJ...768...14W,2018AJ....155..214T}, as we can see in Fig.~\ref{fig:radius_inflation}. In line with this trend, \citet{2016AJ....152..185G,2017AJ....154..254G,2019AJ....158..227G} found that warm Jupiters can reinflate as the host stars evolve off the main sequence and their equilibrium temperatures surpass the 1000 K threshold. Figure~\ref{fig:radius_inflation} also shows an evolution model for a 4.5 Gyr Jupiter-mass planet subjected to different irradiation levels (red dashed line). Many Jupiters hotter than 1000 K lie well above this curve, a phenomenon often termed \textbf{as} the radius anomaly. Overall, even though a radius expansion due to stellar heating was predicted early on, the magnitude of this expansion has been observed to be several times larger than expected, a feature that is not yet well understood \citep[e.g.][]{2013ApJ...763...13W,2016ApJ...819..116G,2021A&A...645A..79S}.

Given their large masses and close-in orbits, HJs have also been found to show different signs of interaction with their host stars. \citet{2000ApJ...533L.151C} first suggested that massive close-in planets could increase and modulate their host star's activity levels through tidal and magnetic interactions, similar to what was previously observed in the comparable cases of RS CVn binary systems \citep[e.g.][]{1996IAUS..176...45P}. Pioneer works by \citet{2003ApJ...597.1092S,2005ApJ...622.1075S} found the first pieces of evidence of periodic chromospheric activity (through Ca II H $\&$ K variability) modulated by the orbital motion of close-in giant planets. In the HD~179949 planetary system, the authors found a two-component activity signal that reflects the rotation period of the star \citep[of about 7 days;][]{2012MNRAS.423.1006F} and the orbital period of its orbiting planet HD~179949~b \citep[3.1 days;][]{2001ApJ...551..507T}, being the later interpreted as the result of magnetic star-planet interactions (MSPIs). After the success achieved by the Ca II K line variability, subsequent MSPI detections through broad band photometry \citep[e.g.][]{2008A&A...482..691W,2009EM&P..105..373P} and X-ray observations \citep[e.g.][]{2011ApJ...741L..18P,2015ApJ...805...52P,2013A&A...552A...7S,2015ApJ...811L...2M} were reported. Interestingly, planet-induced variability is often found not to be constant over time. This was first noted by \citet{2003A&A...406..373S}, who detected photometric activity synchronised with the planetary orbit of HD~192263~b, showing alternation between variable and stable moments. \citet{2008ApJ...676..628S} confirmed this phenomenon and designated it as the `on/off' nature of MSPIs. The authors state that this recurrent behaviour is likely related to the changing stellar magnetic field structure throughout its activity cycle. 

\begin{figure}
    \centering
    \includegraphics[width=0.49\textwidth]{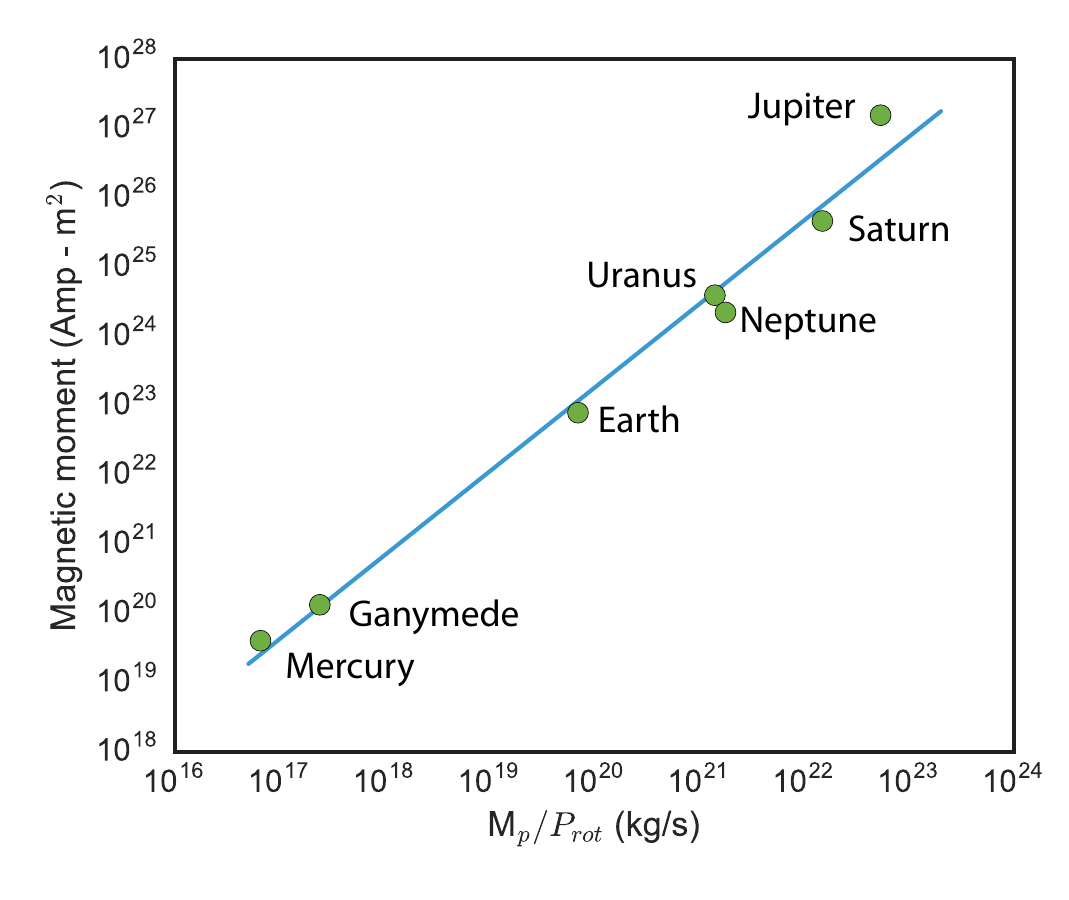}
    \includegraphics[width=0.473\textwidth]{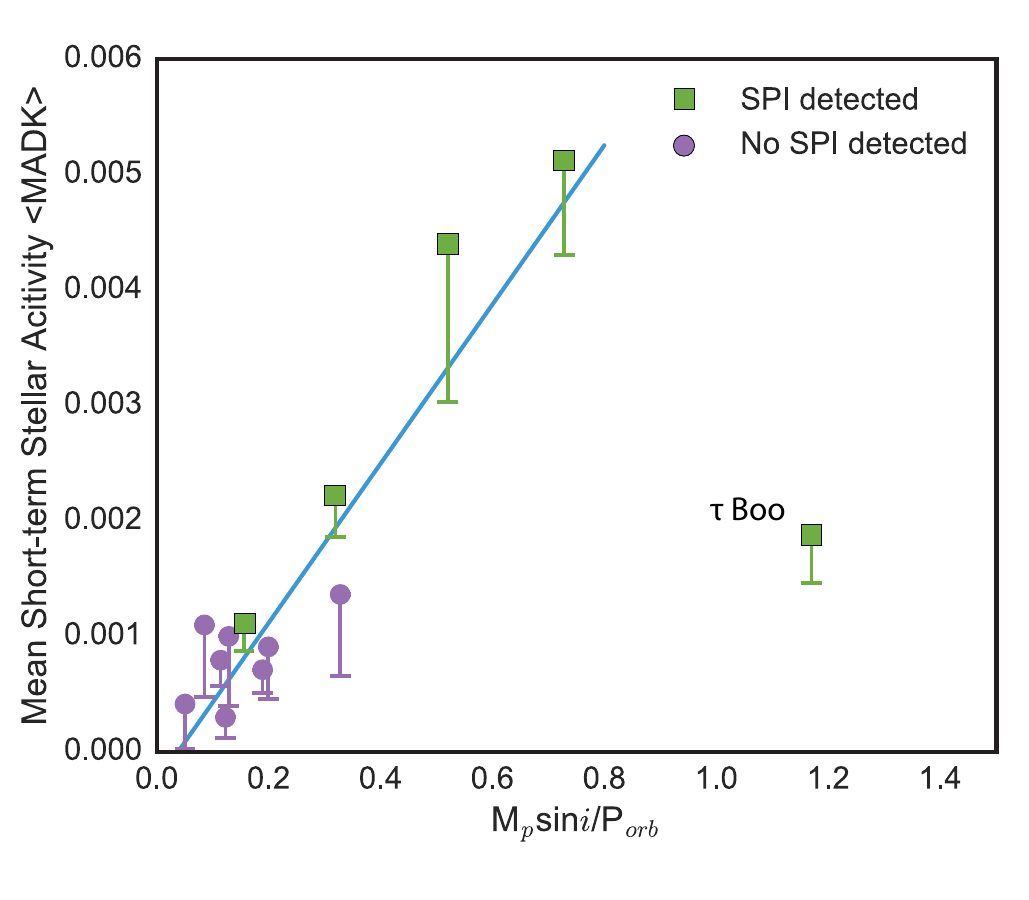}
    \caption[Chromospheric activity as a function of the ratio of planet mass to rotation period.]{Left: Measured magnetic moments for the six magnetised Solar System planets plus Ganymede as a function of the ratio of body mass to rotation period. Right: Mean Ca II K chromospheric activity of different planet-hosting stars with evidence (green squares) and no signs (purple circles) of MSPIs against $M_{\rm p} \, \textrm{sin} \, i / P_{\rm orb}$. We note that $P_{\rm orb}$ = $P_{\rm rot}$ under the assumption of tidally locked planets. Source: \citet{2018haex.bookE..20S}.}
    \label{fig:magmom}
\end{figure}

The increasing number of MSPI detections enabled population studies and comparisons with the Solar System planets, providing relevant insights into the magnetic fields triggering the interactions. For example, the magnetic moments of different Solar System bodies are known to be correlated with the ratio between their masses and rotational periods, as shown in Fig.~\ref{fig:magmom} (left panel). Similarly, \citet{2008ApJ...676..628S} computed the average of the Mean Absolute Deviation of the Ca II K line variability $\rm <MADK>$ of different targets with detections and non-detections of MSPIs, and found a similar correlation with $M_{\rm p} \, \textrm{sin} \, i / P_{\rm orb}$ (see Fig.~\ref{fig:magmom}, right panel). We note that due to the strong tidal forces, close-in gas giants are expected to be tidally locked with their host stars \citep{1981A&A....99..126H,2018AJ....155..118W}, that is, $P_{\rm orb}$ = $P_{\rm rot}$. Therefore, the results shown in Fig.~\ref{fig:magmom} suggest that the strength of the planet-induced stellar activity correlates with the magnetic moment of the planet. We note that the planetary system $\tau$ Boo significantly deviates from the general trend. $\tau$ Boo is known to be one of the few cases of a tidally locked star (i.e. it rotates with the same periodicity as the planetary orbit), which has been extensively discussed to weaken the power of MSPIs given that it is expected to scale with the relative velocity between the two interacting bodies \citep[e.g.][]{2008A&A...482..691W,2009A&A...505..339L,2012A&A...544A..23L,2013MNRAS.435.1451F}.

In addition to phase-resolved techniques, statistical studies based on single-epoch observations have also been used to gain insight into MSPIs. The ultimate goal of these studies is to search for a significant difference in the activity levels of stars with and without close-in giant planets. \citet{2010ApJ...717L.138H} measured the log $R'_{HK}$ activity indicator for 23 planetary systems with HJs and found a correlation with the planet surface gravities and a tentative correlation with the planet mass. In line with these results, \citet{2012A&A...540A..82K} confirmed that both the log $R'_{HK}$ and the equivalent width of the Ca II K line correlate with the planetary mass and also with the inverse of the planetary semi-major axis, as would be expected for magnetic and tidal SPIs. \citet{2010A&A...515A..98P} reported a correlation between the stellar X-ray luminosity and the ratio $M_{\rm p} \, \textrm{sin} \, i/a$, suggesting that massive close-in planets tend to orbit more X-ray luminous stars. The authors warn, however, that this trend could also be explained by observational biases of the RV method since it favours the detection of less massive and further-out planets around less-active X-ray faint stars.  Similar tentative correlations between $M_{\rm p} / a$ and high-energy radiation were found by \citet{2013ApJ...766....9S} and \citet{2016ApJ...820...89F}. All these correlations found in single-epoch observations of large planet samples are promising indicators that MSPIs are common and detectable. However, a careful consideration of possible observational biases is imperative to reach reliable conclusions. In addition, there might be other effects provoked by the presence of planets that could affect the stellar activity levels through other channels.  Massive planets can alter the angular momentum evolution of their stars, which might increase the stellar rotation through tidal spin-up, or, alternatively, decrease the efficiency of the stellar magnetic braking \citep[e.g.][]{2010A&A...520A..53L,2011ApJ...733...67C,2022MNRAS.513.4380I}. In both cases, the star would be more active than expected. Therefore, to actually disentangle the possible causes of the observed increased activity of HJ host, it is necessary to monitor the stellar activity along planetary orbit and stellar rotation period, as it was first done at a population level in \citet{2008ApJ...676..628S}.

\begin{figure}
    \centering
    \includegraphics[width=0.45\textwidth]{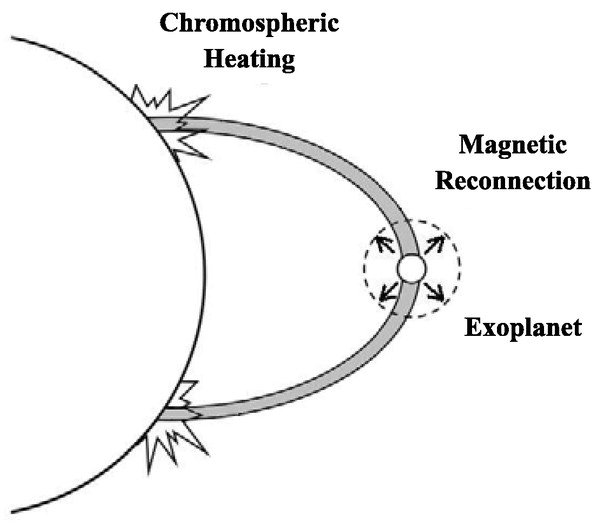}
    \includegraphics[width=0.45\textwidth]{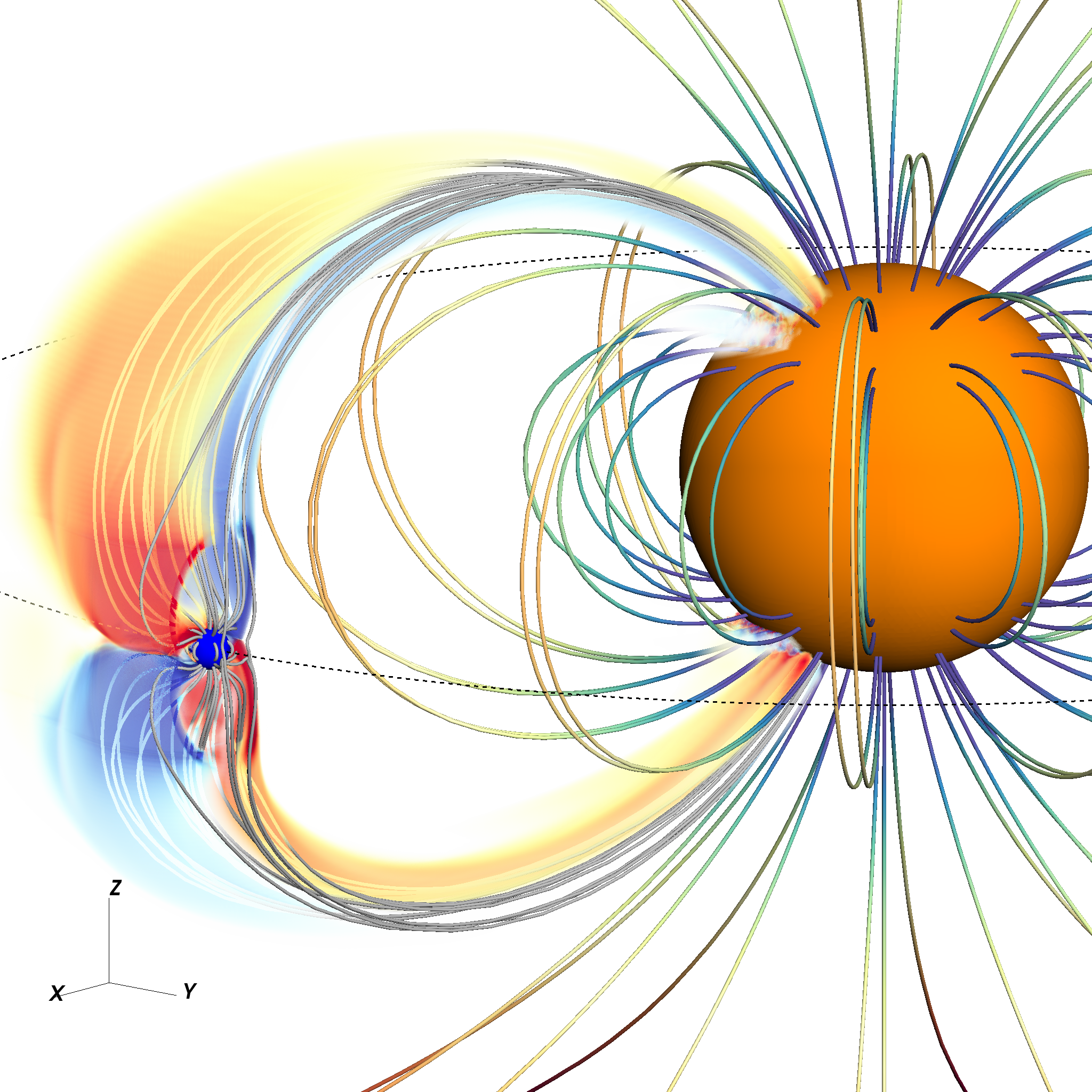}
    \caption[Energy channelling due to MSPIs and a MSPI model.]{Left: 2D representation of the energy channelling due to MSPIs. Source: \citet{2004ApJ...602L..53I}. Right: 3D representation of an MSPI model showing the Alfvén wings connecting a star-planet system. The stellar magnetic field lines are colour-coded by the strength of the magnetic field, and the planetary field is shown by the grey lines. Source: \citet{2015ApJ...815..111S}.}
    \label{fig:mspi}
\end{figure}

Both the phase-resolved and single-epoch results opened a new field of research and stimulated the development of theoretical models to explain the observations. MSPIs can affect the emission, heating, atmospheric escape, and migration of the planet, and can generate magneto-hydrodynamic shocks and channelling of magnetic energy \citep[e.g.][]{2018haex.bookE..25S}. The latter effect consists on the existence of an energy flux in the form of two wings (often referred to as Alfvén wings), which, depending on the relative motion between the ambient plasma and the orbiting planet and the interplanetary magnetic field can connect onto the star as schematized in Fig.~\ref{fig:mspi}. These wings would create a local activity enhancement at the stellar impact points, which would be displaced as the planet orbits its star, generating the observed phased modulations mentioned above. Overall, the formalisms of the MSPIs models are complex and have undergone a significant progress over the last years, but still have many challenges to face \citep[e.g.][]{2009A&A...505..339L,2012A&A...544A..23L,2013A&A...552A.119S,2018JGRA..123.9560S,2015ApJ...815..111S,2018haex.bookE..25S,2023spi..conf....1S}. Current and future efforts to develop improved models will allow to provide key insights for the upcoming MSPI observations of close-in planetary systems. In chapters~\ref{ch:mspis} and \ref{ch:toi_5005}, we delve into the MSPI field through the detection and study of two synchronised photometric signals within the HD 118203 and TOI-5005 planetary systems, respectively.

\subsection{Intermediate-size planets}
\label{sec:intermediate}

\begin{figure}
    \centering
    \includegraphics[width=\textwidth]{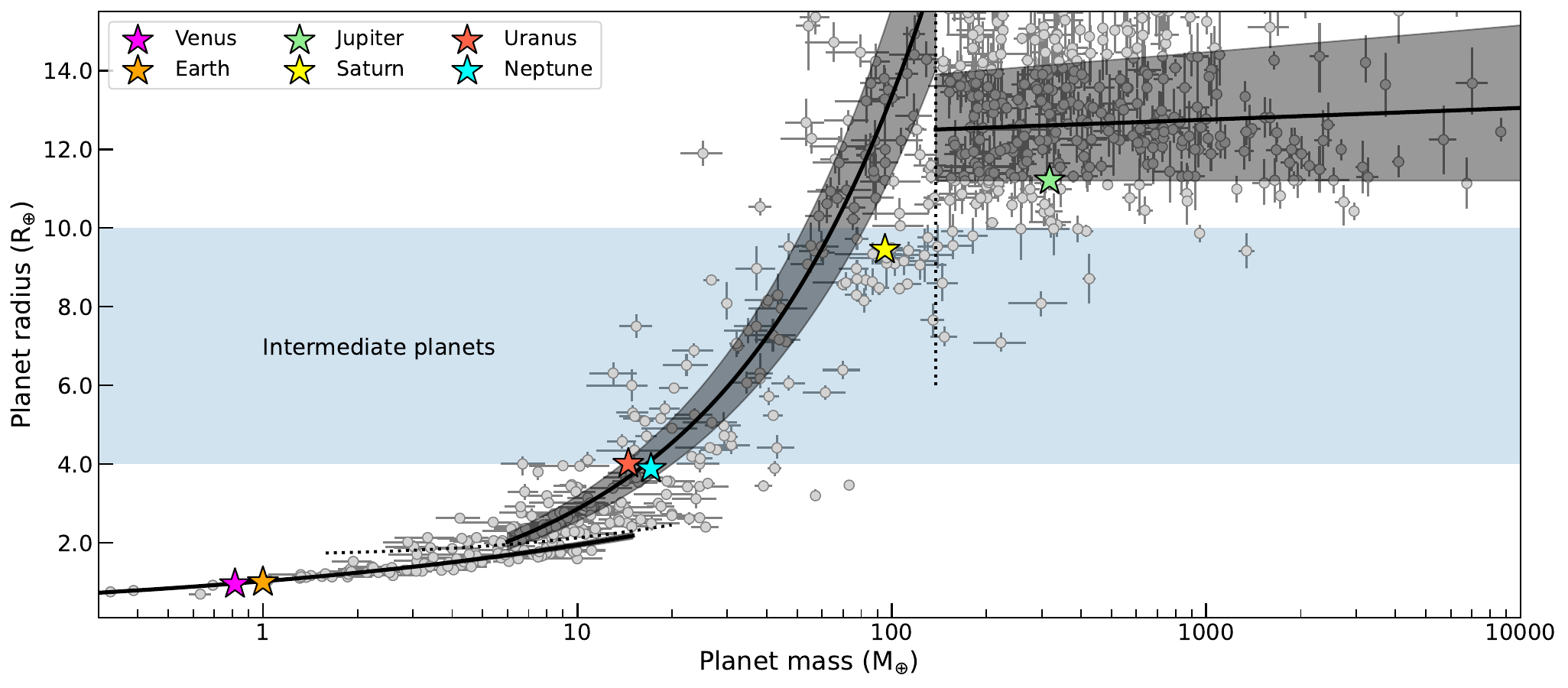}
    \caption[Mass-radius diagram of known exoplanets.]{Mass-radius diagram of known exoplanets with masses and radii constrained to a precision better than 25$\%$ and 8$\%$, respectively. Planet data was obtained from the PlanetS catalogue \citep{2020A&A...634A..43O,2024A&A...688A..59P}. Solar System planets are highlighted with coloured stars. The solid black lines and grey shades correspond to the empirical mass-radius relationships and associated 1$\sigma$ uncertainties from \citet{2024A&A...688A..59P}. This diagram was made through the \texttt{mr-plotter} package (\url{https://github.com/castro-gzlz/mr-plotter}).}
    \label{fig:mr_intermediate}
\end{figure}

Intermediate-size planets (4 $\rm R_{\rm \oplus}$ $<$ $R_{\rm p}$ $<$ 10 $\rm R_{\oplus}$) are very scarce in our current sample of known planets and are by far the least studied planet population. This is mainly because of two factors: exponential improvement of instrumental precision and a strong interest in smaller planets. With the exception of a few instruments (e.g. HARPS), an important fraction of intermediate-size planets were out of reach from the ground during the first decade of exoplanet exploration. The launch of \textit{Kepler} in 2009 allowed, for the first time, large-scale detections of planets smaller than Jupiter and Saturn, with an outstanding precision improvement that allowed detecting planets down to the sub-Neptune domain, where most resources were invested. Intermediate-size planets are thus the `great unknowns' of the exoplanet plethora. While poorly explored, these planets occupy a very wide range of radii and masses with no analogue in the Solar System, making them relevant targets to probe planet formation and evolution theories. In Fig.~\ref{fig:mr_intermediate}, we show the mass-radius diagram of precisely characterized exoplanets together with the Solar System planets, where we can appreciate the scarcity of intermediate-size planets when compared to the giant and small planets populations as as well as their large radius and mass span with no representation in the Solar System. 

Intermediate-size planets are believed to have undergone similar formation and evolution processes to those of gas giant planets (see Sect.~\ref{sec:giant} for further details and discussions). However, the reason for their smaller size is not yet well understood. These planets might once have been gas giants that lost a fraction of their atmospheres, or they might never have reached a Jupiter-size wingspan. Thus, whether their present-day properties are primarily driven by formation or evolution processes is a major question in exoplanet research. Some theoretical works have proposed that gas giant planets receiving large irradiations from their host stars could be eroded into smaller planets \citep[e.g.][]{2014ApJ...783...54K,2019A&A...624A.101L}. However, most theoretical works and observational constraints indicate that these planets, even under very high levels of irradiation, are too massive to significantly decrease their sizes through evaporation \citep[e.g.][]{2004A&A...418L...1L,2008ASPC..398..275L,2009ApJ...693...23M}. In consequence, the most accepted view is that intermediate-size planets do not originate from gas giants. While different values have been proposed, the most accepted mass threshold from which a protoplanetary rocky core can trigger runaway gas accretion (see Sect.~\ref{sec:formation}) is 10 $\rm M_{\oplus}$ \citep[e.g.][]{2006ApJ...648..666R,2015ApJ...811...41L,2019ApJ...878...36L}. Once this mass is reached, the accretion process is expected to be very effective and quick. This value is, however, well below the expected core masses of intermediate-size planets, and thus raises an important question about why the gas accretion was not completed for these planets.  \citet{2004ApJ...604..388I} predicted that the core accretion mechanism would generate a deficit of intermediate-size planets (known as `planet desert') with semi-major axes lower than 3 au. However, such a deficit has not been observed in the planet distribution\footnote{We warn that in the literature, one can find claims of observational evidence of the planet desert. Indeed, the current sample of known exoplanets shows a considerably larger fraction of gas giants compared to intermediate-size planets (e.g. Fig.~\ref{fig:mr_intermediate}). However, this is due to an observational bias $-$i.e. early all-sky RV and transit surveys overpopulated the gas giant region, and their low occurrence rates together with the fixed (and limited) Kepler FoV did not allow for compensation for this early bias.} and instead, the frequency of intermediate-size planets is comparable to that of giant planets in close-in distances \citep[e.g.][]{2012ApJS..201...15H, 2021AJ....162..243B}. Indeed, microlensing surveys suggest that the abundance of intermediate-size planets could be even larger than that of gas giant planets beyond the ice line \citep[e.g.][]{2016ApJ...833..145S}. Theoretical works such as those from \citet{2011A&A...526A.111M} and \citet{2016ApJ...829..114B} suggest that for the intermediate-size planet population, the core accretion process might have been halted because of a later formation of the planetary core or an early dissipation of the protoplanetary disk. However, it is not clear the reason why any of those processes would happen so frequently during the formation of giant planets. 

Some observational trends and correlations of intermediate-size planets are similar to those observed in the giant planet population, hinting at a similar origin, but others suggest the existence of different mechanisms shaping the particularities of each population. The typical metallicity of stars hosting intermediate-size planets is known to be as high as that of stars hosting gas giant planets \citep[for an extensive review on different correlations between the exoplanet occurrence and stellar metallicity, we refer to][]{2019Geosc...9..105A}. Indeed, \citet{2018AJ....155...89P} found that stars hosting intermediate-size planets have the highest mean stellar metallicities among all planet hosts. This can lead us to naively discard the possibility that intermediate-size planets would systematically originate in protoplanetary disks with little solid material, which would lead to a less massive or slower forming proto-planetary core that subsequently would have accreted less gas from the disk. Overall, the metallicity trend mentioned above highly supports that intermediate-size planets were formed through core accretion, akin to the gas giant population. However, the reason why the gas accretion process is stopped relatively frequently is not yet well understood.

\begin{figure}
    \centering
    \includegraphics[width=0.5\textwidth]{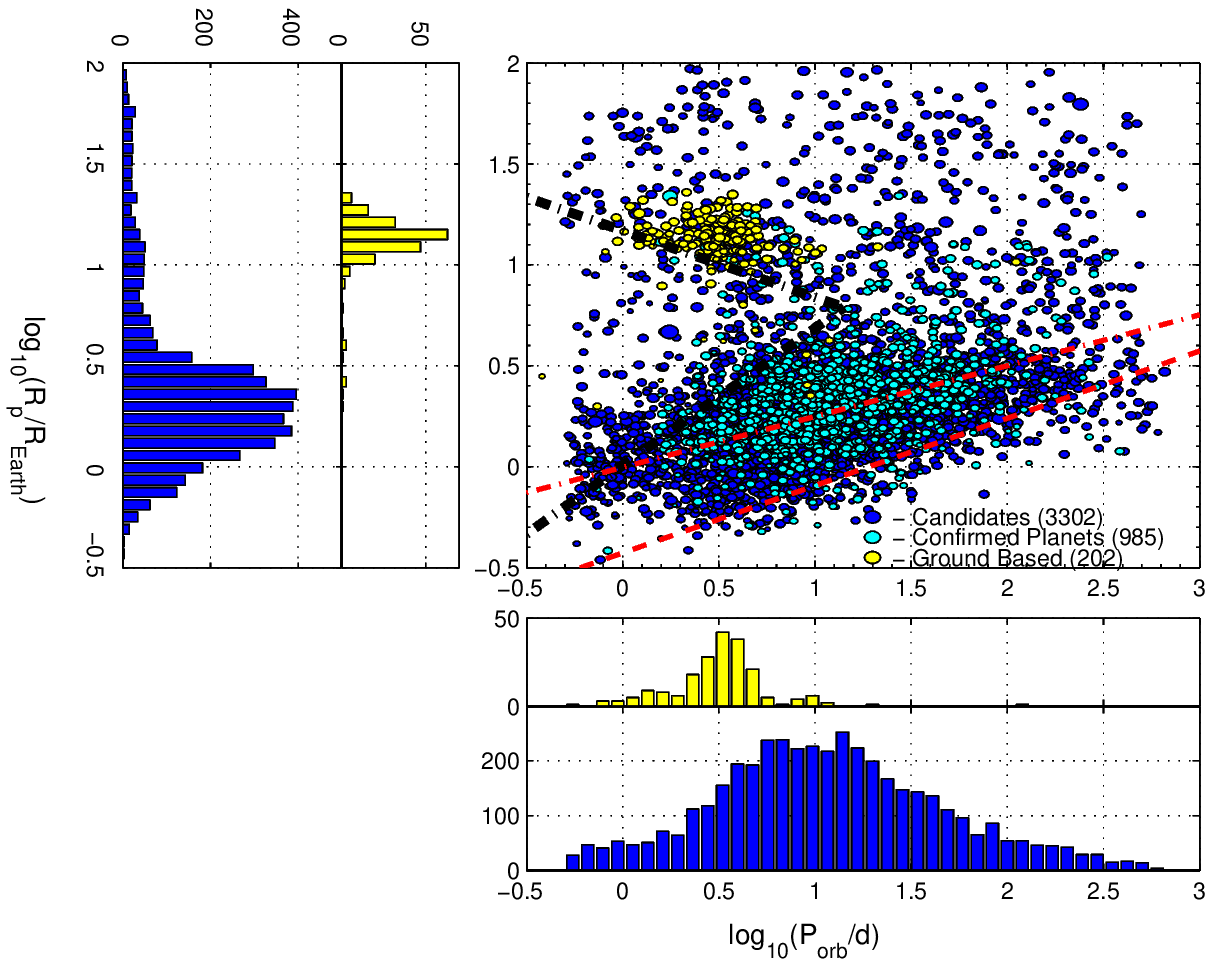}
    \includegraphics[width=0.456\textwidth]{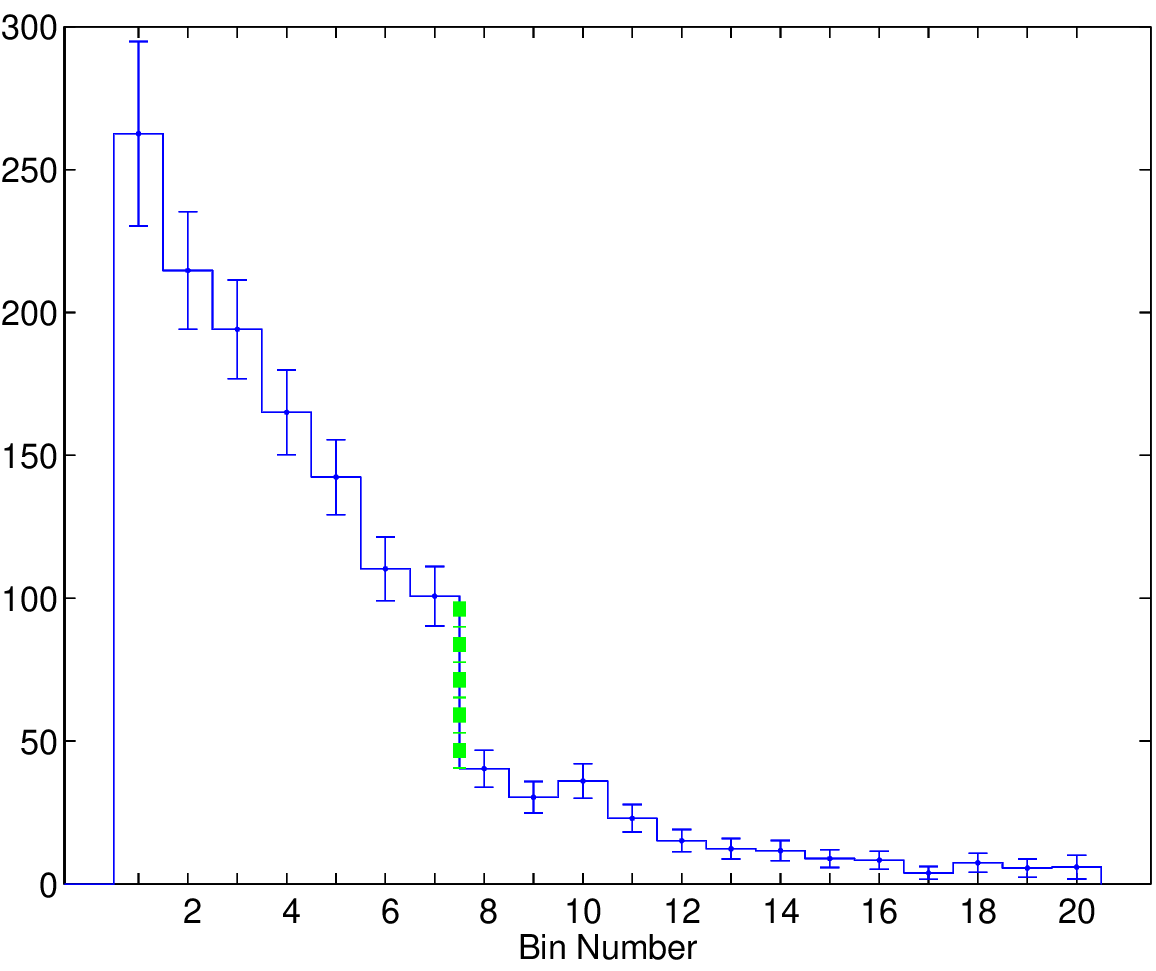}
    \caption[Planetary radii as a function of planetary orbital period for planets detected through ground-based transit searches as well as \textit{Kepler} confirmed and candidate planets.]{Left: Planetary radii as a function of planetary orbital period for planets detected through ground-based transit searches as well as \textit{Kepler} confirmed and candidate planets. The two dash-dotted black lines are the estimated desert boundaries from \citet{2016A&A...589A..75M}. Right: Density contrast between the planets inside and outside the desert (i.e. green dotted vertical line) that was used to define the lower boundary. Source: \citet{2016A&A...589A..75M}.}
    \label{fig:mazeh_desert}
\end{figure}

A notable difference between the intermediate-size and gas giant population is the short-period Neptunian (or sub-Jovian) desert. Soon after the first \textit{Kepler} results, the community realized that there was a scarcity of planets of intermediate masses and sizes in the shortest-period orbits \citep[e.g.][]{2011A&A...528A...2B,2011ApJ...727L..44S,2011ApJ...742...38Y,2013ApJ...763...12B,2016MNRAS.455L..96H,2016NatCo...711201L}. This scarcity cannot be due to observational biases since closer-in planets are easier to detect through the transit and RV techniques (see Sects.~\ref{subsec:transit_method} and \ref{subsec:rv_method}), and thus it must reflect the existence of a process or set of processes that either prevent intermediate-size planets to reach these locations or that remove them in short time scales. The boundaries of the Neptunian desert were first mapped by \citet{2016A&A...589A..75M} based on the observed distribution of \textit{Kepler} candidates and ground-based confirmed planets. The authors assumed a geometry composed of two straight lines in the period-radius and period-mass spaces and performed an iterative process to search for the maximum contrast. In Fig.~\ref{fig:mazeh_desert}, we show the Neptunian desert in the period-radius parameter space together with the maximum planet density contrast used to define the lower boundary \citep{2016A&A...589A..75M}. 

The origin of the Neptunian desert is today an open question in exoplanet research. As discussed above, giant planets are thought to form at large orbital distances, so the lack of planets is not expected to be shaped by formation mechanisms. Instead, the most commonly accepted theory is that, having less massive H/He envelopes than gas giants, intermediate-size planets are more prone to loose their atmospheres in close-in orbits due to evaporation \citep[e.g.][]{2003ApJ...598L.121L,2003Natur.422..143V,2004A&A...418L...1L,2011A&A...529A.136E,2012MNRAS.425.2931O,2014ApJ...792....1L,2015AREPS..43..459T,2019AREPS..47...67O}. This hypothesis is reinforced by recent observations of huge atmospheric mass-loss rates of close-in intermediate-size planets \citep[e.g.][]{2015Natur.522..459E,2018A&A...620A.147B,2022NatAs...6..141B,2022AJ....164..234V}. Alternatively, the desert has been also proposed to be shaped by the migration mechanisms that bring intermediate-size planets to their close-in orbits, which could potentially stop the process before entering the desert \citep[e.g. through the tidal disruption limit; see Sect.~\ref{sec:migration} and ][]{2016A&A...589A..75M,2016ApJ...820L...8M,2018MNRAS.479.5012O,2022AJ....164..234V}. As discussed in Sect.~\ref{sec:migration}, the orbital eccentricity is one of the best tracers of orbital migration. Warm intermediate-size planets show less excited eccentricities (i.e. typically between 0 and 0.2) than those of warm gas giants discussed in Sect~\ref{sec:migration} \citep[see][]{2020A&A...635A..37C}. This could suggest that HEM and disk-driven migration play different roles in the two populations, or that circular warm intermediate-size planets were gas giants that evaporated and circularised, although the second option is thought not to be physically possible, as discussed above. Another striking difference is the relative abundance of very close-in Neptunes (at the edge of the desert) with non-zero eccentricities, in contrast with the typically circular orbits of the closer-in HJs \citep[][]{2020A&A...635A..37C}. This result is especially striking when considering that lower-mass planets are easier (i.e. should take less time) to circularise. This has been proposed that could be related with a late HEM followed by a delayed atmospheric evaporation, so that these eccentric super-Neptunes and sub-Jovians could have arrived recently to their current orbits and are thus still evaporating and circularizing \citep{2018Natur.553..477B,2021A&A...647A..40A,2023A&A...669A..63B}. 

In summary, there are different observational constraints such as the metallicity-occurrence correlation, the mean bulk planet densities and inferred core masses larger than 10 $\rm M_{\oplus}$, and their frequent location in close-in orbits, that suggest that intermediate-size planets underwent similar formation and evolution histories to those of the gas giant counterparts. However, the distributions of some of their properties are sometimes unique (i.e. the existence of the Neptunian desert) or significantly differ from those of the gas giant planet population (i.e. the eccentricity distribution), which evidence the existence of different processes affecting the two populations. A major question in exoplanet research is to uncover which processes are those and why they affect the two populations differently. In chapters~\ref{ch:nep_des} and \ref{ch:toi_5005}, we delve into the evolution processes affecting close-in intermediate-size planets through the study of their underlying distribution at a population level and the confirmation and precise characterisation of the short-period super-Neptune planet TOI-5005~b, respectively.

\subsection{Small planets}
\label{sec:small}

Small planets ($R_{\rm p}$ $<$ 4 $\rm R_{\oplus}$) are one of the topics of greater interest in exoplanet research. This is, to a greater extent, due to their astrobiological value, since the solid surfaces of small rocky planets are considered the best locations for finding life outside the Earth \citep[e.g.][]{1993Icar..101..108K,2013ApJ...767L...8K,2013ApJ...778..109Z,2014AsBio..14...50H}. In addition, their peculiar properties embed 
an intrinsic scientific interest. In Sects. \ref{sec:giant} and \ref{sec:intermediate}, we showed that, even though in some regions of the parameter space there are strong differences between the gas giant and intermediate-size planet populations, overall they seem to share similar formation and evolution histories. In contrast, the small planet population shows particularities that hint at completely different evolutionary pathways. 

\begin{figure}
    \centering
    \includegraphics[width=0.45\textwidth]
    {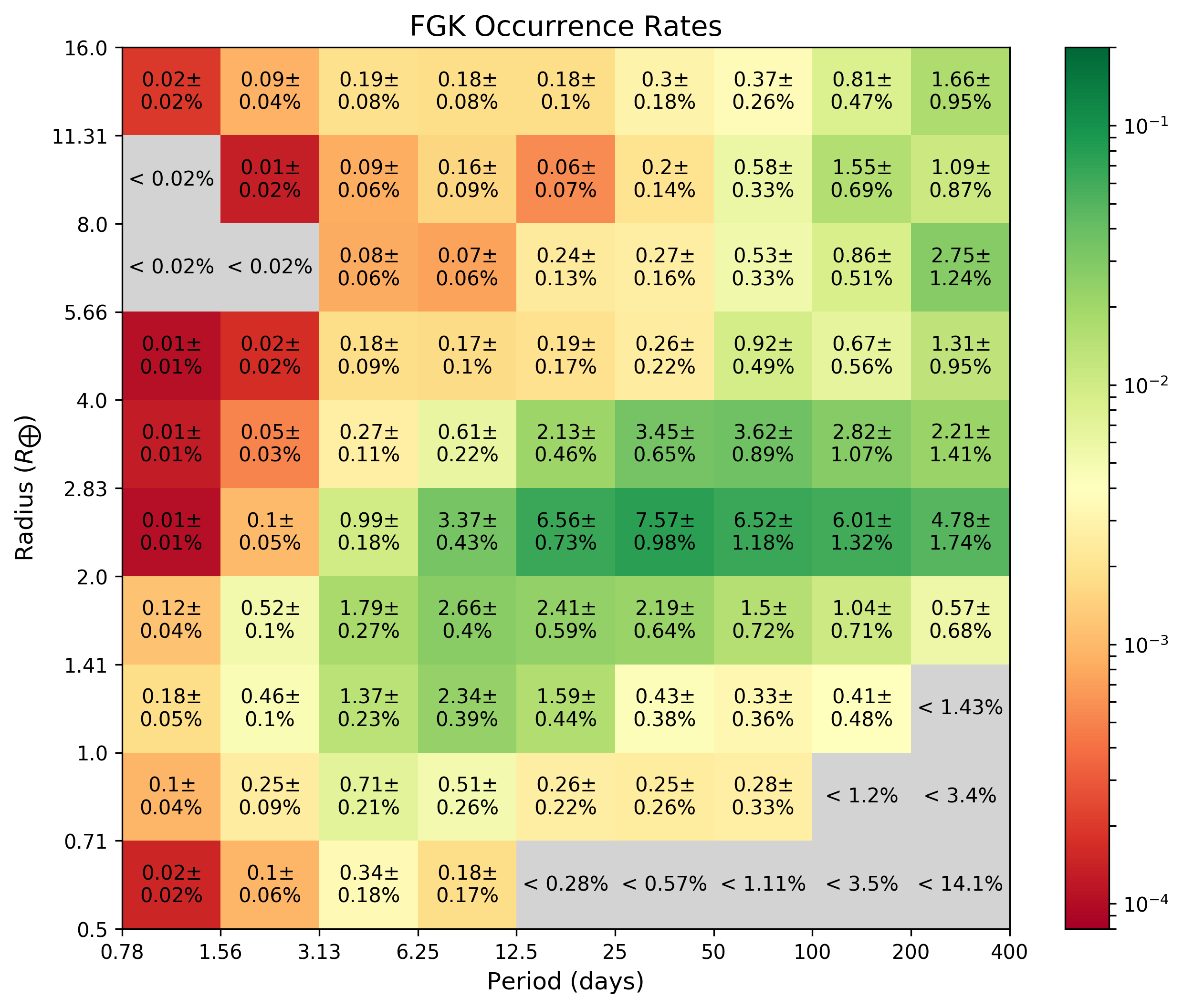}
    \includegraphics[width=0.54\textwidth]{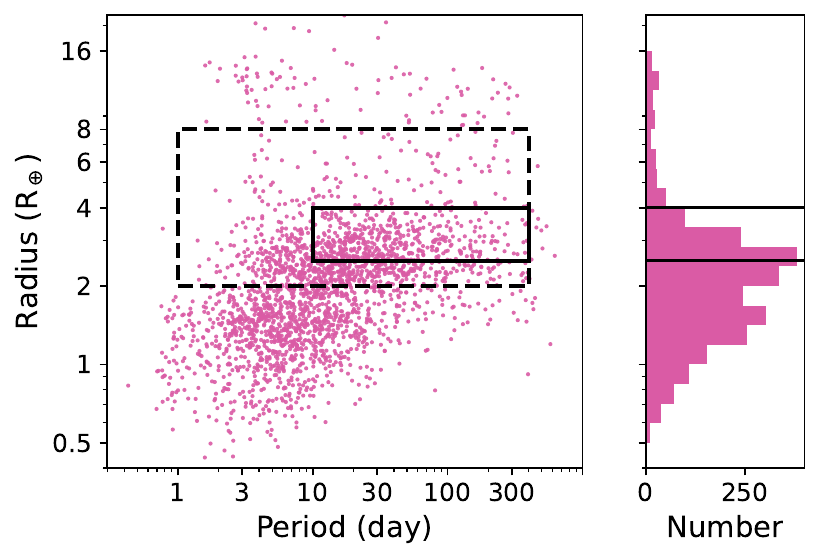}
    \caption[Planet occurrence in the period-radius parameter space and period-radius distribution of \textit{Kepler} detections showing the location of the radius cliff.]{Left: Planet occurrence in the period-radius parameter space showing the abundance of planets with radii lower than 4~$\rm R_{\oplus}$. The percentages represent the fraction of planets per star in each period-radius bin. Source: \citet{2020AJ....159..248K}. Right: Period-radius distribution together with a horizontal histogram showing the location of the radius cliff. Source: \citet{2024AJ....167..288D}. Both panels are based on \textit{Kepler} data.}
    \label{fig:radius_cliff}
\end{figure}

Small planets are the most abundant type of planet within 1 AU of main-sequence stars. This fact was first suggested by \citet{2011arXiv1109.2497M} based on HARPS RV data, and confirmed by \citet{2012ApJS..201...15H} based on \textit{Kepler} photometric data. In the left panel of Fig.~\ref{fig:radius_cliff}, we show the planet occurrence rates in different period-radius bins as computed by \citet{2020AJ....159..248K}, and in the right panel, we show the planet occurrence drop at $\sim$ 4 $\rm R_{\oplus}$, commonly referred to as the `radius cliff' \citep[e.g.][]{2019ApJ...887L..33K,2024AJ....167..288D}. The radius cliff thus marks a clear distinction between small and giant planets, evidencing that formation and/or evolution processes in the closer-in orbits are much more effective for the small planet population. Another relevant feature that differentiates small planets from giant planets is the stellar metallicity of their host stars. Small planets have been found around stars with a wide rage of metallicities, with sub-solar and super-solar values roughly in a similar proportion \citep[e.g.][]{2008A&A...487..373S,2012Natur.486..375B,2014Natur.509..593B}, in contrast to the metallicity-occurrence positive trend found for giant planets (see Sects.~\ref{sec:giant} and \ref{sec:intermediate}). This observational evidence further reinforces the idea of the existence of formation differences between small and giant planets. In particular, it suggests that small planetary cores can be effectively assembled in disks with limited solid material, in contrast to the more massive cores of giant planets. In the following, we introduce the main proposed formation and evolution mechanisms of small planets (Sect.~\ref{sec:form_evo_small}) and describe the main observational results constraining these theories based on population studies (Sect.~\ref{constraints_small}).  

\subsubsection{Formation and evolution mechanisms}
\label{sec:form_evo_small}

Several models have been proposed to explain the origins of small planets \citep[see][]{2008MNRAS.384..663R}. Some of these models require the existence of dynamical interactions with giant planets \citep[e.g.][]{2005A&A...441..791F,2006Sci...313.1413R}, which were recently ruled out as the main formation mechanisms given the higher occurrence of small planets than giant planets \citep[][]{2011arXiv1109.2497M,2012ApJS..201...15H}. Models proposing an `in situ' accretion $-$i.e. planets would grow close to their stars by successive impacts between planetary embryos; e.g.~\citealt{2013MNRAS.431.3444C}$-$ have been found to not be self-consistent, given that in protoplanetary disks massive enough to allow this process, orbital migration is expected to act very effectively for newborn small planets \citep[e.g.][]{2014ApJ...793....3B,2015MNRAS.448.1751I,2015A&A...578A..36O}. Today, there are two main mechanisms to explain the formation and evolution of small planets, the so-called \textit{drift} and \textit{migration} models \citep[see e.g.][]{2021JGRE..12606639B}. 

The \textit{drift} model proposes that dust grains (often referred to as `pebbles') coagulate and grow until becoming large enough to partially decouple from the gas and migrate inwards \citep[see][]{2017AREPS..45..359J,2017ASSL..445..197O}. Hence, the majority of the planetesimal growth would occur close-in within the protoplanetary disk from the dust that drifted inwards. There is observational evidence of the existence of pebbles in gas-rich disks around young stars \citep{2007prpl.conf..767N,2015ApJ...813...41P}, and the dust disks are found to be more compact than the gas disk, which is commonly interpreted as an evidence of the drift process \citep[e.g.][]{2012ApJ...744..162A,2016ApJ...832..110C,2020A&A...638A..38T}. The ring-shaped structures found in many protoplanetary disks \cite[e.g.][]{2015ApJ...808L...3A,2016PhRvL.117y1101I,2018ApJ...869L..41A,2018MNRAS.475.5296D,2019A&A...622A..75L,2021ApJ...909..212K} are also interpreted to be produced by growing and drifting pebbles \citep[e.g.][]{2018ApJ...869L..45B,2018ApJ...869L..46D} that would be trapped in pressure bumps in the inner parts of the disk \citep[][]{2014ApJ...792L..27B,2014ApJ...780...53C,2017ApJ...835..230F,2019A&A...630A.147F,2019MNRAS.484.2296J}. The final phases of the process are thought to involve mutual collisions of planetary embryos, as well as some degree of orbital migration \citep{2012ApJ...751..158H,2015MNRAS.453.1471D,2016ApJ...832...34M}. 

The \textit{migration} model proposes that planetary cores are assembled at large distances from the star and subsequently migrate inwards, similarly to the disk-driven migration mechanism proposed for giant planets (see Sect.~\ref{sec:giant}, Sect.~\ref{sec:intermediate}). Thus, while both the \textit{drift} and \textit{migration} models involve the inward transportation of mass through the proto-planetary disk, they fundamentally differ in the configuration and size of the transported mass (pebbles vs proto-planets). As introduced in Sect.~\ref{sec:formation}, planetary cores preferentially form at large orbital distances, beyond the ice line, where pebble accretion is more efficient \citep{2014A&A...572A.107L,2015Icar..258..418M,2017ASSL..445..197O}. Once they become massive enough, the cores can migrate inwards until reaching the inner regions of the protoplanetary disk \citep[e.g.][]{2010ApJ...719..810I,2014A&A...569A..56C,2016MNRAS.457.2480C}. The final steps of this model are thought to involve giant collisions between cores, as in the \textit{drift} model.  

Both the \textit{drift} and \textit{migration} models predict that small planets accrete gaseous envelopes directly from the disk, which would thus be primarily composed of $\rm H_{2}$ and He. Their present-day sizes are hypothesized to be determined by different processes occurring during planet formation $-$i.e. the cooling atmospheric process \citep[e.g.][]{2014ApJ...797...95L}, dissolution of $\rm H_{2}$ in magma oceans \citep[e.g.][]{2019ApJ...887L..33K}, photo-evaporation of the disk \citep[e.g.][]{2016ApJ...825...29G,2020ApJ...892..124O,2016ApJ...817..107O}, or giant impacts \citep{2019MNRAS.485.4454B}$-$  and evolution $-$i.e. atmospheric escape through photo-evaporation  \citep[e.g.][]{2013ApJ...775..105O,2014ApJ...795...65J,2016ApJ...831..180C}, core-powered mass loss \citep[e.g.][]{2018MNRAS.476..759G,2019MNRAS.487...24G}, or Jeans escape \citep[e.g.][]{1982P&SS...30..773H,1987tpaa.book.....C,1989GeoRL..16.1225M,2011ApJ...729L..24V}. Both the photo-evaporation and core-powered mass loss hypotheses rely on the physical basis that heating the upper atmosphere of a planet would produce a hydrodynamic outflow that would lead to a process of atmospheric escape, but the energy source that causes the atmospheric heating differs.  In the atmospheric escape model, the heating is triggered by the high-energy X-ray and UV radiation (XUV) emitted by the stellar corona, which is absorbed by the planet's upper atmosphere. The high-energy photons are known to be capable of ionizing atoms and molecules, causing the upper atmosphere to heat up to high temperatures ($\sim 10^{3}-10^{4}$~K) that would trigger the outflow \citep[e.g.][]{2004Icar..170..167Y,2007P&SS...55.1426G,2009ApJ...693...23M,2012MNRAS.425.2931O}. In contrast, in the core-powered mass loss model, the heating process is provoked by the infrared (IR) irradiation coming from the cooling planetary interior \citep[e.g.][]{2019MNRAS.487...24G}. The Jeans escape mechanism can occur when the kinetic energy of the lighter atmospheric particles overcomes the gravitational pull of the planet, and hence their velocities surpass the escape velocity. The effectiveness of this process highly depends on the gravitational attraction of the body, and thus it is most efficient in the less massive planets \citep[e.g.][]{2017aeil.book.....C,2017A&A...598A..90F}. While the three atmospheric mass-loss processes may be simultaneously acting, it is not yet clear which is the predominant process shaping the observed properties of small planets at a population level (see Sect.~\ref{constraints_small}). 

Rock-dominated cores surrounded by $\rm H_{2}$/He envelopes are a common output of the \textit{drift} and \textit{migration} formation and evolution models. However, these models differ in the planetary retention capacity of other volatile elements (i.e. chemical elements with low boiling points). Pebbles are expected to loose their volatiles as they migrate inwards \citep[e.g.][]{2011ApJ...738..141O,2019A&A...624A..28I}, but large migrating planetary cores are expected to be capable of retaining the majority of them \citep[e.g.][]{2017A&A...598L...5A,2017A&A...602A..21S}. After $\rm H_{2}$ and He, water ($\rm H_{2}$O) is the most abundant volatile element in protoplanetary disks \citep[e.g.][]{2003ApJ...591.1220L,2007ApJ...667..303T}. Therefore, according to the \textit{drift} model, small planets should have rocky cores with little water in their structure or atmosphere, while the \textit{migration} model predicts that small planets could have abundant reservoirs of water. The amount of water that these planets can hold is today an open question, with simulations suggesting values ranging from a few per cent \citep{2019A&A...624A.109B} up to 50$\%$ in mass \citep{2021A&A...650A.152I}. A key question yet to be answered is where all this water would be placed within the planetary structure. It has been proposed that it could form an independent layer in liquid or solid state \citep[e.g.][]{2019PNAS..116.9723Z}, or be in the planetary atmosphere forming an extensive envelope of steam \citep[e.g.][]{2020A&A...638A..41T}, or could also be mixed with the rocks in the planetary interiors \citep[e.g.][]{2024NatAs...8.1399L}. In all, regardless of the location of water and other volatiles, the \textit{migration} model predicts water-rich low-density planets, while the \textit{drift} model predicts both high-density and low-density planets, depending on their capabilities to preserve their primordial $\rm H_{2}$/He atmospheres. This degenerated scenario complicates the inference of the internal structure of individual planets based on their densities alone (see Sect.~\ref{constraints_small}).

After formation, the dynamical evolution predicted by the \textit{drift} and \textit{migration} models is similar. In both cases, small planets in multi-planetary systems are expected to build up configurations where each neighbouring planet is in mean motion resonance \citep[e.g.][]{2007ApJ...654.1110T,2008A&A...482..677C,2018MNRAS.479L..81R}, as we are starting to observe through ALMA observations of young protoplanets \citep[e.g.][]{2024A&A...685L...1C}. However, after the dissipation of the disk, resonant chains typically become unstable, leading to a period of late giant impacts \citep{2009ApJ...699..824O,2014A&A...569A..56C} and the braking of the chains \citep{2017MNRAS.470.1750I,2021A&A...650A.152I}.

\subsubsection*{Constraints from population studies}
\label{constraints_small}

\begin{figure*}
    \centering
    \includegraphics[width=0.48\textwidth]{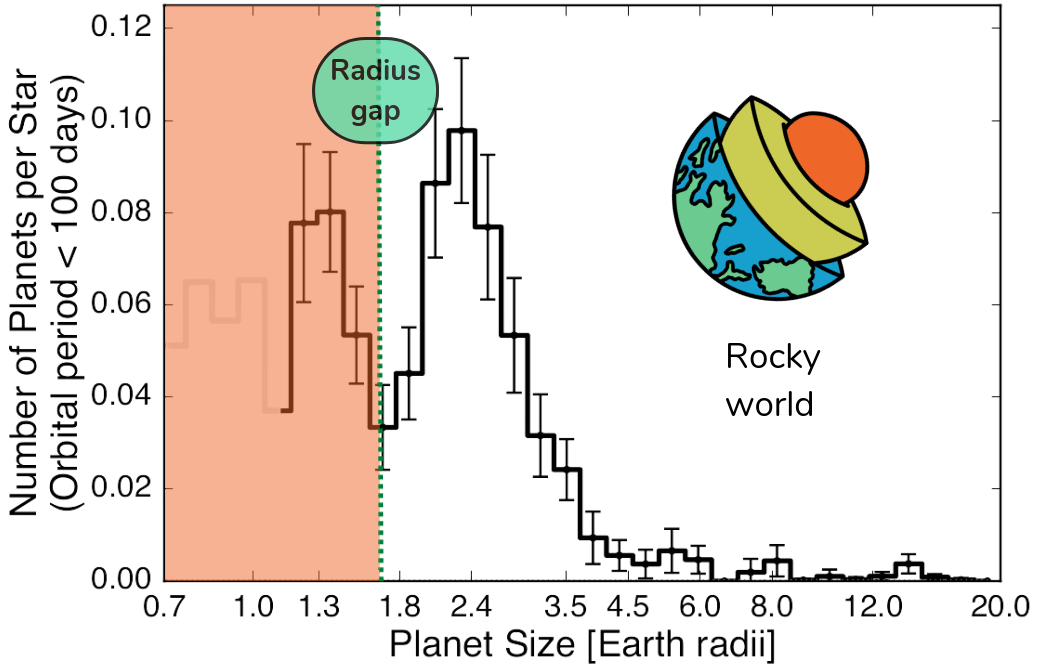}
    \includegraphics[width=0.48\textwidth]{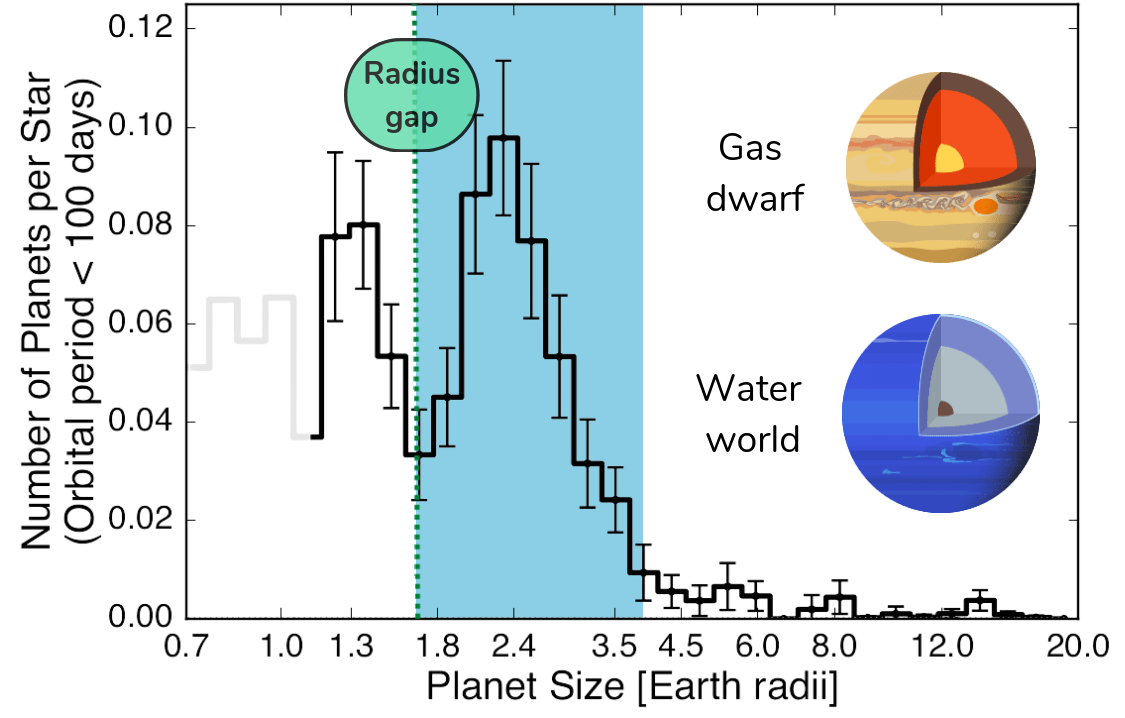}
    \caption[Bi-modal distribution of radii for planets with orbital periods shorter than 100 days.]{Bias-corrected distribution of planet radii for planets with orbital periods shorter than 100 days. The light grey region of the histogram for radii smaller than 1.14 $R_{\oplus}$ suffers from low completeness. The left and right panels highlight the lower and higher models of the bi-modal distribution, and the planet structures overplotted represent the most widely accepted hypotheses. This figure has been adapted from \citet{2017AJ....154..109F}.}
    \label{fig:radius_valley}
\end{figure*}

\begin{figure}
    \centering
    \includegraphics[width=\textwidth]{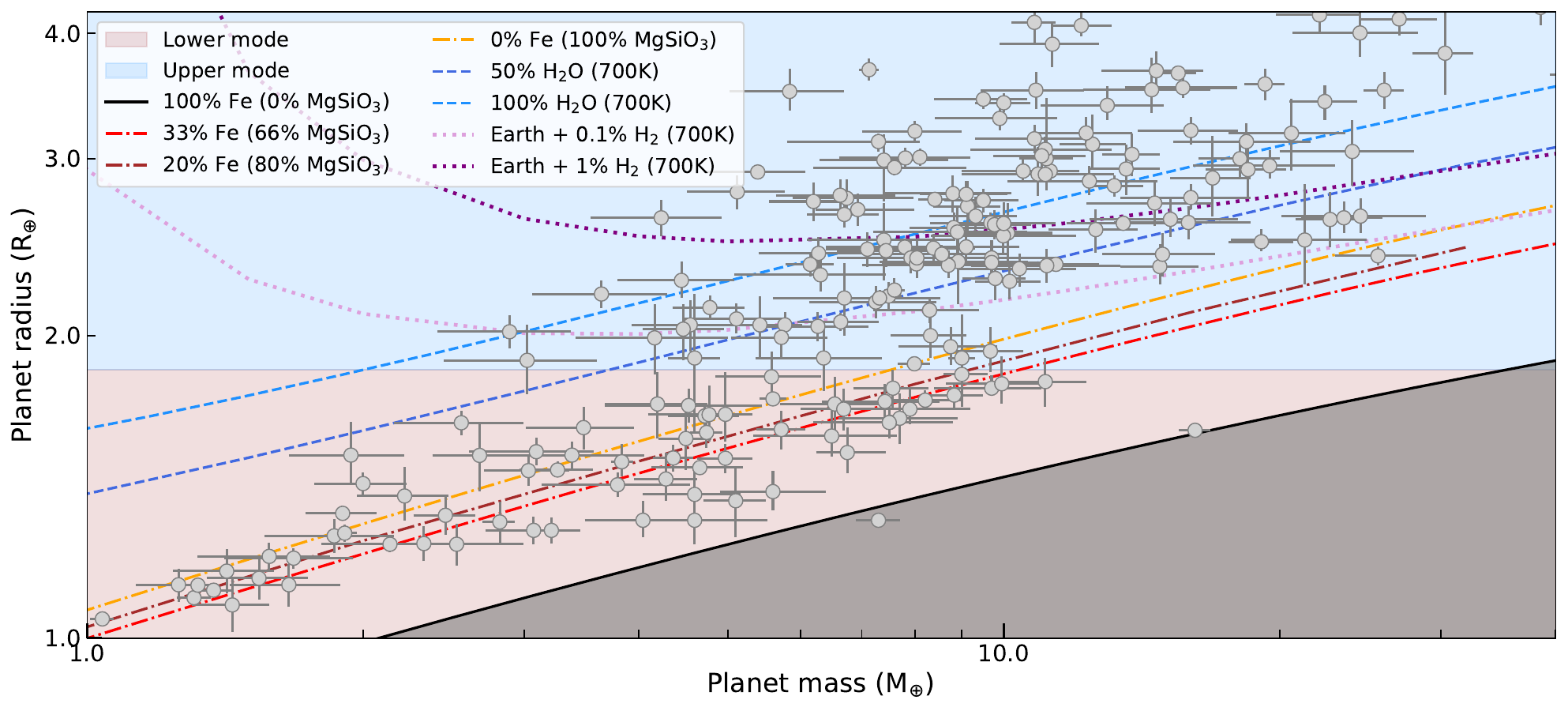}
    \caption[Mass-radius diagram of small planets with precise masses and radii.]{Mass-radius diagram of small planets ($R_{\rm p}$ $<$ 4 $\rm R_{\oplus}$) with precise masses and radii constrained to a precision better than 20$\%$. The red and blue regions indicate the lower and upper modes of the bi-modal radius distribution, respectively. The iso-composition curves for rocky planets, water worlds, and gas dwarfs are from \citet{2016ApJ...819..127Z,2019PNAS..116.9723Z}. The planet data was collected from the NASA Exoplanet Archive \citep{2013PASP..125..989A} and the diagram was made through the \texttt{mr-plotter} package (\url{https://github.com/castro-gzlz/mr-plotter}).}
    \label{fig:mr_small}
\end{figure}

Recent observational results are starting to shed some light on the nature of small planets. One of the most relevant findings is the existence of a drop in the occurrence rate of planets between $R_{p}$ $\simeq$ 1.6~$\rm R_{\oplus}$ and $R_{p}$ $\simeq$ 1.9~$\rm R_{\oplus}$, reaching the minium occurrence at $R_{p}$ $\simeq$ 1.75 $\rm R_{\oplus}$, which is often referred to as the radius gap or radius valley \citep{2017AJ....154..109F,2018AJ....156..264F,2018MNRAS.479.4786V}. Figure~\ref{fig:radius_valley} shows the bias-corrected radii distribution of small planets based on \textit{Kepler} data, where we can appreciate a significant bimodality separated by the radius gap. This occurrence drop is interpreted as the existence of two sub-populations of small planets that trace different formation and/or evolution histories. Interestingly, in addition to being separated by the radius valley, these sub-populations have been found to show different bulk densities, reinforcing the idea of having followed different formation or evolutionary pathways. In Fig.~\ref{fig:mr_small}, we show the mass-radius diagram of small planets known to date. Planets in the lower mode of the radius distribution typically have densities compatible with an Earth-like composition. That is, a core composed of iron (Fe) and nickel (Ni) comprising $\simeq$ 33 $\%$ of the planetary mass, and a mantle composed of silicates (e.g. $\rm MgSiO_{3}$) comprising $\simeq$ 67 $\%$ of the planetary mass \citep[][]{2007ApJ...669.1279S,2016ApJ...819..127Z,2019PNAS..116.9723Z}. These planets are often referred to as rocky planets, referring to their most likely rocky-dominated composition. They are also referred to as super-Earths, given that they are slightly larger than the Earth while seemingly showing the same iron-rock structure \citep[e.g.][]{2005ApJ...634..625R,2009Natur.462..891C,2007ApJ...669.1279S,2009ApJ...696...75H}. In contrast, in the upper mode, the vast majority of known planets have low densities, incompatible with compositions based on iron and rock alone. These planets are often referred to as mini-Neptunes or sub-Neptunes, given their larger radii and puffier nature than super-Earths \citep[e.g.][]{2014ApJ...792....1L,2014ApJS..210...20M,2015ApJ...801...41R}. Unfortunately, the internal compositions of planets with densities lower than Earth's are difficult to infer based on radius and mass data alone due to compositional degeneracies. The two most popular hypotheses propose rock-dominated structures surrounded by primordial $\rm H_{2}$/He envelopes, commonly known as gas dwarfs, or rock-dominated structures surrounded by a layer of condensed (i.e. solid or liquid) water, commonly known as water worlds. In Fig.~\ref{fig:mr_small}, we plot different iso-composition curves for gas dwarfs and water worlds from \citet{2019PNAS..116.9723Z}, illustrating such degeneracy. Unveiling the typical compositions of sub-Neptunes is one of the main goals of exoplanet research, which, in addition to the composition itself, can provide critical clues on the formation of these planets, given that the water world hypothesis would strongly favour the \textit{migration} model over the \textit{drift} model, as introduced in Sect.~\ref{sec:form_evo_small}. 

\begin{figure}
    \centering
    \includegraphics[width=0.502\textwidth]{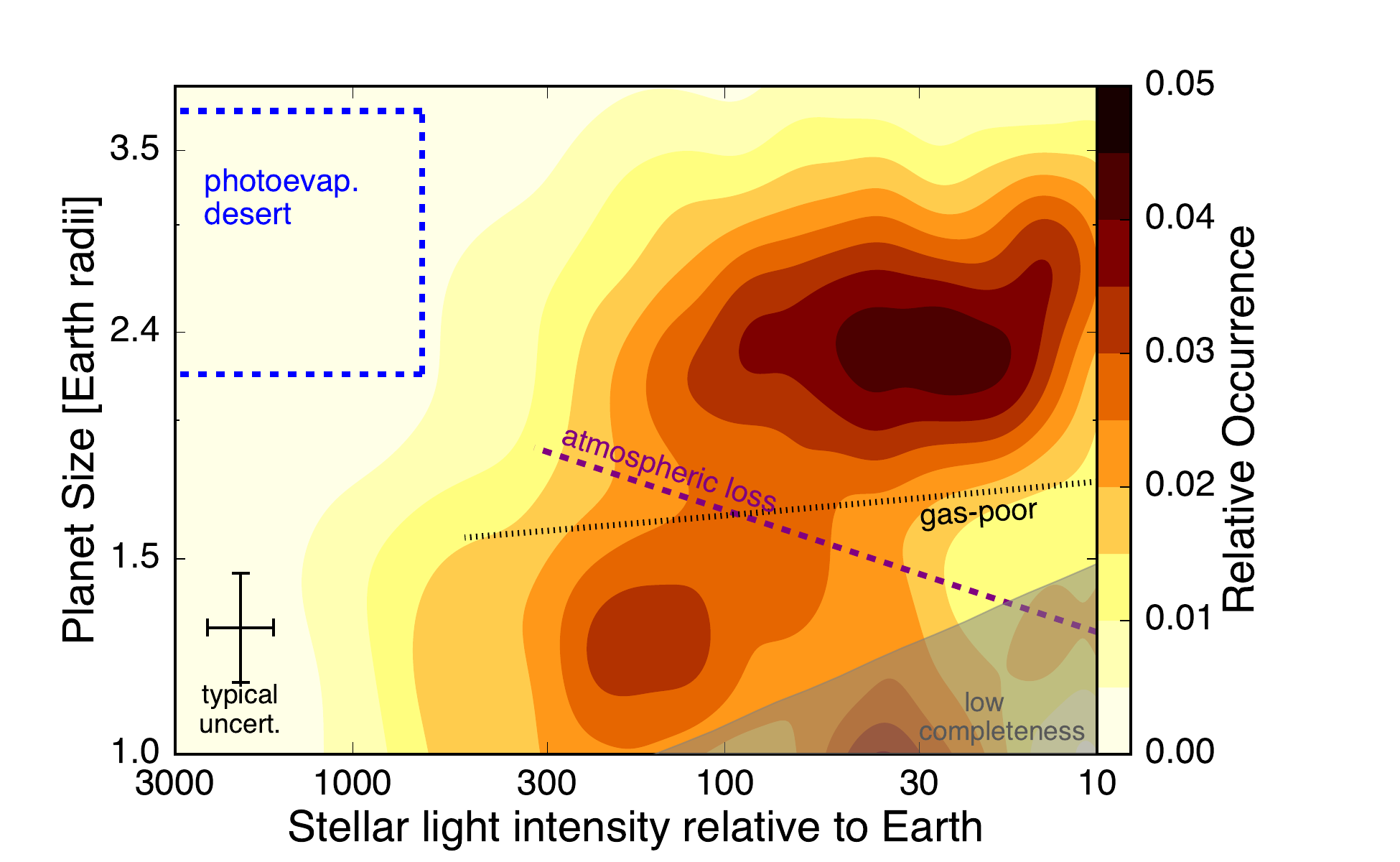}
    \includegraphics[width=0.48\textwidth]{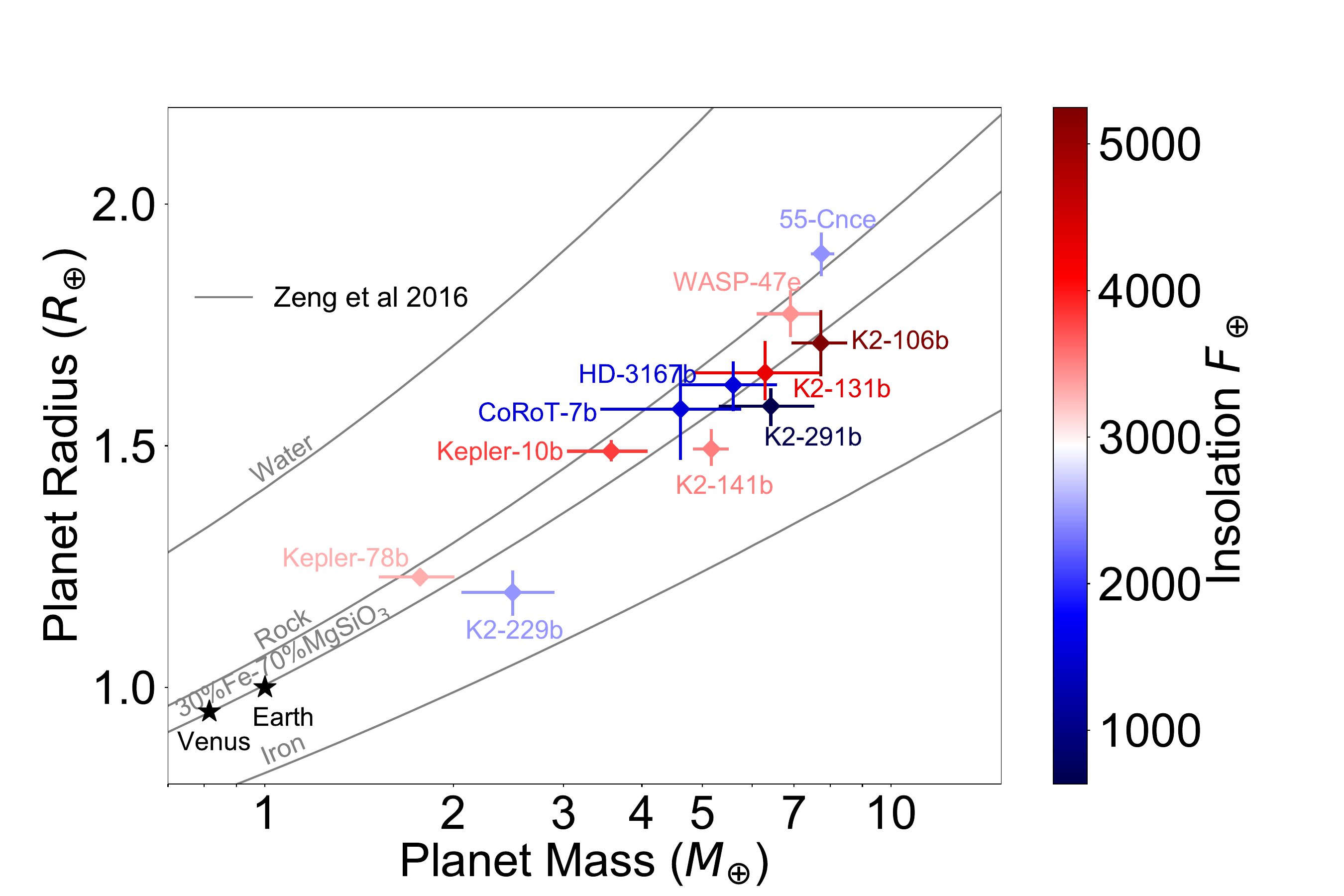}
    \caption[Dependency of the radius gap with insolation flux and mass-radius diagram of ultra-short period planets homogeneously characterised.]{Left: Iso-density contours of the \textit{Kepler} planets in the radius-insolation space showing the dependency of the radius gap with insolation flux reported in \citet{2017AJ....154..109F,2018MNRAS.479.4786V}. The models for atmospheric loss and gas-poor planets represented with dotted lines are from \citet{2013ApJ...776....2L}. Source: \citet{2017AJ....154..109F}. Right: Mass-radius diagram of ultra-short period (USP; $P_{\rm orb}$ $<$ 1 day) planets homogeneously characterised. Iso-composition curves for rocky planets are from \citet{2016ApJ...819..127Z}. Source: \citet{2019ApJ...883...79D}.}
    \label{fig:support_gas_dwarf}
\end{figure}

Today, different observational results support either the gas dwarf or water world hypotheses to describe the compositions of sub-Neptune planets, but no firm conclusions can be stated. \citet{2017AJ....154..109F} identified a dependency of the radius gap with orbital period (or stellar insolation flux). This trend has been studied by several authors and is interpreted to support the gas dwarf hypothesis \citep[e.g.][]{2018MNRAS.479.4786V,2023MNRAS.519.4056H}. Indeed, the slope of the gap with insolation matches the predictions of theoretical models where all small planets originally formed with $\rm H_{2}$/He envelopes but only the more highly irradiated ones lost them (see the dashed magenta line in Fig.~\ref{fig:support_gas_dwarf}; \citealt{2013ApJ...776....2L}). Based on this and other similar early results, it was widely accepted that sub-Neptunes would be mainly composed of rocks and $\rm H_{2}$/He envelopes (i.e. gas dwarfs), and that super-Earths would be the naked rocky cores of these gas dwarfs whose atmospheres would have been stripped out through atmospheric mas-loss mechanisms (see Sect.~\ref{sec:form_evo_small} for a description of the three main proposed mechanisms). This hypothesis is further strengthened by the measured masses and radii of ultra-short period (USP; $P_{\rm orb}$ $<$ 1 day) planets. These planets are so highly irradiated that it is very unlikely that they have preserved $\rm H_{2}$/He envelopes. In line with this theoretical prediction, \citet{2019ApJ...883...79D} performed a homogeneous characterisation of USP planets and found that the vast majority of them are consistent with Earth-like compositions (see Fig.~\ref{fig:support_gas_dwarf}, right panel). 

\begin{figure}
    \centering
    \includegraphics[width=0.485\textwidth]{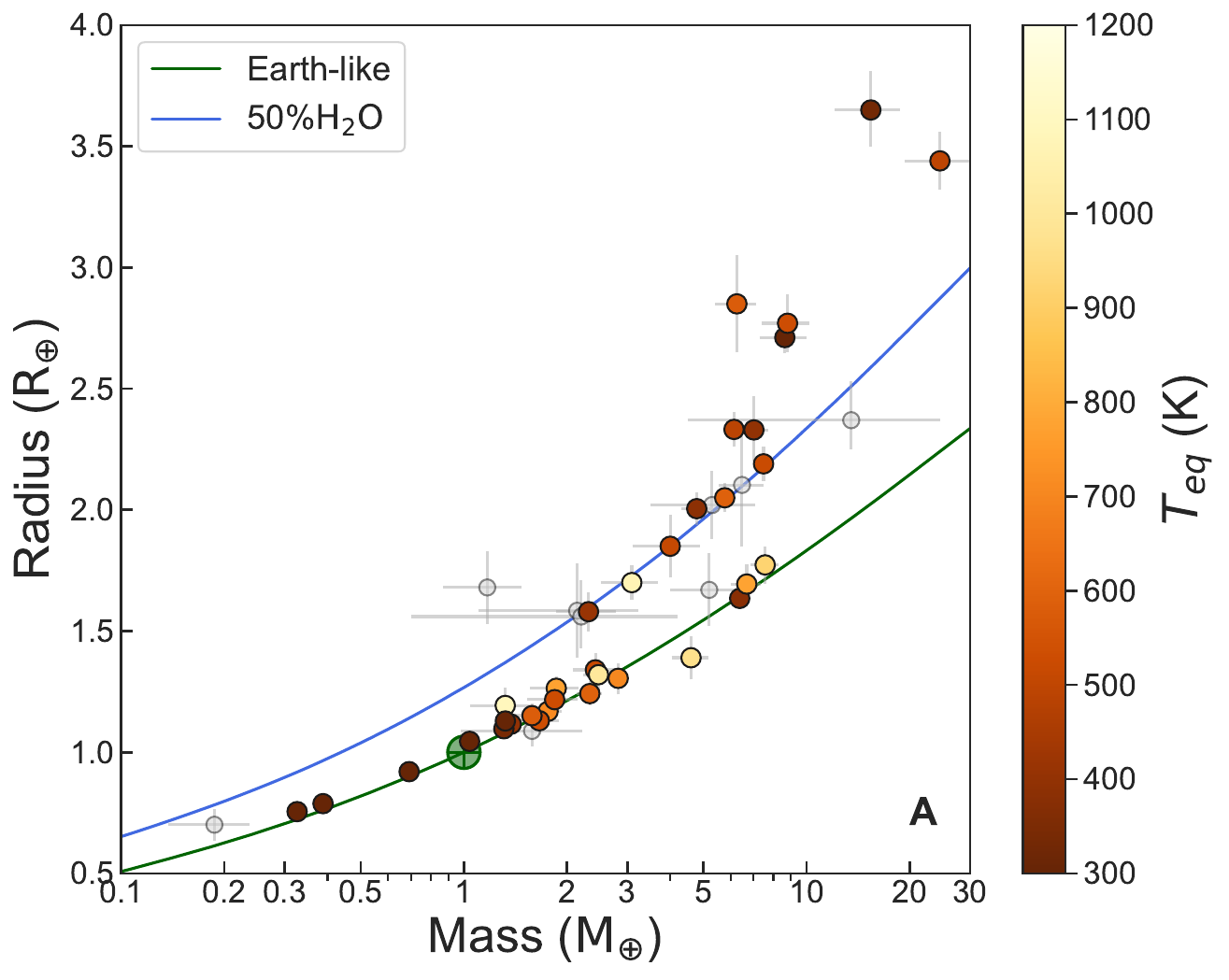}
    \includegraphics[width=0.48\textwidth]{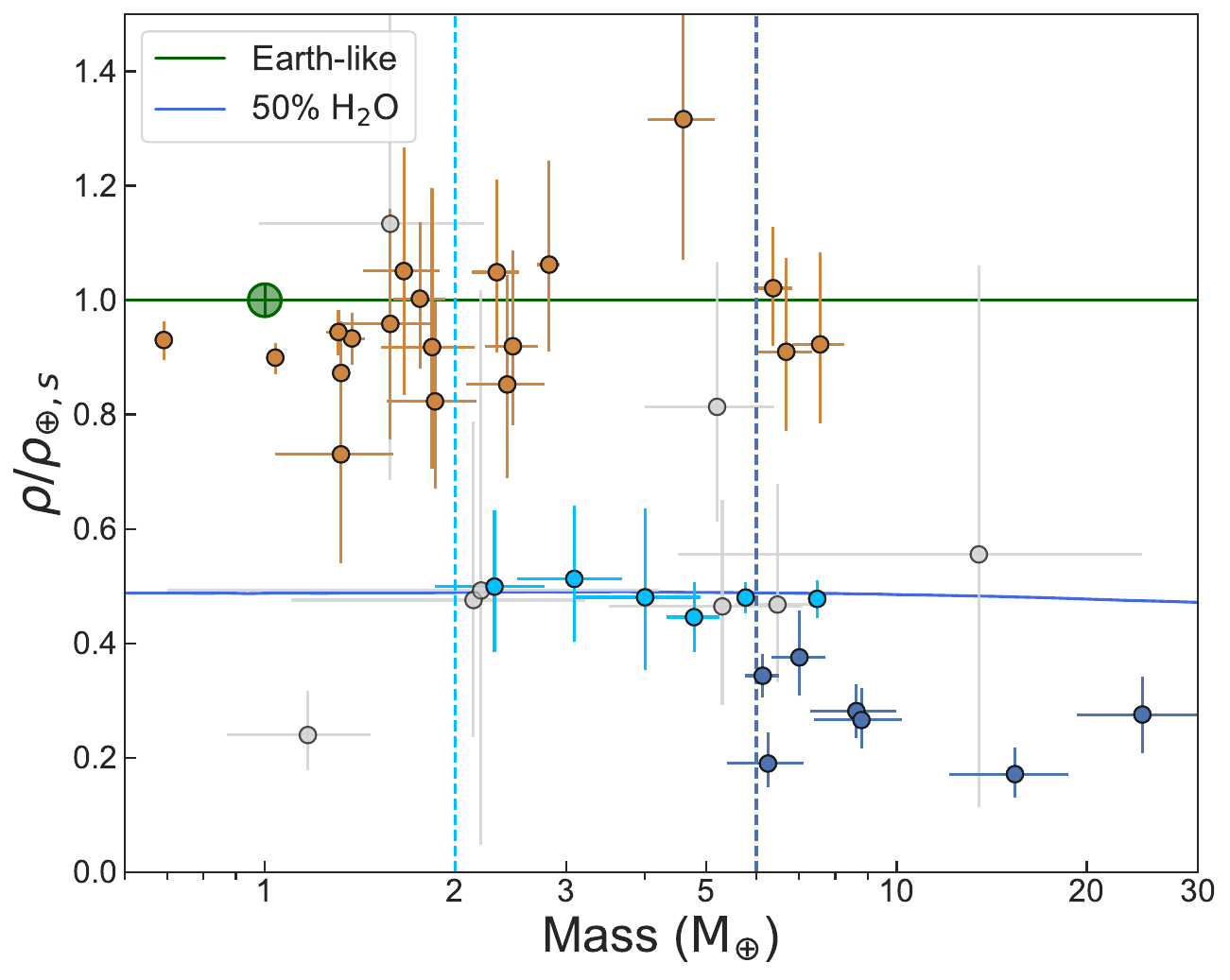}
    \caption[Mass-radius and mass-density diagrams for small transiting planets around M-dwarfs.]{Mass-radius (left panel) and mass-density (right panel) diagrams for small transiting planets around M-dwarfs. The theoretical iso-composition curves for the Earth-like and water world (i.e. 50$\%$ water-dominated ices, and 50$\%$ silicates) models \citep{2019PNAS..116.9723Z} are plotted. In the left panel, planets are colour coded by their equilibrium temperature $T_{\rm eq}$. In the right panel, densities are normalised by the Earth-like model, and planets are colour coded according to their characteristic bulk densities: rocky planets (brown), water worlds (light blue), and puffy sub-Neptunes (dark blue). Source: \citet{2022Sci...377.1211L}.}
    \label{fig:luque_palle}
\end{figure}

Despite the early results pointing towards the gas dwarf hypothesis, the water world hypothesis has not been completely abandoned. Indeed, recent results have been discussed to support it. \citet{2022Sci...377.1211L} homogeneously characterised all small transiting planets around M-dwarf stars and found that, similarly to super-Earths, which closely follow the Earth-like iso-composition curve in the mass-radius diagram, sub-Neptunes around M-dwarfs follow the water world iso-composition curve computed by \citet{2019PNAS..116.9723Z}, consisting of solid water and rocks in a 1:1 proportion (see Fig.~\ref{fig:luque_palle}). The authors argue that material condensing beyond the ice line is expected to have a 1:1 water-to-rock ratio if it has the same composition as the Solar System \citep[e.g.][]{2003ApJ...591.1220L}, which would explain this trend. In addition, if sub-Neptunes were $\rm H_{2}$/He rich, a small change in the gas fraction would drastically change the density of these planets, which is not empirically observed. This powerful observational evidence has provoked the resurgence of the water world hypothesis (and hence the \textit{migration} model) over the last few years. We note, however, that some discrepant voices have already arisen on how this trend has been interpreted. \citet{2023ApJ...947L..19R} argue that it could also be explained through the gas dwarf + $\rm H_{2}$/He mass-loss hypothesis, so that conclusive evidence for a population of water worlds around M-dwarfs remains elusive. Some other authors propose that for small planets more irradiated than the runaway greenhouse limit (i.e. S $>$ 1.1 $S_{\oplus}$), if they are water-rich, such water would be forming an extensive hydrosphere of steam and supercritical water, which, given the lower density of steam water than condensed water, would not be consistent with the proposed 1:1 rock-water ratio \citep[e.g.][]{2020A&A...638A..41T,2020ApJ...896L..22M,2021ApJ...914...84A,2022A&A...660A.102A}. In a very recent study, \citet{2024A&A...688A..59P} used the PlanetS catalogue\footnote{Available at \url{https://dace.unige.ch/exoplanets/}.} to discuss that the transition between super-Earths and sub-Neptunes around M-dwarfs is continuous, which comes into tension with the results from \citet{2022Sci...377.1211L}.

In all, there are currently different observational results based on the measure of fundamental planetary properties such as the mass, radius, and orbital period of sub-Neptune planets that point towards the gas dwarf or water world hypotheses. The gas dwarf hypothesis was the first one proposed, and probably the most accepted one, but the water world hypothesis is progressively gaining more traction, mainly due to the results presented in \citet{2022Sci...377.1211L}. However, to date, there is no conclusive evidence for the nature of sub-Neptune planets due to the observational and theoretical discrepancies presented above.

Direct observations of the planetary atmospheres (mainly through the transmission spectroscopy technique) may provide critical insight into the conundrum of the nature of small planets. Unveiling the compositions of the planetary atmospheres is considered to be critical to break the degeneracies in interior structure models, potentially allowing us to unambiguously unveil the bulk planetary structures \citep[e.g.][]{2008ApJ...673.1160A,2009ApJ...690.1056M,2013ApJ...775...10V}. Unfortunately, a large fraction of transmission spectroscopy measurements of sub-Neptune planets have resulted in featureless spectra \citep[commonly referred to as `flat' spectra; e.g.][]{2010Natur.468..669B,2011ApJ...743...92B,2014Natur.505...66K,2020AJ....159...57L}. Some of the successful atmospheric observations revealed hydrogen-dominated atmospheres for sub-Neptunes, in line with the gas dwarf hypothesis \citep[e.g.][]{2016ApJ...820...99T,2019NatAs...3..813B,2023ApJ...956L..13M}. An intriguing case difficult to interpret is 55~Canc~e \citep{2004ApJ...614L..81M,2018A&A...619A...1B}, the only USP planet in the sample of \citet{2019ApJ...883...79D} found to have a lower density than that expected for a rocky composition (see Fig.~\ref{fig:support_gas_dwarf}, right panel), and for which a hydrogen-dominated atmosphere has been detected. At the time of writing these lines, \citet{2024ApJ...974L..10P} used the James Webb Space Telescope \citep[JWST;][]{2006SSRv..123..485G} to detect, for the first time, a water-rich `steam world' atmosphere in GJ 9827~d \citep{2017AJ....154..266N,2024A&A...684A..22P}, evidencing that water can be abundant in sub-Neptunes, and that it can be allocated in their atmospheres. 

Today, the exploration of sub-Neptune atmospheric composition is still taking its first steps. The recent detections of both hydrogen-dominated and water-dominated atmospheres suggest that both gas dwarfs and `steam' water worlds may co-exist in the upper mode of the bi-modal radius distribution (i.e in the sub-Neptune sub-population). Therefore, unveiling whether there is a predominance of hydrogen-rich or water-rich atmospheres in the sub-Neptune population and whether water could also be allocated in other regions of the planetary structure requires of additional atmospheric, mass and radius measurements at a population level, for which the exoplanet community is investing a lot of resources \citep[e.g. the THIRSTEE program, PI Luque; see also][]{2024A&A...692A.238L}. In chapters~\ref{ch:k2_ojos} and \ref{ch:toi-244}, we delve into the small planet's internal structure problem through the validation of K2 transit-like signals, and the precise characterisation of the super-Earth TOI-244~b and its contextualization at a population level, respectively. 

\section{Motivations and objectives}

The primary motivation of this thesis aligns with one of the main goals of exoplanet research: advancing our understanding of the formation and evolution of planetary systems. Our comprehension of how planets form and evolve has long been limited to the Solar System. Since the discovery of 51~Peg~b, the plethora of planet properties has apparently increased the complexity of the problem. However, this wealth of data allowed us to put relevant constraints on the processes leading to the observed planet populations. As introduced in Sect.~\ref{exoplanet_populations}, to reach a deep understanding of this fundamental problem, both theoretical and observational efforts are necessary. This thesis focuses on the latter.

Constraining planet formation and evolution theories requires a large number of well-characterised planets. Large-scale planet detections are feasible thanks to the new generation of space-based photometers (Sect.~\ref{sec:space_based_photometers}). However, in-depth planet characterisations (e.g. mass measurement, and atmospheric composition) typically require large investments of observing time, more specific analyses, and are limited to a selected number of relatively bright targets, preventing us from making these characterisations on a large-scale basis (Sect.~\ref{sec:ground_based_spectrographs}). Notably, both high-volume detections (with a limited characterisation) and in-depth characterisations (with a volume-limited yield) are critical to constrain different formation and evolution theories (see Sect.~\ref{exoplanet_populations}). As introduced in Sect.~\ref{sec:space_based_photometers}, the transition between \textit{Kepler}/K2 and TESS also marked a transition in the main focus of exoplanet research.

This thesis has been developed in the transition between \textit{Kepler}/K2 and TESS and aims to take advantage of the excellent opportunities offered by these surveys. In this regard, the main objective consisted of delving into the planet formation and evolution problem through the exploitation of \textit{Kepler}/K2 and TESS data in combination with ground-based follow-up observations, serving as a preparation for the PLATO mission. This endeavour was done both through large-scale planet detections (based on K2) and precise characterisations of planets in poorly explored niches (based on TESS and ultra-stable high-resolution spectrographs). The exploitation of K2 data was carried out in the context of the K2-OjOS collaboration (Sect.~\ref{sec:intro_kepler}), and the exploitation of TESS and ground-based spectroscopic data was carried out in the course of the ESPRESSO consortium (Sect.~\ref{espresso-consortium}) and the HARPS-NOMADS collaboration (Sect.~\ref{harps-nomads}). This general objective is divided into different specific goals:

\begin{itemize}

  \item Searching for transit-like signals in K2 C18 through an automated algorithm and conducting injection-recovery tests to compare with the K2-OjOS members' recovery performance
  
  \item Building a sample of high-quality planet candidates from the signals found based on the vetting diagnostics prepared by the K2-OjOS members
  \item Characterising the candidate K2 hosts through archival photometric and spectroscopic data
  \item Modelling the K2 transits through all available photometry (i.e. C18, and C5 and C16 when available), aiming at the maximum possible precision of the system parameters and ephemeris
  \item Studying the presence of contaminant stars within the K2 photometric apertures
  \item Using the relevant data to subject the K2 candidates to statistical validation analyses, and subsequently labelling them as validated planets (VP), planet candidates (PC), or false positives (FP)
  \item Contextualising the new K2 validated planets (if any) in the currently known exoplanet populations and identifying interesting targets for follow-up observations

  \item Characterising the TESS host star's properties and rotation-induced activity signals, which could be contaminating the measured RVs (especially in the M-dwarf TOI-244)

   \item Performing RV model comparison analyses through Bayesian Inference to select the models that best represent our datasets, and use them to infer the orbital and physical properties of the systems 

   \item Contextualising the inferred properties of the studied TESS planets in the known sample of small rocky planets (in the case of TOI-244~b) and hot Neptunes (in the case of TOI-5005~b), and discuss how these properties agree or disagree with the current formation and evolution theories  
  
\end{itemize}

In addition to the original goals enumerated above, during our investigations, a set of new objectives emerged. The super-Earth TOI-244~b resulted in having a bulk density lower than expected for an Earth-like rocky composition. This unexpected feature motivated us to further explore the properties of the currently known sample of unusually low-density super-Earths and try to gain insight into their internal compositions. In particular, we aimed at investigating whether the properties of these planets are correlated with the chemical compositions of their stellar hosts (which would suggest a formation origin) or the stellar irradiation conditions (which would suggest an evolutionary origin). 

Regarding the super-Neptune TOI-5005~b, its orbital properties placed it inside the original boundaries of the Neptunian desert. However, we identified that the current exoplanet sample in this region of the parameter space is far from being a desert, which led us to put into question the accuracy of the original boundaries. Determining whether a planetary system lies within the Neptunian desert (and any other desert of the parameter space) is critical to properly interpret its possible formation and evolutionary pathway (see Sect.~\ref{sec:intermediate}). Therefore, this situation motivated us to delineate the boundaries of the desert by following a population-based approach. This allowed us to mitigate the biases associated with the observed planet distribution and, in turn, be able to properly contextualise TOI-5005~b (and any other close-in Neptune) within the underlying exo-Neptunian landscape. In this process, we identified an overabundance of planets in a particular region of the parameter space. This result further motivated us to unveil whether these planets have particular properties (e.g. densities, orbital eccentricities, and obliquities) that would hint at specific formation or evolutionary processes. Another by-product of the initial investigation was the identification of a periodic photometric TESS signal synchronised with the orbital period of the hot Jupiter HD 118203~b, which motivated us to investigate its origins through testing the MSPI hypothesis (see Sect.~\ref{sec:inflation_mspi}). In all, the original objectives of this thesis were successfully addressed, and led to different unexpected challenges that motivated us to define new objectives and investigations with the ultimate goal of maximising our understanding of both the particular properties of the studied systems and the populations they belong to. 

In the following chapters \ref{ch:k2_ojos}-\ref{ch:toi_5005}, we present the main results of this thesis. These chapters are based on peer-reviewed articles and are self-consistent, as they contain brief introductions to the specific studied problems, descriptions of the data and their analysis, as well as the results and conclusions. In particular, in Chapter~\ref{ch:k2_ojos}, we present the results of the K2-OjOS collaboration. In Chapter~\ref{ch:toi-244}, we present the detection characterisation of the TOI-244 planetary system and a study of the growing population of low-density super-Earths. In Chapter~\ref{ch:mspis}, we present the detection of signs of MSPIs in the HD~118203 planetary system. In Chapter~\ref{ch:nep_des} and Chapter~\ref{ch:toi_5005}, we present a planet occurrence study across the exo-Neptunian landscape and the detection and characterisation of the close-in super-Neptune TOI-5005~b, respectively. Finally,  Chapter~\ref{ch:int_discussion} contains a general discussion that integrates the results and discussions of the individual chapters, and we conclude in Chapter~\ref{ch:conclusions}.



\newpage
\chapter{
The K2-OjOS Project: New and revisited planets and candidates in K2}
\label{ch:k2_ojos}
\vspace{2cm}
\pagestyle{fancy}
\fancyhf{}
\lhead[\small{\textbf{\thepage}}]{\small{\textbf{\nouppercase{\leftmark}}}}
\rhead[\small{\textbf{\nouppercase{\rightmark}}}]{\small{\textbf{\thepage}}}

\bigskip

\bigskip


\newcommand{\totalinspectedlightcurves}{20\,427 }


\newcommand{\totalsignals}{42 }
\newcommand{\totaltargets}{37 }

\newcommand{\newsignals}{18 }
\newcommand{\knownsignals}{24 }



\newcommand{\newVP}{four }
\newcommand{\newPC}{14 }


\newcommand{\medianfactorP}{30}
\newcommand{\medianfactorTo}{1.6 }
\newcommand{\medianfactorrprs}{1.4 }

\newcommand{\planetsandcandidatestoimprove}{22}
\newcommand{\targetstoimprove}{22}

\newcommand{\updatedsignals}{24}

\newcommand{\signalsKOJOSpercentage}{70}

\newcommand{\starsCfive}{25\,137 }
\newcommand{\starsCsixteen}{29\,888 }
\newcommand{\starsCeigtheen}{20\,427 }

\newcommand{\starsCfiveeighteen}{11\,444 }
\newcommand{\starsCfivesixteeneighteen}{3261 }


\begin{figure}
    \centering
    \includegraphics[width = 0.70\textwidth]{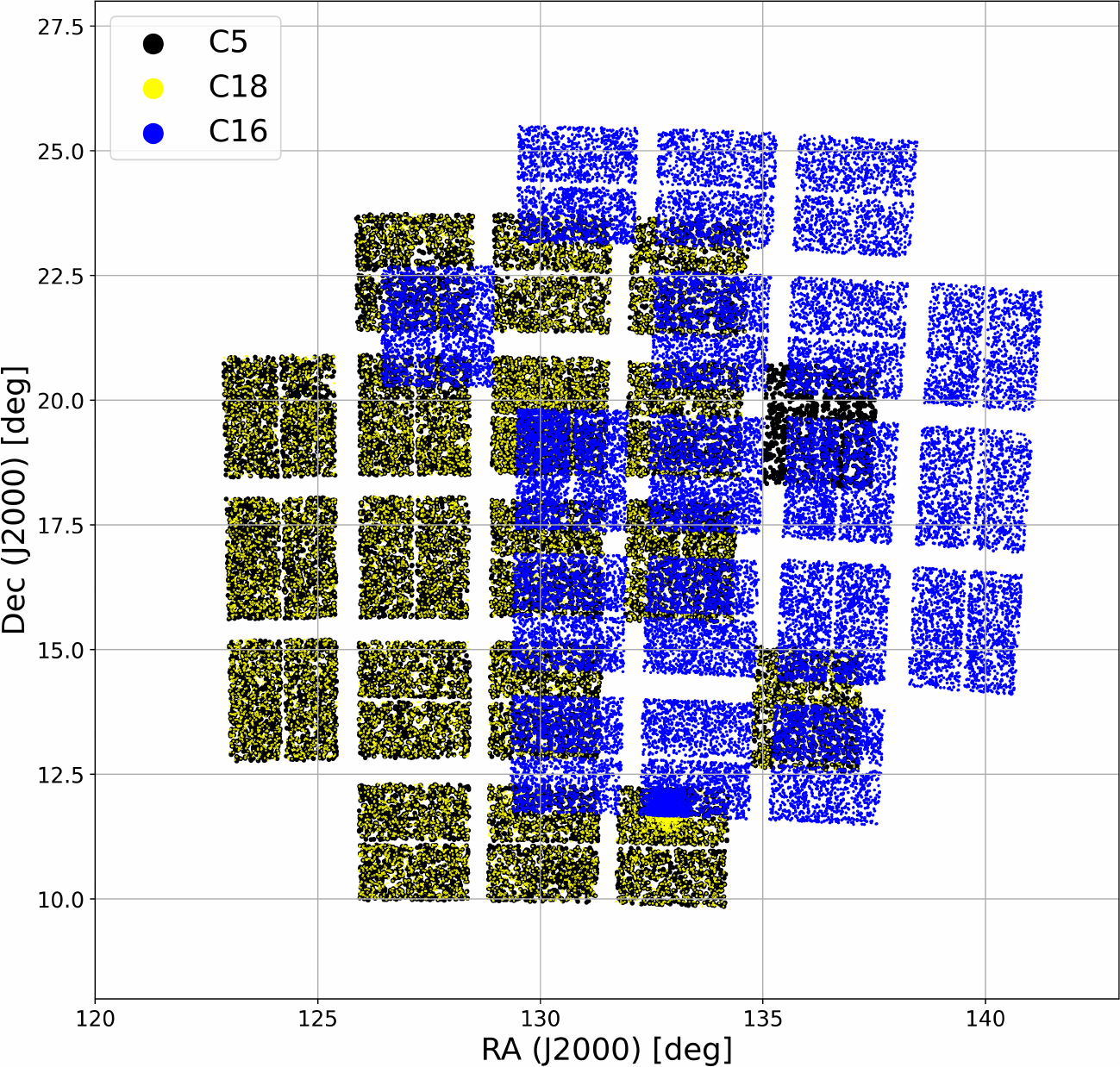}
    \caption[Sky position of the K2-OjOS stellar targets observed in K2 C5, C16, and C18.]{Sky position of the stellar targets observed in C5, C16, and C18.}
    \label{fig:C5_C16_C18_fields}
\end{figure}

According to the NASA Exoplanet Archive \citep{2013PASP..125..989A}, 426 validated and confirmed planets have been detected using K2 data \citep[e.g.][]{2016ApJS..226....7C,2018A&A...620A..77L,2018MNRAS.476L..50D,2018MNRAS.480L...1D,2019MNRAS.489.5928D,2019MNRAS.482.1807K,2020MNRAS.499.5416C,2021MNRAS.508..195D}, representing 10$\%$ of the total known planets. There are also nearly a thousand K2 candidates, which either do not meet the imposed validation criteria \citep[e.g. candidates reported in][]{2015ApJ...809...25M,2018AJ....156...78L,2018AJ....155..136M}, or are awaiting for validation analyses \citep[e.g. many candidates reported in][]{2016A&A...594A.100B,2016MNRAS.461.3399P,2016ApJS..222...14V,2018AJ....156...22Y,2019ApJS..244...11K}. Even though a fairly extensive analysis of K2 data has been carried out so far, several campaigns still can be studied in greater detail.

In general, K2 fields are uniformly distributed along the ecliptic, so most of the targets were observed in just one campaign  (i.e. $\sim$80 d; see Sect.~\ref{sec:intro_kepler}). Nevertheless, certain fields partially overlap, providing unique science opportunities due to the longer duty cycles and temporal baselines.
The most optimal campaigns to take advantage of the existing overlaps in K2 fields are C5 (observations from 27 April 2015 to 10 July 2015), C16 (observations from 7 December 2017 to 25 February 2018), and C18  (observations from 12 May 2018 to 2 July 2018), which observed \starsCfive (C5), \starsCsixteen (C16) and \starsCeigtheen (C18) stellar targets in the  K2 long-cadence mode (30 min). C18 field covers 95$\%$ of C5 field (see Fig. \ref{fig:C5_C16_C18_fields}) and both campaigns have \starsCfiveeighteen stellar targets in common, which were observed with a 3-yr temporal baseline and a 4-month duty cycle. The C16 field covers 30$\%$ of C5 and C18, and the three campaigns have \starsCfivesixteeneighteen stellar targets in common, increasing their duty cycle up to nearly 7 months. Joining photometric data from these three campaigns allows us to search for long-period planets not being identified in data from just one campaign, search for long-term transit timing variations (TTVs), and precisely measure the transit ephemeris, which is essential for future follow-up studies scheduled by the next ground- and space-based telescope generation. Although several works have been published reporting candidates, validated and confirmed transiting planets starting from C5 and C16 data \citep[e.g.][]{2017AJ....154..207D,2017AJ....153...64M,2018AJ....156..277L,2018AJ....155..136M,2018AJ....155...21P,2018AJ....156...22Y}, there are still no works aimed to an exhaustive analysis of C18.

In this context, we started K2-OjOS\footnote{ \url{https://sites.google.com/view/k2-ojos/english}} \citep[][]{2020sea..confE..97C}, a Professional-Amateur (Pro-Am) project in which 10 amateur astronomers visually inspected the light curves of each C18 star in order to: (1) detect, characterize, and validate new extrasolar planets, (2) revisit the orbital and physical parameters of planets detected in C5 with new data from C16 and C18, (3) detect variable stars, and (4) compare visual and automated detection procedures (see more details on the collaboration in Sect.~\ref{sec:intro_kepler}).

\begin{figure*}
    \centering
    \includegraphics[width=0.88\textwidth]{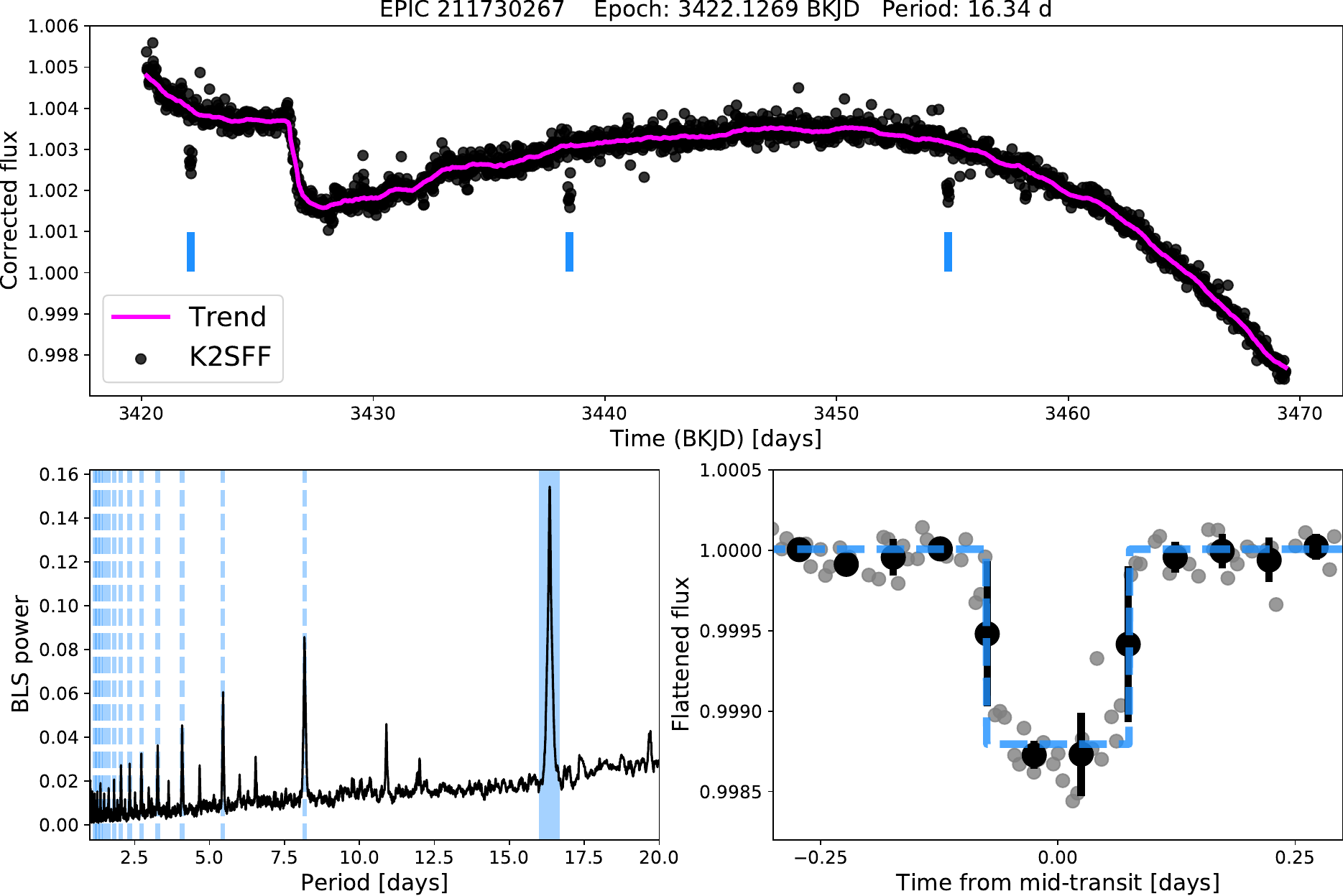}
    \caption[Summary plot of the preliminary vetting carried out by the K2-OjOS members for the newly detected planet K2-357 b (EPIC 211730267.01).]{Summary plot of the preliminary vetting carried out by the K2-OjOS members for the newly detected planet K2-357 b (EPIC 211730267.01). Top: Long cadence K2SFF-corrected light curve along with the trend line. Blue vertical lines indicate the transit locations. Bottom left-hand panel: Box Least Squares Periodogram in which the main period ($P$ = 16.34 d, blue vertical line) and its harmonics (blue vertical dashed lines) are highlighted. Bottom right-hand panel: Phase-folded transit together with the BLS square model (0.12$\%$ depth and 3.6 h duration). Black dots correspond to 1.2-h binned data.}
    \label{fig:K2-OjOS_summary_plot}
\end{figure*}

In Sect.~\ref{project}, we describe the K2-OjOS project, including the signal detection methodology, preliminary vetting, independent search through the BLS algorithm, and the transit injection and recovery. In Sect.~\ref{sec:Observations}, we describe the observations and data reduction for both the K2 photometry and the high-resolution imaging. In Sect.~\ref{analysis}, we present the analysis, in which we describe the stellar characterisation of the host stars, transit modelling, search for transit timing variations and planet validation. In Sect.~\ref{results_discussion}, we present the results, in which we quantify the transit ephemeris refinement of the already known planets, compare the recovery rates of both the K2-OjOS and BLS searches, contextualize and discuss the characteristics of the host stars, planets and candidates in our sample, and highlight interesting features of five individual systems. We summarise our results and conclude in Sect.~\ref{summary}.

\section{The K2-OjOS project}
\label{project}
\subsection{Visual search for planetary signals}
\label{visual_search}

The K2-OjOS members had previous knowledge about the typical photometric features of stars hosting
transiting exoplanets (i.e. periodicities, transit shapes, depths, and durations), as well as about the most recognisable types of variable stars. Furthermore, we developed tutorials with guidelines for proper identification, which are available on the K2-OjOS website. We also established an online chat and scheduled frequent meetings to allow communication between K2-OjOS members and professional astronomers. 

We distributed the C18 20\,427 target stars in 21 batches (20 with 1000 light curves and 1 with 427). Then, we assigned one batch to each K2-OjOS member, and they systematically inspected the K2 Self Flat Fielding (K2SFF) corrected light curves \citep{2014PASP..126..948V} available on the Center for Astrophysics of the Harvard University and Smithsonian Institution website.\footnote{\url{https://www.cfa.harvard.edu/~avanderb/k2.html}} As they finished inspecting their batches, we assigned them more, until reaching all 21 batches. As a result, six members analysed one batch and four members analysed more than one (2, 3, 4, and 6 batches).  The members carried out a preliminary classification of the stars with detected variability, discerning between the following main categories: \texttt{planets}, \texttt{eclipsing binaries}, \texttt{rotatings}, \texttt{pulsatings}, \texttt{irregulars}, and \texttt{artefacts}. Finally, they did a double check, exchanging the targets found with a different member. The results were the following: 216 \texttt{planets}, 427 \texttt{eclipsing binaries}, 374 \texttt{rotatings}, 288 \texttt{pulsatings}, 195 \texttt{irregulars}, and 101 \texttt{artefacts}. The amateur astronomers, jointly with professional astronomers, subjected the 216 \texttt{planets} to a thorough vetting in order to avoid false detections or misclassifications and thus obtain the final sample of planet candidates (Sect.~\ref{preliminary_vetting}).

\subsection{Preliminary vetting and signal selection}
\label{preliminary_vetting}
The K2-OjOS members performed a preliminary vetting of the 216 targets classified as \texttt{planet} in the visual searching step, in order to create, together with professional astronomers, a high-quality planet candidate sample. In the following, we briefly summarise the procedure, which is fully available on the K2-OjOS website. (1) The members checked if the targets were observed in the overlapping C5 and C16, and if so, they checked that the signals found were also present in those campaigns. (2) They removed outliers and long-term trends from the light curves through the \textsc{wotan} package \citep[][]{2019AJ....158..143H}, being cautious of not overfitting nor removing relevant data. (3) They assessed if the signals were periodic, and if so, computed relevant signal features as the orbital periods, mid-transit times, transit durations, and transit depths through the BLS algorithm \citep[][]{2002A&A...391..369K}. (4) They produced phase-folded light curves to check their shapes. During this process,  we removed from the \texttt{planet} category 174 targets (80$\%$ of the  \texttt{planet} targets), which were mostly misclassified eclipsing binaries or artefacts. As a result, we ended up with 42 planet candidates in 37 stars, of which 18 are new detections. In Fig. \ref{fig:K2-OjOS_summary_plot}, we show a summary plot of this process created by the K2-OjOS team for the newly detected planet K2-357 b.  

\subsection{Independent search through BLS algorithm}

We carried out an independent search by applying the BLS algorithm to the whole C18 sample, in order to cross-check the search. We used the BLS implementation of the \textsc{lctools} software \citep[][]{2019arXiv191008034S} considering low-restrictive signal properties: signal-to-noise ratio $>$ 6, planet size $>$ 0.5 $\rm R_{\oplus}$, orbital period $>$ 0.25 d and transit duration $>$ 1 h. As a result, the BLS search retrieved all the findings made by the K2-OjOS members except single transits, and no extra candidate was found. 

\subsection{Transit injection and recovery}
\label{sec:injection_recovery}

To quantify and compare the K2-OjOS and BLS detection efficiencies, we simulated transit signals and injected them into real K2 light curves. In the following, we detail the transit injection procedure. 

First, we randomly selected 200 K2SFF corrected light curves of C18 stars that were analysed by \citet{2020ApJS..247...28H}, in order to acquire homogeneous estimates of their stellar parameters. Secondly, we exchanged the light curves with real planetary signals or eclipsing binaries for light curves without any hint of having eclipses. Thirdly, we randomly injected one simulated transit in 40$\%$ of the 200 selected light curves generated through the \citet{2002ApJ...580L.171M} quadratic transit model as implemented in the \textsc{batman} package \citep{2015PASP..127.1161K}. We adopted stellar masses and radii from the \citet{2020ApJS..247...28H} catalogue and fixed the orbital inclination to $i=90^{\circ}$ and eccentricity to $e=0$. We randomly generated the orbital periods between 1 d and the 49-d temporal baseline of C18, the $R_{p} / R_{\star}$ ratios between 1$\%$ and 10$\%$, and the mid-transit times between the C18 starting time and one orbital period later. We repeated the process, generating different batches of 200 light curves with different injected transits. 

Each K2-OjOS member analysed the same number of batches as the number of original analysed batches with real C18 data.  Regarding the BLS search, the necessary condition for a signal to be considered as retrieved is that any of the 10 main BLS periods have to be within 1$\%$ of the injected orbital period. The results of both independent searches are shown in Sect.~\ref{sec:detection_efficiency}. 


     

\section{Observations and data reduction}
\label{sec:Observations}

\subsection{K2 photometry extraction and processing}
\label{K2_photometry}

\begin{figure*}
    \centering
    \includegraphics[scale=0.38]{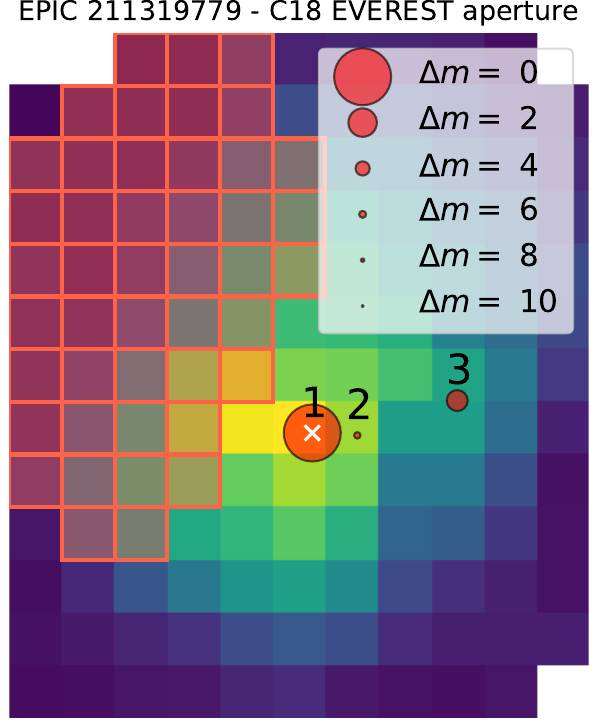}
    \includegraphics[scale=0.38]{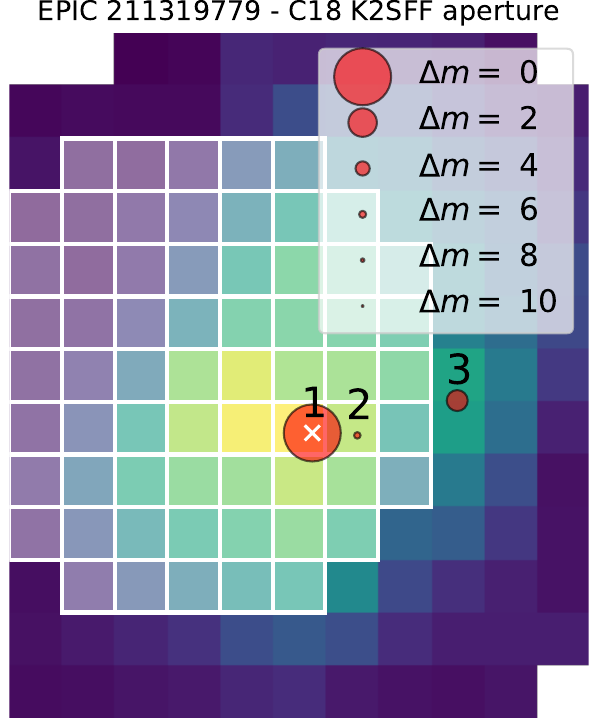}
    \includegraphics[scale=0.14]{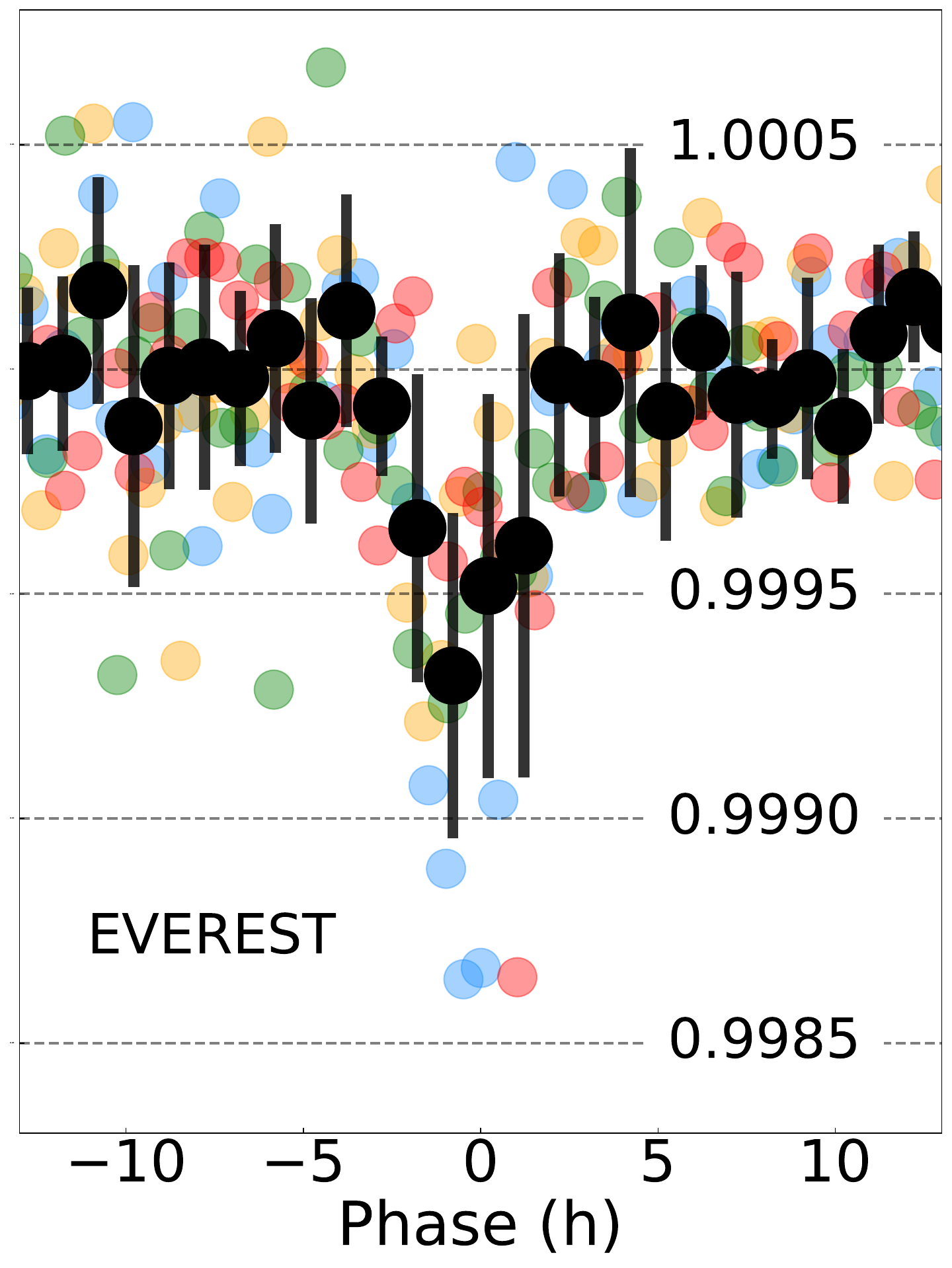}
    \includegraphics[scale=0.14]{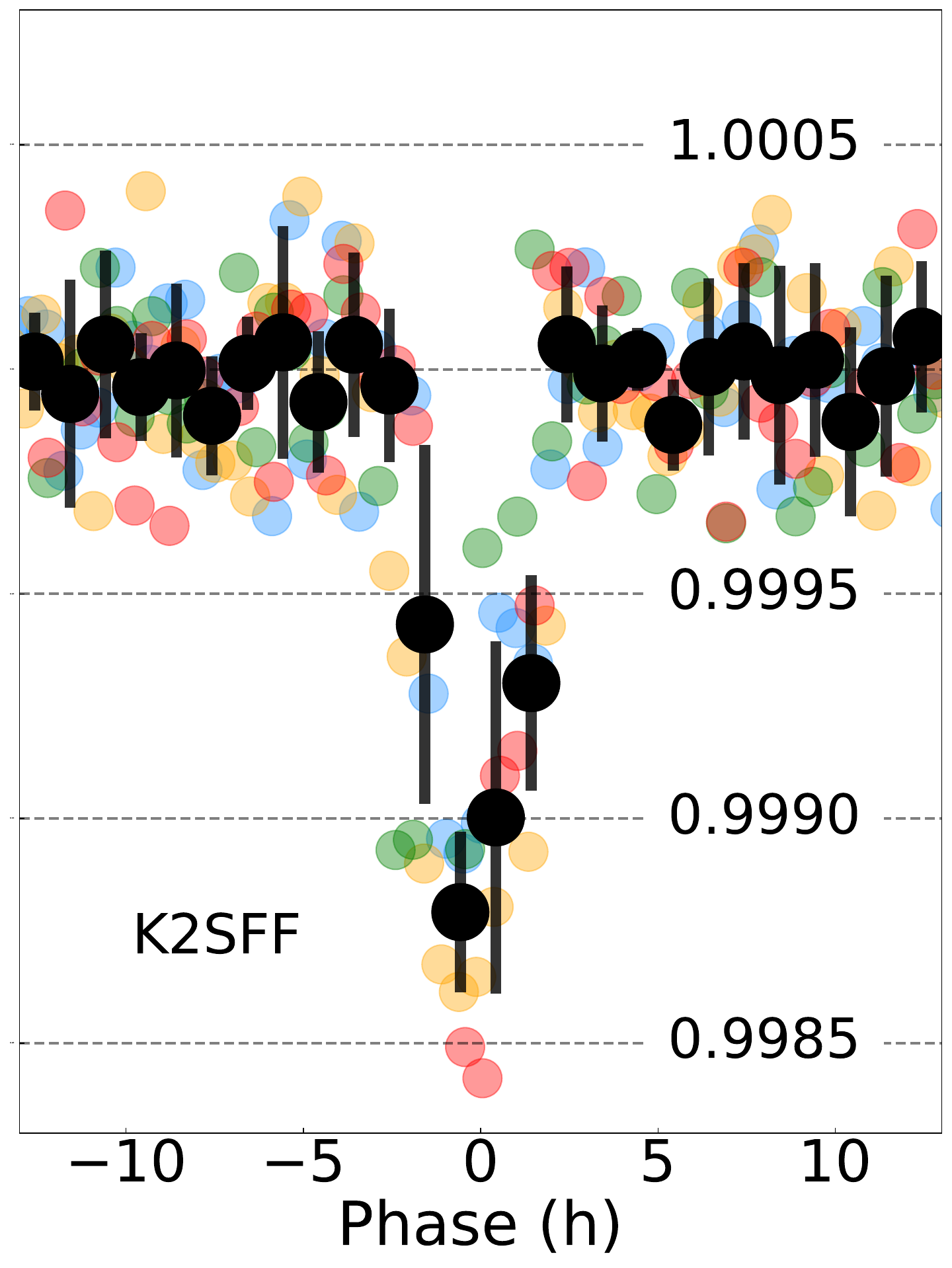}
    \caption[Photometric apertures for EPIC 211319779 corresponding to the EVEREST and K2SFF pipelines, and phase-folded light curves obtained with each pipeline.]{Left panels: Photometric apertures for EPIC 211319779 (star $\#$1) corresponding to the EVEREST and K2SFF pipelines. Both plots have been made through \textsc{tpfplotter}. Right panels: Phase-folded light curves obtained with each pipeline. For both plots, the first transit is plotted in blue, and the second, third, and fourth are in orange, green, and red, respectively. The black dots correspond to 1-h binned data.}
    \label{fig:EVEREST_vs_K2SFF}
\end{figure*}

We downloaded from the Mikulsky Archive for Space Telescopes (MAST) the C18 (and C5 and C16 when available) light curves processed with the EVEREST pipeline \citep{2016AJ....152..100L,2018AJ....156...99L} for the \totaltargets targets found in the K2SFF light curves by the K2-OjOS team, to compare both pipelines and decide which one to use in the subsequent analysis. We checked that the C18 transit-like signals found starting from the K2SFF light curves were also detectable in the overlapping campaigns and the EVEREST light curves. We found three cases (EPICs 211319779, 211606790, and 211407755) for which the EVEREST photometric apertures are not suitable to collect completely the stellar fluxes, as part of the edges of these apertures are located over the target PSFs. Figure \ref{fig:EVEREST_vs_K2SFF} illustrates this phenomenon for EPIC 211319779, for which the unsuitable EVEREST aperture decreases the C18 phase-folded transit depth by a factor of two. In this particular case, the first EVEREST transit corresponds to a situation in which the target is located inside the aperture and skimming its edge. However, over the course of the campaign, the aperture is gradually separating from the target, causing a great flux dimming for the remaining three transits. Besides, we found one target (EPIC 212008766, $G_{\rm mag}$ = 12.87) whose C5 EVEREST photometry is contaminated, as its aperture, unlike that of K2SFF, encompasses the nearby star Gaia DR2 664406705976755840 ($G_{\rm mag}$ = 15.16). This shows the importance of a thorough cross-comparison between different photometric pipelines in order to assess the quality of the signals. Due to the smaller out-of-transit scatter obtained with the EVEREST pipeline, we used its light curves for the subsequent analysis, except for the aforementioned four targets. For EPICs 211319779, 211606790, and 212008766, we used the K2SFF, and for EPIC 211407755, whose K2SFF aperture edge is also too close to the target, we applied the SFF corrector to the raw flux collected in a bigger aperture.

We removed long-term trends and normalised each light curve by using the \textsc{wotan} package \citep{2019AJ....158..143H}. Choosing a suitable detrending procedure is highly important as it has a direct impact on the transit depth, and therefore on the derived planetary radius. First, we removed upper outliers that are more than 5-$\sigma$ above the running mean. Then, we used the robust time-windowed bi-weight method while cutting off the extremes of each time series in order to avoid edge effects. We note that this method is not suitable for some of our targets, whose transit signals would disturb the trend (i.e. there is no window-length capable of reproducing the long-term trend while not being distorted by the transits). This phenomenon typically occurs in stars with strong stellar rotation modulations (e.g. EPICs 211335816, 211418290, and 211424769). In consequence, for these targets we used the cosine de-trending method \citep{2019AJ....158..143H} by following a two-step process. First, we performed a preliminary detrending in which we used the Transit Least Squares algorithm \citep[TLS;][]{2019A&A...623A..39H} to identify transit-like signals that could disturb the trend. Secondly, we detrended the original time series, masking the transit data points found in the first step.

After the de-trending process we joined the light curves from different campaigns generating a single light curve per target, and then used the TLS to obtain constraints on $P$ and $T_{0}$ to be used in the posterior transit modelling, as well as to search for additional shallow transit-like signals within the joined data.

\subsection{High-resolution imaging}
\label{sec:AstraLux}

We observed the four stars that meet all the validation criteria described in Sect.~\ref{validation} (EPICs 211730267, 211914998, 211525753, and 211537087) with the high-spatial resolution camera AstraLux \citep{hormuth08}, located at the 2.2\,m telescope of the Calar Alto Observatory (Almería, Spain) on the nights of 22 and 23 of 2021 March. This camera uses the lucky-imaging approach to obtain diffraction-limited images based on the observation of a large number of frames with very short exposures below the coherence time. We used the  Sloan Digital Sky Survey z filter (SDSSz) and obtained 60\,000 frames with 20\,ms exposure times for EPIC\,211730267, 60\,000 $\times$ 30\,ms for EPIC\,211914998, 18\,400 $\times$ 30\,ms for EPIC\,211525753, and 23\,400 $\times$ 20\,ms for EPIC\,211537087. In order to focus on the closer regions, we restricted the field-of-view by windowing to $6\times6$ arcseconds.

\begin{figure}
    \centering
    \includegraphics[width = 0.6\textwidth]{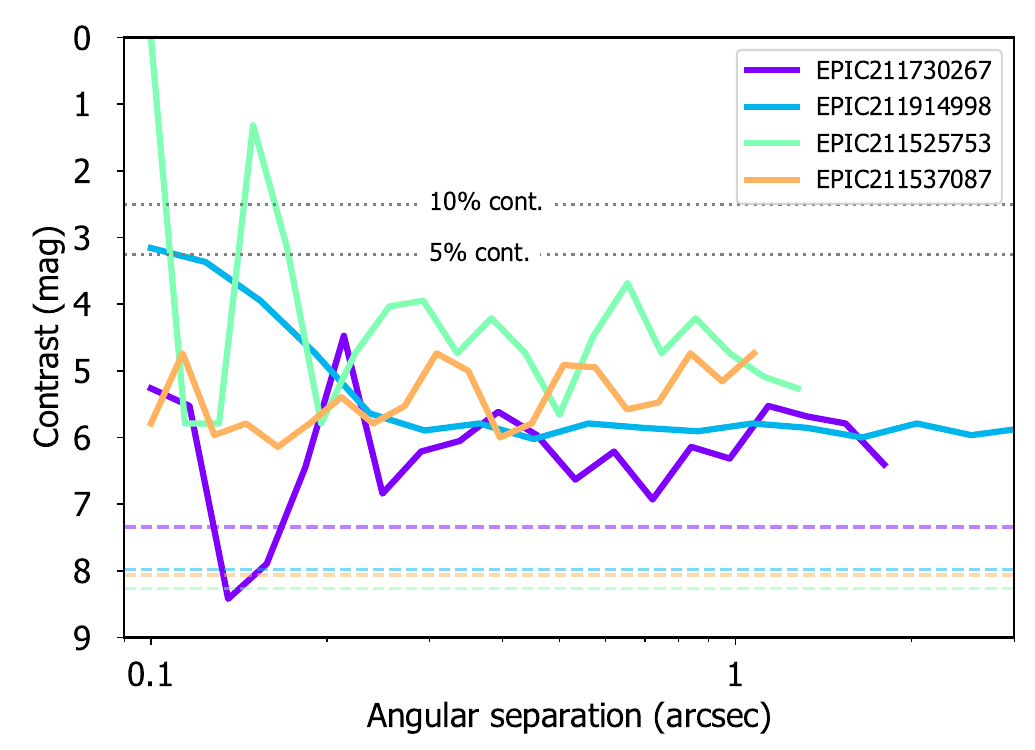}
    \caption{Contrast curves obtained from the AstraLux combined frames.}
    \label{fig:astralux}
\end{figure}

We used the instrument pipeline to select the 10$\%$ frames with the highest Strehl ratio \citep{strehl1902} and combine them into a final high-spatial resolution image. Based on this final image, we computed the sensitivity curve by using our own developed \textsc{astrasens} package\footnote{\url{https://github.com/jlillo/astrasens}} with the procedure described in \cite{lillo-box12,lillo-box14b}. We find no evidence of additional sources within this field of view and within the computed sensitivity limits, shown in Fig. ~\ref{fig:astralux}. 

We use these contrast curves to estimate the probability of contamination from blended sources in the K2 aperture and undetectable from the public images. This probability is called the blended source confidence (BSC), and the steps for estimating it are fully described in \cite{lillo-box14b}. We use a \textsc{python} implementation of this approach (\textsc{bsc}) which uses the \textsc{trilegal}\footnote{\url{http://stev.oapd.inaf.it/cgi-bin/trilegal}} Galactic model \citep[v1.6;][]{girardi12} to retrieve a simulated source population of the region around the corresponding target.\footnote{This is done in \textsc{python} by using the \textsc{astrobase} implementation by \cite{astrobase}.} For instance, the transit signal in EPIC\,211730267 could be mimicked by a blended chance-aligned binary with a magnitude contrast up to $\Delta m=6.8$~mag. However, given the high-resolution image, we estimate using the \textsc{bsc} code that the probability of having an undetected source in the AstraLux image within this contrast range is 0.15$\%$. Similarly, we find such probability to be 2.83$\%$ for EPIC\,211914998, 0.34$\%$ for EPIC\,211525753, and 0.15$\%$ for EPIC\,211537087.

\section{Analysis}
\label{analysis}

\subsection{Stellar characterization}
\label{stellar_charact}

Obtaining reliable stellar parameters is highly important since planetary parameters and statistical validation analysis depend on them. The Ecliptic Plane Input Catalogue \citep[EPIC; ][]{2016ApJS..224....2H}, whose stellar parameters were based on photometry, proper motions, and models of the distribution of stars in the Milky Way, misclassifies between 56 and 72$\%$ of subgiants as dwarfs, and 9$\%$ of dwarfs as subgiants \citep{2016ApJS..224....2H}. Besides, it underestimates the radii for low-mass stars, as a result of the choice of the isochrones from the Padova data base \citep{2008A&A...482..883M}, which tend to under-predict the radii of these stars \citep{2012ApJ...757..112B,2016ApJS..224....2H}. For M dwarf stars, this bias has been empirically estimated to be 39$\%$ by \citet{2017ApJ...836..167D} and 43$\%$ by \citet{2020MNRAS.499.5416C}. Given that $\sim$40$\%$ of selected K2 targets are low-mass M and K dwarfs \citep{2016ApJS..224....2H}, improving the stellar radii estimates of these targets is crucial to accurately characterise the planets observed by K2. In this section, we detail our procedure to infer the radii ($R_{\rm \star}$), masses ($M_{\rm \star}$), effective temperatures ($T_{\rm eff}$), surface gravities (log $g$), and metallicities ([Fe/H]) for the stars in our sample.

Among the \totaltargets targets, 21 have published spectroscopic parameters derived from different spectrographs and pipelines, and 16 lack spectra. Due to this heterogeneity, we performed an independent and uniform stellar characterisation utilising the \textsc{isochrones} package \citep{2015ascl.soft03010M}. The package is an interpolation tool that fits photometric and/or spectroscopic parameters to the MIST (MESA Isochrones Stellar Tracks) stellar models \citep{2015ApJS..220...15P,2016ApJ...823..102C,2016ApJS..222....8D} by using \textsc{multinest} \citep{2008MNRAS.384..449F,2009MNRAS.398.1601F,2019OJAp....2E..10F}, and thus it can predict the value of any physical property derived by the models.
We ran \textsc{isochrones} by using the following data for all our \totaltargets target stars: 2MASS \textit{JHK} photometry \citep{2006AJ....131.1163S} and \textit{Gaia} DR2 parallaxes \citep{2018A&A...616A...1G}, accounting for the systematics reported in \citet{2018ApJ...862...61S} and \citet{2018A&A...616A...9L}. Including \textit{Gaia} parallaxes within the \textsc{isochrones} analysis has proven to be of crucial importance in order to remove most of the potential for misclassifying dwarfs and subgiants, as well as to correct the aforementioned radii underestimation, specially for stars lacking spectroscopy \citep{2018AJ....156..277L}. We included additional priors of inferred spectroscopic parameters $T_{\rm eff}$, [Fe/H], and log $g$ for the 21 targets with published spectra, and for three more targets (EPICs 211335816, 211407755, 211816003) without spectra but with $T_{\rm eff}$, [Fe/H], and log $g$ available from \citet{2020ApJS..247...28H}; the authors derived stellar parameters for 195\,250 K2 targets by using random forest regression on photometric colours, trained on a sample of 26\,838 K2 stars with spectroscopic measurements from the Large Sky Area Multi-Object Fibre Spectroscopic Telescope \citep[LAMOST;][]{2012RAA....12.1197C} DR5.  The resulting \textsc{isochrones}-derived stellar parameters, together with the references of the spectroscopic priors, are shown in Table \ref{tab:stellar_parameters}. 

\begin{figure}
\centering
    \includegraphics[width = 0.6\textwidth]{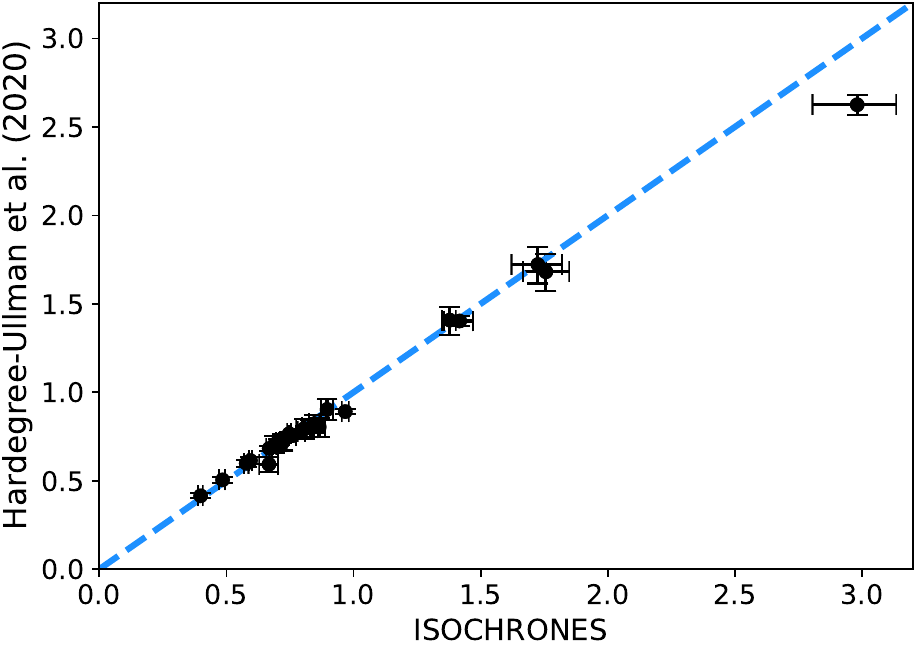}
    \caption[Comparison between \textsc{isochrones} and Hardegree-Ullman et al. (2020) stellar radii.]{Comparison between \textsc{isochrones}-derived and \citet{2020ApJS..247...28H} reported stellar radii ($R_{\star}$). The dashed line represents the 1:1 relation.}
    \label{fig: R_Isoc_vs_HU}
\end{figure}

In order to check the consistency of our results and search for possible outliers, we compared the \textsc{isochrones}-derived stellar parameters with those obtained by \cite{2020ApJS..247...28H} for the 23 common targets in both samples.  In Fig. \ref{fig: R_Isoc_vs_HU}, we plot the comparison between stellar radii, which are typically consistent at the 1-$\sigma$ level, showing a strong agreement between both independent analyses. There is however an outlier (EPIC 211418290, the largest star of our sample), for which we derive $R_{\star}$ = $3.02^{+0.17}_{-0.16}$ $\rm R_{\odot}$ by using \citet{2018AJ....155..136M} spectroscopic values as priors, while \citet{2020ApJS..247...28H} derive $R_{\star}$ = $2.625^{+0.058}_{-0.055}$ $\rm R_{\odot}$, which are consistent at the 2-$\sigma$ level.

\renewcommand{\arraystretch}{2.2}
\begin{table*}
\fontsize{7.7pt}{7.7pt}\selectfont
\caption[Stellar parameters of the K2-OjOS sample.]{\textsc{isochrones}-derived parameters.}
	\begin{center}
		\begin{tabular}{c|c|c|c|c|c|c|c|c}
		    \hline \hline
			EPIC ID & $C$ & $K_p$ (mag) & $T_{\rm eff}$ (K) & [Fe/H] (dex) & log $g$ (cgs) & $R_{\star}(\rm R_{\odot})$ & $M_{\star}(\rm M_{\odot})$ & Spectroscopic parameters \\
			\hline
			211309648 & 18 & 13.848 & $6283^{+235}_{-232}$ & $0.05^{+0.15}_{-0.16}$ & $4.18^{+0.07}_{-0.07}$ & $1.50^{+0.15}_{-0.12}$ & $1.24^{+0.10}_{-0.08}$ & -- \\
			211317649 & 18 & 13.214 & $5693^{+62}_{-63}$ & $0.20^{+0.07}_{-0.07}$ & $4.38^{+0.05}_{-0.04}$ & $1.08^{+0.05}_{-0.05}$ & $1.03^{+0.04}_{-0.04}$ & \cite{Boisse_2013} \\
			211319617 & 5,18 & 12.393 & $5306^{+45}_{-40}$ & $-0.50^{+0.07}_{-0.06}$ & $4.57^{+0.03}_{-0.03}$ & $0.72^{+0.01}_{-0.01}$ & $0.72^{+0.04}_{-0.03}$ & \cite{2018AJ....155..136M} \\
			211319779 & 18 & 12.640 & $5163^{+161}_{-127}$ & $-0.20^{+0.14}_{-0.17}$ & $4.58^{+0.02}_{-0.03}$ & $0.75^{+0.02}_{-0.02}$ & $0.78^{+0.04}_{-0.05}$ & -- \\
			211335816 & 5,18 & 11.929 & $6223^{+105}_{-134}$ & $0.03^{+0.09}_{-0.12}$ & $4.07^{+0.04}_{-0.04}$ & $1.73^{+0.10}_{-0.10}$ & $1.27^{+0.07}_{-0.06}$ & \cite{2020ApJS..247...28H} \\
			211359660 & 5,18 & 11.742 & $5181^{+48}_{-42}$ & $-0.02^{+0.06}_{-0.07}$ & $4.56^{+0.02}_{-0.03}$ & $0.80^{+0.01}_{-0.01}$ & $0.84^{+0.03}_{-0.04}$ & \cite{2018AJ....155..136M} \\
			211393988 & 18 & 13.862 & $6972^{+456}_{-381}$ & $0.02^{+0.15}_{-0.16}$ & $4.10^{+0.10}_{-0.11}$ & $1.82^{+0.29}_{-0.22}$ & $1.53^{+0.15}_{-0.14}$ & -- \\
			211407755 & 5,16,18 & 14.573 & $4726^{+82}_{-74}$ & $-0.20^{+0.14}_{-0.18}$ & $4.63^{+0.03}_{-0.03}$ & $0.67^{+0.04}_{-0.04}$ & $0.68^{+0.04}_{-0.04}$ & \cite{2020ApJS..247...28H} \\
			211418290 & 5,18 & 11.504 & $5041^{+55}_{-46}$ & $-0.43^{+0.08}_{-0.09}$ & $3.64^{+0.06}_{-0.04}$ & $3.02^{+0.17}_{-0.16}$ & $1.47^{+0.12}_{-0.12}$ & \cite{2018AJ....155..136M} \\
			211418729 & 5,18 & 14.286 & $5039^{+52}_{-55}$ & $0.38^{+0.04}_{-0.04}$ & $4.51^{+0.03}_{-0.02}$ & $0.87^{+0.02}_{-0.02}$ & $0.89^{+0.03}_{-0.02}$ & \cite{2017AJ....154..188S} \\
			211424769 & 5,18 & 9.438 & $6217^{+46}_{-41}$ & $-0.06^{+0.06}_{-0.07}$ & $4.21^{+0.02}_{-0.02}$ & $1.38^{+0.03}_{-0.02}$ & $1.12^{+0.05}_{-0.04}$ & \cite{2018AJ....155..136M} \\
			211480861 & 18 & 9.961 & $7461^{+587}_{-401}$ & $0.03^{+0.14}_{-0.16}$ & $4.03^{+0.06}_{-0.05}$ & $2.11^{+0.12}_{-0.12}$ & $1.74^{+0.15}_{-0.11}$ & -- \\
			211525389 & 5,18 & 11.687 & $5636^{+49}_{-52}$ & $0.23^{+0.04}_{-0.03}$ & $4.49^{+0.01}_{-0.01}$ & $0.97^{+0.01}_{-0.01}$ & $1.05^{+0.02}_{-0.02}$ & \cite{2018AJ....155...21P} \\
			211525753 & 18 & 13.746 & $5790^{+177}_{-147}$ & $-0.12^{+0.15}_{-0.18}$ & $4.49^{+0.03}_{-0.04}$ & $0.92^{+0.04}_{-0.03}$ & $0.96^{+0.05}_{-0.06}$ & -- \\
			211537087 & 18 & 13.438 & $5568^{+178}_{-150}$ & $-0.11^{+0.15}_{-0.17}$ & $4.52^{+0.03}_{-0.04}$ & $0.86^{+0.03}_{-0.03}$ & $0.90^{+0.05}_{-0.06}$ & -- \\
			211590050 & 18 & 13.317 & $6414^{+310}_{-280}$ & $0.03^{+0.13}_{-0.16}$ & $4.16^{+0.07}_{-0.06}$ & $1.57^{+0.14}_{-0.12}$ & $1.29^{+0.10}_{-0.09}$ & -- \\
			211594205 & 5,18 & 10.680 & $5245^{+34}_{-35}$ & $-0.10^{+0.04}_{-0.04}$ & $4.62^{+0.00}_{-0.00}$ & $0.75^{+0.01}_{-0.01}$ & $0.84^{+0.01}_{-0.01}$ & \cite{2018AJ....155...21P} \\
			211606790 & 5,16,18 & 12.673 & $5456^{+49}_{-50}$ & $0.07^{+0.07}_{-0.07}$ & $4.12^{+0.03}_{-0.03}$ & $1.76^{+0.09}_{-0.09}$ & $1.47^{+0.07}_{-0.05}$ & \cite{2018AJ....155..136M} \\
			211644764 & 18 & 13.105 & $6688^{+377}_{-269}$ & $0.04^{+0.14}_{-0.14}$ & $4.05^{+0.08}_{-0.07}$ & $1.90^{+0.21}_{-0.18}$ & $1.48^{+0.12}_{-0.11}$ & -- \\
			211705502 & 18 & 13.216 & $6471^{+314}_{-269}$ & $0.01^{+0.15}_{-0.16}$ & $4.24^{+0.06}_{-0.06}$ & $1.39^{+0.12}_{-0.09}$ & $1.26^{+0.10}_{-0.10}$ & -- \\
			211724246 & 18 & 13.242 & $6486^{+290}_{-302}$ & $0.07^{+0.15}_{-0.14}$ & $3.97^{+0.07}_{-0.07}$ & $2.10^{+0.23}_{-0.19}$ & $1.50^{+0.13}_{-0.11}$ & -- \\
			211730267 & 18 & 13.459 & $5794^{+210}_{-169}$ & $-0.02^{+0.14}_{-0.16}$ & $4.43^{+0.04}_{-0.04}$ & $1.00^{+0.05}_{-0.05}$ & $0.98^{+0.07}_{-0.07}$ & -- \\
			211733267 & 5,16,18 & 12.150 & $5342^{+48}_{-47}$ & $0.02^{+0.06}_{-0.07}$ & $4.46^{+0.03}_{-0.02}$ & $0.90^{+0.02}_{-0.02}$ & $0.85^{+0.03}_{-0.02}$ & \cite{2018AJ....155..136M} \\
			211791178 & 5,18 & 13.648 & $4648^{+88}_{-76}$ & $-0.31^{+0.12}_{-0.10}$ & $4.50^{+0.03}_{-0.02}$ & $0.81^{+0.03}_{-0.03}$ & $0.74^{+0.02}_{-0.02}$ & \cite{2017ApJ...836..167D} \\
			211816003 & 5,16,18 & 13.654 & $5397^{+105}_{-96}$ & $-0.09^{+0.12}_{-0.12}$ & $4.53^{+0.03}_{-0.04}$ & $0.83^{+0.03}_{-0.03}$ & $0.86^{+0.05}_{-0.05}$ & \cite{2020ApJS..247...28H} \\
			211818569 & 5,18 & 12.935 & $4690^{+43}_{-41}$ & $-0.16^{+0.05}_{-0.06}$ & $4.63^{+0.02}_{-0.02}$ & $0.67^{+0.01}_{-0.01}$ & $0.69^{+0.03}_{-0.03}$ & \cite{2018AJ....155..136M} \\
			211822797 & 5,16,18 & 14.568 & $4057^{+37}_{-39}$ & $0.20^{+0.05}_{-0.07}$ & $4.70^{+0.01}_{-0.01}$ & $0.58^{+0.01}_{-0.01}$ & $0.62^{+0.02}_{-0.02}$ & \cite{2017ApJ...836..167D} \\
			211904310 & 18 & 13.636 & $6404^{+284}_{-261}$ & $0.05^{+0.15}_{-0.16}$ & $4.12^{+0.08}_{-0.07}$ & $1.65^{+0.19}_{-0.16}$ & $1.33^{+0.10}_{-0.11}$ & -- \\
			211913977 & 5,16,18 & 12.619 & $4927^{+47}_{-46}$ & $0.07^{+0.07}_{-0.07}$ & $4.59^{+0.01}_{-0.02}$ & $0.76^{+0.01}_{-0.01}$ & $0.82^{+0.02}_{-0.03}$ & \cite{2018AJ....155..136M} \\
			211914998 & 18 & 13.587 & $5650^{+193}_{-152}$ & $-0.07^{+0.15}_{-0.16}$ & $4.49^{+0.03}_{-0.04}$ & $0.91^{+0.04}_{-0.04}$ & $0.93^{+0.06}_{-0.06}$ & -- \\
			211916756 & 5,18 & 15.498 & $3566^{+34}_{-37}$ & $0.10^{+0.06}_{-0.06}$ & $4.85^{+0.01}_{-0.01}$ & $0.40^{+0.01}_{-0.01}$ & $0.41^{+0.01}_{-0.01}$ & \cite{2018AJ....156..277L} \\
			211919004 & 5,16,18 & 13.131 & $5161^{+53}_{-47}$ & $0.20^{+0.04}_{-0.04}$ & $4.53^{+0.03}_{-0.03}$ & $0.85^{+0.02}_{-0.02}$ & $0.89^{+0.03}_{-0.03}$ & \cite{2018AJ....155...21P} \\
			211969807 & 5,16,18 & 15.149 & $3712^{+54}_{-76}$ & $0.18^{+0.09}_{-0.10}$ & $4.78^{+0.02}_{-0.02}$ & $0.48^{+0.01}_{-0.01}$ & $0.51^{+0.02}_{-0.02}$ & \cite{2017ApJ...836..167D} \\
			212006344 & 5,16,18 & 12.466 & $4009^{+47}_{-44}$ & $0.32^{+0.06}_{-0.07}$ & $4.69^{+0.01}_{-0.01}$ & $0.59^{+0.01}_{-0.01}$ & $0.63^{+0.02}_{-0.02}$ & \cite{2017ApJ...836..167D} \\
			212008766 & 5,18 & 12.802 & $5044^{+43}_{-42}$ & $-0.16^{+0.03}_{-0.04}$ & $4.64^{+0.01}_{-0.01}$ & $0.70^{+0.01}_{-0.01}$ & $0.79^{+0.01}_{-0.01}$ & \cite{2018AJ....155...21P} \\
			212012119 & 5,18 & 11.753 & $4841^{+39}_{-37}$ & $-0.06^{+0.04}_{-0.03}$ & $4.65^{+0.00}_{-0.00}$ & $0.69^{+0.01}_{-0.01}$ & $0.77^{+0.01}_{-0.01}$ & \cite{2018AJ....155...21P} \\
			212110888 & 5,16,18 & 11.441 & $6168^{+46}_{-76}$ & $0.01^{+0.03}_{-0.04}$ & $4.20^{+0.03}_{-0.03}$ & $1.42^{+0.06}_{-0.05}$ & $1.16^{+0.04}_{-0.04}$ & \cite{2016PASP..128l4402B} \\
			\hline
		\end{tabular}
	\end{center}
	\label{tab:stellar_parameters}
\end{table*}
 
\subsection{Transit modelling}
\label{transit_modeling}

\begin{figure*}
    \centering
    \includegraphics[width=4.8cm]{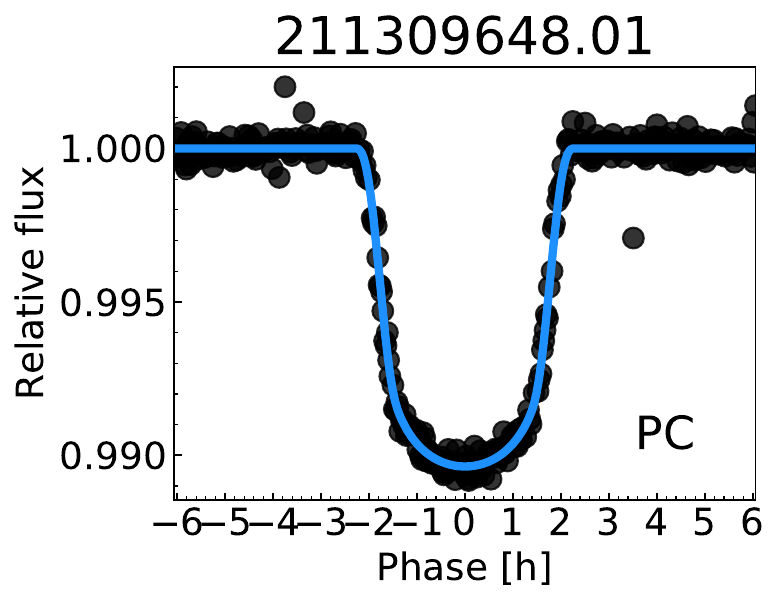}
    \includegraphics[width=4.8cm]{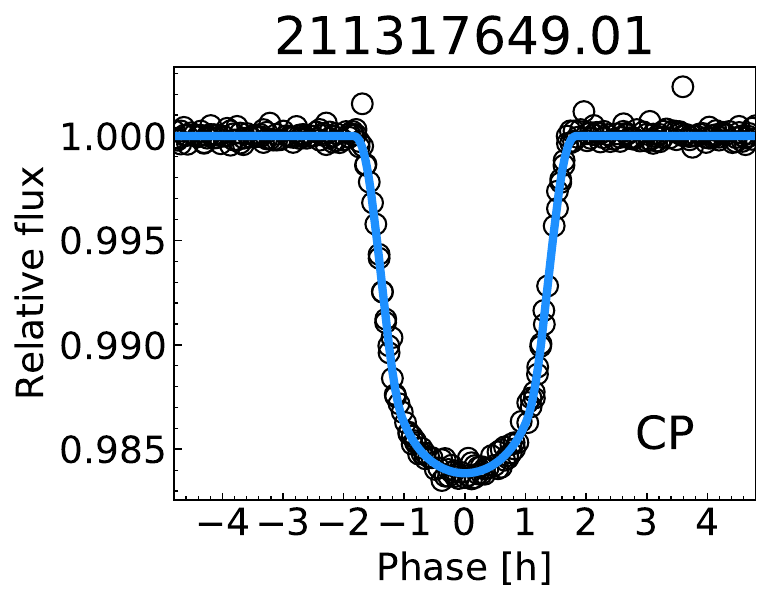}
    \includegraphics[width=4.8cm]{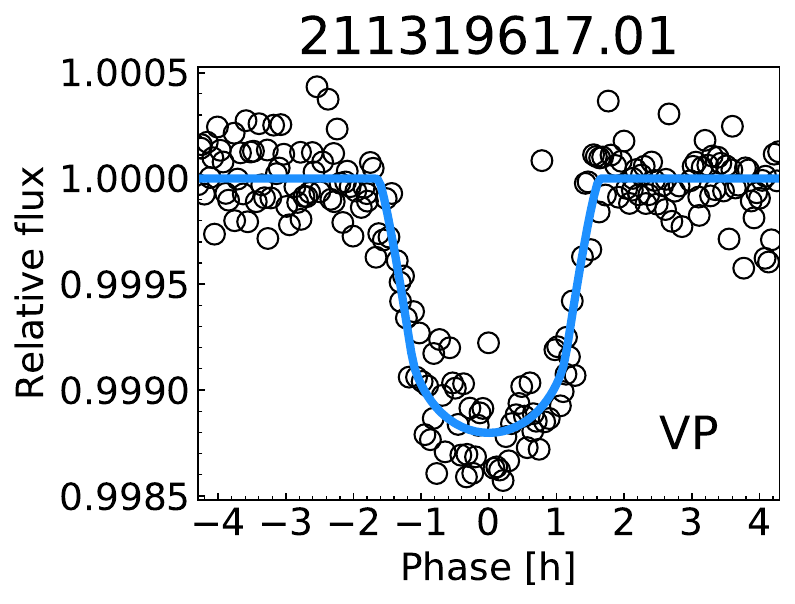}
    \includegraphics[width=4.8cm]{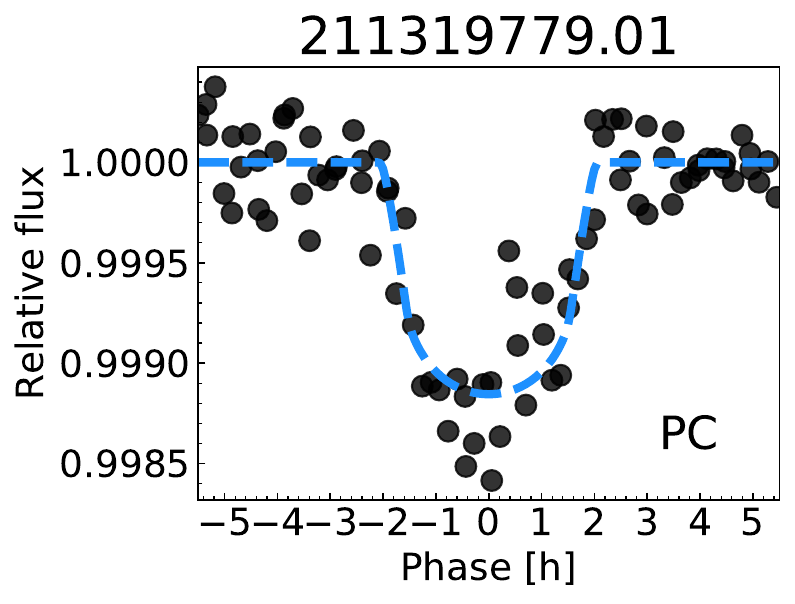}
    \includegraphics[width=4.8cm]{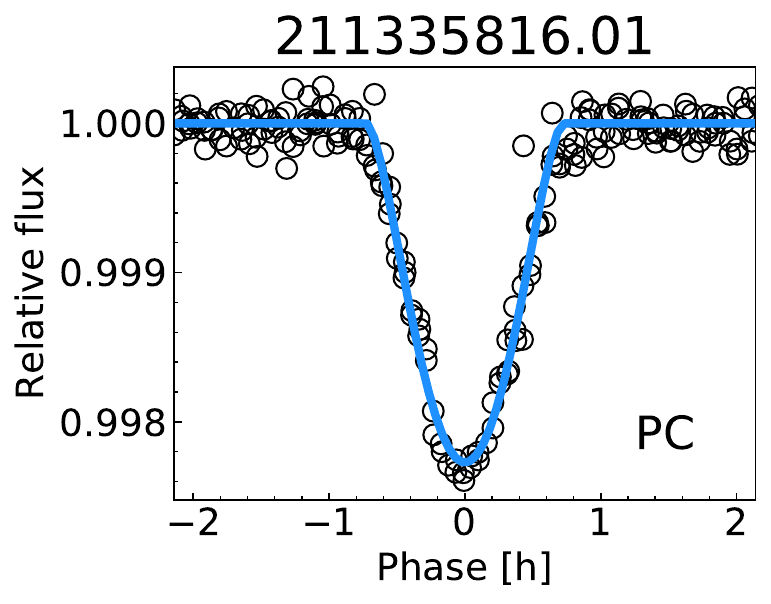}
    \includegraphics[width=4.8cm]{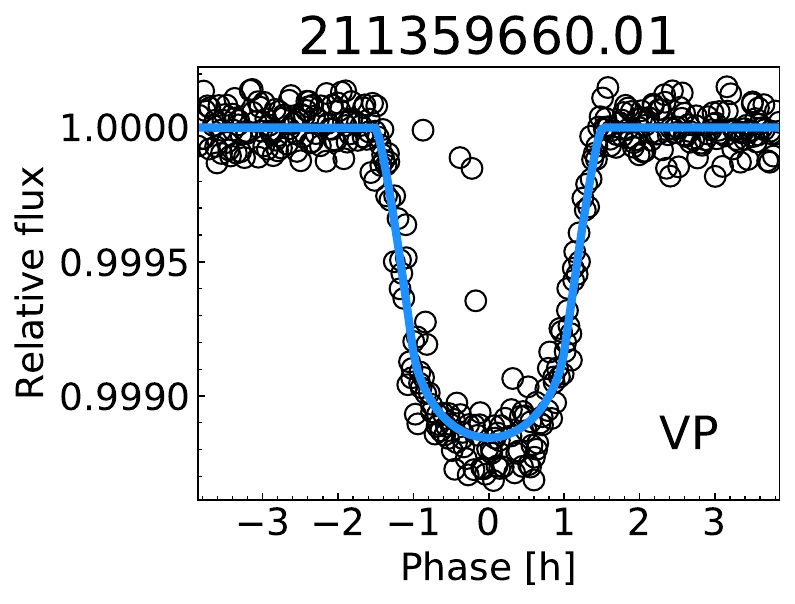}
    \includegraphics[width=4.8cm]{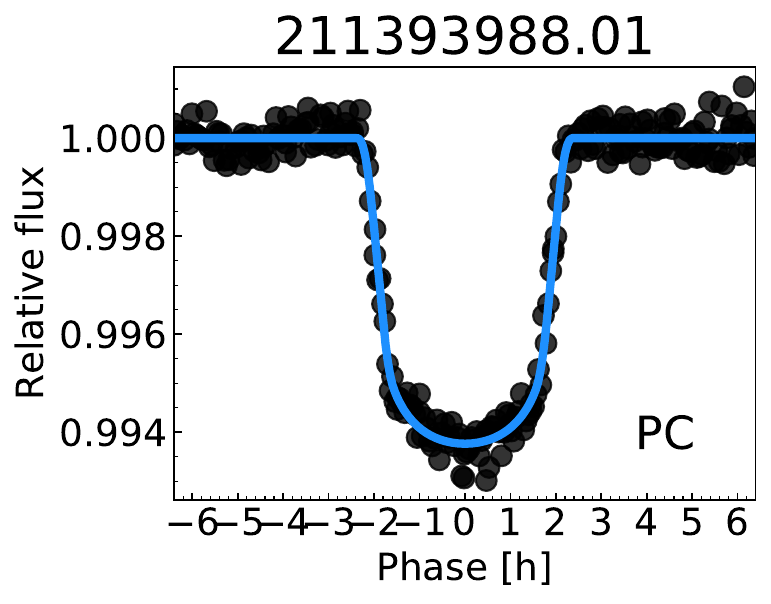}
    \includegraphics[width=4.8cm]{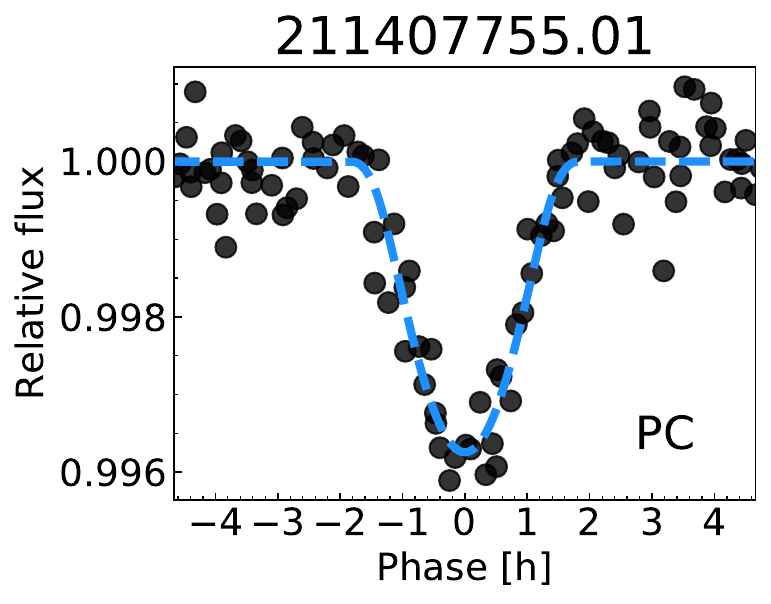}
    \includegraphics[width=4.8cm]{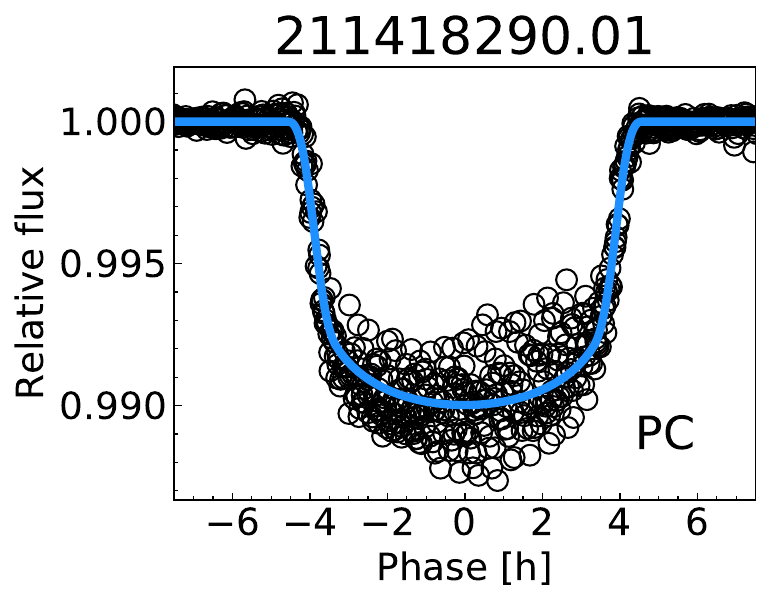}
    \includegraphics[width=4.8cm]{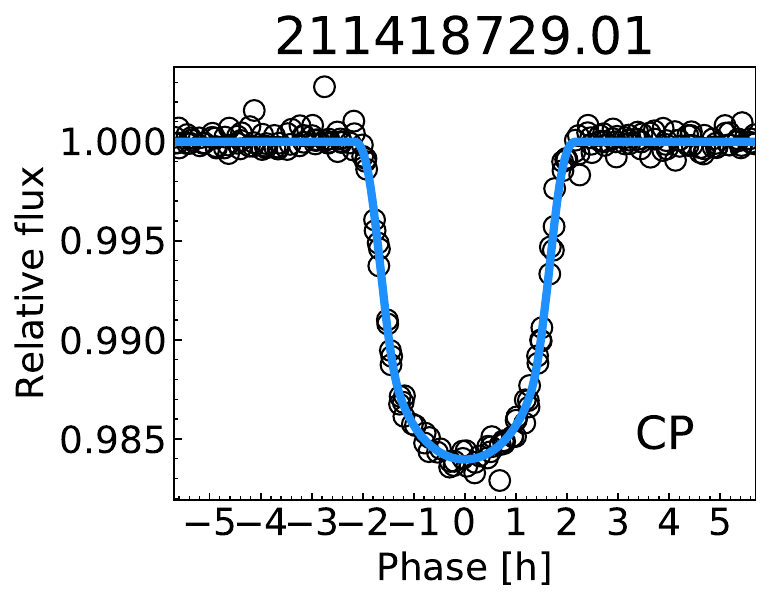}
    \includegraphics[width=4.8cm]{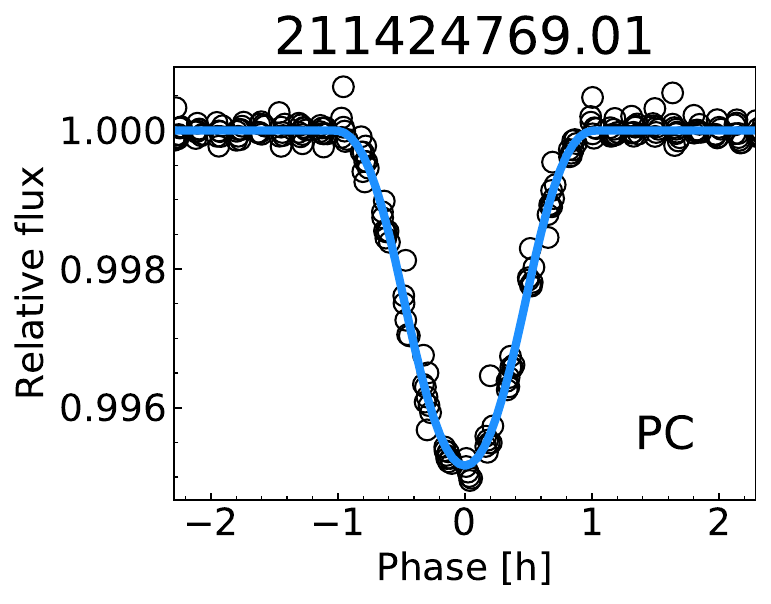}
    \includegraphics[width=4.8cm]{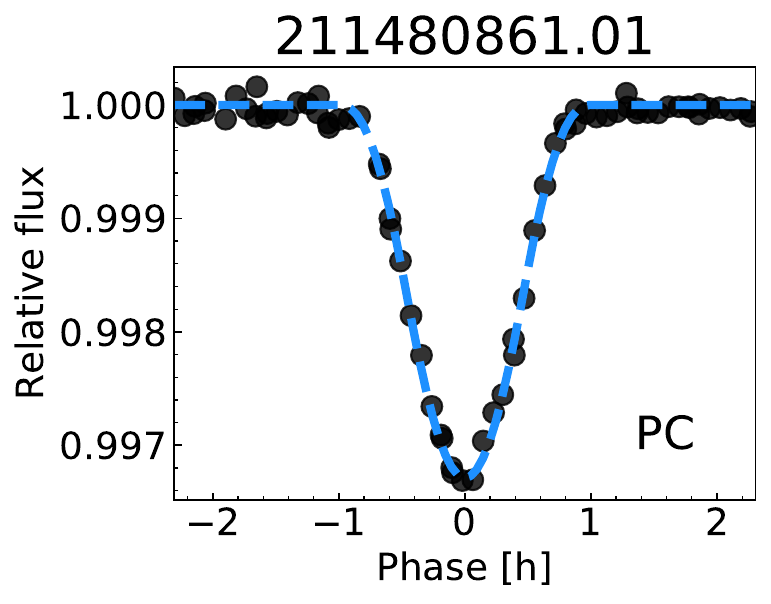}
    \includegraphics[width=4.8cm]{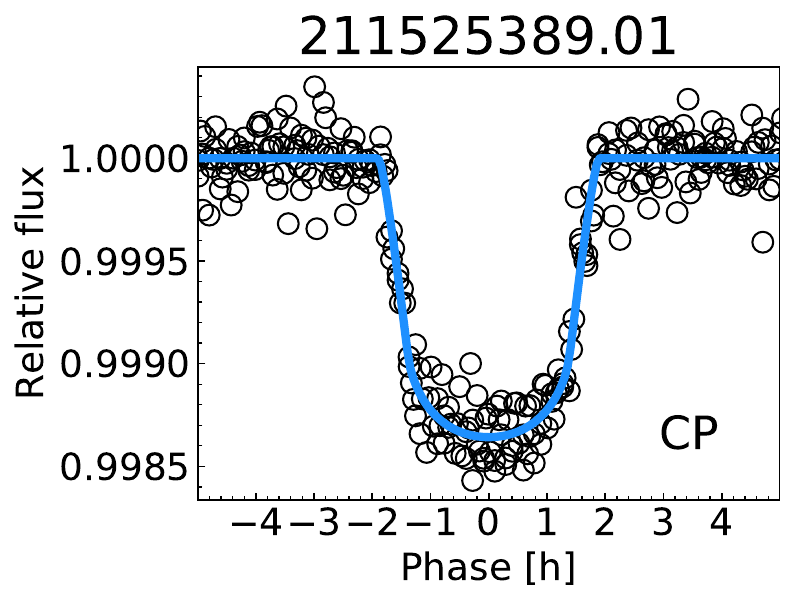}
    \includegraphics[width=4.8cm]{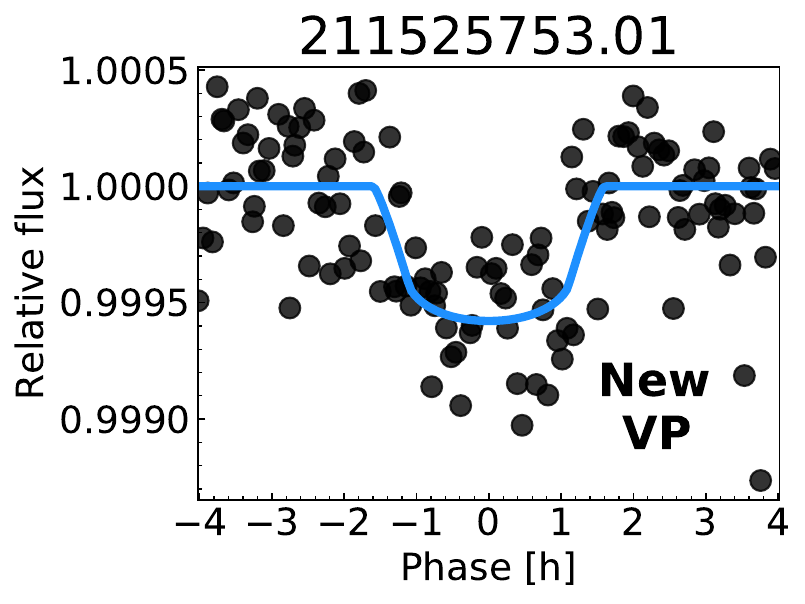}
    \includegraphics[width=4.8cm]{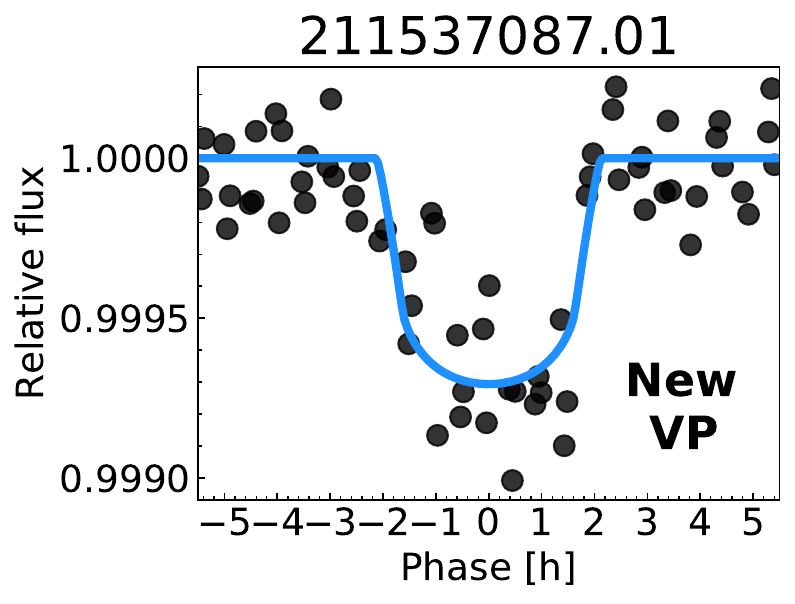}
    \includegraphics[width=4.8cm]{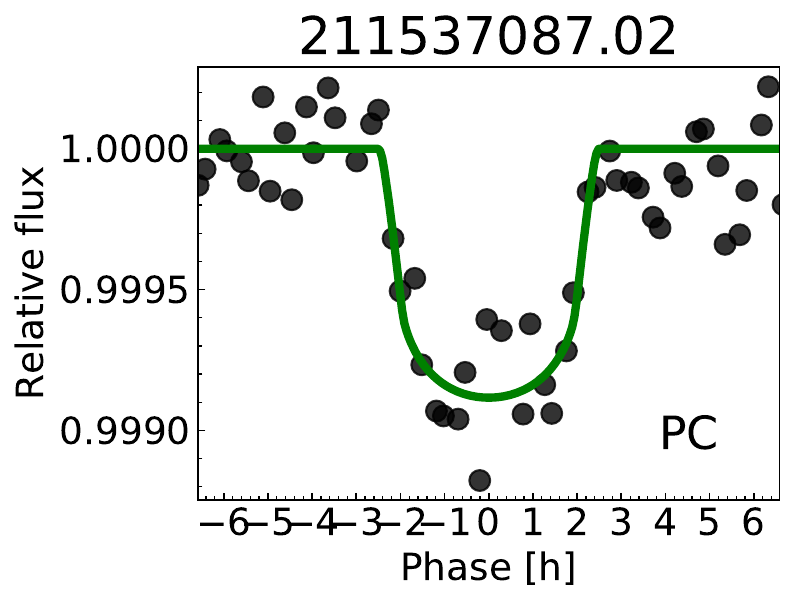}
    \includegraphics[width=4.8cm]{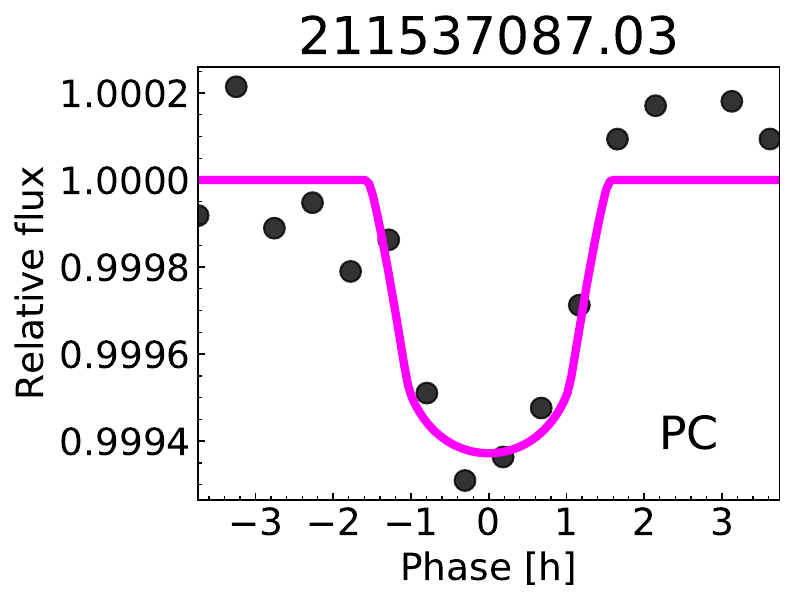}
    \includegraphics[width=4.8cm]{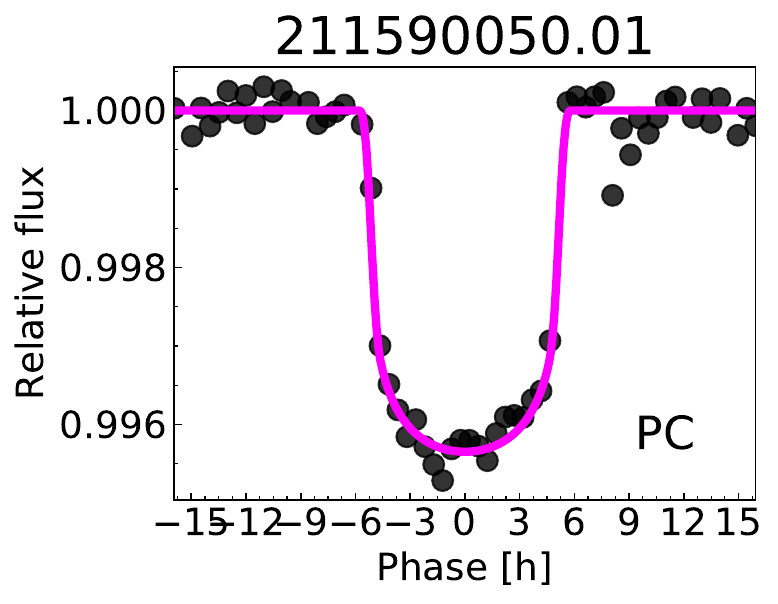}
    \caption[Phase-folded transits of confirmed planets, validated planets, and planet candidates.]{Phase-folded transits of CP, VP, and PC analysed in this chapter. Data are plotted with solid symbols for new detections and with open circles for known ones. The posterior models are over-plotted in blue for signals with three or more transits, in green for two-transit signals, and in magenta for single transits. The model for EPIC 211594205.01 is over-plotted in red indicating strong TTVs. The dashed models for EPICs 211319779.01, 211407755.01, 211480861.01, and 211791178.01 indicate that their photometry is contaminated.}
    \label{fig:phase_folded_1}
\end{figure*}

\begin{figure}
    \centering
    \includegraphics[width=4.8cm]{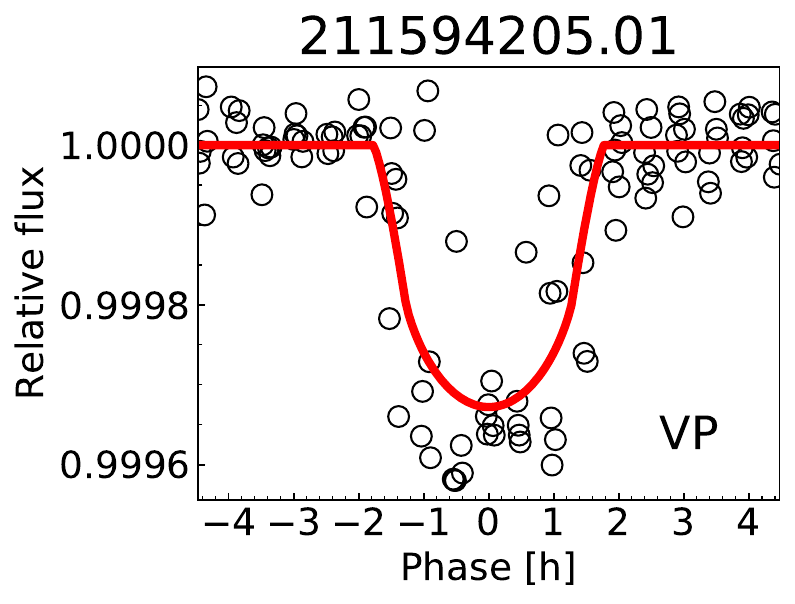}
    \includegraphics[width=4.8cm]{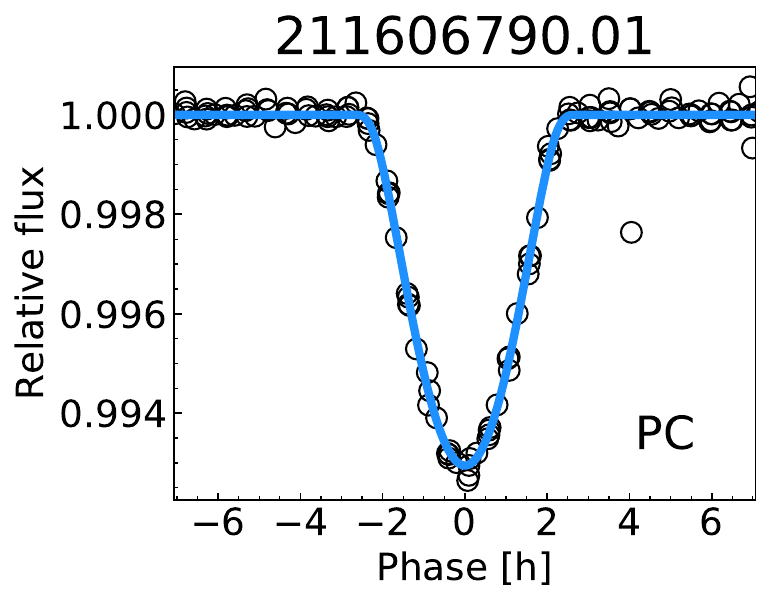}
    \includegraphics[width=4.8cm]{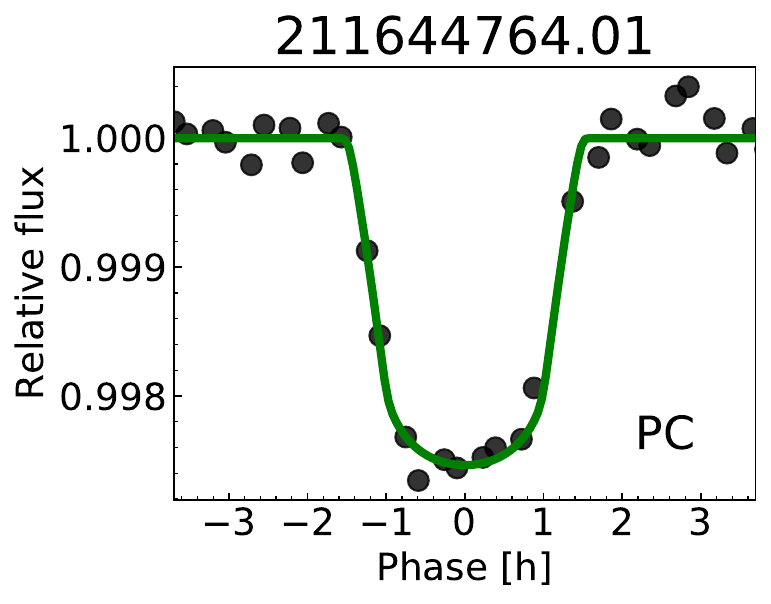}
    \includegraphics[width=4.8cm]{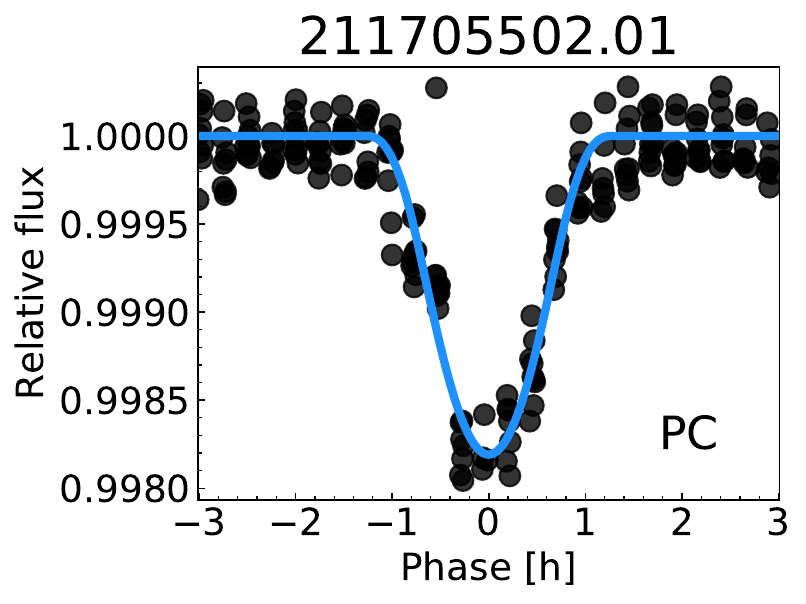}
    \includegraphics[width=4.8cm]{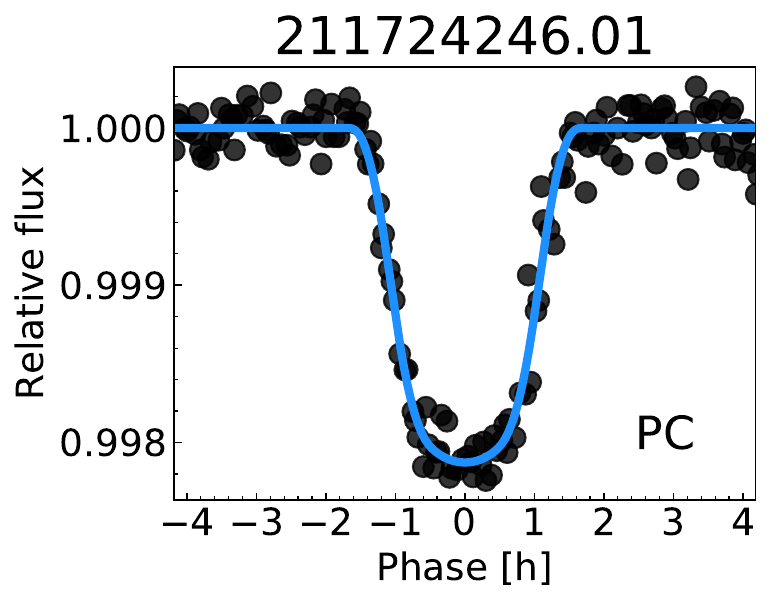}
    \includegraphics[width=4.8cm]{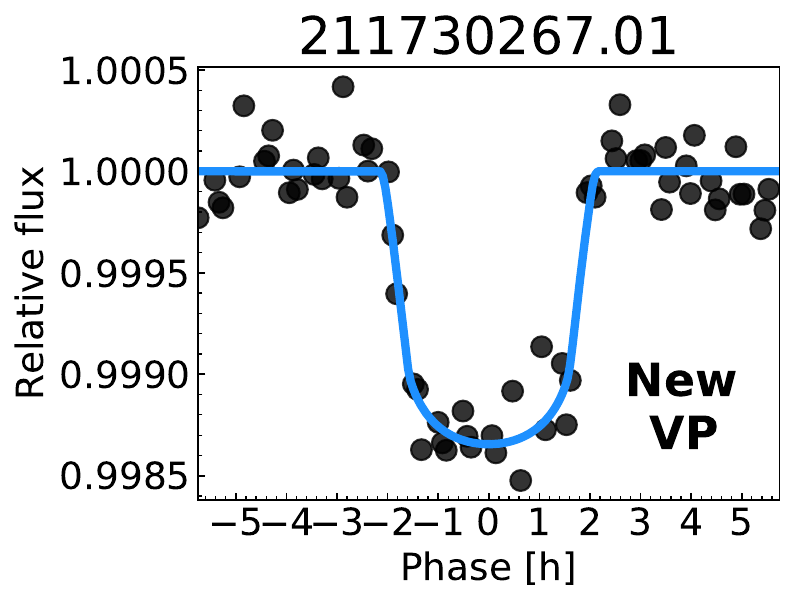}
    \includegraphics[width=4.8cm]{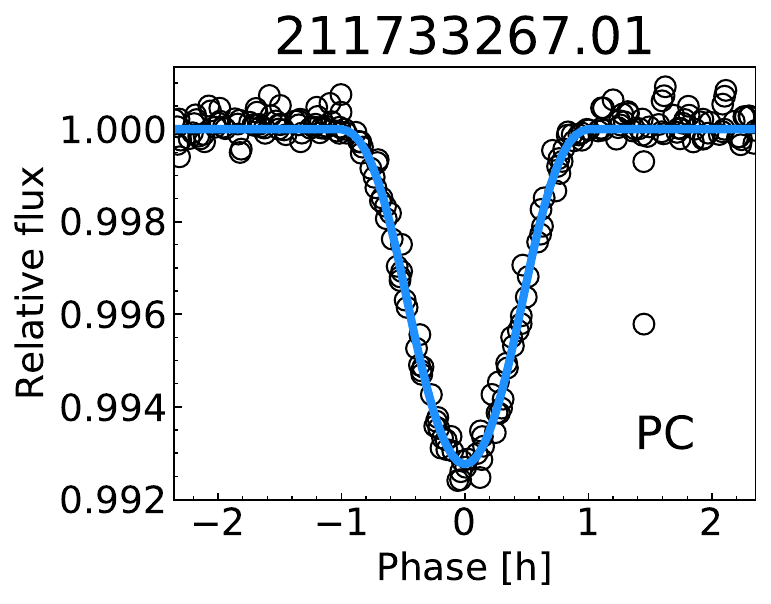}
    \includegraphics[width=4.8cm]{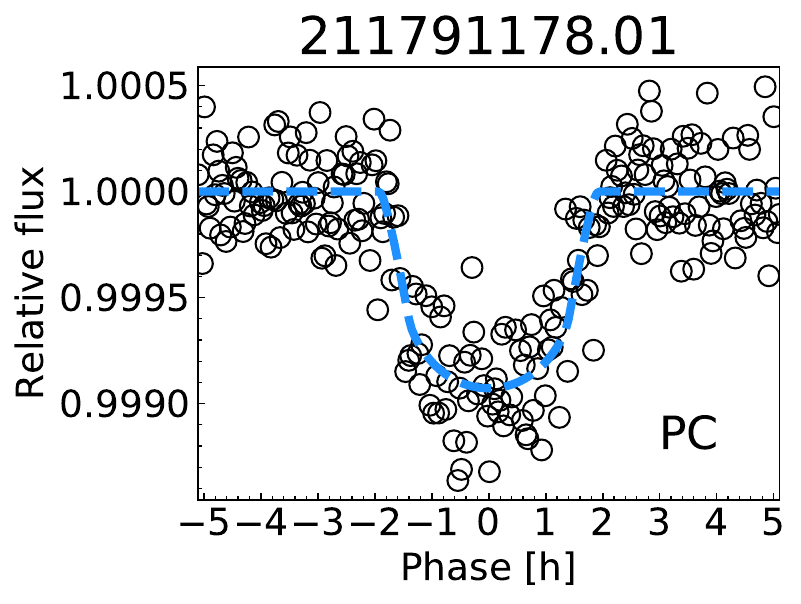}
    \includegraphics[width=4.8cm]{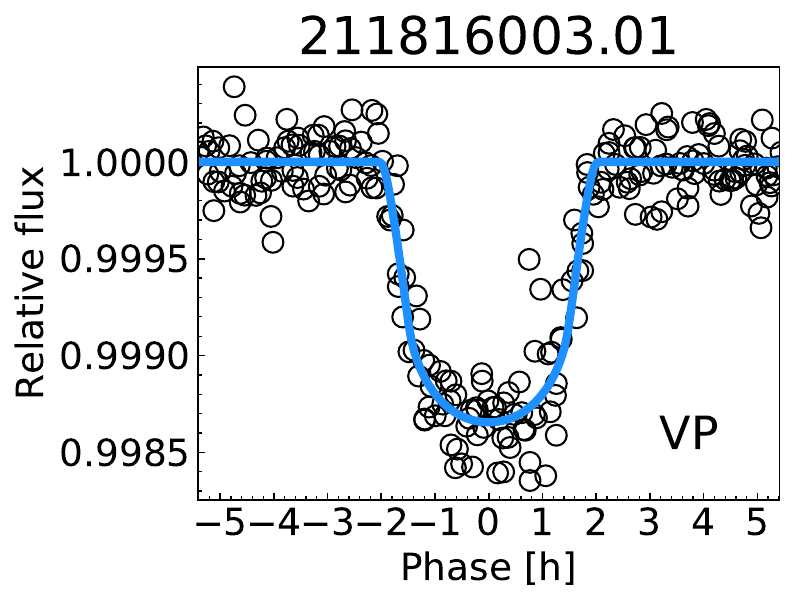}
    \includegraphics[width=4.8cm]{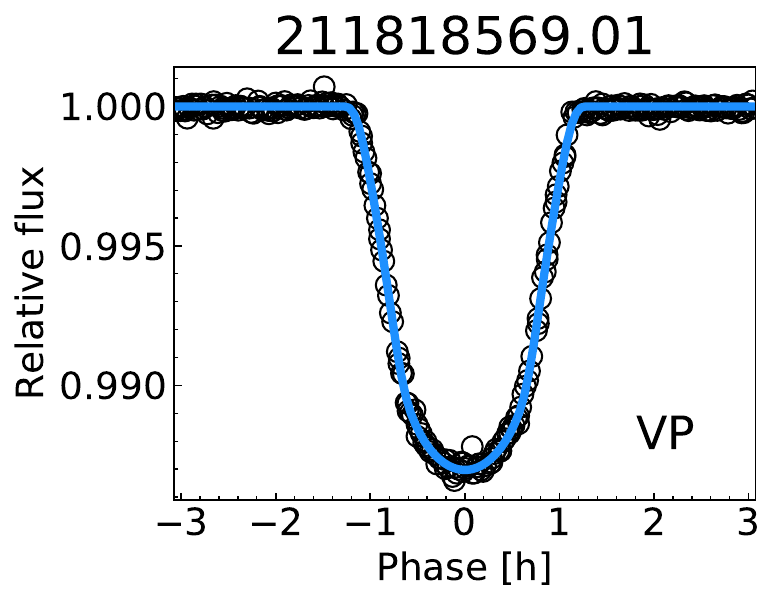}
    \includegraphics[width=4.8cm]{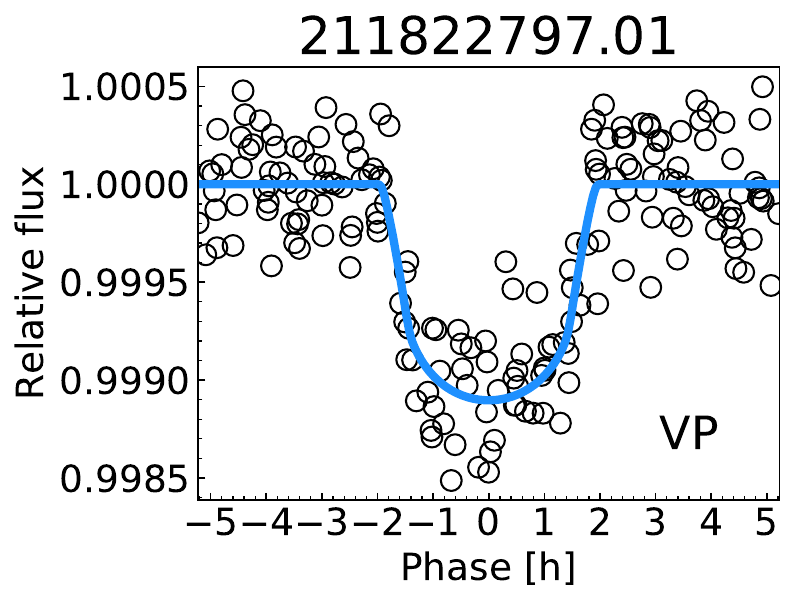}
    \includegraphics[width=4.8cm]{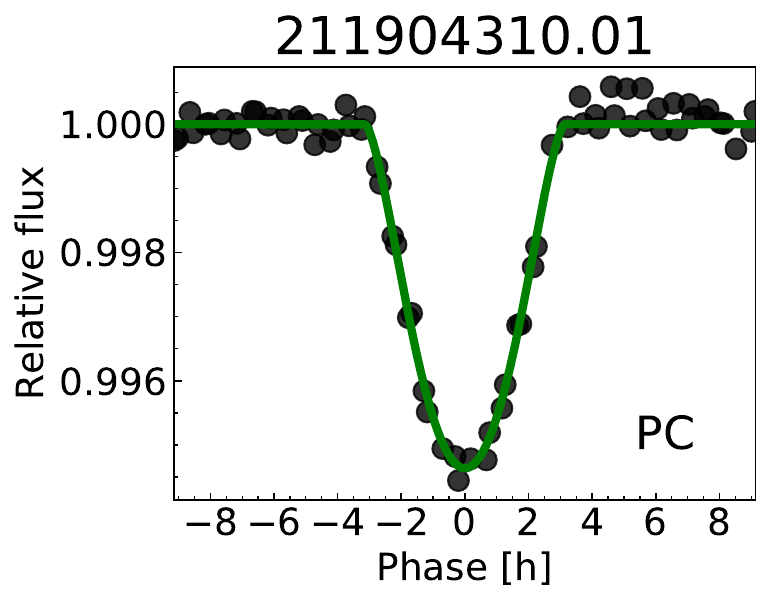}
    \includegraphics[width=4.8cm]{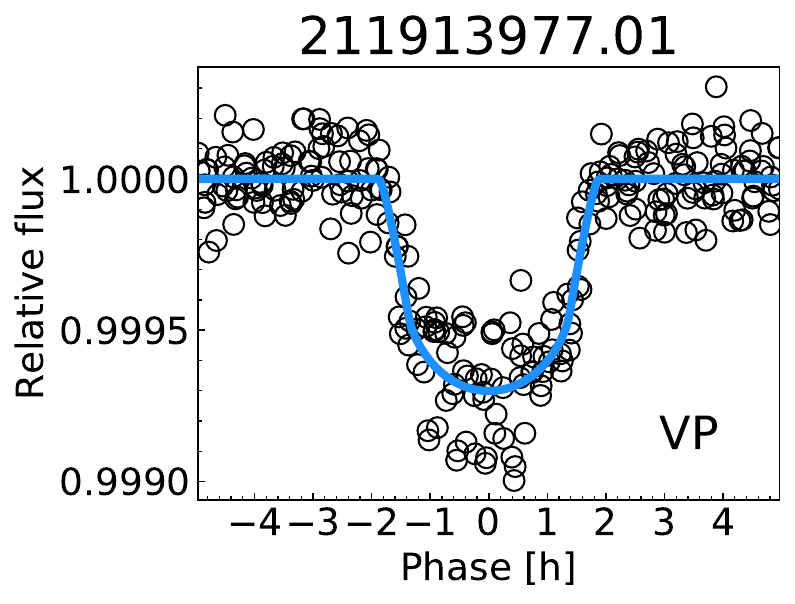}
    \includegraphics[width=4.8cm]{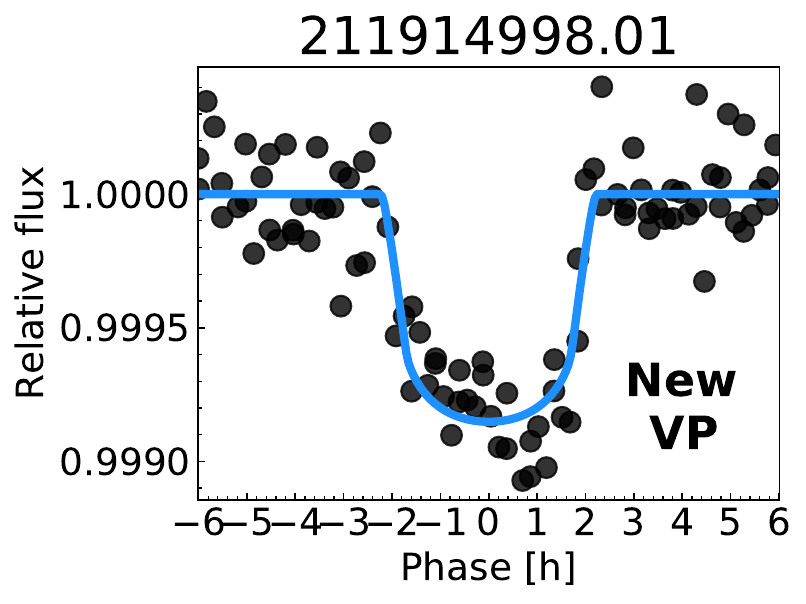}
    \includegraphics[width=4.8cm]{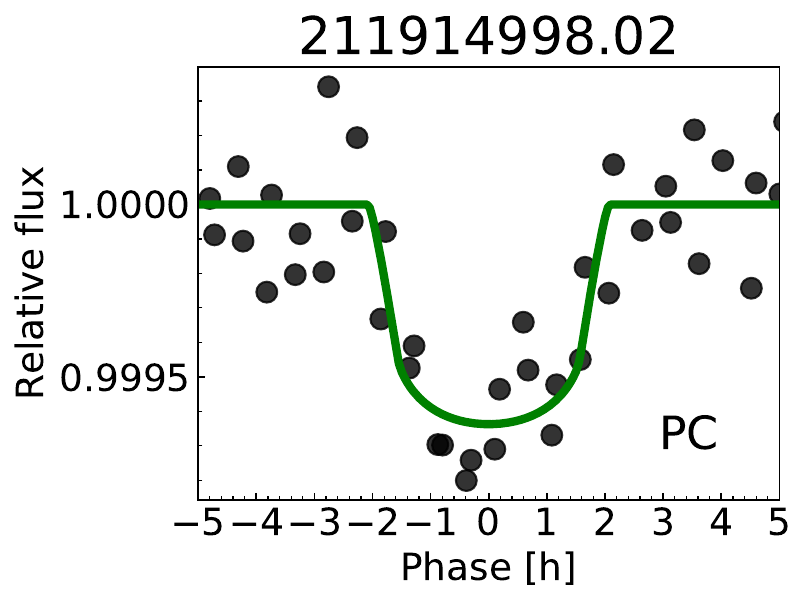}
    \includegraphics[width=4.8cm]{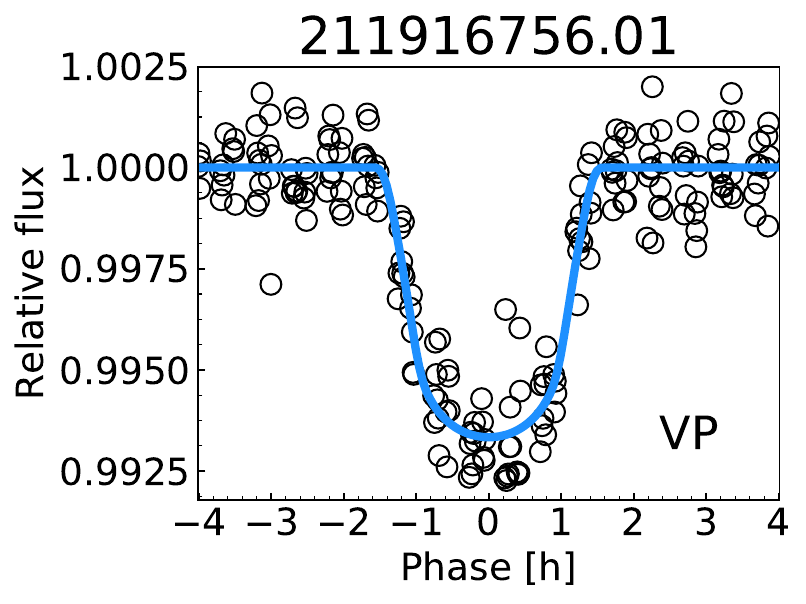}
    \includegraphics[width=4.8cm]{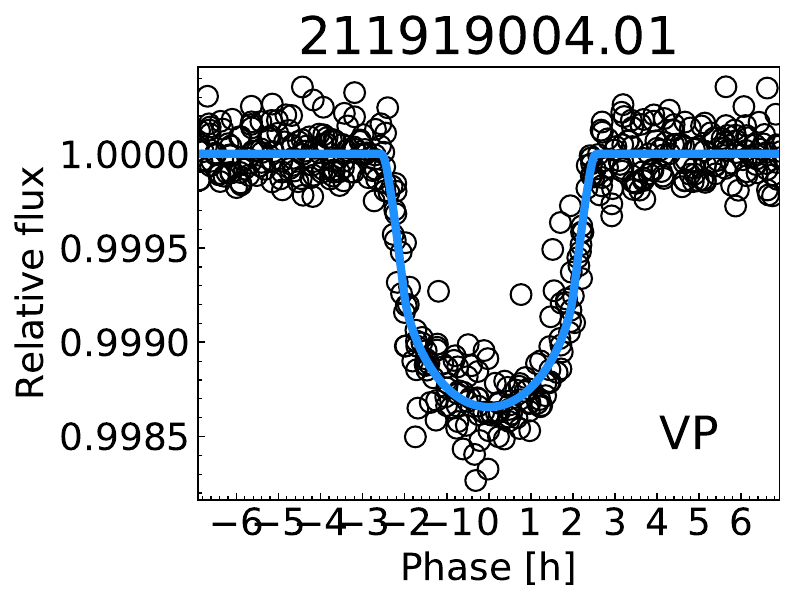}
    \includegraphics[width=4.8cm]{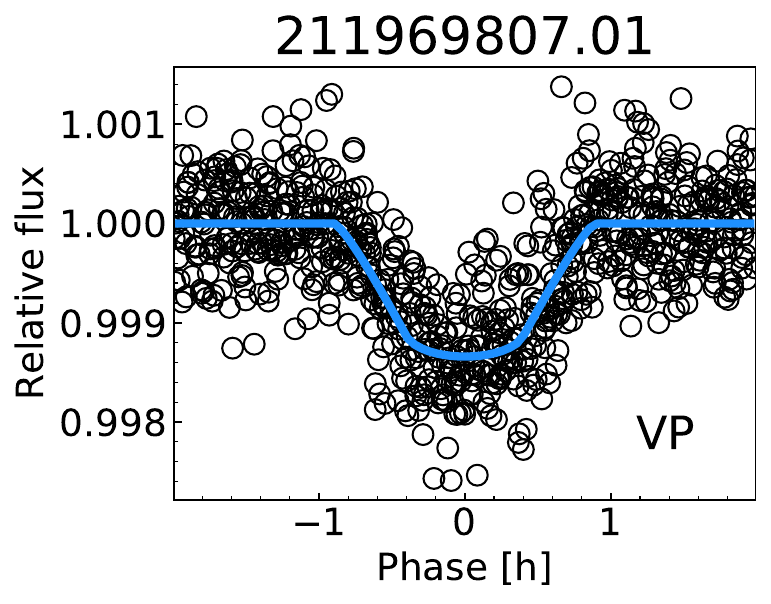}
    \caption[Phase-folded transits of confirmed planets, validated planets, and planet candidates.]{Phase-folded transits of CP, VP, and PC analysed in this chapter. Data are plotted with solid symbols for new detections and with open circles for known ones. The posterior models are over-plotted in blue for signals with three or more transits, in green for two-transit signals, and in magenta for single transits. The model for EPIC 211594205.01 is over-plotted in red indicating strong TTVs. The dashed models for EPICs 211319779.01, 211407755.01, 211480861.01, and 211791178.01 indicate that their photometry is contaminated.}
    \label{fig:phase_folded_2}
\end{figure}

\begin{figure}[t]
    \centering
    \includegraphics[width=4.8cm]{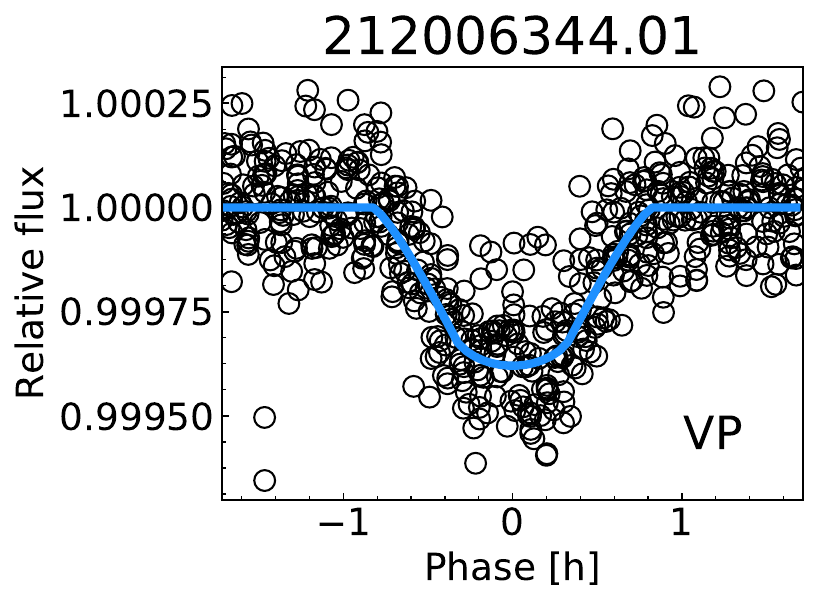}
    \includegraphics[width=4.8cm]{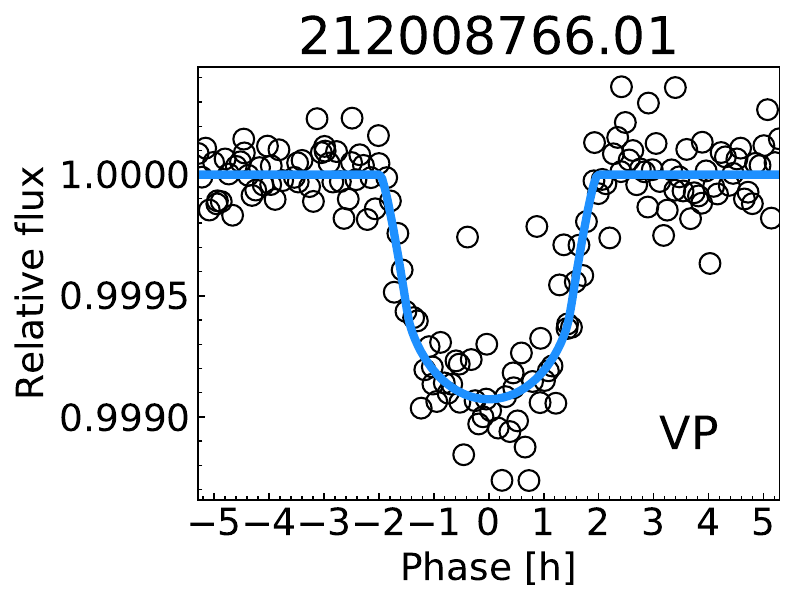}
    \includegraphics[width=4.8cm]{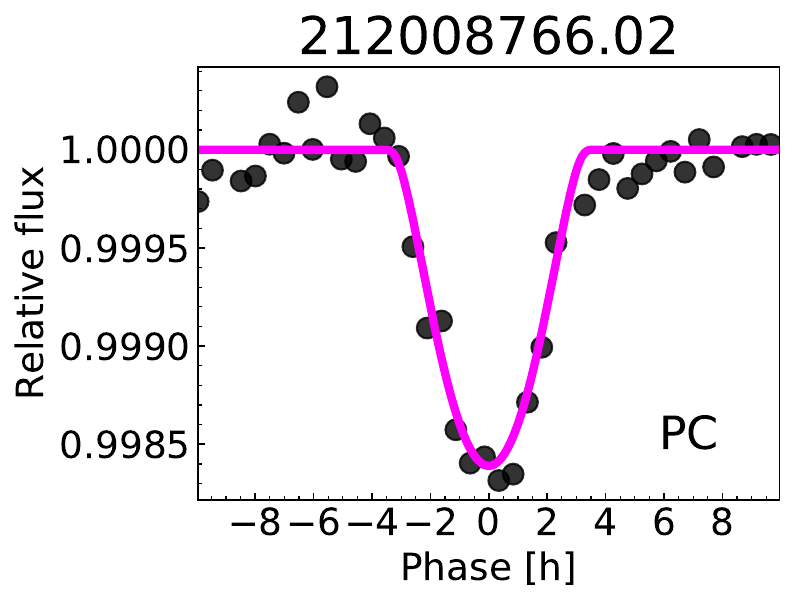}
    \includegraphics[width=4.8cm]{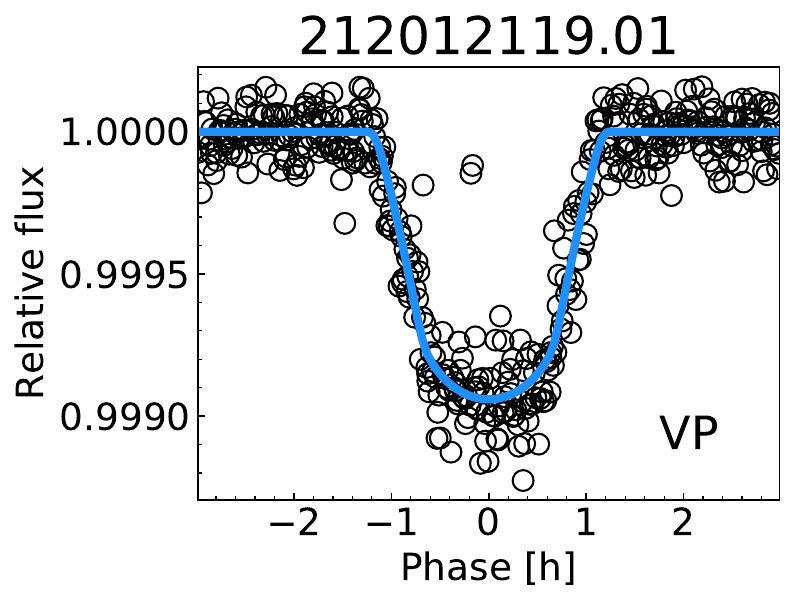}
    \includegraphics[width=4.8cm]{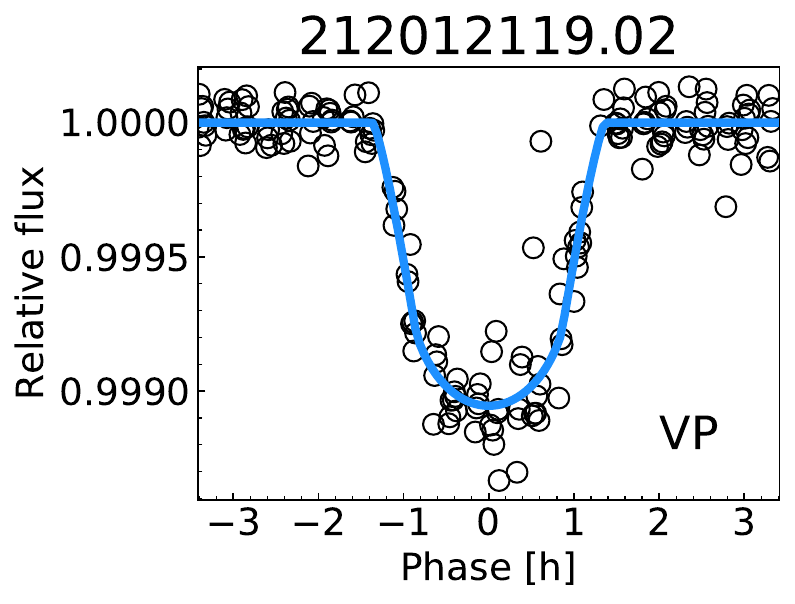}
    \includegraphics[width=4.8cm]{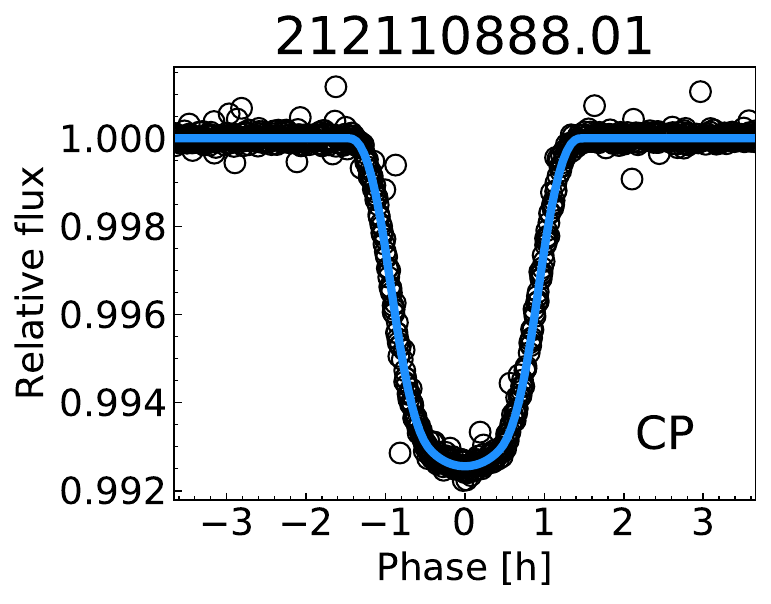}
    \caption[Phase-folded transits of confirmed planets, validated planets, and planet candidates.]{Phase-folded transits of CP, VP, and PC analysed in this chapter. Data are plotted with solid symbols for new detections and with open circles for known ones. The posterior models are over-plotted in blue for signals with three or more transits, in green for two-transit signals, and in magenta for single transits. The model for EPIC 211594205.01 is over-plotted in red indicating strong TTVs. The dashed models for EPICs 211319779.01, 211407755.01, 211480861.01, and 211791178.01 indicate that their photometry is contaminated.}
    \label{fig:phase_folded_3}
\end{figure}

To derive the planetary parameters, we used the pre-processed light curves as described in Sect.~\ref{K2_photometry} and modelled the transits with the \citet{2002ApJ...580L.171M} quadratic limb darkened transit model. For that, we used the \textsc{pyaneti} package \citep{2019MNRAS.482.1017B}, accounting for the K2 30-min cadence by super-sampling the transit model with 10 subsamples per cadence \citep{2010MNRAS.408.1758K}. We assumed circular orbits by fixing the eccentricity to 0. We set wide uniform priors  for the impact parameter ($b$), scaled planet radius ($R_{p} / R_{\star}$) and semimajor axis ($a/R_{\star}$). We set narrow uniform priors for the orbital period ($P$) and the mid-transit time ($T_{0}$) by using the values and uncertainties obtained from the TLS algorithm. We settled Gaussian priors on the quadratic limb darkening coefficients in the q-space \citep{2013MNRAS.435.2152K}, which we estimated from the \textsc{limb-darkening} package \citep{2015MNRAS.450.1879E} by adopting the ATLAS models for the stellar atmospheres \citep{1979ApJS...40....1K}. We also included a photometric jitter term in order to account for underestimated white noise. 

For the single-transit candidates, we fit the same parameters as for the multi-transit candidates, except for $P$ and $a/R_{\star}$, which cannot be determined. Instead, we obtained a lower bound for $P$ as the time between the transit and the farthest edge of the light curve, and we also estimated a lower bound for $a/R_{\star}$, making use of the bound for $P$ and Kepler's third law.

Table \ref{tab:planet_params} shows the main derived parameters together with their uncertainties (i.e. the median and 68.3$\%$ credible intervals of the posterior distribution) for the new \newVP planets and \newPC candidates presented in this chapter, as well as for the 24 already published planets and candidates. Figure \ref{fig:phase_folded_1} shows the light curves folded to the orbital periods of each planet and candidate, together with the median transit model. 

\begin{landscape} \renewcommand{\arraystretch}{1.18}
\begin{table}
\fontsize{8.2pt}{8.2pt}\selectfont
\caption[Planetary parameters and dispositions of the K2-OjOS sample.]{Planetary parameters and dispositions (CP = confirmed planet, VP = validated planet, PC = planet candidate) for the planets and candidates first detected in this work (Detection: New) and for the previously published planets and candidates (Detection: Known). }
	\begin{center}
		\begin{tabular}{c|c|c|c|c|c|c|c|c|c|c|c}
		    \hline \hline
			ID & Name & $T_{0}$ (BKJD) & $P$ (d) & $R_{p}/R_{\star}$ (\%) & $a/R_{\star}$ & $b$ & $T_{eq}$ [A = 0] (K) & $a$ (AU) & $R_{\rm p}$ ($\rm R_{\oplus}$) & Detection & Disp \\
			\hline
			211309648.01 &   & $3421.4841636^{+0.0002984}_{-0.0002959}$ & $3.1393226^{+0.0000325}_{-0.0000316}$ & $9.42^{+0.14}_{-0.11}$ & $6.25^{+0.26}_{-0.33}$ & $0.31^{+0.13}_{-0.20}$ & $1780^{+80}_{-74}$ & \textbf{$0.0434^{+0.0047}_{-0.0047}$} & $15.43^{+1.56}_{-1.56}$ & New & PC \\
			211317649.01 & HAT-P-43 b & $3420.5965728^{+0.0001280}_{-0.0001230}$ & $3.3326414^{+0.0000151}_{-0.0000153}$ & $11.58^{+0.05}_{-0.05}$ & $8.86^{+0.05}_{-0.07}$ & $0.07^{+0.07}_{-0.05}$ & $1353^{+16}_{-16}$ & $0.0444^{+0.0021}_{-0.0021}$ & $13.63^{+0.64}_{-0.63}$ & Known & CP \\
			211319617.01 & K2-180 b & $2310.3959512^{+0.0012311}_{-0.0011967}$ & $8.8656511^{+0.0000151}_{-0.0000152}$ & $3.20^{+0.31}_{-0.11}$ & $21.93^{+2.36}_{-6.55}$ & $0.46^{+0.33}_{-0.31}$ & $802^{+55}_{-41}$ & $0.0734^{+0.0080}_{-0.0219}$ & $2.52^{+0.24}_{-0.10}$ & Known & VP \\
			211319779.01 &   & $3422.1814025^{+0.0039344}_{-0.0038465}$ & $13.8253157^{+0.0021214}_{-0.0020261}$ & -- & -- & -- & -- & -- & -- & New & PC \\
			211335816.01 &   & $2312.0099067^{+0.0001884}_{-0.0001853}$ & $4.9898611^{+0.0000014}_{-0.0000014}$ & $8.21^{+0.39}_{-0.47}$ & $11.30^{+0.22}_{-0.19}$ & $0.99^{+0.00}_{-0.01}$ & $1308^{+31}_{-31}$ & $0.0911^{+0.0055}_{-0.0055}$ & $15.44^{+1.22}_{-1.23}$ & Known & PC \\
			211359660.01 & K2-182 b & $2312.9420111^{+0.0002707}_{-0.0002661}$ & $4.7369729^{+0.0000020}_{-0.0000020}$ & $3.08^{+0.11}_{-0.04}$ & $13.98^{+0.55}_{-1.81}$ & $0.28^{+0.27}_{-0.20}$ & $982^{+9}_{-22}$ & $0.0519^{+0.0022}_{-0.0067}$ & $2.70^{+0.09}_{-0.05}$ & Known & VP \\
			211393988.01 &   & $3421.7347162^{+0.0006113}_{-0.0006035}$ & $4.8718955^{+0.0001146}_{-0.0001140}$ & $7.35^{+0.15}_{-0.09}$ & $8.94^{+0.43}_{-0.86}$ & $0.33^{+0.20}_{-0.21}$ & $1663^{+129}_{-119}$ & $0.0740^{+0.0133}_{-0.0130}$ & $14.64^{+2.33}_{-2.39}$ & New & PC \\
			211407755.01 &   & $2327.2333494^{+0.0017168}_{-0.0017028}$ & $36.0861607^{+0.0000913}_{-0.0000926}$ & -- & -- & -- & -- & -- & -- & New & PC \\
			211418290.01 &   & $2308.8080774^{+0.0002891}_{-0.0002927}$ & $5.0321716^{+0.0000020}_{-0.0000021}$ & $9.25^{+0.04}_{-0.03}$ & $4.88^{+0.03}_{-0.06}$ & $0.10^{+0.10}_{-0.07}$ & $1615^{+20}_{-19}$ & $0.0684^{+0.0039}_{-0.0039}$ & $30.48^{+1.71}_{-1.73}$ & Known & PC \\
			211418729.01 & K2-114 b & $2318.7149031^{+0.0001778}_{-0.0001781}$ & $11.3909344^{+0.0000039}_{-0.0000039}$ & $11.44^{+0.18}_{-0.12}$ & $24.38^{+0.69}_{-1.04}$ & $0.26^{+0.13}_{-0.16}$ & $722^{+7}_{-13}$ & $0.0983^{+0.0037}_{-0.0044}$ & $10.88^{+0.29}_{-0.28}$ & Known & CP \\
			211424769.01 &   & $2311.4983392^{+0.0002221}_{-0.0002172}$ & $5.1762331^{+0.0000015}_{-0.0000015}$ & $10.23^{+1.23}_{-1.27}$ & $12.94^{+0.55}_{-0.36}$ & $0.97^{+0.02}_{-0.02}$ & $1222^{+20}_{-26}$ & $0.0832^{+0.0039}_{-0.0030}$ & $15.39^{+1.88}_{-1.92}$ & Known & PC \\
			211480861.01 &   & $3421.5574015^{+0.0004884}_{-0.0004842}$ & $6.3508198^{+0.0001244}_{-0.0001214}$ & -- & -- & -- & -- & -- & -- & New & PC \\
			211525389.01 & K2-105 b & $2314.9895717^{+0.0004439}_{-0.0004516}$ & $8.2669928^{+0.0000070}_{-0.0000067}$ & $3.39^{+0.10}_{-0.05}$ & $18.48^{+0.85}_{-2.57}$ & $0.31^{+0.26}_{-0.21}$ & $928^{+1}_{-23}$ & $0.0833^{+0.0039}_{-0.0116}$ & $3.59^{+0.11}_{-0.07}$ & Known & CP \\
			211525753.01 & K2-355 b & $3419.1918298^{+0.0049327}_{-0.0055620}$ & $5.7385647^{+0.0013939}_{-0.0011048}$ & $2.23^{+0.19}_{-0.12}$ & $14.85^{+2.05}_{-4.15}$ & $0.43^{+0.35}_{-0.30}$ & $1066^{+190}_{-77}$ & $0.0632^{+0.0096}_{-0.0176}$ & $2.25^{+0.22}_{-0.16}$ & New & VP \\
			211537087.01 & K2-356 b & $3423.0408399^{+0.0045164}_{-0.0045890}$ & $21.0267366^{+0.0036912}_{-0.0038993}$ & $2.44^{+0.13}_{-0.10}$ & $40.15^{+4.07}_{-7.66}$ & $0.36^{+0.30}_{-0.25}$ & $624^{+0}_{-37}$ & $0.1599^{+0.0178}_{-0.0306}$ & $2.29^{+0.15}_{-0.12}$ & New & VP \\
			211537087.02 &   & $3420.0192397^{+0.0039220}_{-0.0038371}$ & $42.3822512^{+0.0059266}_{-0.0055157}$ & $2.74^{+0.18}_{-0.10}$ & $67.60^{+6.47}_{-15.11}$ & $0.40^{+0.31}_{-0.27}$ & $482^{+3}_{-29}$ & $0.2688^{+0.0288}_{-0.0596}$ & $2.58^{+0.19}_{-0.14}$ & New & PC \\
			211537087.03 &   & $3426.9974440^{+0.0055336}_{-0.0063669}$ & $>$ 41.9 & $2.43^{+0.46}_{-0.26}$ & $>$ 57.0 & $0.70^{+0.23}_{-0.44}$ & $<$ 82 & $>$ 0.228 & $2.28^{+0.43}_{-0.26}$ & New & PC \\
			211590050.01 &   & $3442.4009415^{+0.0021095}_{-0.0017790}$ & $>$ 26.0 & $6.11^{+0.11}_{-0.77}$ & $>$ 25.6 & $0.27^{+0.21}_{-0.18}$ & $<$ 222 & $>$ 0.187 & $10.48^{+0.96}_{-0.95}$ & New & PC \\
			211594205.01 & K2-184 b & $2315.5253369^{+0.0009324}_{-0.0009632}$ & $16.9780102^{+0.0000237}_{-0.0000242}$ & $1.80^{+0.13}_{-0.05}$ & $48.36^{+4.39}_{-11.75}$ & $0.42^{+0.31}_{-0.29}$ & $533^{+10}_{-23}$ & $0.1685^{+0.0155}_{-0.0409}$ & $1.47^{+0.11}_{-0.05}$ & Known & VP \\
			211606790.01 &   & $2317.0889820^{+0.0004416}_{-0.0004583}$ & $37.2470875^{+0.0000218}_{-0.0000211}$ & $12.56^{+1.16}_{-1.16}$ & $33.57^{+0.92}_{-0.64}$ & $0.97^{+0.02}_{-0.02}$ & $665^{+0}_{-10}$ & $0.2755^{+0.0156}_{-0.0148}$ & $24.07^{+2.52}_{-2.44}$ & Known & PC \\
			211644764.01 &   & $3435.7199786^{+0.0019066}_{-0.0018808}$ & $23.5306712^{+0.0020516}_{-0.0020952}$ & $4.71^{+0.21}_{-0.12}$ & $65.70^{+6.01}_{-13.98}$ & $0.42^{+0.29}_{-0.29}$ & $591^{+10}_{-47}$ & $0.5304^{+0.0901}_{-0.1078}$ & $9.25^{+1.03}_{-1.02}$ & New & PC \\
			211705502.01 &   & $3421.3479060^{+0.0008176}_{-0.0008147}$ & $2.5844972^{+0.0000782}_{-0.0000779}$ & $6.85^{+0.55}_{-0.72}$ & $4.03^{+0.16}_{-0.14}$ & $0.99^{+0.01}_{-0.01}$ & $2277^{+118}_{-116}$ & $0.0260^{+0.0025}_{-0.0024}$ & $10.29^{+1.31}_{-1.32}$ & New & PC \\
			211724246.01 &   & $3423.1623021^{+0.0006878}_{-0.0006778}$ & $4.7464962^{+0.0001276}_{-0.0001258}$ & $5.06^{+0.11}_{-0.10}$ & $6.64^{+0.40}_{-0.31}$ & $0.91^{+0.01}_{-0.01}$ & $1775^{+96}_{-97}$ & $0.0651^{+0.0082}_{-0.0077}$ & $11.60^{+1.30}_{-1.28}$ & New & PC \\
			211730267.01 & K2-357 b & $3422.1162073^{+0.0017974}_{-0.0017414}$ & $16.3488637^{+0.0013811}_{-0.0014022}$ & $3.40^{+0.13}_{-0.08}$ & $31.62^{+1.98}_{-5.24}$ & $0.35^{+0.28}_{-0.24}$ & $735^{+8}_{-39}$ & $0.1450^{+0.0136}_{-0.0235}$ & $3.72^{+0.23}_{-0.21}$ & New & VP \\
			211733267.01 &   & $2311.9322437^{+0.0002781}_{-0.0002844}$ & $8.6580805^{+0.0000030}_{-0.0000030}$ & $12.22^{+1.84}_{-1.66}$ & $24.56^{+1.23}_{-0.78}$ & $0.96^{+0.03}_{-0.03}$ & $762^{+4}_{-19}$ & $0.1030^{+0.0055}_{-0.0041}$ & $11.99^{+1.82}_{-1.64}$ & Known & PC \\
			211791178.01 &   & $2313.9228292^{+0.0015524}_{-0.0015507}$ & $9.5607076^{+0.0000408}_{-0.0000362}$ & -- & -- & -- & -- & -- & -- & Known & PC \\
			211816003.01 & K2-272 b & $2311.8541321^{+0.0014315}_{-0.0015034}$ & $14.4536756^{+0.0000243}_{-0.0000240}$ & $3.43^{+0.44}_{-0.18}$ & $26.16^{+5.13}_{-9.27}$ & $0.58^{+0.28}_{-0.39}$ & $747^{+82}_{-65}$ & $0.1007^{+0.0199}_{-0.0354}$ & $3.13^{+0.38}_{-0.21}$ & Known & VP \\
			211818569.01 & K2-121 b & $2310.5605812^{+0.0000785}_{-0.0000763}$ & $5.1857539^{+0.0000006}_{-0.0000005}$ & $10.28^{+0.27}_{-0.19}$ & $20.19^{+1.04}_{-1.35}$ & $0.34^{+0.15}_{-0.22}$ & $738^{+7}_{-19}$ & $0.0629^{+0.0033}_{-0.0043}$ & $7.52^{+0.22}_{-0.18}$ & Known & VP \\
			211822797.01 & K2-103 b & $2311.4052567^{+0.0022025}_{-0.0022530}$ & $21.1701963^{+0.0000662}_{-0.0000642}$ & $3.03^{+0.19}_{-0.10}$ & $44.37^{+3.58}_{-9.35}$ & $0.38^{+0.30}_{-0.27}$ & $431^{+4}_{-17}$ & $0.1195^{+0.0100}_{-0.0251}$ & $1.92^{+0.12}_{-0.07}$ & Known & VP \\
			211904310.01 &   & $3432.6937079^{+0.0012236}_{-0.0012639}$ & $24.3998690^{+0.0019024}_{-0.0019239}$ & $10.29^{+1.32}_{-1.18}$ & $15.59^{+0.65}_{-0.47}$ & $0.97^{+0.02}_{-0.02}$ & $1144^{+56}_{-54}$ & $0.1201^{+0.0148}_{-0.0142}$ & $18.43^{+3.17}_{-2.76}$ & New & PC \\
			211913977.01 & K2-101 b & $2319.6849877^{+0.0011205}_{-0.0011677}$ & $14.6762429^{+0.0000201}_{-0.0000203}$ & $2.40^{+0.17}_{-0.07}$ & $31.82^{+2.67}_{-6.79}$ & $0.41^{+0.29}_{-0.28}$ & $618^{+9}_{-25}$ & $0.1124^{+0.0095}_{-0.0239}$ & $2.00^{+0.14}_{-0.07}$ & Known & VP \\
			211914998.01 & K2-358 b & $3426.0094216^{+0.0031628}_{-0.0029238}$ & $11.2510285^{+0.0018188}_{-0.0017892}$ & $2.69^{+0.12}_{-0.07}$ & $20.61^{+1.41}_{-3.89}$ & $0.35^{+0.30}_{-0.24}$ & $888^{+0}_{-47}$ & $0.0863^{+0.0079}_{-0.0159}$ & $2.68^{+0.17}_{-0.14}$ & New & VP \\
			211914998.02 &   & $3435.5995863^{+0.0051114}_{-0.0047980}$ & $24.6237203^{+0.0090182}_{-0.0079234}$ & $2.35^{+0.20}_{-0.12}$ & $47.18^{+6.26}_{-14.07}$ & $0.43^{+0.35}_{-0.30}$ & $585^{+12}_{-43}$ & $0.1984^{+0.0291}_{-0.0587}$ & $2.33^{+0.23}_{-0.16}$ & New & PC \\
			211916756.01 & K2-95 b & $2307.7422121^{+0.0007600}_{-0.0007697}$ & $10.1346454^{+0.0000121}_{-0.0000123}$ & $7.56^{+0.45}_{-0.22}$ & $28.28^{+3.06}_{-5.82}$ & $0.46^{+0.25}_{-0.31}$ & $474^{+8}_{-24}$ & $0.0525^{+0.0059}_{-0.0108}$ & $3.31^{+0.20}_{-0.14}$ & Known & VP \\
			211919004.01 & K2-273 b & $2316.0961619^{+0.0009530}_{-0.0009911}$ & $11.7195663^{+0.0000134}_{-0.0000135}$ & $3.27^{+0.25}_{-0.08}$ & $18.49^{+1.24}_{-3.62}$ & $0.37^{+0.30}_{-0.26}$ & $849^{+7}_{-29}$ & $0.0729^{+0.0053}_{-0.0141}$ & $3.05^{+0.23}_{-0.12}$ & Known & VP \\
			211969807.01 & K2-104 b & $2307.3799790^{+0.0009780}_{-0.0009972}$ & $1.9741954^{+0.0000024}_{-0.0000024}$ & $3.40^{+0.19}_{-0.09}$ & $10.89^{+0.95}_{-2.50}$ & $0.39^{+0.32}_{-0.27}$ & $798^{+9}_{-39}$ & $0.0242^{+0.0022}_{-0.0055}$ & $1.79^{+0.10}_{-0.06}$ & Known & VP \\
			212006344.01 & K2-122 b & $2308.8300469^{+0.0006420}_{-0.0006263}$ & $2.2193023^{+0.0000020}_{-0.0000020}$ & $1.80^{+0.15}_{-0.06}$ & $13.61^{+1.35}_{-3.45}$ & $0.43^{+0.32}_{-0.30}$ & $769^{+20}_{-37}$ & $0.0373^{+0.0038}_{-0.0095}$ & $1.16^{+0.10}_{-0.05}$ & Known & VP \\
			212008766.01 & K2-274 b & $2312.1112611^{+0.0013858}_{-0.0014376}$ & $14.1330314^{+0.0000282}_{-0.0000282}$ & $2.73^{+0.16}_{-0.07}$ & $29.68^{+1.73}_{-5.06}$ & $0.34^{+0.29}_{-0.23}$ & $655^{+0}_{-20}$ & $0.0964^{+0.0059}_{-0.0163}$ & $2.08^{+0.12}_{-0.07}$ & Known & VP \\
			212008766.02 &   & $3443.8646162^{+0.0033673}_{-0.0032444}$ & $>$ 74.8 & $5.69^{+1.06}_{-8.70}$ & $>$ 98.40 & $0.97^{+0.02}_{-0.02}$ & $<$ 86 & $>$ 0.320 & $4.34^{+0.81}_{-0.66}$ & New & PC \\
			212012119.01 & K2-275 b & $2309.1339592^{+0.0003369}_{-0.0003349}$ & $3.2809626^{+0.0000016}_{-0.0000016}$ & $3.11^{+0.32}_{-0.36}$ & $8.61^{+0.42}_{-2.54}$ & $0.77^{+0.13}_{-0.56}$ & $1166^{+0}_{-216}$ & $0.0277^{+0.0141}_{-0.0082}$ & $2.34^{+0.25}_{-0.27}$ & Known & VP \\
			212012119.02 & K2-275 c & $2309.4867161^{+0.0004575}_{-0.0005019}$ & $8.4388385^{+0.0000075}_{-0.0000067}$ & $2.93^{+0.11}_{-0.06}$ & $27.50^{+1.45}_{-3.26}$ & $0.33^{+0.23}_{-0.23}$ & $653^{+3}_{-17}$ & $0.0881^{+0.0049}_{-0.0104}$ & $2.21^{+0.08}_{-0.06}$ & Known & VP \\
			212110888.01 & K2-34 b & $2308.3514834^{+0.0000689}_{-0.0000699}$ & $2.9956348^{+0.0000002}_{-0.0000002}$ & $8.87^{+0.05}_{-0.05}$ & $6.68^{+0.10}_{-0.10}$ & $0.83^{+0.01}_{-0.01}$ & $1687^{+24}_{-25}$ & $0.0442^{+0.0020}_{-0.0020}$ & $13.75^{+0.58}_{-0.59}$ & Known & CP \\
			\hline
		\end{tabular}
	\end{center}
\label{tab:planet_params}
\end{table} \end{landscape} 


\subsection{Transit timing variations}
\label{TTVs}

We searched for TTVs produced by additional non-transiting planets in the light curves of our sample. For this, we took for each target the pre-processed and combined light curves as described in Section~\ref{K2_photometry}, and searched for TTVs using the Python Tool for Transit Variations \citep[\textsc{pyttv;}][]{Korth2020}.

The procedure is as follows: the transits from all the planets in a system are fitted together simultaneously by modelling them with the quadratic \citet{2002ApJ...580L.171M} transit model implemented in \textsc{pytransit} \citep{Parviainen2015} via a Taylor-series expansion \citep{2020MNRAS.499.3356P}, and fitted for all the transit centres $t_{c}$, impact parameter $b$, and planet-to-star radius ratio $R_{p}/R_{\star}$ for all planets, for the quadratic limb darkening coefficients $(u,v)$ and mean stellar density {\ensuremath{\rho_{\star}}}. The search for TTVs is carried out by fitting a linear, quadratic or sinusoidal model to the transit times, which is subtracted afterwards and evaluated through the GLS periodogram from \citet{2009A&A...496..577Z}, where best-fitting parameters and their uncertainties are calculated. The model with the lowest Bayesian Information Criterion (BIC) is chosen as the best model, and the significance of the other models with respect to the best model is calculated via the $\Delta \mathrm{BIC}$. 

Significant TTVs were detected for EPIC 211594205 that hint at the existence of an additional non-transiting planet (See Sect.~\ref{sec:EPIC_211594205} for a further discussion on the system). Besides, weak TTVs were detected for EPIC 211418290 and EPIC 211816003. The latter ones are most likely produced by stellar activity and spots, which is visible by the high scatter of the in-transit residuals compared to the out-of-transit residuals in their phase-folded transits (Fig.~\ref{fig:phase_folded_1}). 

As TTVs are most sensitive to planets near resonant orbits, we checked the period ratios of the planets in our sample with more than one planet in a system. K2-356 b (EPIC 211537087.01) and EPIC 211537087.02 have a period ratio close to 2, which hints at strong perturbations that can lead to significant TTVs for both of them. In Sect.~\ref{epic_211537087}, we include a further discussion on this system and estimate the TTV periods and amplitudes for both the planet and the candidate.


\subsection{Statistical validation}
\label{validation}

We carried out a statistical validation analysis for the \newsignals new planet candidates found. First, we computed the false positive probabilities (FPPs), which are the probabilities of the signals being astrophysical false positives (Sect.~\ref{sec:FPP_calculation}). Secondly, we assessed the reliability of the FPP calculation in order to obtain the final disposition of each new planet candidate (Sect.~\ref{sec:FPP_reliability}).

\subsubsection{FPP calculation}
\label{sec:FPP_calculation}

We obtained the FPPs by using the \textsc{vespa} package \citep{2012ApJ...761....6M,2015ascl.soft03011M}, which computes the likelihood of the main astrophysical false positives scenarios: eclipsing binaries (EBs), background eclipsing binaries (BEBs), and hierarchical triple systems (HEBs), taking into account the target coordinates and relying on simulations of the Galaxy from the \textsc{trilegal} population synthesis code \citep{2005A&A...436..895G}. Briefly, to assign the probability for each scenario, \textsc{vespa} starts from \textsc{isochrones} to carry out single-, binary-, and triple-star model fits to the observed photometric,  spectroscopic and parallax constraints. Then \textsc{vespa} simulates thousands of planetary and non-planetary scenarios to be compared with the observed phase-folded light curve, which is modelled through a trapezoidal transit fit. Finally, the FPP is computed as the posterior probability of the non-planetary scenarios.

We ran \textsc{vespa} starting from the aforementioned constraints and several additional constraints that help to assess the different scenarios. We used the orbital period and the planet-to-star radius ratio as derived from \textsc{pyaneti}. We computed the maximum aperture radius for which the signal is expected to come from (\texttt{maxrad} parameter) as $\sqrt{A \pi^{-1}}$, being $A$ the area of the aperture. We also constrained the maximum allowed depth of a potential secondary eclipse (\texttt{secthresh}) to be thrice the standard deviation of the out-of-transit region. This constraint is quite conservative, as any secondary eclipse with that depth would be noticed even by eye. We show the obtained FPP broken down by scenario in Table~\ref{tab:FPPs}.

\subsubsection{Reliability of FPPs and final dispositions}
\label{sec:FPP_reliability}

The most commonly adopted criteria to consider a candidate as statistically validated planet (VP) is to have a false positive probability lower than 1$\%$  \citep[FPP $<$ 0.01; e.g.][]{2014ApJ...784...45R, 2015ApJ...809...25M, 2016ApJ...822...86M, 2019A&A...627A..66H}, while a candidate with a false positive probability greater than 90$\%$ (FPP $>$ 0.9) is considered as a false positive \citep[FP; e.g.][]{2015ApJ...809...25M, 2016ApJ...822...86M}. For the rest of the cases (0.01 $<$ FPP $<$ 0.9), the planet candidate disposition (PC) prevails. However, it has been widely discussed \citep[e.g.][]{2018AJ....156..277L,2018AJ....155..136M} and proved \citep[e.g.][]{2017A&A...606A..75C,2017ApJ...847L..18S} that only relying on the FPP can lead to misclassifications related to several factors not being taken into consideration by the validation packages.  In the next paragraphs, we detail the conditions that any target and candidate need to meet before assigning it a disposition based on its FPP, as well as which crucial factors can prevent validation, independently of the FPP.

As planetary signals must be periodic, we do not validate any candidate with fewer than three consecutive transits within the light curve. Therefore, the signals found from EPICs 211537087.03 (1 dip), 211590050.01 (1 dip), 211644764.01 (2 dips), 211904310.01 (2 dips), and 212008766.02 (1 dip) are considered as planetary candidates. For those targets with three or more transit signals, we searched for odd-even transit depth mismatches in order to identify possible secondary transits. For that, we modelled separately the odd and even transit events and compared their depths, avoiding validating any signal with a transit depth mismatch higher than 3-$\sigma$.  We also avoid validating noisy signals in order to discard possible non-physical origins. Quantitatively, we do not validate any signal with a signal-to-noise ratio (S/N)\footnote{We computed the signal-to-noise ratio as S/N = $ d  \sqrt{N_{p}} \sigma^{-1}$, where $d$ is the transit depth, $N_{p}$ the number of data points in transit, and $\sigma$ the standard deviation of the de-trended light curve.} lower than 10, which is a conservative threshold adopted by several authors \citep[e.g.][]{2012ApJS..201...15H,2016ApJ...822...86M,2020MNRAS.499.5416C}.  Besides, similarly to previous works \citep[e.g.][]{2018AJ....155..136M,2021AJ....161...24G}, we do not validate any signal with $R_{\rm p}$ $>$ 8 $\rm R_{\oplus}$ in order to avoid validating brown dwarfs and low-mass eclipsing stellar companions.   In Table \ref{tab:FPPs} we include the number of observed transits, odd-even mismatches, S/N and $R_{\rm p}$ of each new signal. 

\begin{landscape}
\begin{table*}
\renewcommand{\arraystretch}{1.7}
\fontsize{10.2pt}{10.2pt}\selectfont
\caption[Summary of the K2-OjOS statistical validation analysis.]{Summary of the statistical validation analysis for the new targets reported. From left to right: ID, final probabilities (computed from \textsc{vespa}) that the signal is due to a BEB, EB, HEB, probability that the signal comes from a planet, FPP, number of measured transits, signal-to-noise ratio, odd-even mismatch, Astrometric Goodness of Fit of the astrometric solution for the star in the Along-Scan direction, Astrometric Excess Noise significance, candidate radius, fulfilment of the condition $\delta'$ $>$ $\gamma^{-1}$, and disposition assigned (VP = validated planet, PC = planet candidate).}
	\begin{center}
		\begin{tabular}{c|c|c|c|c|c|c|c|c|c|c|c|c|c}
		    \hline \hline
			ID & P(BEB) & P(EB) & P(HEB) & P(Pl) & FPP & $\#$tr & S/N & mismatch [$\sigma$] & GOF\_AL & D & $R_{p}$ [$R_{\oplus}$] & $\delta'$ $>$ $\gamma^{-1}$ & Disp \\
			\hline
			211309648.01 & $\rm 2.8 \times 10^{-25}$ & $\rm 2.1 \times 10^{-2}$ & $\rm 4.1 \times 10^{-4}$ & $\rm 9.8 \times 10^{-1}$ & $\rm 2.1 \times 10^{-2}$ & $\geq 3$ & 101 & {0.43} & {-3.41} & {0.00} & {$\geq 8$} & Yes & PC \\
			211319779.01 & $\rm 1.0 \times 10^{-2}$ & $\rm 2.8 \times 10^{-6}$ & $\rm 1.9 \times 10^{-10}$ & $\rm 9.9 \times 10^{-1}$ & $\rm 1.0 \times 10^{-2}$ & $\geq 3$ & 18 & {0.25} & {1.43} & {0.00} & {2.4} & No & PC \\
			211393988.01 & $\rm 5.2 \times 10^{-15}$ & $\rm 5.7 \times 10^{-2}$ & $\rm 7.9 \times 10^{-4}$ & $\rm 9.4 \times 10^{-1}$ & $\rm 5.8 \times 10^{-2}$ & $\geq 3$ & 62 & {1.20} & {0.97} & {0.00} & {$\geq 8$} & Yes & PC \\
			211407755.01 & $\rm 7.6 \times 10^{-1}$ & $\rm 2.1 \times 10^{-1}$ & $\rm 1.1 \times 10^{-3}$ & $\rm 2.5 \times 10^{-2}$ & $\rm 9.7 \times 10^{-1}$ & $\geq 3$ & 22 & {0.42} & {-2.62} & {0.00} & {6.1} & No & PC \\
			211480861.01 & $\rm 2.8 \times 10^{-2}$ & $\rm 8.3 \times 10^{-1}$ & $\rm 7.0 \times 10^{-2}$ & $\rm 6.9 \times 10^{-2}$ & $\rm 9.3 \times 10^{-1}$ & $\geq 3$  & 48 & {0.04} & {96.54} & {185.60} & {$\geq 8$} & Yes & PC \\
			211525753.01 & $\rm 7.6 \times 10^{-3}$ & $\rm 4.5 \times 10^{-5}$ & $\rm 1.4 \times 10^{-9}$ & $\rm 9.9 \times 10^{-1}$ & $\rm 7.7 \times 10^{-3}$ & $\geq 3$ & 13 & {0.15} & {-5.52} & {1.04} & {5.7} & Yes & \textbf{VP}  \\
			211537087.01 & $\rm 2.8 \times 10^{-3}$ & $\rm 7.2 \times 10^{-5}$ & $\rm 2.5 \times 10^{-9}$ & $\rm 1.0 \times 10^{0}$ & $\rm 2.9 \times 10^{-3}$ & $\geq 3$ & 15 & {0.63} & {-3.51} & {0.00} & {2.3} & Yes & \textbf{VP} \\
			211537087.02 & $\rm 7.8 \times 10^{-3}$ & $\rm 2.8 \times 10^{-3}$ & $\rm 4.5 \times 10^{-9}$ & $\rm 9.9 \times 10^{-1}$ & $\rm 1.1 \times 10^{-2}$ & 2 & 20 & {-} & {-3.51} & {0.00} & {2.6} & Yes & PC \\
			211537087.03 & - & - & - & - & - & 1 & 11 & {-} & {-3.51} & {0.00} & {2.3} & Yes & PC \\
			211590050.01 & - & - & - & - & - & 1 & 105 & {-} & {-0.60} & {0.00} & {$\geq 8$} & Yes & PC  \\
			211644764.01 & $\rm 5.9 \times 10^{-2}$ & $\rm 3.6 \times 10^{-1}$ & $\rm 7.7 \times 10^{-2}$ & $\rm 5.1 \times 10^{-1}$ & $\rm 4.9 \times 10^{-1}$ & 2 & 56 & {-} & {4.64} & {0.64} & {$\geq 8$} & Yes & PC \\
			211705502.01 & $\rm 7.9 \times 10^{-1}$ & $\rm 2.0 \times 10^{-1}$ & $\rm 2.8 \times 10^{-3}$ & $\rm 4.9 \times 10^{-3}$ & $\rm 9.9 \times 10^{-1}$ & $\geq 3$ & 46 & {3.70} & {0.57} & {0.00} & {$\geq 8$} & Yes & PC \\
			211724246.01 & $\rm 4.1 \times 10^{-2}$ & $\rm 2.1 \times 10^{-2}$ & $\rm 1.2 \times 10^{-3}$ & $\rm 9.4 \times 10^{-1}$ & $\rm 6.3 \times 10^{-2}$ & $\geq 3$ & 58 & {0.33} & {-3.50} & {0.00} & {$\geq 8$} & Yes & PC \\
			211730267.01 & $\rm 2.9 \times 10^{-4}$ & $\rm 3.5 \times 10^{-4}$ & $\rm 1.1 \times 10^{-16}$ & $\rm 1.0 \times 10^{0}$ & $\rm 6.3 \times 10^{-4}$ & $\geq 3$ & 28 & {0.33} & {-4.86} & {0.00} & {3.7} & Yes & \textbf{VP} \\
			211904310.01 & $\rm 3.3 \times 10^{-1}$ & $\rm 2.1 \times 10^{-1}$ & $\rm 2.0 \times 10^{-2}$ & $\rm 4.4 \times 10^{-1}$ & $\rm 5.6 \times 10^{-1}$ & 2 & 42 & {-} & {-2.13} & {0.00} & {$\geq 8$} & Yes & PC  \\
			211914998.01 & $\rm 8.9 \times 10^{-4}$ & $\rm 4.7 \times 10^{-5}$ & $\rm 1.3 \times 10^{-11}$ & $\rm 1.0 \times 10^{0}$ & $\rm 9.3 \times 10^{-4}$ & $\geq 3$ & 20 & {1.58} & {-6.55} & {0.00} & {2.5} & Yes & \textbf{VP} \\
			211914998.02 & $\rm 1.8 \times 10^{-2}$ & $\rm 4.0 \times 10^{-3}$ & $\rm 3.5 \times 10^{-6}$ & $\rm 9.8 \times 10^{-1}$ & $\rm 2.2 \times 10^{-2}$ & 2 & 8 & {-} & {-6.55} & {0.00} & {2.1} & Yes & PC \\
			212008766.02 & - & - & - & - & - & 1 & 44 & {-} & {-1.15} & {0.00} & {2.0} & Yes & PC \\
			\hline
		\end{tabular}
	\end{center}
	\label{tab:FPPs}
\end{table*}
\end{landscape}

\begin{table}
\centering
\renewcommand{\arraystretch}{1.45}
\caption[Dilution factors of the K2-OjOS sample.]{Dilution factors and magnitudes differences for each nearby \textit{Gaia} DR2 star located inside or outside the aperture at a distance $r$ $<$ 40 arcsecond. }

\begin{tabular}{cccccc}
\hline \hline
EPIC & Aperture & $r$ (arcsecond) & $\Delta_{m}$ & $\gamma_{\rm pri}$ & $\gamma_{\rm sec}$ \\ \hline
211309648 & Outside & 31.86 & 5.93 & 1.004 & 236.548 \\
 & Outside & 39.52 & 5.85 & 1.005 & 219.675 \\ 
211319779 & Inside & 3.41 & 6.24 & 1.003 & 312.918 \\
 & Outside & 11.22 & 2.88 & 1.070 & 15.232 \\ 
211393988 & Outside & 32.18 & -0.96 & 3.424 & 1.412 \\
 & Outside & 34.51 & 3.40 & 1.044 & 23.985 \\ 
\multicolumn{1}{l}{211407755} & Inside & 4.73 & 3.50 & 1.040 & 26.223 \\
\multicolumn{1}{l}{} & Inside & 11.04 & 6.62 & 1.002 & 446.985 \\
\multicolumn{1}{l}{} & Outside & 21.97 & 2.17 & 1.135 & 8.410 \\
\multicolumn{1}{l}{} & Outside & 35.39 & 3.91 & 1.027 & 37.749 \\ 
211480861 & Inside & 12.87 & 6.52 & 1.002 & 407.481 \\
 & Outside & 28.77 & 3.91 & 1.027 & 37.789 \\
 & Outside & 32.41 & 7.27 & 1.001 & 807.789 \\
 & Outside & 34.93 & 6.94 & 1.002 & 598.97 \\ 
211525753 & Outside & 28.45 & 2.72 & 1.082 & 13.242 \\ 
211537087 & Outside & 17.66 & 7.36 & 1.001 & 881.968 \\ 
211590050 & $-$ & $-$ & $-$ & $-$ & $-$ \\
211644764 & Outside & 16.85 & 7.23 & 1.001 & 779.825 \\
 & Outside & 35.27 & 5.62 & 1.006 & 178.812 \\ 
\multicolumn{1}{l}{211705502} & Inside & 5.88 & 6.47 & 1.003 & 389.830 \\
\multicolumn{1}{l}{} & Outside & 31.91 & 4.41 & 1.017 & 58.932 \\
\multicolumn{1}{l}{} & Outside & 35.12 & 0.12 & 1.898 & 2.114 \\ 
211724246 & Outside & 21.68 & 7.24 & 1.001 & 789.497 \\
 & Outside & 26.78 & 5.68 & 1.005 & 188.241 \\
 & Outside & 30.04 & 7.07 & 1.001 & 673.419 \\
 & Outside & 32.11 & 5.76 & 1.005 & 202.131 \\
 & Outside & 36.57 & 5.11 & 1.009 & 111.235 \\ 
211730267 & Outside & 16.56 & 6.47 & 1.003 & 386.549 \\
 & Outside & 30.35 & 6.09 & 1.004 & 272.894 \\
 & Outside & 32.65 & 6.85 & 1.002 & 552.47 \\ 
211904310 & Outside & 30.95 & 5.57 & 1.006 & 169.749 \\ 
211914998 & Outside & 34.94 & 4.40 & 1.017 & 58.671 \\ 
\hline
\end{tabular}
\label{tab:dilution_factors}
\end{table}

\begin{figure*}
\centering
    \includegraphics[scale = 0.35]{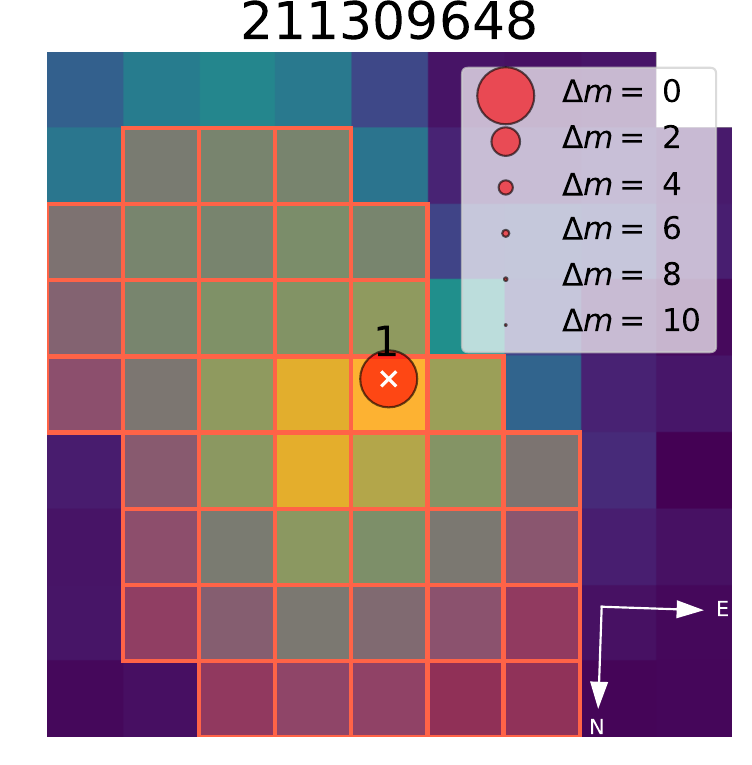}
    \includegraphics[scale=0.35]{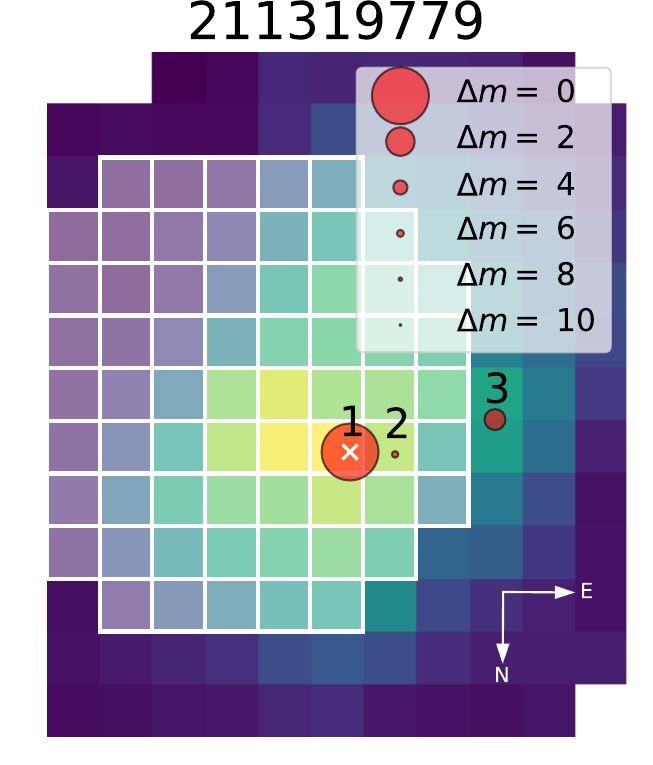}
    \includegraphics[scale=0.34]{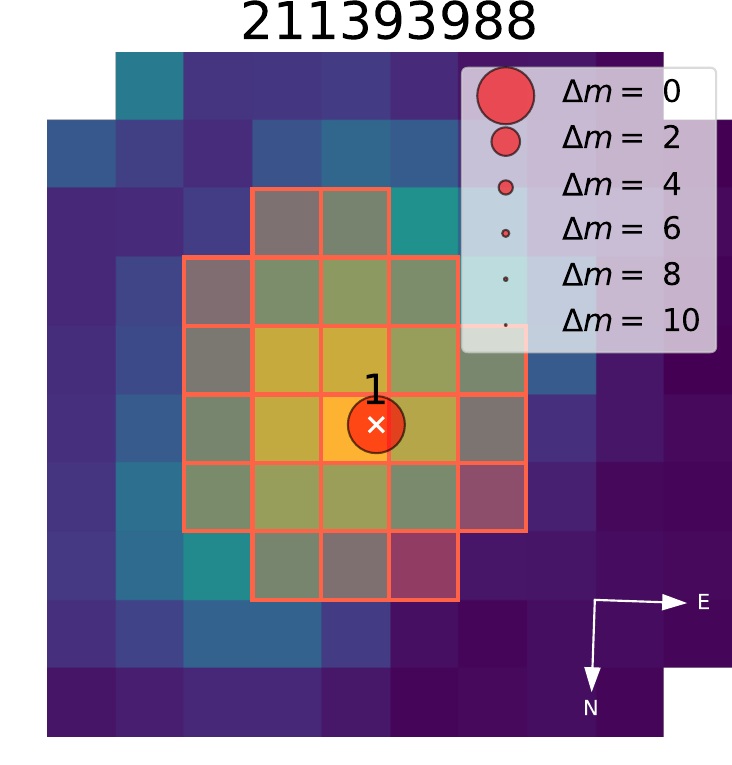}
    \includegraphics[scale=0.34]{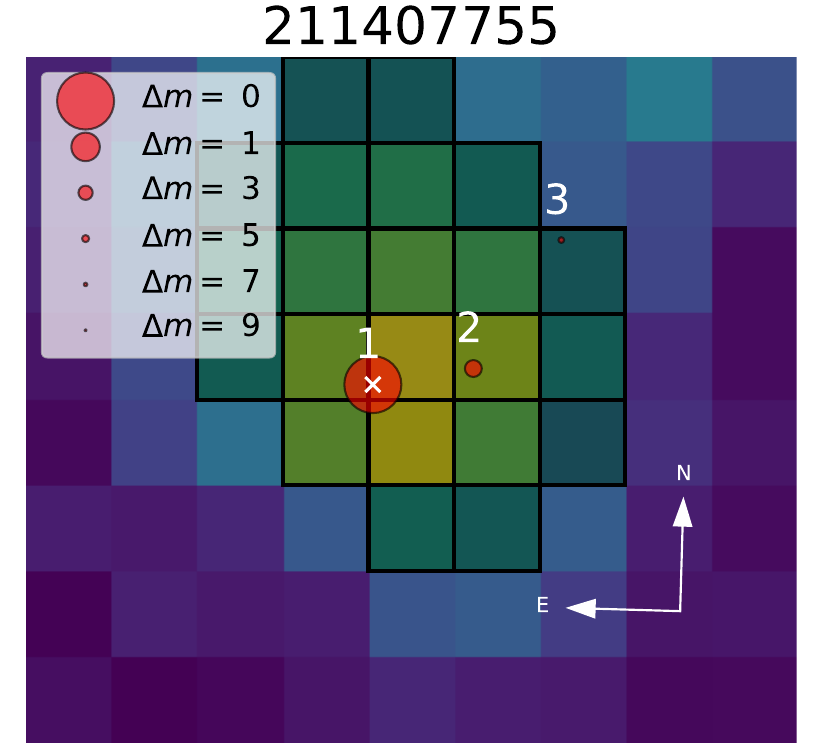}
    \includegraphics[scale=0.33]{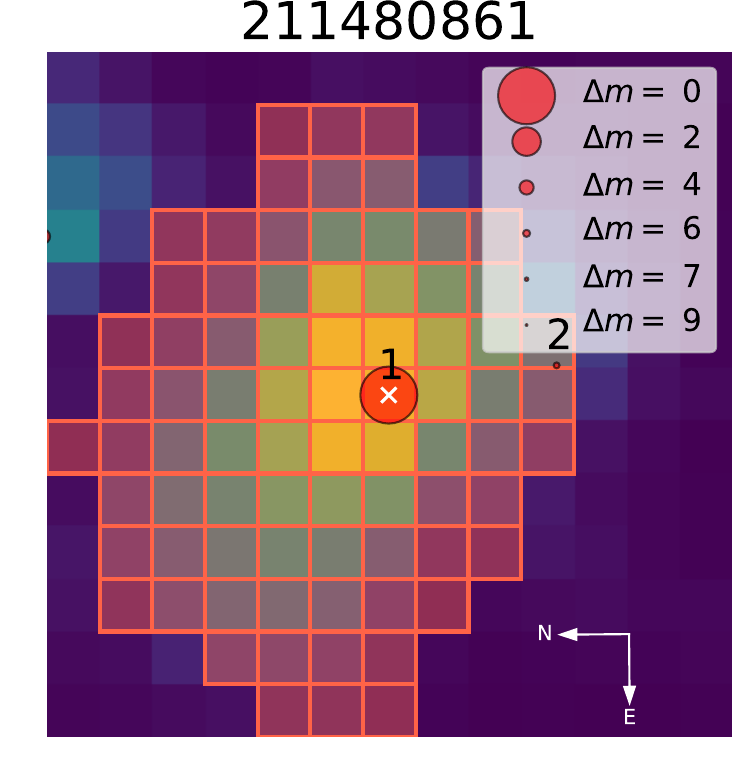}
    \includegraphics[scale=0.33]{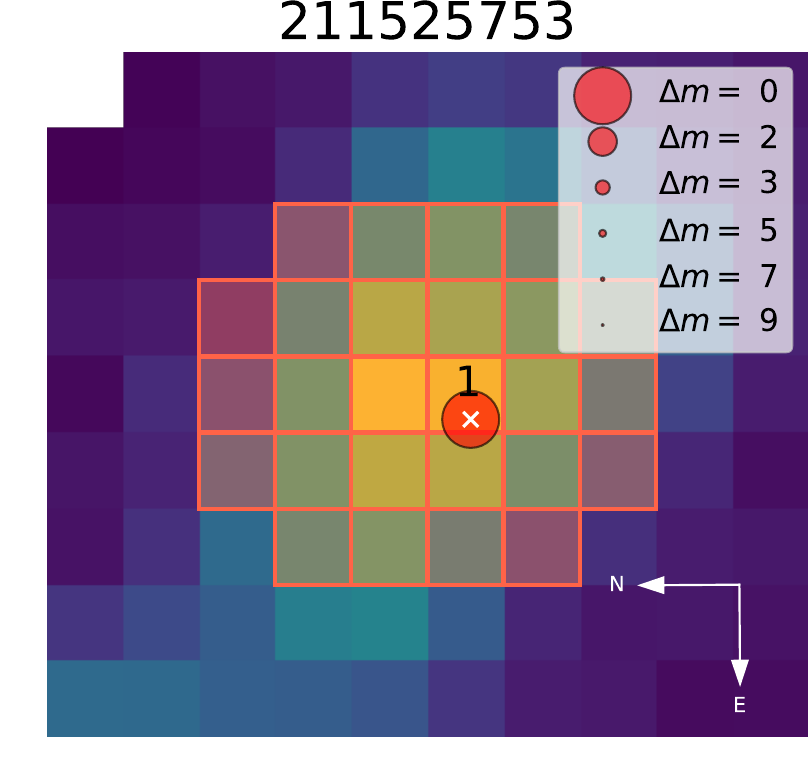}
    \includegraphics[scale=0.33]{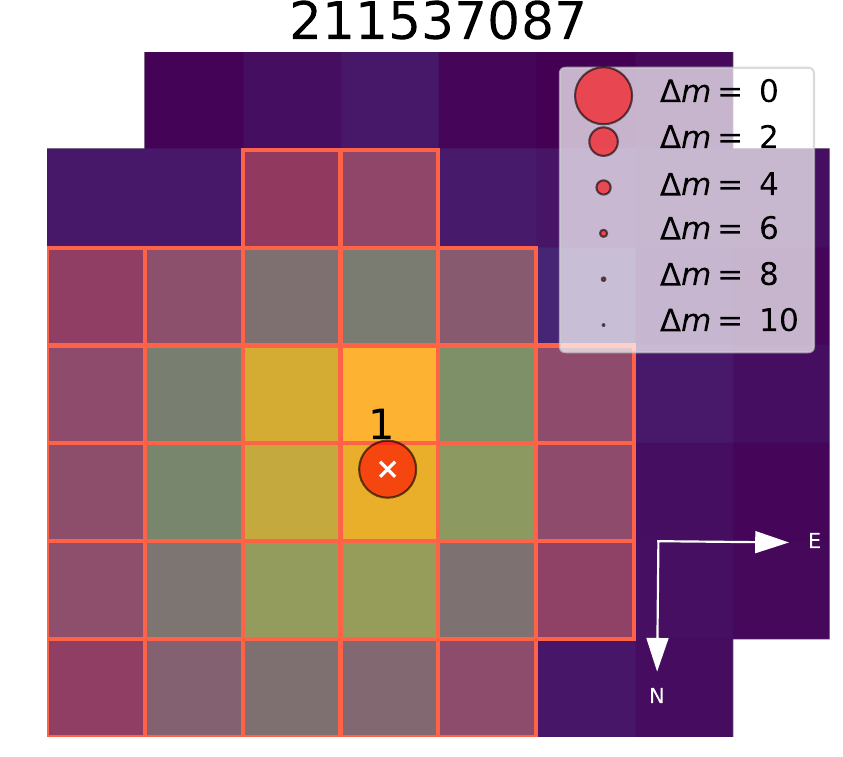}
    \includegraphics[scale=0.33]{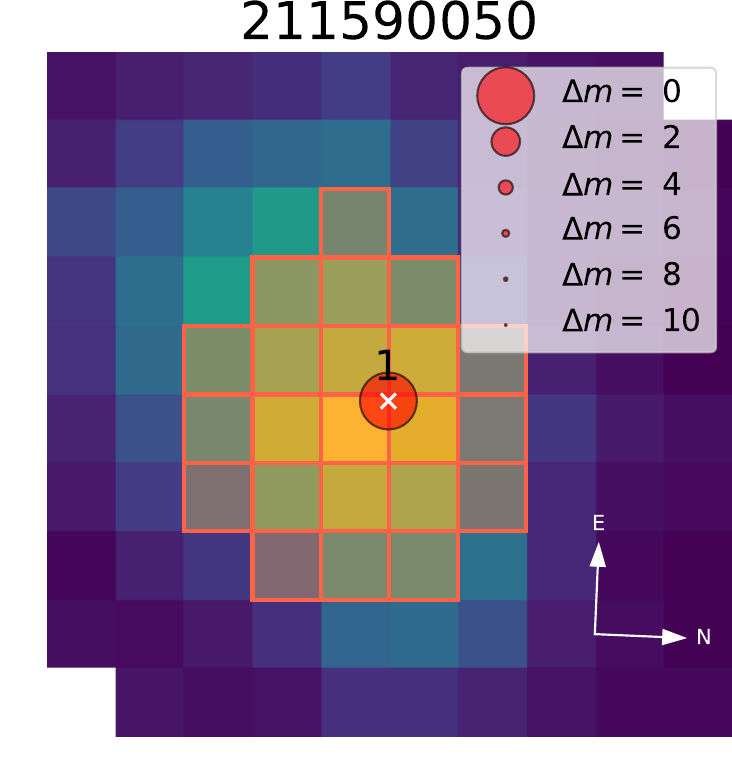}
    \includegraphics[scale=0.33]{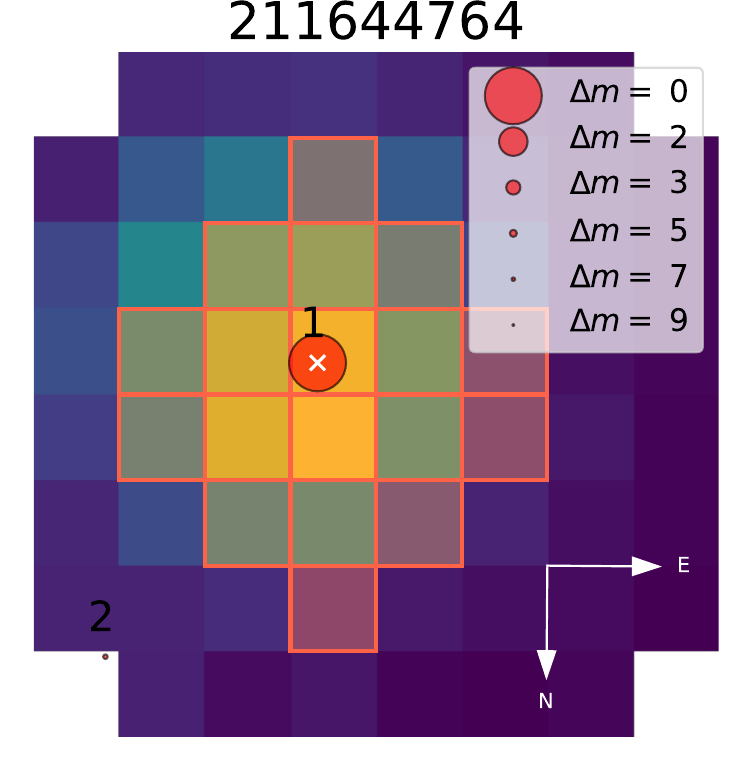}
    \includegraphics[scale=0.33]{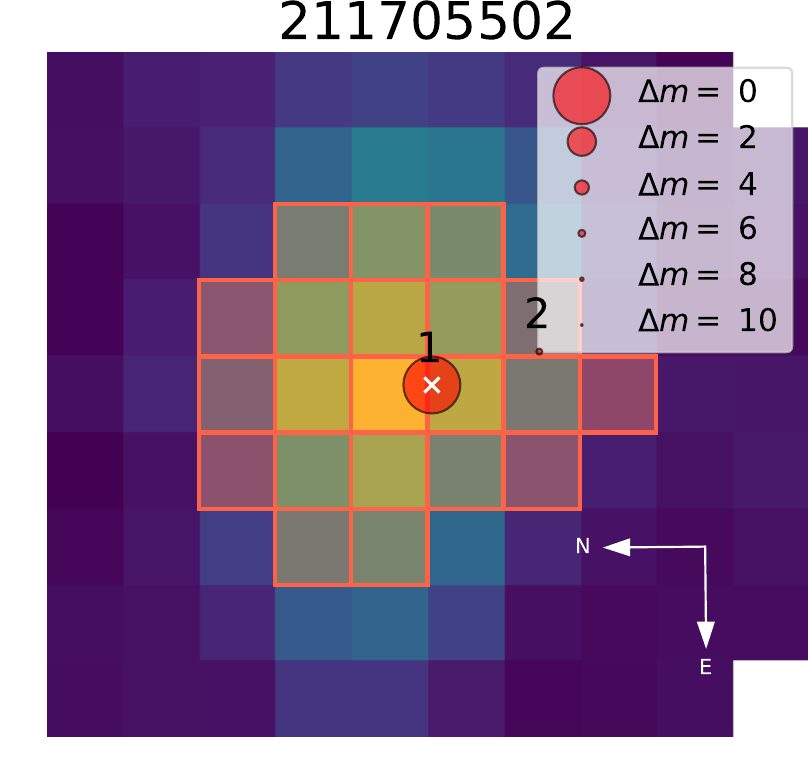}
    \includegraphics[scale=0.33]{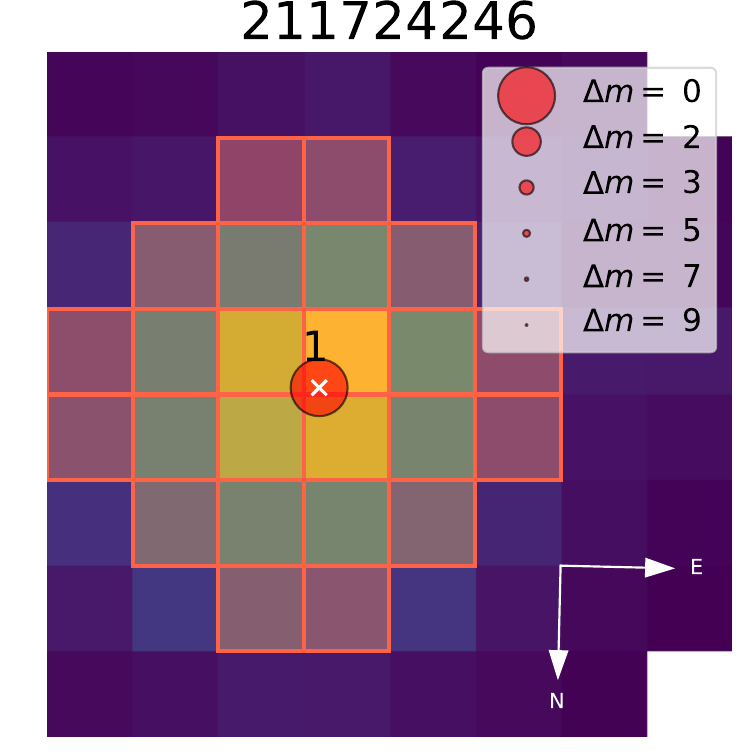}
    \includegraphics[scale=0.33]{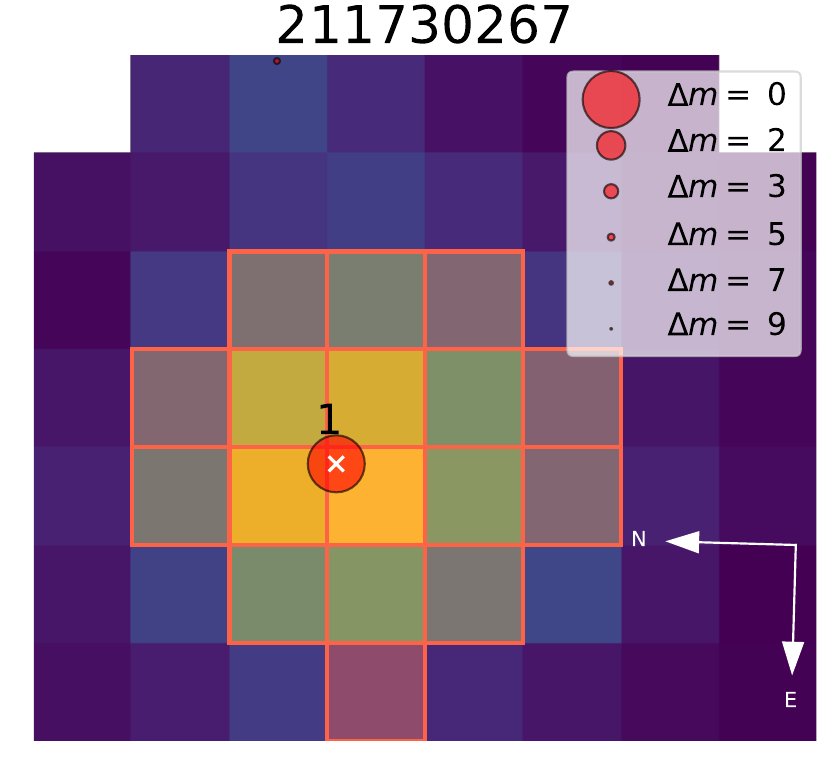}
    \includegraphics[scale=0.34]{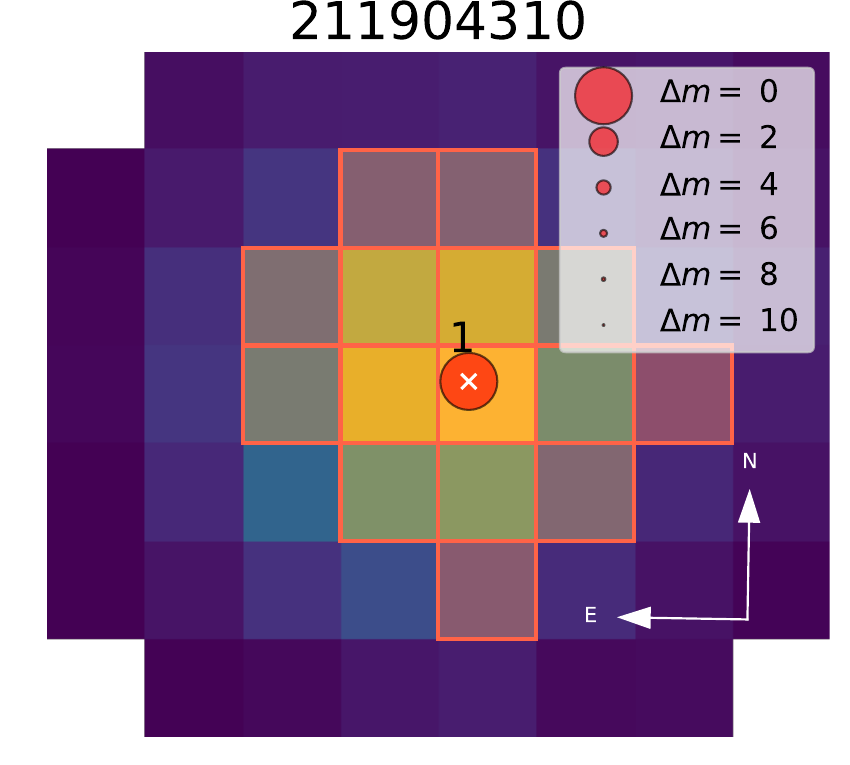}
    \includegraphics[scale=0.34]{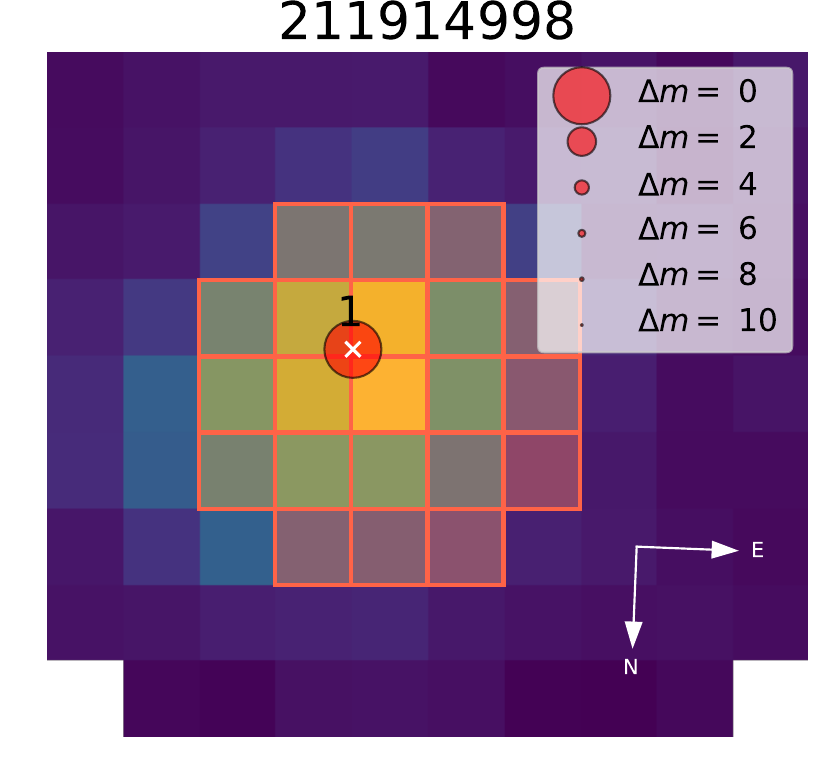}

    \caption[Target pixel files of the new K2-OjOS targets.]{Target pixel files (TPFs) of each new target reported in this chapter. The red circles are the \textit{Gaia} DR2 sources in the field, which are scaled according to the magnitude difference between the target (highlighted with a white cross) and each nearby star. The over-plotted apertures are those considered to obtain the photometry, being the red ones corresponding to the EVEREST pipeline, and the white one to the K2SFF pipeline. The black aperture was defined manually.}
    \label{fig:tpfplotter_1}
\end{figure*}

We searched for hints of binarity by using \textit{Gaia} DR2. Systems with large Astrometric Goodness of Fit of the astrometric solution for the source in the Along-Scan direction (\texttt{GOF\_AL} $>$ 20) and Astrometric Excess Noise significance (\texttt{D} $>$ 5) are plausibly poorly-resolved binaries \citep{2018RNAAS...2...20E}. There is one target in our sample that meets these two conditions (EPIC 211480861.01). Given the possible presence of multiple stars in the system, we designate this target as PC regardless of its FPP, and we do not report its planetary parameters. We include \texttt{GOF\_AL} and \texttt{D} in Table \ref{tab:FPPs} for each target.

Another important consideration before relying on the FPP consists of searching for nearby stars inside or surrounding the aperture, which could be contaminating the photometry. A fairly common false positive scenario involves the presence of a fainter contaminant eclipsing binary, whose deep dips are diluted by the target star, so what we observe appears to be more similar to the typical shallower planetary transits. In another possible scenario, if a planetary signal comes from the target star and there is also a bright nearby contaminant star, the transit depths will be shallower, thus systematically causing an underestimation of the planet radius. In any case in which we find contaminant stars causing that the origin of the signal cannot be determined, we prevent the candidate from validation, and we do not report its parameters. 

In order to search for nearby contaminant stars in our new targets, we updated the \textsc{tpfplotter} package \citep{2020A&A...635A.128A}, which, in addition to \textit{TESS}, is now compatible with \textit{Kepler} and K2 data. The package overlaps the \textit{Gaia} DR2 catalogue to the Target Pixel Files (TPFs), computing and plotting the location of potential contaminant sources relative to the photometric aperture (see Fig. \ref{fig:tpfplotter_1}). Note that the apertures of some targets occupy almost the entire TPF, so in these cases there can be stars outside the TPF but still contaminating the photometry due to the broad point spread function (PSF) of the \textit{Kepler} telescope, which has a typical full-width at half-maximum of FWHM $\approx$ 6 arcseconds. For this reason, we conservatively looked for all the \textit{Gaia} DR2 sources within a search radius of 40 arcsec and with up to 10 magnitudes fainter than the target star. We quantified the photometric contamination by computing the dilution factor as $\gamma = 1+10^{0.4\Delta m}$ \citep[equation 1,][]{2018AJ....156..277L}, which defines the relationship between the observed transit depth ($\delta'$) and the true transit depth ($\delta$) as $\delta'$ = $\gamma^{-1}$ $\delta$, being $\Delta m$ the magnitude of the contaminant star minus the magnitude of the star where the signal comes from, in the \textit{Kepler} bandpass. We use the notation $\gamma_{\rm pri}$ and $\gamma_{\rm sec}$ to indicate that the dilution factor is computed considering that the signal comes from the target (primary) star with a true transit depth $\delta_{\rm pri}$ or from a nearby (secondary) star with a true transit depth $\delta_{\rm sec}$. 

We show in Table \ref{tab:dilution_factors} both $\delta_{\rm pri}$ and $\delta_{\rm sec}$ for all the sources found at a distance $<$ 40 arcsec of our newly detected targets, as well as their separation and magnitude differences. We do not know a priori where the signal comes from, so we followed a procedure to assess whether we can discard the nearby star origin. The procedure consisted of assuming that the signal comes from any of the nearby stars located inside the aperture or outside but separated by up to 6 arcsec from the nearest edge. So, as their hypothetical eclipses cannot be greater than 100$\%$ (i.e. $\delta_{\rm sec}$ $\leq$ 1), if the condition $\delta'$ $>$ $\gamma_{\rm sec}^{-1}$ is met, we can ensure that the observed depth $\delta'$ is too deep to be caused by the nearby secondary star. Otherwise, the origin of the signal is uncertain, so the FPP is not reliable, and we consider it a PC until its origin is determined.

The condition $\delta'$ $>$ $\gamma_{\rm sec}^{-1}$ is met for all our new targets except for EPIC 211319779 and EPIC 211407755. The dip observed in EPIC 211319779 could be caused by a $\sim$30$\%$ dip coming from star $\#2$, and by a $\sim$1.5$\%$ dip coming from star $\#3$. Similarly, the dip observed in EPIC 211407755 could be caused by a  $\sim$1$\%$ dip coming from star $\#2$ (see Fig. \ref{fig:tpfplotter_1} and Table \ref{tab:dilution_factors}). For these two cases, we performed pixel-level multi-aperture analysis in order to figure out the actual origin of the signals found. In some cases, when the target star and the potential contaminant faint star are located several pixels apart, assessing the photometry created with different photometric apertures can solve the signal origin uncertainty, being decisive to unveil possible FP scenarios \citep[e.g.][found that two K2 validated planets were background eclipsing binaries, and hence FPs]{2017A&A...606A..75C}. Unfortunately, our targets are not suitable for reaching such decisive conclusions through multi-aperture analysis, because of the great closeness between the stars. Even though there are clear hints of that the signal from EPIC 211319779 does not come from star $\#3$ (e.g. different apertures both enclosing and excluding it do not alter transit depths, and the transits are still present considering the EVEREST aperture, which is 14 arcsec away from star $\#3$ as shown in Fig. \ref{fig:EVEREST_vs_K2SFF}), the stars $\#1$ and $\#2$ for both  EPIC 211319779 and EPIC 211407755 are 1 pixel away, causing the PSFs to be completely blended. As the signal origin cannot be ascertained for these two targets, we consider them as PCs.

Of the \newsignals new candidates found, four satisfy all the aforementioned conditions. Besides, these candidates have FPP $<$ 0.01. Before relying on the FPP to assign the final disposition, we searched for close sources by observing these stars with the high-spatial resolution camera AstraLux, located at the 2.2 m telescope of the Calar Alto Observatory. The presence of close contaminant sources, identically to nearby sources, implies a potential misidentification in the origin of the signal. Besides, the \textsc{isochrones} stellar characterisation would be unreliable because of the photometric contamination between both sources. For each target, we found no evidence of additional sources within a 6 × 6 arcsec field of view and the computed sensitivity limits (see Sect.~\ref{sec:AstraLux} for further details).  After meeting all the conditions imposed, we consider these four candidates as validated planets. To sum up, the statistical validation analysis carried out over the \newsignals new signals found resulted in four validated planets and 14 planet candidates, three of which we report without planetary parameters due to the presence of photometric contamination that prevents the determination of the origin of the signals.


\section{Results and Discussion}
\label{results_discussion}

The K2-OjOS search in K2 C18 gave rise to 42 planet candidates, of which 24 were published in previous works (four confirmed planets, 14 validated planets, and six planet candidates), and 18 are new detections (four validated planets and 14 planet candidates). In Sect.~\ref{sec:refining_ephemeris}, we quantify the refinement of transit ephemeris and planetary parameters that we achieve by modelling C5, C16 (when available) and C18 photometric data jointly for the previously known planets and candidates. In Sect.~\ref{sec:detection_efficiency}, we compute and compare the detection efficiencies of both the K2-OjOS and BLS searches.  In Sections \ref{sec:charact_stellar_sample} and \ref{sec:charact_planet_sample}, we contextualize our results by comparing the K2-OjOS detections to the population of known host stars and exoplanets,\footnote{All data for known host stars, planets, and candidates were obtained from the NASA Exoplanet Archive (NEA).} and discuss about the possible internal structure of the K2-OjOS planets and candidates. We also compute the habitable zone (HZ) boundaries for our target stars based on both conservative and more optimistic climate models. In Sect.~\ref{sec:highlights_individual_systems}, we highlight interesting features of five individual systems.

\subsection{Refining transit ephemeris and planetary parameters}
\label{sec:refining_ephemeris}

Obtaining long temporal baselines of photometric data for targets hosting transiting planets allows us to measure transit ephemerides very precisely. Besides, increasing the number of observed transits allows us to better constrain the planetary parameters due to a greater in-transit coverage. The latter is especially important for long-period planets observed by K2, whose long 30-min cadence corresponds to few data points per transit (typically between 4 and 10). For deep enough dips, transit follow-up can be carried out from ground-based facilities, but for many targets, these observations need to be done from space. 

K2 C18 observed a field that covers 30$\%$ of C16 and 95$\%$ of C5. The targets observed in both C5 and C18 have a 3-year temporal baseline with a 4-month duty cycle, while for those targets observed in C5, C16, and C18, their duty cycle increases up to 7 months. In our sample of targets with planets and/or candidates with a full characterisation, 12 of them were observed in C18 alone, 13 were observed in both C5 and C18, and 9 were observed in C5, C16, and C18. All the 22 targets with observations in multiple campaigns host published planets or candidates, whose planetary parameters were derived starting from C5 data alone, since when their corresponding papers were in preparation, C16 and C18 had not started yet. In this work, we modelled for the first time the light curves of these \targetstoimprove \, targets by joining photometric data from C5, C16 (when available), and C18, managing to refine their published transit ephemeris and planetary parameters. For the orbital period, we obtain uncertainty improvement factors between 10 and 88, with a median value of \medianfactorP. We also obtain more precise $T_{0}$ and $R_{p}/R_{\star}$, decreasing their uncertainties by a median factor of \medianfactorTo and \medianfactorrprs respectively. The significant orbital period refinement is to be expected given previous similar studies, in which for example \citet[][]{2018AJ....156..277L} decreased the period uncertainty by a median value of 26 for a subset of targets observed in both C5 and C16, and \citet[][]{2021MNRAS.508..195D} obtained a maximum improvement factor of 80 for a planetary signal modelled with C5, C16 and C18 data jointly. The great ephemeris refinement obtained for the 22 targets facilitates future planning of follow-up observations by the new telescope generation. A representative example is K2-274 b, for which we compute, for the year 2028 (the second year of the scheduled \textit{PLATO} mission), a propagated uncertainty in the mid-transit time of 4 min, while the uncertainty obtained from the currently published parameters is 1 h and 45 min.

\subsection{Detection efficiency}
\label{sec:detection_efficiency}

\begin{figure*}
    \includegraphics[scale=0.44]{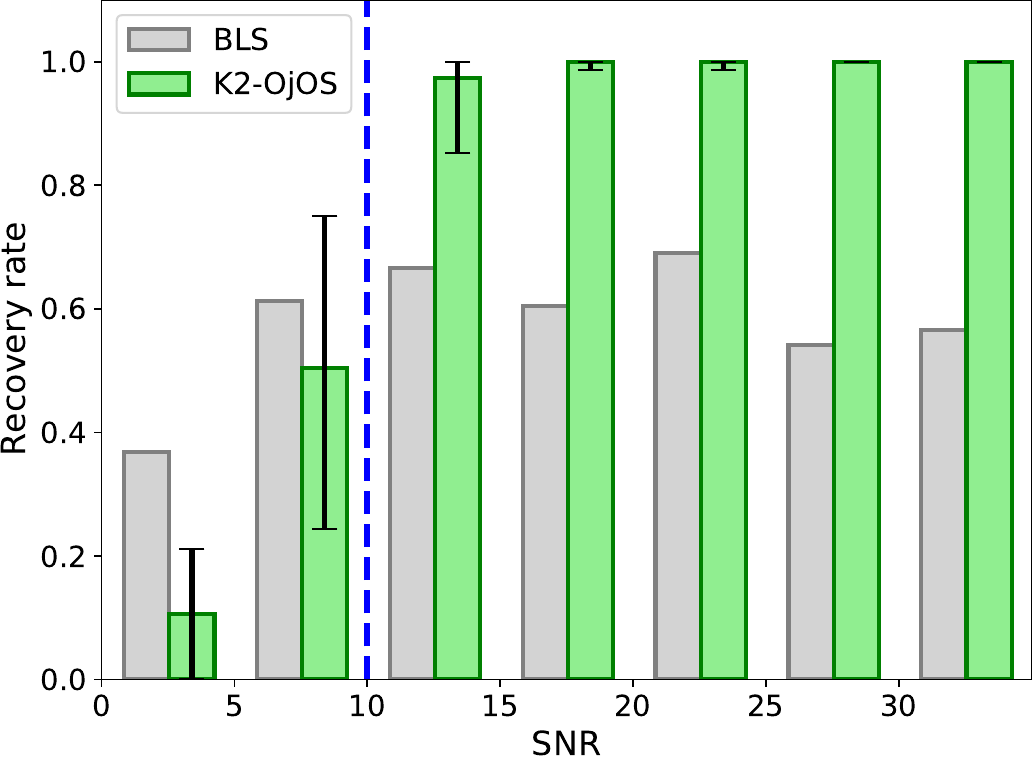}
     \includegraphics[scale=0.44]{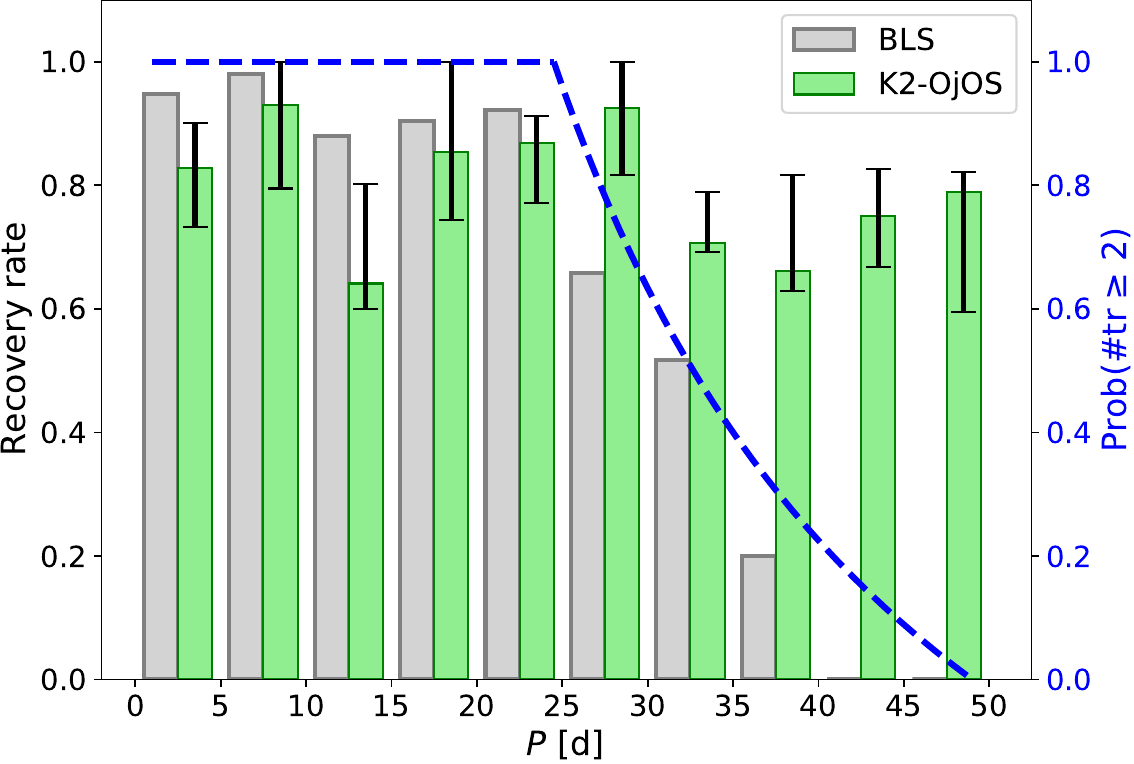}
    \caption[BLS and K2-OjOS recovery rates for different ranges of S/N and orbital period.]{Left: BLS and K2-OjOS recovery rates for different ranges of S/N. The vertical dashed line indicates the minimum required S/N for a signal to be subjected to validation. Right: BLS and K2-OjOS recovery rates for different ranges of $P$. Superimposed, we plot the probability function of having two or more injected transits per signal. The function takes the value 1 for the interval 1 d $<$ $P$ $<$ 24.5 d, and (49-$P$)/$P$ for 24.5 d $<$ $P$ $<$ 49 d.}
    \label{fig:BLS_K2OJOS_recovered}
\end{figure*}

The K2-OjOS members and the BLS algorithm analysed light curves with simulated transit signals in order to quantify the detection efficiencies of both search methods (Sect.~\ref{sec:injection_recovery}). We first computed the recovery rates of each method by dividing the total number of recovered transits by the total number of injected transits. Given the injection and recovery conditions explained in Sect.~\ref{sec:injection_recovery}, we obtain an overall recovery rate of $\sim$78$\%$ for the K2-OjOS members and of $\sim$58$\%$ for the BLS algorithm. We also computed the recovery rate of the K2-OjOS team without being biased by the number of batches analysed by each member in this particular work; that is, we calculated the mean value of the recovery rates of each member. As a result, we obtain $\sim$78$\%$ as well. In the following, we discuss the K2-OjOS and BLS recovery rates broken down by S/N and $P$.

Fig. \ref{fig:BLS_K2OJOS_recovered} (left panel) is a histogram in which we plot the BLS recovery rates as well as the median and 68.3$\%$ credible intervals of the K2-OjOS recovery rates for different ranges of S/N. The K2-OjOS members retrieved 99.5$\%$ of the injected signals with S/N $>$ 10, which is the condition that any signal must meet before being subjected to validation in this work. The 0.5$\%$ not retrieved typically corresponds to short-period signals with many transits at the noise level. As for the BLS, the algorithm retrieved 62$\%$ of the injected signals with S/N $>$ 10. These small recovery rates obtained even for high S/N are related to the inability of BLS to recover single transits, unlike the visual inspection.

In Fig.~\ref{fig:BLS_K2OJOS_recovered} (right panel), we plot the recovery rates as a function of the orbital period for both the BLS and K2-OjOS searches. We also plot the probability function of having two or more injected transits per signal (i.e. of not having a single transit) within the C18 light curve, computed from the injection features explained in Sect.~\ref{sec:injection_recovery}. Similar to previous works, we find that the K2-OjOS detection efficiency remains insensitive to the orbital period. However, for $P>24.5$ d (half of the C18 temporal baseline), the recovery rate of the BLS method drops to zero given the decreasing probability of multi-transit signals being injected. 

From the comparison between the two methods, we can draw two main conclusions. First, the K2-OjOS visual search itself shows great completeness in the search for potentially validatable signals (S/N $>$ 10). Secondly, for signals with S/N $<$ 10, although we obtain higher recovery rates for the BLS algorithm, we highlight the visual inspection as a good complementary method to detect single transits, which are undetectable by the widely used periodicities-based automated transit searchers as BLS.

\subsection{Characteristics of our sample: The host stars}
\label{sec:charact_stellar_sample}

The K2 stars known to host planets have a median magnitude of $K_{\rm p}$  = 12.6 mag, which is two magnitudes brighter than that of the host stars of the primary \textit{Kepler} mission. Thereby, K2 targets can be excellent for precise RV follow-up and atmospheric characterisations, allowing us to unveil planetary masses, densities, and atmospheric properties of the planets found. Our sample has a median magnitude of $K_{\rm p}$ = 13.1 mag, which is slightly fainter than that of K2 host stars. However, we highlight the presence of three bright targets ($K_{\rm p}$ $<$ 11 mag): EPICs 211424769 ($K_{\rm p}$ = 9.4 mag), 211480861 ($K_{\rm p}$ = 10.0 mag), and 211594205 ($K_{\rm p}$ = 10.7 mag). Regarding effective temperatures, most of the planets and candidates in our sample orbit stars that are clustered around 5200 K. If we compare the relative occurrence of stars in our sample with that of K2 hosts, we find a deficit of solar-type and cool-dwarf stars, and a surplus of stars of early K and late G spectral types.

\subsection{Characteristics of our sample: Planets and candidates}
\label{sec:charact_planet_sample}

\subsubsection{Planet radius and orbital period distribution}

\begin{figure*}
    \centering
    \includegraphics[scale=0.297]{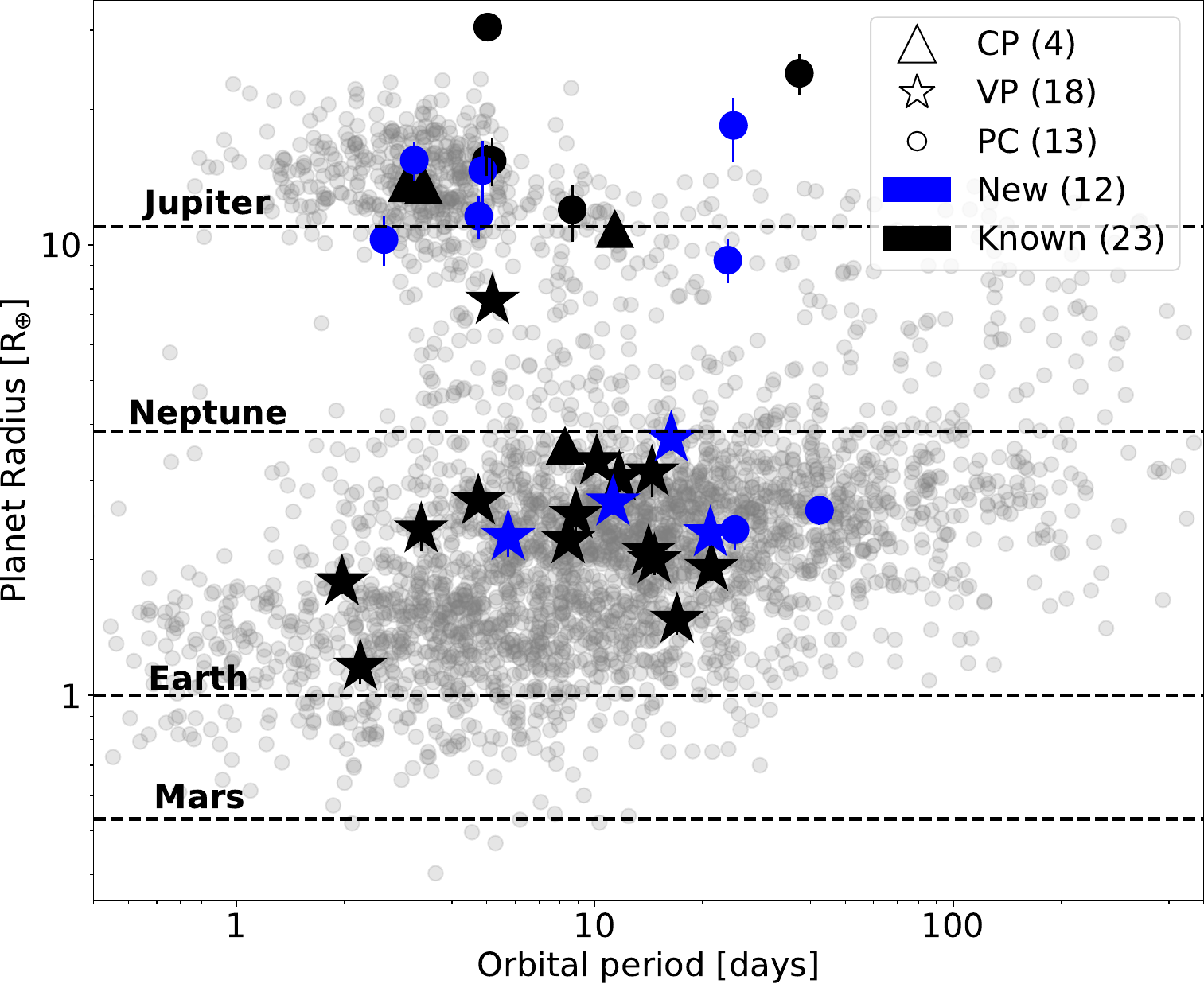}
    \includegraphics[scale=0.24915]{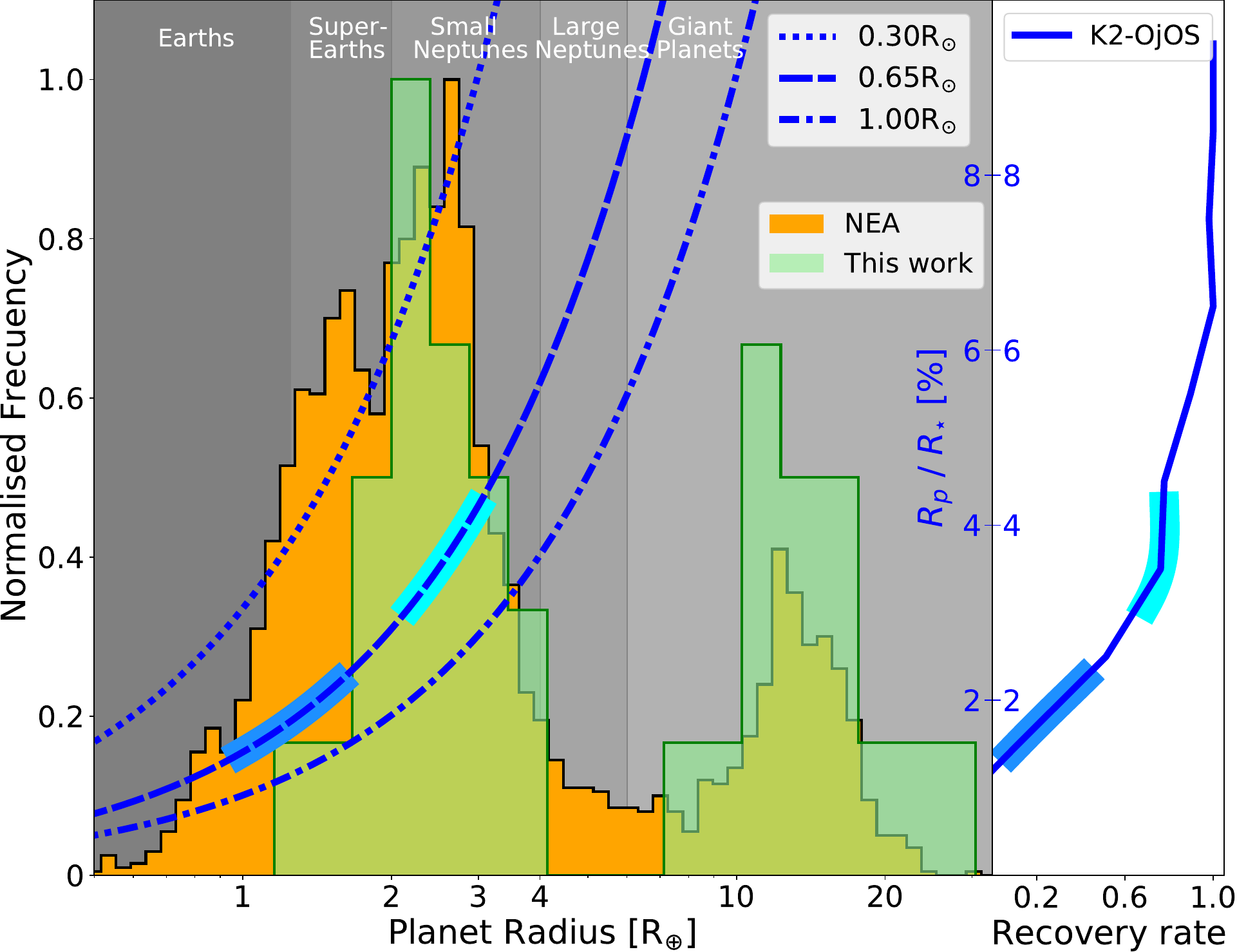}
    \caption[Planet radius as a function of the orbital period for the confirmed planets, validated planets, and planet candidates with measured orbital period and planetary parameters.]{(Left) Planet radius as a function of the orbital period for the confirmed planets (CP), validated planets (VP), and planet candidates (PC) with measured orbital period and planetary parameters. The black markers correspond to previously published detections, while blue markers correspond to the new K2-OjOS detections. The grey data points correspond to the population of known confirmed and validated exoplanets. (Right) Left-hand panel: Histogram of planet radii for both the planets and candidates and the population of known planets. Superimposed we plot $R_{p} / R_{\star}$ (right y-axis) versus $R_{p}$ for three stellar radii. Right-hand panel: K2-OjOS recovery rates as a function of $R_{p} / R_{\star}$. In the left plot we highlight two sections of the  $R_{p} / R_{\star}$ versus $R_{p}$ curve of a 0.65 $\rm R_{\odot}$ star (dodger blue and cyan), each of them being on one side of the radius gap. In the right plot, we can see the K2-OjOS recovery rates corresponding to the highlighted sections. }
    \label{fig:Rp_vs_P}
\end{figure*}

In Fig. \ref{fig:Rp_vs_P} (left), we plot the planet radius as a function of the orbital period for the planets and candidates analysed in this chapter, as well as for the current population of known planets. The latter are mainly grouped into two well differentiated clusters: the small planets cluster, which embraces those planets with $R_{p}$ $<$ 4 $\rm R_{\rm \oplus}$ and orbital periods ranging from less than a day to hundreds of days, and the hot Jupiters cluster, which is composed of large planets ($R_{\rm p}$ $>$ 10 $\rm R_{\rm \oplus}$) with short orbital periods ($P$ $<$ 10~d). Despite the relatively small size of our sample, the K2-OjOS findings alone match quite well with both clusters, especially if we look at the subsample of confirmed and validated planets. In Fig.~\ref{fig:Rp_vs_P} (right, left-hand panel), we plot the distribution of planet radii for both our planet and candidate sample and the population of known planets. The \textit{Kepler} and K2 findings showed that the most common type of planets belong to the small planets cluster \citep[e.g.][]{2012ApJS..201...15H,2013ApJS..204...24B,2013PNAS..11019273P,2015ApJ...809....8B}. Besides, their findings allowed to unveil a bi-modality in the small planets distribution \citep{2017AJ....154..109F}, which shows a lower mode at $\sim$ 1.3 $\rm R_{\oplus}$ and a higher mode at $\sim$ 2.6 $\rm R_{\oplus}$, being both of them separated by the so-called radius gap ($\sim$ 1.9 $\rm R_{\oplus}$). Adopting the same planet size categories as \citet{2013ApJ...766...81F}, our sample of fully characterized planets and candidates contains one Earth (0.8 $\rm R_{\oplus}$ $<$ $R_{\rm p}$ $<$ 1.25 $\rm R_{\oplus}$), four super-Earths (1.25 $\rm R_{\oplus}$ $<$ $R_{\rm p}$ $<$ 2 $\rm R_{\oplus}$), 15 small Neptunes (2 $\rm R_{\oplus}$ $<$ $R_{\rm p}$ $<$ 4 $\rm R_{\oplus}$), and 15 giant planets ($R_{\rm p}$ $>$ 6~$\rm R_{\oplus}$). Focusing on the small planet regime,  when compared to the population of known planets, we find a deficit of planets and candidates with radii smaller than that of the radius gap. We explain this deficit as a consequence of the great difficulty in detecting in K2 data such small planets around stars as large as those in our sample. To illustrate this, in Fig. \ref{fig:Rp_vs_P} (right, left-hand panel) we plot  the $R_{p}$/$R_{\star}$ ratios versus $R_{p}$ for different stellar radii. In the right-hand panel, we plot the obtained K2-OjOS recovery rates as a function of $R_{p}$/$R_{\star}$ (see Sections \ref{sec:injection_recovery} and \ref{sec:detection_efficiency} for further details). We highlight two sections of the $R_{p}$/$R_{\star}$ vs $R_{p}$ curve of a 0.65 $\rm R_{\odot}$ star, which corresponds to the typical stellar radius in our sample. The section located in the lower mode of the small planets distribution corresponds to recovery rates between 8$\%$ and 40$\%$, while the section located in the higher mode corresponds to recovery rates between 70$\%$ and 80$\%$.

\begin{figure*}
    \centering
    \includegraphics[width = 0.65\textwidth]{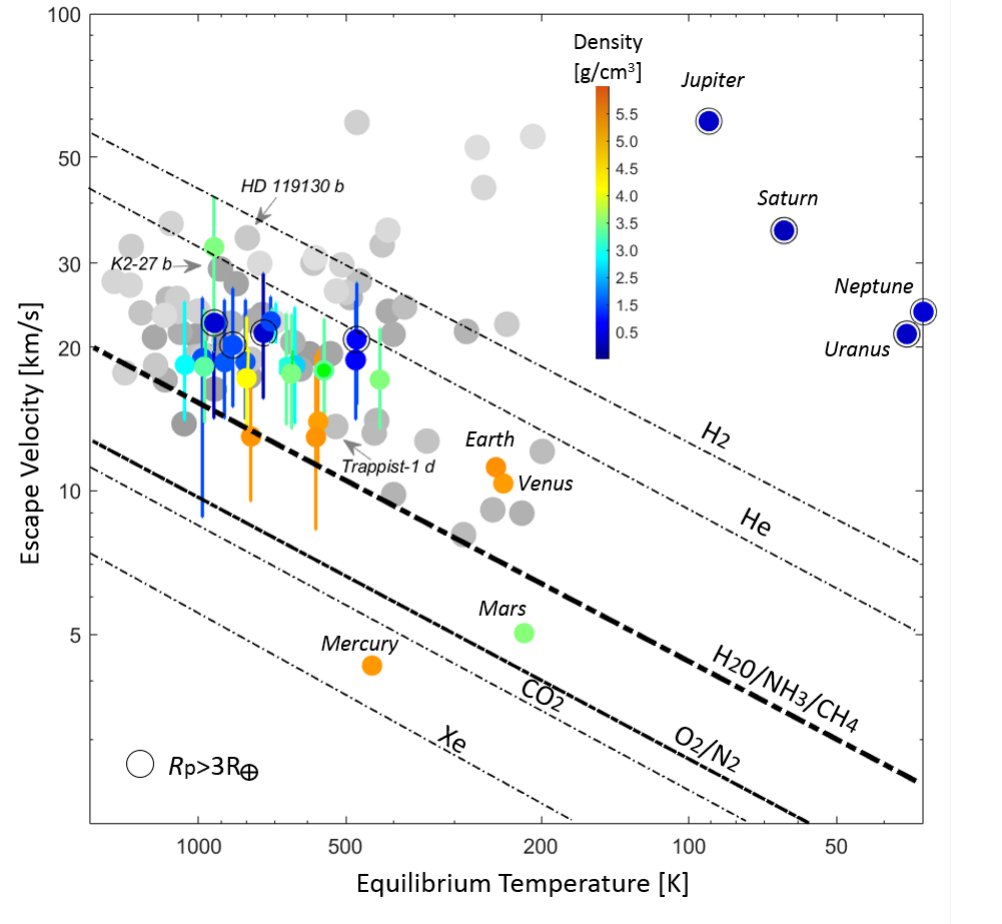}
    \caption[Atmospheric escape velocities as a function of the surface equilibrium temperatures for the small planets and candidates studied in Chapter 2.]{Atmospheric escape velocities as a function of the surface equilibrium temperatures for the small planets and candidates ($R_{\rm p}$ $<$ 4 $\rm R_{\oplus}$) studied in this chapter. 66 NEA confirmed and candidate planets with 0.5 $\rm R_{\oplus}$ $<$ $R_{\rm p}$ $<$ 4.0 $\rm R_{\oplus}$ with uncertainties lower than ±10$\%$ (1-$\sigma$, the average error is about ±7$\%$) and masses determined by the radial velocity method are represented in grey ramp (low bulk density in light grey; high bulk density in dark grey). Dot-dashed straight lines of 0.5 slope stand for threshold velocities of chemicals, labelling each line.}
    \label{fig:esc_velo}
\end{figure*}

\subsubsection{Gas dwarfs versus water worlds: atmospheric escape velocities and retention of volatiles}

The puzzling bi-modality in the population of planets with radii smaller than 4 $\rm R_{\oplus}$ is still a matter of debate \citep[see for example][and references therein]{Zeng9723}. This bimodality is consistent with the existence of two different types of planets. In the range 1-2 $\rm R_{\oplus}$, planets are known to be most likely rocky, whereas the internal compositions of planets between 2 and 4 $\rm R_{\oplus}$ are still an open question. They may either be gas dwarfs or water worlds, being the former planets with a rocky core and a prominent $\rm H_{2}$-He gaseous envelope, and the latter planets with significant amount of multicomponent, $\rm H_{2}O$-dominated ices/fluids in addition to rock and gas \citep[][]{Zeng9723}. Within our dataset, planets below 2 $\rm R_{\oplus}$ are too few to probe the radius gap. However, we can provide some insight into the possible composition of the planets we have detected in the higher mode. We briefly recall two models that explain the observed distribution, leading to different proposals on the composition of planets between 2 and 4 $\rm R_{\oplus}$. 

In the photo-evaporation model, the bi-modality is consistent with the theoretical valley predicted by evaporation numerical analysis \citep[][]{2013ApJ...775..105O,2014ApJ...795...65J,2014ApJ...792....1L,2016ApJ...831..180C}. Along the first 100 Myr of the star's lifetime, high-energy radiation (EUV and X-ray) would have completely stripped the primordial atmosphere of planets that we observe at the lower mode. As a result, gas dwarfs are proposed for planets within the higher mode. On the other side, \citet[][]{Zeng9723} were able to reproduce the two radii subpopulations of small exoplanets through a pebble accretion model independent of the planet growth mechanism. This model involves similar ice and rock contributions to the planet composition, leading to water worlds rather than gas dwarfs for planets in the higher mode. The authors used Monte Carlo simulations to show that the radii's bimodal distribution could arise from the dichotomy of rocky and icy cores.

We now study whether the planets and candidates with radii between 2 and 4 $\rm R_{\oplus}$ analysed in this chapter can keep an atmospheric $\rm H_{2}$-He envelope over a billion-year timescale. Estimations of gas envelopes can be obtained following the strong correlation that atmospheric escape has with the escape velocities of planetary bodies and their atmospheric compositions in the Solar System. The same equations can be applied to estimate which gaseous species a planetary atmosphere can hold \citep[][]{Zeng9723}. 

In Fig. \ref{fig:esc_velo}, we plot the escape velocities ($V_{\rm esc}$) of the planets and candidates with $R$ $<$ 4 $\rm R_{\oplus}$ in our sample, as a function of their equilibrium temperatures ($T_{\rm eq}$). The markers are coloured as a function of the planet densities. Given the difficulty of measuring the masses of our faint host stars via radial velocity measurements, we estimate them from a mass-radius probabilistic algorithm implemented in the widely used program  \textsc{forecaster} \citep[][]{2017ApJ...834...17C}. We use these masses to derive both the escape velocities ($V_{\rm esc}=\sqrt{2GMR^{-1}}$) and planet bulk densities ($\rho = M/V$; $V= 4/3 \pi R^{3}$). The grey dashed dotted lines indicate the thermal escape thresholds of different molecular species. It is interesting to notice that the majority of our planets lie within a specific region with escape velocities of 20 $\pm$ 5 km $\rm s^{-1}$ and equilibrium temperatures between 500 and 1000 K. This particular region of the diagram is characterized by the thermal loss of H$_2$-He gas while the other components are retained. For escape velocities around 20 km $\rm s^{-1}$, one would expect rocky planets to be smaller and denser, while gas dwarfs would be much bigger and with smaller densities than the planets in our sample. Altogether, the escape velocities, density and radii, allow us to tentatively propose, within the uncertainty derived by the use of estimated masses from \textsc{forecaster}, that our subsample of planets and candidates in the higher mode would be composed of water worlds since they would not be able to retain their primordial H$_2$/He envelopes.

\subsubsection{Potentially habitable systems}
\label{sec:habitability}

\begin{figure*}
    \centering
    \includegraphics[width = 0.98\textwidth]{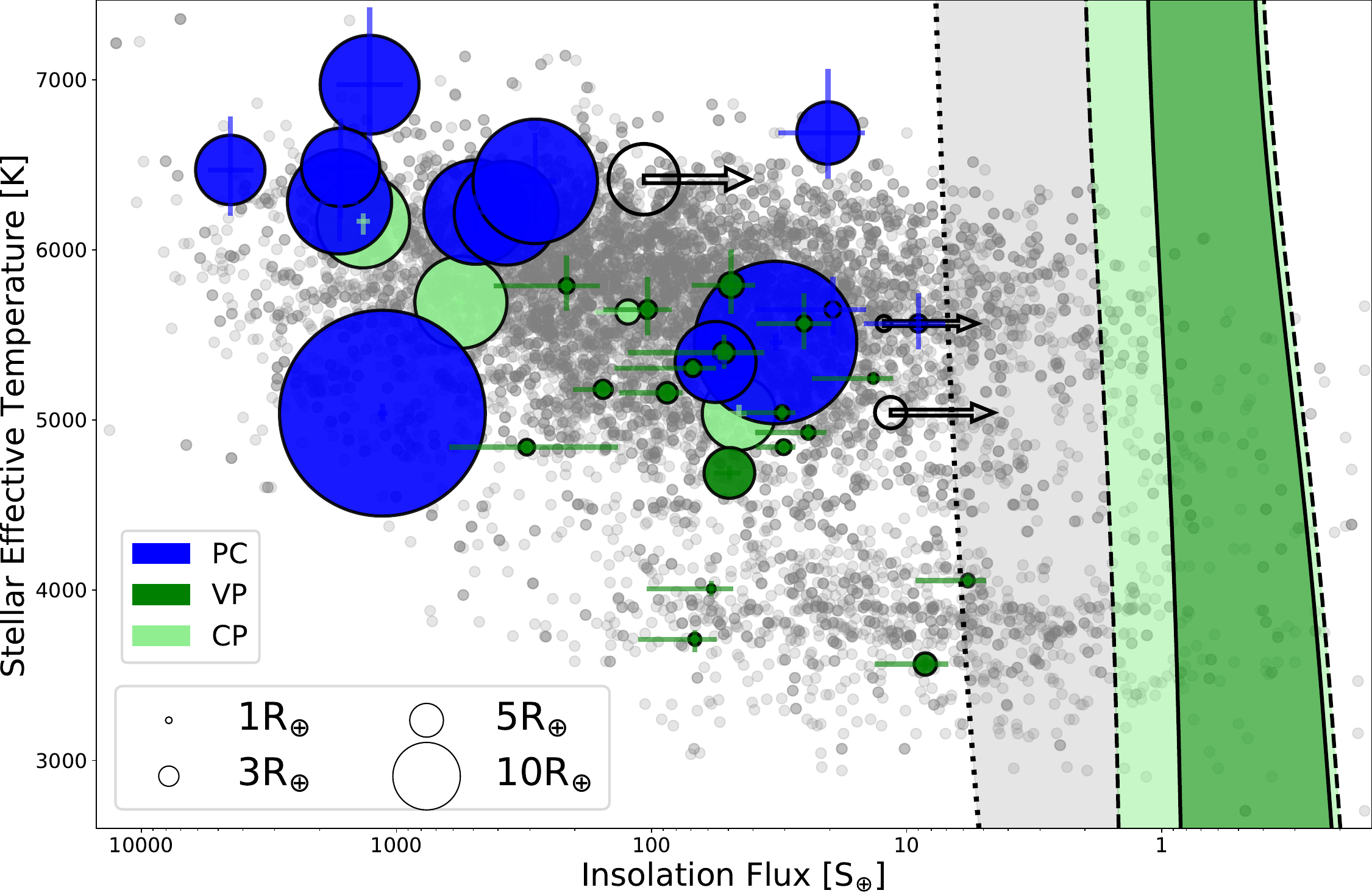}
    \caption[Stellar effective temperatures as a function of the insolation fluxes received by the planets and candidates studied in Chapter 2.]{Stellar effective temperatures as a function of the insolation fluxes received by the planets and candidates. Open circles correspond to the maximum insolation fluxes for single transits, computed from their minimum orbital periods. The grey data points correspond to the population of known planets and candidates. The dark green region is bounded by two solid lines corresponding to the moist greenhouse inner edge and the maximum greenhouse outer edge. The light green region is bounded by two dashed lines corresponding to the recent Venus inner limit and early Mars outer limit. The \citet{2013ApJ...778..109Z} inner edge corresponds to the dotted line. }
    \label{fig:Teff_vs_S}
\end{figure*}

To assess whether any of the low-insolated planets and candidates analysed in this chapter could be inside the Habitable Zone (HZ) we computed the insolation flux boundaries derived from the \citet{2013ApJ...765..131K} climate model, as well as from the more optimistic \citet{2013ApJ...778..109Z} model. For the Solar System, the former define a conservative HZ between 0.99 AU and 1.70 AU, whereas the latter argues that in certain particular conditions ($\rm N_{2}$-dominated atmosphere, surface gravity of $g_{\rm surf}$ = 25 $\rm m\,s^{-2}$, surface pressure of $P_{\rm surf}$ = 1 bar, relative humidity of $\Phi$ = 1$\%$, surface albedo of $A$ = 0.8, and $\rm C0_{2}$ mixing ratio of $X_{\rm C0_{2}}$ = $\rm 10^{-4}$), the HZ inner edge (IHZ) can be as close as 0.38 AU. We used the polynomial relations from \citet{2013ApJ...765..131K} to determine for a wide range of effective temperatures (2600  K $<$ $T_{\rm eff}$ $<$ 7200 $\rm K$) the moist greenhouse inner edge and the maximum greenhouse outer edge, as well as the more optimistic limits of recent Venus inner limit and early Mars outer limit (see Fig. \ref{fig:Teff_vs_S}). The \citet{2013ApJ...778..109Z} model analytical expression is defined within the distance-luminosity parameter space. To transform the model into the $S_{\rm eff}$-$T_{\rm eff}$ parameter space we used the semi-analytical formulas for the Hertzsprung-Russell diagram from \citet{2008SerAJ.177...73Z}.  None of the analysed planets or candidates belong to the \citet{2013ApJ...765..131K} HZ, but there is one validated planet (K2-103 b, EPIC 211822797.01) whose orbit is located slightly further than \citet{2013ApJ...778..109Z} IHZ. We discuss the habitability of this planet in Sect.~\ref{sec:EPIC_211822797}.

\subsection{Highlights of five individual systems}
\label{sec:highlights_individual_systems}

\subsubsection{A 2:1 Period commensurability on the new planetary system K2-356 (EPIC 211537087)}
\label{epic_211537087}

\begin{figure*}
    \centering\includegraphics[width=0.67\textwidth]{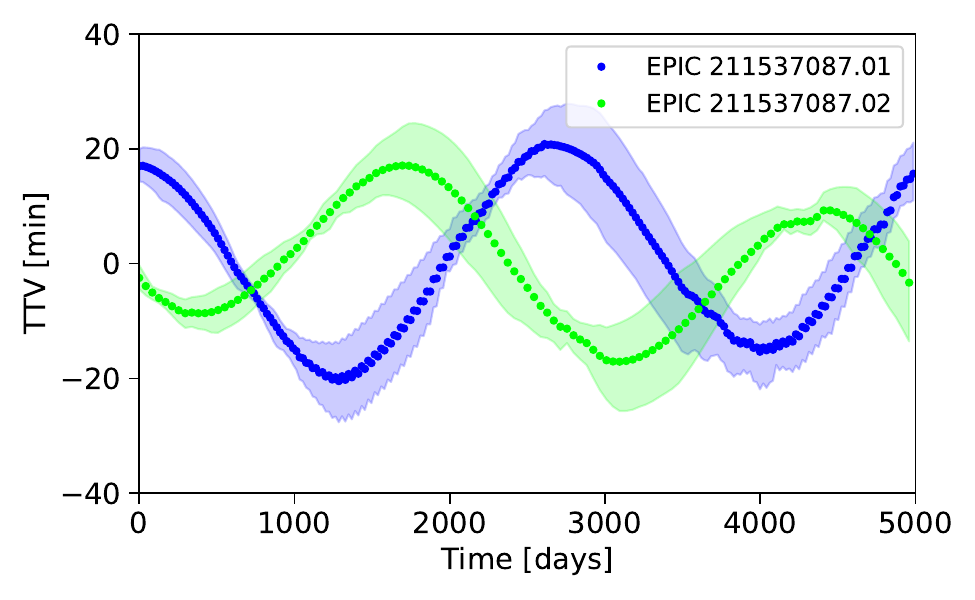}
    \caption[Predicted TTVs for EPIC 211537087.01 (K2-356~b) and EPIC 211537087.0.]{Predicted TTVs ranges with an amplitude of around 20 min for EPIC 211537087.01 (K2-356~b) and EPIC 211537087.02 estimated via forward modelling with \textsc{rebound} assuming circular orbits and planet masses of $M_\mathrm{01}$\,=\,6.1$^{+4.6}_{-2.7}\,\mathrm{M_{\oplus}}$ and $M_\mathrm{02}$\,=\,7.3$^{+5.5}_{-3.2}\,\mathrm{M_{\oplus}}$ derived from the probabilistic mass-radius relation from \textsc{forecaster}. The period of the interaction, the so-called TTV-period or cycle period, is of $\sim$2780\,d, which agrees with the theoretically estimated value of $\sim$2700\,d. The coloured shaded area marks the uncertainties in the expected TTVs based on uncertainties in the planetary masses, orbital periods, and $T_{\rm 0}$.}
    \label{fig:ttv_sim}
\end{figure*}

The K2-OjOS team detected a new planetary system of three planet candidates transiting around K2-356 (EPIC 211537087), a G-type star with $K_{p}$ = 13.44, $T_{\rm eff}$ = $5568$ K, $R$ = $0.86$ $\rm R_{\odot}$, and $M$ = $0.90$ $\rm M_{\odot}$. The three candidates are small Neptunes; EPIC 211537087.01 has $R_{\rm p}$ = 2.29 $\rm R_{\oplus}$ with a 21.03-d orbital period, EPIC 211537087.02 has $R_{\rm p}$ = 2.58 $\rm R_{\oplus}$ with a 42.38-d orbital period, and EPIC 211537087.03 is a single transit with $R_{\rm p}$ = 2.28 $\rm R_{\oplus}$ and orbital period greater than 41.9 d. EPIC 211537087.01 and EPIC 211537087.02 orbit near a 2:1 period commensurability. The analysis of the AstraLux high-resolution image results in a very low probability of the target having a BEB: 0.15$\%$ (see Sect.~\ref{sec:AstraLux} for further details). Although both EPIC 211537087.01 and EPIC 211537087.02 show false positive probabilities lower than the required threshold to validate a planet, we only validate EPIC 211537087.01 (K2-356 b) as EPIC 211537087.02 shows only two transits within the light curve. However, we argue that the origin of EPIC 211537087.02 as well as EPIC 211537087.03 must be planetary, given the extremely low probability of finding multiple false positive signals \citep[][]{2010arXiv1006.3727R,2011ApJS..197....8L}. Besides, for planet candidates which have a period ratio near a first-order mean motion resonance, the probability of both signals being true planets is even higher, since such resonances would not be seen for random eclipsing binaries \citep{2011ApJS..197....8L}.

As TTVs are more sensitive to planets near resonant orbits and the number of transits (three and two for EPIC 211537087.01 and EPIC 211537087.02, respectively) is not sufficient to detect any TTVs, we computed the theoretical TTVs for this system. We estimated a TTV period of $\sim$2700 d using the analytical formula described in \citet{2012ApJ...761..122L}. To have an idea of the expected TTV amplitude, we carried out n-body simulations using \textsc{rebound} \citep{2012A&A...537A.128R}, assuming circular orbits and adopting the values for the orbital periods and mid-transit times from Table \ref{tab:planet_params}. The values for the masses ($M_\mathrm{01}$\,=\,6.1$^{+4.6}_{-2.7}\,\mathrm{M_{\oplus}}$ and $M_\mathrm{02}$\,=\,7.3$^{+5.5}_{-3.2}\,\mathrm{M_{\oplus}}$) were estimated using the mass-radius relation implemented in \textsc{forecaster} \citep{2017ApJ...834...17C}, starting from the planet radii (2.29\,$\pm$\,0.2\,$\mathrm{R_{\oplus}}$ and 2.58\,$\pm$\,0.2\,$\mathrm{R_{\oplus}}$) for EPIC 211537087.01 (K2-356~b) and EPIC 211537087.02.
The simulations predict a TTV period of $\sim$2780 d, confirming the analytical estimations, and TTV amplitudes of $\sim$20 min (Fig.~\ref{fig:ttv_sim}). The influence of the third planet candidate EPIC 211537087.03 (for which the orbital period is unknown) was not considered, which could affect our predictions of both the TTV amplitude and the TTV period.

\subsubsection{Transit timing variations on the bright star (V = 10.35 mag) K2-184 (EPIC 211594205)}
\label{sec:EPIC_211594205}

\begin{figure*}
    \includegraphics[scale=0.49]{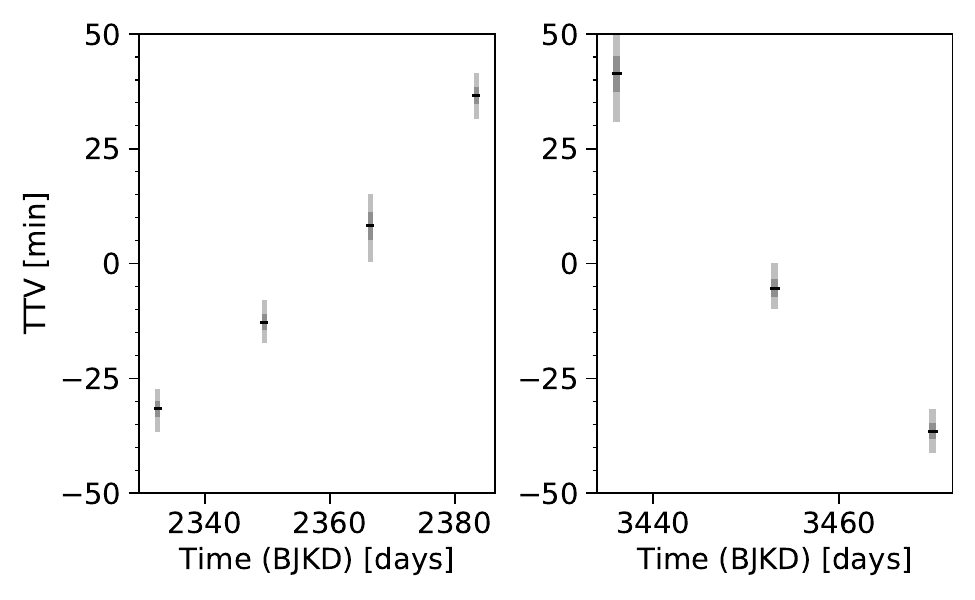}
    \includegraphics[scale=0.47]{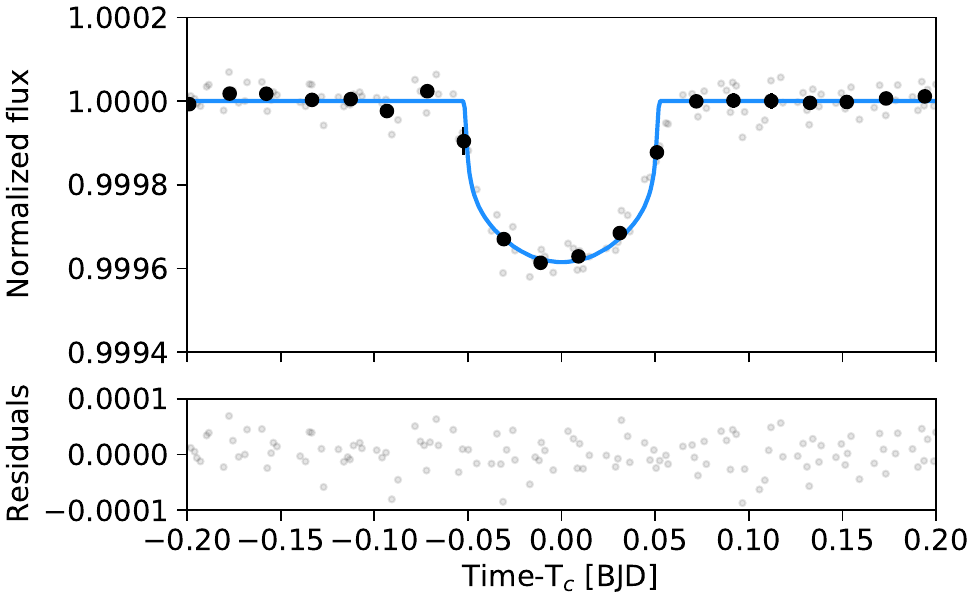}
    
    \caption[Transit timing variations and phase-folded transits of K2-184 b (EPIC 211594205.01).]{Left: Transit timing variations for K2-184~b observed in C5 (left-hand panel) and C18 (right-hand panel). The black lines mark the median values, and the 68$\%$ and 99$\%$ central posterior percentiles are indicated by the dark and light shaded areas, respectively. Right: Phase-folded transits of K2-184 b (EPIC 211594205.01). Each transit is shifted by its mid-transit time estimated by \textsc{pyttv}. The black points are the binned data points with a 30-min binning. The blue line is the transit model. Note that in-transit and out-of-transit data show the same noise properties, indicating that the data are well fitted.}
    \label{fig:ttv}
\end{figure*}

K2-184 b (EPIC 211594205.01)  is a super-Earth with $R_{\rm p}$ = 1.5 $\rm R_{\oplus}$, which orbits a late G-type star ($K_{p}$ = 10.68, $T_{\rm eff}$ = $5245$ K, $R$ = $0.75$ $\rm R_{\odot}$, and $M$ = $0.84$ $\rm M_{\odot}$) with an orbital period of 16.98 d. It was published as a candidate by \citet{2016A&A...594A.100B} and \citet{2016MNRAS.461.3399P}, and latter validated by \citet{2018AJ....155..136M} and \citet{2018AJ....156..277L}. Other works that studied this planet are \citet{2018AJ....155...21P} and \citet{2019ApJS..244...11K}. In this work, the TTV analysis on the joined C5 and C18 data found significant sinusoidal TTVs (Fig.~\ref{fig:ttv}, left) that hint at the existence of an additional non-transiting planet, since no other planet was detected in the light curve that could produce the detected TTVs. Unfortunately, there is no chance to derive characteristics of the TTV signal, e.g. TTV amplitude or TTV period, because of an insufficient coverage of the TTV period due to the lack of transit observations. The phase-folded transit accounting for the TTVs together with the best-fitting model is shown in Fig.~\ref{fig:ttv} (right). In Table \ref{tab:planet_params} we report the median and 68.3$\%$ credible intervals of the main planetary parameters.

The great brightness of the target (V = 10.35 mag) makes it very appropriate for photometric and/or RV follow-up. We predicted the planet mass through \textsc{forecaster} and then estimated the RV semi-amplitude, obtaining $K \sim$ 1.3 m $\rm s^{-1}$, which is achievable by the current precision spectrographs.

\subsubsection{A new single transit on K2-274 (EPIC 212008766)}

K2-274 b (EPIC 212008766.01) is a planet with $R_{p}$ = 2.1 $\rm R_{\oplus}$, which orbits an early K-type star ($K_{p}$ = 12.80, $T_{\rm eff}$ = $5044$ K, $R$ = $0.70$ $\rm R_{\odot}$, and $M$ = $0.79$ $\rm M_{\odot}$) with an orbital period of 14.13 d. It was published as a candidate in 2016 November by \citet{2016A&A...594A.100B} and \citet{2016MNRAS.461.3399P}, and in 2016 December by \citet{2016MNRAS.463.1780L}. In 2018, it was first studied for validation by \citet{2018AJ....155..136M}, who did not validate the planet with a FFP = 0.15$\%$ due to their more conservative considered threshold (FPP $<$ 0.1$\%$). Later, \citet{2018AJ....156..277L} validated the planet with a computed FPP of 0.03$\%$. Interestingly, the authors found that the photometric pipeline they used (k2phot) includes a nearby contaminant star within the aperture, the same one that we found for the EVEREST pipeline, which made us choose the smaller K2SFF aperture for the subsequent analysis. Other works that studied this planet are \citet{2018AJ....155...21P} and \citet{2019ApJS..244...11K}. 

All the aforementioned works detected and analysed the planet starting from C5 data alone. For this work, the K2-OjOS team retrieved the signal in C18, allowing us to model the photometry by joining both C5 and C18 datasets. Besides, the team detected a new single-transit event at 3443.86 BKJD with a 0.15$\%$ dip, hinting at the existence of an additional long-period planet in the system. Our transit modelling corresponds to a planet candidate of 4.3 $\rm R_{\oplus}$. Remarkably, the V-shape of the transit is consistent with a grazing transit. This is to be expected for planets with large semimajor axes, as for a given orbital inclination $b$ $\propto$ $a$. Given the non-presence of such a deep dip within the C5 continuous photometry, we constrain the orbital period for EPIC 212008766.02 to be greater than 74.8 days. If confirmed, it would be the second-longest-period planet detected by K2.

\subsubsection{Habitability of K2-103 b (EPIC 211822797.01)}
\label{sec:EPIC_211822797}

K2-103 b (EPIC 211822797.01)  was first detected as a candidate by \citet{2016A&A...594A.100B} and later validated by \citet{2017AJ....153...64M} and \citet{2017AJ....154..207D} starting from C5 data alone. Other works that analysed the planet are \citet{2017AJ....154..224R} and \citet{2019ApJS..244...11K}. The planet orbits around EPIC 211822797, a late K-type dwarf with $K_{p}$ = 14.57,  $T_{\rm eff}$ = $4057$ K, $R$ = $0.58$ $\rm R_{\odot}$, and $M$ = $0.62$ $\rm M_{\odot}$. Our planet transit modelling of C5, C16, and C18 photometry results in an orbital period of 21.17 d and a radius of $R_{\rm p}$ = 1.92 $\rm R_{\oplus}$, which locates the planet inside the radius gap. The late spectral type of the host star, together with the relatively long orbital period, causes this planet to receive the least amount of insolation flux of the planets and candidates in our sample: 5.75 $\rm S_{\oplus}$. Although this flux is not low enough to consider the planet within the HZ according to \citet{2013ApJ...765..131K} model, if we consider the optimistic conditions proposed by \citet{2013ApJ...778..109Z} (Sect.~\ref{sec:habitability}), this planet would be within the Habitable Zone of its star (see Fig. \ref{fig:Teff_vs_S}). In terms of distances, the \citet{2013ApJ...778..109Z} inner edge is located at 0.116 AU, while the semimajor axis of the planet is $a$ = 0.120 AU. 

\subsubsection{Disposition of K2-120 b (EPIC 211791178.01)}

\begin{figure}
    \centering
    \includegraphics[width=0.5\textwidth]{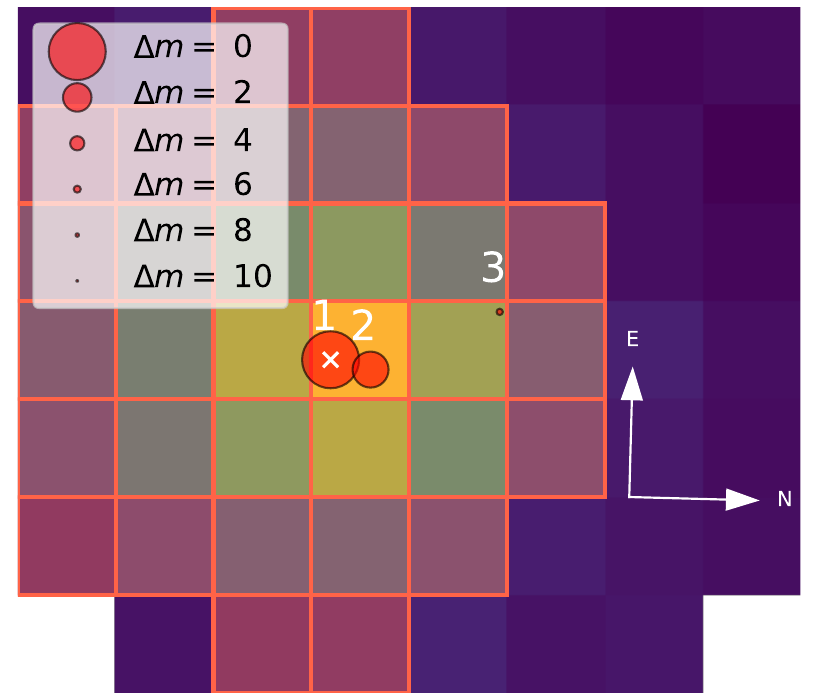}
    \caption[Target pixel file of EPIC 211791178]{TPF of EPIC 211791178 (star $\#$1) together with the EVEREST aperture and nearby \textit{Gaia} DR2 sources.}
    \label{fig:tpfplotter_K2_120}
\end{figure}

K2-120 b (EPIC 211791178.01) was a validated planet retrieved by the K2-OjOS visual searching. As detailed in Sect.~\ref{validation}, we carried out a complete statistical validation analysis for our newly detected candidates, but not for the retrieved known planets and candidates, which were already subjected to validation. This is mainly motivated by the fact that simply adding photometry does not crucially affect the transit shape as for perturbing \textsc{vespa} dispositions. However, we performed some routine checks for all 37 analysed targets (e.g. revisit photometry with different apertures and/or pipelines, search for nearby contaminant stars, and search for secondary eclipses). As a result, we detected for this target a bright contaminant star within the photometric aperture. In Fig. \ref{fig:tpfplotter_K2_120} we show the locations of EPIC 211791178 (source $\#$1, $G_{mag}$ = 13.9) and \textit{Gaia} DR2 659785145072281600 (source $\#$2, $G_{mag}$ = 15.3) within the TPF. These sources are separated 1.67 arcsec from each other, with $\Delta m$ = 1.33, making any type of multi-aperture analysis pointless. Their \textit{Gaia} DR2 measured parallaxes and proper motions are almost identical: $\pi = 3.45 \pm 0.03$ mas, $\mu_{\alpha}$ = $-31.30 \pm 0.06$ $\rm mas \times yr ^{-1}$, $\mu_{\delta}$ = $-20.02 \pm 0.03$ $\rm mas \times yr ^{-1}$ for source $\#$1, and $\pi = 3.28 \pm 0.05$ mas, $\mu_{\alpha}$ = $-31.51 \pm 0.09$ $\rm mas \times yr ^{-1}$, $\mu_{\delta}$ = $-19.19 \pm 0.05$ $\rm mas \times yr ^{-1}$ for source $\#$2, which indicates that both sources form a binary system.

Given the observed transit depth of $\sim$0.077$\%$, if the validated signal comes from source $\#$1, the flux would be diluted by $\gamma_{\rm pri}$ = 1.29, whereas if the signal comes from source $\#$2, the dilution factor would be $\gamma_{\rm sec}$ = 4.40. This yields a dilution-corrected planet radius of 2.80 $\rm R_{\oplus}$ for the first case, and 5.16 $\rm R_{\oplus}$ for the second case, in contrast to the current value of 2.48 $\rm R_{\oplus}$. Therefore, following the criteria in Sect.~\ref{sec:FPP_reliability}, as the origin of the signal cannot be ascertained, we consider EPIC 211791178.01 as a planet candidate. 

\section{Summary and conclusions}
\label{summary}

We have presented the first results of K2-OjOS, a Pro-Am project primarily dedicated to searching, characterising, and validating new extrasolar planets. In this chapter, a group of 10 amateur astronomers visually inspected the 20\,427 light curves of K2 C18 and performed a preliminary vetting of the signals found jointly with professional astronomers, resulting in 42 planet candidates in 37 systems. We characterised homogeneously all the host stars starting from published spectroscopic parameters, photometry, and parallaxes. We modelled the transit signals by joining K2 photometry from C18 and the overlapping C5 and C16 when available, and searched for TTVs in the joined dataset. An exhaustive search revealed that 24 of the findings had been previously published in several works devoted to analysing C5 data alone, while the remaining 18 are new detections. For the former, we refined their ephemeris by joining C5, C16, and C18 data, managing to decrease their uncertainties by a median factor of \medianfactorP\, for $P$, and \medianfactorTo for $T_{0}$. For the latter, we carried out a careful statistical validation analysis that resulted in four new validated planets (K2-355~b, K2-356~b, K2-357~b, K2-358~b) and 14 planet candidates. For the planet sample with 2 $\rm R_\oplus$ $<$ $R_{p}$ $<$ 4 $\rm R_\oplus$, their escape velocities and densities computed from estimated masses, suggest a composition compatible with water worlds.  Regarding individual systems, we highlight the presence of a 2:1 period commensurability in the new system K2-356, the detection of significant TTVs in the bright star K2-184 (V = 10.35 mag), the location of K2-103 b inside the HZ according to optimistic models, the detection of a new single transit in the known system K2-274, and the disposition reassignment of K2-120 b,
which we consider a planet candidate, as the origin of the signal cannot be ascertained. 

Although exoplanetary research is moving from mass detections towards a more comprehensive characterisation and understanding of individual systems, works aimed at detecting large numbers of planets and candidates are of great value. These works greatly increase the statistical information of the population of planets in the Galaxy, and allow follow-up studies to have a greater diversity of systems from which to choose to invest telescope and economic resources. In this context, well-coordinated Pro-Am projects in which amateur and professional astronomers join forces can play an important role.
\newpage
\chapter{TOI-244 b: A low-density super-Earth transiting the bright early-type M-dwarf GJ 1018}
\label{ch:toi-244}
\vspace{2cm}
\pagestyle{fancy}
\fancyhf{}
\lhead[\small{\textbf{\thepage}}]{\small{\textbf{\nouppercase{\leftmark}}}}
\rhead[\small{\textbf{\nouppercase{\rightmark}}}]{\small{\textbf{\thepage}}}

\bigskip

Recent discoveries have revealed the existence of rocky planets inconsistent with Earth-like compositions. For example, planets K2-229 b \citep{2018NatAs...2..393S}, Kepler-107 c \citep{2019NatAs...3..416B}, K2-233 c \citep{2020A&A...640A..48L}, L 168-9 b \citep{2020A&A...636A..58A}, and K2-38 b \citep{2020A&A...641A..92T} have been found to have unusually high densities, similar to that of Mercury. These densities are thought to be caused by the presence of elevated iron content in the planetary cores, which is usually explained through the existence of high iron abundances in the initial protoplanetary disks \citep[e.g.][]{2020MNRAS.493.4910S,2020ApJ...901...97A,2022A&A...662A..19J,2023A&A...670A...6B}, or through external factors such as mantle stripping caused by collisions during planetary formation \citep[e.g.][]{2010ApJ...712L..73M}. On the other hand, planets TOI-561~b \citep{2021MNRAS.501.4148L,2021AJ....161...56W,2023AJ....165...88B}, L 98-59~c and d \citep{2021A&A...653A..41D}, HD 260655 c \citep{2022A&A...664A.199L}, and TOI-4481~b \citep{2023arXiv230106873P} have been found to have lower densities than expected for an Earth-like composition. These densities could be explained by a scarcity or total absence of iron in the planet's structure, by the presence of a significant amount of volatile elements, or by a mixture of both.

The identification of rocky planets inconsistent with Earth-like compositions has only been possible recently thanks to the measurement of very precise planet masses. Today, only 24$\%$ of exoplanets have a true dynamical mass measured, and the percentage is reduced down to 9$\%$ for small planets (i.e. $R_{\rm p}$~$<$~4~$\rm~R_{\oplus}$). Measuring an accurate mass for transiting exoplanets is of crucial importance in order to perform internal structure analysis \citep[e.g.][]{2020A&A...640A..48L,2021NatAs...5..775D} as well as atmospheric characterisation through transmission spectroscopy. \citet{2019ApJ...885L..25B} conducted retrievals on simulated transmission spectra from the James Webb Space Telescope, concluding that a 20$\%$ mass precision or better is required so that inferences of atmospheric properties are not limited by the mass precision of the planet. However, the percentage of small planets meeting this threshold is lower than 5$\%$.

In this chapter, we confirm and characterize the small ($R_{p}$ = 1.5 $\rm R_{\oplus}$) and close-in ($P$ = 7.49 d) transiting planet TOI-244 b orbiting the bright ($K$ = 7.97 mag) and nearby ($d$ = 22 pc) M 2.5V star GJ 1018. Based on the TESS transit signal, we carried out an intensive radial velocity campaign with ESPRESSO in order to confirm its planetary nature, obtain a precise mass measurement, as well as to search for additional planets. In addition to TESS and ESPRESSO data, we used complementary spectroscopic and photometric data sets from HARPS and ASAS-SN to maximise the information for this system. 

In Sect.~\ref{sec:observations},  we describe the TESS, ESPRESSO, HARPS, and ASAS-SN observations. In Sect.~\ref{sec:stellar_charact}, we present our stellar characterisation based on precise photometry and the ESPRESSO spectra. In Sect.~\ref{sec:analysis_results}, we describe our analysis of photometric and spectroscopic data and present the derived planetary parameters. In Sect.~\ref{sec:discussion}, we discuss the results, and we conclude in Sect.~\ref{conclusions}.

\section{Observations}
\label{sec:observations}

\subsection{TESS photometry}
\label{sec:obs_tess}

\begin{figure}
    \centering
    \includegraphics[width = 0.6\textwidth]{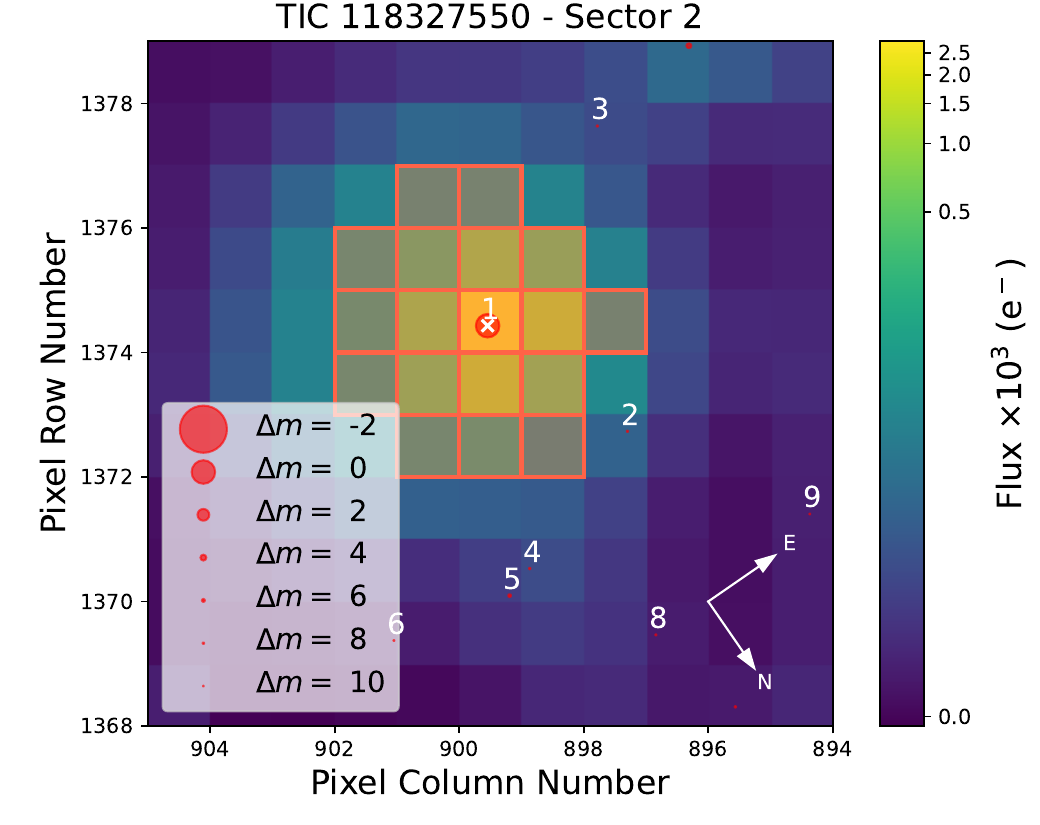}
    \caption[Target pixel file of GJ 1018.]{Target pixel file of GJ 1018. The orange grid is the selected aperture, and the red circles correspond to the nearby \textit{Gaia} sources. Symbol sizes scale to their $G$ magnitudes. This plot has been prepared through \texttt{tpfplotter} \citep{2020A&A...635A.128A}.}
    \label{fig:tpfplotter}
\end{figure}

\begin{figure*}
    \centering
    \includegraphics[width=0.49\textwidth]{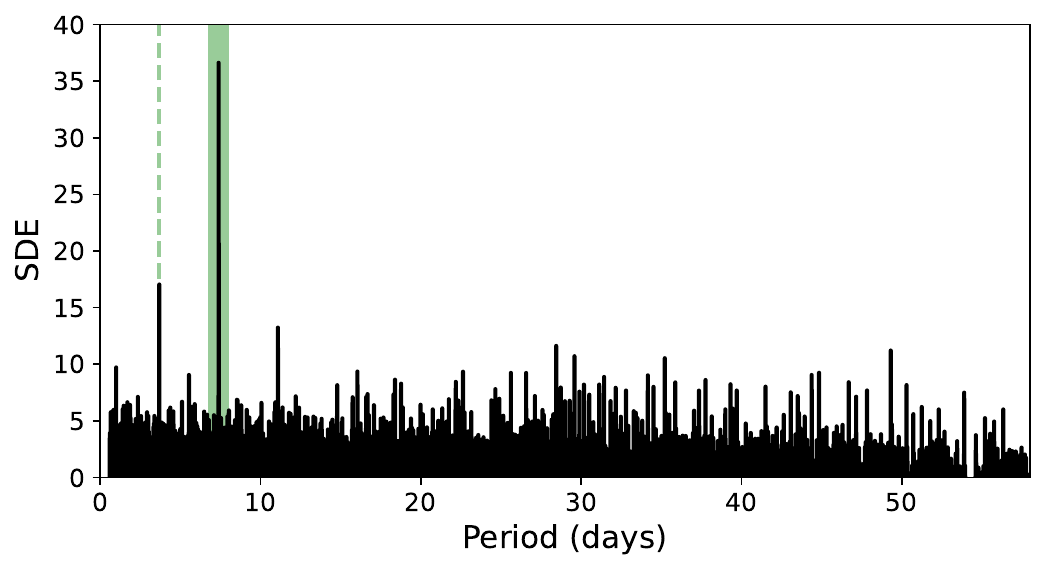}
    \includegraphics[width=0.49\textwidth]{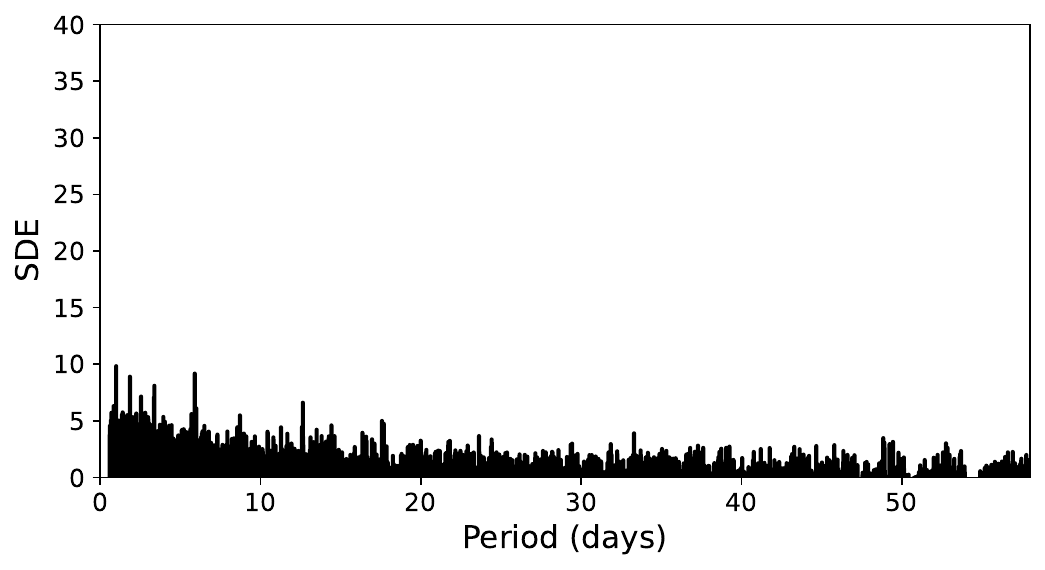}
    \caption[TLS periodogram of the TESS light curve of GJ 1018.]{Left: TLS periodogram of the complete and flattened TESS light curve of GJ 1018, where the highest peak (highlighted with the green broad line) corresponds to the TOI-244.01 transit signal. The green, thin dashed line corresponds to the second harmonic. Right: TLS periodogram after masking the TOI-244.01 signal. } 
    \label{fig:tls_periodograms}
\end{figure*}

The star GJ 1018 (TOI-244, TIC 118327550) was observed by TESS (camera $\#$2, CCD $\#$3) at a 2-min cadence in sector 2 (S2) from 22 August 2018 to 20 September 2018, and in S29 from 26 August 2020 to 22 September 2020, resulting in a total of 36$\,$229 target pixel files (TPFs) spanning a temporal baseline of two years. The central time of each sector shows a short gap ($\sim$1-2 days) due to the satellite repointing toward the Earth to downlink the data, resulting in a total duty cycle of 50 days.

\begin{figure}
    \centering
    \includegraphics[width=\textwidth]{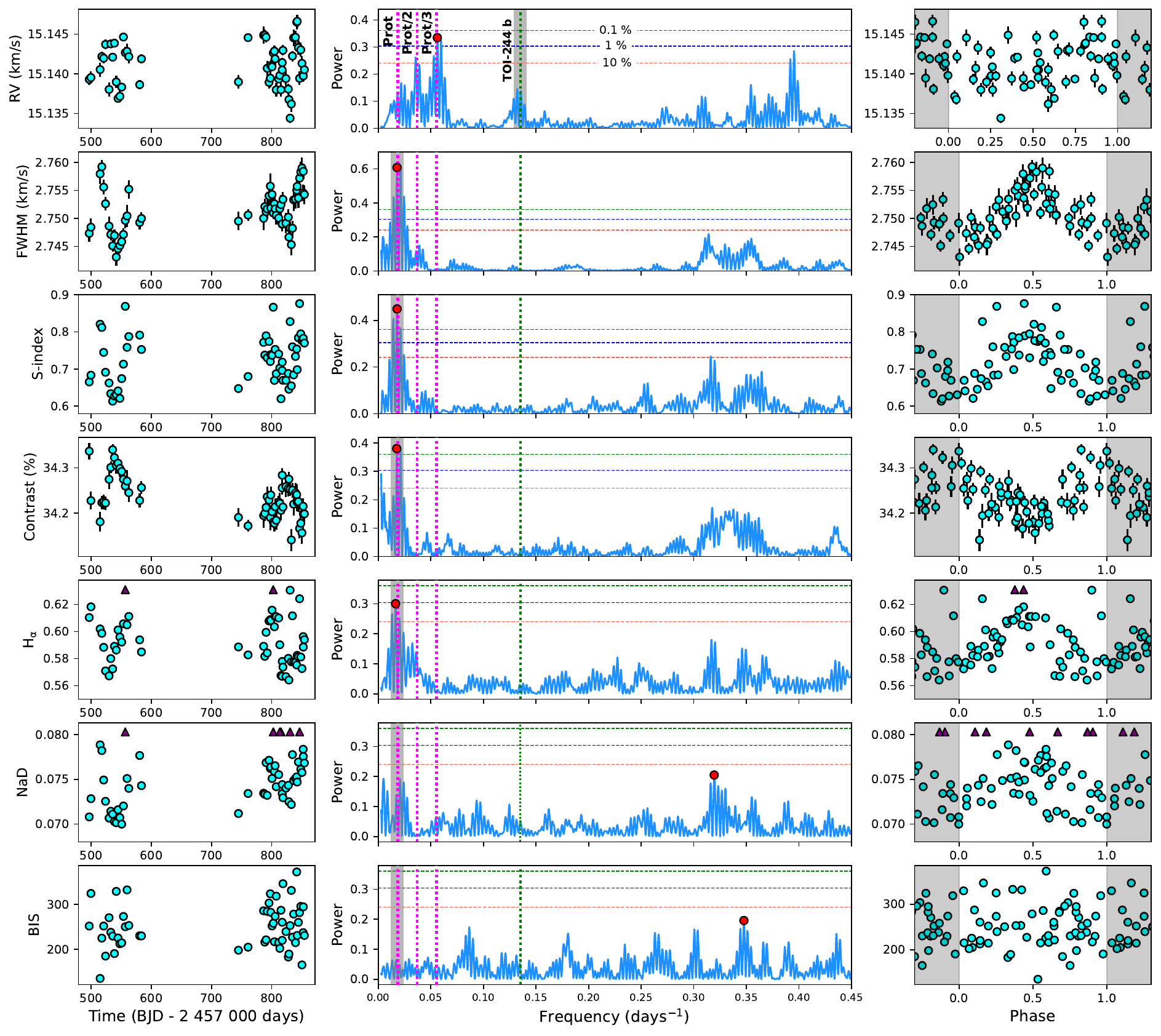}
    \caption[Time series and GLS periodograms of the ESPRESSO RVs and indicators of GJ 1018.]{Left: Time series of the ESPRESSO RVs and activity indicators. Centre: GLS periodograms of the time series. The red circles highlight the maximum power frequencies. The green dotted vertical lines indicate the location of the orbital period of TOI-244~b ($P_{\rm orb}$~=~7.4 days). The magenta dotted vertical lines indicate the rotation period of the star ($P_{\rm rot}$ $\sim$ 56 days) and its second and third harmonics. The grey vertical bands indicate the periods by which the time series in the right panel are folded. Right: ESPRESSO time series folded to the grey bands periods. The triangle markers within the left and right panels indicate the location of data points outside the boundaries of the plot.}
    \label{fig:gls_espresso}
\end{figure}

The observations were processed by the Science Processing Operation Center (SPOC) pipeline \citep{2016SPIE.9913E..3EJ} and are publicly available in the TESS archive of the Mikulski Archive for Space Telescopes (MAST)\footnote{\url{https://mast.stsci.edu/portal/Mashup/Clients/Mast/Portal.html}}. The TESS SPOC data products include simple aperture photometry (SAP) and pre-search data-conditioned simple aperture photometry (PDCSAP), being the latter the SAP processed by the PDC algorithm, which corrects the photometry of instrumental systematics that are common to all stars in the same CCD \citep{2012PASP..124.1000S,2012PASP..124..985S,2014PASP..126..100S}. The photometric aperture was automatically selected by the SPOC module Create Optimal Apertures (COA), which maximises the signal-to-noise ratio of the flux measurement \citep{10.1117/12.857625,2020ksci.rept....7S}. Besides, COA estimates the fraction of flux inside the photometric aperture that comes from the target star and uses it to correct for contamination within the PDCSAP light curve. This fraction (also called CROWDSAP) is 0.9996 for GJ 1018. In Fig. \ref{fig:tpfplotter}, we plot the selected aperture over a TPF of GJ 1018, together with all the nearby stars detected in the \textit{Gaia} Data Release 3 \citep[DR3;][]{2022arXiv220800211G}. There are no additional sources within the aperture, and the nearby sources surrounding the aperture have a magnitude difference $\Delta G$ $>$ 6 mag with GJ 1018 in the \textit{Gaia} passband. These large contrasts ensure negligible contamination \citep[e.g.][]{2018AJ....156..277L,2022MNRAS.509.1075C}.

In October 2018, the SPOC pipeline identified a periodic flux decrease (known as a threshold crossing event; TCE) of 7.4 days through the Transiting Planet Search (TPS) algorithm. The algorithm first characterises the power spectral density of the observation noise, and then estimates the likelihood of the existence of a transit-like signal over a wide range of trial transit durations and orbital periods \citep{2002ApJ...575..493J,10.1117/12.856764,2020ksci.rept....9J}. A transit model fit was performed \citep{Li:DVmodelFit2019}, and a suite of diagnostic tests was carried out to help make or break confidence in the planetary hypothesis \citep{Twicken:DVdiagnostics2018}, all of which the candidate signal passed. Finally, the TESS Science Office examined the light curve and additional information to designate TIC 118327550.01 as a TESS Object of Interest (TOI-244.01) that would benefit from follow-up observations \citep{2021ApJS..254...39G}. 

We downloaded from MAST the PDCSAP light curves and removed all data points with a quality flag different from zero. As a result, we removed one data point in S2 (1359.6474 TJD) due to an argabrightening event (bit 5, value 16). In sector S29, we removed 3005 data points (located between 2098.7624 TJD and 2101.3791 TJD, and between 2112.6041 TJD and 2114.4374 TJD) due to scattered light caused by the Earth or Moon (bit 13, value 4096), and we also removed four data points catalogued as impulsive outliers (bit 10, value 512) which have fluxes deviated 5.2$\sigma$ (2097.5194 TJD), 6.5$\sigma$ (2103.2500 TJD), 7.4$\sigma$ (2103.3000 TJD), and 7.1$\sigma$ (2105.0583 TJD) from the PDCSAP light curve. The final TESS light curves are presented in Table \ref{tab:tess_phot}.

We computed the generalised Lomb-Scargle periodogram \citep[GLS,][]{2009A&A...496..577Z} of the TESS light curve (for S2 and S29 separately and jointly) and found no significant periodicities. We also computed the transit least squares periodogram \citep[TLS,][]{2019A&A...623A..39H} in order to unveil the significance of the TOI-244.01 signal and to determine whether there are additional transit signals. To do so, we first produced a flattened version of the PDCSAP light curve in order to mitigate possible trends caused by the stellar activity or uncorrected systematics. For that matter, we filtered the photometric data of each sector separately using the robust time-windowed bi-weight method implemented within the \texttt{wotan} package \citep{2019AJ....158..143H} with a 0.5-day window length. We detrended the PDCSAP and joined the two flattened light curves to create a long time series. We present the de-trended light curve in Table~\ref{tab:tess_phot}. In Fig.~\ref{fig:tls_periodograms} (left panel), we plot the TLS periodogram of the flattened light curve, which shows a strong peak at 7.4 days with a signal detection efficiency (SDE) of 36.6. In the right panel, we plot the periodogram over the same light curve with the TOI-244.01 transits masked. This periodogram shows no prominent peaks. However, the six highest peaks have SDEs between 6 and 10, which are slightly above some empirical thresholds for transit detection\footnote{Empirical thresholds for transit detection range from SDE $>$ 6 \citep{2015ApJ...807...45D}, SDE $>$ 6.5 \citep{2018AJ....156...78L}, SDE $>$ 7 \citep{2012ApJ...761..123S}, to SDE $>$ 10 \citep{2018MNRAS.473L.131W}.}. To check the reliability of those peaks, we folded the flattened light curve to the corresponding periodicities and inspected a binned version of the folded data. We found no hints of any transit signal, so we conclude those peaks are spurious.

\subsection{ESPRESSO spectroscopy}
\label{sec:espresso_obs}

We observed GJ 1018 with the ESPRESSO high-resolution echelle spectrograph \citep{2021A&A...645A..96P}, which is mounted on the Very Large Telescope (VLT) located at ESO's Paranal Observatory (Chile) and has been operational since 2018. The observations were performed in the course of the ESPRESSO Guaranteed Time Observations (GTOs) under the programs IDs 108.2254.002, 108.2254.005, and 108.2254.006, whose main aim is to determine precise mass measurements of transiting planet candidates. We obtained a total of 57 spectra between 9 October 2021 and 2 October 2022 with a typical cadence of 2-3 days and a typical exposure time of 1200 s, resulting in a mean S/N of 57 at 650 nm. All observations were made in the slow-readout single UT high resolution mode (HR21; 2$\, \times \,$1 binning), which has a spectral resolution power of 140$\,$000 and embraces a wavelength range from 380 nm to 788~nm. For each exposure, we simultaneously illuminated the second fibre with the Fabry-Pérot interferometer, which allows the calibration of the instrumental drift with a precision better than 10 cm$ \rm s^{-1}$ \citep{2010SPIE.7735E..4XW}.

We reduced the data through the ESPRESSO Data Reduction Software (DRS) pipeline version 3.0.0\footnote{The ESPRESSO Data Reduction Software is publicly available at \url{https://www.eso.org/sci/software/pipelines/index.html}.} \citep{2021A&A...645A..96P}. The DRS also extracts the radial velocities (RVs) based on a modified implementation of the original cross-correlation technique presented by \citet{1996A&AS..119..373B}, in which the different spectral lines of a numerical mask are weighed as a function of their RV information \citep{2002A&A...388..632P}. In particular, we used an M3 mask to obtain the cross-correlation function (CCF) of each observation, which was later fitted to a Gaussian profile in order to compute the RVs (centre of the Gaussian). Finally, all the RVs were corrected for secular acceleration. The standard deviation of the RV observations is 2.7 m/s, and the median uncertainty per data point is 0.7 m/s. The DRS also computes activity indicators such as the full width at half maximum (FWHM) and the amplitude or contrast of the CCF, the bisector span (BIS), the $\rm H_{\alpha}$ and Sodium doublet (NaD) line depths, and the S-index.  We present the complete ESPRESSO data set in Table \ref{tab:espresso_rvs}.

In Fig.~\ref{fig:gls_espresso}, we show the GLS periodogram of the ESPRESSO RVs and activity indicators time series. The FWHM, S-index, Contrast of the CCF, and $\rm H_{\alpha}$ show maximum power periods of 56.0, 55.9, 56.8, and 60.6 days, with false alarm probabilities (FAPs)\footnote{All the FAPs in this chapter have been computed analytically following Eq. 24 of \citet{2009A&A...496..577Z}.}  of  $5 \times 10^{-6}~\%$, $1.8 \times 10^{-3}~\%$, $1.2 \times 10^{-1}~\%$, and 1.1$~\%$ respectively, thus unveiling the presence of a significant activity-related signal that most likely corresponds to the rotation period of the star. The remaining two indicators BIS and NaD show no significant peaks within the periodogram. Their maximum power periods correspond to 2.9 days (39.0$\%$ FAP), and 3.1 days (30.4$\%$ FAP) respectively, being most likely related to the mean observing cadence of $\sim$2-3 days. It is remarkable that, when folded to the $\sim$56-day periodicity found in most indicators, the NaD time series shows moderate signs of a sinusoidal behaviour, suggesting that the activity signal might be manifested in this indicator as well. The RV periodogram shows a peak coinciding with the orbital period of TOI-244.01 that cannot be seen in any of the indicators, which suggests a planetary origin for the signal. However, the signal is not significant. The maximum power period of the RV time series is 17.8 days ($2.9 \times 10^{-1} \% $ FAP), which is compatible with the third harmonic of the $\sim$56-day signal present in most activity indicators. Also, in this region of the periodogram, there is another peak at 28.3 days with a 9.5$\%$ FAP that is compatible with the second harmonic of the $\sim$56-day signal. The presence of these two prominent peaks indicates that the RV time series is significantly affected by the stellar activity. Finally, there is another relevant peak at 2.5 days (2.2$\%$ FAP). This periodicity is compatible with the mean observing cadence, but we also explore the possibility of it being a planetary signal in Sect.~\ref{subsec:rv_modeling}.

\subsection{HARPS spectroscopy}
\label{sec:harps_obs}
GJ 1018 was observed by the High Accuracy Radial velocity Planet Searcher (HARPS), which is mounted on the 3.6~m telescope located at ESO's La Silla Observatory (Chile) and has been operational since 2003. HARPS is a fibre-fed cross-dispersed echelle spectrograph located in a vacuum vessel that protects the instrument from temperature and refractive index variations. It has a spectral resolution power of 115$\,$000 and covers a wavelength range between 378 and 691 nm. A total of 15 spectra were acquired between 15 December 2018 and 8 January 2019 under the program 1102.C-0339 (PI X. Bonfils). We downloaded the reduced spectra (DRS version 3.8), which are publicly available in the ESO's Science Portal\footnote{\url{http://archive.eso.org/scienceportal/home}}. Eleven spectra were acquired with 1800 s exposure time, resulting in a median S/N of 14.6 per resolution element at 650 nm, and the remaining four spectra were acquired with 1500 s exposure time, resulting in a median S/N of 11.9.   We present the complete HARPS data set in Table~\ref{tab:harps_rvs}. The standard deviation of the RV observations is 4.3 m/s, and the median uncertainty per data point is 2.1 m/s. The S/N values obtained with relatively long exposures are remarkably low, leading to a photon-noise limited RV precision (i.e. three times worse than ESPRESSO).

\subsection{ASAS-SN photometry}
\label{sec:asas_sn}

\begin{table*}[]
\fontsize{8.6pt}{8.6pt}\selectfont
\caption{Summary of the ASAS-SN observations of GJ 1018.}
\renewcommand{\arraystretch}{1.6}
\begin{tabular}{lllllll}
\hline \hline
Start date & End date & Camera & Band & Station & Location & $\#$Obs \\ \hline
11 May 2014 & 15 Sep 2018 & bf & V & Cassius & Cerro Tololo International Observatory (Chile) & 263 \\
18 Sep 2017 & Operational & bj & g & Henrietta Leavitt & McDonald Observatory (USA) & 385 \\
10 Nov 2017 & Operational & bn & g & Cecilia Payne & South African Astrophysical Observatory & 196 \\
7 Oct 2018 & Operational & bF & g & Cassius & Cerro Tololo International Observatory (Chile) & 266 \\ \hline
\end{tabular}
\label{tab:asas_sn}
\end{table*}

The sky region encompassing GJ 1018 is being observed by three different stations of the All-Sky Survey for Supernovae \citep[ASAS-SN,][]{2014ApJ...788...48S}. In Table \ref{tab:asas_sn}, we include the observing periods, the number of observations, the observing bands, and the location of the different stations. Each station consists of four Nikon telephoto lenses of 14 cm aperture equipped with a 2048 $\times$ 2048 pixels CCD camera. The pixel scale is 8.0 arc seconds, which corresponds to a 4.5 $\rm deg^{2}$ field of view per camera. Every night, the cameras take three consecutive exposures of 90 s exposure time each, which are combined afterwards to increase the signal-to-noise of the light curves by a factor of $\sqrt{3}$. Images obtained in poor weather conditions, that are out of focus (FWHM $>$ 2.5 pixels), or where the studied source is near the detector edge (at a distance $<$ 0.2 deg) are discarded by the survey pipeline. The pipeline performs aperture photometry at a selected location through \texttt{IRAF} \citep{1986SPIE..627..733T}, considering a 2 pixel radius aperture and 7-10 pixel radius annulus for the target star and for the reference stars, which are selected from the Photometric All-Sky Survey \citep[APASS,][]{2012JAVSO..40..430H,2019JAVSO..47..130H} following \citet{2017PASP..129j4502K}.

We computed the light curves through the ASAS-SN Sky Patrol web interface\footnote{\url{https://asas-sn.osu.edu/}}. Given that GJ 1018 is a high proper motion star ($\mu_{\alpha}$ = $-154.88 \pm 0.02$ $\rm mas/yr$, $\mu_{\delta}$ = $45.63 \pm 0.03$ $\rm mas/ yr$) and the typical FWHMs are comparable to the radius of the aperture ($\sim$ 2 pixels), we shifted the aperture location over time in order to ensure the target centring and thus avoid flux loses. GJ 1018 is visible from the three observatories from mid-May to mid-February of the following year, so we computed the photometry on a year-by-year basis, introducing the coordinates corrected for proper motion corresponding to the central time of each observing window (1 October). After the light curves computation, we discarded those epochs in which the flux is below the estimated 5-$\sigma$ detection limit for the target location, as well as those data points with a deviation greater than 5-$\sigma$ of a flattened version of the photometric time series. In Table \ref{tab:asassn_phot}, we present the complete photometric data set.

In Fig.~\ref{fig:asassn}, we show the ASAS-SN photometric time series of GJ 1018. The complete data set shows a long-term trend that could be caused by instrumental systematics, or more likely by the magnetic cycle of the star, given the similarity of the trends in each individual camera. We also show the GLS periodograms of the photometry corrected for the above-mentioned trends; that is, we divided the time series of each camera by a degree three polynomial that was previously fit to the data. For cameras bf, bj, and bF, we obtain maximum power periods of 58.3, 60.9, and 60.0 days, with FAPs of 0.011$\%$, 0.0019$\%$, and 2.79$\%$ respectively. These periodicities are consistent with the ones observed in most of the ESPRESSO activity indicators (Sect.~\ref{sec:espresso_obs}). Hence, they most likely reflect the rotation period of GJ 1018 (see Sect.~\ref{subsec:joint_analysis} for an accurate determination). The maximum power period for camera bn is 30.7 days (FAP = 0.70$\%$), which coincides with the second harmonic of the aforementioned rotation period.

\section{Stellar characterization}
\label{sec:stellar_charact}

\subsection{General description of GJ 1018}
\label{sec:general_description}

\begin{table}[]
\fontsize{11pt}{11pt}\selectfont
\centering
\caption[Stellar properties of GJ 1018.]{Stellar properties of GJ 1018. Results based on (1) \texttt{SteParSyn} (2) \texttt{ODUSSEAS} analysis. \textbf{References.} [1] \citet{2021ApJS..254...39G}; [2] \citet{2019AJ....158..138S}; [3] \citet{1991adc..rept.....G}; [4] \citet{1979nlcs.book.....L}; [5] \citet{skrutskie2006}; [6] \citet{2022arXiv220800211G}; [7] This work; [8] \citet{2019AJ....158..173A};  [9] \citet{2019JAVSO..47..130H}.}
\label{tab:stellar_params}
\renewcommand{\arraystretch}{1.22}
\setlength{\tabcolsep}{1.5pt}
\begin{tabular}{llc}
\hline \hline
Parameter & Value & Reference \\ \hline
\multicolumn{3}{l}{Identifiers} \\ \hline
TOI, TIC & 244, 118327550 & [1], [2] \\
GJ & 1018 & [3] \\
LP & 937-95 & [4] \\
2MASS & J00421695-3643053 & [5] \\
Gaia DR3 & 5001098681543159040 & [6] \\ \hline
\multicolumn{3}{l}{Coordintates, parallax and kinematics} \\ \hline
RA, DEC & 00:42:16.74, -36:43:04.71 & [6] \\
$\rm \mu_{\alpha}$ (mas/yr) & -154.880 $\pm$ 0.020 & [6] \\
$\rm \mu_{\delta}$ (mas/yr) & 45.632 $\pm$ 0.029 & [6] \\
Parallax (mas) & 45.300 $\pm$ 0.027 & [6] \\
Distance (pc) & 22.075 $\pm$ 0.013 & [6] \\
RV (km/s) & 15.1410 $\pm$ 0.013 & [7], Sect. \ref{subsec:joint_analysis} \\
U (km/s) & -23.30 $\pm$ 0.02 & [7], Sect. \ref{sec:abundances} \\
V (km/s) & 17.09 $\pm$ 0.02 & [7], Sect. \ref{sec:abundances} \\
W (km/s) & -8.62 $\pm$ 0.03 & [7], Sect. \ref{sec:abundances} \\
Gal. population & Thin disk & [7], Sect. \ref{sec:abundances} \\ \hline
\multicolumn{3}{l}{Atmospheric parameters and spectral type} \\ \hline
$T_{\rm eff}$ (K) & 3433 $\pm$ 100 & [7], Sect. \ref{sec:steparsyn} \\
log $g$ (dex) & 4.66 $\pm$ 0.07 & [7], Sect. \ref{sec:steparsyn} \\
$\rm [Fe/H]$ (dex) (1) & -0.39 $\pm$ 0.07 & [7], Sect. \ref{sec:steparsyn} \\
$\rm [Fe/H]$ (dex) (2) & -0.03 $\pm$ 0.11 & [7], Sect. \ref{sec:steparsyn} \\
SpT & M2.5 V & [7], Sect. \ref{sec:steparsyn} \\ \hline
Physicall parameters &  &  \\ \hline
$\rm R_{\star}$ ($\rm R_{\odot}$) & 0.428 $\pm$ 0.025 & [7], Sect. \ref{sec:stellar_radius_mass} \\
$\rm M_{\star}$  ($\rm M_{\odot}$) & 0.427 $\pm$ 0.029 & [7], Sect. \ref{sec:stellar_radius_mass} \\
L ($\rm L_{\odot} \times 10^{-2}$) & $2.277\pm 0.042$ & [7], Sect. \ref{sec:luminosity} \\
$\rm P_{rot}$ (days) & $53.3^{+1.2}_{-1.1}$ & [7], Sect. \ref{subsec:joint_analysis} \\
Age (Gyr) & 7 $\pm$ 4 & [8] \\ \hline
Chemical abundances &  &  \\ \hline
$\rm [Mg/H]$ (dex) (1) & -0.28 $\pm$ 0.07 & [7], Sect. \ref{sec:abundances} \\
$\rm [Si/H]$ (dex) (1) & -0.33 $\pm$ 0.07 & [7], Sect. \ref{sec:abundances} \\
$\rm [Mg/H]$ (dex) (2) & 0.00 $\pm$ 0.10 & [7], Sect. \ref{sec:abundances} \\
$\rm [Si/H]$ (dex) (2) & -0.01 $\pm$ 0.10 & [7], Sect. \ref{sec:abundances} \\ \hline
Magnitudes &  &  \\ \hline
$\rm M_{bol}$ (mag) & 8.847 $\pm$ 0.020 & [7], Sect. \ref{sec:luminosity} \\
TESS (mag) & 10.3475 $\pm$ 0.0073 & [2] \\
G (mag) & 11.553 $\pm$ 0.001 & [6] \\
B (mag) & 14.199 $\pm$ 0.033 & [9] \\
V (mag) & 12.684 $\pm$ 0.047 & [9] \\
g' (mag) & 13.426 $\pm$ 0.063 & [9] \\
r' (mag) & 12.097 $\pm$ 0.045 & [9] \\
i' (mag) & 10.831 $\pm$ 0.062 & [9] \\
J (mag) & 8.827 $\pm$ 0.023 & [5] \\
H (mag) & 8.251 $\pm$ 0.033 & [5] \\
$\rm K_{s}$ (mag) & 7.970 $\pm$ 0.029 & [5] \\ 
\hline
\end{tabular}

\end{table}





GJ 1018 is a bright ($K$ = 7.97 mag) early-type M-dwarf star located in the solar neighbourhood. The \textit{Gaia} DR3 provides a  parallax of $\pi$ =  45.300 $\pm$ 0.027 mas, which corresponds to a distance of $d$ = 22.075 $\pm$ 0.013 pc. The TESS Input Catalogue \citep[TIC v8.0.1,][]{2019AJ....158..138S} estimates an effective temperature of $T_{\rm eff}$ = 3407 $\pm$ 157 K, surface gravity of log $g$ = 4.820 $\pm$ 0.004 dex, stellar radius of $R$ = 0.41 $\pm$ 0.01 $\rm R_{\odot}$, and stellar mass of $M$ = 0.40 $\pm$ 0.02 $\rm M_{\odot}$. In the following, we describe our stellar characterisation based on precise photometry and a high S/N spectrum obtained from the combination of all ESPRESSO spectra. In Table \ref{tab:stellar_params}, we summarise the general properties of GJ~1018 and our derived parameters.

\subsection{Stellar atmospheric parameters}
\label{sec:steparsyn}

We computed the stellar atmospheric parameters of GJ 1018 by means of the {\sc SteParSyn} code\footnote{\url{https://github.com/hmtabernero/SteParSyn/}} \citep{tab22}. The code implements the spectral synthesis method with an MCMC sampler to retrieve the stellar atmospheric parameters. We employed a grid of synthetic spectra computed with the {\sc Turbospectrum} code \citep{ple12} alongside BT-Settl stellar atmospheric models \citep{all12} and atomic and molecular data of the Vienna atomic line database \citep[VALD3;][]{rya15}. We considered a selection of $\rm Fe_{I}$, $\rm Ti_{I}$ lines, and TiO molecular bands that are well-suited to analyse M-dwarf stars \citep{mar21}. In all, {\sc SteParSyn} has allowed us to compute the following stellar parameters: $T_{\rm eff}$~$=$~3433~$\pm$~10~K, $\log{g}$~$=$~4.66~$\pm$~0.07 dex, and [Fe/H]~$=$~$-0.39$~$\pm$~0.07~dex. In order to account for systematics associated with $T_{\rm eff}$ when it is determined by spectral fitting techniques (e.g. \citealt{marfil21} found a dispersion between 40 and 100 K among different model atmospheres valid for M-dwarfs), we increased the $T_{\rm eff}$ uncertainty up to 100 K.

We performed an independent computation of the atmospheric parameters through the newly machine learning tool \texttt{ODUSSEAS}\footnote{\url{https://github.com/AlexandrosAntoniadis/ODUSSEAS/}} \citep{Antoniadis-Karnavas2020}, which has been designed to compute the $T_{\rm eff}$ and [Fe/H] of M-dwarfs. The method is based on measuring the pseudo Equivalent Widths (pEWs) of specific absorption and blended lines in the wavelength range 5300\,{\AA}-6900\,{\AA} \citep{neves14}. Briefly, \texttt{ODUSSEAS} compares the measured EWs to the machine learning models generated from reference HARPS spectra, convolved to the resolution of the observed spectrum. Although \texttt{ODUSSEAS} was initially developed for spectra with resolution powers from 48\,000 to 115\,000, it has been applied successfully to ESPRESSO spectra with a resolution power of 140\,000. For the very high resolutions of ESPRESSO and HARPS, the results derived by \texttt{ODUSSEAS} are essentially the same, either by using directly the original highest-resolution spectra of ESPRESSO, or by convolving first the ESPRESSO spectra to the lower, but still high, resolution of HARPS. We ran \texttt{ODUSSEAS} by considering a reference data set composed of 47 stars with interferometry-based $T_{\mathrm{eff}}$ \citep{khata21, rabus19}, and [Fe/H] derived through the photometric calibration by \citet{neves12} using the \textit{Gaia} DR3 parallaxes. We obtain $T_{\mathrm{eff}}$~=~3419~$\pm$~92~K and [Fe/H]~=~-0.03~$\pm$~0.11 dex.

The \texttt{ODUSSEAS} effective temperature is in very good agreement with the value computed by {\sc SteParSyn}. However, there exists a strong discrepancy between the metallicities. Obtaining accurate metallicities for M-dwarfs is a complicated task, given the strong blending of spectral lines and molecular bands. For example, \citet{2022A&A...658A.194P} found mean deviations of around 0.1-0.3 dex between different techniques, showing that any uncertainty below those values is underestimated. Overall, we adopt the $T_{\rm eff}$ and $\log{g}$ estimations from the spectral synthesis method. However, given the strong discrepancy in [Fe/H], we decide not to adopt either estimate, and instead discuss both results independently.

\subsection{Stellar bolometric luminosity}
\label{sec:luminosity}

\begin{figure}
    \centering
    \includegraphics[width=0.6\textwidth]{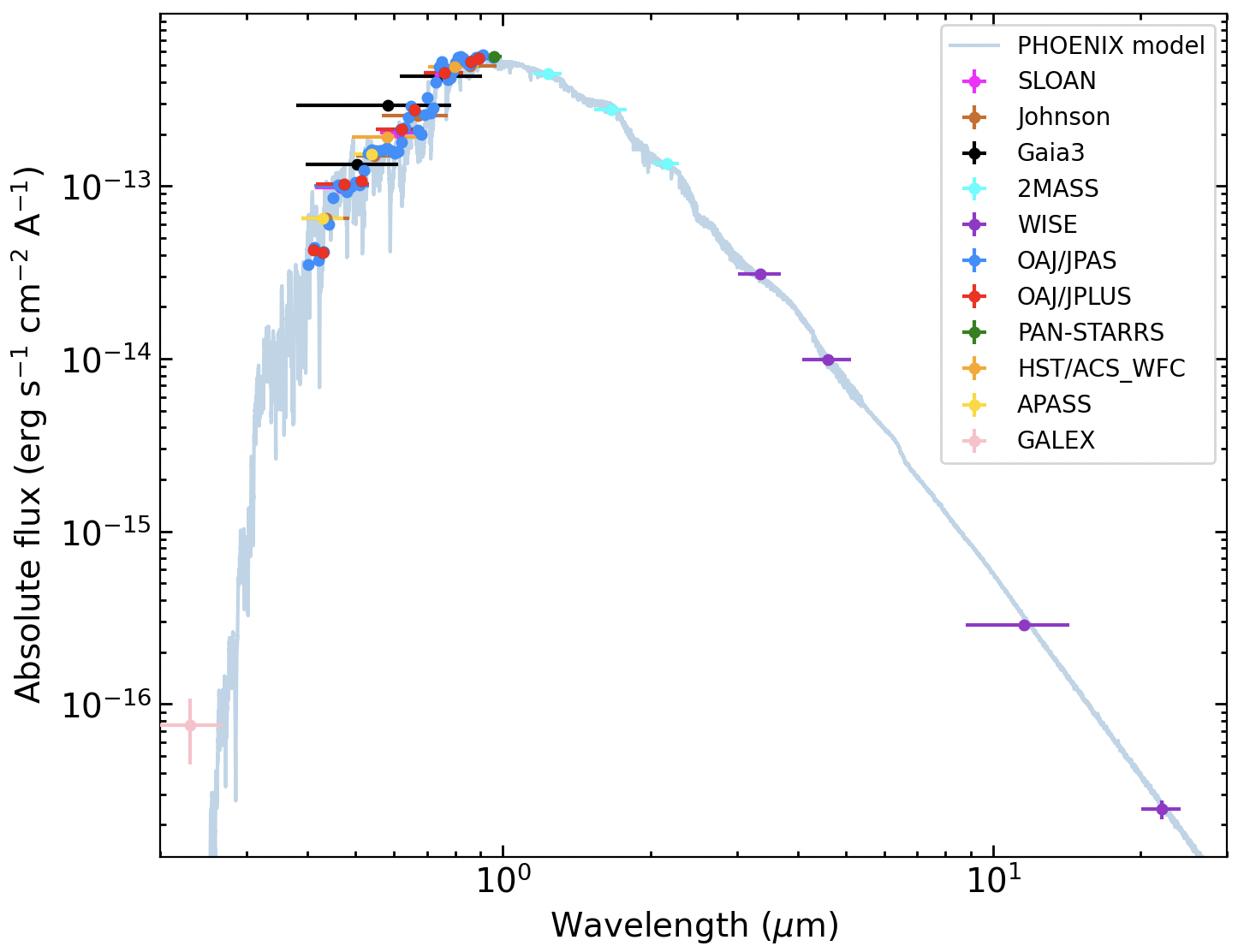}
    \caption[Photometric spectral energy distribution of GJ 1018.]{Photometric spectral energy distribution of GJ 1018 from 0.23 to $\sim$25 $\mu$m. The PHOENIX model corresponding to 3300 K, solar metallicity and high gravity is added \citep{2003IAUS..211..325A} in order to show that most fluxes are photospheric in origin and that there are no infrared flux excesses at long wavelengths. Vertical error bars denote flux uncertainties, and horizontal error bars account for the width of the passbands. The fluxes, effective wavelengths and widths of all passbands were taken from the Virtual Observatory SED Analyser database.}
    \label{fig:phot_sed}
\end{figure}

To determine the bolometric luminosity of GJ 1018, we first built the photometric spectral energy distribution (SED) of the star using broadband and narrowband photometry from the literature. The stellar SED is shown in Fig.~\ref{fig:phot_sed}, which includes the Galaxy Evolution Explorer (GALEX) near-ultraviolet photometry \citep{2017ApJS..230...24B}, the Johnson $BVRI$ photometry \citep{2002yCat.2237....0D}, the $BV$ data from the AAVSO Photometric All-sky Survey \citep[APASS,][]{2019JAVSO..47..130H}, the $griz$ data from the Sloan Digital Sky Survey \citep{2000AJ....120.1579Y}, the {\sl Gaia} Early Data Release 3 photometry \citep{2021A&A...649A...1G}, the $y$ data from the Panoramic Survey Telescope and Rapid Response System (Pan-STARRS, \citealt{2020ApJS..251....7F}), the Two Micron All Sky Survey (2MASS) near-infrared $JHK_s$ photometry \citep{skrutskie2006}, the Wide-field Infrared Survey Explorer ({\sl WISE}) $W1$, $W2$, $W3$, and $W4$ data \citep{wright2010}, and the optical multi-photometry of the Observatorio Astrofísico de Javalambre (OAJ) Physics of the Accelerating Universe Astrophysical Survey (JPAS) and Photometric Local Universe Survey (JPLUS) catalogues accessible through the Spanish Virtual Observatory \citep{bayo2008}. In total, there are 89 photometric data points defining the SED of GJ 1018 between 0.23 and $\sim$25 $\mu$m. The OAJ/JPAS data cover very nicely the optical region in the interval 0.40--0.95 $\mu$m with a cadence of one measurement per 0.01 $\mu$m. We used the {\sl Gaia} trigonometric parallax to convert all observed photometry and fluxes into absolute fluxes, which we present in Table \ref{tab:svo_phot}. The SED of GJ 1018 clearly indicates its photospheric origin for wavelengths longer than 0.4 $\mu$m; there are no mid-infrared flux excesses up to 25 $\mu$m.

We integrated the SED over wavelength to obtain the absolute bolometric flux ($F_{\rm bol}$) using the trapezoidal rule. We did not include the $\sl Gaia$ $G$-band flux in the computations because the filter's large passband width encompasses various redder and bluer filters. We then applied $M_{\rm bol} = -2.5~{\rm log}\,F_{\rm bol} - 18.988$ \citep{cushing2005}, where $F_{\rm bol}$ is in units of W$\,$m$^{-2}$, to derive the absolute bolometric magnitude $M_{\rm bol}$ = 8.8465 $^{+0.0203}_{-0.0200}$ mag for GJ 1018, from which we obtain a bolometric luminosity of $L$ = $2.277 \, \pm \, 0.042 \times 10^{-2}$ L$_{\odot}$. The quoted error bar accounts for the photometric uncertainties in all observed bands and the trigonometric distance error.

\subsection{Stellar radius and mass}
\label{sec:stellar_radius_mass}
We determined the radius of GJ 1018 from the well-known Stefan-Boltzmann law. Based on the effective temperature from Sect.~\ref{sec:steparsyn} and the bolometric luminosity from Sect.~\ref{sec:luminosity}, we obtained a radius of $R_{\star} = 0.428 \pm 0.025$ R$_\odot$. This radius determination is independent of any evolutionary model and depends only on distance (well known from ${\sl Gaia}$), bolometric luminosity (well determined from the SED, see Sect.~\ref{sec:luminosity}), and the model atmospheres used to fit the observed spectra.

The mass of GJ 1018 can be derived following different approaches. One is given by the empirical mass-luminosity relationship of \citet{mann19}, which is derived from 62 nearby, late-type binaries with known orbits. The authors calibrated the stellar mass as a function of the absolute $K$-band magnitude of the stars, finding that metallicity has little impact on the mass of the stars for approximately solar composition. According to the \citet{mann19} relationship, GJ 1018 has a mass of $M_{\star}$ = 0.400 $\pm$ 0.025 M$_\odot$, where the error budget includes the photometric error and the dispersion of the mass-luminosity relation. Another widely used relation valid for M-dwarfs is the mass-radius equation given in \citet{2019A&A...625A..68S}, obtained from 55 detached, double-lined, double-eclipsing, and main-sequence M-dwarf binaries from the literature. Following this relation, we obtain a mass of $M_{\star}$ = $0.427\pm0.029$~$\rm M_{\odot}$. Both mass determinations are compatible at the 1$\sigma$ level. Given that the relation by \citet{2019A&A...625A..68S} makes use of the Stefan-Boltzmann law, its mass derivation is consistent with our method for obtaining the stellar radius, thus we adopted $R_{\star} = 0.428 \pm 0.025$ R$_\odot$ and $M_{\star} = 0.427 \pm 0.029$ M$_\odot$.

\subsection{Galactic membership and Mg and Si abundances}
\label{sec:abundances}

Because of heavy line blending, the determination of the individual elemental abundances of M-dwarfs from the visible spectra is very hard  \citep[e.g.][]{2020A&A...644A..68M}. In this chapter, we estimated the abundance of Mg and Si closely following the procedure presented in \citet{2021A&A...653A..41D}. In brief, we used the systemic RV, parallax, ra/dec coordinates and proper motions from \textit{Gaia} DR3 to derive the Galactic space velocity UVW of GJ~1018. We obtain U  = -23.30 $\pm$ 0.02 km/s, V = 17.09 $\pm$ 0.02 km/s, and W = -8.62 $\pm$ 0.03 km/s with respect to the local standard of rest (LSR) adopting the solar peculiar motion from \citet{2003A&A...409..523R}. Based on these velocities, we used the widely-known kinematic approach from \citet{2003A&A...410..527B} and the kinematic characteristics for stellar components in the Solar neighbourhood from \citet{2003A&A...409..523R} to estimate the probability that GJ 1018 belongs to the thin disk (D), the thick disk (TD), and the halo (H), obtaining 99.11$\%$, 0.889$\%$, and 0.001$\%$ respectively. Hence, it is very likely that GJ 1018 is a member of
the Galactic thin-disk population. Then, from the APOGEE DR17 \citep{2022ApJS..259...35A} we selected cool stars with metallicities similar to that of GJ 1018 and belonging to the chemically defined Galactic thin disk. We obtained a sample of several thousand stars for which we calculated the mean abundances of Mg and Si and their standard deviation. Considering the {\sc SteParSyn} metallicity, we obtain [Mg/H] = -0.28 $\pm$ 0.07 dex and [Si/H] = -0.33 $\pm$ 0.07 dex. Considering the \texttt{ODUSSEAS} metallicity, we obtain [Mg/H] = 0.00 $\pm$ 0.10 dex and [Si/H] = -0.01 $\pm$ 0.10 dex. 

\section{Analysis and results}
\label{sec:analysis_results}

\subsection{TESS photometry analysis}
\label{subsec:TESS_analysis}

We first analysed the TESS PDCSAP photometry described in Sect.~\ref{sec:obs_tess} through a model that consists of two components: a transit model, and a Gaussian process (GP) that models the correlated photometric noise \citep{2006gpml.book.....R,2012RSPTA.37110550R}. 

We implemented the \citet{2002ApJ...580L.171M} quadratic limb darkened transit model through \texttt{batman} \citep{2015PASP..127.1161K}. The model is defined by the orbital period of the planet ($P_{\rm orb}$), the time of inferior conjunction ($T_{\rm 0}$), the orbital inclination ($i$), the quadratic limb darkening (LD) coefficients $u_{\rm 1}$ and $u_{\rm 2}$, the planet-to-star radius ratio ($R_{\rm p}/R_{\star}$), and the semimajor axis scaled to the stellar radius, which we parametrized through $P_{\rm orb}$ and the stellar mass ($M_{\rm \star}$) and radius ($R_{\rm \star}$) following the Kepler's Third Law. We modelled the TESS correlated noise through a GP with an approximate Matérn-3/2 kernel \citep{2017AJ....154..220F,2018RNAAS...2...31F}, which has been successfully used to model the unknown
mixture of stellar variability and residual systematics of TESS SPOC photometry \citep[e.g.][]{2022arXiv220113274M,2022AJ....163..298M,2023arXiv230409220M}. This kernel is especially suitable to model TESS light curves where the stellar rotation modulation is not observable (as is the case of GJ 1018, see Sect.~\ref{sec:obs_tess}), and instead, residual systematics are a significant component of the photometric variability, given that it has covariance properties that are especially well matched to short-term instrumental red-noise structures \citep{2017AJ....153..177P,2020AJ....160..259S}. The approximate Matérn-3/2 kernel can be written in terms of the temporal separation between two data points $\tau = t_{i} - t_{j}$ as

\begin{equation}
    K_{3/2} = \eta_{\sigma}^{2} \left[ \left(1 + \frac{1}{\epsilon} \right) e^{-(1-\epsilon) \sqrt{3} \tau / \eta_{\rho}} \, \left(1 - \frac{1}{\epsilon} \right) e^{-(1+\epsilon) \sqrt{3} \tau / \eta_{\rho}} \right],
\end{equation}

\noindent where the hyperparameters $\eta_{\sigma}$ and $\eta_{\rho}$ are the characteristic amplitude and timescale of the correlated variations, respectively, and $\epsilon$ controls the approximation to the exact Matérn-3/2 kernel. Given that the amplitudes and timescales of the TESS systematics can vary from one sector to another, we fit those parameters independently ($\eta_{\sigma_{i}}$ and $\eta_{\rho_{i}}$, where \textit{i} denotes the sector), while $\epsilon$ was fixed to its default value of $10^{-2}$ \citep{2017AJ....154..220F}. We also included a jitter term for each sector, which we added quadratically to the TESS flux uncertainties in order to model the uncorrelated noise not taken into account in our model. 

We used a Markov Chain Monte Carlo (MCMC) affine-invariant ensemble sampler \citep{2010CAMCS...5...65G} as implemented in \texttt{emcee} \citep{2013PASP..125..306F} in order to sample the posterior probability density function of the different parameters involved in our model. To do so, we used four times as many walkers as the number of parameters, and performed two consecutive runs. The first run (or burn-in) consisted of 200\,000 iterations. After this run, we reset the sampler and initialised the second run (or production) with 100\,000 iterations while considering the initial values from the last iteration of the burn-in phase. To ensure the convergence of the chains, we estimated the autocorrelation time for each parameter and checked that it is at least 30 times smaller than the chain length. 

We ran an MCMC fit starting from wide uniform priors for all the parameters involved in the model except for those for which we have prior information, which we constrained through Gaussian priors. These parameters are the $T_{0}$ and $P$ of TOI-244.01 (from the TLS periodogram, Sect. \ref{sec:obs_tess}), the stellar radius and mass (from our spectroscopic analysis, Sect. \ref{sec:stellar_radius_mass}), and the quadratic LD coefficients, which we computed from the \texttt{ldtk} package \citep{2015MNRAS.453.3821P}. The package infers the coefficients of a given LD law relying on the \citet{2013A&A...553A...6H} synthetic spectra library, spectroscopic $T_{\rm eff}$, log $g$, and [Fe/H], and the instrument transmission curve. We used the \texttt{ldtk} uncertainties as the widths of the Gaussian distributions. In order to account for possible systematics in the estimated LD coefficients \citep[e.g.][]{2022AJ....163..228P}, we also ran the MCMC fit by considering conservative widths of 0.2, obtaining identical results for the planetary parameters. We show the prior distributions for this analysis and subsequent ones in Table~\ref{table:priors}.

\begin{figure*}
    \centering
    \includegraphics[width = 0.99\textwidth]{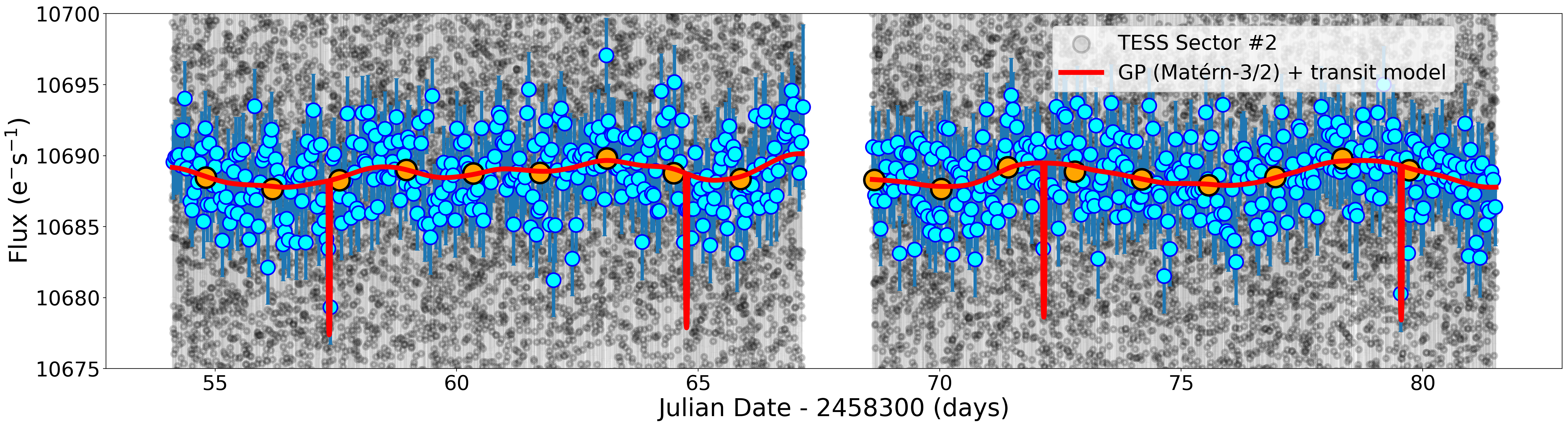}
    \includegraphics[width = 0.99\textwidth]{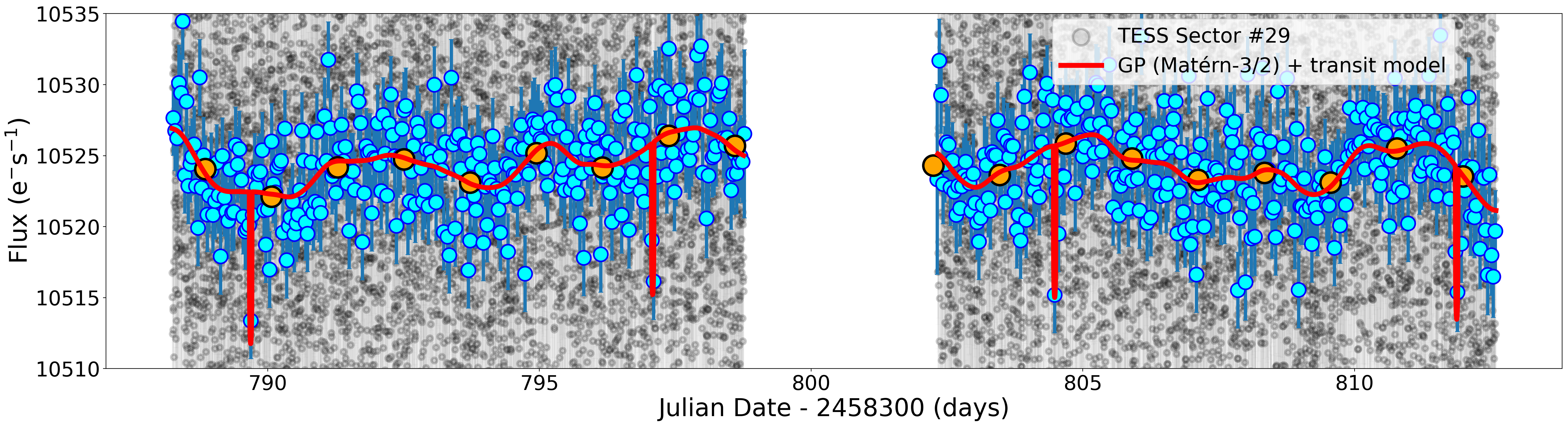}
    \caption[TESS light curve of GJ 1018 with the median posterior global model (transit + GP).]{TESS light curve of GJ 1018 with the median posterior global model (transit + GP) superimposed, in which we can see the different locations of the TOI-244.01 transit events over the smooth photometric modulation. The grey data points correspond to the SPOC 2-minute cadence PDCSAP photometry, and the blue and orange data correspond to 50-minute and 1.5-day binned data, respectively. }
    \label{fig:tess_photometry}
\end{figure*}

In Table \ref{tab:final_derived_params}, we include the median and 1$\sigma$ (68.3$\%$ credible intervals) of the posterior distributions of the fit parameters. In Fig.~\ref{fig:tess_photometry}, we show the complete TESS light curve together with the global model (transit + GP) evaluated on the fit parameters.

\begin{table}[]
\centering
\small
\renewcommand{\arraystretch}{1.175}
\setlength{\tabcolsep}{8pt}
\caption[Prior distribution of the parameters used in our models to describe TOI-244~b.]{Prior distribution of the parameters used in our models. The $\mathcal{U}(a,b)$ and $\mathcal{LU}(a,b)$ symbols indicate uniform and log-uniform distributions, $a$ and $b$ being the lower and upper limits. The $\mathcal{G}(\mu, \sigma)$ and $\mathcal{ZTG}(\mu, \sigma)$ symbols indicate Gaussian and zero-truncated Gaussian distributions, being $\mu$ and $\sigma$ the mean and width of the distributions.}
\label{table:priors}
\begin{tabular}{llc}
\hline \hline
Symbol                             & Parameter         & Distribution                                       \\ \hline 
\multicolumn{3}{l}{Orbital parameters}                                                                                                 \\ \hline
$P$ {[}days{]}                              & Orbital period             & $\mathcal{G} \, (7.397, 0.002)$                                 \\
$T_{\rm 0}$ {[}RJD{]}                       & Time of inf. conj.         & $\mathcal{G} \, (58357.36, 0.02)$                         \\
$i$ {[}degrees{]}                           & Orbital inclination        & $\mathcal{U} \, (50, 90)$                                   \\
$cos(w) \sqrt{e}$                           & Ecc. parametrization       & $\mathcal{U} \, (- \frac{\sqrt{2}}{2}, \frac{\sqrt{2}}{2})$ \\
$sin(w) \sqrt{e}$                           & Ecc. parametrization       & $\mathcal{U} \, (- \frac{\sqrt{2}}{2}, \frac{\sqrt{2}}{2})$ \\ \hline
\multicolumn{3}{l}{Planet parameters}                                                                                                  \\ \hline
$R_{\rm p}$/$R_{\rm \star}$                 & Planet-to-star radius ratio & $\mathcal{U} \, (0.0, 0.1)$                                 \\
$K$ {[}$\rm m/s${]}                        & RV semi-amplitude          & $\mathcal{U} \, (1, 10)$                         \\ \hline
\multicolumn{3}{l}{Stellar parameters}                                                                                                  \\ \hline
$M_{\rm \star}$ $[\rm M_{\odot}]$           & Stellar Mass               & $\mathcal{ZTG} \, (0.427, 0.029)$                           \\
$R_{\rm \star}$ $[\rm R_{\odot}]$           & Stellar Radius             & $\mathcal{ZTG} \, (0.428, 0.025)$                           \\
$u1$                                        & LD coefficient             & $\mathcal{ZTG} \, (0.2204, 0.0011)$                         \\
$u2$                                        & LD coefficient             & $\mathcal{ZTG} \, (0.4112, 0.0017)$                         \\ \hline
\multicolumn{3}{l}{Matérn-3/2 GP hyper-parameters}                                                                                      \\ \hline
$\eta_{\sigma_{S2}}$ {[}$\rm e^{-}/s${]}    & S2 amplitude               & $\mathcal{U} \, (0, 10^{2})$                                \\
$\eta_{\rho_{S2}}$ {[}days{]}               & S2 timescale               & $\mathcal{U} \, (0, 10^{2})$                                \\
$\eta_{\sigma_{S29}}$ {[}$\rm e^{-}/s${]}   & S29 amplitude              & $\mathcal{U} \, (0, 10^{2})$                                \\
$\eta_{\rho_{S29}}$ {[}days{]}              & S29 amplitude              & $\mathcal{U} \, (0, 10^{2})$                                \\ \hline
\multicolumn{3}{l}{Quasiperiodic GP hyper-parameters}                                                                                  \\ \hline
$\eta_{\rm 1,RV}$ {[}$\rm m/s${]}          & Amplitude (RV)             & $\mathcal{LU} \, (10^{-2}, 10)$                        \\
$\eta_{\rm 2,RV}$ {[}days{]}                & Decay timescale            & $\mathcal{U} \, (0, 5 \, 10^{2})$                        \\
$\eta_{\rm 3,RV}$ {[}days{]}                & Correlation period         & $\mathcal{U} \, (40, 80)$                                   \\
$\eta_{\rm 4,RV}$                           & Periodic scale             & $\mathcal{LU} \, (10^{-3}, 10)$                             \\
$\eta_{\rm 1,FWHM}$ {[}$\rm m/s${]}        & Amplitude (FWHM)           & $\mathcal{LU} \, (10^{-2}, 10)$                        \\ \hline
\multicolumn{3}{l}{Instrument-dependent parameters (TESS)}                                                                             \\ \hline
$F_{\rm 0,S2}$ {[}$\rm e^{-}/s${]}          & TESS S2 flux offset        & $\mathcal{U} \, (-10^{2}, 10^{2})$                          \\
$F_{\rm 0,S29}$ {[}$\rm e^{-}/s${]}         & TESS S29 flux offset       & $\mathcal{U} \, (-10^{2}, 10^{2})$                          \\
$\sigma_{\rm TESS,S2}$ {[}$\rm e^{-}/s${]}  & TESS S2 jitter             & $\mathcal{U} \, (0, 10^{2})$                                \\
$\sigma_{\rm TESS,S29}$ {[}$\rm e^{-}/s${]} & TESS S29 jitter            & $\mathcal{U} \, (0, 10^{2})$                                \\ \hline
\multicolumn{3}{l}{Instrument-dependent parameters (ESPRESSO and HARPS)}                                                               \\ \hline
$\gamma_{\rm ESPR,RV}$ {[}$\rm m/s${]}     & ESPR. sys. RV              & $\mathcal{U} \, (13000, 16000)$                                   \\
$\gamma_{\rm HAR,RV}$ {[}$\rm m/s${]}      & HARPS sys. RV              & $\mathcal{U} \, (13\,10^{3}, 16\,10^{3})$                                   \\
$\gamma_{\rm ESPR,FWHM}$ {[}$\rm m/s${]}   & ESPR. sys. FWHM            & $\mathcal{U} \, (0, 10^{4})$                                    \\
$\gamma_{\rm HAR,FWHM}$ {[}$\rm m/s${]}    & HARPS sys. FWHM            & $\mathcal{U} \, (0, 10^{4})$                                    \\
$\sigma_{\rm ESPR,RV}$ {[}$\rm m/s${]}     & ESPR. RV jitter            & $\mathcal{U} \, (0, 10)$                               \\
$\sigma_{\rm HAR,RV}$ {[}$\rm m/s${]}      & HARPS RV jitter            & $\mathcal{U} \, (0, 10)$                               \\
$\sigma_{\rm ESPR,FWHM}$ {[}$\rm m/s${]}   & ESPR. FWHM jitter          & $\mathcal{U} \, (0, 10)$                               \\
$\sigma_{\rm HAR,FWHM}$ {[}$\rm m/s${]}    & HARPS FWHM jitter          & $\mathcal{U} \, (0, 10)$                               \\ \hline
\multicolumn{3}{l}{Orbital and planet parameters of an additional Keplerian}                                                           \\ \hline
$P_{\rm 2}$ {[}days{]}                      & Orbital period             & $\mathcal{U} \, (0, 5 \, 10^{2})$                        \\
$T_{\rm 0,2}$ {[}RJD{]}                     & Time of inf. conj.         & $\mathcal{U} \, (58467, 59855)$                             \\
$cos(w_{2}) \sqrt{e_{2}}$                   & Ecc. parametrization       & $\mathcal{U} \, (- \frac{\sqrt{2}}{2}, \frac{\sqrt{2}}{2})$ \\
$sin(w_{2}) \sqrt{e_{2}}$                   & Ecc. parametrization       & $\mathcal{U} \, (- \frac{\sqrt{2}}{2}, \frac{\sqrt{2}}{2})$ \\
$K_{2}$ {[}$\rm m/s${]}                    & RV semi-amplitude          & $\mathcal{U} \, (1, 10)$                         \\ \hline
\end{tabular}

\end{table}

\subsection{ESPRESSO and HARPS radial velocity analysis}
\label{subsec:rv_modeling}

We analysed the ESPRESSO and HARPS RV data sets described in Sects.~\ref{sec:espresso_obs} and \ref{sec:harps_obs} through a model composed of three components: a Keplerian, which models planetary-induced RV signals, an instrumental component, which models the systemic velocity as measured by each instrument, and a GP, which models the RV correlated noise induced by the stellar activity. 

We implemented the Keplerian component through the Python package \texttt{radvel} \citep{2018PASP..130d4504F} by using the parametrization $\left\lbrace P, T_{0}, K, \sqrt{e} cos(w), \sqrt{e} sin(w)\right\rbrace$, being $P$ the orbital period of the planet, $T_{0}$ the time of inferior conjunction, $K$ the semi-amplitude, $e$ the orbital eccentricity, and $w$ the planetary argument of periastron. The instrumental component of our model consists of an offset that corresponds to the systemic radial velocity of the star as measured by each instrument ($\gamma_{ins})$. The periodograms of the spectroscopic data show that the stellar rotation induces significant activity-related RVs (Sect.~\ref{sec:espresso_obs}). We modelled this correlated noise through a GP with a quasiperiodic kernel \citep{2015ITPAM..38..252A,2016A&A...588A..31F}. The choice of this kernel is motivated by the fact that although stellar rotation is a periodic phenomenon, the activity-induced RV signals are quasiperiodic, given that active regions evolve; that is, they move on the stellar surface and appear and disappear throughout the magnetic cycle timescale. The quasiperiodic kernel depends on four hyper-parameters ($\eta_{1}$, $\eta_{2}$, $\eta_{3}$, and $\eta_{4}$) and can be written in terms of the separation between data points $\tau = t_{i} - t_{j}$ as

\begin{equation}
    K_{QP} (\tau) = \eta_{1}^{2} \rm{exp} \left[ - \frac{\tau^{2}}{2\eta_{2}^{2}} - \frac{2sin^{2} \left(  \frac{\pi \tau}{\eta_{3}} \right)}{\eta_{4}^{2}}  \right].
\end{equation}

\noindent The hyper-parameter $\eta_{1}$ scales with the amplitude of the stellar activity signal. $\eta_{3}$ corresponds to the main periodicity of the signal, and it is considered a measure of the stellar rotation period \citep[e.g.][]{2018MNRAS.474.2094A}. $\eta_{2}$ is the length-scale of exponential decay. For two data points in the X-axis far from each other, the larger $\eta_{2}$ is, the more closely correlated those data points are \citep[e.g.][]{2018MNRAS.474.2094A}; therefore, $\eta_{2}$ is considered as a measure of the timescale of growth and decline of the active regions \citep{2014MNRAS.443.2517H,2016A&A...588A..31F}. $\eta_{4}$ controls the amplitude of the $\rm sin^{2}$ term. The smaller $\eta_{4}$ is, data points separated by one orbital period will be much more closely correlated than those data points separated by a different period of time; thus, $\eta_{4}$ indicates the complexity of the harmonic content of the activity signal. Finally, in order to model the white noise not taken into account in our model, we included a jitter term per instrument ($\sigma_{ins}$) that we added quadratically to the uncertainties of our RV measurements. 

\begin{figure}
    \centering
    \includegraphics[width=0.65\textwidth]{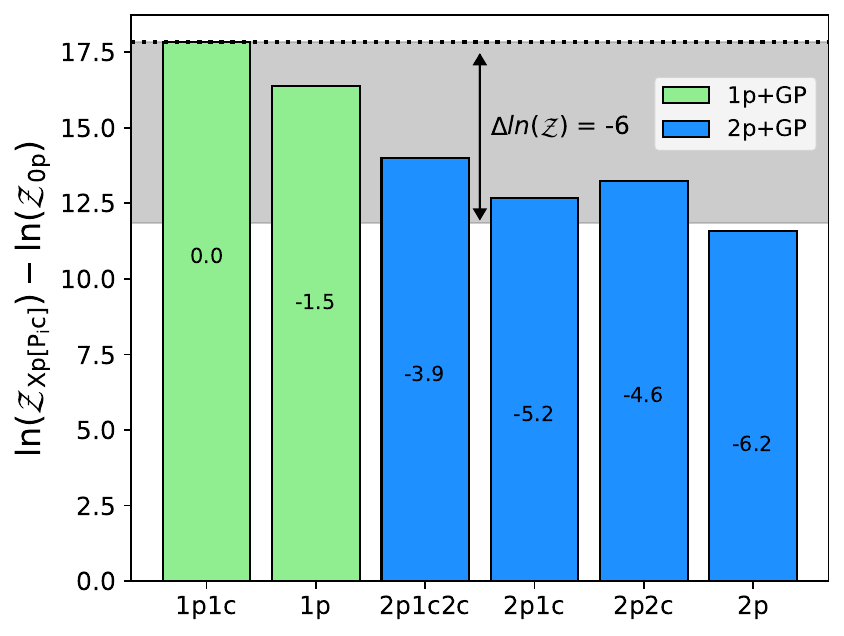}
    \caption[Differences of the log-evidences of different models and the 0 planet model tested on the ESPRESSO+HARPS data set.]{Differences of the log-evidences of different models and the 0 planet model, tested on the ESPRESSO+HARPS data set. The grey shaded region indicates the $\rm 0 \geq \Delta ln \mathcal{Z} \geq$ -6 region from the largest evidence model (1p1c).} 
    \label{fig:log-evidenes}
\end{figure}

In order to obtain a more robust estimation of the model parameters, we included in our data set an activity indicator to be modelled jointly with the HARPS and ESPRESSO RVs. This approach consists of using two GP kernels with shared hyperparameters except the signal amplitude (hereafter we differentiate between $\eta_{\rm 1,RV}$ and $\eta_{\rm 1,indicator}$). This approach has been used in previous works and relies on the assumption that the variations on the activity indicators are caused by the stellar activity alone, and that their periodicity and coherence are the same as those of the activity component of the RV \citep[e.g.][]{2020A&A...639A..77S,2020A&A...642A.121L,2022A&A...665A.154B}. From the time series and periodograms of Fig.~\ref{fig:gls_espresso}, we can perceive a certain correlation between the RVs and the activity indicators. We computed the strength of these correlations by means of the Pearson product-moment correlation coefficient \citep{doi:10.1080/00031305.1988.10475524}, obtaining the following results: 0.35 for the correlation with the FWHM, 0.20 with S-index, -0.07 with $\rm H_{\alpha}$, -0.22 with BIS, 0.02 with NaD, and 0.21 with the contrast of the CCF. Showing a moderate degree of correlation, we decided to use the FWHM of the CCF in our modelling.

We explored the constraining capacity of the HARPS data, motivated by their shorter time coverage and smaller precision than ESPRESSO data. To do so, we ran an MCMC fit considering a circular Keplerian model ($e$ = 0, $w$ = 0) to the ESPRESSO and ESPRESSO+HARPS data sets. Besides, for each data set, we performed a complementary fit with zero planets in order to obtain the significance against the null hypothesis. To do so, we used the Perrakis algorithm through the \texttt{bayev} implementation \citep{2016A&A...585A.134D} to compute the logarithm of the Bayesian evidence of the model ln($\mathcal{Z}$) based on the 15$\%$ of the final flattened chain. In the same way as in the previous section, we used wide uniform priors for all the parameters, except for those for which we have prior information (see Table \ref{table:priors} for the detailed prior distributions). As a result, the inferred parameters are compatible in both tested data sets (ESPRESSO and ESPRESSO+HARPS). However, we obtain a larger Bayesian evidence against the null hypothesis when using the ESPRESSO+HARPS data set: $\rm \Delta ln(\mathcal{Z})_{ESPR} $ = +16.5, $\rm \Delta ln(\mathcal{Z})_{ESPR+HARPS} $ = +17.7. This way, the inclusion of HARPS data slightly increases the significance of the detection. Consequently, due to the higher evidence, together with the advantage of having a longer time span to search for additional signals, we used the ESPRESSO+HARPS data set for the subsequent analysis. 

We tested six different models in order to assess whether we can detect additional planetary signals and to select the simplest model that best represents our data. The models involve one and two planets in all the possible combinations of circular and eccentric orbits. To identify the models, we use the nomenclature Xp[$\rm P_{i}c$], where X is the total number of planets considered in the system, and $\rm P_{i}$ indicates which planets have assumed circular orbits. In this chapter, planet "1" corresponds to TOI-244.01, and planet "2" to an additional planet without prior orbital information (see Table \ref{table:priors}). This way, we have tested the following models: 1p1c, 1p, 2p1c2c, 2p, 2p1c, and 2p2c. In Fig.~\ref{fig:log-evidenes}, we compare the obtained difference of the log-evidence of each model and the log-evidence of the 0 planet model $\rm  ln(\mathcal{Z}_{\rm Xp[\rm P_{i}c]}) - ln(\mathcal{Z}_{0p})$. We assume that a difference of +6 (i.e. $\rm \Delta ln \mathcal{Z}$ $>$ 6) indicates strong evidence in favour of the largest evidence model \citep{2008ConPh..49...71T}. For our data set, there is no model meeting the condition when compared to a simpler model,  so we chose the simplest model as the one that best represents our data set, which, in this case, coincides with the one with the largest evidence: the 1 planet model in a circular orbit (1p1c). 

\subsection{Joint analysis}
\label{subsec:joint_analysis}

\begin{table}[]
\centering
\caption[Main parameters of TOI-244 b derived from the joint analysis.]{TOI-244 b main parameters obtained from the fitted parameters in the joint analysis (Sect. \ref{sec:analysis_results}, Table \ref{tab:final_derived_params}) and the computed stellar parameters (Sect. \ref{sec:stellar_charact}, Table \ref{tab:stellar_params}).}
\label{tab:planet_params}
\renewcommand{\arraystretch}{2.03}
\setlength{\tabcolsep}{5pt}
\begin{tabular}{lll}
\hline \hline
Symbol                            & Parameter          & Value                     \\ \hline
$P$ {[}days{]}                             & Orbital period              & $7.397225^{+0.000026}_{-0.000023}$ \\
$T_{0}$ {[}RJD{]}                           & Time of inf. conj.    & 58357.3627 $\pm$ 0.0020          \\
$R_{\rm p}$ {[}$\rm R_{\oplus}${]}         & Planet radius               & 1.52 $\pm$ 0.12                    \\
$M_{\rm p}$ {[}$\rm M_{\oplus}${]}         & Planet mass                 & 2.68 $\pm$ 0.30                    \\
$\rho_{\rm p}$ {[}$\rm g \cdot cm^{-3}${]} & Planet density              & 4.2 $\pm$ 1.1                      \\
$a / R_{\star}$                            & $a$ relative to $R_{\star}$ & 28.1 $\pm$ 1.8                     \\
$a$ {[}AU{]}                               & Orbit semi-major axis       & 0.0559 $\pm$ 0.0013                \\
$S$ {[}$\rm S_{\oplus}${]}                 & Insolation flux             & 7.3 $\pm$ 0.4                      \\
$T_{\rm eq}$ {[}K{]}                       & Equilibrium temperature     & 458 $\pm$ 20                       \\
$g$ {[}$\rm m \cdot s^{-2}${]}             & Surface gravity             & 11.3 $\pm$ 2.2                     \\
$b$                                        & Impact parameter            & 0.61 $\pm$ 0.31                    \\
$\delta$ {[}ppt{]}                         & Transit depth               & 1.06 $\pm$ 0.12                    \\
$T_{\rm 14}$ {[}hours{]}                   & Transit duration            & 1.7 $\pm$ 0.5                      \\ \hline 
\end{tabular}
\end{table}

\begin{figure*}
    \centering
    \includegraphics[width = 0.223\textwidth]{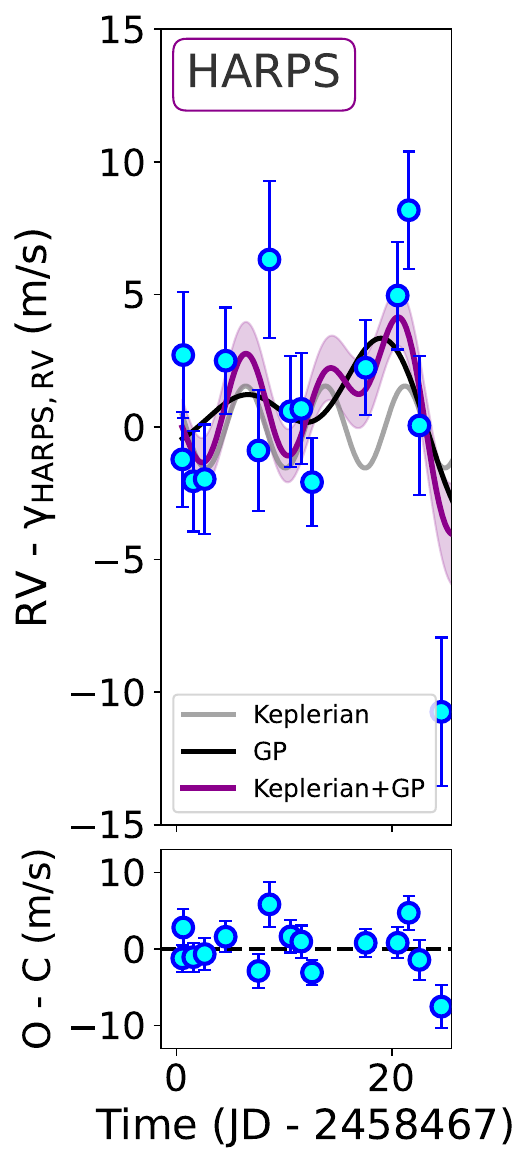}
    \includegraphics[width = 0.76\textwidth]{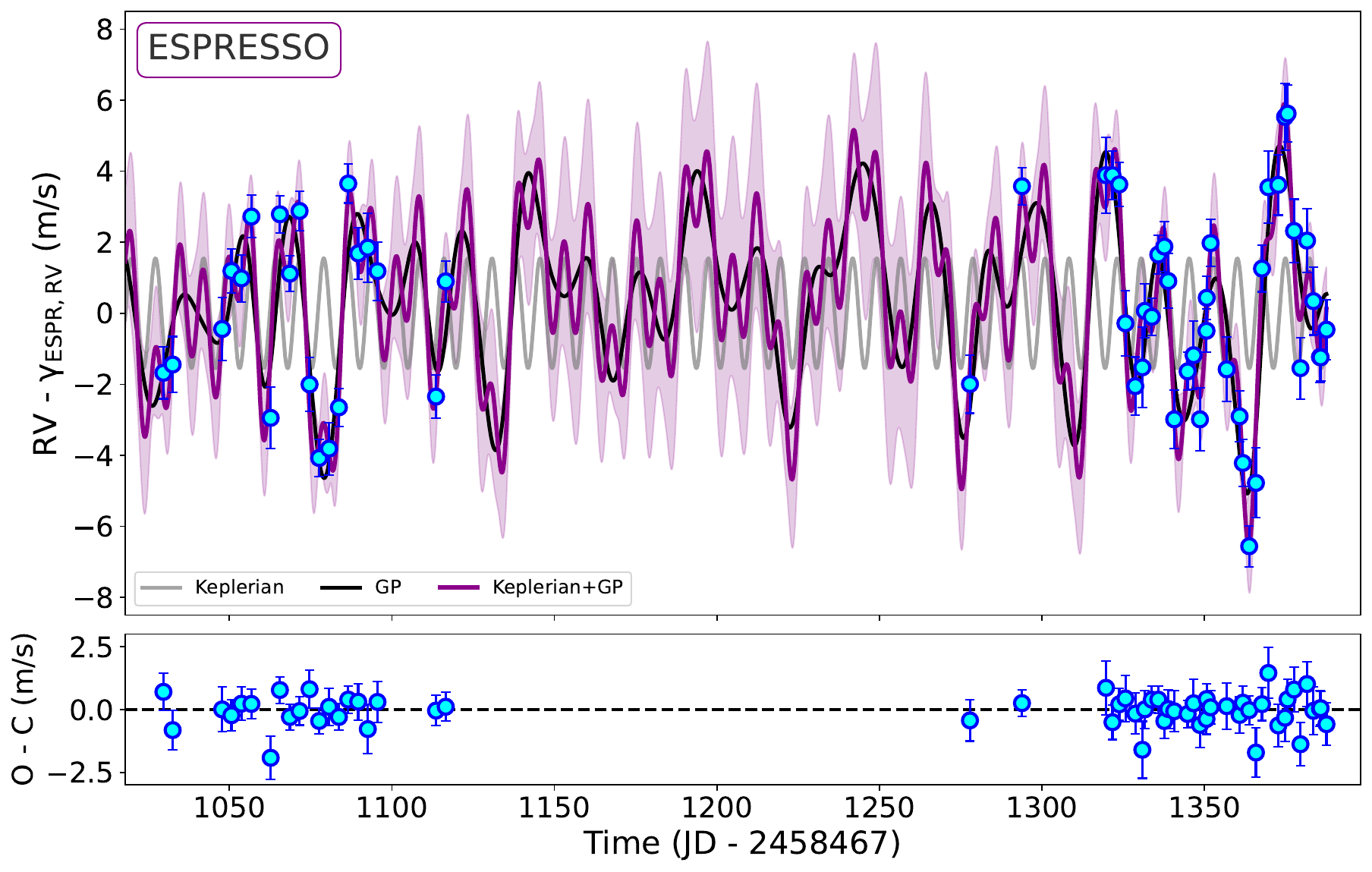}
    \caption[ESPRESSO and HARPS RVs of GJ 1018 with the global, GP, and Keplerian models.]{ESPRESSO and HARPS radial velocities of GJ 1018. The purple solid line indicates the median posterior global model, and the black and grey solid lines indicate the median posterior GP and Keplerian models, respectively. The shaded purple region indicates the 68.7$\%$ confidence interval of the global model. }
    \label{fig:final_plots}
\end{figure*}

\begin{figure*}
\centering
    \includegraphics[width=0.4675\textwidth]{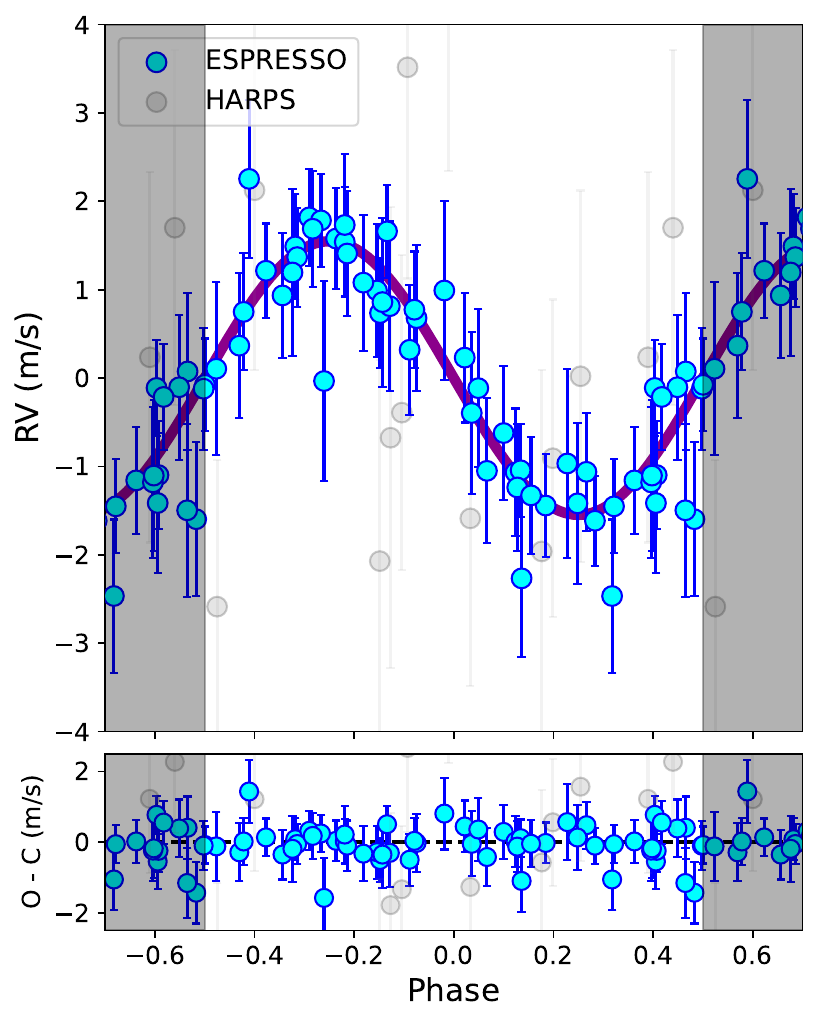}
    \includegraphics[width=0.495\textwidth]{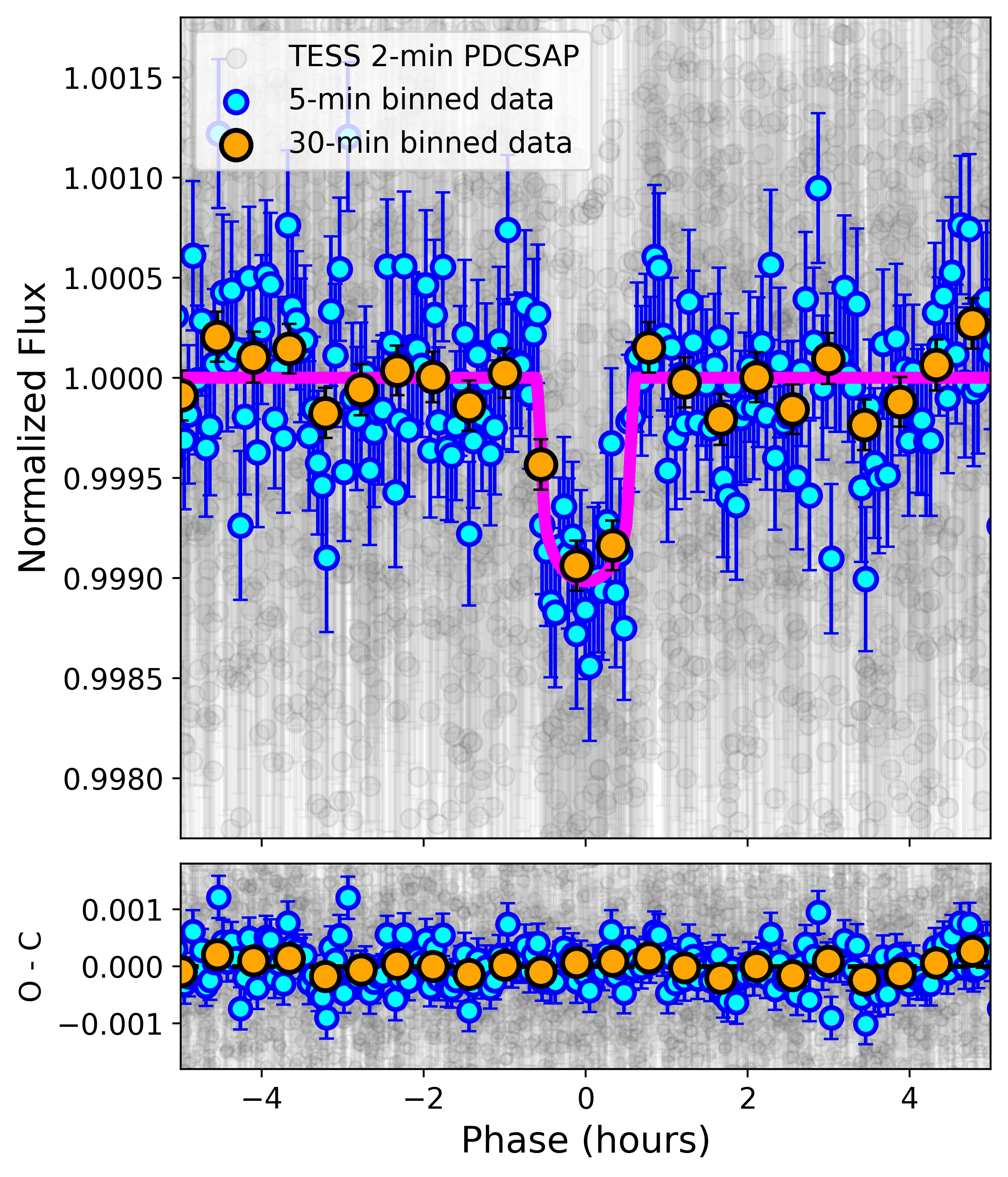}
    \caption[ESPRESSO RVs, HARPS RVs, and TESS photometry subtracted from their corresponding GP components and folded to the orbital period of TOI-244 b.]{ESPRESSO and HARPS RVs (left) and TESS photometry (right) subtracted from their corresponding GP components and folded to the orbital period of TOI-244 b. The solid lines indicate the median posterior models.}
    \label{fig:final_plots_2}
\end{figure*}

We inferred the final parameters of the system by modelling jointly the TESS photometry (same way as in Sect.~\ref{subsec:TESS_analysis}) and ESPRESSO+HARPS radial velocities (same way as in Sect.~\ref{subsec:rv_modeling}; considering the 1p1c model). In Fig.~\ref{fig:final_plots}, we show the complete RV data set together with the median posterior Keplerian+GP model. In Fig.~\ref{fig:final_plots_2}, we include the ESPRESSO+HARPS RVs and TESS photometry folded to the inferred orbital period, together with the median posterior model, being both subtracted from their corresponding GP components.  In Table \ref{tab:planet_params}, we present the main parameters of TOI-244 b. The complete list of fitted parameters can be found in Table \ref{tab:final_derived_params}. 

\section{Discussion}
\label{sec:discussion}

\subsection{Detectability of additional planets in the RV data set}

\begin{figure}
    \centering
    \includegraphics[width = 0.65\textwidth]{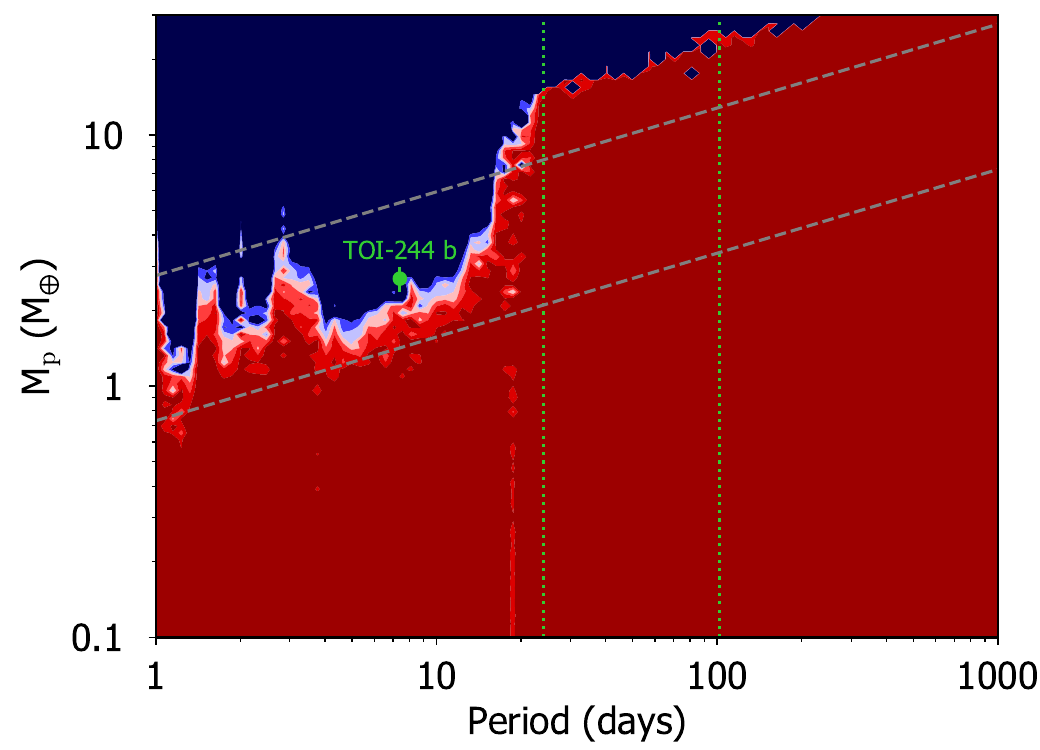}
    \caption[Detectability map for the ESPRESSO+HARPS data set of GJ~1018.]{Detectability map for the ESPRESSO+HARPS data set obtained from the injection-recovery test. The redder values correspond to false alarm probabilities (FAPs) larger than 1\% while the blue colour indicates that the signal is recovered with FAP $<1$\%. The location of TOI-244\,b is indicated by the green symbol. The green dotted vertical lines indicate the HZ around GJ 1018 according to \citet{2013ApJ...765..131K}. The bottom dashed line indicates the limit imposed by the median radial velocity uncertainty of the original data set (i.e. before removing the planet and activity), while the upper dashed line indicates the limit corresponding to the rms of the data.}
    \label{fig:IR_RV}
\end{figure}

We constrained the possibility of additional planets in the radial velocity data set by following the injection-recovery procedure described in \citet{2022arXiv221207332S}. In brief, we first removed the radial velocity contribution from the confirmed planet TOI-244\,b. The residuals from this include the activity of the star. We then injected a Keplerian model from a grid of periods (from 1  to 1000 days, 100 bins in log-space) and planet masses (from 0.1 to 30 $\rm M_{\oplus}$, 100 bins in log-space) and random phases. We assumed coplanar orbits with TOI-244\,b. From the resulting RVs we computed the activity model using the hyperparameters determined in the joint analysis section and removed its contribution. We then computed the false alarm probability of the power at the injected period. We consider the injected planet signal is detected if the FAP is below 1\%. Figure~\ref{fig:IR_RV} shows the result of this exercise. All injections implying a root mean square (rms) larger than 1.5 times the original rms are assumed as detected. We repeated the process five times and averaged all the iterations to obtain the final detectability matrix shown in Fig.~\ref{fig:IR_RV}. The current ESPRESSO+HARPS data set allows us to detect planets down to 2~$\rm M_{\oplus}$ at orbital periods up to around 22 days.

It is of special interest to constrain the presence of planets in the habitable zone (HZ) of GJ 1018, given that late K-dwarfs and early M-dwarfs represent an ideal trade-off between detectability and true habitability of their planets \citep[see the KOBE experiment;][]{2022A&A...667A.102L}. For the HZ of GJ 1018, which is located in the period range 24-102 days \citep{2013ApJ...765..131K}, the current data allows us to discard planets with masses above 20~$\rm M_{\oplus}$.

\subsection{Internal structure of TOI-244 b}
\label{sec:internal_structure}

\subsubsection{TOI-244 b in the mass-radius diagram}
\label{sec:M-R}

In Fig.~\ref{fig:M-R}, we plot the radius versus the mass for all the known planets from the NASA Exoplanet Archive with measured dynamical masses with a precision better than $20\%$. TOI-244 b is located in an unpopulated region, significantly separated (1.8$\sigma$ and 6.7$\sigma$ in radius and mass, respectively) from the Earth-like composition curve \citep[33$\%$ Fe and 66$\%$ $\rm MgSiO_{3}$ in mass;][]{2019PNAS..116.9723Z}, where rocky planets typically reside. Being located above that curve, TOI-244 b has a lower density than expected. We highlight other five planets in this particular region of the parameter space: TOI-561\,b \citep{2021MNRAS.501.4148L,2021AJ....161...56W,2023AJ....165...88B}, L 98-59~c and d \citep{2021A&A...653A..41D}, HD 260655 c \citep{2022A&A...664A.199L}, and TOI-4481\,b \citep{2023arXiv230106873P}. The existence of these planets might be explained by the presence of lighter elements than those expected for Earth-like compositions. This translates into three possible scenarios: planets with a scarcity or complete absence of iron in their cores (being thus practically composed of silicates), planets with a significant amount of volatile elements, or planets in which both scenarios coexist and have a significant effect on the density of the planet. In the following, we discuss each scenario separately.

\subsubsection{TOI-244 b as an iron-free planet}
\label{sec:iron_free}

\begin{figure}
\centering
\includegraphics[width = 0.75\textwidth]{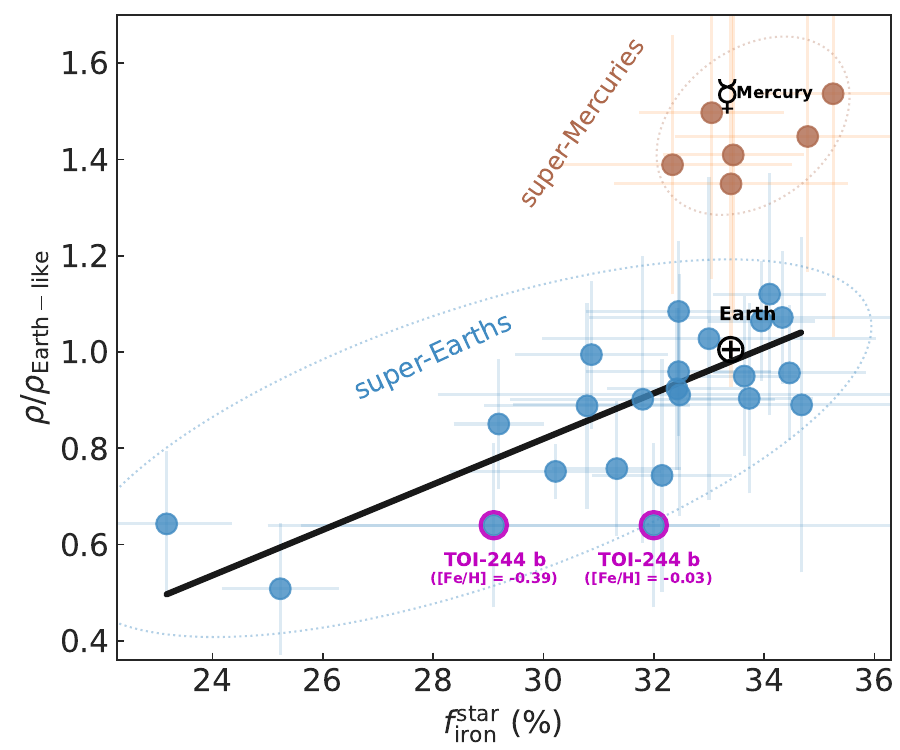}
    \caption[Planet density versus the estimated iron-to-silicate mass fraction of the original protoplanetary disk for the sample studied in Adibekyan et al. (2021) and TOI-244~b.]{Planet density normalized by the expected density of an Earth-like composition \citep{2017A&A...597A..37D} versus the estimated iron-to-silicate mass fraction of the original protoplanetary disk \citep{2015A&A...580L..13S,2017A&A...608A..94S} for the sample studied in \citet{2021Sci...374..330A} and for TOI-244~b considering the metal-poor and solar metallicity characterizations (see Sect.~\ref{sec:steparsyn}).  The Earth and Mercury are indicated with their respective symbols in black.  All error bars
show 1$\sigma$ uncertainties.}

    \label{fig:density_vs_iron}
\end{figure}

The abundance of refractory elements such as Mg, Si, and Fe of solar-type stars is considered a proxy of the composition of the initial protoplanetary disk from which stars and planets were formed  \citep[e.g.][]{2015A&A...577A..83D,Unterborn_2016}. In a recent study, \citet{2021Sci...374..330A} used stellar abundances of 22 stars to estimate the iron-to-silicate mass fractions of their initial protoplanetary disks ($f^{\rm star}_{\rm iron}$)  based on the stoichiometric model presented by \citet{2015A&A...580L..13S,2017A&A...608A..94S}. The authors found a relationship between $f^{\rm star}_{\rm iron}$ and the densities and iron mass fractions of their hosted planets, disclosing a compositional link between the rocky planets and their host stars. In other words, stars with lower $f^{\rm star}_{\rm iron}$ have lighter planets with lower iron-to-silicate mass fractions\footnote{The sample analysed in \citet{2021Sci...374..330A} is composed of highly irradiated planets, which are assumed to have negligible atmospheres, similar to the scenario we are discussing.}. In Fig.~\ref{fig:density_vs_iron}, we plot the planet density normalized by the expected density of an Earth-like composition \citep[$\rho$/$\rho_{Earth-like}$;][]{2017A&A...597A..37D} as a function of $f^{\rm star}_{\rm iron}$ for the same planet sample as in \citet{2021Sci...374..330A} and for TOI-244~b. For the metal-poor stellar characterization, we estimate $f^{\rm star}_{\rm iron}$ = 29.2 $\pm$ 4.2 $\%$, which would place TOI-244 b in agreement with the correlation. For the solar metallicity characterization, we estimate $f^{\rm star}_{\rm iron}$ = 32.0 $\pm$ 6.5 $\%$, which would place TOI-244~b slightly more deviated from the trend.

Overall, the low density of the planet together with the possibility of having a relatively low $f^{\rm star}_{\rm iron}$ (in the metal-poor scenario) suggests that TOI-244~b might have an iron mass fraction smaller than that expected for an Earth-like planet. However, according to Fig.~\ref{fig:M-R} (left), TOI-244~b has a density below what is expected for a 100$\%$ silicate composition at the 1$\sigma$ level. Hence, the most likely scenario to explain the low density of TOI-244~b is that it has a non-negligible amount of volatile elements in its composition. The colour coding of Fig.~\ref{fig:M-R} (left) represents the metallicity of the stellar hosts, showing that the growing population of low-density super-Earths tend to be formed around metal-poor stars (see Sect.~\ref{sec:density_metallicity} for a more extended discussion). Interestingly, the presence of volatiles in planets orbiting metal-poor stars is very expected, given that the building blocks of the original protoplanetary disks that form metal-poor stars are expected to have lower iron and higher water content than the expected for the disks that form stars with solar metallicities \citep{2017A&A...608A..94S}.

\subsubsection{TOI-244 b as a volatile-rich planet}
\label{volatile_rich}

\begin{figure*}
    \centering
    \includegraphics[width=0.48\textwidth]{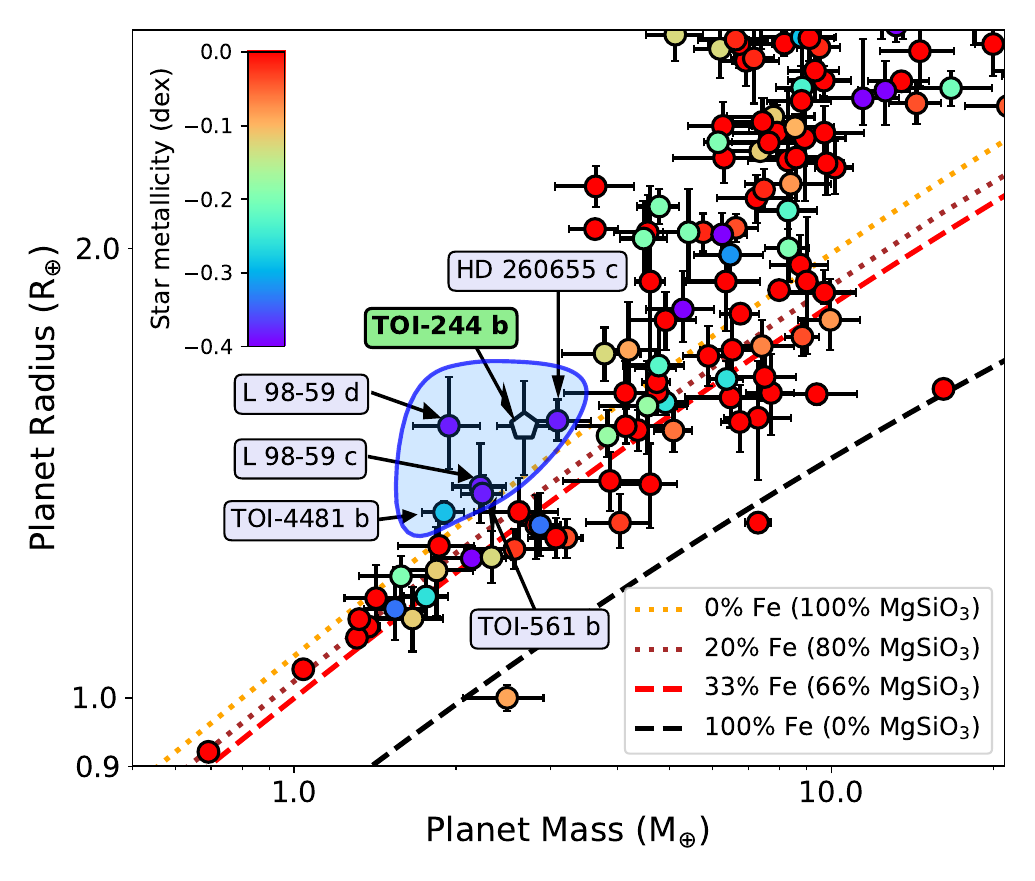}
    \includegraphics[width=0.48\textwidth]{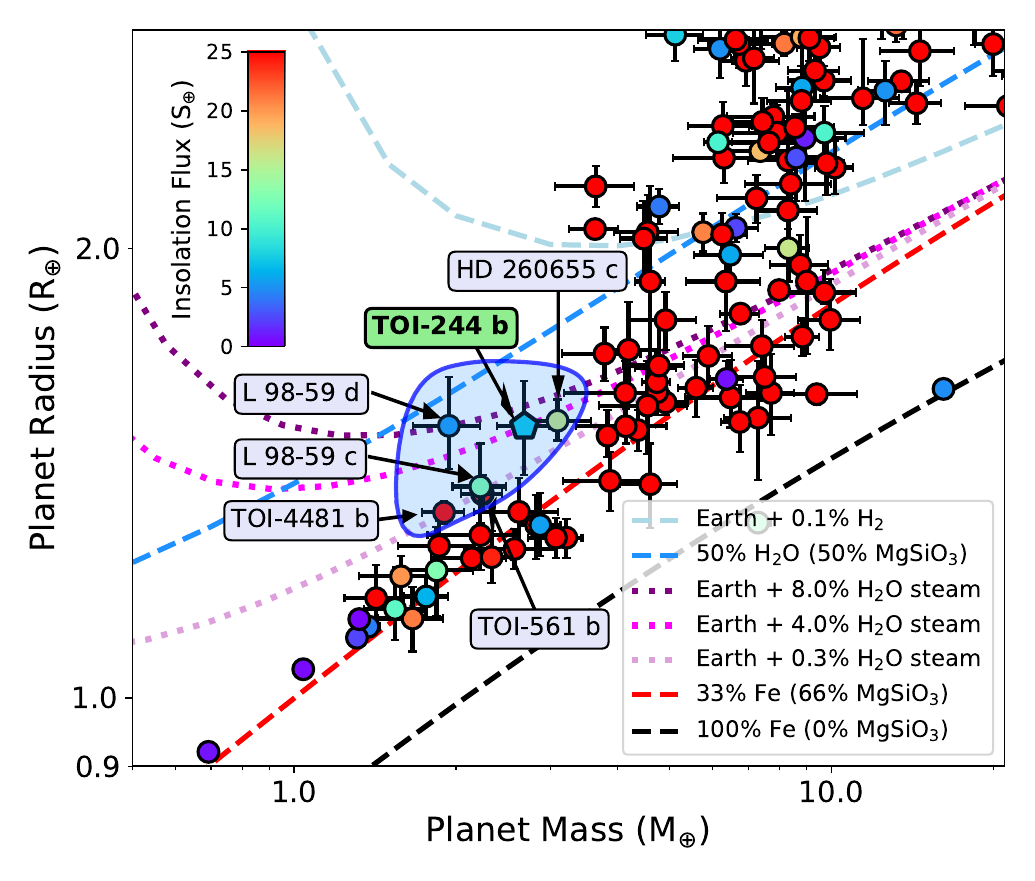}
    \caption[Mass-radius diagram for the small planet sample with measured dynamical masses with a precision better than 20$\%$ and TOI-244~b.]{Mass-radius diagrams for the small planet sample with measured dynamical masses with a precision better than 20$\%$. Left: colour coding indicates the stellar host metallicities. TOI-244~b is not filled with any colour since we get inconsistent metallicities with the different methods used (see Sect.~\ref{sec:steparsyn}). The dashed and dotted lines correspond to theoretical interior models that consider different mass percentages of Fe and $\rm MgSiO_{3}$ \citep{2016ApJ...819..127Z,2019PNAS..116.9723Z}. Right: colour coding indicates the stellar insolation received by the planets. The dashed lines correspond to theoretical models from \citet{2019PNAS..116.9723Z}, which consider planets without a significant amount of volatiles (back and red lines), planets with condensed water (dark blue), and planets with $\rm H_{2}$ atmospheric envelopes (light blue). The dotted lines correspond to the \citet{2020A&A...638A..41T} theoretical models for Earth-like planets with $\rm H_{2}O$-dominated atmospheres. These plots have been prepared using \texttt{mr-plotter}, which is available at \url{https://github.com/castro-gzlz/mr-plotter}.}
    \label{fig:M-R}
\end{figure*}

The most abundant volatiles on protoplanetary disks are $\rm H_{2}$/He and $\rm H_{2}O$  \citep[e.g.][]{2003ApJ...591.1220L,2007ApJ...667..303T}. In interior modelling, it is common to consider a hydrogen-rich envelope, which is representative of a primordial atmosphere. However, water is commonly considered in condensed form (either liquid or solid), based on traditional mass-radius relationships \citep[e.g.][]{2007ApJ...669.1279S, 2012ApJ...744...59S,2016ApJ...819..127Z}. In recent work, \citet{2020A&A...638A..41T} computed new mass-radius relationships taking into account that for $\rm H_{2}O$-rich rocky planets more irradiated than the runaway greenhouse limit (i.e $S$ $>$ 1.1 $\rm S_{\oplus}$), water is unstable in condensed form, so it would form a thick $\rm H_{2}O$-dominated atmosphere. Hence, given the much lower density of water vapour than liquid or solid water, the water content of highly irradiated $\rm H_{2}O$-rich planets is expected to be dramatically lower than the commonly computed ones \citep{2020ApJ...896L..22M}. In the right-hand mass-radius diagram of Fig.~\ref{fig:M-R}, we include a theoretical composition model consistent with Earth-like planets with $\rm H_{2}$-dominated atmospheres contributing 0.1$\%$ to the total planetary mass \citep{2019PNAS..116.9723Z}, showing that a small amount of hydrogen would significantly increase the radius of a rocky planet. We also include three theoretical models of Earth-like planets with an additional $\rm H_{2}O$-dominated atmosphere \citep{2020A&A...638A..41T}. In this case, a 4$\%$ water mass fraction would reproduce the observed density of TOI-244 b.  In addition, we include the 50$\%$ $\rm H_{2}O$-condensed and 50$\%$ $\rm MgSiO_{3}$ theoretical model from \citet{2019PNAS..116.9723Z}, which is widely used to describe planets in the higher mode of the bimodal radius distribution (2 $\rm R_{\oplus}$ $<$ $R_{\rm p}$ $<$ 4 $\rm R_{\oplus}$). This model allows us to illustrate the huge amount of condensed water ($\sim$20$\%$ in mass) that would be needed to explain the low density of TOI-244 b, in contrast with the 4$\%$ mass fraction when considering steam. In this diagram, the colour coding indicates the isolation fluxes received by the planets. Interestingly, TOI-244~b receives a lower insolation than the typically received by planets over the Earth-like composition curve, which could be interpreted as a hint that it could have maintained an atmospheric envelope (we discuss a possible density-insolation relation on a population level in Sect.~\ref{sec:density_insolation}). To infer whether the atmosphere of TOI-244~b could be a primary \ce{H2}-dominated atmosphere, or a secondary outgassed, potentially water-dominated atmosphere \citep{Kite2020pnas}, we estimated the amount of material that could be removed by atmospheric loss processes such as photo-evaporation and Jeans escape.

Planets close to their stars are exposed to high levels of X-ray and extreme UV radiation (XUV), which can lead to partial or total loss of a hydrogen envelope by photo-evaporation. If the escape is energy-limited, the mass loss rate can be written as \citep{Erkaev2007,2013ApJ...775..105O}:
\begin{equation}
    \dot{M} = \epsilon \frac{\pi F_\mathrm{XUV} R_\mathrm{p}^3}{G M_\mathrm{p}}, \label{eq:atmospheric-loss}
\end{equation}
where $F_\mathrm{XUV}$ is the XUV flux received by the planet, $G$ the gravitational constant and $\epsilon$ 
is an efficiency parameter. Stars emit the most XUV at a young age, during the so-called saturation regime \citep{Pizzolato2003}, which is particularly long for M-dwarfs. This makes photo-evaporation extremely efficient in the vicinity of M-dwarfs at a young age. Using the analytical fits of \cite{Sanz-Forcada2011}, we estimated that GJ 1018 emitted $3.9\times 10^{29}$ erg$\,$s\textsuperscript{-1} of XUV during its saturation regime, that lasted for $\tau_\mathrm{sat}=350$ Myr. This translates into a mass loss rate during the saturation regime of \rm $\rm 0.074~M_{\oplus}$/Gyr, during which the planet lost $\rm 0.026~M_{\oplus}$ of hydrogen. Following the approach of \cite{Aguichine2021}, we integrated Eq. \ref{eq:atmospheric-loss} in time assuming that, at first order, the mass and radius of the planet remain roughly constant, and only $F_{\mathrm{XUV}}$ varies in time, following the fit of \cite{Sanz-Forcada2011}. We then find that, in total, TOI-244 b could have lost $\rm 0.1~M_{\oplus}$ of hydrogen by photo-evaporation. This value is most likely a lower estimate, since an initially greater \ce{H2} content would produce a larger planetary radius, greatly increasing the mass loss rate (see Eq. \ref{eq:atmospheric-loss}). This estimate is consistent with more refined models of hydrogen mass loss. For example, the work by \cite{2023ApJ...947L..19R} also predicts that planets with core masses $\rm <3~M_{\oplus}$ at $T_\mathrm{eq}=500$ K are entirely stripped of their envelopes, if the latter are made of pure \ce{H2}.

Despite the efficient loss of hydrogen, recent studies indicate that secondary atmospheres, made of a mixture of volatile gases, can be restored by outgassing from the magma after the photo-evaporation phase \citep{Kite2020pnas,Tian2023}. The equilibrium temperature of TOI-244 b is greater than that of the Earth, but so is its surface gravity. The characteristics of TOI-244 b are such that its Jeans escape parameter $\Lambda_\mathrm{p} = GM_\mathrm{p}m_\mathrm{H}/(k_\mathrm{B}T_\mathrm{eq}R_\mathrm{p})$ \citep{2017A&A...598A..90F} has a value consistent with the Jeans escape parameter of the Earth within the measurements uncertainty range: $\Lambda_\mathrm{p}/\Lambda_\oplus = 1.07^{+0.28}_{-0.23}$. 

Our computations suggest that even if hydrogen was initially present in the secondary atmosphere of TOI-244 b, it has possibly been removed by Jeans escape, as it happened on Earth \citep[see][]{Catling2017book}. Both Jeans escape and photo-evaporation appear to be efficient mechanisms to entirely remove hydrogen from the atmosphere of TOI-244~b. This favours the modelling of its interior with an atmosphere of high mean molecular weight volatiles, such as water vapour. However, we cannot rule out the possibility that its atmosphere contains other molecules, for instance carbon-rich species such as \ce{CH4}, CO or \ce{CO2}. Spectroscopic measurements are necessary to break the compositional degeneracy, but the bulk water content can be used as a proxy for all heavy volatile species (\ce{CH4}, \ce{O2}, CO, \ce{CO2}, etc).

\subsubsection{Internal structure modelling}
\label{sec:final_internal_strucutre}

\begin{figure}
    \centering
    \includegraphics[width = 0.75\textwidth]{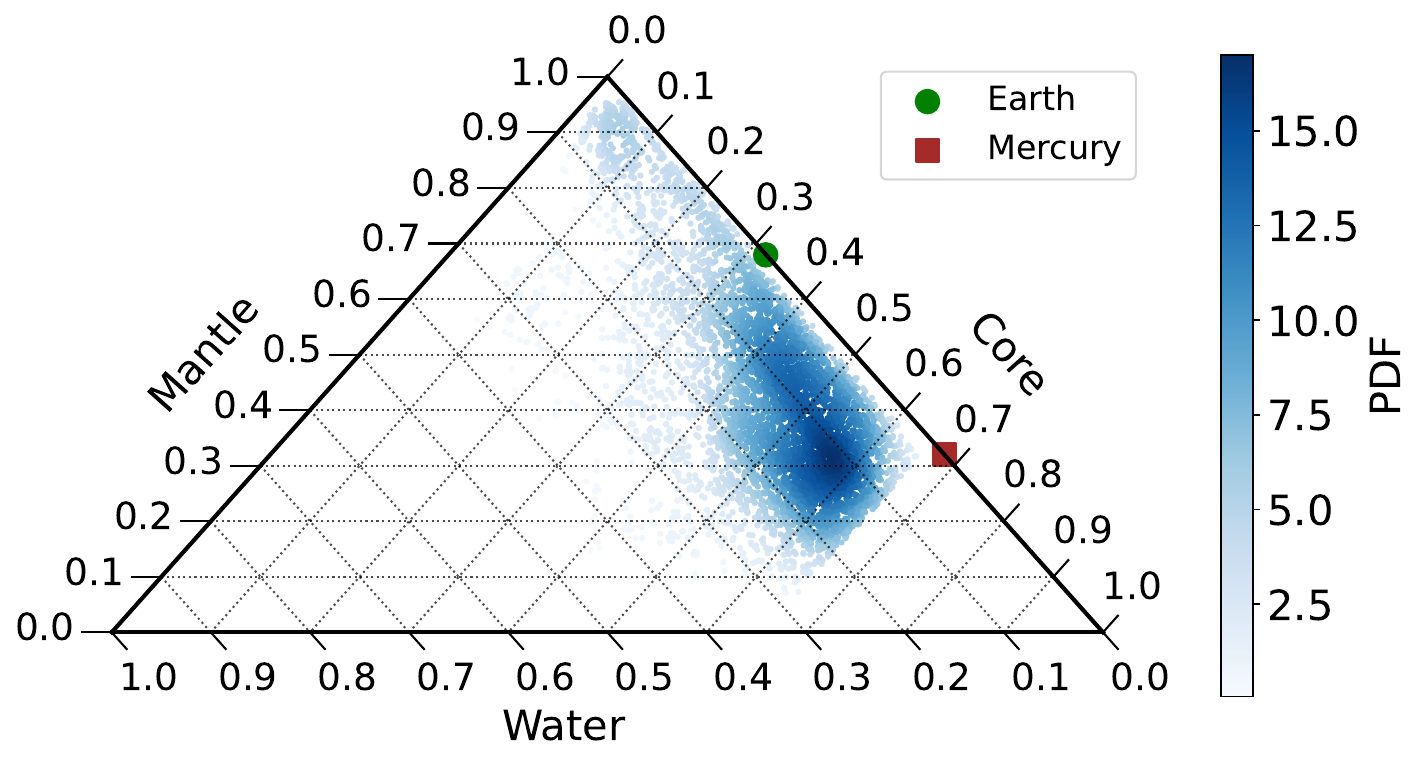}
    \caption[Ternary diagram of the sampled 2D marginal posterior distribution for the CMF and WMF of TOI-244 b in our interior structure retrieval.]{Ternary diagram of the sampled 2D marginal posterior distribution for the CMF and WMF of TOI-244 b in our interior structure retrieval. The colour code displays the probability density function (PDF).}
    \label{fig:ternary_interior}
\end{figure}

We performed a retrieval on the mass and radius data shown in Table \ref{tab:planet_params} with a 1D interior-atmosphere model to estimate TOI-244~b's compositional parameters. Our interior model is stratified in three layers: a Fe-rich core, a silicate-dominated mantle \citep{Brugger16,Brugger17}, and a water layer (hydrosphere). Given the irradiation conditions of TOI-244 b, volatiles such as water cannot condense out, and are therefore in gaseous and supercritical phases \citep{Mousis2020}. Our interior structure model takes into account the low density of these phases of water by using an equation of state adequate in this region of the water phase diagram. In addition, we establish the coupling interface between the interior and the atmosphere at 300 bar. The temperature at this pressure constitutes the boundary condition for our interior model, and it is calculated self-consistently by a k-correlated atmospheric model. Furthermore, the atmospheric thickness is calculated by the atmospheric model by integrating the hydrostatic equilibrium equation, and then added to the interior model's radius to obtain the planetary radius \citep{Acuna21}.

The two compositional parameters that are free in our MCMC retrieval are the core mass fraction (CMF) and water mass fraction (WMF). The mean and uncertainties of their 1$\sigma$ intervals are shown in Table \ref{tab:mcmc_interior}. Our observable parameters are the mass and the radius. The Fe/Si mole ratio is not considered as an observable, not only because of the difficulty of obtaining an accurate metallicity for GJ 1018, but also because the assumption that the composition of rocky planets reflects the stellar one
is being questioned \citep{2021Sci...374..330A}. Our CMF is compatible within uncertainties with that of the Earth, as well as with the derived CMF distribution of a sample of Super-Earths, between 0.1 and 0.5 \citep{Plotnykov20}. The 1$\sigma$ interval of the WMF of TOI-244 b spans from 0.04 to 0.20, being in the transition between super-Earths (WMF $<$ 0.05) and sub-Neptunes (WMF $>$ 0.20). Overall, our internal structure modelling suggests that TOI-244~b has a $479^{+128}_{-96}$ km thick hydrosphere over a 1.17~$\pm$~0.09~$\rm R_{\oplus}$ structure composed of a Fe-rich core and a silicate-dominated mantle compatible with that of the Earth.

\begin{table}
\centering
\renewcommand{\arraystretch}{1.3}
\setlength{\tabcolsep}{25pt}
\caption[Inferred internal composition of TOI~244~b.]{\label{compo} Composition of TOI~244 b assuming a Fe-rich core, a silicate-dominated mantle, and a water layer (hydrosphere). Errors are the 1$\sigma$ confidence intervals of the interior and atmosphere output parameters.}
\begin{tabular}{lc}\hline \hline
Parameters &  \\
\hline 
Core Mass fraction (CMF)  & 0.43 $\pm$ 0.16 \\ 
Water mass fraction (WMF) & 0.12 $\pm$ 0.08  \\
Temperature at 300 bar [K] &  2427$\pm$37 \\ 
Thickness at 300 bar [km] & 479$_{-96}^{+128}$  \\ 
Albedo & 0.28 $\pm$ 0.27 \\
Core + Mantle radius [$\rm R_{\oplus}$]  & 1.17 $\pm$ 0.09 \\
\hline 
\end{tabular}
\label{tab:mcmc_interior}
\end{table}

\subsection{Emerging trends in the growing population of low-density super-Earths}

\begin{table*}[]
\centering
\fontsize{10pt}{10pt}\selectfont
\caption[Main characteristics of low-density super-Earths.]{\label{tab:low_density} Normalized density to the density expected for a planet composed 100$\%$ of silicates \citep[$\rho$/$\rho_{rock}$;][]{2019PNAS..116.9723Z}, stellar host metallicity ([Fe/H]), insolation flux ($S$), deviation from the expected density of an Earth-like composition ($\sigma_{Earth}$), spectral type (SpT), and reference papers for all the known planets with $R_{\rm p}<2\,\rm R_{\oplus}$, $M_{p}<3.5\,\rm M_{\oplus}$, and a mass precision better than~30$\%$. Metallicity from (1) \texttt{SteParSyn} (2) \texttt{ODUSSEAS}. \textbf{References.} [1] \citet{2019AJ....158...32K}; [2] \citet{2019A&A...629A.111C}; [3] \citet{2021A&A...653A..41D}; [4] \citet{2022A&A...664A.199L}; [5] \citet{2023arXiv230106873P}; [6] \citet{2019A&A...622L...7T}; [7] \citet{2022A&A...665A.154B}; [8] \citet{2021MNRAS.501.4148L}; [9] \citet{2021AJ....161...56W}; [10] \citet{2022MNRAS.511.4551L}; [11] \citet{2023AJ....165...88B}; [12] \citet{2013ApJ...779..188M}; [13] \citet{2014ApJ...784...45R}; [14] \citet{2014ApJ...787...80H}; [15] \citet{2015Natur.522..321J}; [16] \citet{2016ApJ...822...86M}; [17] \citet{2017AJ....153..267M}; [18] \citet{2018MNRAS.478..460A}; [19] \citet{2018ApJ...866...99B}; [20] \citet{2023NatAs...7..206P}; [21] \citet{2012ApJ...750L..37M}; [22] \citet{2014ApJ...784...28K}; [23] \citet{2016ApJS..225....9H}; [24] \citet{2019RAA....19...41G}; [25] \citet{2020ApJ...890L...7S}; [26] \citet{2022ApJ...937L..17C}; [27] \citet{2014ApJS..210...25X}; [28] \citet{2015ApJ...807...45D}.}
\setlength{\tabcolsep}{7.3pt}
\renewcommand{\arraystretch}{1.5}
\begin{tabular}{lclcccc}
\hline \hline
Planet & $\rho$/$\rho_{rock}$ & \multicolumn{1}{c}{{[Fe/H] (dex)}} & $S$ ($\rm S_{\oplus}$) & $\sigma_{Earth} \, (\sigma)$ & SpT & Ref.               \\ \hline 
L 98-59 c       & 0.90 $\pm$ 0.20               & -0.46 $\pm$ 0.26                        & $12.8^{+2.6}_{-2.1}$               & 1.4                       & M3           & {[}1{]}-{[}3{]}             \\
L 98-59 d       & 0.61 $\pm$ 0.15               & -0.46 $\pm$ 0.26                        & $5.0^{+1.0}_{-0.8}$                & 3.3                       & M3           & {[}1{]}-{[}3{]}             \\
HD 260655 c     & 0.87 $\pm$ 0.16               & -0.43 $\pm$ 0.04                        & 16.1 $\pm$ 0.3                     & 1.9                       & M0           & {[}4{]}                     \\
TOI-4481 b      & 0.89 $\pm$ 0.12               & -0.28 $\pm$ 0.07                        & 130 $\pm$ 6                        & 3.3                       & M1.5         & {[}5{]}                     \\
HD 23472 c      & 0.52 $\pm$ 0.17               & -0.20 $\pm$ 0.05                        & $8.0^{+0.9}_{-0.8}$                & 3.5                       & K4           & {[}6{]}, {[}7{]}            \\
TOI-561 b       & 0.63 $\pm$ 0.15               & -0.40 $\pm$ 0.05                        & 4745 $\pm$ 269                     & 3.4                       & G9           & {[}8{]}-{[}11{]}            \\
Kepler-138 b    & 0.43 $\pm$ 0.08               & -0.18 $\pm$ 0.10                        & 9.9 $\pm$ 0.7                      & 8.0                       & M1           & {[}12{]}-{[}20{]}           \\
Kepler-138 c    & 0.71 $\pm$ 0.18               & -0.18 $\pm$ 0.10                        & 6.8 $\pm$ 0.5                      & 2.3                       & M1           & {[}12{]}, {[}14{]}-{[}24{]} \\
Kepler-138 d    & 0.66 $\pm$ 0.18               & -0.18 $\pm$ 0.10                        & 3.4 $\pm$ 0.2                      & 2.6                       & M1           & {[}12{]}, {[}15{]}-{[}23{]} \\
GJ 1252 b       & 0.93 $\pm$ 0.23               & \multicolumn{1}{c}{+0.10 $\pm$ 0.10}    & $233^{+48}_{-41}$                  & 1.0                       & M3           & {[}25{]}, {[}26{]}          \\
Kepler-114 c    & 0.71 $\pm$ 0.27               & -0.20 $\pm$ 0.10                        & 29 $\pm$ 7                         & 1.6                       & M0           & {[}27{]}, {[}28{]}          \\
TOI-244 b (1)      & 0.79 $\pm$ 0.19               & \multicolumn{1}{c}{-0.39 $\pm$ 0.07}                  & 7.3 $\pm$ 0.4                      & 1.8                       & M2.5         & This work   \\
TOI-244 b (2)      & 0.79 $\pm$ 0.19               & \multicolumn{1}{c}{-0.03 $\pm$ 0.11}                  & 7.3 $\pm$ 0.4                      & 1.8                       & M2.5         & This work   \\
\hline  
\end{tabular}
\end{table*}

We now discuss the possible existence of trends in the growing population of low-density super-Earths. In particular, we investigated the possibility of a density-metallicity and a density-insolation relation within the super-Earth population. Our sample consists of all known planets with $R_{\rm p}<2\,\rm R_{\oplus}$, $M_{p}<3.5\,\rm M_{\oplus}$, and a mass precision better than 30$\%$, making a total of 40 well-characterized super-Earths. The threshold in mass aims at discarding puffy sub-Neptunes above the 0.1$\%$~$\rm H_{2}$ theoretical model from \citet{2019PNAS..116.9723Z}. In Table \ref{tab:low_density}, we show the metallicities of the stellar hosts and the received insolation fluxes of the low-density planets in our sample; that is, those with a measured density lower than expected for a planet composed entirely of silicates (i.e. $\rho$/$\rho_{rock}$ $<$ 1).

\begin{figure*}
    \centering
    \includegraphics[width=0.5\textwidth]{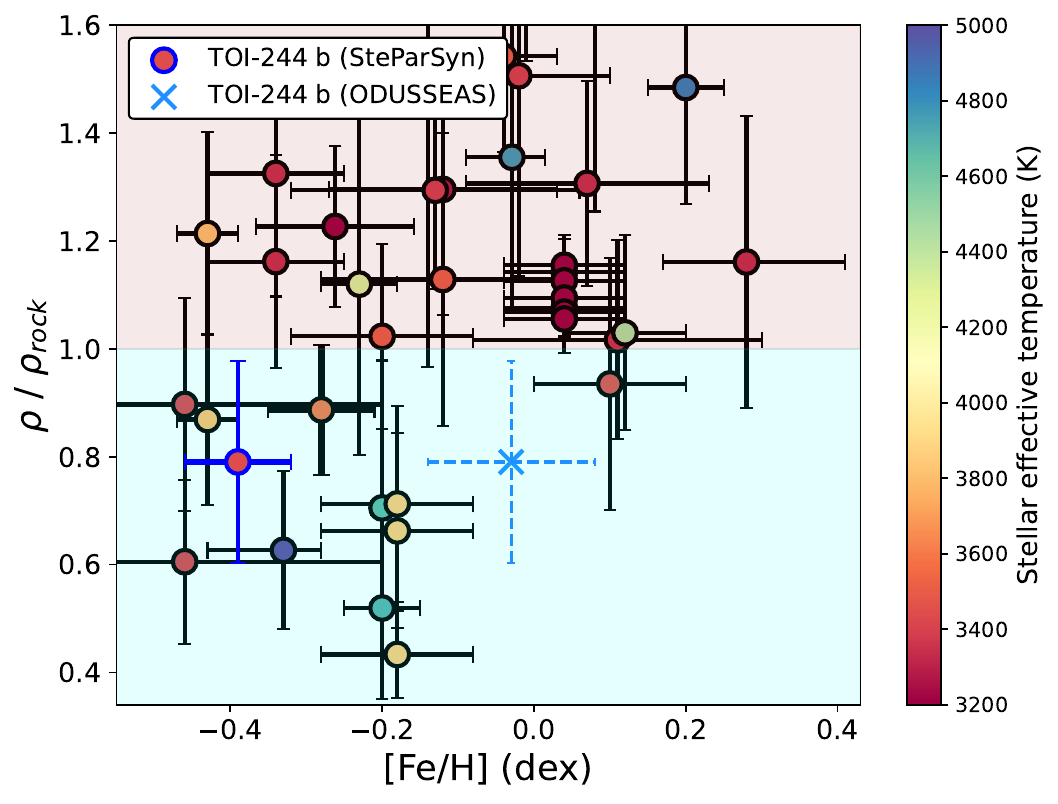}\includegraphics[width=0.5\textwidth]{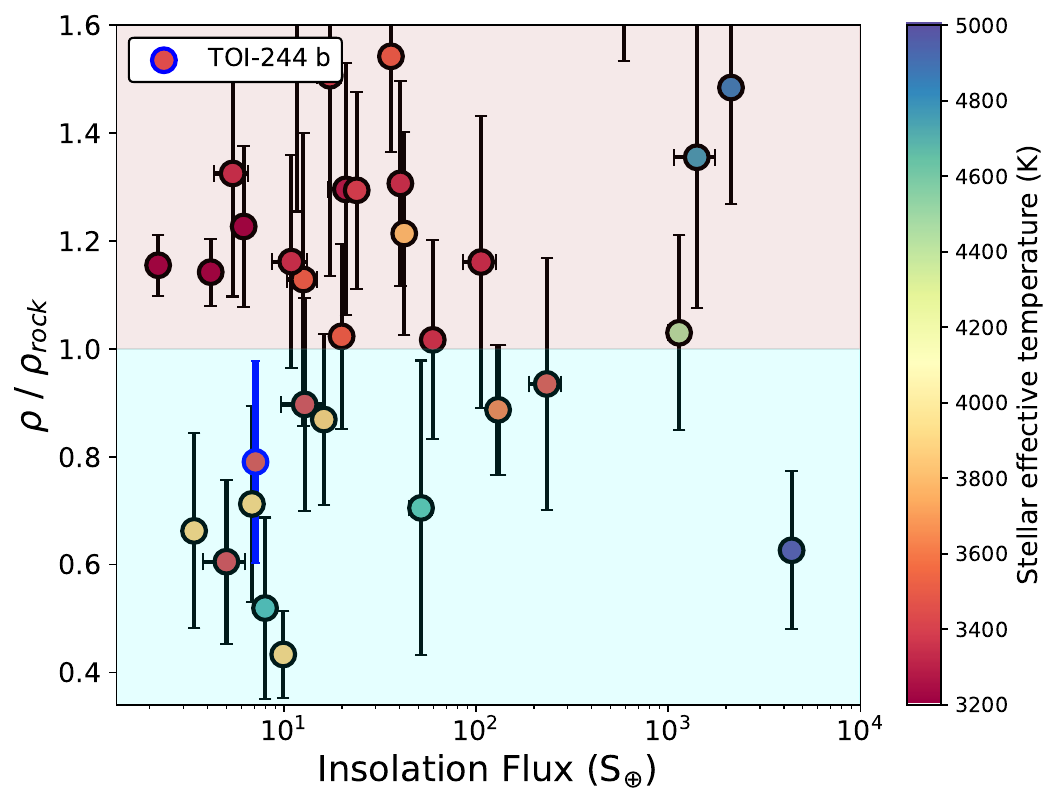}
    \caption[Planet density versus stellar metallicity and insolation flux for planets with $R_{\rm p}<2\,\rm R_{\oplus}$, $M_{p}<3.5\,\rm M_{\oplus}$, and mass precisions better than~30$\%$.]{Normalized density to the density expected for a planet composed 100$\%$ of silicates \citep{2019PNAS..116.9723Z} versus the stellar metallicity of the stellar hosts (left) and the stellar insolation flux (right) for all the confirmed planets with $R_{\rm p}<2\,\rm R_{\oplus}$, $M_{p}<3.5\,\rm M_{\oplus}$, and a mass precision better than~30$\%$. The colour coding indicates the effective temperature of the host stars.}
    \label{fig:trends}
\end{figure*}

\subsubsection{Density-metallicity}
\label{sec:density_metallicity}

All the stars hosting published low-density super-Earths are metal-poor ([Fe/H] between -0.20 and -0.45 dex) except GJ~1252, which has a slightly super-solar metallicity ([Fe/H] = +0.10 $\pm$ 0.10 dex). In Fig.~\ref{fig:trends}, we plot $\rho$/$\rho_{rock}$ versus [Fe/H] for all the planets in our sample and for TOI-244~b, considering our two estimated metallicities (see Sect.~\ref{sec:steparsyn}). We see that both Earth-like and low-density planets are found around metal-poor stars. However, there is a scarcity of low-density planets around stars with super-solar metallicities (corresponding to the lower-right parameter space in Fig.~\ref{fig:trends}). In order to quantify this possible trend, we carried out a Kolmogorov–Smirnov (KS) statistical test between the super-solar and subsolar populations, which allowed us to determine how likely it is that we would see both populations if they were drawn from the same probability distribution. We obtain a D statistic of $0.43\pm0.05$, which corresponds to a p-value of $0.0013^{+0.0055}_{-0.0011}$. This result allows us to confidently reject the null-hypothesis (i.e. both samples are drawn from the same distribution). Although statistically significant, we only propose this trend as a possible or tentative trend. This is because the sample of well-characterised rocky planets is still small, and most of those planets orbit M-dwarf stars, whose computed metallicities are less reliable than those of FGK stars.

According to several previous works, this possible trend is expected to exist. Metal-poor stars are known to have an enhancement of Mg and Si \citep[e.g.][]{2003A&A...410..527B}, and, as discussed in Sect.~\ref{sec:iron_free}, \citet{2021Sci...374..330A} found that stars with higher Mg/Fe and Si/Fe ratios host lighter planets with higher Mg/Fe and Si/Fe ratios. However, the correlation from \citet{2021Sci...374..330A}  would not explain the possible metallicity-density trend for planets with $\rho$/$\rho_{rock}$ $<$ 1, since the densities of those planets must be affected not only by the core and mantle composition, but also by the presence of volatile elements. Interestingly, the water mass fractions of protoplanetary disks forming metal-poor stars are known to be higher than the ones expected for stars with solar metallicities \citep{2017A&A...608A..94S}, so metal-poor stars are expected to host iron-poor and water-rich planets. Overall, both the observed densities and the previous knowledge of stellar composition point toward a possible compositional connection between the volatile content of rocky planets and their host stars. In order to try to understand why metal-poor stars host both low-density and Earth-like super-Earths, in the following, we study the possible influence of the insolation flux on the population of low-density super-Earths, which could hint at the capability of those planets to have retained their original abundant water reservoirs.

\subsubsection{Density-insolation}
\label{sec:density_insolation}
In Fig.~\ref{fig:trends} (right-hand panel), we plot $\rho$/$\rho_{rock}$ versus the stellar insolation flux ($S$) received by the planets in our sample. Most of the planets receive insolations lower than 300 $\rm S_{\oplus}$. In this region, we see that the density distribution of planets with $\rho$/$\rho_{rock}$~$>$~1 is homogeneous across different stellar insolations. This result is to be expected, given that the density of those planets is only determined by the iron and silicate content of their interiors. In contrast, for planets with  $\rho$/$\rho_{rock}$ $<$ 1, their low densities must be explained through the presence of a significant amount of volatile elements. Interestingly, the lowest density planets (i.e. 0.4 $<$ $\rho$/$\rho_{rock}$ $<$ 0.8) tend to be clustered in the low-insolation region of the diagram (i.e. $S$ $<$ 10 $\rm S_{\oplus}$), while the less dense planets near the rocky planet domain tend to receive higher insolations. Hence, this trend might indicate an influence of the stellar irradiation on the volatile content of super-Earths. 

In the highly irradiated region (i.e. 300-10\,000 $\rm S_{\oplus}$), we highlight the presence of the low-density planet TOI-561~b. With $\rho$/$\rho_{rock}$ = 0.63 $\pm$ 0.15, it might require volatile elements in its structure. This planet is an outlier in the density-insolation relation discussed above. With an insolation flux of $S$~=~4745~$\pm$~269~$\rm S_{\oplus}$, it is unlikely that it has managed to maintain an envelope. Hence, other hypotheses are needed to explain the volatile storage mechanism, such as magma oceans \citep[e.g.][]{2021ApJ...922L...4D}.  

\subsubsection{Synthesis}

In Sect. \ref{sec:density_metallicity} we have seen that low-density super-Earths tend to be hosted by metal-poor stars, which we argue is to be expected since those stars are already known to host planets with iron-poor cores \citep{2021Sci...374..330A} and are also known to have been formed in water-rich protoplanetary disks \citep{2017A&A...608A..94S}. However, there also exist planets with Earth-like densities (i.e. with a negligible water content in mass) around metal-poor stars, which were also presumably formed in water-rich environments. Interestingly, in Sect. \ref{sec:density_insolation} we found that the lowest-density planets (i.e. 0.4 $<$ $\rho$/$\rho_{rock}$ $<$ 0.8) tend to receive low insolation fluxes (i.e. $S$ $<$ 10 $\rm S_{\oplus}$). These two emerging trends combined indicate that low-irradiated super-Earths around metal-poor stars tend to have the lowest densities, and that such low densities are not typically reached when only one of those two factors is met (there is only the exception of TOI-561, which consists of a metal-poor star with a highly irradiated planet). On a compositional level, the aforementioned trend suggests that while many super-Earths were formed on iron-poor and water-rich protoplanetary disks, only the less irradiated ones could have been able to preserve a significant amount of water from their original reservoirs.

\section{Summary and conclusions}
\label{conclusions}

We carried out an intensive radial velocity campaign with ESPRESSO in order to confirm and precisely characterize the close-in ($P$ = 7.4 days) transiting super-Earth-sized planet candidate TOI-244.01, which had been detected by TESS orbiting the bright ($K$ = 7.97 mag) and nearby ($d$ = 22 pc) star GJ 1018. The powerful combination of ESPRESSO and TESS data allowed us to determine a very precise planetary radius and mass (8$\%$ and 11$\%$ relative uncertainties, respectively), thus confirming the planetary nature of TOI-244~b. With a radius of $R_{\rm p}$~=~1.52~$\pm$~0.12~$\rm R_{\oplus}$, TOI-244~b is located amidst the lower mode of the bimodal radius distribution of small planets, where planets are considered to be composed of iron and silicates in a proportion similar to that of the Earth (33$\%$ Fe, and 67$\%$ $\rm MgSiO_{3}$ in mass). However, with a measured dynamical mass of $M_{\rm p}$~=~2.66~$\pm$~0.31~$\rm M_{\oplus}$, TOI-244~b is less dense than the majority of super-Earths of its size ($\rho$ = 4.2 $\pm$ 1.1 $\rm g \, cm^{-3}$). 

We investigated the possibility that a scarcity of iron in the core of TOI-244~b could contribute to the planet's low density. To do so, we estimated the iron-to-silicate mass fraction that would have been present in the original protoplanetary disk ($f_{\rm iron}^{\rm star}$) based on our computed stellar metallicity and estimated [Si/H] and [Mg/H] abundances. Unfortunately, depending on the technique used, we found a strong discrepancy between different metallicity computations for this M-dwarf (-0.39 $\pm$ 0.07 dex and -0.03 $\pm$ 0.11 dex). We found $f_{\rm iron}^{\rm star}$ = 29.2 $\pm$ 4.2 $\%$ and $f_{\rm iron}^{\rm star}$ = 32.0 $\pm$ 6.5 $\%$ for the metal-poor and solar metallicity scenarios, respectively, suggesting that TOI-244~b could have a lower amount of iron in its core than the Earth. However, according to the location of TOI-244~b in the mass-radius diagram, it is more likely that its low density is related to the presence of a significant amount of volatile elements. In that sense, we found that atmospheric loss processes may have been very efficient in removing a potential primordial hydrogen envelope, but high mean molecular weight volatiles such as water could have been retained. Hence, we chose our internal structure model to be composed of an iron-rich core, a silicate-dominated mantle, and a hydrosphere. We ran our model without constraining the Fe/Si mode ratio, not only because of the difficulty of obtaining an accurate metallicity for GJ 1018, but also because the assumption that the composition of rocky planets reflects the stellar one is being questioned. As a result, our internal structure modelling suggests that TOI-244~b has a $479^{+128}_{-96}$ km thick hydrosphere over a 1.17~$\pm$~0.09~$\rm R_{\oplus}$ structure composed of a Fe-rich core and a silicate-dominated mantle compatible with that of the Earth.

On a population level, we found that the current population of well-characterised low-density super-Earths tend to orbit stars with subsolar metallicities, and that the lowest density planets tend to receive lower insolation fluxes than the less dense ones in the $\rho$/$\rho_{rock}$ $<$ 1 regime. However, these possible trends need to be confirmed in the future with a larger sample of well-characterised systems.

Overall, the detection and precise characterisation of the super-Earth TOI-244~b, thanks to ESPRESSO and TESS data, allowed us to make a first detailed description of its composition. Given its unusual properties, the brightness of the host star, and the likely presence of an extended atmosphere of high mean molecular weight volatiles, TOI-244~b will probably become a key target for atmospheric studies. 
\newpage
\chapter{
Signs of magnetic star-planet interactions in HD 118203}
\label{ch:mspis}
\vspace{2cm}
\pagestyle{fancy}
\fancyhf{}
\lhead[\small{\textbf{\thepage}}]{\small{\textbf{\nouppercase{\leftmark}}}}
\rhead[\small{\textbf{\nouppercase{\rightmark}}}]{\small{\textbf{\thepage}}}

\bigskip

To understand planetary magnetic fields, it is important to study under what conditions they appear. In the Solar System, there is a strong correlation between the magnetic moment of a body and the ratio between its mass and its rotation period \citep[e.g.][]{2002Icar..157..507K}. Similarly, \citet{2003ApJ...597.1092S,2005ApJ...622.1075S,2008ApJ...676..628S} found that the stellar variability induced by magnetic star-planet interactions (MSPIs) is also correlated with such a ratio, which suggests that MSPIs are dominated by the planetary magnetic moment. This evidence is supported by several formalisms \citep[e.g.][]{2009A&A...505..339L,2012A&A...544A..23L,2013A&A...552A.119S}. 

The rotation periods of exoplanets are difficult to measure. However, they can be confidently estimated for hot Jupiters since those planets undergo strong and well-known tidal forces that affect both their orbits and rotation velocities. In particular, their orbits tend to circularise and their rotation periods tend to synchronise with the planetary orbital period in very short time scales \citep[i.e. of a few million years;][]{1981A&A....99..126H,2018AJ....155..118W}. To date, all MSPIs have been found in circular planetary systems, which allowed us to assume that the orbital and rotation periods coincide. However, there are a few exceptions in which hot Jupiters managed to preserve an eccentric orbit because of particular conditions, such as planet migration or gravitational interactions induced by an outer companion. In such eccentric systems, the planetary rotation periods are not expected to be synchronised, but rather pseudo-synchronised with the orbital periods \citep{1981A&A....99..126H,2011CeMDA.111..105C}. This pseudo-synchronisation depends on the planet's eccentricity and translates into higher rotation velocities that would generate larger magnetic planetary moments than those in circular systems. Interestingly, eccentric planetary systems also offer a unique opportunity to probe the possible imprint of the orbital geometry into the planet-induced activity signals. Unfortunately, the previous attempts to detect MSPIs in eccentric systems have turned out to be unsuccessful \citep[e.g.][]{2014IAUS..299..291H,2016A&A...592A.143F}. To date, there is only one case of periodic activity \citep{2015ApJ...803...49Q} and one case of sudden activity enhancement \citep{2015ApJ...811L...2M} in eccentric systems that could be explained by MSPIs.

In this chapter, we aim to search for MSPIs in the bright ($V$ = 8.05 $\pm$ 0.03 mag) and slightly evolved star HD~118203, which is known to host the eccentric ($e$ = 0.32 $\pm$ 0.02) and close-in ($a$ = 0.0864 $\pm$ 0.0006~au) transiting Jupiter-sized planet HD~118203~b \citep{2006A&A...446..717D,2020AJ....159..243P}. In Sect.~\ref{sec:observations}, we describe the observations. In Sect.~\ref{sec:analysis_results}, we present our analysis of photometric and spectroscopic time series. In Sect.~\ref{sec:discussion}, we discuss the results found, and we conclude in Sect.~\ref{sec:conclusions}.


\section{Observations}
\label{sec:observations}

\subsection{TESS photometry}
\label{sec:obs_tess}

\begin{figure}
    \centering
    \includegraphics[width = 0.6\textwidth]{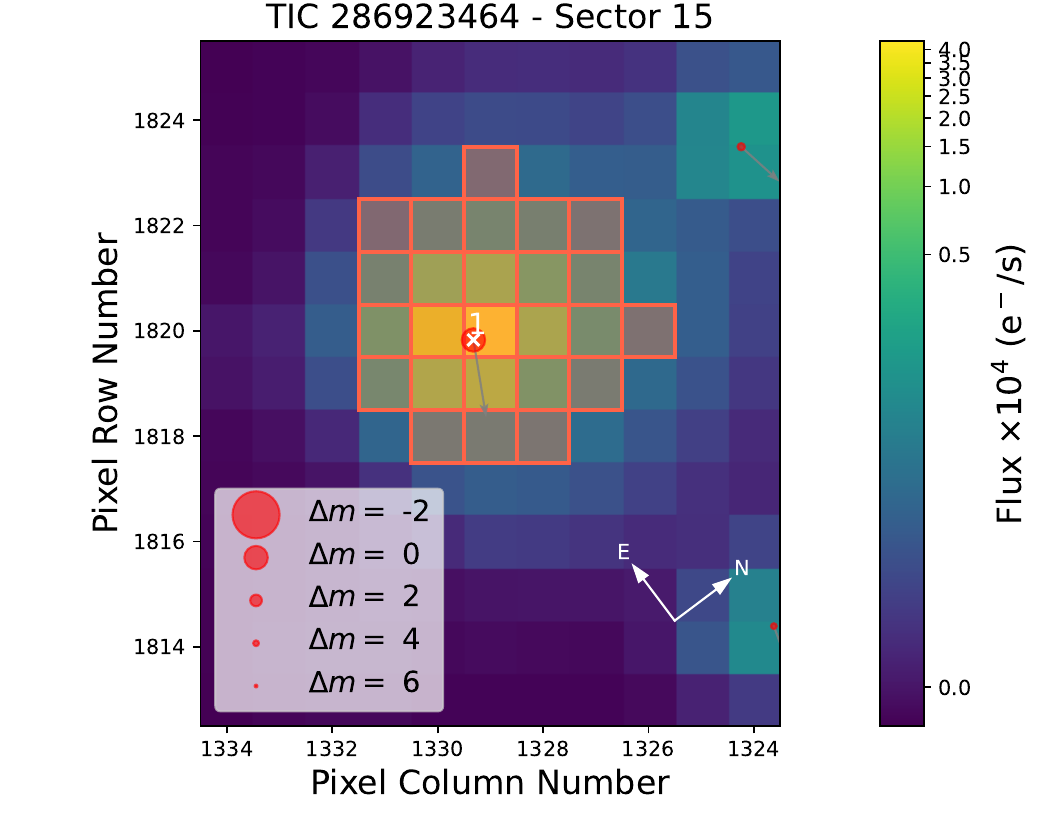}
    \caption[Target pixel file of HD 118203.]{TESS target pixel file of HD 118203. The orange grid is the selected aperture, and the red circles correspond to the nearby \textit{Gaia} DR3 sources scaled to their $G$ magnitudes.}
    \label{fig:tpfplotter}
\end{figure}

The star HD 118203 (TOI-1271, TIC 286923464) has been observed by the Transiting Exoplanet Survey Satellite \citep[TESS;][]{2015JATIS...1a4003R} in sectors 15, 16, 22, and 49 at a 2-min cadence, resulting in a 3-month duty cycle obtained throughout a 2.5-year baseline. In total, 61556 target pixel files (TPFs) were acquired, of which 9502 were discarded as having a non-zero quality flag (QF). In Table \ref{tab:TESS_summary} we summarise the details of the observations. We downloaded the pre-search data conditioned simple aperture photometry \citep[PDCSAP;][]{2012PASP..124.1000S,2012PASP..124..985S,2014PASP..126..100S} from the Mikulski Archive for Space Telescopes (MAST)\footnote{\url{https://mast.stsci.edu/portal/Mashup/Clients/Mast/Portal.html}}, which was processed by the Science Processing
Operation Center (SPOC) pipeline \citep{2016SPIE.9913E..3EJ}. The SPOC module Create Optimal Aperture (COA) selected the photometric aperture that maximised the signal-to-noise of the measured flux, and then estimated the flux fraction inside the aperture that came from the target star (99.96$\%$) in order to correct for possible photometric contamination. In Figure \ref{fig:tpfplotter}, we used \texttt{tpfplotter}\footnote{\url{https://github.com/jlillo/tpfplotter}} \citep{2020A&A...635A.128A} to plot the selected aperture over a TPF of HD~118203, together with all the nearby stars detected in the \textit{Gaia} Data Release 3 \citep[DR3;][]{2022arXiv220800211G}. There are no additional sources within the aperture with a magnitude difference $\Delta$G $<$ 6 mag with HD 118203 in the \textit{Gaia} passband, which implies a negligible contamination \citep[e.g.][]{2022MNRAS.509.1075C}. In October 2019, the Transit Planet Search (TPS) algorithm of the SPOC pipeline detected a periodic flux decrease of 3.86 ppt (parts per thousand) every 6.13 days, which passed all the diagnostic tests described in \citet{Twicken:DVdiagnostics2018}. Both the orbital period and mid-transit time coincide with that of the RV-detected planet HD 118203~b, which confirms its transiting nature. The out-of-transit photometry of this target has a standard deviation of 0.46 ppt, which makes it an ideal target for searching for low-amplitude planet-induced activity signals. 

\begin{table*}[]
\centering
\fontsize{10.2pt}{10.2pt}\selectfont
\caption[Summary of the TESS observations of HD 118203.]{Summary of the TESS observations of HD 118203. (1) NZQF stands for the number of TESS observations with a non-zero quality flag.}
\renewcommand{\arraystretch}{1.4}
\setlength{\tabcolsep}{11pt}
\begin{tabular}{ccccccc}
\hline \hline
Sector & Start date        & End date        & TPFs  & Camera & CCD & $\rm NZQF^{(1)}$ \\ \hline
15     & 15 August 2019    & 9 September 2019 & 15256 & 4      & 4   & 2493   \\
16     & 12 September 2019 & 5 October 2019   & 14678 & 4      & 3   & 2041   \\
22     & 20 February 2020  & 17 March 2020    & 17815 & 3      & 2   & 804    \\
49     & 1 March 2022      & 25 March 2022    & 13807 & 3      & 1   & 4164   \\ \hline
\end{tabular}
\label{tab:TESS_summary}
\end{table*}

\subsection{ELODIE spectroscopy}
\label{sec:obs_elodie}

HD 118203 was observed by the ELODIE spectrograph \citep{1996A&AS..119..373B}, which was in operation between 1993 and 2006 at the 1.93~m telescope located at the Observatoire de Haute-Provence, France. ELODIE was a fibre-fed echelle spectrograph placed in a temperature-controlled room, with a resolution power of 42000 and a wavelength range from 389.5 to 681.5 nm split into 67 spectral orders. A total of 43 spectra were acquired between May 2004 and July 2005 with a typical cadence of 1 day and a typical exposure time of 1200 s. These observations were analysed in the discovery paper \citep{2006A&A...446..717D}. In addition, 13 more spectra were acquired right after the publication of \citealp{2006A&A...446..717D} (between March and June 2006). In this chapter, we analyse all the 56 ELODIE spectra, which have a median signal-to-noise ratio of 43.5 at 555 nm, and are publicly available on the Data $\&$ Analysis Center for Exoplanets (DACE)\footnote{\url{https://dace.unige.ch/dashboard/}}. The spectra were reduced through the ELODIE Data Reduction Software (TACOS), which also extracted the RVs and activity indicators such as the full width at half maximum (FWHM) and the contrast of the cross-correlation function (CCF) through the cross-correlation technique \citep{1996A&AS..119..373B}. We used these indicators to search for planet-induced and rotation-related activity signals. 

\subsection{ASAS-SN photometry}
\label{sec:asas_sn}

HD 118203 has been monitored by four stations of the All-Sky Survey for Supernovae \citep[ASAS-SN;][]{2014ApJ...788...48S,2017PASP..129j4502K}. Each station consists of four Nikon telephoto lenses of 14 cm aperture, which are equipped with a 2048 $\times$ 2048 pixels CCD camera with a pixel scale of 8.0 arc seconds. The survey pipeline performs aperture photometry through \texttt{IRAF} \citep{1986SPIE..627..733T}, considering a 2-pixel radius aperture and 7-10 pixel radius annulus for the target star and for the reference stars, which are selected from the Photometric All-Sky Survey \citep[APASS,][]{2012JAVSO..40..430H,2019JAVSO..47..130H}.

We used the ASAS-SN Sky Patrol web interface\footnote{\url{https://asas-sn.osu.edu/}} to compute and download the light curves of HD 118203. This target is a high proper motion star ($\mu_{\alpha}$ = $-85.88 \pm 0.05$ $\rm mas\,yr^{-1}$, $\mu_{\delta}$ = $-78.91 \pm 0.04$ $\rm mas\,yr^{-1}$; \citealp{2022arXiv220800211G}). Hence, given that the ASAS-SN photometry is extracted in a fixed location of the detector, and the typical FWHMs are comparable to the radius of the aperture, we shifted the aperture location over time in order to minimise possible flux losses. HD 118203 is visible from the four stations from 15 November to 1 September. Hence, similarly to \citet{2023A&A...675A..52C}, we computed the photometry of HD 118203 considering the coordinates corrected for proper motion corresponding to the central time of each observing window. Finally, we discarded those epochs in which the flux is below the estimated 5$\sigma$ detection limit for the target location, as well as those data points with a deviation greater than 5$\sigma$ of a flattened version of the photometric time series, which could be caused by flares, cosmic rays, or uncorrected instrumental systematics. The median standard deviation of the final ASAS-SN photometry of HD 118203 is 15.5 ppt.


\section{Analysis and results}
\label{sec:analysis_results}

\subsection{Stellar characterization}

\begin{figure}
    \centering
    \includegraphics[width = 0.6\textwidth]{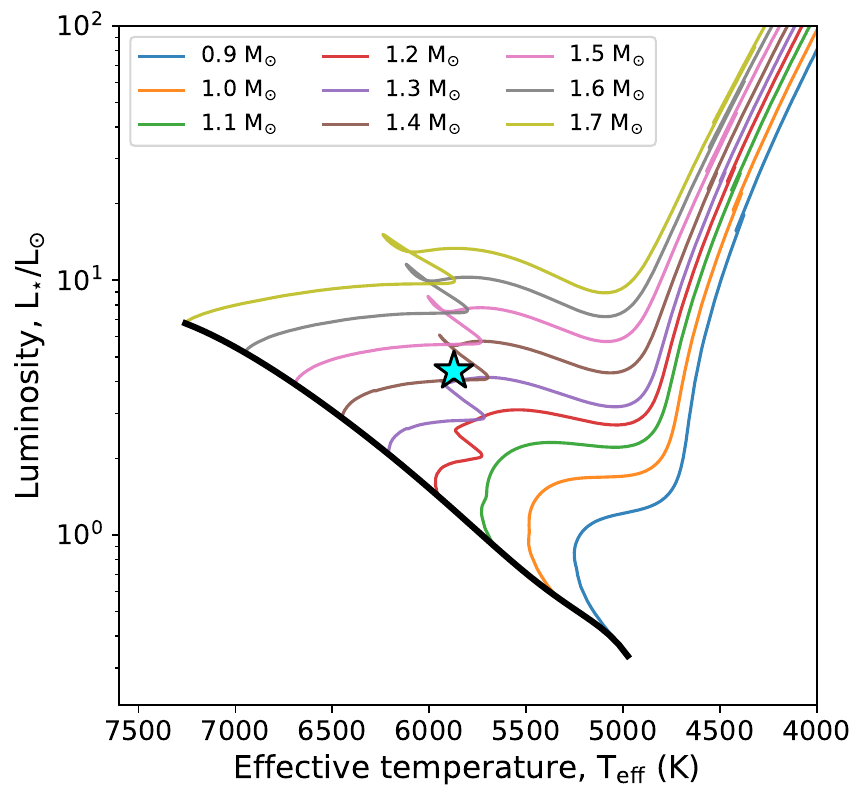}
    \caption[HD~118203 in the Hertzsprung–Russell diagram.]{HD~118203 in the Hertzsprung–Russell diagram. The black line is the main sequence, and the coloured lines are the PARSEC 2.1s tracks from \citet{2012MNRAS.427..127B,2013EPJWC..4303001B} for stars with Z = 0.04 ([M/H] $\approx$ +0.3). }
    \label{fig:HR}
\end{figure}

Given its brightness \citep[V = 8.05 $\pm$ 0.03 mag;][]{2019JAVSO..47..130H} and the early discovery of an orbiting planet \citep{2006A&A...446..717D}, HD~118203 has been extensively studied by several groups which were mainly focussed on determining elemental abundances of planet-hosting stars \citep[e.g.][]{2011ApJ...738...97B,2011A&A...530A..54G,2013A&A...556A.150S,2013A&A...554A..84M,2015A&A...576A..94S,2015A&A...576A..69D,2016A&A...588A..98M,2017AJ....153...21L,2018A&A...612A..93M}. While a comprehensive stellar characterization is out of the scope of this work, we need to adopt a stellar mass, radius, and effective temperature to determine the orbital and physical properties of HD~118203~b. We chose the spectroscopic values from the catalogue of Stars With ExoplanETs \citep[SWEET-Cat;][]{2021A&A...656A..53S}. In Table \ref{tab:stellar_params} we summarise the main properties of the star. The effective temperature and log $g$ indicate that the star is slightly evolved. In Fig.~\ref{fig:HR} we place  HD~118203 within the Hertzsprung–Russell diagram, which illustrates the star ascending the subgiant branch.

\begin{table}[]
\centering
\caption[Stellar properties of HD 118203.]{Properties of HD 118203. [1] \citet{1918AnHar..91....1C}; [2] \citet{2021ApJS..254...39G}; [3] \citet{2019AJ....158..138S}; [4] \citet{skrutskie2006}; [5] \citet{2022arXiv220800211G}; [6] \citet{2021A&A...656A..53S}; [7] \citet{2019JAVSO..47..130H}.}
\label{tab:stellar_params}
\renewcommand{\arraystretch}{1.5}
\setlength{\tabcolsep}{7pt}
\begin{tabular}{llc}
\hline \hline
Parameter & Value & Reference \\ \hline
\multicolumn{3}{l}{Identifiers} \\ \hline
HD & 118203 & [1]  \\
TOI & 1271 & [2] \\
TIC & 286923464 & [3] \\
2MASS & J13340254+5343426 & [4] \\
Gaia DR3 & 1560420854826284928 & [5] \\ \hline
\multicolumn{3}{l}{Coordinates, parallax and kinematics} \\ \hline
RA, DEC & 13:34:02.39, 53:43:41.48 & [5] \\
$\rm \mu_{\alpha}$ ($\rm mas\,yr^{-1}$) & -85.877 $\pm$ 0.052 & [5] \\
$\rm \mu_{\delta}$ ($\rm mas\,yr^{-1}$) & -78.913 $\pm$ 0.038 & [5] \\
Parallax (mas) & 10.864 $\pm$ 0.018 & [5] \\
Distance (pc) & 92.04 $\pm$ 0.15 & [5] \\
RV ($\rm km\,s^{-1}$) & -29.37 $\pm$ 0.13 & [5] \\ \hline
\multicolumn{3}{l}{Atmospheric parameters} \\ \hline
$T_{\rm eff}$ (K) & 5872 $\pm$ 20 & [6] \\
log $g$ (dex) & 4.05 $\pm$ 0.04 & [6] \\
$\rm [Fe/H]$ (dex) & 0.27 $\pm$ 0.02 & [6] \\
\hline
Physical parameters &  &  \\ \hline
$\rm R_{\star}$ ($\rm R_{\odot}$) & 1.993 $\pm$ 0.065 & [6] \\
$\rm M_{\star}$  ($\rm M_{\odot}$) & 1.353 $\pm$ 0.006 & [6] \\
L ($\rm log_{10}$ $\rm L_{\odot}$) & $0.6458\pm 0.0018$ & [5] \\ \hline
Magnitudes &  &  \\ \hline
TESS (mag) & 7.4556 $\pm$ 0.0060 & [3] \\
G (mag) & 7.89255 $\pm$ 0.00031 & [5] \\
B (mag) & 8.746 $\pm$ 0.025 & [7] \\
V (mag) & 8.050  $\pm$ 0.030 & [7] \\
J (mag) & 6.861 $\pm$ 0.021 & [4] \\
H (mag) & 6.608 $\pm$ 0.038 & [4] \\
$\rm K_{s}$ (mag) & 6.543 $\pm$ 0.023 & [4] \\ 
\hline
\end{tabular}
\end{table}

\subsection{Joint radial velocity and transit model}
\label{sec:joint_fit}

\begin{figure*}
    \centering
    \includegraphics[width = \textwidth]{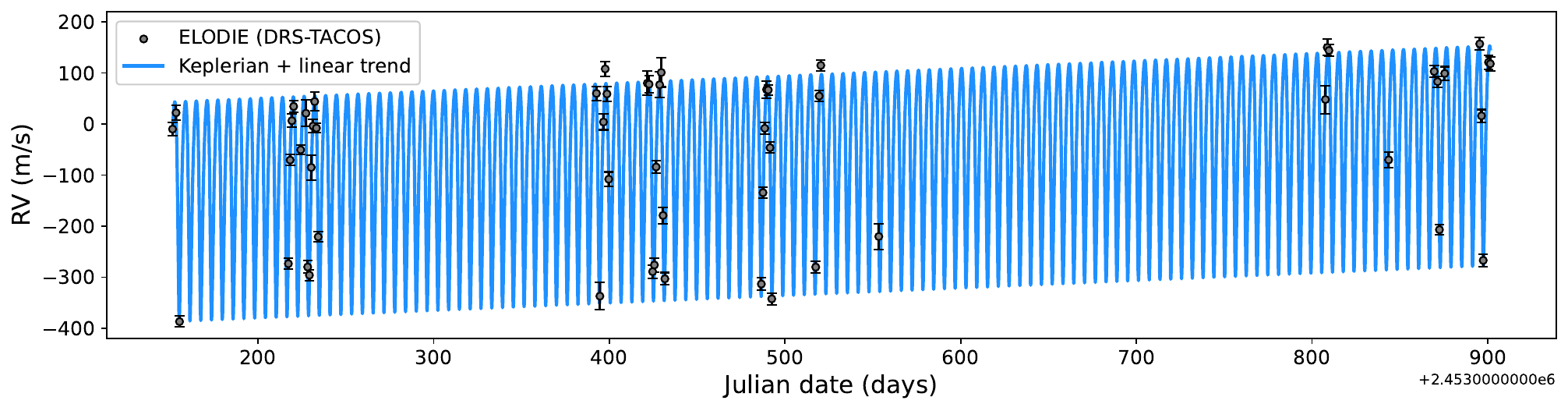}

    \vspace{0.5cm}

    \includegraphics[width = \textwidth]{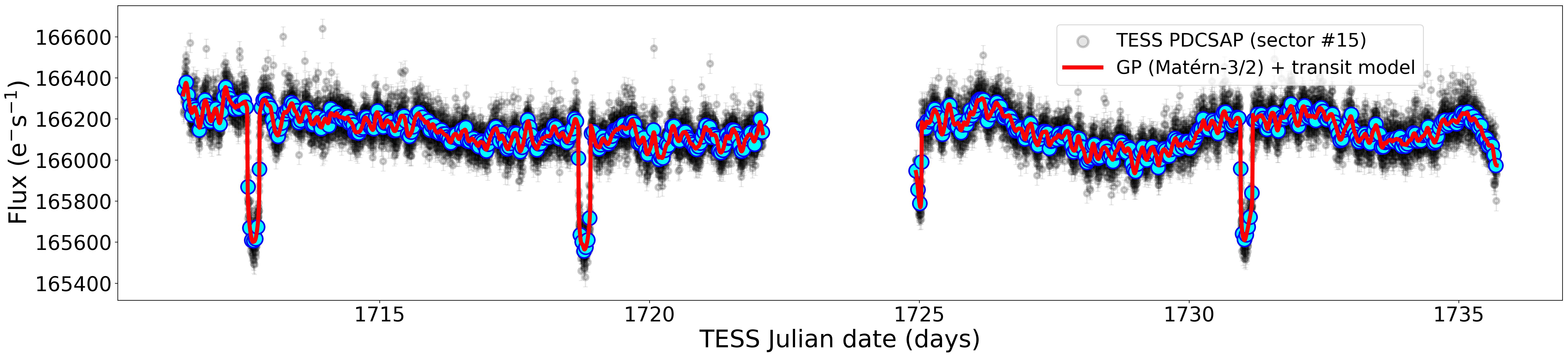}

     \vspace{0.5cm}
     
    \includegraphics[width = \textwidth]{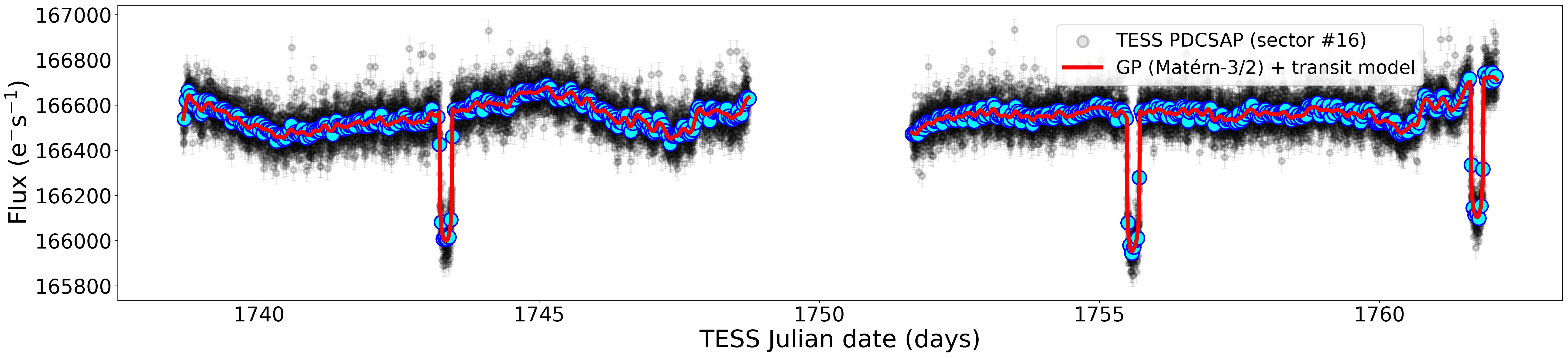}

    \vspace{0.5cm}
    
    \includegraphics[width = \textwidth]{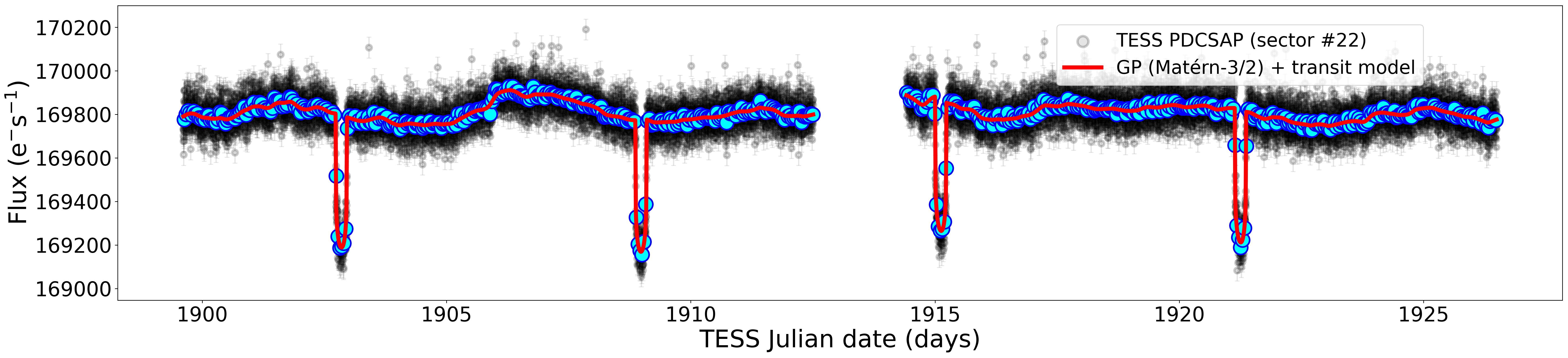}
    
    \vspace{0.5cm}
    
    \includegraphics[width =\textwidth]{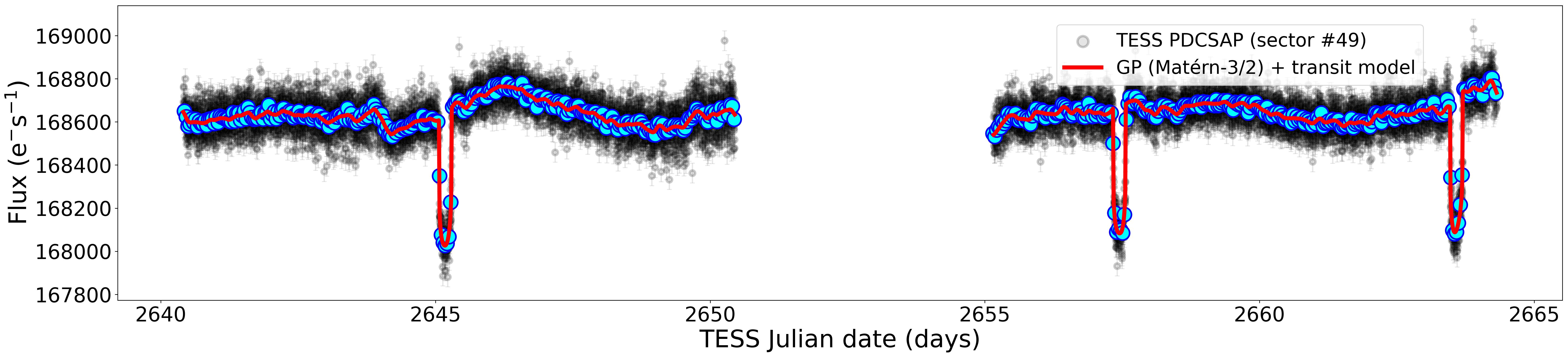}

    \vspace{0.3cm}

     \caption[ELODIE RVs and TESS photometry of HD 118203 with the inferred models.]{Top panel: ELODIE RVs of HD 118203 with the median posterior model (Keplerian + linear trend) superimposed. Lower panels: TESS photometry of HD 118203 with the median posterior model (transit + GP) superimposed. The grey data points correspond to the SPOC 2-minute PDCSAP photometry, and the blue data points correspond to 50-minute binned data.}
     \label{fig:TESS_complete}
\end{figure*}

We inferred the planetary and orbital parameters of HD~118203~b by modelling jointly the TESS photometry and ELODIE RVs described in Sects. \ref{sec:obs_tess} and \ref{sec:obs_elodie}, respectively. The RVs are not significantly influenced by the stellar activity, but the TESS photometry shows a clear modulation in all the observed sectors (see Sects. \ref{sec:var_tess} and \ref{sec:variability_ELODIE} for a detailed analysis of the variability of both datasets). Hence, we modelled the ELODIE RVs with a simple Keplerian, and the TESS photometry with a model composed of a transit model and a Gaussian Process \citep[GP;][]{2006gpml.book.....R,2012RSPTA.37110550R}. Modelling planetary transits together with correlated photometric noise (either of stellar or instrumental origin) has been shown to be the most optimal procedure to preserve the transit shapes \citep[e.g][]{2021A&A...645A..41L,2023A&A...669A.109L,2023A&A...679A..33D} and to properly propagate the uncertainties of the model parameters \citep{2021A&A...649A..26L}. In the following, we describe each model as well as the fitting procedure.

We implemented the Keplerian model based on the \texttt{radvel} package \citep{2018PASP..130d4504F}. We considered the parametrization $\left\lbrace P_{\rm orb}, T_{0}, K, \sqrt{e} \cos(w), \sqrt{e} \sin(w)\right\rbrace$ recommended by \citet{2013PASP..125...83E}, where $P_{\rm orb}$ is the orbital period of the planet, $T_{0}$ the time of inferior conjunction, $K$ the semi-amplitude, $e$ the orbital eccentricity, and $w$ the planetary argument of the periastron. We included a linear trend component in order to account for a possible long-period drift. This component is described by the systemic radial velocity of the star ($v_{\rm sys}$) and a slope ($\gamma$). Finally, in order to model the white noise not taken into account in our model, we included a jitter term ($\sigma_{\rm ELODIE}$) that we added quadratically to the uncertainties of the RV measurements. We implemented the \citet{2002ApJ...580L.171M} quadratic limb darkened transit model through \texttt{batman} \citep{2015PASP..127.1161K}, which is described by the $P_{\rm orb}$, $T_{0}$, $e$, $w$, the orbital inclination ($i$), the quadratic limb darkening (LD) coefficients $u_{1}$ and  $u_{2}$, the planet-to-star radius ratio ($R_{\rm p}/R_{\star}$), and the semimajor axis scaled to the stellar radius, which we parametrized through $P_{\rm orb}$ and the stellar mass ($M_{\rm \star}$) and radius ($R_{\rm \star}$) following the Kepler's Third Law. We implemented a GP defined by an approximate Matérn-3/2 kernel through \texttt{celerite} \citep{2017AJ....154..220F,2018RNAAS...2...31F}, which has been extensively used to model TESS photometry given its simplicity and flexibility. This GP kernel can be written as:

\begin{equation}
    K_{3/2} = \eta_{\sigma}^{2} \left[ \left(1 + \frac{1}{\epsilon} \right) e^{-(1-\epsilon) \sqrt{3} \tau / \eta_{\rho}} \cdot \left(1 - \frac{1}{\epsilon} \right) e^{-(1+\epsilon) \sqrt{3} \tau / \eta_{\rho}} \right],
\end{equation}

\noindent where the hyperparameters $\eta_{\sigma}$ and $\eta_{\rho}$ are the characteristic amplitude and timescale of the correlated variations, respectively, and $\epsilon$ controls the approximation to the exact Matérn-3/2 kernel. Since both the amplitude and timescale of the TESS variability can vary from one sector to another \citep[e.g.][]{2023A&A...677A.182M},  we fitted those parameters independently; that is, we fitted $\eta_{\rm \sigma_{i}}$ and $\eta_{\rm \rho_{i}}$, where \textit{i} indicates the sector. We also included an offset ($F_{\rm 0,i}$) and a jitter term ($\sigma_{\rm TESS,i}$) for each sector in order to model the white noise not taken into account in our model.

\begin{figure*}
    \centering
    \includegraphics[width = 0.45\textwidth]{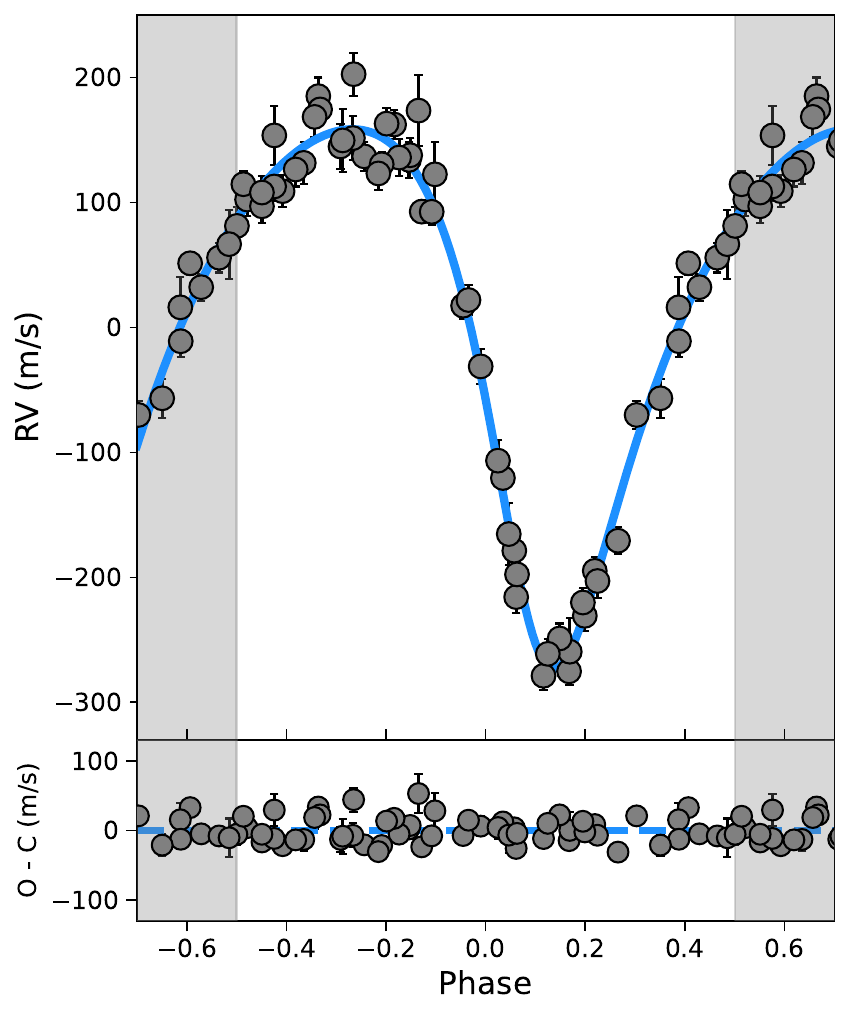}
    \includegraphics[width = 0.463\textwidth]{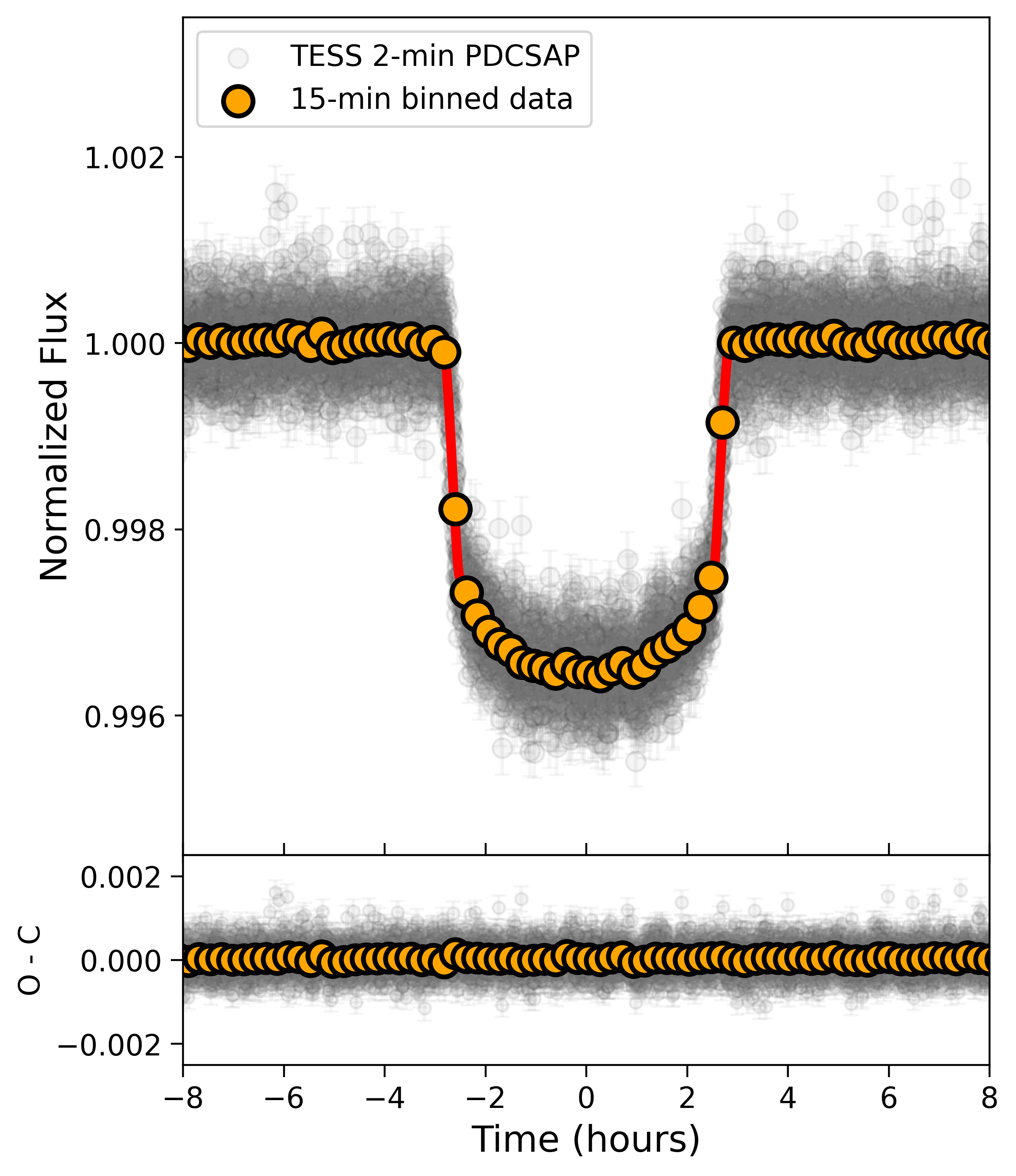}
    \caption[ELODIE RVs and TESS photometry folded to the orbital period of HD~118203~b.]{ELODIE RVs (left) and TESS photometry (right) folded to the orbital period of HD~118203~b. ELODIE data were subtracted from the linear trend, and TESS data were subtracted from the GP component.}
    \label{fig:folded_rvs_and_phot}
\end{figure*}

We sampled the posterior probability density function of the different parameters involved in our global model by using a Markov Chain Monte Carlo (MCMC) affine-invariant ensemble sampler \citep{2010CAMCS...5...65G} as implemented in \texttt{emcee} \citep{2013PASP..125..306F}. We used 240 walkers (eight times as many as the number of parameters), and performed two consecutive runs. The first run (or burn-in) consisted of 200\,000 iterations. After this run, we reset the sampler and initialised the second run (or production) with 100\,000 iterations while considering the initial values from the last iteration of the burn-in phase. We finally estimated the autocorrelation time for each parameter and checked that it is at least 50 times smaller than the chain length, which indicates that we collected thousands of independent samples after discarding the burn-in phase. In Table~\ref{tab:bestfit}, we include the priors used in the MCMC run, together with the median and 1$\sigma$ credible intervals of the posterior distributions of the fitted parameters. In Fig.~\ref{fig:TESS_complete}, we show the ELODIE and TESS datasets together with the median posterior global model. In Fig. \ref{fig:folded_rvs_and_phot}, we show the phase-folded ELODIE and TESS data after being subtracted from their corresponding linear trend and GP components. In Fig.~\ref{fig:corner_plot}, we show the corner plot of the main parameters.

\subsection{Apsidal precession}

As shown in Sect.~\ref{sec:joint_fit}, the RVs of HD~118203 show a long-term linear trend that might be caused by an outer massive companion. This potential companion could perturb the orbit of HD~118203~b and cause apsidal precession. Determining whether the orbit of HD~118203~b precesses is of crucial importance to be able to interpret the possible links between the eccentric orbital motion of HD~118203~b and the photometric variations of its host star. We used the RVs to study whether there is any change in the argument of the periastron of the HD~118203~b orbit. To that end, we modelled the RVs following the procedure described in Sect. \ref{sec:joint_fit}, but parametrizing the argument of the periastron as $w_{0}$ + $dw \times (t-t_{0})$, being $w_{0}$ and $dw$ free parameters, $t$ the observing times, and $t_{0}$ the dataset mid-point. We obtain $w_{0}$ = 152.7 $\pm$ 3.1 deg and $dw$ = -0.01 $\pm$ 0.15 deg. Hence, being $dw$ consistent with zero, we conclude that the orbit of HD~118203~b does not significantly precess throughout a 750-day time span.

\subsection{Stellar variability in TESS photometry}
\label{sec:var_tess}

The TESS photometry of HD 118203 shows variability in the four observed sectors (Fig.~\ref{fig:TESS_complete}). To analyse such variations, we removed the transits of HD 118203 b. Given that their durations are $\simeq$5.7 hours, we masked all data points located three hours before and after each mid-transit time. 

\subsubsection{Search for periodic signals}
\label{sec:periodic_signals}

\begin{figure*}
    
    \centering
    \includegraphics[width=\textwidth]{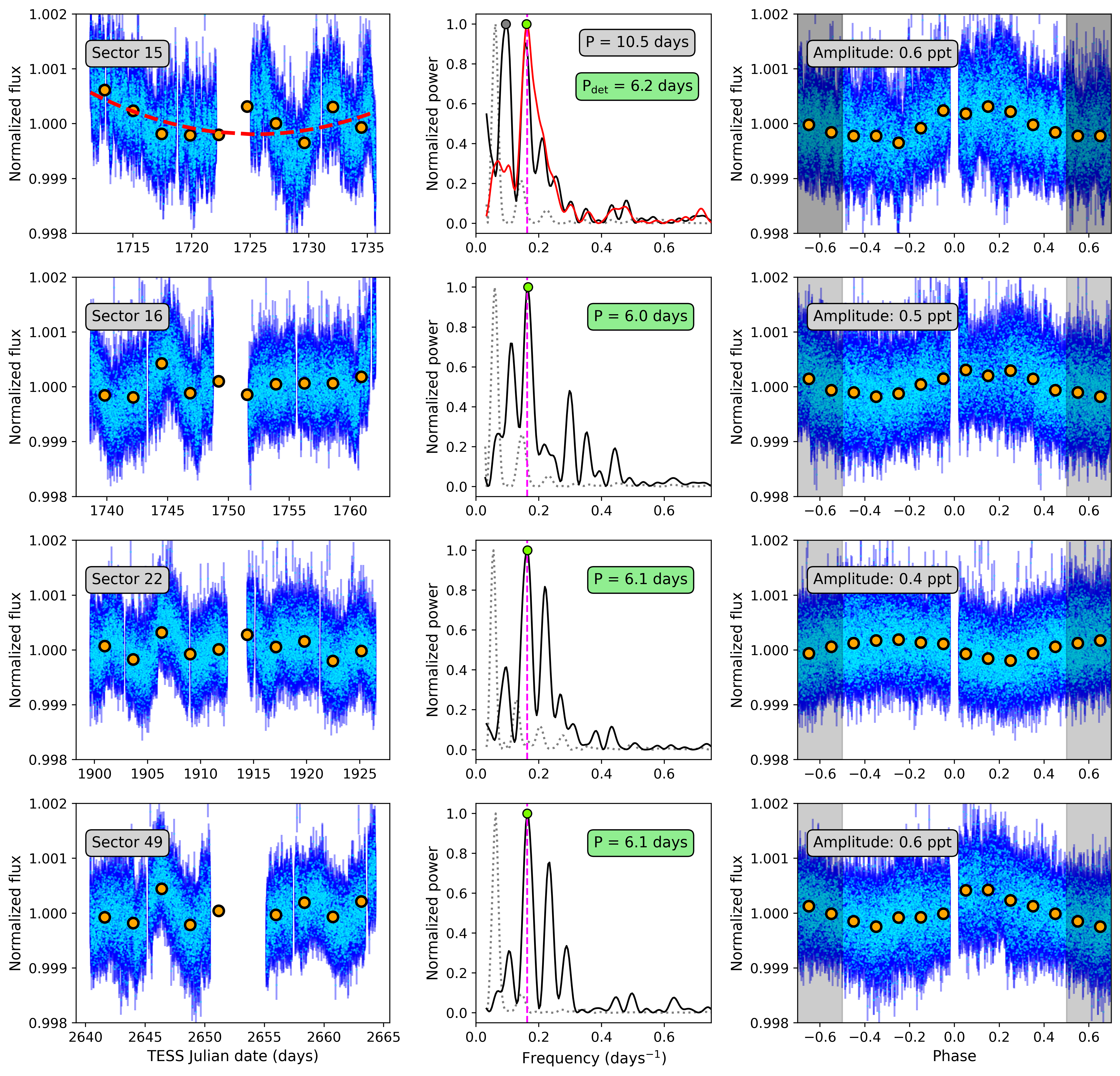}
    \caption[GLS periodograms of the TESS photometry of HD 118203.]{GLS periodograms of the TESS photometry time series and their window functions. Left panels: TESS photometry time series of HD 118203. The red dashed line within the S15 panel indicates the degree 2 polynomial fit used to detrend the data. The orange circles correspond to 1.8-day binned data. Centre panels: GLS periodograms of the time series (solid black) and their window functions (dotted grey). In the S15 panel, the periodogram over the detrended data is represented in red. The green circles and boxes indicate the maximum power frequencies, and the vertical magenta dashed lines indicate the orbital period of HD 118203~b. Right panels: TESS photometry time series folded to the HD 118203~b orbital period. The phase refers to the planet's time of inferior conjunction.  The orange circles correspond to binned data of 10$\%$ the orbital phase.}
    \label{fig:tess_individual_periodograms}
\end{figure*}

We computed the generalised Lomb-Scargle periodogram \citep[GLS;][]{2009A&A...496..577Z}\footnote{We used \texttt{PyAstronomy} \citep[][]{pya}, which is available at \url{https://github.com/sczesla/PyAstronomy}.} of the masked photometry in order to identify possible periodic sinusoidal-like signals. As a result, we obtain a maximum power period of 10.5 days in S15, 6.0 days in S16, 6.1 days in S22, and 6.1 days in S49, all of them with a False Alarm Probability (FAP)\footnote{All the FAPs in this chapter have been computed analytically following Eq. (24) of \citet{2009A&A...496..577Z}.} lower than 0.1$\%$. Thus, in three of the four observed sectors, the TESS photometry shows a periodic sinusoidal-like signal with the same periodicity as the HD 118203~b orbit ($\simeq$6.1 days; see Table \ref{tab:bestfit}). Regarding S15, the first half of the light curve (just before the downlink gap) is dominated by a parabolic trend, but the second half shows a clear $\simeq$6-day periodicity, which stands out in the periodogram as the second highest peak. In order to know if this signal is significant, we detrended the S15 photometry for the parabolic trend, which is most likely the cause of the 10.5-day periodicity. To do so, we modelled the photometry with a second-order polynomial. The periodogram of the de-trended photometry reveals a maximum power period of 6.2 days, also with FAP $<$ 0.1$\%$. This confirms that all the observed sectors show a sinusoidal-like signal with a $\simeq$6.1-day periodicity, which coincides with the orbital period of the confirmed planet. We also ran the GLS over the complete TESS dataset and different combinations of sectors, and in all cases, we detected the signal.

We studied whether the GLS-detected $\simeq$6.1-day signal describes a true periodic photometric variability, or if it could be an alias of the masked transits. To do so, we computed the GLS periodograms of the window functions of each sector. We find no peaks at the $\simeq$6.1-day periodicity detected within the original periodograms, and instead, we find maximum power periods of 16.1 days in S15, 15.8 days in S16, 18.0 days in S22, and 16.0 days in S49. Those periodicities correspond to 67$\%$ (i.e. two-thirds) of their corresponding sector lengths, which suggests that they could be related to the observing baselines. Alternatively, they could also be related to the beat period between the $\simeq$6.1-day transit gap separations and the length of each observing chunk before and after downlink. We filled the masked transit regions with mock data and checked whether the $\simeq$6.1-day periodicity and the 16-to-18-day periodicities of the window functions remain. To do so, we first performed a cubic spline interpolation over the TESS photometry filtered through a median filter with a 5-hour kernel size. Then, we filled the masked regions with 2-min cadence data generated according to the white noise properties of each sector; that is, we used Gaussian distributions with mean values centred on the interpolation models and standard deviations obtained from the flattened photometry.

In Fig.~\ref{fig:mock_data_example}, we show an example of the generated mock photometry. The GLS periodograms of the filled datasets show maximum power periods of 6.1 days (S15), 6.0 days (S16), 6.1 days (S22), and 6.1 days (S49), all of them with FAP $<$ 0.1$\%$. Similarly, the GLS periodograms of the window functions of the filled datasets show the same maximum power periods as those of the window functions of the masked datasets (see Fig.~\ref{fig:wf_comparison}). This analysis confirms that the $\simeq$6.1-day signal found is not related to the data sampling, and instead, it describes a true periodic photometric variability. In Fig.~\ref{fig:tess_individual_periodograms}, we show the sector-by-sector TESS photometry, the GLS periodograms of the original time series and their window functions, and the photometry folded in phase to the orbital period of HD 118203~b. 

\subsubsection{Persistence and evolution of the $\simeq$6.1-day signal}
\label{sec:persistence_evolution}

\begin{figure*}
    
    \centering
    \includegraphics[width = 0.45\textwidth]{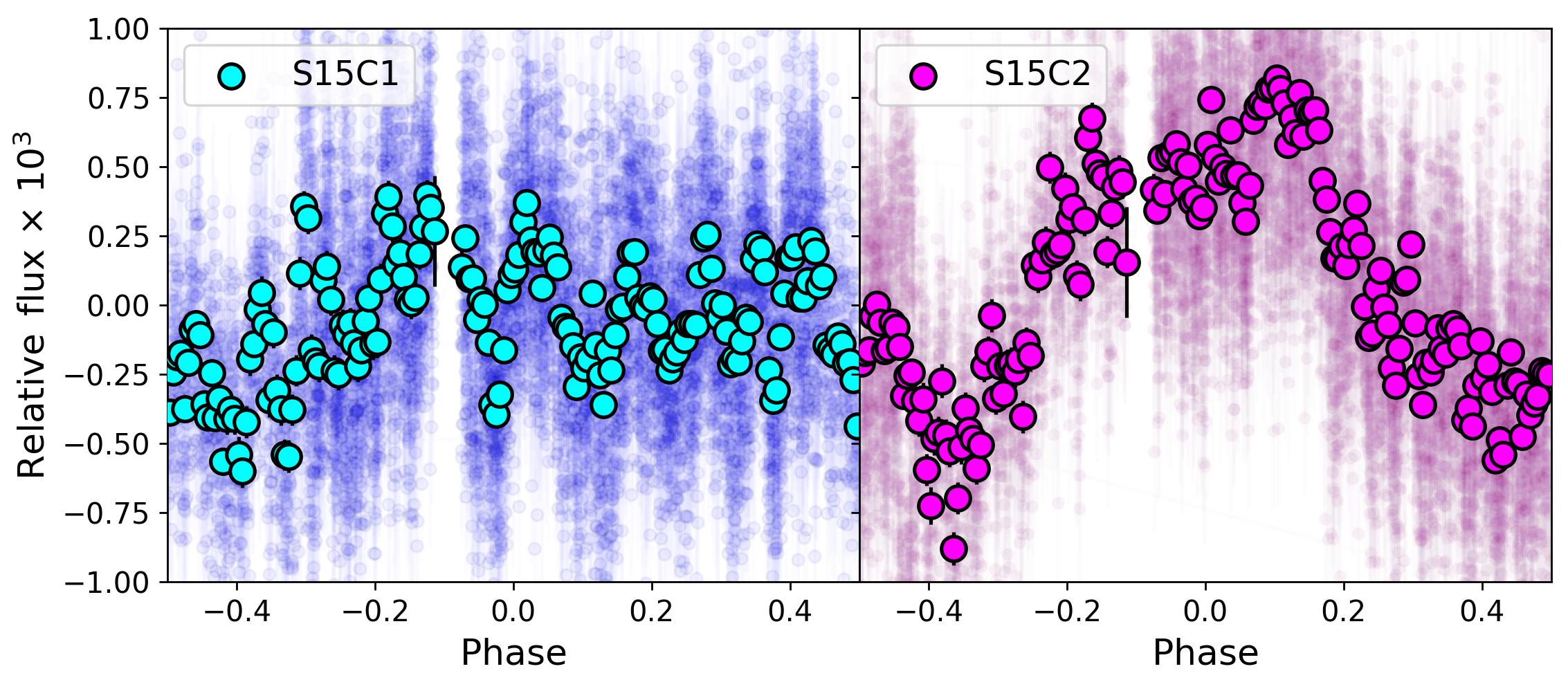}
    \includegraphics[width = 0.45\textwidth]{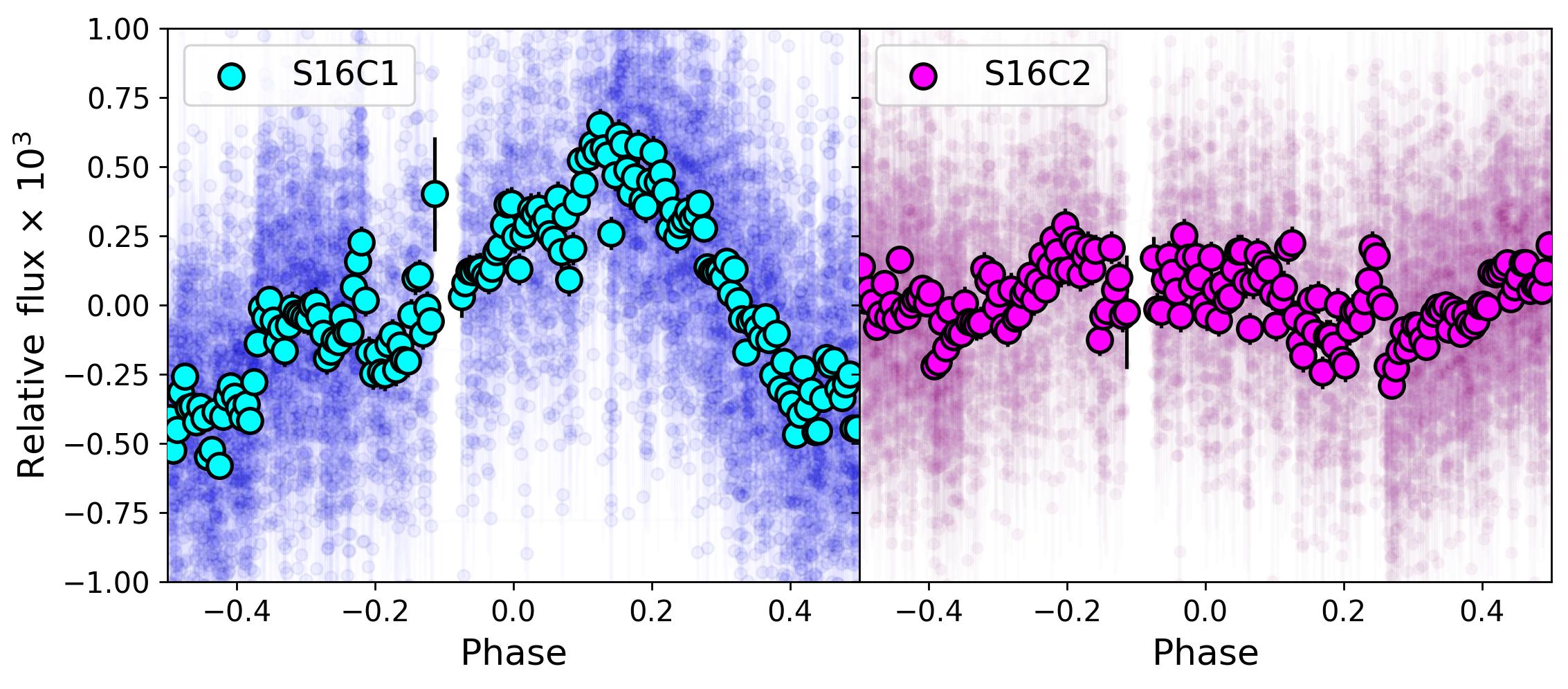}
    \includegraphics[width = 0.45\textwidth]{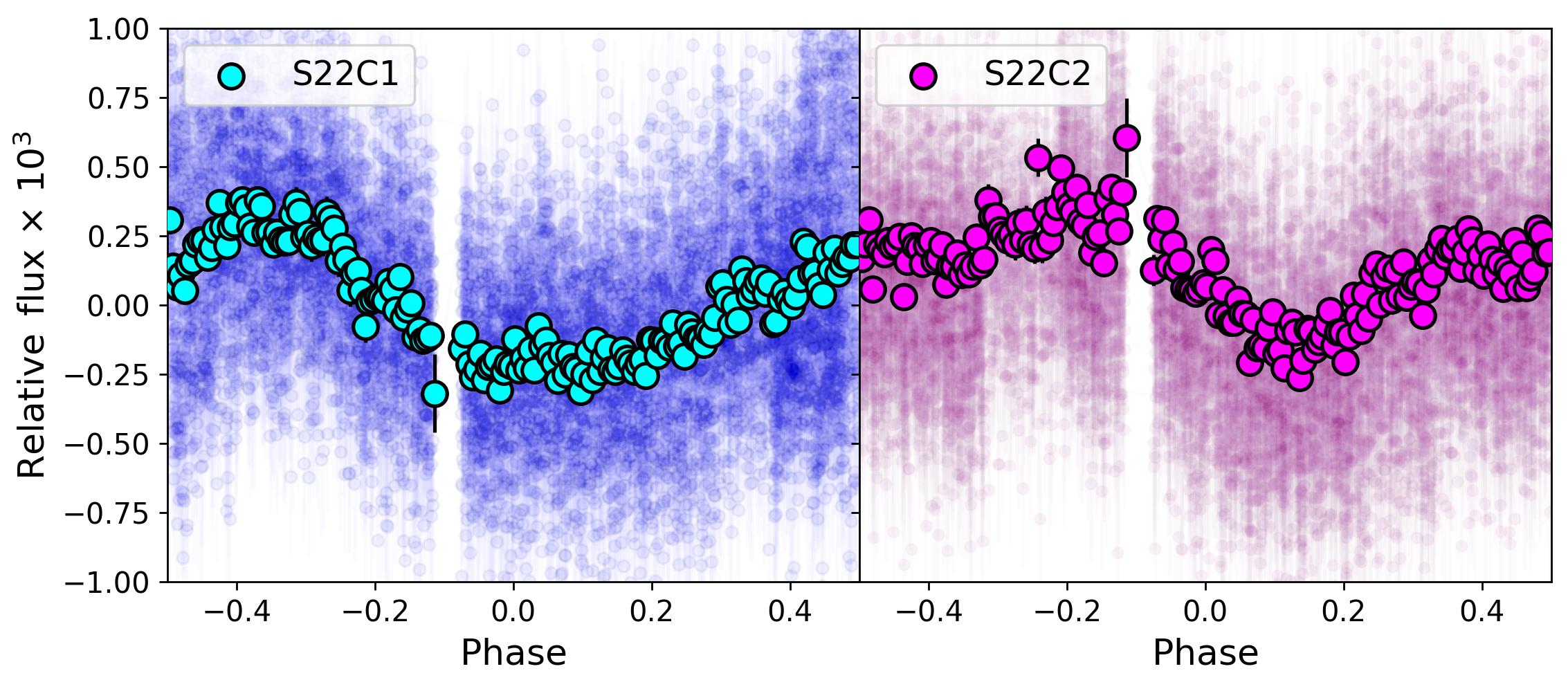}
    \includegraphics[width = 0.45\textwidth]{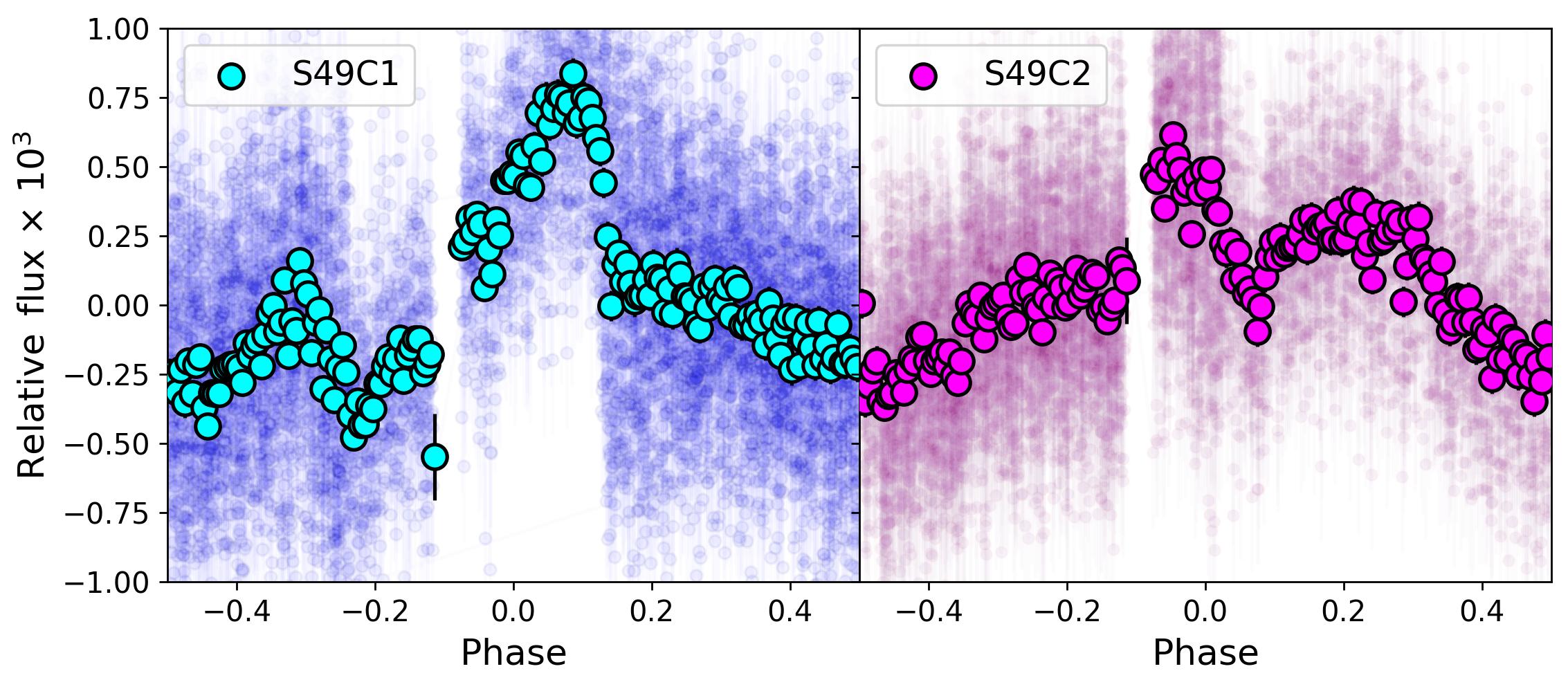}
    \caption[TESS PDCSAP split into eight chunks and folded in phase with HD~118203~b.]{TESS PDCSAP split into eight chunks and folded in phase with the orbital period of HD~118203~b. Large circles correspond to 50-minute binned data for better visualisation of the photometric variations.}
    \label{fig:phase_with_planet}

\end{figure*}

We studied whether the $\simeq$6.1-day signal remains invariant or suffers changes from one orbit to another. To do so, we split the photometry of each sector into two chunks (i.e. before and after the downlink gap), which correspond to $\simeq$1.5 orbits. Hereinafter, we refer to them as S$\mathcal{X}$C$\mathcal{Y}$, where $\mathcal{X}$ denotes the sector and $\mathcal{Y}$ denotes its corresponding chunk (i.e. 1 or 2). We repeated the GLS analysis over each chunk and detected the $\simeq$6.1-day signal in six of them. No significant signals were detected either in S15C1 or in S16C2. In Fig.~\ref{fig:phase_with_planet}, we plot each chunk folded in phase with the orbital period of HD 118203 b, in which we can appreciate how the $\simeq$6.1-day signal appears and disappears from one orbit to another. 

Figure~\ref{fig:phase_with_planet} also reveals that the $\simeq$6.1-day signal experiences a strong and quick evolution. The magnitude of such evolution can be quantified in terms of the signal amplitude and shape. For example, the signal amplitude in S22C1 and S22C2 is decreased by $\simeq$50$\%$ with respect to that of S15C2 or S16C1. Also, the signal shape in S22 is inverted with respect to that of the other three sectors (i.e. in S15C2, S16C1, S49C1, and S49C2, the periodic transits of HD 118203 b occur before reaching the maximum flux emission, but in S22 the transits occur in the descendant region). Hence, in summary, the $\simeq$6.1-day signal appears and disappears suddenly, and when it shows up, it undergoes strong changes in both amplitude and shape. 

\subsubsection{Links with the orbital motion of HD 118203~b}
\label{sec:links}

\begin{table*}[]
\centering
\caption[Properties of the TESS photometric variations of HD 118203.]{Properties of the TESS photometric variations studied in Sect.~\ref{sec:links}. $f_{\rm inc}$ is the orbital phase fraction in which the TESS photometric flux increases. $(f_{\rm inc})_{\rm vis}$ and $(f_{\rm inc})_{\rm hid}$ represent the same fraction but limited to the phases where a hypothetical co-rotating and small active region would be visible and hidden from Earth, respectively. $\phi_{\rm max}$ and $\phi_{\rm min}$ are the phase offsets between the subplanetary point and the maximum and minimum flux emission, respectively. R and p-value measure the correlations between the TESS flux derivative of HD~118203 and the orbital angular velocity of HD~118203~b.}

\renewcommand{\arraystretch}{1.3}
\setlength{\tabcolsep}{9.7pt}

\begin{tabular}{cccc|cc|cc}
\hline \hline
Sector (Chunk) & $f_{\rm inc}$ & $(f_{\rm inc})_{\rm vis}$ & $(f_{\rm inc})_{\rm hid}$ & $\phi_{\rm max}$ & $\phi_{\rm min}$ & R     & p-value \\ \hline
S15 (C2)       & 0.53      & 0.83              & 0.34              & +0.20        & -0.31        & -0.24 & 0.017   \\
S16 (C1)       & 0.56      & 0.88              & 0.38              & +0.25        & -0.36        & -0.18 & 0.070    \\
S22 (C1)       & 0.52      & 0.15              & 0.78              & -0.29        & +0.08        & -0.20 & 0.049    \\
S22 (C2)       & 0.52      & 0.46              & 0.55              & -0.08        & +0.20        & +0.33 & $7.3 \, \times \,10^{-4}$    \\
S49 (C1)       & 0.56      & 0.79              & 0.43              & +0.17        & -0.14        & +0.34 & $6.3 \, \times \,10^{-4}$     \\
S49 (C2)       & 0.54      & 0.70              & 0.41              & +0.30        & -0.36        & -0.13 & 0.18    \\ \hline
\end{tabular}
\label{tab:properties_variations}
\end{table*}

\begin{figure*}
     \centering

     \includegraphics[width = 0.3\textwidth]{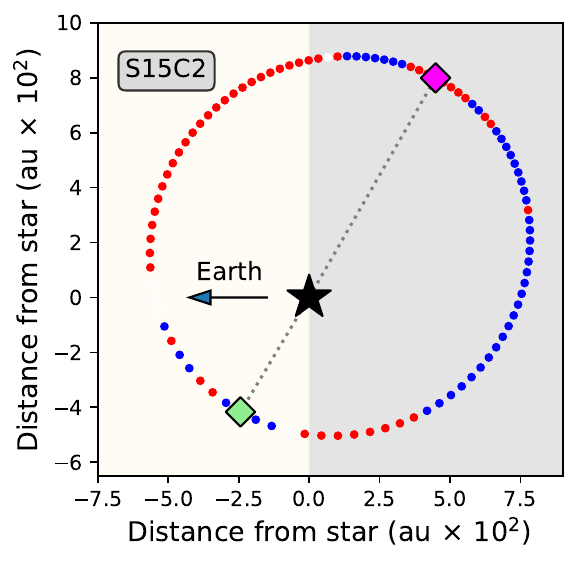}
     \includegraphics[width = 0.3\textwidth]{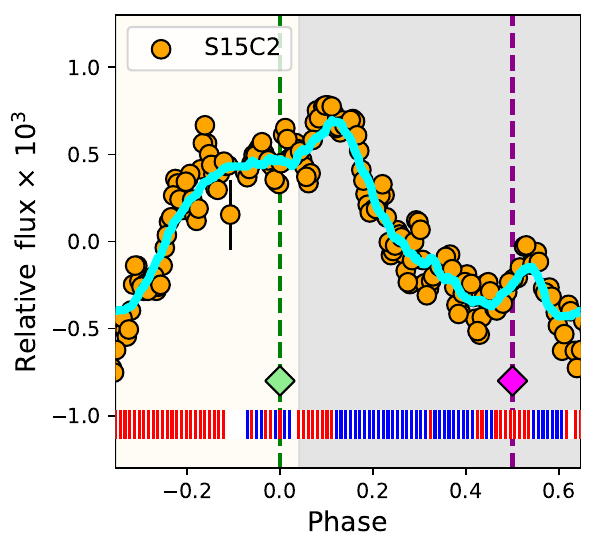}
     \includegraphics[width = 0.3\textwidth]{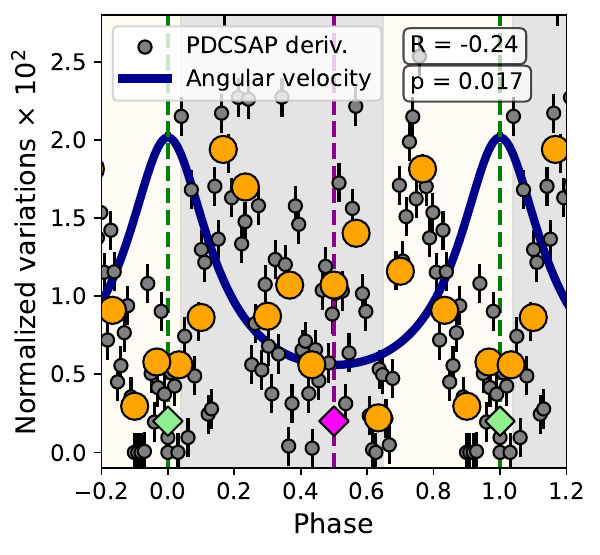}

     \includegraphics[width = 0.3\textwidth]{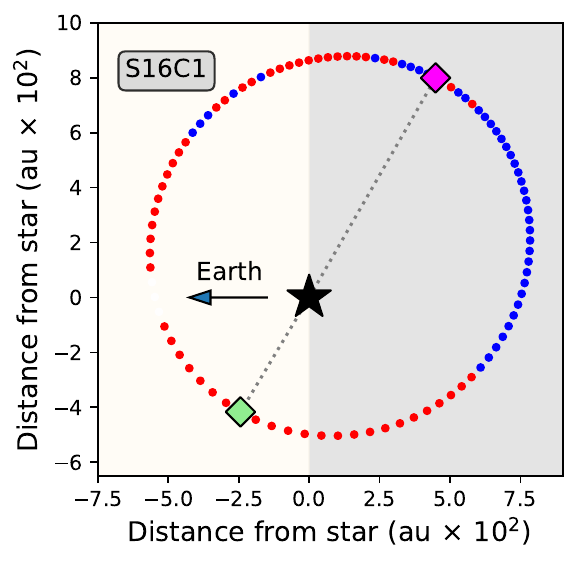}
     \includegraphics[width = 0.3\textwidth]{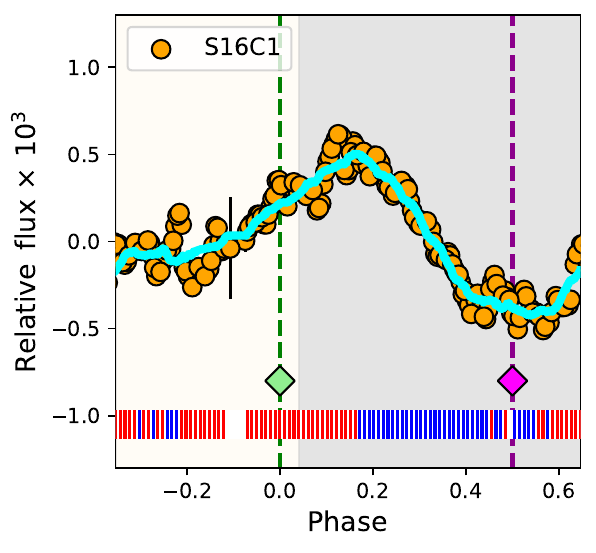}
     \includegraphics[width = 0.3\textwidth]{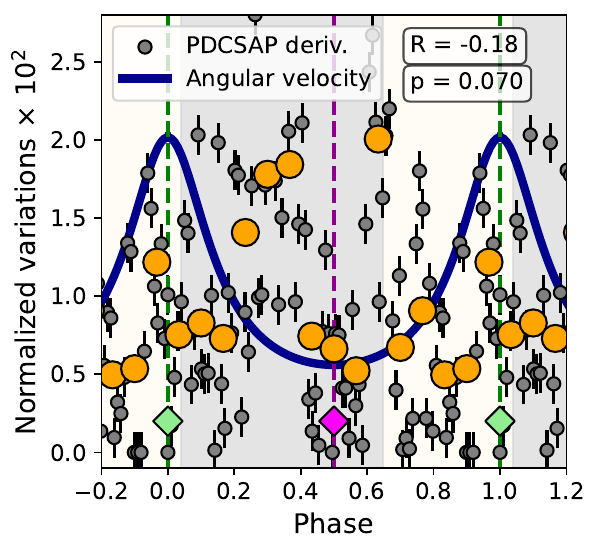}

     \includegraphics[width = 0.3\textwidth]{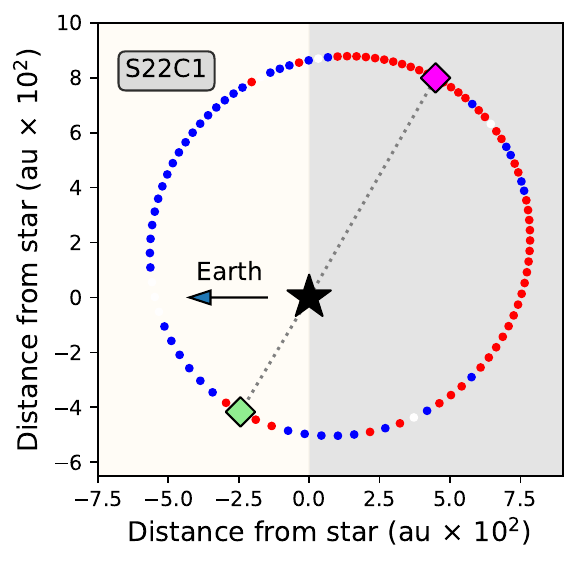}
     \includegraphics[width = 0.3\textwidth]{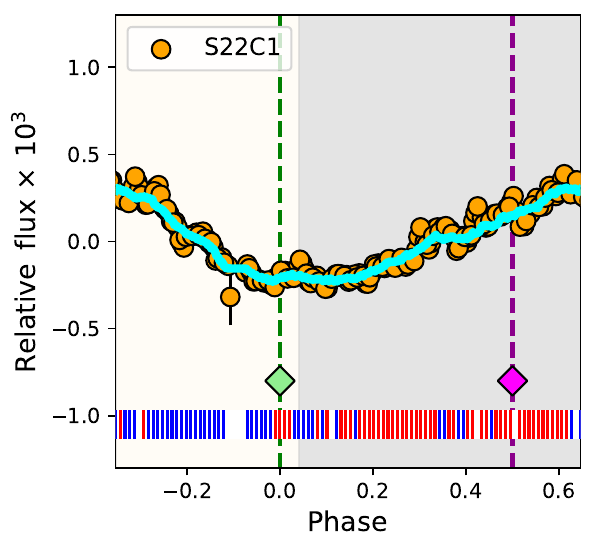}
     \includegraphics[width = 0.3\textwidth]{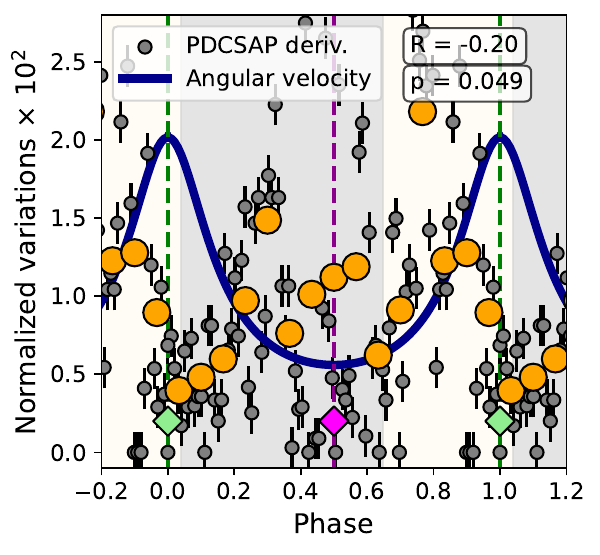}

     \includegraphics[width = 0.3\textwidth]{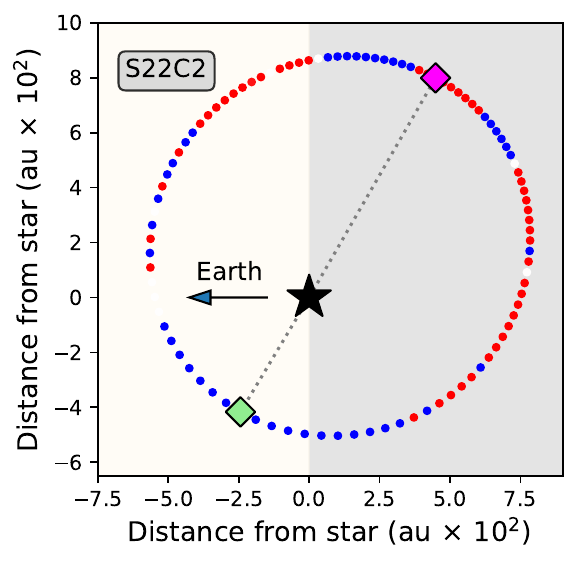}
     \includegraphics[width = 0.3\textwidth]{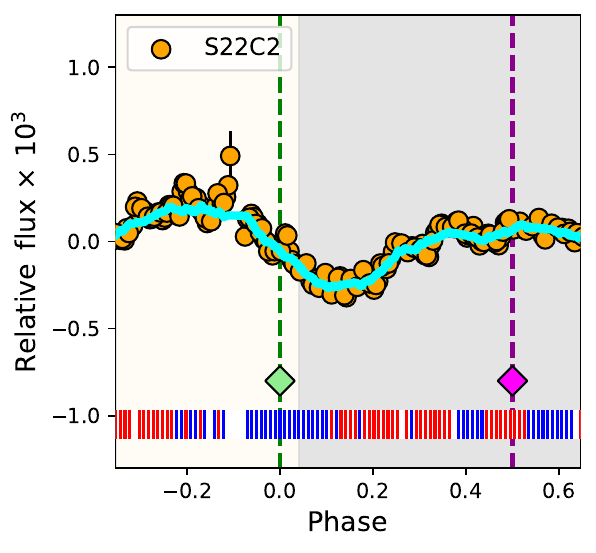}
     \includegraphics[width = 0.3\textwidth]{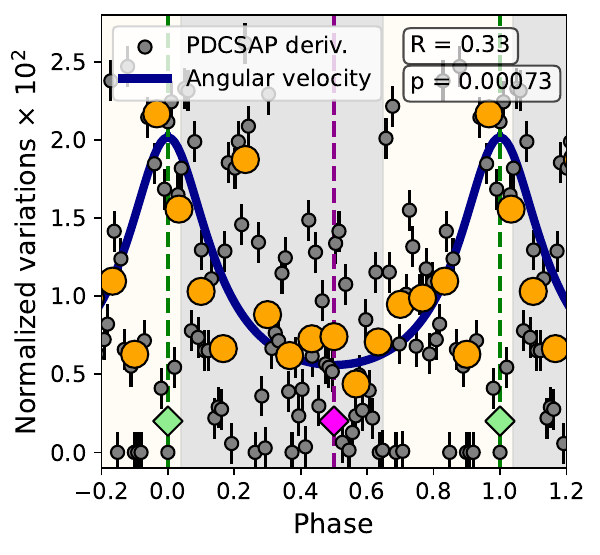}

    \caption[Links between the photometry of HD 118203 and the orbital motion of HD~118203~b.]{Links with the orbital motion of HD~118203~b. Left panels: Orbital path of HD~118203~b, which follows the anticlockwise direction. The circles represent the location of HD~118203~b every $\simeq$1.5 hours and are coloured in red, blue, or white, depending on whether the photometric flux increases, decreases, or remains stable during that time lapse, respectively. Centre panels: TESS PDCSAP photometry of HD~118203 folded in phase with the period of HD~118203~b and binned with $\simeq$5$\%$ phase bins. The cyan line corresponds to the filtered photometry through a median filter. The blue, red, and white vertical lines indicate whether the photometry increases, decreases, or remains stable in time lapses of $\simeq$1.5 hours. Right panels: Comparison between the derivative of the phase-folded PDCSAP photometric variations of HD~118203 and the angular velocity of HD~118203~b. The orange circles correspond to binned data of 7$\%$ of the orbital phase. In all panels: The green and magenta squares represent the periapsis and apoapsis of the orbit, respectively, and the white and grey backgrounds represent the orbital regions in which a hypothetical stellar co-rotating active region would be visible and not visible from Earth, respectively.}

    \label{fig:main_plot}
        
\end{figure*}

\begin{figure*}

     \includegraphics[width = 0.32\textwidth]{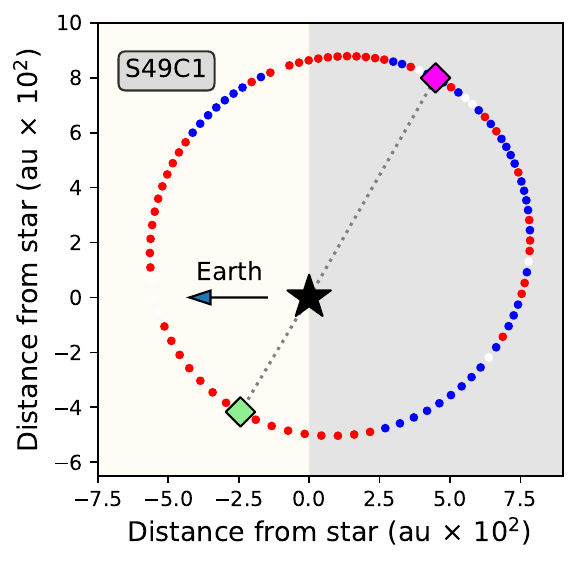}
     \includegraphics[width = 0.32\textwidth]{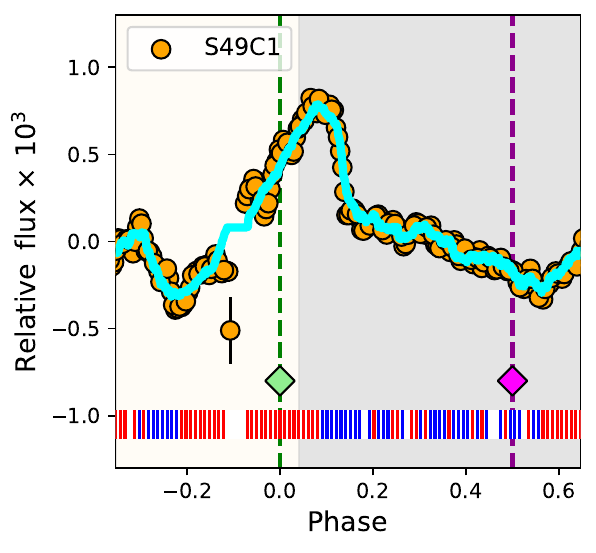}
     \includegraphics[width = 0.32\textwidth]{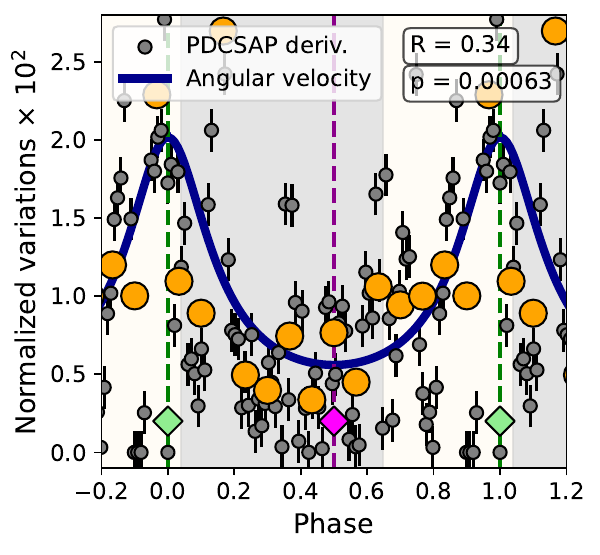}

     \includegraphics[width = 0.32\textwidth]{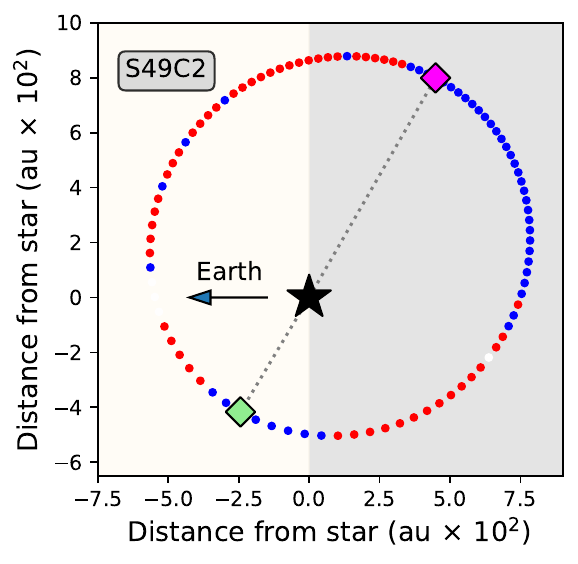}
     \includegraphics[width = 0.32\textwidth]{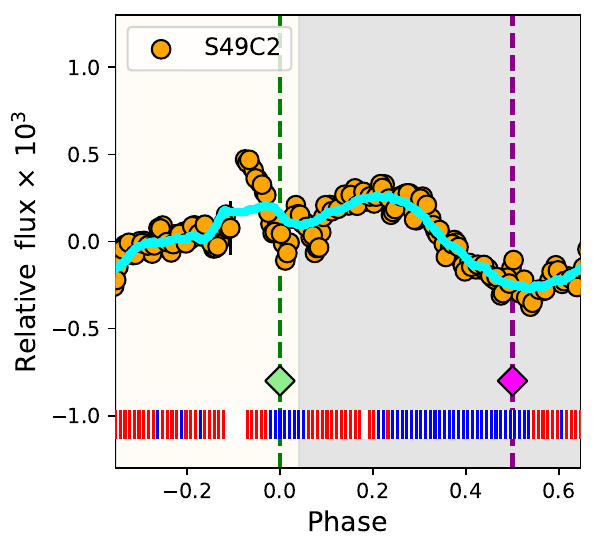}
     \includegraphics[width = 0.32\textwidth]{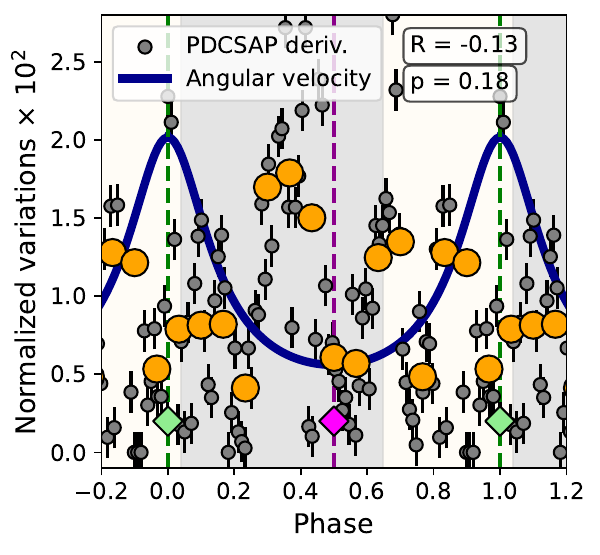}

     \caption[Continuation of Fig.~\ref{fig:main_plot}.]{Continuation of Fig.~\ref{fig:main_plot}. Left panels: Orbital path of HD~118203~b, which follows the anticlockwise direction. Centre panels: TESS PDCSAP photometry of HD~118203 folded in phase with the period of HD~118203~b and binned with $\simeq$5$\%$ phase bins. Right panels: Comparison between the derivative of the phase-folded PDCSAP photometric variations of HD~118203 and the angular velocity of HD~118203~b.}

     \label{fig:main_plot_2}

\end{figure*}

We studied whether the $\simeq$6.1-day photometric signal is linked to the orbital motion of HD~118203~b. That is, we compared the observed photometric variations with the location and angular velocity of HD~118203~b throughout its eccentric orbit. 

We divided the complete phase-folded PDCSAP photometry into 100 bins of $\simeq$1.5 hours long, for which we estimated the flux derivative. In order to mitigate the short-term variations induced by the photometric scatter, we previously filtered the phase-folded photometry through a median filter with a kernel size of 701 cadences (i.e. $\simeq$10$\%$ of the orbital phase) and then computed the flux differences of the filtered photometry. In Table~\ref{tab:properties_variations}, we include the orbital phase fraction in which the photometric flux increases ($f_{\rm inc}$)\footnote{We note that in the transit regions, there are three cadences in which we do not have information on the flux derivative. In those cases, we considered that the flux increases or decreases when there are at least four cadences before and after with the same derivative sign; that is, in S16C1, S22C1, S49C1, and S49C2.}. We find that those fractions roughly correspond to half an orbit. In Table~\ref{tab:properties_variations} we also include the orbital fraction of increasing photometry limited to the phases where a hypothetical co-rotating and small active region would be visible $(f_{\rm inc})_{\rm vis}$ and hidden $(f_{\rm inc})_{\rm hid}$ from Earth. In this case, we find a clear imbalance between the increasing and decreasing regions. In S15C2, S16C1, S49C1, and S49C2, the flux keeps increasing throughout 70-90$\%$ of the time that the hypothetical co-rotating active region would be visible from Earth. On the contrary, in S22C1 the flux decreases during 85$\%$ of the visible phase, and in S22C2 the increasing and decreasing regions last similar.  We also computed the phase offset between the subplanetary point (i.e. time of mid-transit) and the maximum ($\phi_{\rm max}$) and minimum ($\phi_{\rm min}$) flux emission based on the fitted orbital parameters and the filtered photometry. The results are included in Table~\ref{tab:properties_variations}. No flux extreme coincides with the subplanetary point, and instead, there are offsets between -0.35 and 0.25 orbital phases. In order to visualize the aforementioned results, in the left panels of Figs.~\ref{fig:main_plot} and \ref{fig:main_plot_2} we represent the orbital motion of HD 118203~b based on the fitted parameters in Sect.~\ref{sec:joint_fit}, and in the centre panels of the same figures we show the phase-folded binned and filtered photometry. 

Based on Kepler's second law, we computed the angular velocity of HD 118203~b as:
\begin{equation}
    w(r) = \frac{d\theta}{dt} = \frac{\pi}{P_{\rm orb} r^{2}} \left( d_{a} + d_{p} \right) \sqrt{d_{a} d_{p}}
\end{equation}

\begin{figure*}
    \centering
    \includegraphics[width=0.9\textwidth]{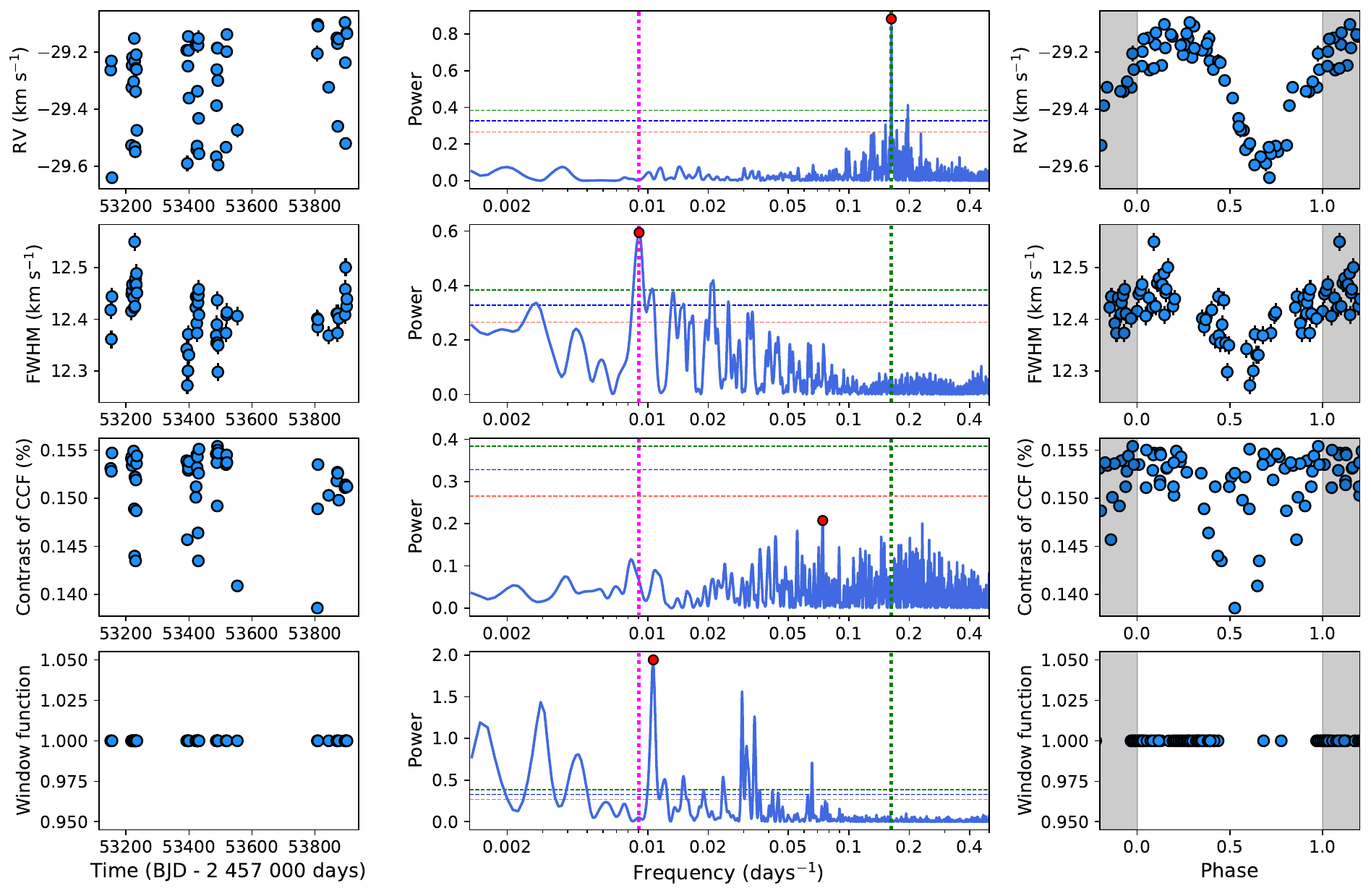}
    \caption[GLS periodograms of the ELODIE time series of HD 118203.]{GLS periodograms of the ELODIE time series and window function. Left panels: Time series of the ELODIE RVs, activity indicators described in Sect.~\ref{sec:obs_elodie}, and window function of the observations. Centre panels: GLS periodogram of the time series and the window function. The red circle indicates the maximum power frequencies. The green dotted vertical lines highlight the $\simeq$6.1-day orbital period of HD~118203~b. The magenta dotted vertical lines indicate the $\simeq$110-day periodicity found in the ELODIE FWHMs. The horizontal dotted lines correspond to the 10 (orange), 1 (blue), and 0.1$\%$ (green) FAP levels. Right panels: ELODIE time series and window function folded to the maximum power periods.}
    \label{fig:gls_to_ELODIE}
\end{figure*}

where $r$ is the distance between the star and the planet, and $d_{p}$ and $d_{a}$ are the distances between the star and the periapsis and apoapsis, respectively. We studied whether this quantity is correlated with the absolute photometric derivative. A certain degree of correlation would be expected in a scenario in which the active regions are moving over the stellar surface synchronised with the eccentric orbital motion of HD 118203 b. We note that there is a geometric factor that also affects the flux derivative. However, we cannot take it into account since it depends on the unknown size, location, evolution, and amount of active regions generating the variations. Hence, we warn the reader to just take this analysis as a first approach to quantify a possible link between the eccentric orbital motion of this planet and the photometric variations generated by hypothetical active regions co-rotating with it. In order to make the comparison, we previously normalised each quantity through the total accumulated variations, which we estimated by using the trapezoidal rule. We quantified the existence of possible correlations by means of the Pearson product-moment correlation coefficient and its associated p-value \citep{doi:10.1080/00031305.1988.10475524}, which we include in Table \ref{tab:properties_variations}. We find a low degree of correlation (R $<$ 0.3) for S15C2, S16C1, S22C1, and S49C2, and we find a moderate degree of correlation (0.30 $<$ R $<$ 0.50) for S22C2 and S49C1, with R coefficients of 0.33 and 0.34, and p-values of $7.3\,\times\,10^{-4}$ and $6.3\,\times\,10^{-4}$, respectively. In the right panels of Figs.~\ref{fig:main_plot}~and~\ref{fig:main_plot_2}, we show the absolute photometric derivatives and the angular velocity of HD~118203~b resolved in phase, where we can visualize the moderate correlation found in S22C2 and S49C1 as well as the lack of correlation in the remaining sectors.

\subsection{Stellar variability in ELODIE spectroscopic data}
\label{sec:variability_ELODIE}

We computed the GLS periodogram of the ELODIE RVs and activity indicators described in Sect.~\ref{sec:obs_elodie}. The RVs periodogram shows a maximum power period of 6.1 days with a FAP of $10^{-20}\%$ that corresponds to the orbital period of HD~118203~b. The FWHMs periodogram shows a maximum power period of 110.3 days with a FAP of $10^{-6}\%$ that suggests the existence of a periodic activity signal. This signal has no counterpart either in the RVs or in the contrasts of the CCFs, the latter being devoid of any signal with FAP $<$ 10$\%$. We also computed the GLS periodogram
of the window function of the observations in order to determine whether the detected peaks could be related to the sampling of the data. We find a maximum power period of 93.8
days, which is relatively close to the periodicity detected in
the FWHMs. In Fig.~\ref{fig:gls_to_ELODIE}, we show the computed periodograms together with the ELODIE time series folded in phase to their corresponding maximum power periods.

\subsection{Stellar variability in ASAS-SN photometry}
\label{sec:variability_asassn}

\begin{figure*}
    \centering
    \includegraphics[width=0.95\textwidth]{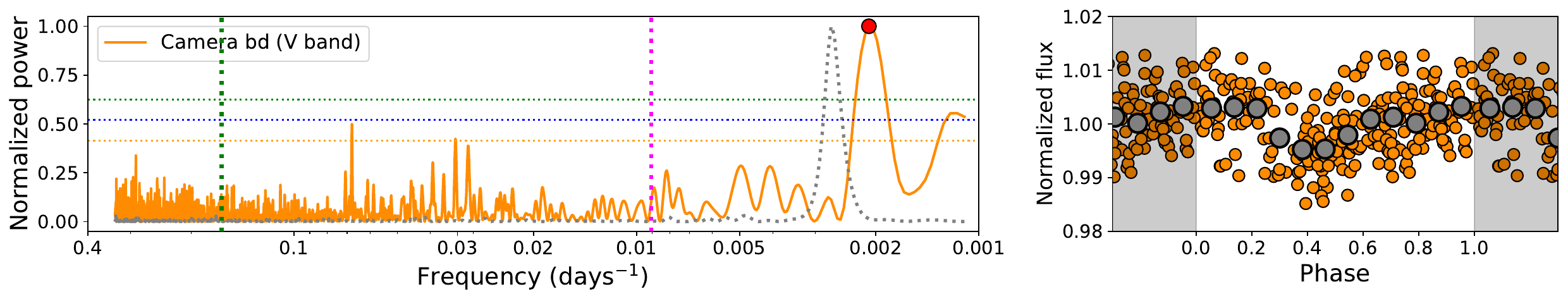}
    \caption[GLS periodogram of ASAS-SN photometry of HD 118203.]{GLS periodogram of the ASAS-SN photometry acquired by the camera bd (solid orange) and its window function (dotted grey). The red circle indicates the maximum power frequency. The green dotted vertical line highlights the $\simeq$6.1-day orbital period of HD~118203~b, and the magenta dotted vertical line indicates the $\simeq$110-day periodicity found in the ELODIE FWHMs. The horizontal dotted lines correspond to the 10 (orange), 1 (blue), and 0.1$\%$ (green) FAP levels. The right panel illustrates the phase-folded photometry to the maximum power period. The grey data points correspond to a 35-day binning.}
    \label{fig:asas-sn}
\end{figure*}

We computed the GLS periodograms of the ASAS-SN photometry acquired by the four cameras. The periodograms of three of them show no significant periodic signals; that is, there are no peaks with FAP $<$ 10$\%$. The periodogram of the camera \textit{bd}, however, shows a peak at 477 days with a FAP of $\simeq$ $10^{-5}$$\%$, but no peak appears at the 110.3-day signal found in the ELODIE FWHMs.  We also computed the GLS periodogram of the window function. It shows a maximum power period of 366 days, which is possibly related to a yearly alias due to the seasonal visibility of the target. In Fig.~\ref{fig:asas-sn}, we show the periodograms together with the ASAS-SN photometry folded in phase to its maximum power 477-day period.

\section{Discussion}
\label{sec:discussion}
\subsection{Origin of the $\simeq$6.1-day TESS photometric signal}

Close-in giant planets can modify the activity levels of their host stars through MSPIs \citep[][]{2000ApJ...533L.151C,2003ApJ...597.1092S,2005ApJ...622.1075S,2008ApJ...676..628S,2008A&A...482..691W,2009EM&P..105..373P,2011ApJ...741L..18P,2015ApJ...811L...2M,2019NatAs...3.1128C}. The commonly considered criterion to confirm those interactions consists of detecting a periodic activity signal that matches the orbital period of the hosted planet, as well as another periodic signal induced by the rotation of the star. In this chapter, we have found evidence of periodic stellar variability in the TESS photometry of HD 118203 that matches the orbital period of its hosted giant planet HD 118203~b (i.e. $\simeq$6.1 days; Sect.~\ref{sec:var_tess}). We searched for possible rotation-induced activity in the ELODIE activity indicators (Sect.~\ref{sec:variability_ELODIE}) and ASAS-SN photometry (Sect.~\ref{sec:variability_asassn}) and found significant signals at $\simeq$110 and $\simeq$477 days, respectively. However, given the disparity of results in both datasets, the non-repeatability of the signals found, and the relative closeness of the ELODIE periodicity to the highest peak of the window function periodogram, we consider that we do not have enough evidence to confidently confirm the rotation period of HD~118203.

With a mass of $\simeq$1.3 $\rm M_{\odot}$, HD 118203 must have been a F-type star during its main sequence (MS) phase. Such a mass coincides with the Kraft break \citep{1967ApJ...150..551K}. Stars with lower masses are known to have large convective envelopes that provoke angular momentum loss due to magnetised stellar winds. In contrast, stars with higher masses are known to have very thin convective envelopes, so they can keep rotating quickly during the MS phase. As stars leave the MS, their cores contract and their envelopes expand, resulting in an increase in their moments of inertia that leads to a decrease in the surface rotation rate as the star evolves across the subgiant branch. Besides, stars with masses above the Kraft break develop convective envelopes so that wind-driven angular momentum loss affects both MS rotational regimes in evolved stages \citep[e.g.][]{2013ApJ...776...67V}. This has been empirically shown by \citet{2012A&A...548L...1D}, who found that subgiants with masses both above and below the Kraft break show rotation periods of several tens of days. Therefore, whether HD 118203 underwent significant magnetic braking during its MS phase or remained rotating rapidly because of a thin convective envelope, both models and theory indicate that its rotation period would have been increased during its subgiant phase.

Another factor that might also affect the rotation velocity of this star is the presence of a close-in massive giant planet. However, the energy dissipation due to tidal forcing is much lower in stars than in their hosted planets \citep[e.g.][]{Fabrycky_Tremaine_2007,2011CeMDA.111..105C,Beauge_Nesvorny_2012}. This has been observationally proved for G-type stars with tightly orbiting giant planets, which show similar rotation periods to those of G-type stars without close-in companions \citep{2013PASP..125..989A}. F-type stars, however, are less studied in this regard, but we know they should be more prone to reach a synchronization state since radiative envelopes dissipate more energy than convective envelopes \citep[e.g.][]{2011A&A...529A..50L}, and, as discussed above, they are much less slowed down by magnetized winds. Interestingly, to date, the only star thought to be tidally locked to the orbit of its hosted hot Jupiter is an MS F-type star, $\tau$ Boo \citep{1997ApJ...474L.115B,2000ApJ...529L..41H}. Therefore, the natural question that arises here is whether HD~118203 may be a new case of a tidally locked star to its hosted hot Jupiter. If so, the tidal locking configuration must have been strong enough to avoid the influence of the braking forces experienced by the star when ascending the subgiant branch. 

The tidal theory shows that the equilibrium state of planetary systems with gas giants in eccentric orbits is not a synchronous state, but rather a pseudo-synchronous state that depends on the orbital eccentricity \citep{1981A&A....99..126H,2011CeMDA.111..105C}. According to Eq. (42) in \citet{1981A&A....99..126H}, the pseudo-synchronisation in the HD 118203 planetary system corresponds to a rotation period of 3.73 days. Such pseudo-synchronised rotation is expected to have occurred quickly for the hosted giant planet HD 118203~b (i.e. a few million years), but much more slowly for the host star HD 118203 (in the case that this rotational state was actually reached). Therefore, being HD 118203 a subgiant star, either coming from an MS star with a thick or narrow convective envelope, we expect that it underwent the well-studied rotation period increase when ascending the subgiant branch. In a hypothetical and unusual case in which the rotation of HD 118203 could have been dominated by the tidal forces exerted by its hot Jupiter during its MS phase, the star would be pseudo-synchronised at 3.73 days, being a hypothetical 6.1-day rotation a non-equilibrium state for this system. Hence, although we do not have observational evidence of the rotation period of HD 118203, we argue that it being 6.1 days is very unlikely, which makes MSPIs the most likely scenario to explain the observed TESS activity signal. This argument agrees with literature estimates of the spectroscopic projected rotational velocity in the case that the stellar rotation axis is orthogonal to the line of sight. That is, \citet{2006A&A...446..717D} and \citet{2017AJ....153...21L} computed values of 5 and 7 km$\rm \,s^{-1}$, respectively, while a projected velocity of 17 km$\rm \,s^{-1}$ would be required for the rotation of the star to be the 6.1-day signal found. In the case of a pseudo-synchronised 3.73-day rotation, a 27 km$\rm \,s^{-1}$ projected rotational velocity would be required. In the following, we discuss how the TESS signal properties unveiled in Sects. \ref{sec:persistence_evolution} and \ref{sec:links} match the MSPIs and rotation scenarios. 

\subsection{TESS signal properties}

In Sect.~\ref{sec:persistence_evolution}, we found that the $\simeq$6.1-day photometric signal is not a persistent phenomenon (see Fig.~\ref{fig:phase_with_planet}). Interestingly, this is an expected behaviour for MSPIs. \citet{2008ApJ...676..628S} found evidence of synchronous stellar activity in HD 179949 and $\upsilon$ And during 75$\%$ of the monitored time, while in the remaining 25$\%$ only the rotation of the star showed up. This on/off nature of MSPIs had been previously predicted by the models of \citet{2007astro.ph..2530C}. The authors found that the complex nature of the multipole fields may cause MSPI-induced activity not to repeat exactly from one orbit to another, and sometimes it can even disappear completely. We note, however, that while being expected for MSPIs, this behaviour does not completely rule out the rotation of the star as the source of the variability. Stellar rotation is a periodic phenomenon, but the rotation-induced activity signals are quasiperiodic since the active regions move on the stellar surface and appear and disappear throughout the magnetic cycle timescale. The typical stellar magnetic cycles last several years \citep[e.g.][]{2016A&A...595A..12S}, but some stars have also been found to have a quick spot evolution, showing variations throughout time scales of a few days \citep[e.g.][]{2019ApJ...871..187N}. Hence, the strong changes in amplitude and shape experienced by the $\simeq$6.1-day TESS signal are most likely explained by MSPIs, although we cannot completely discard the possibility that they could be caused by the rotation of the star with quickly evolving spots.

In Sect.~\ref{sec:links}, we studied the possible existence of links between the $\simeq$6.1-day photometric signal and the orbital motion of HD~118203~b. We found that the flux variations span the complete orbit, which suggests that the total active area generating those variations occupies a considerable amount of the stellar surface. Hence, the observed photometric properties do not allow us to identify a link with the orbital location of the planet under the assumption of a small co-rotating spot on the stellar surface. However, we also found that in five of the six analysed chunks, the flux mostly increases (S15C2, S16C1, S49C1, and S49C2) or decreases (S22C1) when such a hypothetical big active area would be visible from Earth. Hence, if we consider an MSPI scenario, the photometry of S15C2, S16C1, S49C1, and S49C2 would be best explained by the presence of bright and extensive spots co-rotating with HD~118203~b, while in S22 such co-rotating spots would be dark. This possible alternation caused by MSPIs has been previously noticed by \citet{2008A&A...482..691W}. Previous works reporting MSPIs also found that the stellar activity extremes tend to be shifted with respect to the subplanetary point, which is interpreted as the co-rotating active regions advancing or delaying with respect to the planet location \citep[e.g.][]{2008ApJ...676..628S}. Depending on the system, those offsets can vary significantly  (e.g. \citealp{2019NatAs...3.1128C} found a 0.46 phase offset for $\upsilon$, a 0.92 phase offset for HD 189733, and offsets in between for other two systems). Interestingly, we have found similar offsets for HD~118203.

In Sect.~\ref{sec:links}, we also studied whether the eccentric orbital motion of HD~118203~b has an imprint in the observed TESS variability. To date, only two eccentric systems have been reported to possibly have MSPIs, Kepler-432 \citep{2015ApJ...803...49Q,2015A&A...573L...5C} and  HD 17156 \citep{2015ApJ...811L...2M}. However, the periodic variability of Kepler-432 was found to be invariant during the whole photometric observations, so that both MSPIs and tidal interactions could explain them \citep{2015ApJ...803...49Q}. Regarding HD~17156, the occasional nature of the activity enhancement also makes it compatible with accretion onto the star of material tidally stripped from the planet \citep{2015ApJ...811L...2M}. Therefore, HD~118203~b represents the best evidence of MSPIs in eccentric planetary systems, which makes it an ideal target to probe a possible eccentricity imprint in the observed photometric variations. We searched for a possible correlation between the stellar flux derivative ($\frac{df}{dt}$) and the planetary angular velocity ($\frac{d\theta}{d t}$), given that the projected area of active regions connected to, and co-moving with the planet, would change according to the eccentric orbital motion of the planet. Unfortunately, we could not confirm the existence of clear links between the photometric variations of HD~118203 and the eccentric orbital motion of HD~118203~b. 

\subsection{Orbital eccentricity as a possible booster of MSPIs}

As discussed before, given the closeness of this gas giant to its host star and the significant orbital eccentricity, the rotation period of HD~118203~b is expected to be pseudo-synchronised at $\simeq$3.73 days. This prediction indicates that the eccentric orbit of this hot Jupiter could be responsible for a higher planetary magnetic moment than for a similar circular system. In particular, HD~118203~b  would have a $\simeq 40\%$ larger magnetic moment than if it were in a circular, synchronised orbit. \citet{2005ApJ...622.1075S,2008ApJ...676..628S} found that circular planetary systems with hot Jupiters with $M_{\rm p}$sin(i) / $P_{\rm rot}$ $>$ 0.4 $M_{J}$ $\rm day^{-1}$ tend to undergo MSPIs, while below that value only 20 $\%$ of their studied systems showed MSPIs. For HD~118203~b, we estimate a $M_{\rm p}$ / $P_{\rm rot}$ of 0.61 $M_{J}$ $\rm day^{-1}$, while in a hypothetical circular orbit, the ratio would be 0.37 $M_{J}$ $\rm day^{-1}$. Hence, the estimated ratio for the eccentric case is well above the 0.4 $M_{J}$ $\rm day^{-1}$ value where MSPIs are expected to be found, while in a hypothetical circular orbit, the estimated ratio corresponds to a region of the parameter space where MSPIs are not expected to be always found. This suggests that the unusually high eccentricity of HD~118203~b could be critical for the generation of the potential MSPIs.

\section{Summary and conclusions}
\label{sec:conclusions}

The close-in ($a$ = 0.0864 $\pm$ 0.0006 au) and eccentric ($e$ = 0.32 $\pm$ 0.02) Jupiter-sized planet HD~118203~b was discovered by \citet{2006A&A...446..717D} through 43 RVs acquired with the ELODIE spectrograph. Recently, the TESS satellite revealed that HD~118203~b transits its bright host star \citep{2020AJ....159..243P}. Having a magnitude of V = 8.05 $\pm$ 0.03, HD~118203 is among the 10 brightest stars to have a transiting giant planet. Close-in giant planets such as HD~118203~b may influence the activity levels of their host stars through magnetic star-planet interactions (MSPIs), which may induce stellar variability modulated by the orbital period of the interacting planet. Since the first claims \citep{2003A&A...406..373S,2003ApJ...597.1092S}, dozens of signs of MSPIs have been reported \citep[e.g.][]{2005ApJ...622.1075S,2008ApJ...676..628S,2008A&A...482..691W,2009EM&P..105..373P,2011ApJ...741L..18P,2015ApJ...811L...2M}. However, a key regime remained unexplored: eccentric planetary systems. Eccentric systems hosting close-in giant planets such as HD~118203 are very scarce since at those close distances gas giants suffer a quick circularisation process \citep{1981A&A....99..126H}. Previous studies focused on detecting MSPIs in eccentric systems were unsuccessful, and to date, there are only two possible cases of MSPIs in eccentric systems \citep{2015ApJ...803...49Q,2015ApJ...811L...2M}. However, eccentric systems can give us important insights on MSPIs.

In this chapter, we analysed the complete ELODIE dataset (43 RVs from the discovery paper and 13 more public RVs) and four TESS sectors of HD~118203 with the primary objective of searching for activity signals potentially induced by its hosted close-in giant planet, and the secondary objective of refining the orbital and physical properties of the planetary system. 

We found evidence of an activity signal within the TESS photometry that matches the orbital period of HD~118203~b (i.e.~$\simeq$6.1 days), which could be the result of magnetic interactions between the planet and its host star. In order to confirm such interactions, the commonly considered criterion consists of independently detecting the rotation period of the star. To do so, we analysed the ELODIE activity indicators and complementary ASAS-SN photometry but found no compelling evidence of an additional rotation-induced activity signal. However, given the evolved nature of the star and the significant orbital eccentricity, we argue that MSPIs is the most likely scenario to explain the observed TESS variability. This argument agrees
with literature estimates of the spectroscopic projected rotational velocity. 

We analysed the persistence and evolution of the signal and found that it appears and disappears on time scales comparable to a planetary orbit. Also, when active, it experiences strong changes in both amplitude and shape. We interpret this behaviour as a possible manifestation of the on/off nature of MSPIs found by \citet{2008ApJ...676..628S}. However, it does not completely rule out the rotation scenario since the star could have a quick spot evolution that could cause those sudden changes \citep[e.g.][]{2019ApJ...871..187N}. We also studied the existence of links between the TESS variability and the orbital motion of HD~118203~b. Unfortunately, the observed variations are complex and cannot be interpreted as a single small spot moving over the stellar surface, which implies that there is a geometric factor that we cannot consider in our analysis. Overall, we found a moderate degree of correlation between the flux derivative of HD~118203 and the angular velocity of HD~118203~b in two chunks of the TESS photometry, but found no correlations in the remaining four chunks.  We also found that the rotation period of the planet is expected to be pseudo-synchronised with the planetary orbital period with a 3.73-day periodicity. Interestingly, such a spin velocity is expected to generate a planetary magnetic moment able to produce strong MSPIs, but in a hypothetical circular scenario in which the planetary rotation and orbital periods were synchronized, the generated magnetic moment could not be enough to produce such interactions \citep{2005ApJ...622.1075S,2008ApJ...676..628S}. This suggests that the unusually high eccentricity of HD~118203~b could be critical for the generation of the potential MSPIs found.

Regarding the secondary objective of the work, we significantly improved the transit ephemeris of this planetary system thanks to the inclusion of two new sectors of TESS data separated by 2.5 years from the last observations. As an example, for the year 2028, we compute a propagated uncertainty in the mid-transit time of just 1 minute, in contrast with the 27-minute uncertainty reported in the current most accurate characterisation \citep{2020AJ....159..243P}. Obtaining accurate ephemerides for very bright targets is very valuable for the community since those targets will most likely be scheduled for atmospheric studies.

Similar to previous studies such as \citet{2008A&A...482..691W} and \citet{2009EM&P..105..373P}, we here report new signs of MSPIs detected photometrically in a planetary system hosting a hot Jupiter, whose confirmation via rotation period determination should be attempted through precise photometric or spectroscopic follow-up observations. In this regard, the PLATO mission will play a major role thanks to the expected long-term and continuous precise photometric monitoring of the sky. Also, state-of-the-art high-resolution spectrographs such as HARPS-N and CARMENES could be very useful to determine the rotation period of this star, either through direct observations of the stellar chromospheric activity or through the Rossiter-McLaughlin effect.  To date, HD 118203 represents the best evidence that magnetic star-planet interactions can be found in eccentric planetary systems, and it opens the door to dedicated searches in such systems that will allow us to better understand the interplay between close-in giant planets and their host stars.
\newpage
\chapter{Mapping the exo-Neptunian landscape: A ridge between the desert and savanna}
\label{ch:nep_des}
\vspace{2cm}
\pagestyle{fancy}
\fancyhf{}
\lhead[\small{\textbf{\thepage}}]{\small{\textbf{\nouppercase{\leftmark}}}}
\rhead[\small{\textbf{\nouppercase{\rightmark}}}]{\small{\textbf{\thepage}}}

Several efforts are underway to disentangle the complex processes shaping the distribution of close-in Neptunian planets. These are mostly based on obtaining a large census of atmospheric escape rates \citep[e.g. the NIGHT spectrograph;][]{2024MNRAS.527.4467F} and system obliquities (e.g. the ATREIDES collaboration; Bourrier et al., in prep). Properly interpreting the results of such large-scale surveys requires accurate characterisation of the intrinsic planet distribution, which itself can also provide useful constraints. Determining where and how the transition between the Neptunian desert and savanna occurs \citep[i.e. a moderately populated region at larger orbital distances;][]{2023A&A...669A..63B}, as well as the relative planet occurrence between different features, is key to understanding the overall evolution of Neptunian worlds. In this chapter, using a population-based approach, we aim to map the boundaries of the Neptunian desert and study its transition into the savanna. In Sect.~\ref{sec:sample_selection_biases_mitigation}, we describe the sample selection and bias mitigation. In Sect.~\ref{sec:oc}, we study the distribution of Neptunes across the orbital period space. In Sect.~\ref{sec:boundaries}, we extend our analysis to a wider radius range and explain how we computed new desert boundaries in the 2D period-radius space. In Sect.~\ref{sec:discussion}, we discuss the new Neptunian landscape as a tracer of close-in planets' origins, and we conclude in Sect.~\ref{sec:conclusion}.

\section{Sample selection and bias mitigation}
\label{sec:sample_selection_biases_mitigation}

\begin{table}[]
\centering
\renewcommand{\arraystretch}{1.3}
\setlength{\tabcolsep}{11pt}
\caption[Transit and detection probabilities of the \textit{Kepler} DR25 catalogue.]{Transit and detection probabilities of the \textit{Kepler} DR25 catalogue. The complete table is accessible at the Centre de Données astronomiques de Strasbourg (\url{https://cdsarc.cds.unistra.fr/viz-bin/cat/J/A+A/689/A250}).}
\label{tab:weights}
\begin{tabular}{cccc}
\hline \hline
KOI Name        & $P_{\textrm{transit}}^{-1}$ & $P_{\rm detection}^{-1}$ & $w$   \\ \hline
K00001.01    & 7.978060                      & 1.034912              & 8.256587   \\
K00002.01          & 4.133503                        & 1.035139              & 4.278752  \\
K00003.01 & 14.807579                        & 1.036065              & 15.341618  \\
        K00004.01  & 3.892077                    & 1.037361              & 4.037490  \\
K00005.01  & 7.179468                     & 1.038220             & 7.453869\\
...           & ...                         & ...                 & ...     \\ \hline
\end{tabular}
\end{table}

\begin{figure*}
    \centering
    \includegraphics[width=0.49\textwidth]{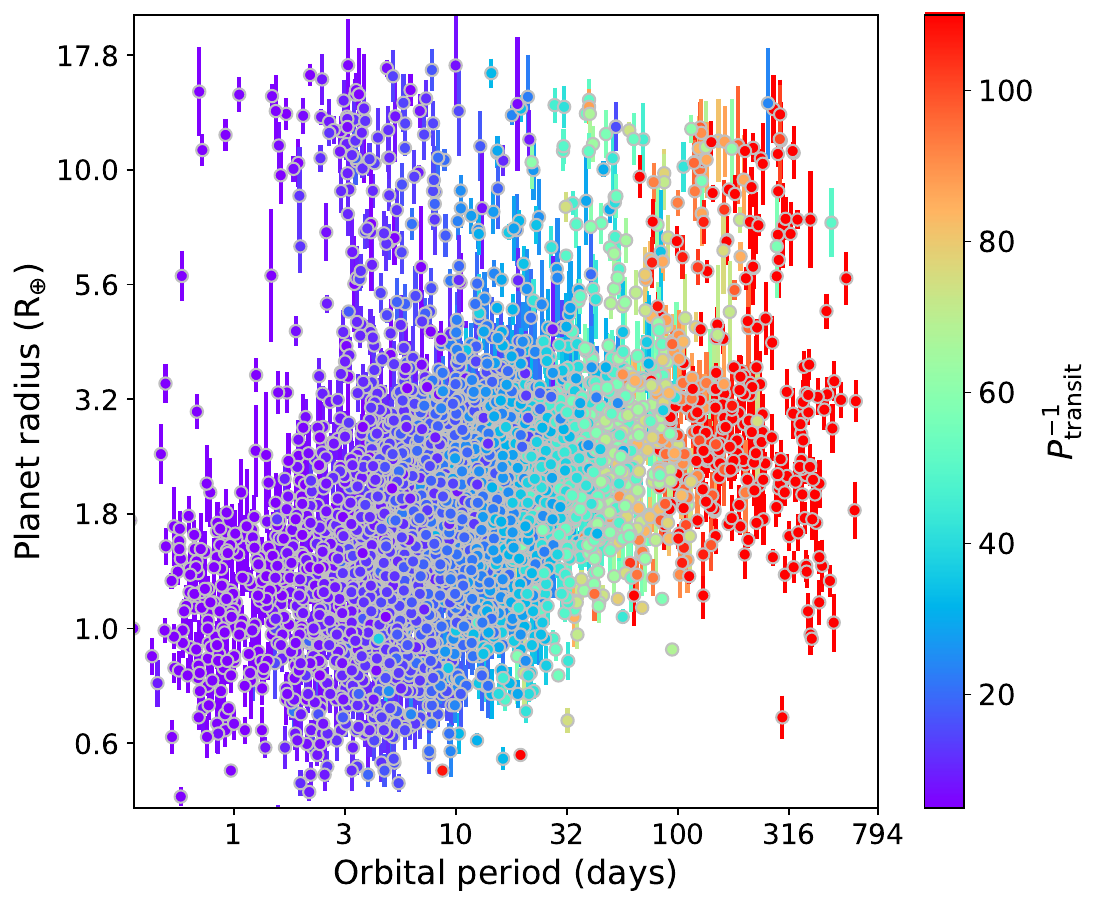}
    \includegraphics[width=0.49\textwidth]{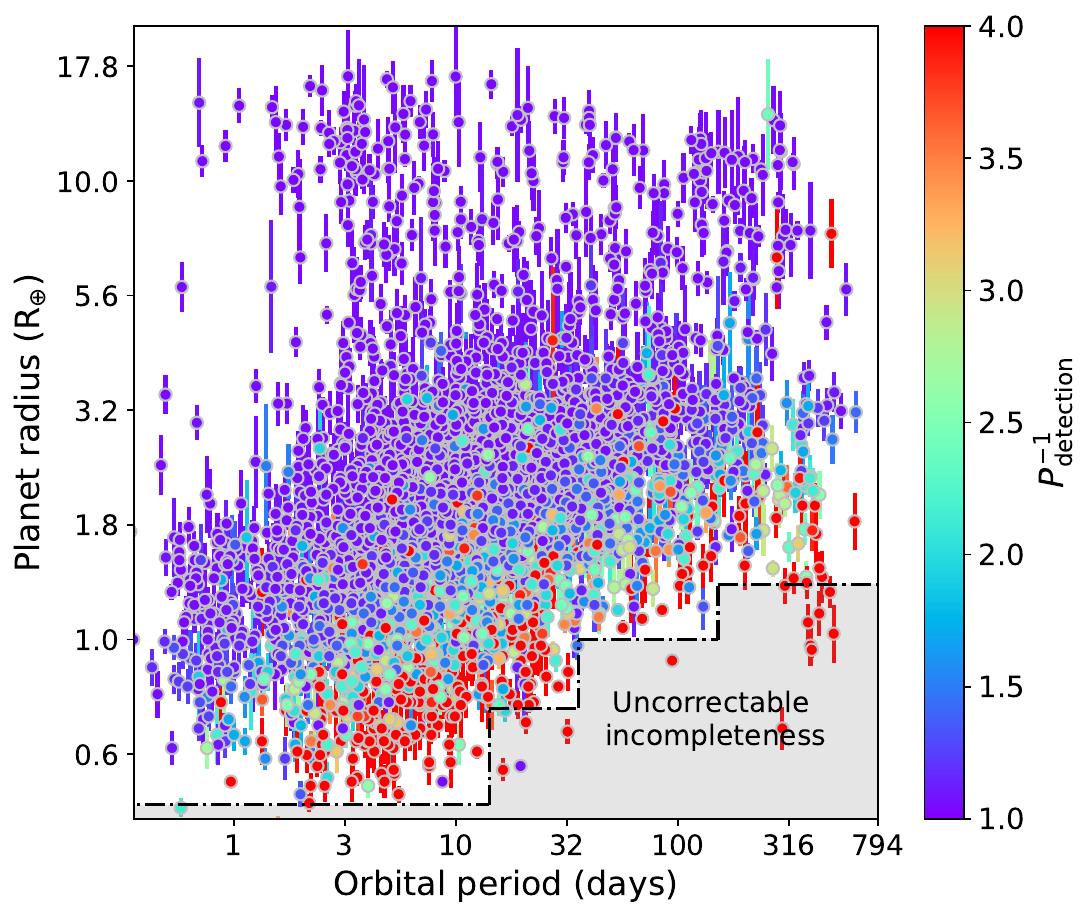}
    \caption[Planet radius as a function of orbital period of the \textit{Kepler} DR25 catalogue, where the colour coding indicates the transit and detection probabilities.]{Planet radius as a function of orbital period of the \textit{Kepler} DR25 final catalogue. The colour coding indicates the transit probabilities and the detection probabilities obtained in Sect.~\ref{sec:sample_selection_biases_mitigation}.}
    \label{fig:probs}
\end{figure*}

We built a sample of planets and candidates based on  \textit{Kepler} \citep{2010Sci...327..977B}, since it is the only survey allowing for a complete planet occurrence study ranging from sub-Earths to Jupiters \citep[e.g.][]{2011ApJ...742...38Y}. In particular, we considered the final \textit{Kepler} catalogue \citep[DR25;][]{2018ApJS..235...38T}. This sample is affected by two main observational biases: non-transiting orbital inclinations and insufficient photometric precision. The correction of these biases has been addressed by several works using the inverse detection efficiency method (IDEM), which estimates the probabilities that a particular planet could have been detected orbiting the observed stars \citep[e.g.][]{2012ApJS..201...15H,2012ApJ...753...90M,2013ApJS..204...24B,2013ApJ...766...81F,2013ApJ...767L...8K,2013PNAS..11019273P,2015ApJ...810...95C,2015ApJ...807...45D,2015ApJ...798..112M}. However, these planet-by-planet probabilities are not typically available, and instead occurrence studies provide the planet occurrences in pre-defined period-radius bins. We followed the IDEM approach to compute such probabilities and made them public to facilitate both the reproducibility of our results and the execution of occurrence studies in other regions of the parameter space. We would like to warn readers, however, that this method has been found to be imprecise near the detection threshold, that is, for small planets at large orbital distances, where there are few detections (see Fig.~\ref{fig:probs}). Hence, we do not recommend using our IDEM probabilities to study occurrences in this region of the parameter space. To that aim, more sophisticated techniques based on extrapolations have been explored \citep[e.g.][]{2014ApJ...795...64F,2014ApJ...791...10M,2018AJ....155..205H,2020AJ....159..248K,2021AJ....161...36B}. 

The geometric probability of observing a transit can be written as a function of the stellar and planetary radii ($R_{\star}$ and $R_{\rm p}$, respectively) and the planetary semi-major axis ($a$):
\begin{equation}
    P_{\textrm{transit}} = \frac{(R_{\star} + R_{\rm p})}{a} \simeq \frac{R_{\star}}{a}.
\end{equation}
In Table~\ref{tab:weights}, we include the inverse of these probabilities $P_{\textrm{transit}}^{-1}$ for the \textit{Kepler} DR25 catalogue. In Figure \ref{fig:probs}, we show the period-radius diagram colour-coded according to $P_{\textrm{transit}}^{-1}$, which indicates the expected dependence on the orbital period. 

The probability of detecting a transiting planet depends on several factors, namely, the instrumental precision, stellar properties (e.g. size, brightness, and activity), and planetary properties (e.g. radius and orbital period). The combination of all of these factors produces a transit signal whose strength is typically quantified through its signal-to-noise ratio (S/N). We adopted the definition introduced by the \textit{Kepler} team, since it allows us to take advantage of the photometric noise measured in different timescales by the \textit{Kepler} pipeline \citep{2010ApJ...713L..87J}:
\begin{equation}
    \rm S/N = \frac{\delta}{\sigma_{CDPP}} \sqrt{N},
\end{equation}
where $\delta$ is the transit depth ($\delta$ = $R_{\rm p}^{2} / R_{\star}^{2}$), $N$ is the number of observed transits, and $\sigma_{\rm CDPP}$ is the combined differential photometric precision (CDPP), which indicates the empirical root-mean-square along the transit duration\footnote{The \textit{Kepler} pipeline computes $\sigma_{\rm CDPP}$ in timescales ranging from 1.5 to 15 hours, so we chose the $\sigma_{\rm CDPP}$ corresponding to the timescale closest to the measured transit duration.}. 

The \textit{Kepler} pipeline considers a transit-like signal as a planet candidate if it meets the criterion S/N $>$ 7.1. However, not all planets that meet this criterion are detected with 100$\%$ probability. \citet{2015ApJ...810...95C,2020AJ....160..159C} performed injection-recovery tests and found that the signal recoverability of the \textit{Kepler} pipeline is well described by a $\Gamma$ cumulative distribution, which has the form
\begin{equation}
    p\big(x|a,b,c\big) = \frac{c}{b^{a} \Gamma \big(a\big)} \int_{0}^{x} t^{a-1} e^{-t/b}.
\end{equation}
We considered the best-fit coefficients $a$ = 33.54, $b$ = 0.2478, and $c$ = 0.9731 as computed by \citet{2020AJ....160..159C} for the DR25 catalogue. This corresponds to a 19.8$\%$ recovery rate of signals with S/N =  7.1, in contrast to the assumed 50$\%$ rate in earlier \textit{Kepler} occurrence works.

For each planet, the fraction of stars around which it would have been detected can be written as
\begin{equation}
    P_{\rm detection} = \frac{\sum_{i=1}^{N_{s}}p_{i}}{N_{s}},
\end{equation}
where $N_{s}$ stands for the number of observed stars. In Table~\ref{tab:weights}, we include the obtained values of $P_{\rm detection}$. In Fig.~\ref{fig:probs}, we plotted the period-radius diagram colour-coded according to $P_{\rm detection}$. 

We corrected each detection for its incompleteness (geometric and detectability) by assigning it a weight,
\begin{equation}
    w = P_{\rm transit}^{-1} \times P_{\rm detection}^{-1} ,
\end{equation}
which we also include in Table~\ref{tab:weights}. We note that this approach allowed us to obtain a complete sample in the regions of the period-radius space where the detection probability is low, yet several detections have been achieved under favourable stellar conditions (i.e. low $\sigma_{\rm CDPP}$). However, the sample remains incomplete in the regions with low detection probabilities and no detections. This is the case for the lower-right region of the period-radius diagram. Building a complete sample involves a trade-off between larger orbital periods and smaller planetary radii. As our aim here is to study the close-in planet population, we selected the \textit{Kepler} detections with $P_{\rm orb}$ $<$ 30 days and $R_{\rm p}$ $>$~1~$\rm R_{\rm \oplus}$, which allowed us to avoid sparsely populated regions near the detection threshold (see Fig.~\ref{fig:probs}, right panel). We note that a certain fraction of candidates in the \textit{Kepler} DR25 catalogue can be false positives (e.g. brown dwarfs or eclipsing binaries). We explored the potential impact of these false detections using two different approaches. On the one hand, we assigned each planet candidate an additional weight based on its false positive probability (FPP) as estimated by \texttt{vespa}\footnote{The FPPs of the candidates were retrieved from the NASA Exoplanet Archive: \url{https://exoplanetarchive.ipac.caltech.edu/cgi-bin/TblView/nph-tblView?app=ExoTbls&config=koifpp}} \citep{2016ApJ...822...86M}. On the other hand, we restricted our sample to candidates statistically validated as planets. In both cases, the computed occurrences are consistent with those from the original DR25 catalogue considered here. This result is in agreement with the work by \citet{2020AJ....159..279B}, who explored the effect of vetting completeness on \textit{Kepler} planet occurrences. The authors found that correcting for reliability can impact the occurrence rates near the detection limit, at orbital periods longer than 200 days and radii smaller than 1.5 $\rm R_{\oplus}$, a parameter space far beyond the range analysed here. 

\section{Occurrence across the orbital period space}
\label{sec:oc}

Transiting planets are commonly classified into three groups according to their radii \citep[e.g.][]{2012ApJS..201...15H,2019PNAS..116.9723Z}: small planets (also known as sub-Neptune planets; $R_{\rm p}$ $<$ 4$\rm R_{\oplus}$), gas giants (also known as Jupiter-size planets; $R_{\rm p}$ $>$ 10$\rm R_{\oplus}$), and intermediate-size planets (also known as Neptunian planets; 4$\rm R_{\oplus}$ $<R_{\rm p}$ $<$ 10$\rm R_{\oplus}$). In this section, we study planet occurrences across the orbital period space, intending to find where and how the Neptunian desert transitions into the savanna. We aim to study specific features of intermediate-size planets, so we focus our analysis on a reduced region of the commonly adopted radius range to minimise potential contamination from the adjacent populations. In Sect.~\ref{sec:oc_nep}, we study the distribution of Neptunian planets with radii 5.5$\rm R_{\oplus}$ $<R_{\rm p}$ $<$ 8.5$\rm R_{\oplus}$. For completeness, in Sect.~\ref{sec:oc_giant_small} we examine the Jupiter-size ($R_{\rm p}$ $>$ 10$\rm R_{\oplus}$) and sub-Neptune planet regimes ($R_{\rm p}$ $<$ 4$\rm R_{\oplus}$), and in Sect.~\ref{oc:trans} we explore the Neptunian regions adjacent to the giant and small planets' regimes, which we refer to as the frontier Neptunian regimes (4$\rm R_{\oplus}$ $<R_{\rm p}$ $<$ 5.5$\rm R_{\oplus}$ and 8.5$\rm R_{\oplus}$ $<R_{\rm p}$ $<$ 10$\rm R_{\oplus}$).

\begin{figure*}
    \centering
    \includegraphics[width=0.4743\textwidth]{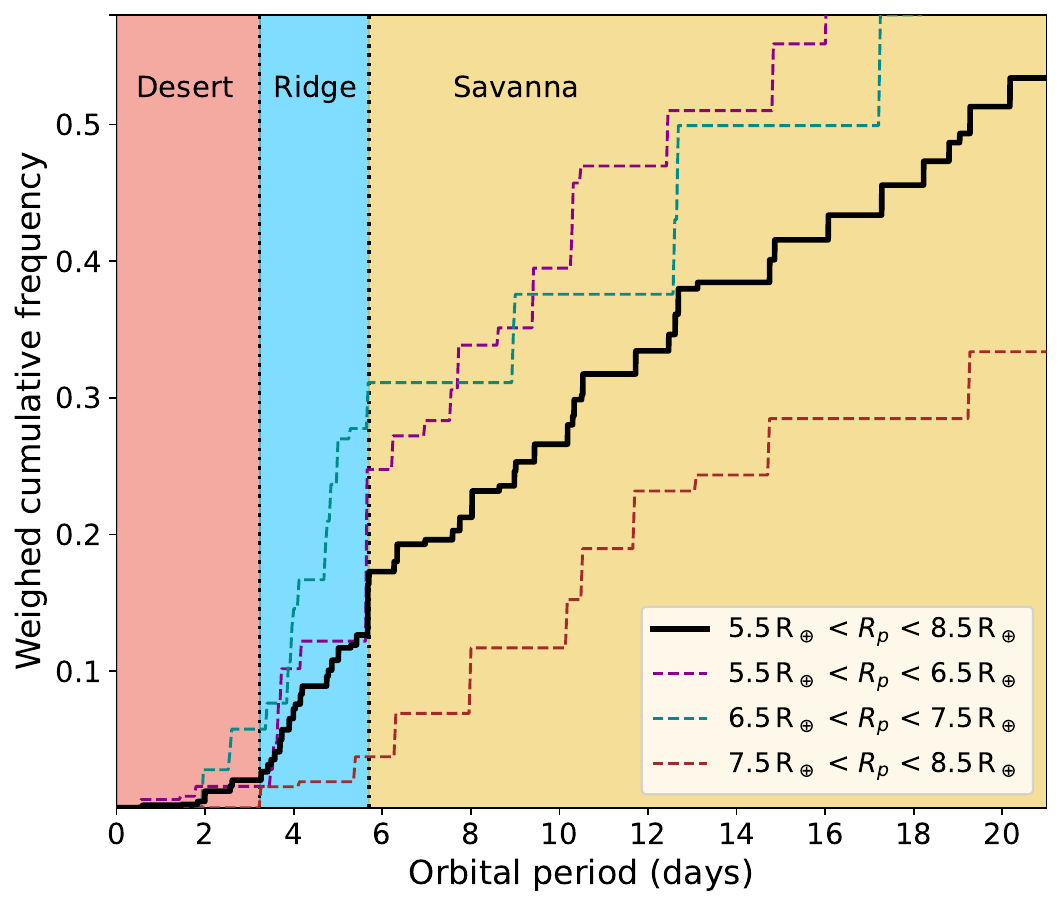}
    \includegraphics[width=0.486\textwidth]{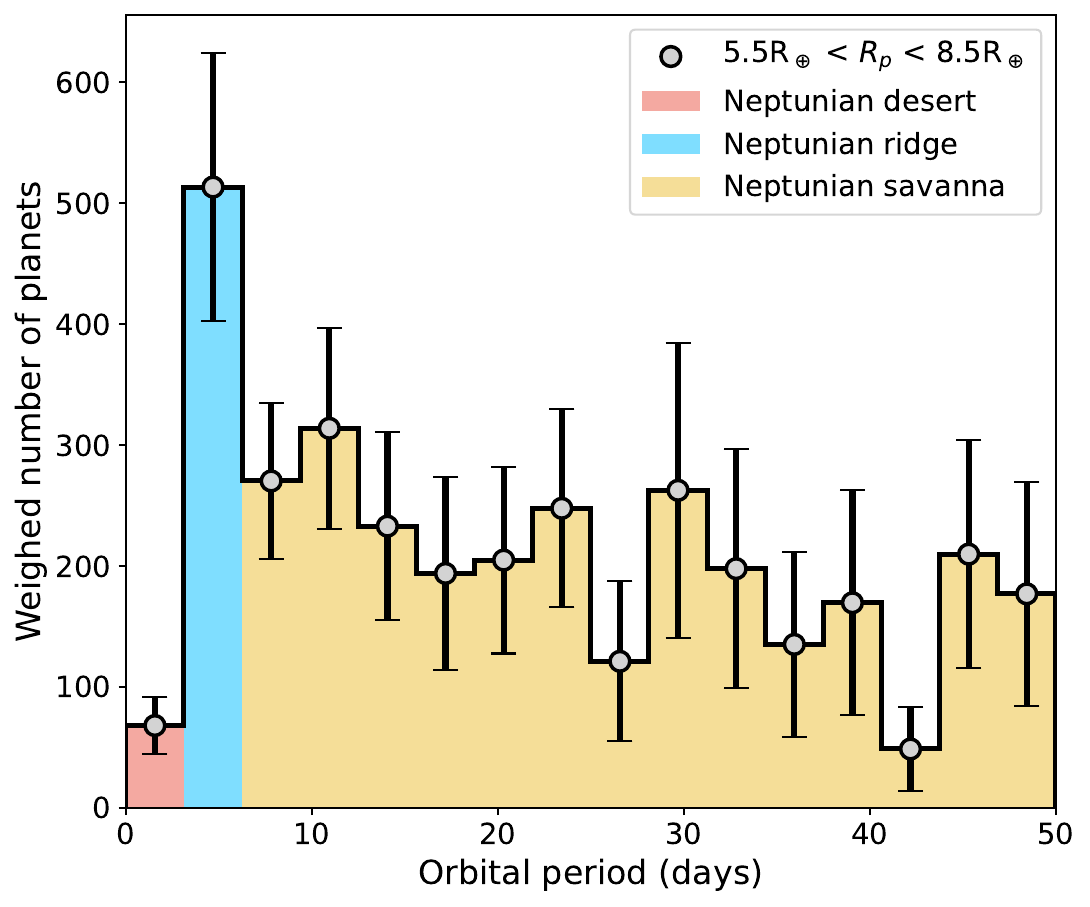}
    \caption[Distribution of Neptunian planets across the orbital period space.]{Distribution of Neptunian planets across the orbital period space, where three regimes are differentiated: a significant deficit of planets at periods $\lessapprox$ 3.2 days (i.e the Neptunian desert), a moderately populated region at periods $\gtrapprox$ 5.7 days (i.e. the Neptunian savanna), and an over-density of planets between these regimes (i.e. the Neptunian ridge). The histogram error bars were computed as the square root of the quadratic sum of the weights. }
    \label{fig:oc:nep}
\end{figure*}

\subsection{Distribution of Neptunian planets}
\label{sec:oc_nep}

In Fig.~\ref{fig:oc:nep}, left panel, we plotted the weighed cumulative frequencies across the orbital period space of Neptunian planets with radii 5.5$\rm R_{\oplus}$ $<R_{\rm p}$ $<$ 8.5$\rm R_{\oplus}$. We find three well-differentiated regimes: a quasi-flat region at $P_{\rm orb} \lessapprox 3.2$ days, a steep frequency increase at 3.2 days $ \lessapprox P_{\rm orb}$ $\lessapprox$ 5.7 days, and a milder increase at $P_{\rm orb} \gtrapprox$ 5.7 days. This distribution validates the existence of the desert and the savanna as two differentiated features and reveals an abrupt transition between both regimes. This transition corresponds to an over-density of planets (with respect to both the desert and the savanna), which we propose to name the Neptunian `ridge'. Based on the cumulative frequencies, we built the weighted histogram by considering bin sizes that allow for differentiation between the three regimes (Fig.~\ref{fig:oc:nep}, right panel). This distribution contrasts with the original desert boundaries of \citet{2016A&A...589A..75M}, which range from $\simeq$6 days to $\simeq$15 days (depending on the planet radius) in the Neptunian domain. We note that \citet{2016A&A...589A..75M} mentioned that even though their boundaries intersected at this large orbital period, it was not clear that the desert extended up to it
since the picture was not clear by that time. We also split the Neptunian occurrence into three adjacent radius chunks of 1$\rm R_{\oplus}$ width. We find that the individual distributions of the two chunks in the 5.5$\rm R_{\oplus}$$<$ $R_{\rm p}$ $<$ 7.5$\rm R_{\oplus}$ range peak at the ridge. However, the distribution of the chunk in the 7.5$\rm R_{\oplus}$$<$ $R_{\rm p}$ $<$ 8.5$\rm R_{\oplus}$ range is more homogeneous and does not peak at the ridge. The lower range dominates the complete distribution, and the upper range does not show relevant features. We warn that the number of planets in these individual chunks is too small to reach statistically significant conclusions, but we highlight a possible fading of the ridge in the upper end of the considered radius range. 

We estimated the significance of the Neptunian desert, ridge, and savanna by means of the t-statistic. We propagated the bin uncertainties as $N_{\rm bins}^{-1} (\sum_{i = 1}^{N_{\rm bins}} \delta_{i}^{2})^{1/2}$, with $N_{\rm bins}$ being the number of bins, and $\delta_{i}$ = ($\sum_{j = 1}^{N_{\rm p}} w_{j}^{2})^{1/2}$, where $N_{\rm p}$ is the number of planets detected in each bin. The Neptunian ridge stands out at a 4.7$\sigma$ level above the desert and at a 3.5$\sigma$ level above the savanna. The savanna stands out above the desert at a 4.7$\sigma$ level. We also computed the occurrence fraction between the regimes and find $f_{\rm ridge/desert}$ = 8 $\pm$ 3, $f_{\rm ridge/savanna}$ = 2.7 $\pm$ 0.5, and $f_{\rm savanna/desert}$ = 3.0 $\pm$ 0.9. Overall, both the occurrence fractions and the t-values indicate that the Neptunian ridge describes a region where close-in Neptunes are preferentially found, and it marks the boundary of the Neptunian desert through an abrupt occurrence drop.
\subsection{Distribution of Jupiter-size and sub-Neptune planets}
\label{sec:oc_giant_small}

The distribution of Jupiter-size planets ($R_{\rm p}$ $>$ 10$\rm R_{\oplus}$) has substantial similarities with that of Neptunes. The weighed cumulative frequencies show a steep increase at 3.2 days $\lessapprox$ $P_{\rm orb}$ $\lessapprox$ 5.8 days, which leads to a mild increase at larger orbital periods (Fig.~\ref{fig:oc_jup}). This over-density in the Jupiter-size domain was noticed shortly after the first discoveries and is commonly known as the hot Jupiter pileup \citep[e.g][]{2007ARA&A..45..397U,2009ApJ...693.1084W}. While similar, the Jupiter-size and Neptunian planet occurrences do show some differences. The occurrence fraction between the hot Jupiter pileup and warm Jupiters at longer orbital periods is larger than for Neptunian planets: $f_{\rm pileup/warm}$ = 5.3 $\pm$ 1.1 versus $f_{\rm ridge/savanna}$ = 2.7 $\pm$ 0.5. Taking the pileup (ridge) density as a reference for warm Jupiters (Neptunes), we find an occurrence ratio between the Neptunian savanna and the warm Jupiter regime of $f_{\rm savanna/warm}$ = 2.0 $\pm$ 0.6.  Another difference is that the hot Jupiter pileup does not abruptly drop to a desert of planets. The distribution of sub-Neptune planets ($R_{\rm p}$ $<$ 4$\rm R_{\oplus}$) is much more homogeneous than that of Neptunian and Jupiter-size planets, and it does not show an over-density in the 3-5 day period range (Fig.~\ref{fig:oc_jup}).

\subsection{Distribution of frontier Neptunian planets}
\label{oc:trans}

In Fig.~\ref{fig:oc_transitional}, we show the occurrence of frontier Neptunian planets located near the giant and small
planet regimes. On the one hand, the distribution in the 4$\rm R_{\oplus}$ $<R_{\rm p}$ $<$ 5.5$\rm R_{\oplus}$ radius range does not show remarkable features, similarly to the distribution of sub-Neptunes in the upper radius end (see Figs.~\ref{fig:oc_jup} and \ref{fig:oc_transitional}). We note that the number of planets detected in this radius range is large enough (i.e. the bin uncertainties are small enough) to have significantly detected a ridge with the occurrence contrast observed in Sect.~\ref{sec:oc_nep}. On the other hand, the distribution in the 8.5$\rm R_{\oplus}$ $<R_{\rm p}$ $<$ 10$\rm R_{\oplus}$ radius range shows a clear over-density of planets at $\simeq$3-5 days, which drops at larger orbital distances (Fig.~\ref{fig:oc_transitional}), similarly to the adjacent Neptunian ridge (see Sect.~\ref{sec:oc_nep}) and hot Jupiter pileup (see Sect.~\ref{sec:oc_giant_small}).

\section{Desert boundaries}
\label{sec:boundaries}

\begin{figure}
    \centering
    \includegraphics[width=0.65\textwidth]{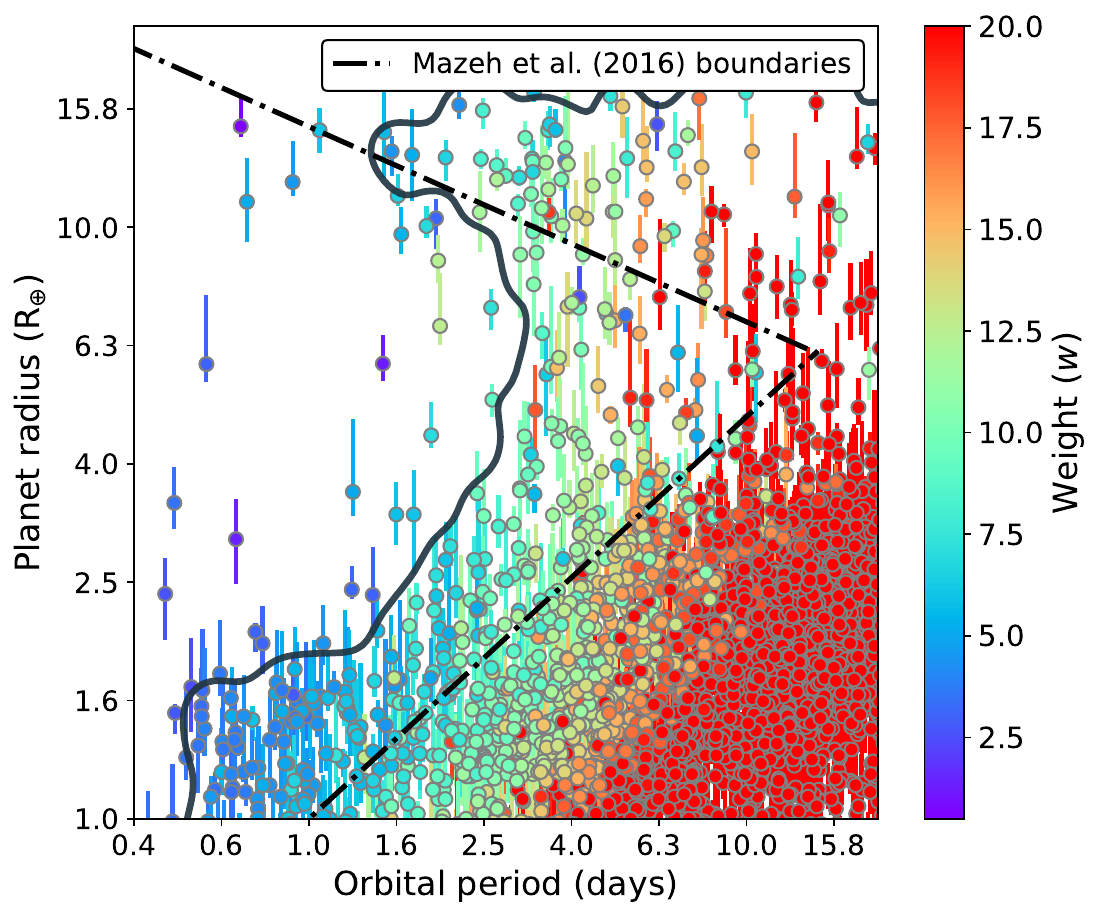}
    \caption[Planet radius versus orbital period of the \textit{Kepler} DR25 catalogue, where the contour line represents the lowest percentile our dataset is sensitive to.]{Planet radius versus orbital period of the \textit{Kepler} DR25 catalogue. Each detection is coloured according to the assigned weight to correct for observational biases. The contour represents the lowest percentile that our dataset is sensitive to.}
    \label{fig:our_desert}
\end{figure}

In Sect.~\ref{sec:oc}, we found that the occurrence of Neptunian planets drops to a desert at $P_{\rm orb} \simeq$ 3.2 days. However, the identification of such a desert in the Jupiter-size and sub-Neptune planet regimes cannot be limited to studying occurrences across the orbital period space. In these regimes, the planet distribution at short orbital distances varies strongly as a function of planet radius, so setting a fixed period boundary is not possible.

We aimed to find a population-based desert that is self-consistent in the entire radius range of our \textit{Kepler} sample. To do so, we estimated the probability density function (PDF) of the 2D period-radius distribution corrected for observational biases. We obtained the PDF through a kernel density estimation \citep[KDE;][]{Parzen_1962}. This method is one of the most widely used non-parametric approaches to estimating the underlying PDF of a dataset, and previously has been used to infer planet occurrence distributions \citep[e.g.][]{2014ApJ...791...10M,2023AJ....166..122D}. The KDE method relies upon a functional form (i.e. kernel) that is associated with each data point, and a bandwidth that controls the PDF smoothness. In most cases, the kernel choice has negligible influence on the computed PDF \citep[e.g.][]{2017arXiv170403924C}. We confirmed that this is true for our dataset by testing several kernel options (e.g. Gaussian, uniform, linear, cosine, and exponential), and arbitrarily selected a Gaussian kernel for our final estimation. We considered the rule of thumb introduced by \citet{1986desd.book.....S} to obtain an optimum bandwidth for our dataset. We also tested varying the optimised bandwidth by different factors between 0.5 and 2, and found compatible PDFs. In Fig.~\ref{fig:our_desert}, we plotted the contour line corresponding to the lowest percentile we are sensitive to. For lower percentiles, we find some close-in planets that, depending on their weights, surpass the percentile threshold. Therefore, we consider this contour a good representation of the desert boundaries. 

We provide the community with these simple, ready-to-use approximations for the desert boundaries:
\begin{equation}
    \label{eq_6}
    \mathcal{L_{R}} = -0.43 \times \mathcal{L_{P}} + 1.14, \,\, \rm{if} \,\, \mathcal{L_{P}} \in [0.12, 0.47] 
\end{equation}
for the upper limit,
\begin{equation}
    \mathcal{L_{R}} = +0.55 \times \mathcal{L_{P}} + 0.36, \,\, \rm{if} \,\, \mathcal{L_{P}} \in [-0.30, 0.47]
\end{equation}
for the lower limit, and
\begin{equation}
    \mathcal{L_{P}} =+0.47, \,\, \rm{if} \,\ \mathcal{L_{R}} \in [0.61, 0.92] 
\end{equation}
\noindent for the Neptunian domain, where $\mathcal{L_{R}}$ = $\rm log_{10}(R_{\rm p}/R_{\oplus})$ and $\mathcal{L_{P}}$ = $\rm log_{10}(P_{\rm orb}/d)$. The upper and lower boundaries also lead to constant-period limits that we approximate as
\begin{equation}
\label{eq_9}
\mathcal{L_{P}} =\begin{cases} 
 +0.12 & \rm{if} \,\, \mathcal{L_{R}} \in [1.08, 1.20] \\  
 -0.30 & \rm{if} \,\, \mathcal{L_{R}} \in [-0.07, 0.20].  
 \end{cases}
\end{equation}

These boundaries show considerable differences with those derived by \citet{2016A&A...589A..75M}. Our desert is drier in all radius ranges, especially in the sub-Neptune and Neptune domains. Accounting for completeness, we find that 2.2$\%$ of the close-in planet population lies within the \citet{2016A&A...589A..75M} desert, while only 0.1$\%$ lies inside our boundaries. Hence, our desert represents the close-in region of the period-radius space where there are no planets at a 3$\sigma$ level. In addition to the desert's dryness, there is also a notable difference regarding the desert's shape. Similar to MA+16, our boundaries narrow towards a smaller radius window as the period increases. However, our revised boundaries do not penetrate into the savanna, as the desert is physically limited by the Neptunian ridge. This sets a constant-period boundary in the Neptunian regime at $\simeq$3 days, as we also discuss in Sect.~\ref{sec:oc_nep}.


\section{The Neptunian landscape as a tracer of close-in planets' origins}
\label{sec:discussion}

\begin{figure}
    \centering
    \includegraphics[width=0.65\textwidth]{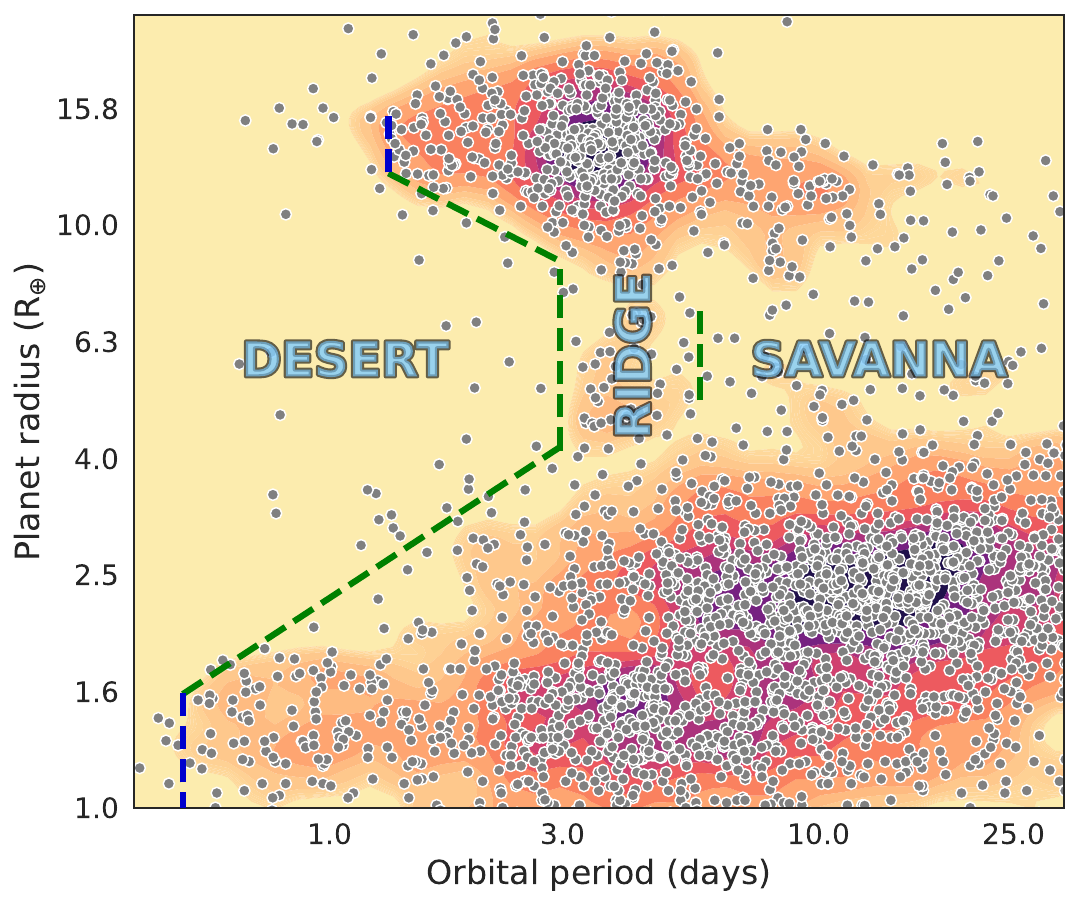}
    \caption[Planet radius as a function of orbital period for all known exoplanets, where we highlight the location of the Neptunian desert, ridge, and savanna.]{Planet radius as a function of orbital period for all known exoplanets, where we highlight the location of the Neptunian desert, ridge, and savanna derived in this work (Eqs. (\ref{eq_6}) to (\ref{eq_9})). The colour code represents the observed density of planets. This plot has been generated with \texttt{nep-des} (\url{https://github.com/castro-gzlz/nep-des}).}

    \label{fig:Per_Rad}
\end{figure}

In Fig.~\ref{fig:Per_Rad}, we highlight the location of the Neptunian desert, ridge, and savanna (Eqs. (\ref{eq_6}) to (\ref{eq_9})) in the context of all known planets. The ridge spans a period range that coincides with that of the hot Jupiter pileup ($\simeq$3-5 days), which suggests that similar evolutionary processes might act on both populations. The current observational constraints suggest that both disk-driven migration and HEM processes are needed to explain the observed properties of hot Jupiters. However, a large number of planets in the pileup have been found in elliptical orbits, and the elliptical/circular fraction increases with orbital period, which is interpreted as HEM processes being the main channel populating the pileup \citep[e.g.][]{2018ARA&A..56..175D,2021JGRE..12606629F}. In a recent study, \citet{2020A&A...635A..37C} found that Neptunian planets at the edge of the desert, with $P_{\rm orb}$ $\lessapprox$ 5 days (corresponding to the newly identified ridge), have moderate ($e$ $\lessapprox$ 0.3) but non-zero eccentricities. This suggests that HEM is also the main channel bringing Neptunes to the ridge, and that this migration process might be the main agent populating the $\simeq$3-5 day over-density observed both in the Jupiter-size and Neptunian populations. While Jupiters in the pileup would remain immune to photo-evaporation and eventually have their orbit circularised, the Neptunes observed in the ridge today may have arrived recently enough through HEM that their orbit has not yet circularised and they have survived evaporation \citep{2018Natur.553..477B,2020A&A...635A..37C,2021A&A...647A..40A}. This picture is consistent with additional dynamical and atmospheric constraints, as many Jupiter-size \citep{2012ApJ...757...18A} and Neptune-size \citep{2023A&A...669A..63B} planets in the over-density have been found on highly misaligned orbits (which is considered a tracer of HEM processes, \citealp[e.g.][]{2012ApJ...754L..36N,2017AJ....154..106N}), and several Neptunian planets within the ridge undergo strong atmospheric escape \citep[e.g.][]{2015Natur.522..459E,2018A&A...620A.147B}. 

We have found that the occurrence fraction between the hot Jupiter pileup and warm Jupiters is about twice ($f_{\rm pileup/warm}$ = 5.3 $\pm$ 1.1)  that between the Neptunian ridge and savanna ($f_{\rm ridge/savanna}$ = 2.7 $\pm$ 0.5). If we assume that migration processes distribute Jupiter-size and Neptunian planets equally, this result could be explained by photo-evaporation, which would be removing Neptunes from the ridge but not from the savanna. However, the reality is probably more complex. The eccentricities of warm Jupiters are preferentially elliptical, while the eccentricities of warm Neptunes in the savanna are preferentially circular\footnote{We refer to Figure 1 of \citet{2020A&A...635A..37C} for an eccentricity comparison between Jupiter-size and Neptunian planets.}. Since photo-evaporation does not affect warm Jupiters and is likely inefficient on warm Neptunes in most of the savanna, this implies that HEM processes act differently on warm Jupiter and Neptunes, and/or that HEM and disk-driven migration bring different fractions of Jupiter- and Neptune-size planets from beyond the ice line into the warm planet regime. The warm-Jupiter regime would be preferentially populated through HEM processes, while the Neptunian savanna would be preferentially populated through disk-driven migration. Thus, the different occurrences between the over-densities and the warm regions of Jupiter-size and Neptunian planets cannot be interpreted through a single process, since the observational constraints hint at different evaporation and migration mechanisms affecting both populations.

\section{Conclusions}
\label{sec:conclusion}

We identified an over-density of intermediate-size planets (5.5$\rm R_{\oplus}$ $<R_{\rm p}$ $<$ 8.5$\rm R_{\oplus}$) at  3.2 days $ \lessapprox P_{\rm orb}$ $\lessapprox$ 5.7 days, which we call the Neptunian ridge, as it separates the desert and savanna of close-in Neptunes. We also determined accurate, population-based boundaries for the Neptunian desert in the radius-period plane, and here provide the community with simple, ready-to-use approximations for these boundaries. 

The period range of the ridge matches that of the well-known hot Jupiter pileup ($\simeq$3-5 days), suggesting the existence of similar evolutionary pathways populating both regimes. The large number of Jupiter- and Neptune-size planets with eccentric, misaligned orbits in this overdensity further suggests that it is populated primarily by HEM processes. In contrast, the larger fraction of warm Jupiters with eccentric orbits, compared to warm Neptunes on circular orbits in the savanna, suggests that HEM and disk-driven migration act differently on both populations. The different relative fraction of Jupiters and Neptunes in the over-density, compared to the warm regime, further suggests that HEM is more efficient at bringing Jupiter-size planets closer in, and/or that we only detect Neptunes in the ridge that have survived evaporation because they migrated recently.

These hypotheses must be further tested through large-scale atmospheric and dynamical surveys. Spin-orbit angle surveys such as ATREIDES (Bourrier et al., in prep) will offer further insight into the relative roles of the different migration processes on the Neptunian population, while atmospheric escape surveys such as NIGHT will allow us to determine how deep into the savanna photo-evaporation remains efficient, or if the ridge also marks the threshold for the onset of hydrodynamical escape. The results of such surveys, coupled with the newly mapped landscape and numerical syntheses of the Neptunian population, will provide a clearer picture of the origins and evolution of close-in giants as a whole. 

\newpage
\chapter{TOI-5005 b: A super-Neptune in the savanna near the ridge}
\label{ch:toi_5005}
\vspace{2cm}
\pagestyle{fancy}
\fancyhf{}
\lhead[\small{\textbf{\thepage}}]{\small{\textbf{\nouppercase{\leftmark}}}}
\rhead[\small{\textbf{\nouppercase{\rightmark}}}]{\small{\textbf{\thepage}}}

\bigskip

Our current understanding of the Neptunian desert, ridge, and savanna is limited by the scarcity of observational constraints. In a first step, obtaining a large sample of confirmed planets with precise radii, masses, and orbital eccentricities is critical to infer possible differences in the formation and evolution mechanisms that gave rise to those features. In a second step, coupling the aforementioned constraints with follow-up observations of the spin-axis angle and atmospheric escape rates will provide a clearer picture of the origins and evolution of close-in Neptunes at a population level. In this regard, TESS is playing a key role. Its photometric precision is high enough to enable the detection of Neptunian planets, which typically show large transit depths of a few parts per thousand. In addition, its focus on bright stars and the monitoring of practically the entire sky is boosting the number of close-in Neptunian candidates around stars amenable to detailed follow-up studies.

This chapter is part of the HARPS-NOMADS collaboration (see Sect.~\ref{sec:ground_based}), where we confirm and characterize the close-in ($P_{\rm orb}$ = 6.3 days) super-Neptune ($R_{\rm p}$ = 6.3 $\rm R_{\rm \oplus}$) TOI-5005~b, which orbits a moderately bright ($\rm V$ = 11.8 mag) solar-type star (G2 V, $T_{\rm eff}$ = 5750 K).  With these properties, TOI-5005~b is located in the Neptunian savanna near the ridge, a poorly populated but key region of the parameter space for understanding the transition between those regimes. In Sect.~\ref{sec:observations}, we describe the TESS, HARPS, and additional PEST and TRAPPIST-South photometric observations. In Sect.~\ref{sec:stellar_charact}, we present our stellar characterisation based on the HARPS spectra. In Sect.~\ref{sec:analysis_results}, we describe our analysis of photometric and
spectroscopic data and present the derived system parameters. In Sect.~\ref{sec:discussion}, we discuss the results, and we conclude in Sect.~\ref{conclusions}.

\section{Observations}
\label{sec:observations}

\subsection{TESS high-precision photometry}
\label{sec:obs_tess}

\begin{table*}[]
\centering
\fontsize{9.2pt}{9.2pt}\selectfont
\caption{TESS observations of TOI-5005.}
\renewcommand{\arraystretch}{1.6}
\setlength{\tabcolsep}{3.5pt}
\begin{tabular}{cccccccccc}
\hline \hline
Sector & Cycle & Start date  & End date     & Camera & CCD & FFIs   & TPFs    & Cadence      & Photometry pipelines \\ \hline
12     & 1     & 21 May 2019 & 18 June 2019 & 1      & 1   & 1234    & 0       & 30 min       & QLP                  \\
39     & 3     & 27 May 2021 & 24 June 2021 & 1      & 1   & 3865 & 3865       & 10 min       & TESS-SPOC, QLP       \\
65     & 5     & 4 May 2023  & 2 June 2023  & 1      & 2   & 11663 & 19515 & 200 s, 120 s & SPOC, TESS-SPOC, QLP \\
\hline
\end{tabular}
\label{tab:TESS_summary}
\end{table*}

The star TOI-5005 (TIC 282485660) has been observed by TESS in sectors 12, 39, and 65 (hereafter S12, S39, and S65).  In Table \ref{tab:TESS_summary}, we summarise the details of the observations. The full-frame images (FFIs) of the three sectors were processed through the Quick Look Pipeline \citep[QLP;][]{2020RNAAS...4..204H}, which computed simple aperture photometry (SAP) for all sources in the TESS Input Catalogue \citep[TIC;][]{2018AJ....156..102S,2019AJ....158..138S} with magnitudes up to T = 13.5 mag. The S39 and S65 FFIs were also processed by the TESS-SPOC pipeline \citep{2020RNAAS...4..201C}, and the S65 TPFs by the SPOC pipeline \citep{2016SPIE.9913E..3EJ}. SPOC and TESS-SPOC operate at the Science Processing Operations Center (NASA Ames Research Center) under the same codebase and provide SAP \citep{twicken:PA2010SPIE,morris:PA2020KDPH} and Presearch Data Conditioned Simple Aperture Photometry (PDCSAP). The PDCSAP is the SAP processed by the PDC algorithm, which corrects
the photometry of instrumental systematics that are common to all stars in the same CCD \citep{2012PASP..124.1000S,2012PASP..124..985S}. The complete QLP, TESS-SPOC, and SPOC datasets are available at the Mikulski Archive for Space Telescopes (MAST).\footnote{\url{https://mast.stsci.edu/portal/Mashup/Clients/Mast/Portal.html}}

In January 2022, the QLP-based faint-star search pipeline \citep{2022ApJS..259...33K} detected in S39 a periodic transit-like flux decrease that was alerted by the TESS Science Office as a TESS Object of Interest (TOI-5005.01) that would benefit from follow-up observations \citep{2021ApJS..254...39G}. The detection yielded a period of $6.30847 \pm 0.00003$ days, a transit duration of $2.9 \pm 0.3$ hours, and a transit depth of $3690 \pm 4$ ppm (parts per million).

\subsubsection{\texttt{TLS} and \texttt{GLS} periodograms}

\begin{figure*}

\includegraphics[width=\textwidth]{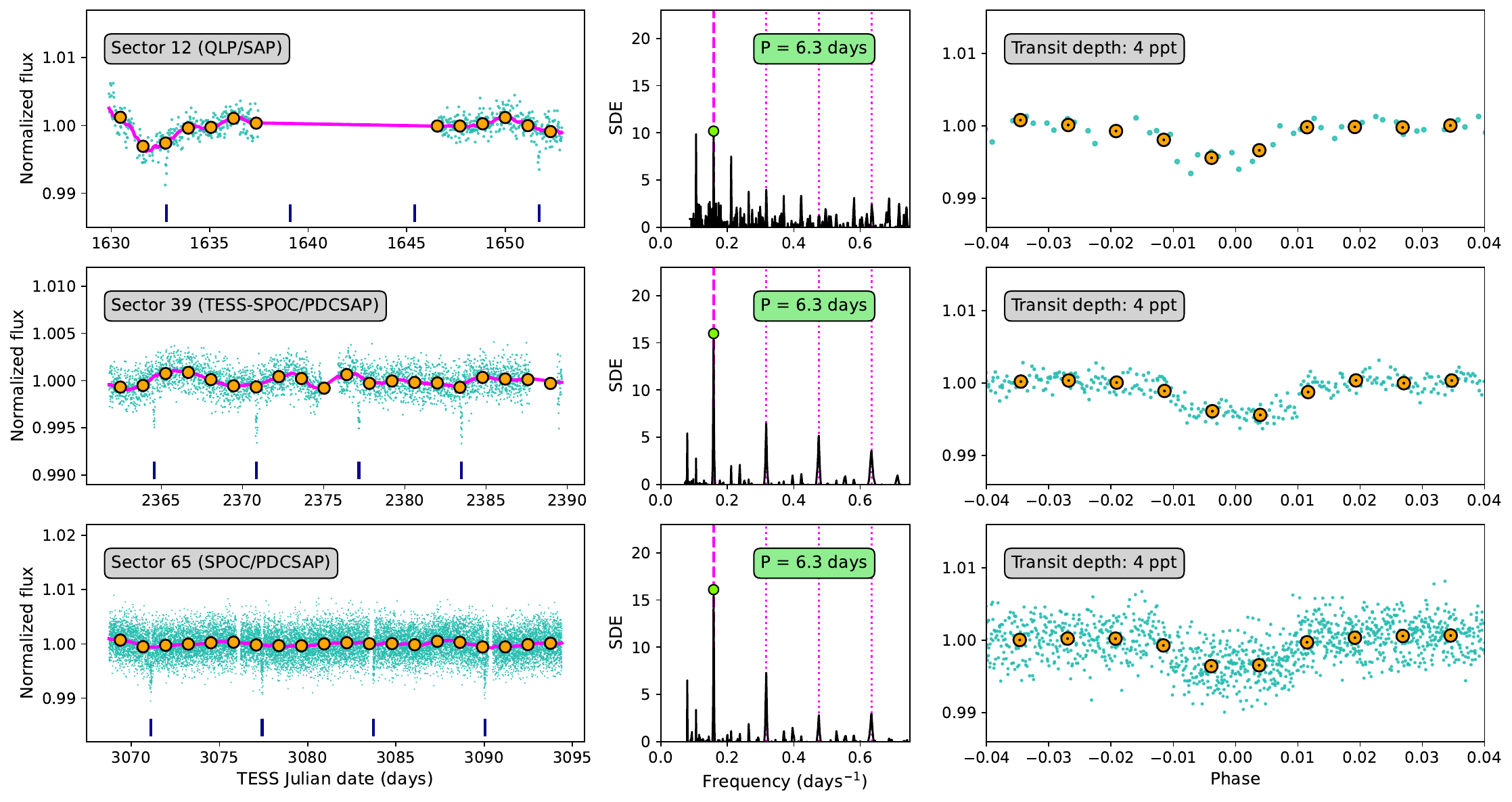}
         \caption[Transit-like signals within the TESS photometry of TOI-5005.]{Transit-like signals within the TESS photometry of TOI-5005. Left panels: Photometric time series. The magenta line is the trend used to flatten the photometry before the periodogram computation, which was obtained through the time-windowed bi-weight method implemented in \texttt{wotan} \citep{2019AJ....158..143H} with a 1-day window length. The orange circles correspond to 1.3-day binned data. Centre panels: Transit Least Squares Periodograms of the corresponding time series. The vertical magenta dashed lines indicate the orbital period of TOI-5005.01, and the vertical magenta dotted lines indicate its second, third, and fourth harmonics. Right panels: Flattened time series folded to the maximum power periods. The orange circles correspond to 1-hour binned data.}
         \label{fig:tls_to_TESS}
\end{figure*}

In Fig.~\ref{fig:tls_to_TESS}, we show the sector-by-sector transit least squares periodograms \citep[\texttt{TLS;}][]{2019A&A...623A..39H} of the QLP/SAP (S12), TESS-SPOC/PDCSAP (S39), and SPOC/PDCSAP (S65) photometry flattened; that is, de-trended from low-frequency trends of stellar or instrumental origin. In the three sectors, the maximum power peak corresponds to a 6.3-day periodicity, which coincides with that of TOI-5005.01. This peak has a signal detection efficiency (SDE) of 10.3 in S12, 16.0 in S39, and 16.5 in S65. Those SDE are above the commonly used empirical thresholds for transit detection, namely, SDE $>$ 6.0 \citep{2015ApJ...807...45D}, SDE $>$ 6.5 \citep{2018AJ....156..277L}, SDE $>$ 7 \citep{2012ApJ...761..123S}, and SDE $>$ 10 \citep{2018MNRAS.473L.131W}. Therefore, although TOI-5005.01 was not announced until S39 was observed, we find that its transit-like signature can be detected in the three sectors, with two transit events in S12, four in S39, and three and a half in S65. We also computed the \texttt{TLS} periodogram of the complete dataset and recovered the 6.3-day signal with an SDE of 51.8. We repeated the process with the TOI-5005.01 transits masked to search for additional transit-like signatures and found no further significant periodicities. We note that the SPOC pipeline also detected the 6.3-day transit signal with a multiple event detection statistic (MES) of 21.7 in S39 and 13.3 in S65 \citep{2002ApJ...575..493J,2020TPSkdph}, which correspond to model-fitted signal-to-noise ratios (S/N) of 24.5 and 18.3, respectively \citep{Twicken:DVdiagnostics2018,Li:DVmodelFit2019}.

\begin{figure*}
    \includegraphics[width=\textwidth]{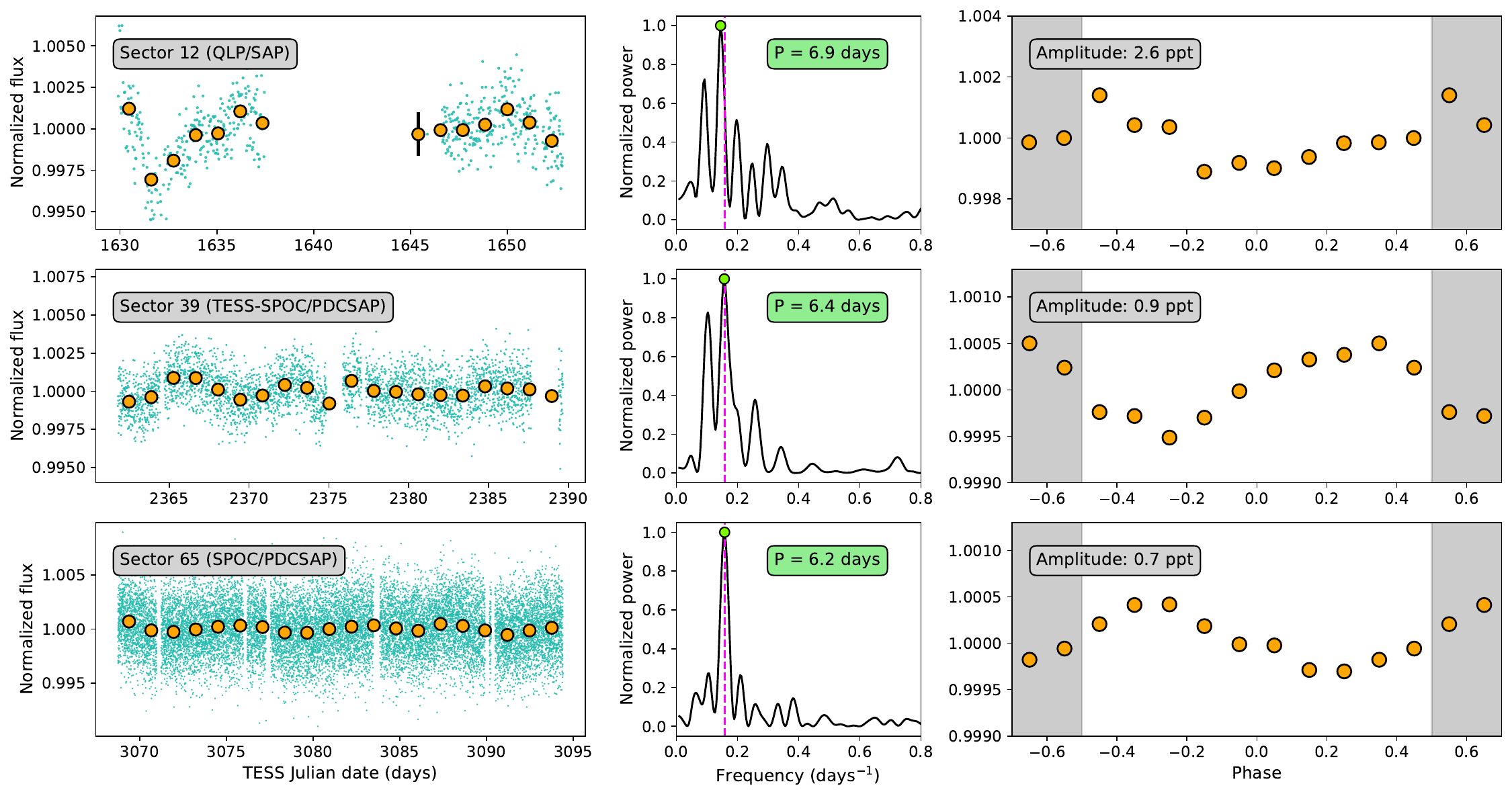}
         \caption[Sinusoidal signals within the TESS photometry of TOI-5005.]{Sinusoidal signals within the TESS photometry of TOI-5005. Left panels: Photometric time series without the TOI-5005.01 transits. The orange circles correspond to 1.3-day binned data. Centre panels: Generalised Lomb-Scargle Periodograms of the time series. The vertical magenta dashed lines indicate the orbital period of TOI-5005.01. The green circles and boxes indicate the maximum power frequencies. Right panels: Photometric time series folded to the orbital period of TOI-5005.01. The orange circles correspond to data binned over phase windows of 0.1 in width.} 
         \label{fig:gls_to_TESS}
\end{figure*}

In Fig.~\ref{fig:gls_to_TESS}, we show the sector-by-sector generalized Lomb-Scargle periodograms \citep[\texttt{GLS;}][]{2009A&A...496..577Z} of the TESS time series after masking the TOI-5005.01 transit-like features. The periodograms of the systematics-corrected time series (i.e. PDCSAP of S39 and S65) show maximum power periods of 6.4 and 6.3 days with False Alarm Probabilities (FAPs) below $10^{-40}\,\%$. These periodicities coincide with the orbital period of TOI-5005.01. In the right panels of Fig.~\ref{fig:gls_to_TESS}, we show the TESS photometry folded in phase to the orbital period of TOI-5005.01, which illustrates the existence of a sinusoidal-like modulation of 0.9 ppt (parts per thousand) in S39 and 0.7 ppt in S65 synchronised with the orbit of the planet candidate. We note, however, that while the signal periodicity persists, the signal phase is modified from one sector to another. The periodogram of the systematics-uncorrected QLP/SAP photometry shows a maximum power period of 6.9 days with a FAP below $10^{-30}\,\%$. This periodicity differs from that of the TOI-5005.01 orbit by 8$\%$, but the photometry folded to its orbital period shows a tentative sinusoidal behaviour. With an orbital period of 6.3 days, TOI-5005.01 is located within the sub-Alfvénic radius of its parent star. Therefore, this potentially synchronized signal suggests a planet-induced origin most likely due to magnetic star-planet interactions \cite[MSPIs; e.g.][]{2000ApJ...533L.151C,2003A&A...406..373S,2003ApJ...597.1092S,2009EM&P..105..373P,2019NatAs...3.1128C,2024A&A...684A.160C}. This signal also seems to be present within the systematics-uncorrected S39 and S65 SAP fluxes. In the S39 SAP fluxes, the 6.3-day periodicity appears as the second-highest peak. In the S65 SAP fluxes, the signal does not appear, but after performing a simple linear detrending, the maximum power period also indicates a 6.0-day periodicity. Consistently, the S39 and S65 SAP fluxes folded in phase to the planetary orbit show a sinusoidal modulation. We illustrate the \texttt{GLS} periodograms of the S39 and S65 SAP fluxes in Fig.~\ref{fig:gls_to_TESS_sap}. In this chapter, we delve into the PDC correction to unveil whether the prominent 6.3-day signal has a true stellar origin or if it could have been generated by residual uncorrected systematics. Before doing such an analysis, it is important to assess the possibility of flux contamination within the photometric aperture. Given the large TESS pixel sizes (21" $\times$ 21"), it is very common that the measured fluxes do not come from the target star exclusively, but also from nearby stars, which could eventually be the sources of the signals found. Indeed, the TESS photometry of TOI-5005 receives flux contributions from several nearby stars.

\subsubsection{The \texttt{TESS-cont} algorithm}
\label{sec:tess-cont}
\begin{figure*}
    \includegraphics[width=0.48\textwidth]{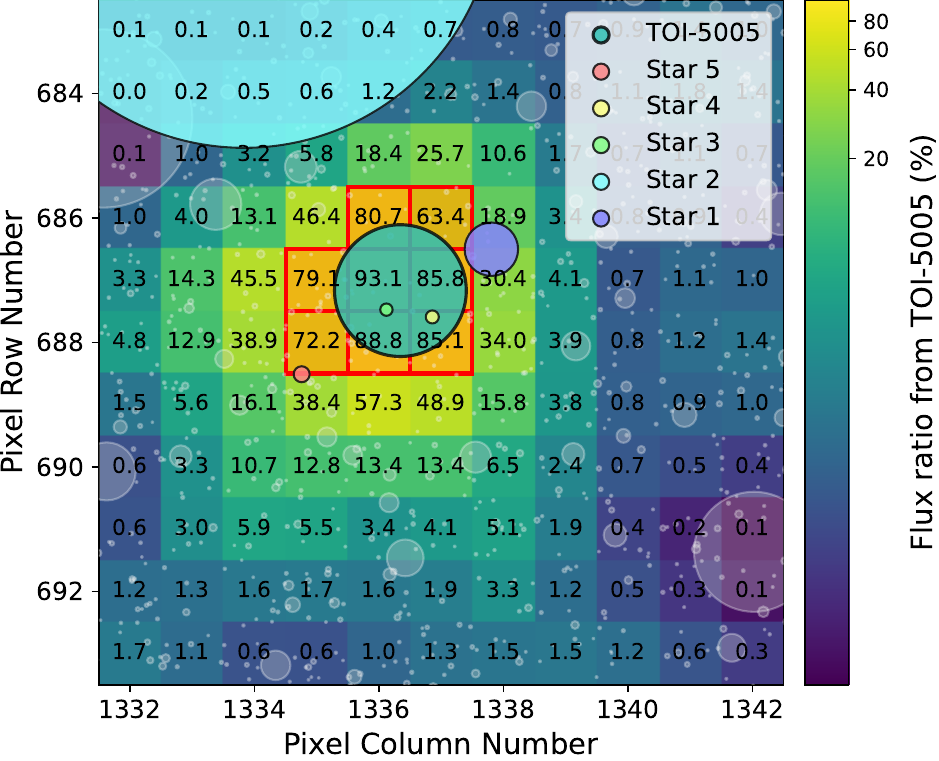}
    \includegraphics[width=0.48\textwidth]{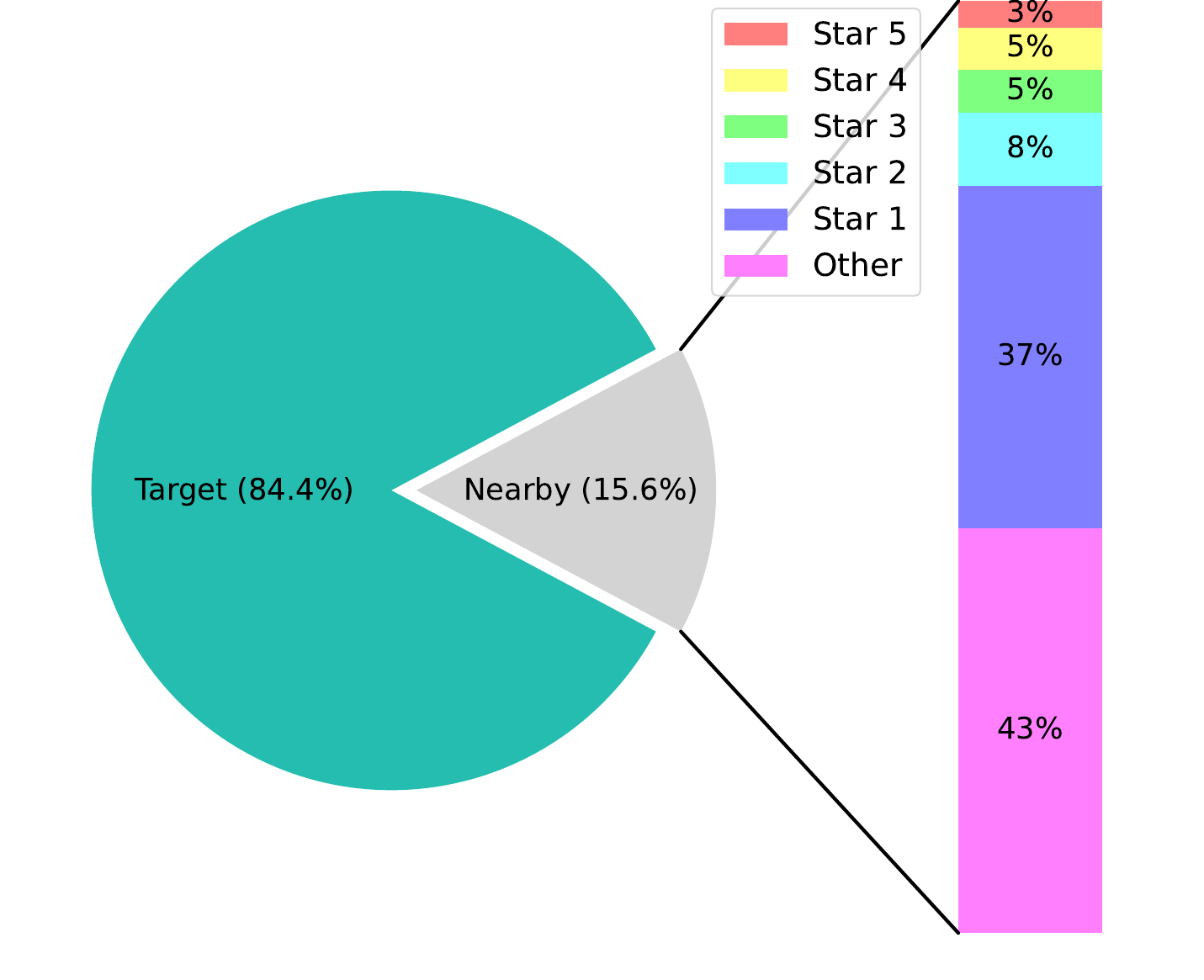}
    \caption[Nearby sources contaminating the photometry of TOI-5005.]{Nearby sources contaminating the TOI-5005 photometry. Left: TPF-shaped heat-map with the pixel-by-pixel flux fraction from TOI-5005 in S65. The red grid is the SPOC aperture. The pixel scale is 21" $\rm pixel^{-1}$. The white disks represent all the \textit{Gaia} sources, and the five sources that most contribute to the aperture flux are highlighted in different colours. The disk areas scale with the emitted fluxes. Right: Flux contributions to the SPOC aperture from the target and most contaminant stars. This plot was created through \texttt{TESS-cont} (\url{https://github.com/castro-gzlz/TESS-cont}).}
    \label{fig:TESS-cont}
\end{figure*}

We developed a Python package to quantify the flux contribution from nearby sources in the TESS photometry: \texttt{TESS-cont}.\footnote{Available at \url{https://github.com/castro-gzlz/TESS-cont}.} In this section, we describe the main aspects of its operation.

The \texttt{TESS-cont} algorithm (1) identifies the main contaminant sources, (2) quantifies their individual and total contributions to the selected aperture (i.e. SPOC or custom), and (3) determines whether any of these sources could be the origin of the observed transit or variability signals. The package first searches for all the nearby \textit{Gaia} DR2 \citep{2018A&A...616A...1G} or DR3 \citep{2023A&A...674A...1G} sources using the \texttt{get\_gaia\_data} function of \texttt{tpfplotter} \citep{2020A&A...635A.128A} and then constructs their Point Spread Functions (PSFs) to estimate the flux distribution across the TPF or FFI of the target star. The TESS PSFs are not Gaussian and vary across the focal plane mainly due to the optics. Therefore, instead of PSFs, TESS has Pixel Response Functions (PRFs) that better represent the flux distribution of the point sources. TESS PRFs were created by the SPOC pipeline based on micro-dithered data taken during the commissioning phase. These PRFs were built for a discrete number of CCD locations, while real sources can appear at any location. Therefore, to better represent the flux distribution of any source, \texttt{TESS-cont} uses the \texttt{TESS\_PRF} module \citep{2022ascl.soft07008B} to perform a bilinear interpolation between the four nearest SPOC PRFs. The obtained PRFs are then scaled to the stellar relative fluxes and placed in a TPF-shaped array. For each TPF pixel, \texttt{TESS-cont} computes the flux contribution from each source, and this is used to compute the total flux contributions within the photometric aperture. 
 
\subsubsection{Contamination analysis}

In Fig.~\ref{fig:TESS-cont}, we illustrate the \texttt{TESS-cont} output for TOI-5005 in S65. The left panel consists of a TPF-shaped heat-map with the pixel-by-pixel flux fraction from the target star, and the right panel shows the aperture flux contributions from the target star and main contaminant sources. The TOI-5005 flux falling inside the SPOC photometric aperture is 86.1$\%$ and 84.4$\%$ in S39 and S65, respectively, which is in agreement with the CROWDSAP metric estimated by SPOC. There is a relatively bright source ($G$ = 13.6 mag) surrounding the aperture (TIC 282476646, labelled as Star~1) that contributes 37$\%$ of the contaminant flux (i.e. 6$\%$ of the total flux). The following most contaminant sources are TIC 282485676, TIC 282476644, TIC 282478138, and TIC 282480455, which are labelled as Star~2, Star~3, Star~4, and Star~5 in Fig.~\ref{fig:TESS-cont}, and contribute 8$\%$, 5$\%$, 5$\%$, and 3$\%$ to the total contaminant flux, respectively. The remaining 43$\%$ of the contaminant flux is contributed by other nearby sources. We studied whether the dilution-corrected SPOC-derived $3690 \pm 4$ ppm transit and the 0.9 ppt sinusoidal modulation could have originated in any of the contaminant sources. To do so, we used the \texttt{TESS-cont} \texttt{DILUTION} feature to dilute the dilution-corrected transit depth assumed to come from TOI-5005, and de-blend it by considering that it comes from the contaminant sources  \citep[][]{2018AJ....156..277L,2020MNRAS.499.5416C,2021MNRAS.508..195D}. We obtain dilution-corrected transit depths of 5$\%$, 26$\%$, 43$\%$, 44$\%$, and 70$\%$ for Star~1, Star~2, Star~3, Star~4, and Star~5, respectively. For the 6.3-day sinusoidal-like signal, we obtain dilution-corrected amplitudes between 1$\%$ (Star~1) and 17$\%$ (Star~5). Therefore, given that these values are lower than 100$\%$, this analysis cannot discard that the transit and sinusoidal signals could have originated in any of the nearby contaminant sources. We note that the ground-based photometry (Sect.~\ref{sec:ground_based}) and high-resolution spectroscopy (Sect.~\ref{sec:obs_harps}) allowed us to discard these sources as the origin of the planetary signal. However, having no counterpart in independent observations, the origin of the 6.3-day photometric modulation remains uncertain. Among the five most contaminant sources, Star~1 and Star~2 have available QLP photometry at MAST. Star~1 shows no signs of stellar variability, but Star~2 exhibits a strong variability with a periodicity of 0.7~days. In addition, it shows a hint of a $\simeq$6.7-day modulation in the second half of S39. Since this modulation is not detected in S12, S65 or the first half of S39, and its amplitude is comparable to that of the 0.7-day variability, it cannot correspond to the 6.3-day persistent signal. However, it makes us suspect that the pixels near TOI-5005 could be affected by uncorrected systematics (see Sect.~\ref{subsubsec:photometric_variability} and Appendix~\hyperref[sec:cbv_correction]{E}). 

\subsubsection{Dataset selection}

In Sects.~\ref{subsec:TESS_analysis}, \ref{subsec:joint_analysis}, and \ref{subsec:stellar_signals_analysis}, we analyse the photometric signals based on SAP fluxes corrected for crowding only. We chose SAP instead of PDCSAP for two main reasons. First, we aim to investigate whether the detected sinusoidal modulation has a stellar origin or if it could have been artificially originated by residual uncorrected systematics. Second, there are no SPOC/PDCSAP fluxes for the S12 FFIs, which prevents us from building a homogeneous dataset based on PDCSAP photometry. In fact, in terms of homogeneity, QLP/SAP photometry was also extracted differently than SPOC/SAP. While SPOC/SAP is simply the sum of the calibrated TPF fluxes within a pixel-based grid aperture, QLP/SAP uses several circular apertures, and it has a higher level of processing \citep[e.g. see][]{2019ApJ...871L..24V,2020RNAAS...4..251F}. Therefore, we decided to analyse a homogeneous dataset composed of SPOC/SAP photometry (S39 and S65) and SAP photometry that we extracted similarly to SPOC (S12). The S12 photometric aperture was automatically selected by \texttt{TESS-cont} to minimise contamination; that is, we only considered pixels with a target flux contribution larger than 60$\%$, similarly to the S39 and S65 SPOC apertures. In Fig.~\ref{fig:S12_aperture}, we show our selected S12 aperture over an 11 $\times$ 11 pixel FFI cutout and a similarly shaped heat-map containing the pixel-by-pixel target flux fractions. In Table~\ref{tab:TESS_SAP}, we present the complete TESS SAP dataset together with its associated quality flags (QFs) determined by SPOC. We discarded those observations with SPOC QFs different from zero, which were mainly flagged because of stray light coming from the Earth or Moon. For completeness, we repeated the analysis in Sect.~\ref{subsec:joint_analysis} by considering the highest level dataset available (QLP/SAP for S12, and SPOC/PDCSAP for S39 and S65; Table~\ref{tab:TESS_QLP_PDCSAP}), and found a consistent solution for the planetary and orbital parameters within 1$\sigma$.

\subsection{Ground-based photometry}
\label{sec:ground_based}

\begin{figure}
    \centering
    \includegraphics[width=0.57\textwidth]{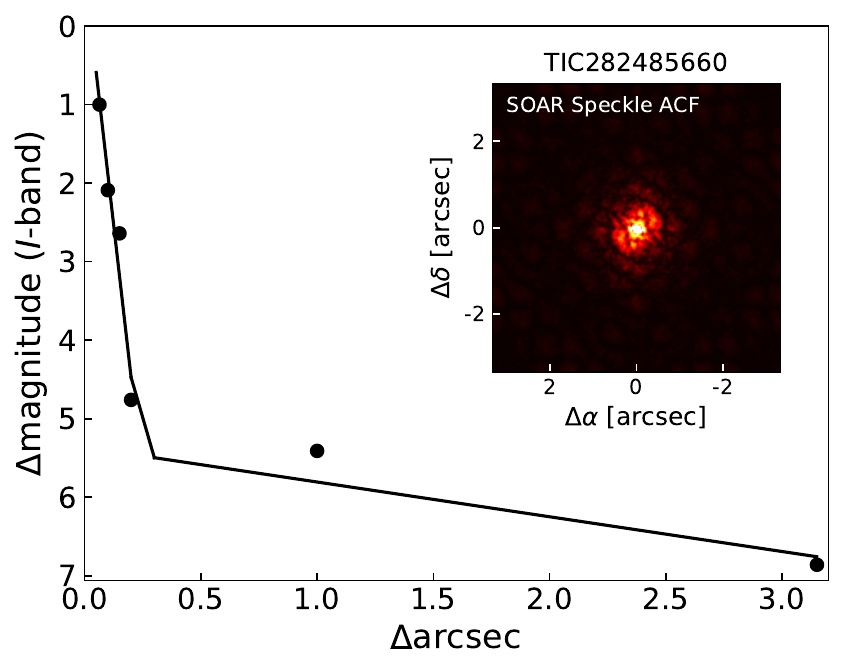}
    \caption[Detection sensitivity to nearby companions for the SOAR observation of TOI-5005.]{Detection sensitivity (5$\sigma$) to nearby companions for the SOAR observation of TOI-5005 as a function of separation from the target star and magnitude difference relative to the target. The inset shows the speckle autocorrelation function.}
    \label{fig:SOAR_speckle}
\end{figure}

We observed two transits of TOI-5005.01 from ground-based facilities to attempt to determine the true source of the TESS detection and to obtain precise ephemerides. 

\begin{figure*}
\centering
    \includegraphics[width=0.97\textwidth]{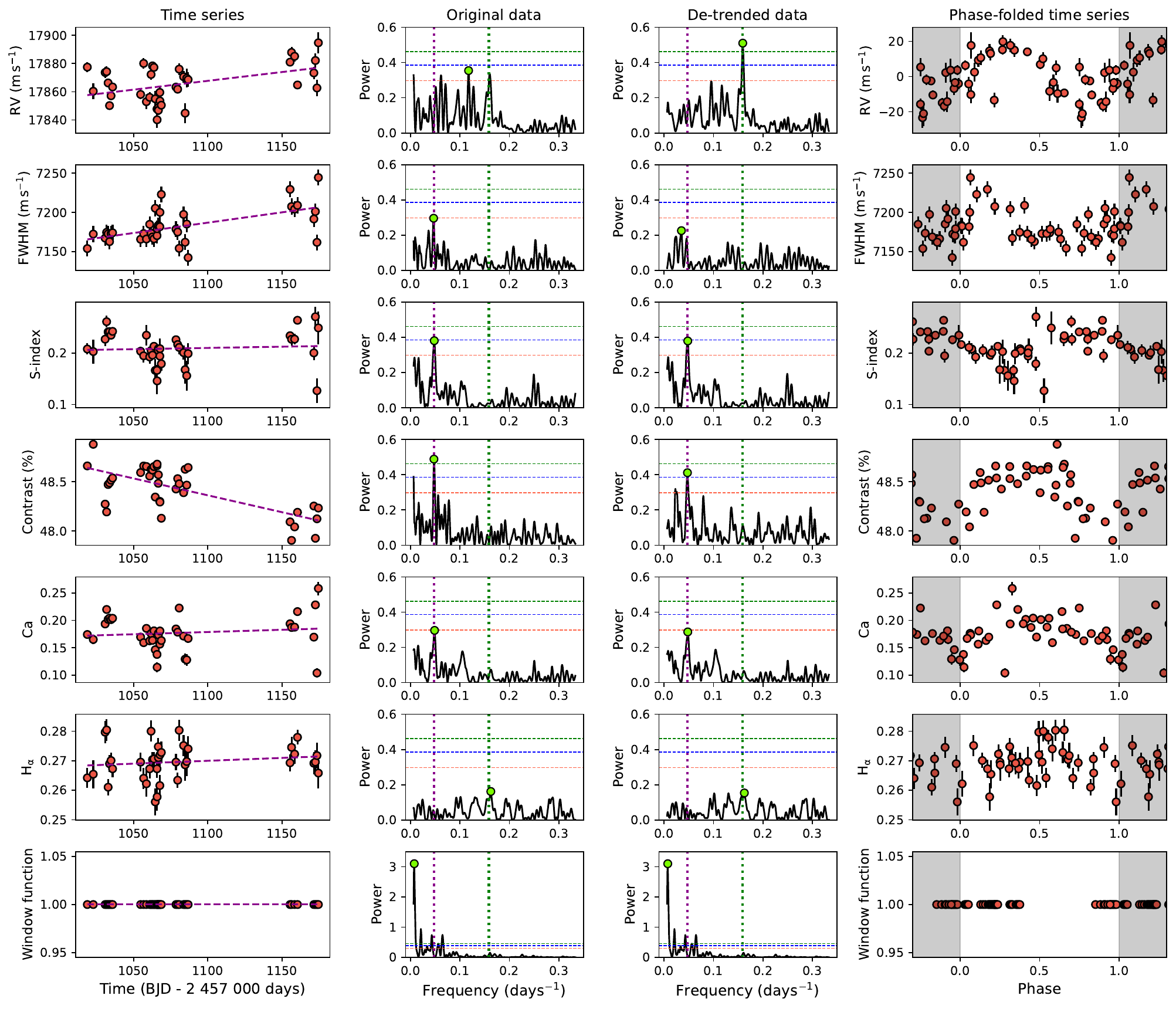}
    \caption[GLS periodograms of the HARPS RVs and activity indicators of TOI-5005.]{Left panels: time series of the HARPS RVs, activity indicators, and the window function. The magenta dashed lines represent the linear trends fit to the data. Centre panels: GLS periodograms of the original time series and de-trended time series. The green dotted vertical lines indicate the location of the orbital period of TOI-5005.01 ($P_{\rm orb}$ = 6.3 days). The magenta dotted vertical lines indicate the $\simeq$21 days activity-related signal that most likely reflects the stellar rotation period. The horizontal dotted lines correspond to the 10 (orange), 1 (blue), and 0.1$\%$ (green) FAP levels. Right panels: HARPS data folded to the maximum power periods; that is, 6.3 days (RVs), 21.3 days (FWHM), 20.8 days (S-index), 21.0 days (Contrast), 20.5 days (Ca), 6.2 days ($\rm H_{\alpha}$), and 129.8 days (window function).}
    \label{fig:gls_to_HARPS}
\end{figure*}


We observed a full transit window of TOI-5005.01 in the Sloan $r'$ filter on 5 May 2022 from the Perth Exoplanet Survey Telescope (PEST) located near Perth, Australia. These observations are part of the TESS Follow-up Observing Program \citep[TFOP;][]{collins:2019}.\footnote{\url{https://tess.mit.edu/followup}} We scheduled the transit observations through the {\tt TESS Transit Finder}, which is a customized version of the {\tt Tapir} package \citep{Jensen:2013}. The 0.3-m telescope is equipped with a $5544\times3694$ QHY183M camera.  Images are binned $2\times2$ in software, giving an image scale of 0.7" pixel$^{-1}$, resulting in a $32'\times21'$ field of view. A custom pipeline based on {\tt C-Munipack}\footnote{Available at \url{http://c-munipack.sourceforge.net}} was used to calibrate the images and extract the differential photometry. We used circular photometric apertures with a radius of 6.4". The target star aperture excluded most of the flux from the nearest known neighbours in the \textit{Gaia} DR3 catalogue, TIC 282476644 (Star~3) and TIC 282478138 (Star~4), which are 7.1" and 15.5" from TOI-5005, respectively. The light curve data are available in Table~\ref{tab:pest_data} and {\tt EXOFOP-TESS}.\footnote{\url{https://exofop.ipac.caltech.edu/tess/}}

We observed one transit of TOI-5005.01 with the TRAPPIST-South telescope \citep{TS_Gillon,TS_Jehin} located at ESO's La Silla Observatory in Chile on 22 March 2024. As the planet was already confirmed on target by the PEST transit and the HARPS radial velocity measurements (see Sect.~\ref{sec:obs_harps}), we obtained this transit observation to keep the ephemeris up to date. TRAPPIST-South is a 0.6-m telescope equipped with an FLI ProLine camera and a back-illuminated CCD which has a pixel size of 0.64" and provides a field of view of $22' \times 22'$. It is a robotic Ritchey-Chrétien telescope with F/8, and it is equipped with a German equatorial mount. The transit was obtained in the Sloan $z'$ filter with an exposure time of 20 s. We reduced the data and performed aperture and differential photometry using a custom pipeline built with the \texttt{prose} package\footnote{Available at \url{https://github.com/lgrcia/prose/}} \citep{prosesoft,2022_prose}. To minimise red and white noise in the transit light curve, we selected four comparison stars and an uncontaminated circular aperture of 4.1" for a full width at half-maximum (FWHM) of 2.6". A meridian flip occurred at 2460392.8576 BJD, which caused an offset in the normalised flux. The light curve data are available in Table~\ref{tab:trappist_south_data}.

\subsection{SOAR high-resolution imaging}

We searched for stellar companions to TOI-5005 with speckle imaging on the 4.1-m Southern Astrophysical Research (SOAR) telescope \citep{2018PASP..130c5002T} on 15 April 2022, observing in Cousins $I$-band, a similar visible bandpass as TESS. This observation was sensitive with 5$\sigma$ detection to a 5.5-magnitude fainter star at an angular distance of 1" from the target. More details of the observations within the SOAR TESS survey are available in \citet{2020AJ....159...19Z}. The 5$\sigma$ detection sensitivity and speckle auto-correlation functions from the observations are shown in Fig.~\ref{fig:SOAR_speckle}. No nearby stars were detected within 3" of TOI-5005 in the SOAR observations.

\subsection{HARPS high-resolution spectroscopy}
\label{sec:obs_harps}

We observed TOI-5005 with the High Accuracy Radial velocity Planet Searcher spectrograph \citep[HARPS;][]{2003Msngr.114...20M}, which is mounted on the ESOS's 3.6m telescope located at La Silla Observatory in Chile. HARPS is a fibre-fed cross-dispersed echelle spectrograph stabilised in a vacuum vessel. It covers a wavelength range between 378 and 691 nm and has a spectral resolution power of 115\,000.

We acquired a total of 38 HARPS spectra between 15 March 2023 and 18 August 2023 under the HARPS-NOMADS programmes 108.21YY.001 and 108.21YY.002 (PI: Armstrong). The nightly seeing conditions ranged from 0.76" to 2.67", with a median value of 1.24". We performed the acquisitions with a typical exposure time of 1800 s, which resulted in S/N per pixel between 14.3 and 36.7, with a median value of 24.1. We used the High Accuracy Mode (HAM) with a 1" science fibre centred on the star and a second fibre in the sky to monitor the sky background.

We used the HARPS Data Reduction Software to reduce the raw spectra and extract the RVs through the cross-correlation technique based on a G2 template \citep{1996A&AS..119..373B,2002A&A...388..632P}. We also used the HARPS DRS to extract several activity indicators such as the full width at half maximum (FWHM) of the
cross-correlation function (CCF), the contrast of the CCF, the $\rm H_{\alpha}$ and sodium doublet (NaD) line depths, and the S-index. We present the complete dataset in Table \ref{tab:harps_rvs}. The HARPS RV uncertainties range from 2.4 to 7.6 $\rm m\,s^{-1}$, with a median value of 3.8 $\rm m\,s^{-1}$ and a root-mean-square (rms) of 4.3 $\rm m\,s^{-1}$. These values contrast with the dispersion of the RV measurements, which have a standard deviation of 13.1 $\rm m\,s^{-1}$. The factor of three between the RV dispersion and uncertainties indicates that our data are not white-noise-dominated but could contain planetary or stellar signals.

In Fig.~\ref{fig:gls_to_HARPS}, we show the HARPS RVs and activity indicators together with their \texttt{GLS} periodograms. Several time series follow upward or downward trends (e.g. RV, FWHM, and Contrast), which could be due to physical (e.g. outer long-period massive companion, stellar magnetic cycle) or instrumental (e.g. night-to-night drifts) effects. The RV periodogram shows a maximum power period of 8.5 days. This periodicity has a FAP of 6$\%$, so we do not consider it significant.\footnote{The most commonly adopted criterion to consider a \texttt{GLS} peak significant is to have a FAP $<$ 0.1 $\%$.}  Interestingly, the second-highest peak coincides with the 6.3-day periodicity of TOI-5005.01, although it is not significant either.  We recomputed the periodogram after subtracting a simple linear trend that we previously fit to the data. The periodogram of the de-trended RVs shows a significant maximum power period of 6.3 days (FAP = 0.05 $\%$), which matches the TOI-5005.01 periodicity and ephemeris. The periodograms of the FWHM, S-index, Contrast, and Ca activity indicators show maximum power periods of 21.3, 20.8, 21.0, and 20.5 days, with FAPs of 10, 1.1, 0.040, and 10 $\%$, respectively. This recurring $\simeq$21-day periodicity indicates the existence of an activity-related signal that most likely corresponds to the rotation period of the star (see Sect.~\ref{subsec:prot}).  The periodogram of the $\rm H_{\alpha}$ indicator shows a maximum power period of 6.2 days with a FAP of 91 $\%$. This periodicity coincides with that of the TOI-5005.01 orbit. As mentioned in Sect.~\ref{sec:obs_tess}, given the closeness of this Neptune-sized planet to its host star, this periodicity could be interpreted as a planet-induced enhancement of chromospheric activity due to magnetic star-planet interactions. However, in this case, the signal significance is extremely small, and the phase-folded data do not show a clear modulation such as that seen within the TESS photometry. Therefore, the weakness of the signal, together with the absence of additional spectroscopic activity signals matching the TOI-5005.01 periodicity, suggests that the possible magnetic star-planet interactions detected within the TESS data cannot be measured in our HARPS data. In Fig~\ref{fig:gls_to_HARPS}, we show the window function of the observations and its periodogram, which shows a maximum power period of 129.8 days. This periodicity does not coincide with any of the previously mentioned planet and activity signals.

\section{Stellar characterization}
\label{sec:stellar_charact}

\begin{table}
\centering
\renewcommand{\arraystretch}{1.9}
\setlength{\tabcolsep}{4pt}
\caption[General properties of TOI-5005 from the literature.]{General properties of TOI-5005. (1) \citealp{2021ApJS..254...39G}; (2) \citealp{2019AJ....158..138S}; (3) \citealp{skrutskie2006}; (4) \citealp{2023A&A...674A...1G}; (5) \citealp{2018AAS...23222306H}.}
\label{tab:stellar_general_prop}
\begin{tabular}{llc}
\hline \hline
Parameter                               & Value                     & Reference \\ \hline
\multicolumn{3}{l}{Identifiers}                                                 \\ \hline
TOI                                     & 5005                      & (1)       \\
TIC                                     & 282485660                 & (2)       \\
2MASS                                   & J15522597-4808419         & (3)       \\
Gaia DR3                                & 5984530842395365248       & (4)       \\ \hline
\multicolumn{3}{l}{Astrometric properties}                                      \\ \hline
RA, Dec                                 & 15:52:25.97, -48:08:42.37 & (4)       \\
$\rm \mu_{\alpha}$ ($\rm mas\,yr^{-1}$) & -7.830 $\pm$ 0.018        & (4)       \\
$\rm \mu_{\delta}$ ($\rm mas\,yr^{-1}$) & -24.642 $\pm$ 0.015       & (4)       \\
Parallax (mas)                          & 4.760 $\pm$ 0.017         & (4)       \\
Distance (pc)                           & 210.08 $\pm$ 0.75         & (4)       \\
RV ($\rm km\,s^{-1}$)                   & 16.58 $\pm$ 0.77          & (4)       \\ \hline
Photometric properties                  &                           &           \\ \hline
TESS (mag)                              & 11.1300 $\pm$ 0.0061      & (2)       \\
G (mag)                                 & 11.63820 $\pm$ 0.00045    & (2)       \\
J (mag)                                 & 10.393 $\pm$ 0.023        & (3)       \\
H (mag)                                 & 10.058 $\pm$ 0.022        & (3)       \\
$\rm K_{s}$ (mag)                       & 10.004 $\pm$ 0.021        & (3)       \\
B (mag)                                 & 12.572 $\pm$ 0.013        & (5)       \\
V (mag)                                 & 11.822 $\pm$ 0.050        & (5)       \\
g' (mag)                                & 12.150 $\pm$ 0.026        & (5)       \\
r' (mag)                                & 11.576 $\pm$ 0.079        & (5)       \\
i' (mag)                                & 11.362 $\pm$ 0.142        & (5)       \\ \hline
\end{tabular}
\end{table}

\subsection{General description}
\label{subsec:general_description}

TOI-5005 is a moderately bright \citep[V = 11.822 $\pm$ 0.050 mag;][]{2018AAS...23222306H} early-type G-dwarf star visible from the southern sky. According to the measured \textit{Gaia} DR3 parallax ($\pi$ = 4.760 $\pm$ 0.017 mas), TOI-5005 is located 210.08 $\pm$ 0.75 pc away from the Sun. In Table \ref{tab:stellar_general_prop}, we show the main astrometric and photometric properties compiled from the literature. The TESS Input Catalogue \citep[TIC v8.2;][]{2019AJ....158..138S} estimates $T_{\rm eff}$ = 5840 $\pm$ 125 K, log $g$ = 4.484 $\pm$ 0.075 dex, $R_{\star}$ = 0.972 $\pm$ 0.045 $\rm R_{\odot}$, and $\rm M_{\star}$ = 1.05 $\pm$ 0.13 $\rm M_{\odot}$. In the next sections, we describe our stellar characterisation based on a high-resolution, high S/N spectrum obtained from the combination of the 38 individual HARPS spectra. 


\subsection{Stellar physical parameters}
\label{subsec:atmospheric_parameters}

We derived the stellar atmospheric parameters ($T_{\mathrm{eff}}$, $\log g$, micro-turbulence, and [Fe/H]) using ARES+MOOG following the methodology described in \citet[][]{Sousa-21, Sousa-14, Santos-13}. We used the ARES code\footnote{The last version of ARES code (ARES v2) can be downloaded at \url{https://github.com/sousasag/ARES}} \citep{Sousa-07, Sousa-15} to consistently measure the equivalent widths (EW) of selected iron lines based on the line list presented in \citet[][]{Sousa-08}. This was done on a 
combined HARPS spectrum of TOI-5005. We then used a minimisation process to find the ionisation and excitation equilibrium and converge to the best set of spectroscopic parameters. This process makes use of a grid of Kurucz model atmospheres \citep{Kurucz-93} and the radiative transfer code MOOG \citep{Sneden-73}. A trigonometric surface gravity was also derived using \textit{Gaia} DR3 data following the same methodology as described in \citet[][]{Sousa-21}. In this process, we derived the stellar mass using the calibration presented in \citet[][]{Torres-2010}: $M_{\star}$ = 0.97 $\pm$ 0.02~$\rm M_{\odot}$. Following a similar calibration, presented in the same work, we obtained the stellar radius: $R_{\star}$ = 0.93 $\pm$ 0.03~$\rm R_{\odot}$. These values are consistent with the photometry-based estimations of the TIC catalogue within 1$\sigma$. We list the stellar atmospheric parameters, mass, and radius of TOI-5005 in Table \ref{tab:stellar_parameters}.

\subsection{Chemical abundances}
\label{subsec:chemical_abundances}

We derived stellar abundances of different elements using the classical curve-of-growth analysis method, assuming local thermodynamic equilibrium. We used the same radiative transfer code (MOOG) and the same model atmospheres that we previously used for the stellar parameters determinations. We followed the methods described in \citet[]{Adibekyan-12, Adibekyan-15, Delgado-17} to derive chemical abundances of refractory elements, and followed the method of \cite{Delgado-21, Bertrandelis-15} to derive abundances of volatile elements such as carbon and oxygen. The oxygen abundances are based on two weak atomic lines which present large uncertainties, especially when the S/N of the spectra are not very high. One of the lines was contaminated by an Earth airglow, and hence we only used 14 spectra to determine its abundance. We obtained all the [X/H] ratios by doing a differential analysis relative to a high S/N solar (Vesta) spectrum from HARPS. 

In addition, we obtained the abundance of lithium by performing spectral synthesis with MOOG, following the same procedure as in \citet{Delgado-14}. We first fixed the macro-turbulence velocity to 3.2 km\,s$^{-1}$ \citep[based on the empirical calibration by][]{doyle-14} to estimate the $v \, \textrm{sin} \, i_{\star}$ from two Fe lines in the region, leading to a value of 1.35 $\pm$ 0.10 km\,s$^{-1}$. We obtained an abundance A(Li)\,=\,1.65 $\pm$ 0.10\,dex, which is a relatively high value for a star of this temperature, suggesting that TOI-5005 is younger than the Sun. We present the abundances of all the elements in Table \ref{tab:stellar_parameters}.

\subsection{Rotation period}
\label{subsec:prot}
The HARPS activity indicators FWHM, S-index, Contrast, and Ca show periodic signals at $\simeq$21 days, which most likely reflect the rotation period ($P_{\rm rot}$) of TOI-5005 (Sect.~\ref{sec:obs_harps}). This signal is significantly detected in the S-index and Contrast indicators (FAPs of 1.1$\%$ and 0.040$\%$, respectively), and is detected with a weaker significance in the FWHM and Ca indicators (FAPs of 10$\%$). Interestingly, it appears not to have a significant effect on the RVs (see Sect.~\ref{subsec:harps_rv_analysis} for a detailed analysis), and it is not detected in the TESS photometry either (probably because of insufficient photometric precision or an incomplete phase coverage). In this section, we study additional observables to compare with the periodicity detected in the HARPS activity indicators. 

The $v \, \textrm{sin} \, i_{\star}$ obtained in Sect.~\ref{subsec:chemical_abundances} (1.35 $\pm$ 0.10 km\,s$^{-1}$) together with the stellar radius derived in Sect.~\ref{subsec:atmospheric_parameters} corresponds to a $P_{\rm rot}  \, \textrm{sin} \, i_{\star}$ of 35 $\pm$ 3 days. Hence, a certain inclination angle relative to our line of sight seems to be required to match the observed $\simeq$21-day activity signal. We note, however, that measuring accurate rotation velocities for slow rotators ($v \, \textrm{sin} \, i_{\star}$ $<$ 3$\rm \, km \, s^{-1}$) is a difficult task and hence our estimate should be taken with care. In particular, the reported $v \, \textrm{sin} \, i_{\star}$ uncertainty might be underestimated. We also used the S-index provided by the HARPS pipeline, which is calibrated to the Mount Wilson scale \citep{1978PASP...90..267V}, to obtain the $\textrm{log} \, R'_{HK}$ of each spectrum. To do so, we considered the B-V colour from the APASS catalogue \citep[][B-V = 0.75 $\pm$ 0.05, see Table \ref{tab:stellar_general_prop}]{2018AAS...23222306H} and used the \texttt{pyrhk} code\footnote{Available at \url{https://github.com/gomesdasilva/pyrhk}.} to obtain an average $\textrm{log} \, R'_{HK}$ of -4.819 $\pm$ 0.052 dex with bolometric corrections from \citet{1982A&A...107...31M}. We used the \citet{2016A&A...595A..12S} activity-rotation empirical calibrations and obtained a rotation period of $P_{\rm rot}$ = 24.9 $\pm$ 5.0 days, which is consistent with the $\simeq$21-day activity signal within 1$\sigma$. In Sect~\ref{subsubsec:HARPS_activity_indicators}, we use an activity model to jointly analyse different HARPS activity indicators. From this analysis, we obtain a precise rotation period of $P_{\rm rot}$ = $21.01^{+0.46}_{-0.60}$ days, which we adopt as our final estimate.

\subsection{Age}
\label{subsec:age}
Chemical abundances can be used for estimating stellar ages. Some chemical elements are related to different astrophysical origin channels and can be used as tracers of the time a star was born. Those elements are called chemical clocks (CCs) and their use for dating stars has been proposed in many works \citep[e.g. see][and references therein]{2024MNRAS.528.3464R}. Almost all those works use linear regressions to describe the relation between the CCs and the stellar age.  \citet{2019A&A...624A..78D} presented a complete set of multidimensional linear regressions using all the chemical clocks that could be made with that dataset. One of the main problems of this technique is that by using different CCs we can obtain slightly different stellar age estimations. To solve this, one option is to combine these estimations to obtain a more robust age estimator, but a simple mean, for example, cannot be used here since all the chemical clocks are very correlated. In a recent work, \citet{2022A&A...660A..15M} constructed a Hierarchical Bayesian model (HBM) for estimating stellar ages, combining the results from different chemical clocks and their multidimensional linear regressions, also properly propagating uncertainties all along the procedure. We used this HBM for estimating the age of TOI-5005. Using [Y/Mg], [Sr/Mg], [Y/Si], [Y/Ti], and [Y/Zn], the stellar $T_{\rm eff}$, $\textrm{log}\,g$, [Fe/H], and the corresponding uncertainties, we obtained the posterior probability distribution for its age. This distribution is compatible with zero and has a 3$\sigma$ upper limit of 3.6 Gyr.

The stellar rotation period has also been used to date stars through gyro-chronology \citep[e.g.][]{2003ApJ...586..464B,2007ApJ...669.1167B,2008ApJ...687.1264M,2015MNRAS.450.1787A}. We used the empirical relations by \citet{2019AJ....158..173A} implemented in \texttt{stardate}\footnote{Available at \url{https://github.com/RuthAngus/stardate}.} to estimate the age of TOI-5005. Based on the \textit{Gaia} parallax (Sect.~\ref{subsec:general_description}), stellar atmospheric parameters (Sect.~\ref{subsec:atmospheric_parameters}), and measured rotation period ($P_{\rm rot}$ = $21.01^{+0.46}_{-0.60}$ days; see Sects.~\ref{subsec:prot} and \ref{subsubsec:HARPS_activity_indicators}), we obtain a stellar age of $2.20^{+0.61}_{-0.28}$ Gyrs, which is compatible with the CC estimate. We include both the CC 3$\sigma$ upper limit ($\rm Age_{CC}$) and the gyro-chronology age ($\rm Age_{gyro}$) in Table~\ref{tab:stellar_parameters}. We note that given its better precision, for subsequent analysis we adopt $\rm Age_{gyro}$.


\subsection{Galactic membership}
\label{subsec:galactic_membership}

We computed the Galactic space velocity (\textit{UVW}) of TOI-5005 based on its radial velocity, parallax, RA/Dec coordinates, and proper motion as measured by \textit{Gaia} (Table~\ref{tab:stellar_general_prop}). We adopted the solar peculiar motion from \citealp{2003A&A...409..523R} ($U_{\odot} = -10.3 \rm \, km\,s^{-1}$, $V_{\odot} = 6.3 \, \rm km\,s^{-1}$, and $W_{\odot} = 5.9 \, \rm km\,s^{-1}$), and derived $U = -15.40 \pm 0.67 \, \rm km\,s^{-1}$, $V = -21.10 \pm 0.37 \, \rm km\,s^{-1}$, and $W = -6.79 \pm 0.08 \, \rm km\,s^{-1}$ with respect to the local standard of rest (LSR). Similarly to \citet{2003A&A...410..527B}, we assumed that the Galactic space velocities of the stellar populations follow a multi-dimensional Gaussian distribution

\begin{equation}
    f(U,V,W) = k \cdot \textnormal{exp}  \left(     
    -\frac{U^{2}}{2\sigma_{U}^{2}}
    -\frac{(V-V_{\rm asym})^{2}}{2\sigma_{V}^{2}}
    -\frac{W^{2}}{2\sigma_{W}^{2}}
    \right)
,\end{equation}

\noindent where $k = \left( 2\pi \right)^{-3/2} \sigma_{U}^{-1} \sigma_{V}^{-1} \sigma_{W}^{-1}$, being $\sigma_{U}$, $\sigma_{V}$, and $\sigma_{W}$ the characteristic velocity dispersions, and $V_{\rm asym}$ the asymmetric drift. We estimated the probabilities that TOI-5005 belongs to the thin disk (TD), the thick disk (D), and the halo (H) by multiplying $f(U,V,W)$ by the probability that a star in our neighbourhood belongs to those three Galactic populations. In order to make our estimation self-consistent, we adopted the relative likelihoods of belonging to each population as well as the characteristics for stellar components
in the solar neighbourhood from \citet{2003A&A...409..523R}. As a result, we obtain TD = 98.865 $\%$, D = 1.133 $\%$, and H = 0.002 $\%$. Therefore, it is very likely that TOI-5005 is a
member of the Galactic thin disk population. This result is consistent with the stellar age estimation (Sect.~\ref{subsec:age}), since most thin disk stars have ages less than 8 Gyr \citep[e.g.][]{1998A&A...338..161F,2001ASPC..245..207B,2003A&A...410..527B}.

\begin{table}[]
\centering
\renewcommand{\arraystretch}{1.3}
\setlength{\tabcolsep}{15.7pt}
\caption{Stellar properties of TOI-5005 derived in this work.}
\label{tab:stellar_parameters}
\begin{tabular}{lll}
\hline \hline
Parameter               & Value  & \multicolumn{1}{c}{Section}      \\ \hline
\multicolumn{3}{l}{Atmospheric parameters and spectral type}                                      \\ \hline
$T_{\rm eff}$ (K)                & $5749 \pm 61$      & Sect. \ref{subsec:atmospheric_parameters} \\
log $g$ (dex)                    & $4.43 \pm 0.10$    & Sect. \ref{subsec:atmospheric_parameters} \\
log $g_{\rm GAIA}$ (dex)             & $4.56 \pm 0.03$    & Sect. \ref{subsec:atmospheric_parameters} \\
{[}Fe/H{]} (dex)                 & $0.15 \pm 0.04$    & Sect. \ref{subsec:atmospheric_parameters} \\
$\xi_{\mathrm{t}}$ ($\rm km\,s^{-1}$)        & $0.97 \pm 0.02$    & Sect. \ref{subsec:atmospheric_parameters} \\
SpT                              & G2 V               & Sect. \ref{subsec:atmospheric_parameters} \\ \hline
\multicolumn{3}{l}{Physical parameters}                                                           \\ \hline
$R_{\star}$ $(\rm R_{\odot})$        & $0.93 \pm 0.03$    & Sect. \ref{subsec:atmospheric_parameters} \\
$M_{\star}$ $(\rm M_{\odot})$        & $0.97 \pm 0.02$    & Sect. \ref{subsec:atmospheric_parameters} \\
$v \, \textrm{sin} \, i_{\star}$ ($\rm km\,s^{-1}$) & $1.35 \pm 0.10$                 & Sect. \ref{subsec:chemical_abundances}    \\
$\textrm{log} \, R'_{HK}$ (dex) & -4.819 $\pm$ 0.052 & Sect.~\ref{subsec:prot}  \\
$P_{\rm rot}$ (days)             & $21.01^{+0.46}_{-0.60}$                &  Sect.~\ref{subsec:prot}                                       \\
$\rm Age_{CC}$ (Gyr)            & $<$ 3.6                & Sect. \ref{subsec:age}                    \\
$\rm Age_{gyro}$ (Gyr)          & $2.20^{+0.61}_{-0.28}$                & Sect. \ref{subsec:age}                                         \\ \hline
\multicolumn{3}{l}{Chemical abundances}                                                           \\ \hline
{[}Mg/H{]} (dex)                 & $0.100 \pm 0.030$  & Sect. \ref{subsec:chemical_abundances}    \\
{[}Si/H{]} (dex)                 & $0.100 \pm 0.040$  & Sect. \ref{subsec:chemical_abundances}    \\
{[}Ni/H{]} (dex)                 & $0.100 \pm 0.020$  & Sect. \ref{subsec:chemical_abundances}    \\
{[}Ti/H{]} (dex)                 & $0.150 \pm 0.030$  & Sect. \ref{subsec:chemical_abundances}    \\
{[}O/H{]} (dex)                  & $0.141 \pm 0.187$  & Sect. \ref{subsec:chemical_abundances}    \\
{[}C/H{]} (dex)                  & $-0.035 \pm 0.034$ & Sect. \ref{subsec:chemical_abundances}    \\
{[}Cu/H{]} (dex)                 & $0.015 \pm 0.051$  & Sect. \ref{subsec:chemical_abundances}    \\
{[}Zn/H{]} (dex)                 & $0.011 \pm 0.025$  & Sect. \ref{subsec:chemical_abundances}    \\
{[}Sr/H{]} (dex)                 & $0.280 \pm 0.077$  & Sect. \ref{subsec:chemical_abundances}    \\
{[}Y/H{]} (dex)                  & $0.237 \pm 0.053$  & Sect. \ref{subsec:chemical_abundances}    \\
{[}Zr/H{]} (dex)                 & $0.194 \pm 0.061$  & Sect. \ref{subsec:chemical_abundances}    \\
{[}Ba/H{]} (dex)                 & $0.267 \pm 0.027$  & Sect. \ref{subsec:chemical_abundances}    \\
{[}Ce/H{]} (dex)                 & $0.246 \pm 0.035$  & Sect. \ref{subsec:chemical_abundances}    \\
{[}Nd/H{]} (dex)                 & $0.192 \pm 0.036$  & Sect. \ref{subsec:chemical_abundances}    \\ \hline
\multicolumn{3}{l}{Galactic space velocities and membership}                                      \\ \hline
U ($\rm km \, s^{-1}$)           & $-15.40 \pm 0.67$  & Sect. \ref{subsec:galactic_membership}    \\
V ($\rm km \, s^{-1}$)           & $-21.10 \pm 0.37$  & Sect. \ref{subsec:galactic_membership}    \\
W ($\rm km \, s^{-1}$)           & $-6.79 \pm 0.08$   & Sect. \ref{subsec:galactic_membership}    \\
Gal. population                  & Thin disk          & Sect. \ref{subsec:galactic_membership}    \\ \hline
\end{tabular}
\end{table}

\section{Analysis and results}
\label{sec:analysis_results}

\subsection{Model inference and parameter determination}
\label{sec:model_sel_param_det}

We analysed the TESS, HARPS, PEST, and TRAPPIST-South datasets through Bayesian inference. Our main goal is to find the model that best represents our data and use it to derive accurate physical parameters. In Appendix~\hyperref[bayesian]{D}, we describe the mathematical framework behind our Bayesian analysis, where we include details on the considered likelihood functions, prior distributions, and their implementation. For each tested model, we used a Markov chain Monte Carlo (MCMC) affine-invariant ensemble sampler \citep{2010CAMCS...5...65G} as implemented in \texttt{emcee}\footnote{Available at \url{https://github.com/dfm/emcee}.} \citep{2013PASP..125..306F} to sample the parameter space and generate marginal posterior distributions associated with each parameter. We used eight times as many walkers as the number of parameters and performed two consecutive runs. The first run consisted of 200\,000 iterations, and the second run consisted of 100\,000 iterations. Between both runs, we reset the sampler and considered the initial values from the last iteration of the first run. Following \citet{2013PASP..125..306F}, we examined the convergence by estimating the autocorrelation times and checking that they are all at least 50 times smaller than the chain length.

To infer the model that best represents our data, we estimated the Bayesian evidence through \texttt{bayev}\footnote{Available at \url{https://github.com/exord/bayev}.} \citep{2016A&A...585A.134D}, which is based on the formalism described in \citet{PERRAKIS201454}. The package uses the marginalized posterior distributions obtained by \texttt{emcee}, the likelihood functions (Eqs. \ref{eq:log-like_uncorrelated} and \ref{eq:log-like_correlated}), and the prior distributions (Eqs. \ref{eq:unifor_priors} and \ref{eq:gaussian_priors}) to obtain the model evidence $\mathcal{Z}$.\footnote{When describing the mathematical framework in Appendix~\hyperref[bayesian]{D}, we refer to the model evidence as $P(D)$. However, for simplicity, in the main text, we adopt the widely used notation $\mathcal{Z}$.} For the model selection, we considered a criterion based on Occam's razor principle. That is, we always selected the simplest model, unless there was a more complex model with significantly more evidence. Following the Jeffreys' scale \citep{jeffreys1961theory}, we consider that a complex model has strong evidence against a simple model if the logarithmic difference $\mathcal{B}$ = ln$(\mathcal{Z}_{\rm complex})$ - ln$(\mathcal{Z}_{\rm simple})$ is larger than five.

\subsection{HARPS radial velocity analysis}
\label{subsec:harps_rv_analysis}

We analysed the HARPS dataset described in Sect.~\ref{sec:obs_harps} through the procedure described in Sect.~\ref{sec:model_sel_param_det}. The RV periodogram showed a significant 6.3-day signal that matches the periodicity and ephemeris of TOI-5005~b, and the periodograms of four activity indicators showed maximum power periods at $\simeq$21 days, which indicated the existence of an activity-related signal that corresponds to the rotation period of the star. This signal, however, did not appear within the RV periodograms, which suggests that the stellar activity did not significantly influence the measured RVs. Under this situation, the main objectives of this section are the following. First, we aim to study whether our RV model should incorporate a stellar noise component. Secondly, we aim to figure out whether there are additional planetary signals in the system. Finally, we aim to select the simplest model that best represents our data. 

We built 21 models with different components aimed at describing the phenomena that might be affecting our dataset. Those components are an instrumental model, a linear drift, a Keplerian, a stochastic process composed of a mean function and an autocorrelation function for describing unknown correlated noise (i.e. a Gaussian process; see Appendix~\hyperref[bayesian]{D}), and a white noise model (i.e. a jitter term, see Appendix~\hyperref[bayesian]{D}).

The instrumental model consists of an offset that describes the systemic RV of the star as measured by HARPS ($v_{\rm HARPS}$). The linear drift adds a slope ($\gamma_{\rm HARPS}$) to the instrumental model to account for possible long-term trends that can be approximated by a straight line. The Keplerian model describes the planetary orbit. We implemented it through \texttt{radvel}\footnote{Available at \url{https://radvel.readthedocs.io}.} \citep{2018PASP..130d4504F} by considering the parametrization $\left\lbrace P_{\rm orb}, T_{0}, K, \sqrt{e} \cos(\omega), \sqrt{e} \sin(\omega)\right\rbrace$, where $P_{\rm orb}$ is the planetary orbital period, $T_{0}$ is the time of inferior conjunction, $K$ is the semi-amplitude, $e$ is the orbital eccentricity, and $\omega$ is the argument of the periastron. To model the unknown correlated noise, we defined a Gaussian process (GP) with a quasi-periodic covariance function that can be written as \citep{2015ITPAM..38..252A,2016A&A...588A..31F}

\begin{equation}
    \label{eq:QP}
    K_{QP} (\tau) = \eta_{1}^{2} \rm{exp} \left[ - \frac{\tau^{2}}{2\eta_{2}^{2}} - \frac{2sin^{2} \left(  \frac{\pi \tau}{\eta_{3}} \right)}{\eta_{4}^{2}}  \right],
\end{equation}where $\tau = x_{i} - x_{j}$ is the separation between two time stamps, and $\eta_{1}$, $\eta_{2}$, $\eta_{3}$, and $\eta_{4}$ are hyper-parameters defining the GP. Our GP covariance choice is motivated by its physical interpretation. The quasi-periodic covariance (hereinafter QP) has been designed so that $\eta_{3}$ can be interpreted as the stellar rotation period, $\eta_{2}$ is a measure of the timescale of appearance and disappearance of the active regions, and $\eta_{4}$ indicates the complexity of the harmonic content of the activity signal (see \citealt{2014MNRAS.443.2517H} and \citealt{2018MNRAS.474.2094A} for further details on the physical interpretation of the QP hyper-parameters). Finally, to account for the existence of uncorrelated noise not taken into account in the error bars, we included a jitter term that we added quadratically to the estimated HARPS uncertainties.

We built different models involving circular and eccentric orbits, one and two planets, linear trends, and correlated noise components. We refer to each model following a similar notation as that in \citet{2021A&A...654A..60L}; that is, $\rm Xp \left[ P_{i}c \right] \left[ Y \right]$, where X is the total number of Keplerians, $\rm P_{i}$ indicates which planets are considered to have a circular orbit, and Y indicates whether the model has a linear drift (L) or a correlated noise component (QP). We tested the following models: 0p, 1p1c, 1p, 2p1c2c, 2p, 2p1c, 2p2c, 0pL, 1p1cL, 1pL, 2p1c2cL, 2pL, 2p1cL, 2p2cL, 0pQP, 1p1cQP, 1pQP, 2p1c2cQP, 2pQP, 2p1cQP, and 2p2cQP. In this chapter, planet `1' corresponds to TOI-5005.01. To ensure this in all cases, we set uninformative but restrictive priors on the orbital period of planet 1; that is, $P_{\rm orb, 1}\,\in$ $\mathcal{U}(5, 7)$ days. To avoid degeneracies in the time of mid-transit due to the infinite possibilities of $T_{0}+n \times P$, being $n$ $\in$ $\mathbb{N}$,  we sampled $T_{0}$ between 0 and $P$; that is, we sampled $T_{0}$ in the orbital phase space. For planet 2, given that we do not have any hint of its existence based on TESS data, we considered a wide, uninformative prior for its orbital period $P_{\rm orb, 2}\,\in$ $\mathcal{U}(0, 1000)$ days. For the remaining parameters, we used wide, uninformative priors large enough to ensure that they do not bias the analysis. 

\begin{figure*}
\centering
\includegraphics[width=0.99\textwidth]{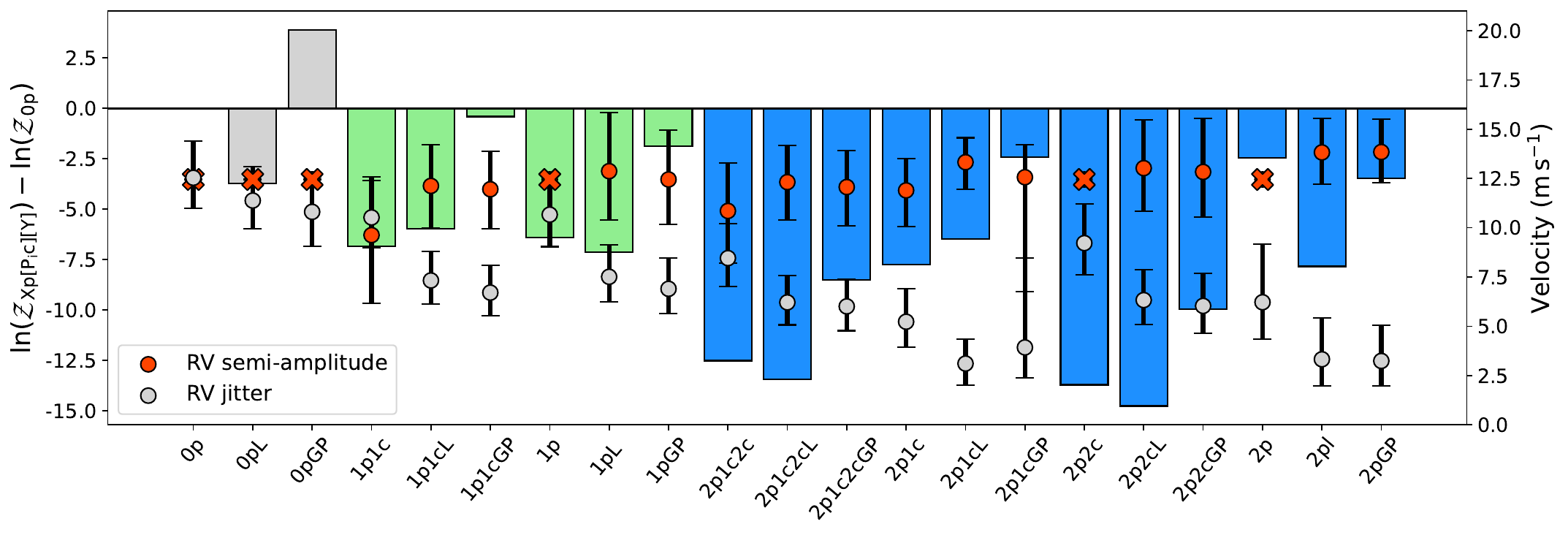}
    \caption[Differences of the log-evidence of the 21 tested models and the 0p model.]{Bar chart showing the differences of the log-evidence of the 21 tested models (labelled on the $X$-axis) and the 0p model. The grey, green, and blue bars correspond to models with zero, one, and two planets, respectively. The red and grey circles indicate the posterior semi-amplitudes and RV jitters of the tested models. The red crosses indicate null semi-amplitude values, either because they correspond to a model without planets or because the MCMC fit did not converge.}
    \label{fig:evidences_harps}
\end{figure*}

In Fig.~\ref{fig:evidences_harps}, we show the log-evidence of each model subtracted from the evidence of the 0p model; that is, $\rm ln(\mathcal{Z}_{Xp[P_{i}c][Y]}) - ln(\mathcal{Z}_{0p})$. In the same figure, we also illustrate the posterior Keplerian semi-amplitudes of TOI-5005 b and the jitter components of each tested model. The main conclusion of this analysis is that there is no model including planets that has a larger Bayesian evidence than the 0p model. Therefore, even though the planetary models converge towards a 6.3-day orbit for planet 1 with a $\simeq$5$\sigma$ detection of the RV semi-amplitude, this signal cannot be confirmed via Bayesian inference and model comparison based on the measured HARPS RVs alone when considering wide, uninformative priors. This result illustrates the well-known difficulties of RV blind searches \citep[e.g.][]{2016SPIE.9908E..12Q,2022A&A...667A.102L}, which typically require large amounts of RV data points to consider planetary candidate signals as confirmed planets \citep[e.g.][]{2020A&A...639A..77S,2022A&A...658A.115F,2024A&A...690A..79G}. To select the model that best represents our dataset, we need to include additional information that constrains the presence of the planetary signal. That is, we need to incorporate the TESS transits in the tested models. In Sect.~\ref{subsec:TESS_analysis}, we describe our TESS transit modelling, and in Sect.~\ref{subsec:joint_analysis} we incorporate it into the RV model comparison analysis.

\subsection{TESS photometric transits analysis}
\label{subsec:TESS_analysis}

\begin{figure*}
    \centering
    \includegraphics[width=0.94\textwidth]{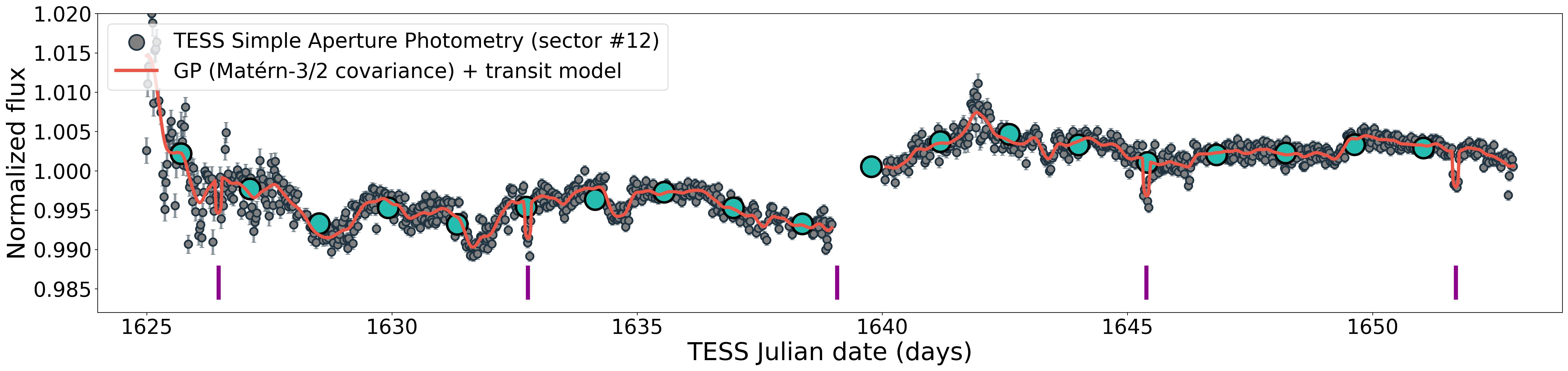}
    \includegraphics[width=0.94\textwidth]{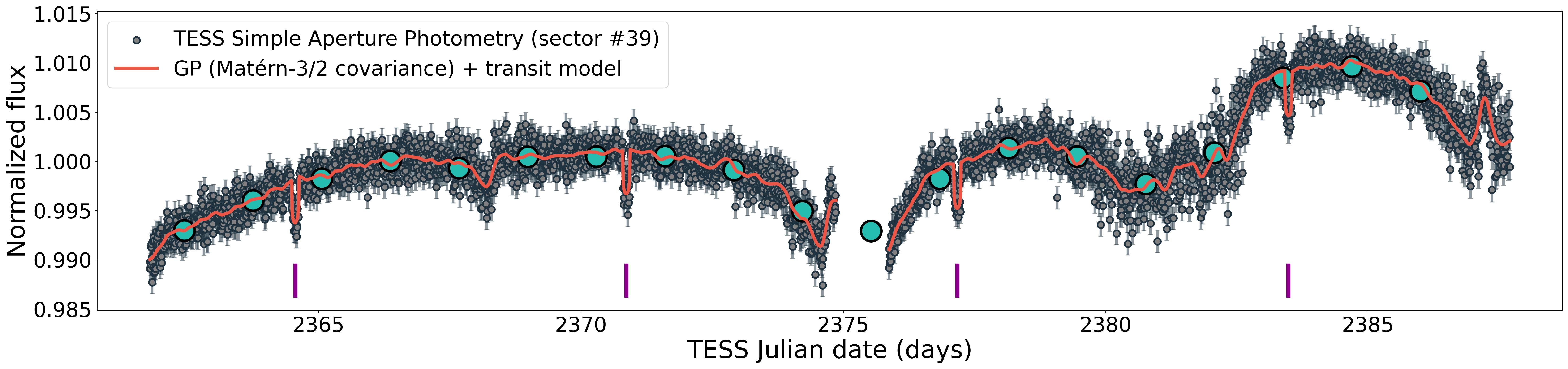}
    \includegraphics[width=0.94\textwidth]{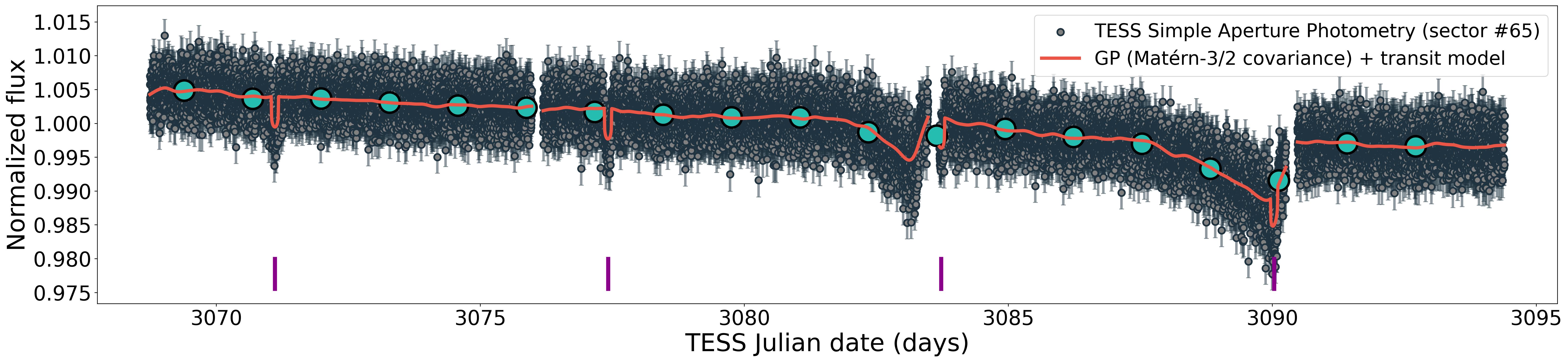}
    \caption[TESS SAP photometry of TOI-5005 with the median posterior model superimposed.]{TESS simple aperture photometry of TOI-5005 with the median posterior model (transit + GP) superimposed. The green data points correspond to 1.5-day binned data. The vertical magenta line indicates the locations of the planetary transits. The TESS Julian date corresponds to Julian date $-$ 2457000 days.}
    \label{fig:transit+GP}
\end{figure*}

We analysed the SAP photometry corrected for contamination described in Sect.~\ref{sec:obs_tess} to infer several orbital and physical parameters of TOI-5005~b. We built a model composed of two components: a transit model and a GP to account for the correlated photometric noise.

We considered the \citet{2002ApJ...580L.171M} quadratic limb darkened model as implemented in \texttt{batman}\footnote{Available at \url{https://github.com/lkreidberg/batman}.} \citep{2015PASP..127.1161K}. The model is described by the orbital period of the planet ($P_{\rm orb}$), the time of inferior conjunction ($T_{0}$), the orbital inclination ($i$), the quadratic limb darkening (LD) coefficients $u_{1}$ and $u_{2}$, the planet-to-star radius ratio ($R_{p} / R_{\star}$), the semi-major axis scaled to the stellar radius ($a/R_{\star}$), the orbital eccentricity ($e$), and the argument of the periastron ($\omega$). We followed the prescription by \citet{2013MNRAS.435.2152K} and parametrized the LD coefficients as $q_{1} =(u_{1} + u_{2})^{2}$ and $q_{2} = 0.5 \, u_{1} (u_{1} + u_{2})^{-1}$. Given that $a/R_{\star}$ is typically poorly constrained based on transit data alone, we parametrized it through the Kepler's third law and the measured stellar mass ($M_{\star}$) and radius ($R_{\star}$) similarly to previous works \citep[e.g.][]{2007ApJ...664.1190S,2020A&A...642A.121L,2022MNRAS.509.1075C}. We also parametrized $e$ and $\omega$ as in Sect.~\ref{subsec:harps_rv_analysis}. We considered uninformative priors for all the parameters except for those for which we have prior independent information, which we constrained through Gaussian priors. Those parameters are the stellar radius and mass, which we derived in our spectroscopic analysis (Sect.~\ref{sec:stellar_charact}), and the LD coefficients, which we computed based on the \texttt{ldtk} package \citep{2015MNRAS.453.3821P}. The package relies upon the TESS transmission curve, $T_{\rm eff}$, log $g$, and [Fe/H], and uses the synthetic spectra library from \citet{2013A&A...553A...6H} to infer the LD coefficients of a given law. To account for possible systematics in the estimated LD coefficients \citep[e.g.][]{2022AJ....163..228P}, we considered uncertainties of 0.2 in $u_{1}$ and $u_{2}$.

The TESS SAP photometry of TOI-5005 is considerably affected by correlated noise. In Sect.~\ref{sec:obs_tess}, we saw that such noise is not dominated by stellar rotation (see also Sect.~\ref{sec:obs_harps}), so it most likely has an instrumental origin. Based on such noise properties, we chose a simple autocorrelation function $k(x_{i}, x_{j})$ (also called a GP kernel; see Appendix~\hyperref[bayesian]{D} for further details) described as

\begin{equation}
    k(x_{i}, x_{j}) = \eta_{\sigma}^{2} \left[ \left(1 + \frac{1}{\epsilon} \right) e^{-(1-\epsilon) \sqrt{3} \tau / \eta_{\rho}} \cdot \left(1 - \frac{1}{\epsilon} \right) e^{-(1+\epsilon) \sqrt{3} \tau / \eta_{\rho}} \right],
\end{equation}where $\tau = x_{i} - x_{j}$ is the temporal separation between two time stamps, and $\eta_{\sigma}$ and $\eta_{\rho}$ are two hyper-parameters that represent the characteristic amplitude and timescale of the correlated variations, respectively. This kernel is called approximate Matérn-3/2, and it has an additional parameter $\epsilon$ that controls the approximation to the exact kernel, which we fixed to its default value of $10^{-2}$ (see \citealt{2017AJ....154..220F} for further details). Given its simplicity and flexibility, this kernel has been extensively used to model TESS photometry where the correlated noise is dominated by unknown instrumental systematics \citep[e.g.][]{2023A&A...679A..33D,2023A&A...677A.182M,2025arXiv250420999C}, as is the case of TOI-5005. Similarly to the transit component, we considered wide, uninformative priors for both hyperparameters since we do not have a priori information about them. Also, given that the amplitudes and timescales of the TESS systematics typically vary from one sector to another, we fit those parameters independently. Hence, we denoted them as $\eta_{\sigma_{i}}$ and $\eta_{\rho_{i}}$, where $i$ refers to the corresponding sector.

We inferred the parameters that best represent the TESS dataset through Bayesian inference as described in Sect.~\ref{sec:model_sel_param_det}. As a sanity check, we repeated the analysis by modelling the highest level public dataset (i.e. QLP/SAP in S12 and SPOC/PDCSAP in S39 and S65), and also by setting wide uninformative priors in the LD coefficients $\mathcal{U}(0,1)$. In both cases, we retrieved consistent planetary parameters at 1$\sigma$. In Fig.~\ref{fig:transit+GP}, we show the complete TESS SAP dataset together with the inferred transit+GP model evaluated on the fit parameters. In Fig.~\ref{fig:individual_transits}, we show the transit+GP model over each individual observed transit event. We note that the obtained orbital and physical parameters based on the TESS data alone are consistent within 1$\sigma$ with the parameters derived in the joint analysis described in the following section. Therefore, for clarity, we only present the final characterisation based on the complete dataset.

\subsection{Joint analysis and system characterisation}
\label{subsec:joint_analysis}

We jointly analysed the HARPS RVs, TESS photometry, and ground-based photometry with the ultimate goal of inferring the orbital and physical parameters of the system. In Sect.~\ref{sec:HARPS_TESS}, we describe the joint analysis of the HARPS and TESS datasets, and in Sect.~\ref{sec:HARPS_TESS_PEST_TRAPPIST} we test the inclusion of the PEST and TRAPPIST-South single transits. 

\subsubsection{HARPS and TESS}
\label{sec:HARPS_TESS}

\begin{figure}
    \centering\includegraphics[width=0.65\textwidth]{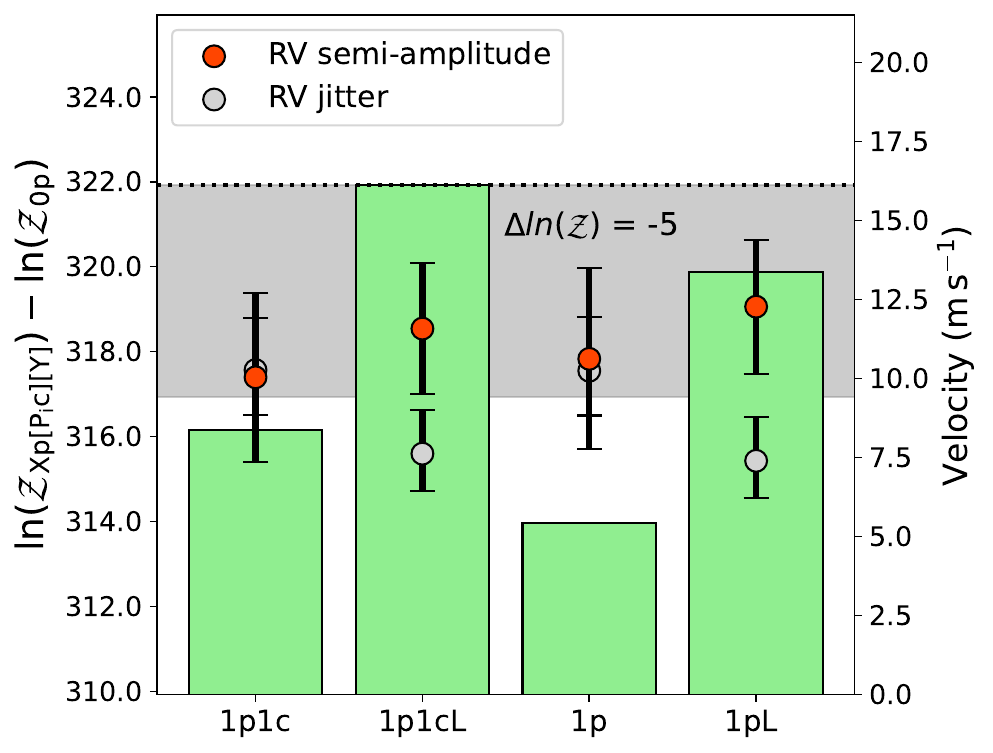}
    \caption[Log-evidence of the four preferred models in the TESS+HARPS joint analysis.]{Bar chart showing the log-evidence of the four preferred models with respect to the 0p model in the TESS+HARPS joint analysis (Sect.~\ref{subsec:joint_analysis}). The grey shaded region indicates the $\rm 0 \geq \Delta ln \mathcal{Z} \geq$ -5 region from the largest evidence model (1p1cL).}
    \label{fig:evidences_tess_harps}
\end{figure}

\begin{figure*}
    \centering
    \includegraphics[width=\textwidth]{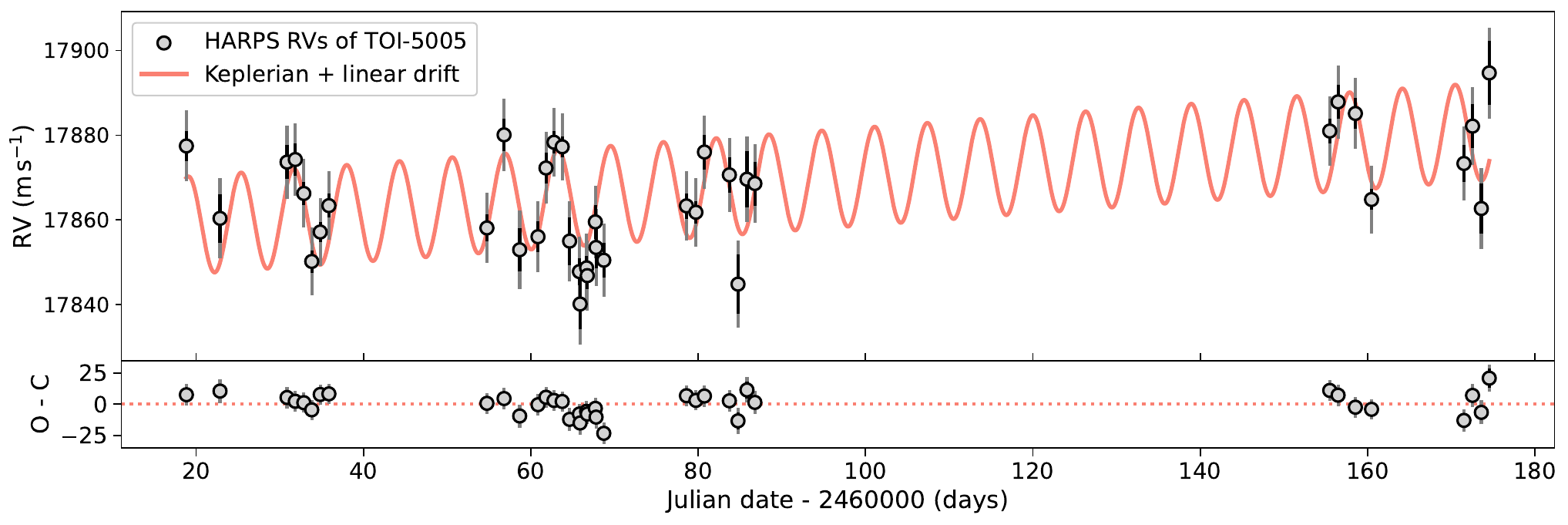}
    \caption[HARPS radial velocities of TOI-5005 together with the simplest model that best represents our data, a Keplerian model plus a linear drift.]{HARPS radial velocities of TOI-5005. The red solid line indicates the simplest model that best represents our data (i.e. a Keplerian model plus a linear drift; 1p1cL), selected based on our joint analysis of the HARPS and TESS datasets (Sect.~\ref{subsec:joint_analysis}). The black error bars represent the uncertainties as estimated by the HARPS DRS ($\sigma_{i,\rm HARPS}$; Sect.~\ref{sec:obs_harps}), and the grey error bars represent the total uncertainties computed as $(\sigma_{i,\rm HARPS}^{2} +  \sigma_{jit,\rm HARPS}^{2})^{0.5}$, where $\sigma_{jit,\rm HARPS}$ is the posterior jitter term. }
    \label{fig:rv_model_complete}. 
\end{figure*}

\begin{figure*}
    \centering
    \includegraphics[width=0.445\textwidth]{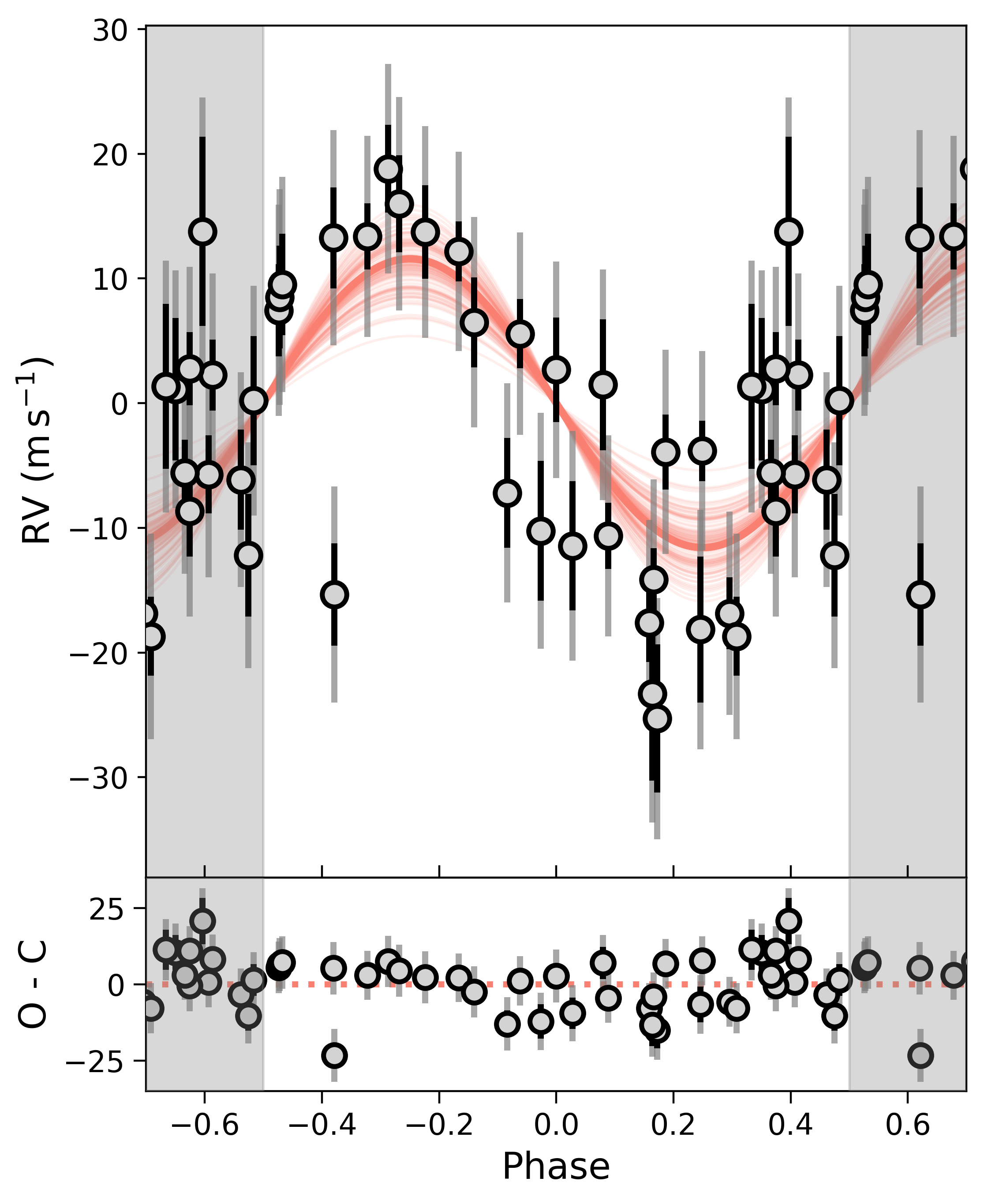}
    \includegraphics[width=0.47\textwidth]{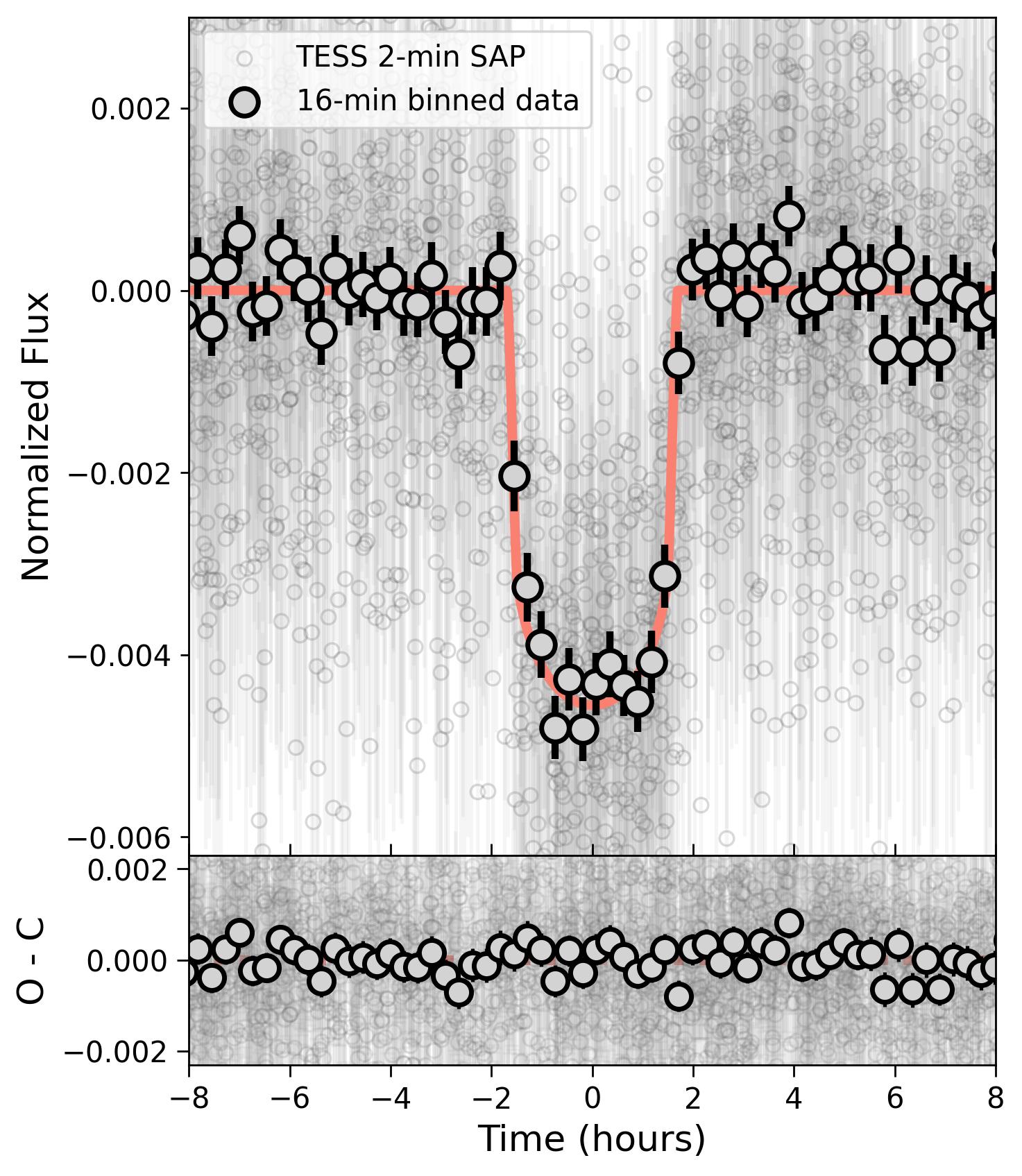}
    \caption[HARPS RVs and TESS photometry of TOI-5005 subtracted from the linear drift and GP component, and folded to the orbital period of TOI-5005~b.]{HARPS RVs (left) and TESS photometry (right) subtracted from the linear drift and GP component, and folded to the orbital period of TOI-5005~b. The black error bars represent the instrument uncertainties, and the grey error bars represent the total uncertainties obtained by quadratically adding the corresponding jitter term. The red solid lines indicate the median posterior models. In the HARPS RVs plot, we also include 100 random posterior models that illustrate the level of uncertainty.}
    \label{fig:phase_folded_plots}
\end{figure*}

As shown in Sect.~\ref{subsec:harps_rv_analysis}, the RV-only analysis is not able to decide which model best represents our data. Therefore, we repeated the model comparison analysis by including the TESS transit analysis described in Sect.~\ref{subsec:TESS_analysis}. Among the 21 tested models, there are four that significantly stand out, showing Bayesian evidence much larger than the remaining 17 models: 1p1c, 1p1cL, 1p, and 1pL. This result discards the detection of additional planets in the system and also discards the need to use a GP component to describe our dataset. In Fig.~\ref{fig:evidences_tess_harps}, we show the log-evidence of those four models. They all meet the condition $\mathcal{B} > 5$ when compared to the 0p model, so we can now confirm the planetary nature of TOI-5005~b via Bayesian inference and model comparison with uninformative priors. Among those four models, the 1p1cL model stands out with the largest evidence and meets the condition $\mathcal{B} > 5$ ($\mathcal{B}$ = 5.8) when compared to the simpler 1p1c model. This result indicates that our dataset is better described when incorporating a linear drift. Also, the models involving eccentric orbits show lower Bayesian evidence than the simpler circular models. Overall, the 1p1cL model is the simplest model that best represents our data when jointly considering the TESS and HARPS observations, so we chose it to infer the orbital and physical parameters of the planetary system following the procedure in Sect.~\ref{sec:model_sel_param_det}.  In Fig.~\ref{fig:rv_model_complete}, we show the HARPS dataset together with the fitted 1p1cL model. In Fig.~\ref{fig:phase_folded_plots}, we show the HARPS RVs and TESS photometry subtracted from the linear drift and GP folded to the period of TOI-5005~b. 

\subsubsection{HARPS, TESS, PEST, and TRAPPIST-South}
\label{sec:HARPS_TESS_PEST_TRAPPIST}

We tested the inclusion of the PEST and TRAPPIST-South single transits (Sect.~\ref{sec:ground_based}) in the joint analysis. For both datasets, we considered a transit model as described in Sect.~\ref{subsec:TESS_analysis} together with a systematics model to account for possible correlations with different parameters. These are the full width at half maximum of the target point spread function (\textit{fwhm}), airmass ($\chi$), position (\textit{x},\textit{y}), displacement (\textit{dx}, \textit{dy}), and distance to the detector centre (\textit{dist}) of the target star, and background flux (\textit{sky}). We assume linear dependencies, so we can write the systematics model as $c_{\rm inst, 0}$ + $\sum_{i = 1}^{N_{\rm inst}} c_{\rm inst, i} \times p_{\rm inst, i}$, where $p_{\rm inst, i}$ are the de-trend parameters described above, $N_{\rm inst}$ is the number of such parameters, and $c_{\rm inst, i}$ are the linear combination coefficients. Similarly to TESS, we considered a jitter term per instrument.  

\begin{figure}
    \centering
\includegraphics [width=0.7\textwidth]{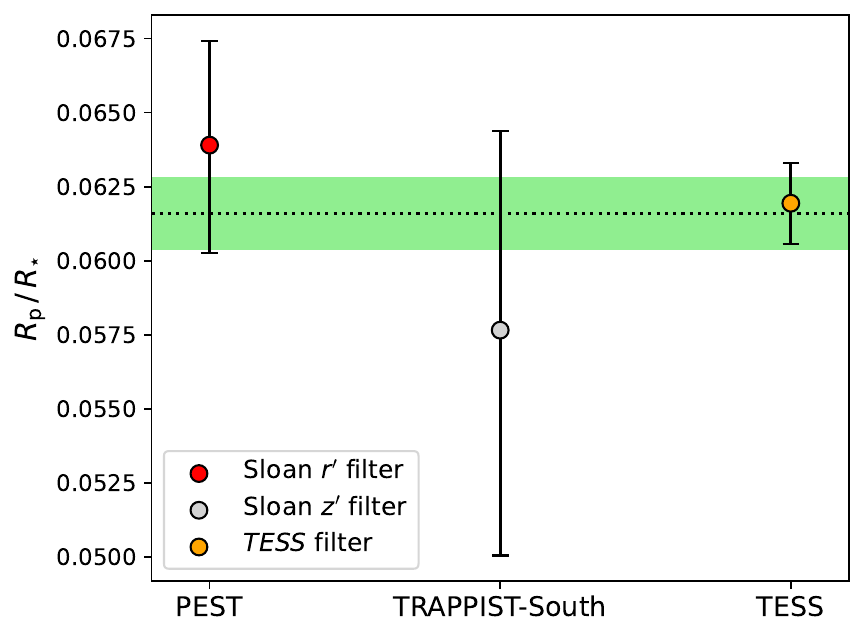}
    \caption[Planet-to-star radius ratios measured by TESS, PEST, and TRAPPIST-South.]{Planet-to-star radius ratios measured by TESS, PEST, and TRAPPIST-South. The horizontal dotted black line and green shadow indicate the median and $\pm$1$\sigma$ interval of the planet-to-star radius ratio measured jointly by the three instruments. }
    \label{fig:rprs_comparison}
\end{figure}

\begin{figure*}
    \centering
    \includegraphics[width=0.49\textwidth]{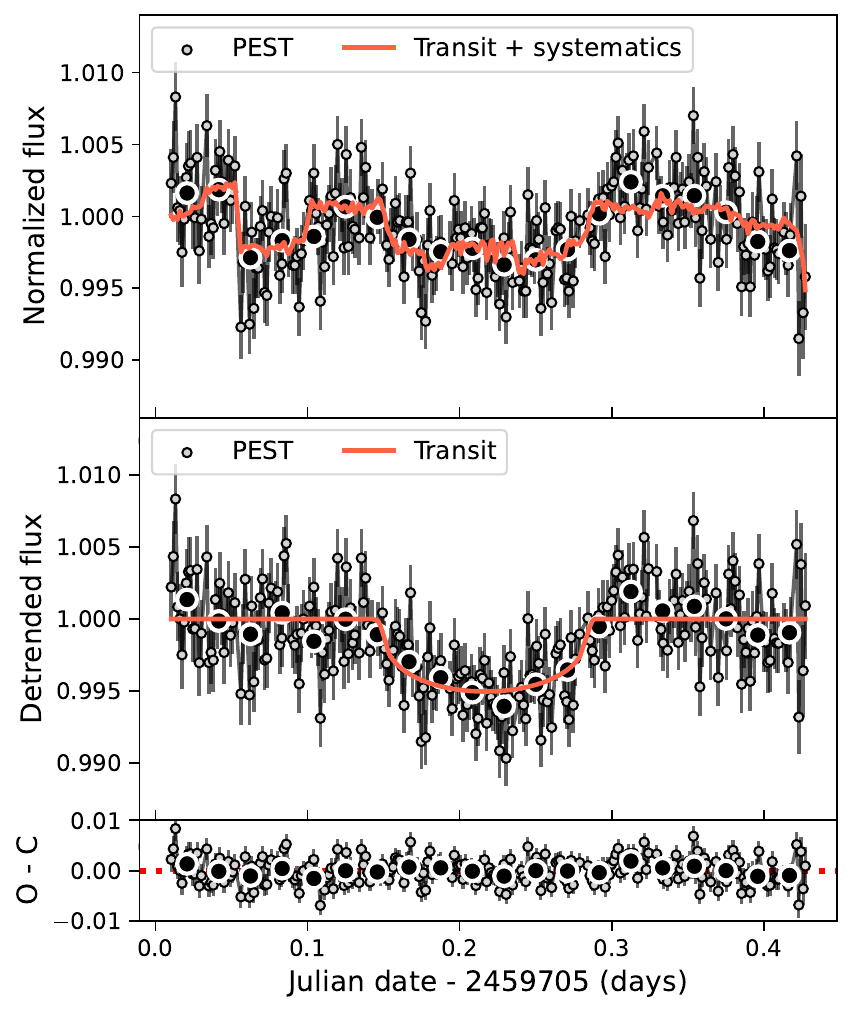}
    \includegraphics[width=0.49\textwidth]{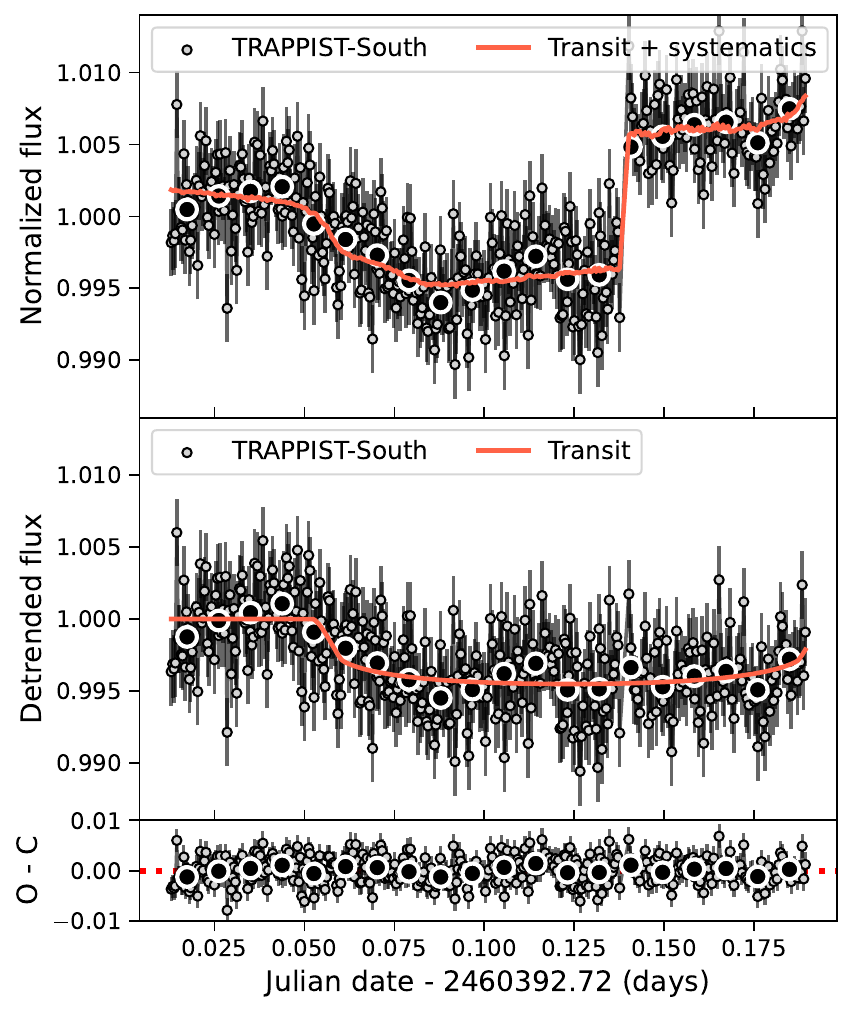}
    \caption[PEST and TRAPPIST-South photometry of TOI-5005~b.]{PEST and TRAPPIST-South photometry of TOI-5005~b acquired on 5 May 2022 and 22 March 2024, respectively. The upper panels show the raw photometry extracted from the instrument pipelines (see Sect.~\ref{sec:ground_based}), the lower panels show the de-trended photometry, i.e. the raw photometry subtracted from the systematics model described in Sect.~\ref{sec:HARPS_TESS_PEST_TRAPPIST}, and the lower panels show the residuals of the joint fit analysis. The red solid lines indicate the median posterior models.}.
    \label{fig:pest_trappist}
\end{figure*}

The TESS, PEST, and TRAPPIST-South photometric data were acquired in different bands: TESS, Sloan \textit{r'}, and Sloan \textit{z'}, respectively. Hence, we considered different $q_{1}$ and $q_{2}$ limb-darkening coefficients for each instrument and set wide Gaussian priors based on \texttt{ldtk} as described in Sect.~\ref{subsec:TESS_analysis}. We first performed a joint analysis that considers different $R_{\rm p}/R_{\rm \star}$ ratios per instrument. We compare the posterior $R_{\rm p}/R_{\rm \star}$ in Fig.~\ref{fig:rprs_comparison}. The obtained values are consistent at the 1$\sigma$ level, which indicates that our photometric dataset is not sensitive to a possible transit depth chromaticity. This allowed us to perform a joint analysis with a common $R_{\rm p}/R_{\rm \star}$ for the three instruments. The derived orbital and physical parameters are consistent at the 1$\sigma$ level with those from the HARPS+TESS analysis (Sect.~\ref{sec:HARPS_TESS}), and several transit parameters show slightly lower uncertainties. We also find that the Bayesian Evidence difference with that of a global model with no planets is larger when including the PEST and TRAPPIST-South data: $(\rm \Delta ln \mathcal{Z})_{HARPS+TESS+PEST+TRAPPIST}$ = +361 against $(\rm \Delta ln \mathcal{Z})_{HARPS+TESS}$ = +322. Therefore, we find the inclusion of the ground-based photometry beneficial in the joint analysis.  In Fig.~\ref{fig:pest_trappist}, we show the PEST and TRAPPIST-South raw and systematics-corrected photometry together with the corresponding posterior models. In Table~\ref{tab:parameters_joint}, we present the median and 1$\sigma$ intervals of the complete parameter set that describes the joint model.  In Fig.~\ref{fig:corner_joint_analysis}, we represent a corner plot with the 1D and 2D posterior distributions of the main fitted orbital and physical parameters. 

\subsection{Stellar signals analysis}
\label{subsec:stellar_signals_analysis}

\subsubsection{HARPS activity indicators}
\label{subsubsec:HARPS_activity_indicators}

Four HARPS activity indicators (FWHM, S-index, Contrast, and Ca) of TOI-5005 show a sinusoidal signal with a periodicity of $\simeq$21 days (Sect.~\ref{sec:obs_harps}, Fig.~\ref{fig:gls_to_HARPS}). This signal most likely reflects the rotation period of the star and is compatible with independent $P_{\rm rot}$ estimates (Sect.~\ref{subsec:prot}). In this section, we follow the procedure described in Sect.~\ref{sec:model_sel_param_det} to obtain an accurate determination of the signal periodicity and assess its significance. We built an activity model based on a GP with a quasi-periodic covariance function (Eq.~\ref{eq:QP}), and fitted it to the individual activity indicators' time series starting from wide, uninformative priors. The MCMC analysis shows that the active regions' timescale ($\eta_{2}$) and harmonic complexity of the signal ($\eta_{4}$) cannot be constrained by our dataset. However, the stellar rotation period ($\eta_{3}$ or $P_{\rm rot}$) converges at $\simeq$21 days. To get a robust estimate and assess how well the activity model describes our dataset, we jointly modelled the four indicators with a common $P_{\rm rot}$. We obtain $P_{\rm rot} = 21.01^{+0.46}_{-0.60}$~days and a Bayesian Evidence against the null hypothesis of $\rm \Delta ln \mathcal{Z}$ = +8.6. We include this value in Table~\ref{tab:stellar_parameters}.

\subsubsection{TESS photometric variability}
\label{subsubsec:photometric_variability}

\begin{table}[]
\centering
\caption[Properties of the TESS variations synchronised with the period of TOI-5005~b]{Main properties of the TESS photometric variations synchronised with the orbital period of TOI-5005~b (Sect.~\ref{subsubsec:photometric_variability}). $f_{\rm inc}$ is the orbital phase fraction in which the TESS photometric flux increases. $(f_{\rm inc})_{\rm vis}$ and $(f_{\rm inc})_{\rm hid}$ represent the same fraction, but limited to the phases where a hypothetical co-rotating active region would be visible and hidden from Earth, respectively. $\phi_{\rm max}$ and $\phi_{\rm min}$ are the phase offsets between the TOI-5005~b transit time and the maximum and minimum flux emissions, respectively.}
\renewcommand{\arraystretch}{1.4}
\setlength{\tabcolsep}{16.6pt}
\begin{tabular}{cccccc}
\hline \hline
Sector & $f_{\rm inc}$ & $(f_{\rm inc})_{\rm vis}$ & $(f_{\rm inc})_{\rm hid}$ & $\phi_{\rm max}$ & $\phi_{\rm min}$ \\ \hline
S39    & 0.56          & 0.83                      & 0.30                      & +0.33            & -0.21            \\
S65    & 0.49          & 0.26                      & 0.72                      & -0.33            & +0.21            \\ \hline
\end{tabular}
\label{tab:properties_variations}
\end{table}

\begin{figure*}
    \centering
    \includegraphics[width=0.25\textwidth]{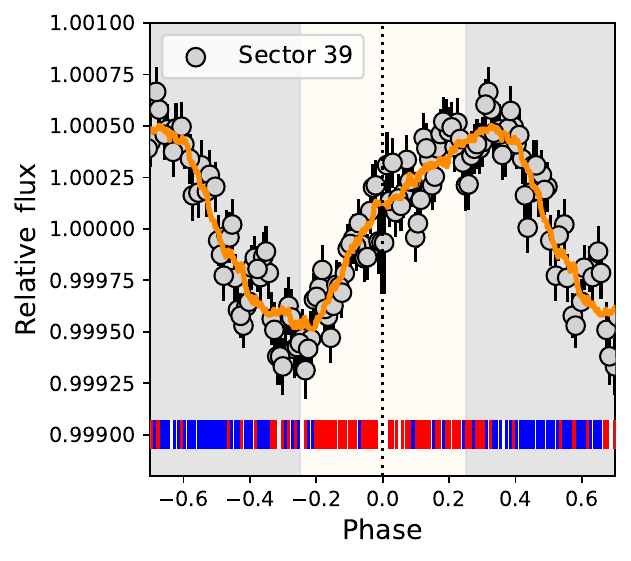}
    \includegraphics[width=0.23\textwidth]{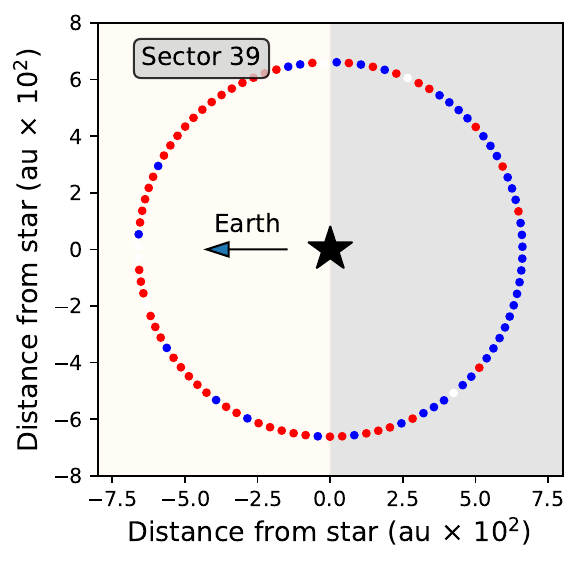}
    \includegraphics[width=0.25\textwidth]{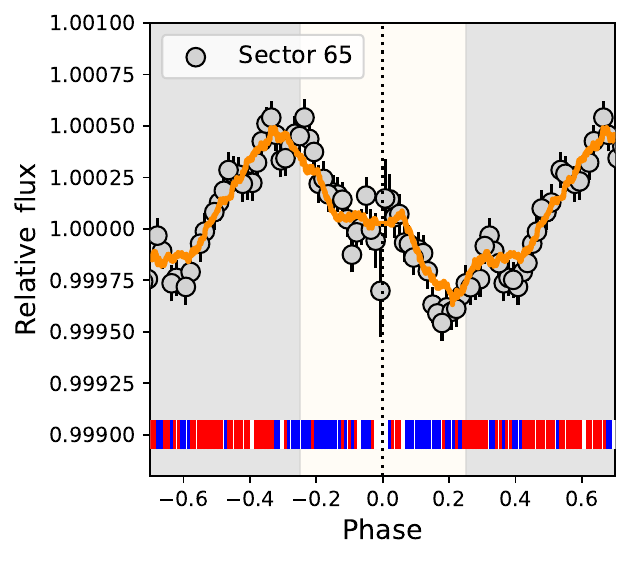}
    \includegraphics[width=0.23\textwidth]{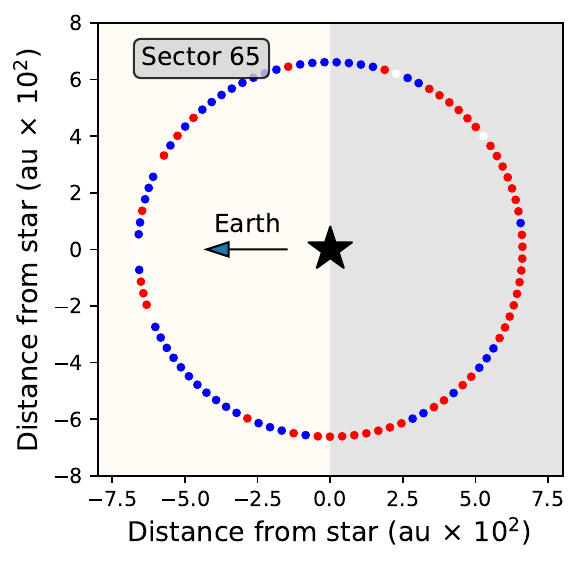}
    \caption[Links between the orbital motion of TOI-5005~b and the TESS variability of TOI-5005.]{Links between the orbital motion of TOI-5005~b and the TESS photometric variability of TOI-5005. Left panels: TESS photometry of TOI-5005 folded in phase with the orbital period of TOI-5005~b (Sect.~\ref{subsec:joint_analysis}, Table \ref{tab:parameters_joint}) and binned with $\simeq$5$\%$ phase bins. The orange line corresponds to the filtered photometry through a median filter with a kernel size of 1401 cadences. The blue, red, and white vertical lines indicate whether the photometry increases, decreases, or remains stable in time lapses of $\simeq$1.5 h. Right panels: Orbital path of TOI-5005 b, which follows the anticlockwise direction. The circles represent the location of TOI-5005 b every $\simeq$1.5 h and are coloured similarly to the vertical lines in the left panels. The white and grey backgrounds represent the orbital regions in which a hypothetical co-rotating active region would be visible and not visible from Earth, respectively.}
    \label{fig:MSPIs}
\end{figure*}

The \texttt{GLS} periodograms of the PDCSAP photometry of TOI-5005 show a significant sinusoidal modulation with a periodicity of $\simeq 6.3$~days (Sect.~\ref{sec:obs_tess}, Fig.~\ref{fig:gls_to_TESS}). This periodicity matches the orbital period of TOI-5005~b, which suggests a planet-induced origin through MSPIs. TESS photometry, however, is frequently affected by significant instrumental systematics. If not modelled properly, they can generate instrumental signals that could be mistakenly interpreted as stellar. We studied whether the observed modulation could have an instrumental origin. To that end, we performed an independent photometric correction, which consists of fitting all the major instrumental systematics observed in the TESS CCDs without being a priori constrained by the observed systematics in nearby stars (see Appendix~\hyperref[sec:cbv_correction]{E} for further details). We find that no combination of the major TESS systematics can explain the observed $\simeq 6.3$-day modulation, which strongly favours a stellar origin. We note that the signal could still be instrumental in the case that the main pixels collecting the stellar flux are affected by particular systematics different from those observed in other regions of the CCD. However, we find this possibility very unlikely since the same signal appears in two different sectors, where TOI-5005 falls in different pixels of different CCDs. Another factor to consider is the flux contamination from nearby sources. Contrary to the planetary signal of TOI-5005~b, we do not have independent observations that confirm that the $\simeq 6.3$-day sinusoidal modulation comes from TOI-5005. However, the observed modulation would require very large photometric variations if it came from the faint contaminant sources (see Sect.~\ref{sec:obs_tess}), which makes it a very unlikely scenario. Taking all the mentioned analyses into account, we consider that the $\simeq 6.3$-day signal most likely has a stellar origin coming from TOI-5005.  

An in-depth analysis of the measured signal is beyond the scope of this work. However, we are interested in studying how the flux variations are related to the orbital motion of TOI-5005~b since it can give us important insight into the potential MSPIs. Similarly to \citet{2024A&A...684A.160C}, we divided the complete S39 and S65 phase-folded photometry into 100 bins of $\simeq$1.5 h long and studied when the flux increases and decreases. Short-term variations induced by the photometric scatter were previously filtered out through a median filter with a kernel size of 1401 cadences; that is, $\simeq$10$\%$ of the orbital phase. We find that the orbital phase fraction in which the photometric flux increases ($f_{\rm inc}$) roughly corresponds to half an orbit, which indicates a high degree of signal symmetry. 

We also computed the orbital fraction of increasing photometry limited to the phases where a hypothetical co-rotating and small active region would be visible $(f_{\rm inc})_{\rm vis}$ and hidden $(f_{\rm inc})_{\rm hid}$ from Earth. Interestingly, we find a strong imbalance between increasing and decreasing regions (see Table~\ref{tab:properties_variations} and Fig.~\ref{fig:MSPIs}). In S39, the TESS flux primarily increases when such a hypothetical region is visible from Earth, while in S65 it primarily increases when this region is hidden. We note that this behaviour discards the possibility that the co-rotating region is small, since, in that case, we would observe a flux increase until TOI-5005~b's transit time (phase = 0), where the flux would decrease until the spot reaches the stellar limb (phase = 0.25), where the flux would become flat until reaching the following limb (phase = 0.75). Instead, the observed photometric variations can be explained by an extensive co-rotating active region (i.e. comparable to the stellar disk) trailing or leading TOI-5005~b's orbit with a bright or dark contribution to the total stellar flux. 

We also computed the phase offset between TOI-5005~b's transit time and the maximum ($\phi_{\rm max}$) and minimum ($\phi_{\rm min}$) flux emission, and found a 0.5 phase difference between S39 and S65 time series (see Table~\ref{tab:properties_variations}). This phase offset shows that the signal shape changes in short time scales, and it can be either interpreted as the potential planet-induced active region changing from bright to dark contrast or alternating its location with respect to the planetary orbit from a trailing to a leading configuration and vice versa. Interestingly, very similar offsets between the planet transit time and the stellar activity extremes, as well as between different planetary orbits have been detected in similar systems with signs of MSPIs \citep[e.g.][]{2005ApJ...622.1075S,2008ApJ...676..628S,2008A&A...482..691W,2019NatAs...3.1128C,2024A&A...684A.160C}.


\section{Discussion}
\label{sec:discussion}

The mass ($M_{\rm p}$ = $32.7\pm 5.9$ $\rm M_{\oplus}$; $M_{\rm p}$ = $0.103\pm 0.018$ $\rm M_{\rm J}$) and radius ($R_{\rm p}$ = $6.25\pm 0.24$~$\rm R_{\rm \oplus}$; $R_{\rm p}$ = $0.558\pm 0.021$ $\rm R_{\rm J}$) derived in the joint analysis place TOI-5005~b approximately halfway between Neptune and Saturn. We adopt the most commonly used convention \citep[e.g.][]{2011ApJ...726...52H,2014A&A...572A...2B,2015ApJ...813..111B} and refer to TOI-5005~b as a super-Neptune. 

There are very few planets with physical properties similar to those of TOI-5005~b. For example, only 119 planets (i.e. 2$\%$ of the known population) have been found within a radius range 5 $\rm R_{\oplus}$ $<$ $R_{\rm p}$ $<$ 7~$\rm R_{\oplus}$. This number is reduced to ten detections if we limit to a period range of 5 days $<$ $P_{\rm orb}$ $<$ 7 days, of which only two have a precise mass measurement (CoRoT-8 b; \citealt{2010A&A...520A..66B}, and TOI-4010 c; \citealt{2023AJ....166....7K}). Interestingly, among the ten detections, TOI-5005~b orbits the brightest host, which makes it the most amenable planet for follow-up observations in this parameter space. Other super-Neptunes similar to TOI-5005~b orbiting relatively bright hosts ($V$ $<$ 12 mag) are K2-39~b \citep{2016AJ....152..143V}, HATS-38~b \citep{2020AJ....160..222J}, K2-334~b \citep{2021MNRAS.508..195D}, TOI-181~b \citep{2023MNRAS.521.1066M}, K2-141~c \citep{2018AJ....155..107M}, TOI-5126~b \citep{2024MNRAS.527.8768F}, and TOI-1248~b \citep{2024ApJS..272...32P}.

In Sects.~\ref{subsec:period-radius} and \ref{subsec:LDSP}, we contextualise TOI-5005~b in different parameter spaces and discuss its observed properties according to different evolutionary hypotheses and additional observational constraints. In Sects.~\ref{sec:internal_structure}, \ref{subsec:mass-loss}, and \ref{subsec:prospects_atm}, we use the system parameters to infer the internal structure of TOI-5005~b, constrain its mass-loss rate, and discuss the prospects for characterising its atmosphere.

\subsection{TOI-5005~b in the period-radius diagram: A new super-Neptune in the savanna near the ridge}
\label{subsec:period-radius}

\begin{figure}
    \centering
    \includegraphics[width=0.65\textwidth]{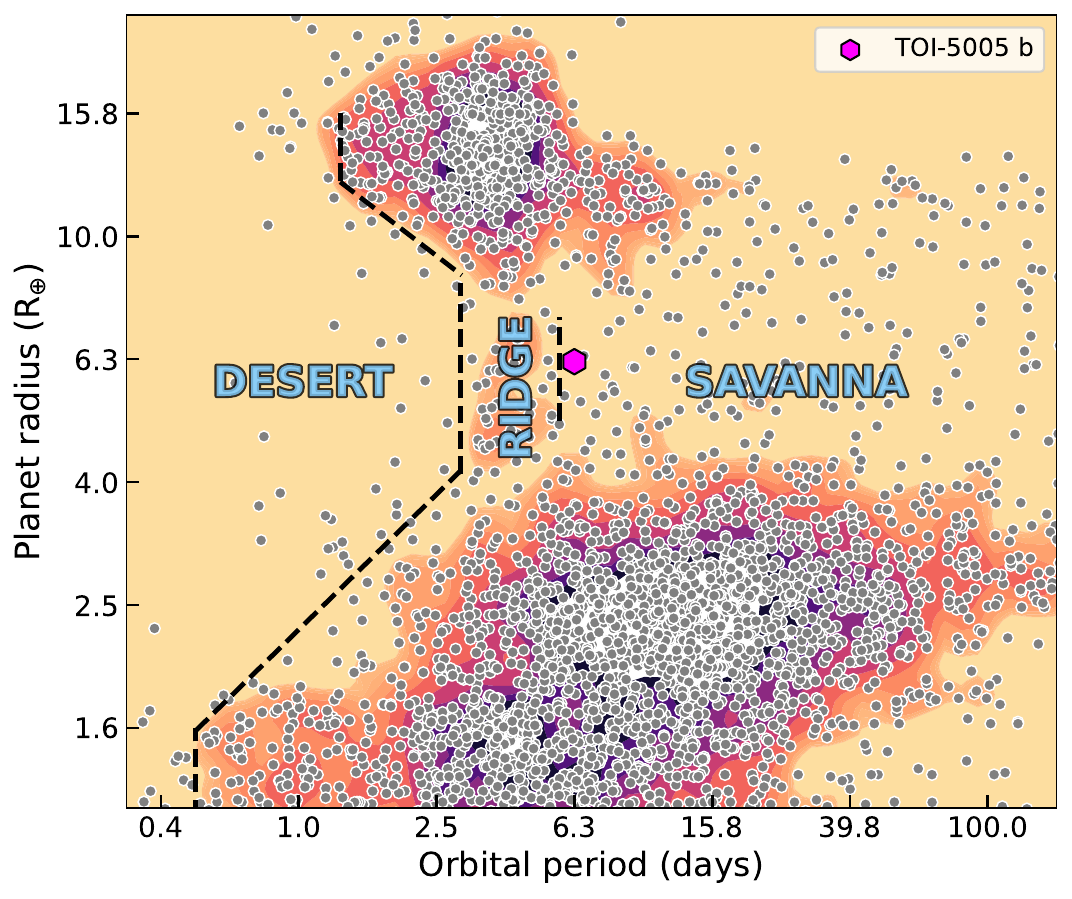}
    \caption[TOI-5005~b in the period-radius diagram of close-in exoplanets, where we highlight the population-based boundaries of the Neptunian desert, ridge, and savanna.]{TOI-5005~b in the period-radius diagram of close-in exoplanets, where we highlight the population-based boundaries of the Neptunian desert, ridge, and savanna derived in \citet{2024A&A...689A.250C}. The data were collected from the NASA Exoplanet Archive \citep{2013PASP..125..989A} on 20/09/2024. This plot was generated with \texttt{nep-des} (\url{https://github.com/castro-gzlz/nep-des}).}
    \label{fig:nep_desert}
\end{figure}

In Fig.~\ref{fig:nep_desert}, we contextualise TOI-5005~b in the period-radius diagram of known close-in planets. Having an orbital period of 6.3 days, TOI-5005~b lies in the Neptunian savanna near the ridge, a recently identified over-density of planets at $\simeq$3-5 days \citep{2024A&A...689A.250C}. The authors argue that the dynamical mechanism that brings planets preferentially to the ridge might be similar to the mechanism that brings larger planets to the $\simeq$3-5 day hot-Jupiter pileup \citep[e.g.][]{2007ARA&A..45..397U,2009ApJ...693.1084W}. Interestingly, different works are revealing a large number of eccentric and misaligned Jupiter- and Neptune-sized planets within and near the $\simeq$3-5 day over-density \citep[e.g.][]{2012ApJ...757...18A,2020A&A...635A..37C,2023A&A...669A..63B}, which suggests that it could be primarily populated by HEM processes \citep[see][]{2012ApJ...754L..36N,2017AJ....154..106N,2018ARA&A..56..175D,2021JGRE..12606629F}. Unfortunately, our HARPS RV dataset did not allow us to constrain the orbital eccentricity of TOI-5005~b. In this hypothesis, planets within the ridge would be expected to have an outer massive companion that triggered such migration processes \citep[e.g.][]{2003ApJ...589..605W,2008ApJ...686..621F,2011CeMDA.111..105C}. Our HARPS RV dataset shows a long-term linear trend that could be caused by an outer companion (see Sects.~\ref{sec:obs_harps} and \ref{subsec:joint_analysis}). We note, however, that several activity indicators of TOI-5005 also show similar trends (Sect.~\ref{sec:obs_harps}), so we cannot discard that they could all reflect the magnetic cycle of the star. Indeed, the RV and FWHM trends are inverse to the Contrast trend, which is what we would observe if the magnetic cycle was causing them \citep{2011arXiv1107.5325L}. Additional long-term high-resolution spectroscopic measurements or high-precision astrometry are needed to resolve this dichotomy. Overall, based on our current data, we cannot infer whether TOI-5005~b reached its location through disk-driven migration or HEM processes.

Atmospheric escape is also thought to shape the close-in period-radius distribution depicted in Fig.~\ref{fig:nep_desert}. While some theoretical works predict that Jupiter-sized planets could be eroded into super-Earths \citep[e.g.][]{2014ApJ...783...54K}, the majority of models and observational constraints indicate that these planets are too massive to significantly evaporate \citep[see][for a review]{2018ARA&A..56..175D,2021JGRE..12606629F}. In contrast, the atmospheres of several Neptunian planets within and near the ridge have been observed to be eroding at very high rates \citep[e.g.][]{2015Natur.522..459E,2018A&A...620A.147B}. Hence, as suggested by \citet[][]{2018Natur.553..477B,Attia2021,2024A&A...689A.250C}, Neptunes in the ridge might survive evaporation for a limited time, so that the ones that we detect today would have arrived relatively recently, presumably through HEM processes. Interestingly, this hypothesis would answer the question of why warm Neptunes present non-zero eccentricity studied in \citet{2020A&A...635A..37C}. Unfortunately, atmospheric escape has not been extensively probed at larger orbital distances (i.e. within the savanna), and evaporation models show important discrepancies in the still poorly explored Neptunian domain (see Sect.~\ref{subsec:mass-loss} for further discussion). Hence, given its unusual location in the period-radius diagram and the brightness of the host star, TOI-5005~b represents a unique opportunity to probe how deep into the savanna atmospheric escape plays a relevant role. This, together with observations of the spin-orbit angle and long-term monitoring to detect massive companions, will provide a clearer picture of the overall evolution of this system. 

\subsection{TOI-5005~b in the $M_{\rm p}-R_{\rm p}$ and $\rho_{\rm p}-P_{\rm  orb}$ diagrams: A new member of the low-density savanna planets}
\label{subsec:LDSP}

\begin{figure*}
    \centering
    \includegraphics[width=0.48\textwidth]{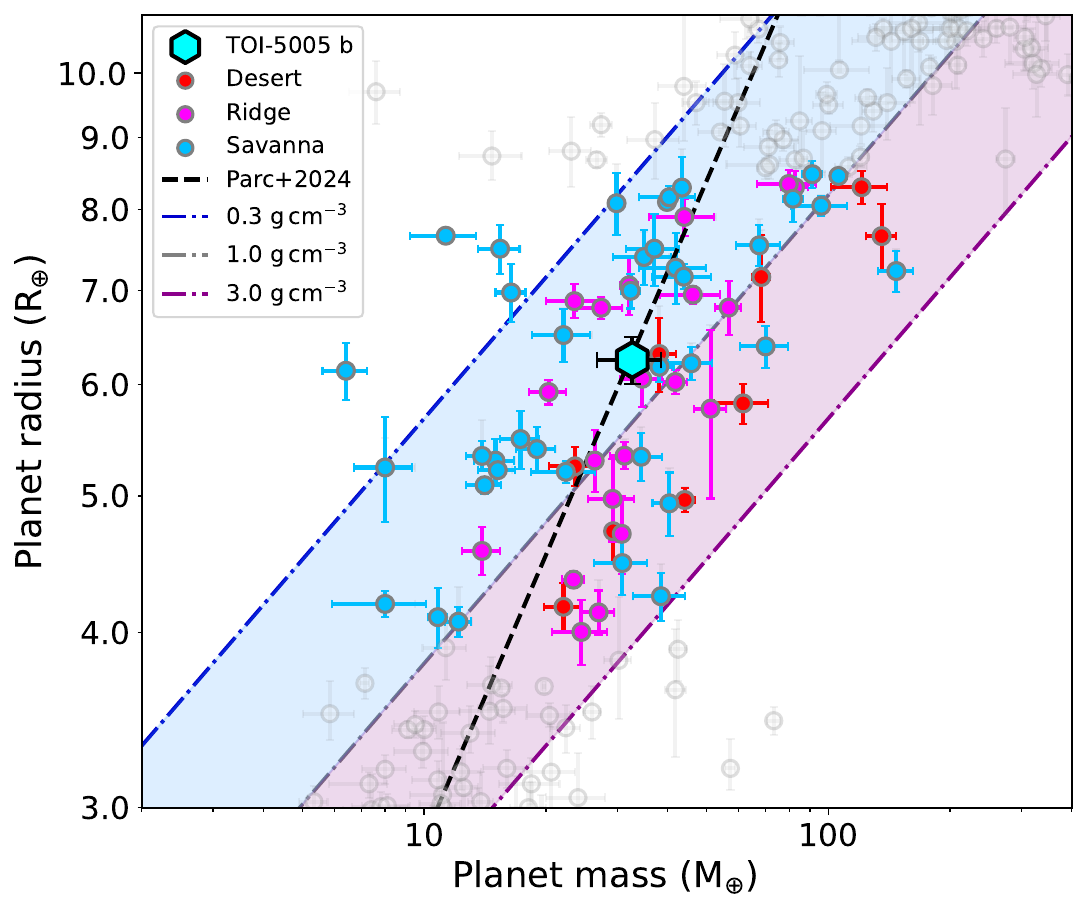}
    \includegraphics[width=0.476\textwidth]{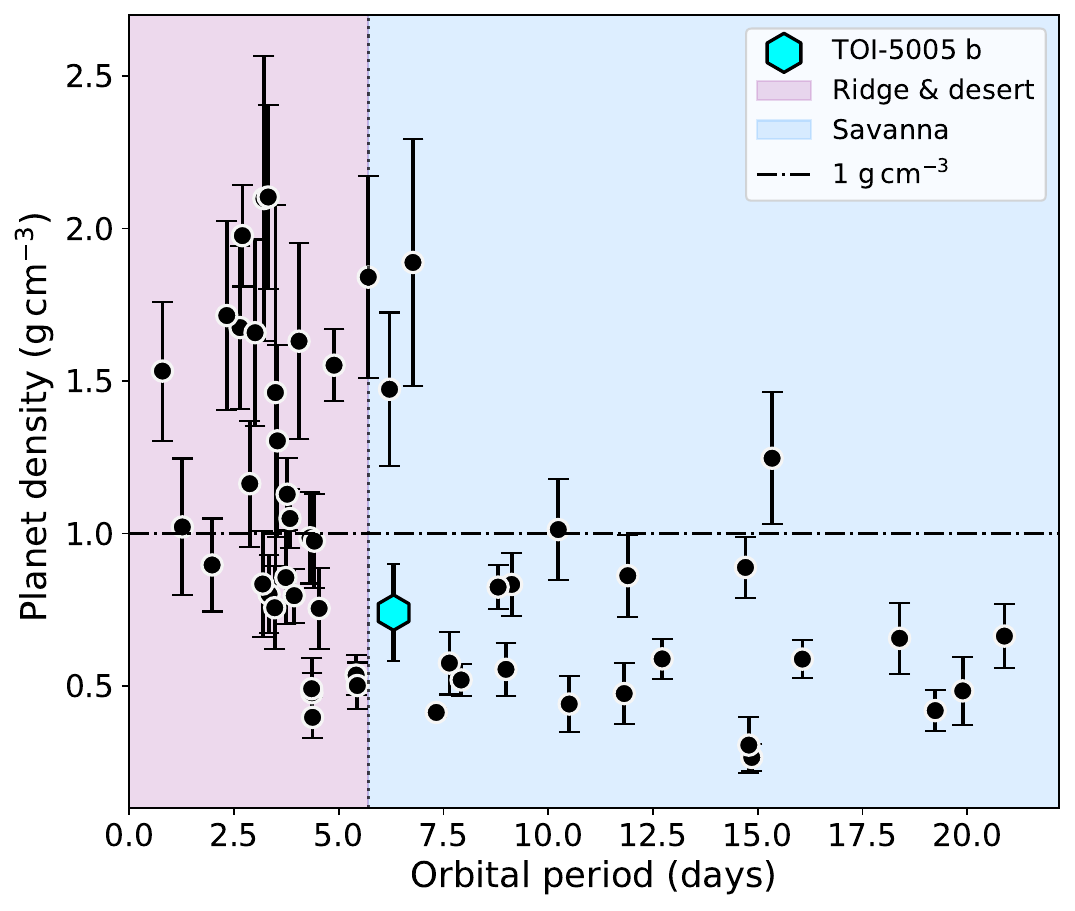}
    \caption[TOI-5005~b in the mass-radius and density-period diagrams.]{TOI-5005~b in the mass-radius and density-period diagrams. Left: Mass-radius diagram of all Neptunian planets with masses and radii with precisions better than 20$\%$. Right: Density-period diagram of the same sample. The data were collected from the NASA Exoplanet Archive on 20/09/2024. This plot was generated with \texttt{mr-plotter} \citep[\url{https://github.com/castro-gzlz/mr-plotter};][]{2023A&A...675A..52C}. }
    \label{fig:mass-radius-density-period}
\end{figure*}

In Fig.~\ref{fig:mass-radius-density-period}, left panel, we contextualise TOI-5005~b in the mass-radius diagram of precisely characterised planets. Its location is consistent with the mass-radius relation for volatile-rich exoplanets derived by \citet{2024A&A...688A..59P}. If we focus on the entire population, we see that there is a large dispersion in this region of the parameter space. Hence, continuing with the discussion in Sect.~\ref{subsec:period-radius}, we aim to study whether there exists any link between the observed planet densities and the new Neptunian landscape presented in \citet{2024A&A...689A.250C}. In the left panel of Fig.~\ref{fig:mass-radius-density-period}, we highlight the planets in the desert, ridge, and savanna in different colours. In the right panel of Fig.~\ref{fig:mass-radius-density-period}, we represent the density-period distribution of Neptunian planets. Interestingly, we find that planets in the ridge (and desert) tend to have densities larger than 1 $\rm g \, cm^{-3}$ (median value of 1.3~$\rm g \, cm^{-3}$), while planets in the savanna show lower densities, typically below 1~$\rm g \, cm^{-3}$ (median value of 0.6~$\rm g \, cm^{-3}$). We note that this difference is unlikely to be attributed to observational biases since denser planets are easier to detect. We also note that denser planets typically show larger uncertainties than low-density planets, which is due to the considered relative precision threshold for the masses and radii. 

We quantified the significance of the trend through a Kolmogorov–Smirnov (KS) statistical test between the ridge and savanna populations, which we performed by taking into account the uncertainties of the densities through bootstrapping. We obtain a $D$-statistic of 0.39 $\pm$ 0.04 ($p$-value of $0.0092^{+0.0184}_{-0.0066}$), which allows us to confidently reject the hypothesis that both samples are drawn from the same distribution. We also searched for possible trends in the mass-period and radius-period spaces (see Fig.~\ref{fig:mass-period-radius-period}). We obtain $D$-statistics 0.33 $\pm$ 0.03 and 0.22 $\pm$ 0.03 ($p$-values of $0.037^{+0.030}_{-0.018}$ and $0.33^{+0.19}_{-0.13}$) for the mass and radius sample, respectively. On the one hand, the mass-period $p$-value is lower than
the most commonly assumed threshold for considering statistical significance (i.e. $p$-value $<$ 0.05), although we note that its upper limit is larger (0.068). On the other hand, the radius-period $p$-value is well above such a threshold. We thus conclude that the radius distribution across the orbital period space is insensitive to the ridge and savanna, but Neptunes in the ridge tend to be more massive and dense than those in the savanna, being the planet density the key parameter that maximises the significance. Having a density of $\rho_{\rm p}$ = $0.74 \pm 0.16$ $\rm g \, cm^{-3}$ and being located in the savanna with an orbital period of 6.3 days, TOI-5005~b is consistent with this newly identified trend.

The density-period trend observed in the Neptunian population might be explained through atmospheric, formation, and dynamical processes. However, determining which is the predominant agent shaping it is not trivial. On the one hand, Neptunian planets could all have been formed with low densities (e.g. $<$ 1 $\rm g \, cm^{-3}$), presumably with extensive H/He atmospheres, and only those receiving high irradiations (i.e. those orbiting at short orbital distances) would be able to undergo strong enough evaporation to increase the bulk planet density. On the other hand, the formation and migration history of different types of Neptunes could also play an important role. While most close-in giant planets are thought to have undergone disk-driven migration soon after their formation, there is increasing evidence that many Neptunes in the ridge would have undergone late HEM processes (see  Sect.~\ref{subsec:period-radius}). In that sense, the density difference between Neptunes in the ridge and the savanna could reflect the existence of two different populations of planets that formed and migrated through different channels. To date, there is not enough observational evidence to discern between the different hypotheses. Hence, confirming and characterising planets located around the boundary between the desert and savanna, such as TOI-5005~b will provide important insight into the transition between two possible different populations of Neptunes.

\subsection{Internal structure} \label{sec:internal_structure}

We performed an MCMC retrieval of the interior structure of TOI-5005~b. We employed GASTLI (GAS gianT modeL for Interiors, \citealp[]{Acuna21,2024A&A...688A..60A}) as a forward interior model, which is a one-dimensional, coupled interior-atmosphere model for warm gas giants. Our interior model is stratified into two layers: a core that consists of a 1:1 rock and water mixture, and an envelope. The mass of the core is computed from the planet's interior mass, $M_{\rm interior}$, and the core mass fraction (CMF), which are our input parameters, $M_{\rm core} =$ CMF $\times \ M_{\rm p}$. The composition of the envelope is determined by the input parameter $Z_{\rm env}$, which is the mass fraction of metals in the envelope, represented by water, whereas (1-$Z_{\rm env}$) is the mass fraction of H/He in the envelope. Hence, the total metal mass fraction in the planet is computed as $Z_{\rm planet}$ = CMF + (1-CMF) $\times \ Z_{\rm env}$. The interior structure models solve the equations of hydrostatic equilibrium, adiabatic temperature profile, mass conservation and Gauss’s theorem to compute the pressure, temperature and gravity profiles, respectively. The density and entropy are obtained with the respective equation of state (EOS) of each material. We adopt state-of-the-art EOS for silicates \citep{sesame,Miguel22}, H/He \citep{Chabrier21,HG23}, and water \citep{Mazevet19,Haldemann20}. Furthermore, we solve the equation of thermal cooling to determine the planetary luminosity as a function of age, $L = 4 \pi \sigma R^{2} T_{\rm int}^4$, where $T_{\rm int}$ is the internal (or intrinsic) temperature \citep{Fortney07,Thorngren16}. The boundary condition of the interior model is determined by coupling it to a grid of atmospheric models obtained with \texttt{petitCODE} \citep{Molliere15,Molliere17}. In the atmospheric models, we consider clear, cloud-free atmospheres. The envelope metal mass fraction,  $Z_{\rm env}$, is estimated using \texttt{easyCHEM} \citep{easychem} assuming chemical equilibrium for a given atmospheric metallicity in $\times$ solar units, log(Fe/H). The interior and the atmosphere are coupled at a pressure of 1000 bar, by using an iterative algorithm to ensure that the planet's interior radius (from the planet centre up to 1000 bar), and boundary temperature converge to constant values \citep{Acuna21}. The total planet radius is the sum of the converged interior radius and the thickness of the atmosphere, which is computed by integrating the gravity and pressure profiles from the bottom of the atmosphere (1000 bar) to the transit pressure \citep[20 mbar, see][]{2014ApJ...792....1L,Grimm18,Mousis20}. The total planet mass, envelope mass, CMF and $Z_{\rm planet}$ are recalculated taking into account the atmospheric mass (i.e $M_{\rm p} = M_{\rm interior} + M_{\rm atm}$), which has a significant effect in planets whose atmosphere is massive. 

For the MCMC retrieval, we consider three observable parameters, which are the planetary mass and radius ($M_{\rm p}$~= $32.7\pm 5.9$, $R_{\rm p}$ = $6.25\pm 0.24$~$\rm R_{\rm \oplus}$), and age ($age \ = \ 2.20^{+0.61}_{-0.28}$ Gyr). We compute the log-likelihood as described in Eqs. 6 and 14 of \citet{Dorn15} and \citet{Acuna21}, respectively. We use the \texttt{emcee} \citep{emcee} MCMC package as a sampler. Our priors are uniform: $\mathcal{U}(0,0.99)$ for CMF; $\mathcal{U}(-2,2.4)$ for log(Fe/H); and $\mathcal{U}(50,250)$ K for the internal temperature. For the interior planet mass, $M_{\rm interior}$, we use a Gaussian prior with a mean and standard deviation equal to the observed mass. This choice eases the convergence of the MCMC retrieval as the difference between the interior mass and the total mass is negligible due to the low atmospheric mass. We assume a constant global equilibrium temperature of 1000 K, which is the maximum limit of our grid of atmospheric models. This is close enough to TOI-5005 b's equilibrium temperature (1040 K) to have a negligible effect on the metal content estimate.

\begin{figure}
    \centering
    \includegraphics[width=0.6\textwidth]{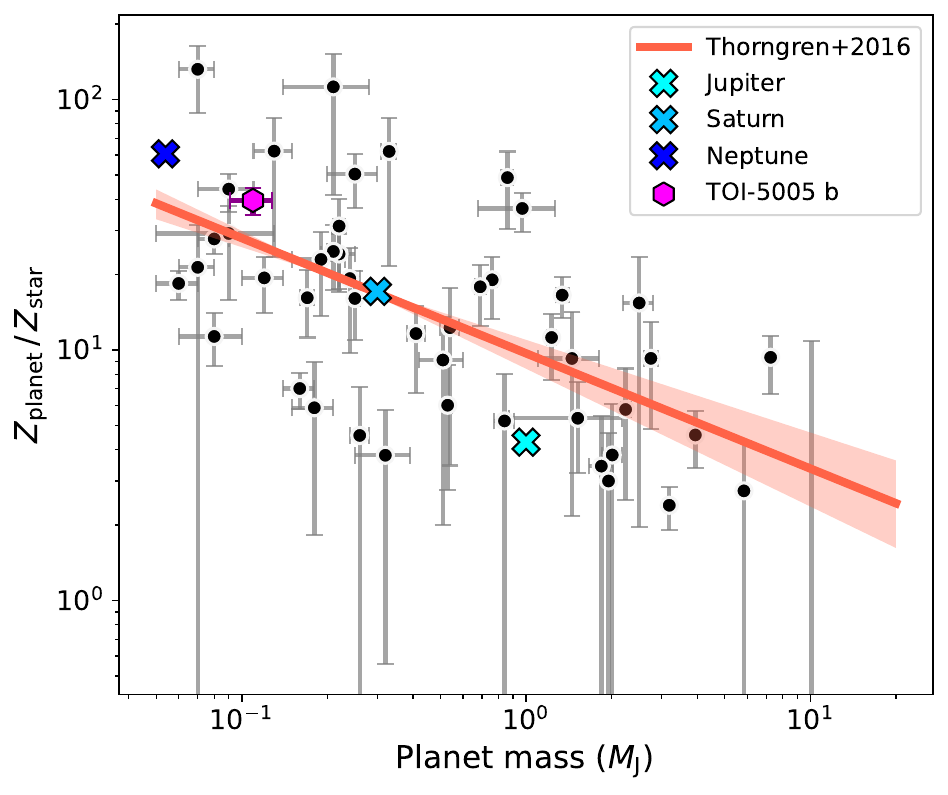}
    \caption[Heavy element enrichment of planets relative to their host stars as a function of mass.]{Heavy element enrichment of planets relative to their host stars as a function of mass. The black data points are from \citet{Thorngren16}, and the planet mass and enrichment of TOI-5005~b were inferred in Sects.~\ref{subsec:joint_analysis}  and \ref{sec:internal_structure}.}
    \label{fig:enrichment_vs_mass}
\end{figure}

We summarise the results of the interior structure retrieval in Table \ref{tab:interior_retrieval}, and show the 1D and 2D posterior distributions in Fig.~\ref{fig:interior_retrieval}. TOI-5005 b's mean metal mass fraction is $Z_{\rm planet} = $ 0.76, with 1$\sigma$ estimates ranging between 0.65 and 0.80. The metal content is distributed between the core and the envelope. In the latter, the envelope mass fraction is compatible with atmospheric metallicities ranging from subsolar (log(Fe/H) = - 1.17, 0.12 $\times$ solar) to 81 $\times$ solar within 1$\sigma$, although a metallicity of $\times$ 250 solar cannot be ruled out (2$\sigma$). For reference, Neptune and Uranus have metal mass fractions $Z_{\rm planet} > $ 0.80, and atmospheric metallicities between 60 and 100 $\times$ solar, whereas Saturn presents a $Z_{\rm planet} = 0.2 $ and an atmospheric metallicity of $\times$ 10 solar \citep[][and references therein]{MV23}. Thus, TOI-5005 b has an overall metal mass fraction slightly lower than that of the Solar System ice giants. To break the degeneracy between the CMF and $Z_{\rm env}$, atmospheric characterisation data is necessary to constrain the atmospheric metallicity, log(Fe/H). This could also help unveil the extent of mixing in the interior: if the atmospheric metallicity is found to be high ($>$ 10 $\times$ solar) it would suggest that metals are fairly well mixed in the interior \citep{Thorngren19}. In addition, we calculate the stellar metal mass fraction as $Z_{\star} = 0.014 \times 10^{\rm [Fe/H]_{\star}}$ \citep{Thorngren16}, and used it to infer the heavy element enrichment relative to the host star: $Z_{\rm planet}/Z_{\star}$ = 37.5 $\pm$ 5. In Fig.~\ref{fig:enrichment_vs_mass}, we plot such ratio versus the measured planet mass for the TOI-5005 system and include the sample and mass-metallicity relations obtained by \citet{Thorngren16}. TOI-5005 b is compatible with such a relation, which indicates that it could have formed via core accretion \citep[see][for further discussion]{Thorngren16}.

\begin{table}[]
\centering
\renewcommand{\arraystretch}{1.35}
\setlength{\tabcolsep}{20pt}
\caption[Main physical parameters of the interior structure retrieval of TOI-5005~b.]{Main physical parameters of the interior structure retrieval of TOI-5005~b (Sect.~\ref{sec:internal_structure}).}
\label{tab:interior_retrieval}
\begin{tabular}{ll}
\hline \hline
Parameter               & Value                  \\ \hline
Core mass fraction, CMF & $0.74^{+0.05}_{-0.45}$ \\
Atmospheric metallicity, log(Fe/H)            & $0.73^{+1.18}_{-1.64}$ \\
Envelope metal mass fraction, $Z_{\rm env}$           & $0.08^{+0.41}_{-0.06}$ \\
Total metal mass fraction, $Z_{\rm planet}$        & $0.76^{+0.04}_{-0.11}$ \\
Internal temperature, $T_{\rm int}$ ($K$)          & $97^{+15}_{-17}$ \\

\hline  
\end{tabular}
\end{table}


\subsection{Atmospheric mass-loss rate}
\label{subsec:mass-loss}
TOI-5005 b presents an intriguing case for the study of atmospheric evolution due to its unique combination of parameters. Despite its relatively high equilibrium temperature ($1040 \pm 20$~K), the planet's heavy mass contributes to a Jeans escape parameter $\Lambda_\mathrm{p} = R_\mathrm{p}/H$ of $38 \pm 7$, where $H$ is the scale height of the planet's atmosphere. Sub-Neptunes with Jeans escape parameters greater than 20--25 are not expected to experience significant atmospheric mass-loss \citep{Owen2016,Cubillos2017,Fossati2017,Vivien2022}. However, given its uniqueness, TOI-5005 b deserves a more detailed analysis.

The photo-evaporation rate of super-Neptunes is expected to be in the energy-limited regime \citep{Owen2016-energy-limited}, and can be computed with \citep{Erkaev2007,Owen2013}
\begin{equation}
    \dot{M}=\epsilon \frac{\pi F_{\mathrm{XUV}} R_{\mathrm{p}}^3}{G M_{\mathrm{p}}}, \label{eq:mass-loss-rate}
\end{equation}
where $F_\mathrm{XUV}$ is the XUV flux received by the planet, $G$ is the gravitational constant, and $\epsilon$ is an efficiency parameter. The XUV flux from the star is not constant in time, and its value is approximated by the analytical fit of XUV luminosity as a function of age obtained by \cite{Sanz-Forcada2011}. The value of $\epsilon$, sometimes noted $\eta$ in the literature, is computed using Eq. (31) from \cite{Owen2017}. We perform a simple Monte Carlo analysis by assuming Gaussian priors on all stellar and planetary parameters from Tables \ref{tab:stellar_parameters} and \ref{tab:parameters_joint}, except for the stellar age, for which we use the posterior distribution from the \texttt{stardate} analysis ($\rm age \ = \ 2.20^{+0.61}_{-0.28}$ Gyr). We find a present-day photo-evaporation rate of $(2.5 \pm 1.4) \times 10^{9}$ g\,s$^{-1}$ ($0.013\pm 0.008$~$\rm M_{\oplus}$~Gyr$^{-1}$). Since the XUV flux emission by the host star is decreasing with time, most of the photo-evaporation happens within the first Gyr of the planet's evolution if the planet migrates early on and stays on its present-day orbit \citep{Bourrier2018-GJ436b,Owen2018}.

To quantify this, we estimate the total mass of H/He lost by TOI-5005 b following the approach of \cite{Aguichine2021}. We integrate the mass-loss rate $\dot{M}$ from $t=0$ to the planet's present-day age, assuming that $M_\mathrm{p}$, $R_\mathrm{p}$ and $T_\mathrm{eq}$ remained roughly constant and that variations in $\dot{M}$ are mainly caused by the changing stellar XUV luminosity as a function of star's age. In this case, we find that TOI-5005 b could have lost $0.36 \pm 0.18$ $\rm M_{\oplus}$~ of H/He, that is, $\sim 1.1\%$ of its current mass. This is not sufficient to change the bulk composition of the planet, but it can alter the composition of its upper atmosphere.

The procedure described above is repeated by replacing Eq.~\ref{eq:mass-loss-rate} by mass-loss rates computed from the publicly available grid of models from \cite{Kubyshkina2018,Kubyshkina2021}, which is adapted to planets with masses 1--40 $\rm M_{\oplus}$. Mass-loss rates are computed by linear interpolation in this grid with respect to five parameters: stellar mass, incident XUV flux, planet equilibrium temperature, planet mass, and planet radius. This approach still uses the stellar XUV luminosity as a function of age from \cite{Sanz-Forcada2011}. The present-day mass loss rate computed from their model is $(3.3 \pm 2.3) \times 10^{10}$ g\,s$^{-1}$ ($0.17\pm 0.12$ ~$\rm M_{\oplus}$~Gyr$^{-1}$), and the total mass lost since formation, obtained by integrating $\dot{M}$, is $1.8 \pm 0.7$~$\rm M_{\oplus}$, which is comparable to the total H/He  in the envelope ($\sim 7.3$ $\rm M_{\oplus}$, see Sect. \ref{sec:internal_structure}).

These results are one order of magnitude greater than the estimate from the model of \cite{Owen2016}. We also computed mass-loss rates with the analytical fits to the model from \cite{Salz2016}, and found results compatible with those produced by the model from \cite{Kubyshkina2018}. However, the model of \cite{Salz2016} is only calibrated to moderate XUV fluxes and is therefore outside of its validity range during the stellar saturation regime, so we do not present its results here. The discrepancy between the values from Eq. \ref{eq:mass-loss-rate} and the models of \cite{Salz2016} and \cite{Kubyshkina2018} is most likely due to the fact that Eq. \ref{eq:mass-loss-rate} computes the mass loss rate at the surface. In reality, XUV photons can be absorbed much higher in the atmosphere, up to nano-bar levels, so that the surface to collect XUV is much bigger and the gravitational potential is weaker, resulting in much greater mass loss rates.

\subsection{Prospects for atmospheric characterisation}
\label{subsec:prospects_atm}

We study the feasibility of atmospheric characterisation of TOI-5005~b through the transmission and emission spectroscopy metrics, as well as modelling the spectrum of a primary and secondary atmosphere in both emission and transmission geometries. We determine the transmission spectroscopy metric \citep[TSM;][]{Kempton2018} using the planet's mass, radius, equilibrium temperature and J-band magnitude of the star. We find a TSM of 84 using the assumption of an H$_2$- and He-rich atmosphere. We also consider a secondary atmosphere with 500$\times$ metallicity, dominated by heavier species such as H$_2$O, CO and CO$_2$. For such a case, the higher mean molecular mass of the atmosphere reduces the TSM to 27. The emission spectroscopy metric (ESM) is 15, but is not strongly affected by the change in the mean molecular mass between the two cases.

\begin{figure*}
\centering
        \includegraphics[width=\textwidth,trim={0cm 0cm 0cm 0},clip]{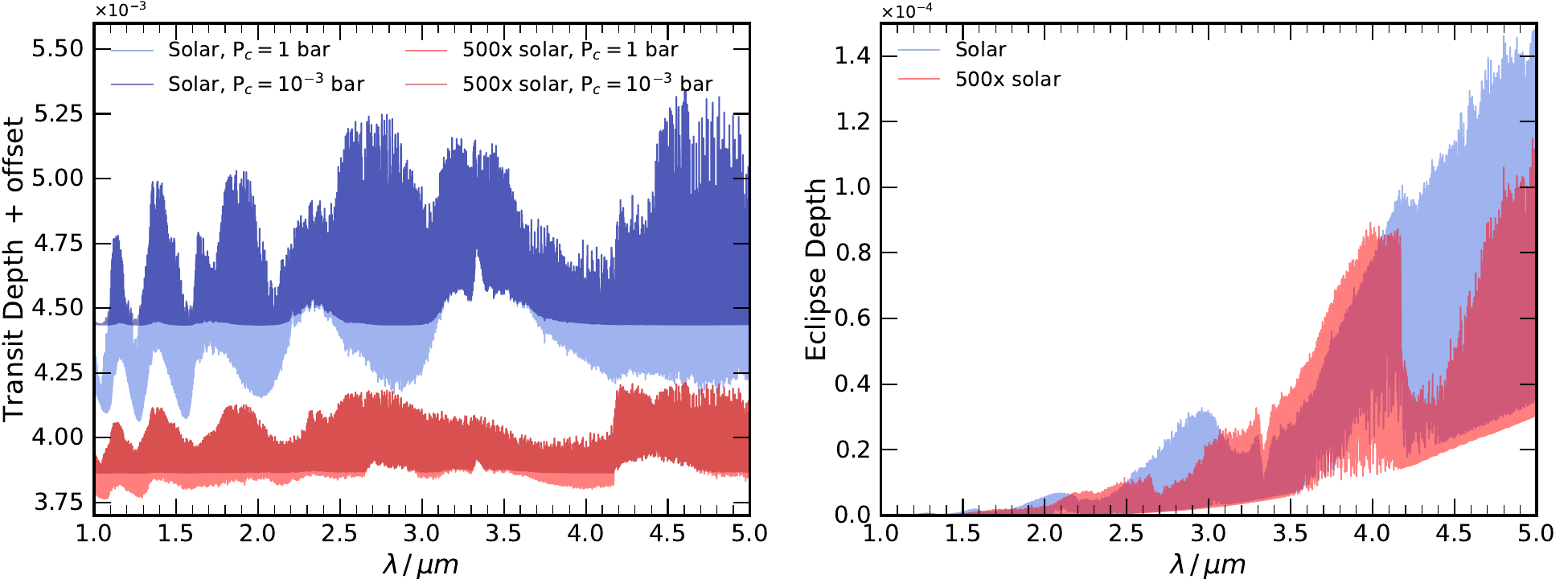}
    \caption[Modelled atmospheric spectra of TOI-5005~b.]{Modelled atmospheric spectra of TOI-5005~b assuming a solar and 500$\times$ solar metallicity atmosphere. Left panel: Transmission spectra, with cloud decks for each model at 1~bar and 1~mbar for each of the cases. Right panel: Eclipse depths of the dayside atmosphere.   Both show similar feature strengths and absorption from spectroscopically active species.}     
\label{fig:atm_model}
\end{figure*}

We generated model spectra in both transmission and emission geometries of TOI-5005~b using GENESIS \citep{Gandhi2017, Gandhi2020}. These are shown in Fig.~\ref{fig:atm_model} for both the primary atmosphere at solar metallicity as well as a secondary atmosphere at 500$\times$ solar metallicity. For all of the models, we assumed a temperature profile with 1.1 $\times$ T$_\mathrm{eq}$ (1140~K) in the deep atmosphere which monotonically cooled to 0.6 $\times$ T$_\mathrm{eq}$ (620~K) in the upper layers, in order to be as realistic as possible with potential spectroscopic signatures. It is likely that any clouds present in the atmosphere will reduce the signal from the planet's atmosphere further in transmission. Therefore, we also modelled the atmosphere with clouds at 1~mbar level for both the solar and high-metallicity cases. The model spectra show that despite the clouds, features do still appear due to absorption from H$_2$O, CO, CH$_4$ and CO$_2$ in the atmosphere. For the high-metallicity case, some features from strong lines originate at the $\sim \mu$bar level, indicating we may also be sensitive to photochemistry if the planet's atmosphere has a high metallicity. In emission in the infrared, we show the model spectra for the solar and 500$\times$ solar metallicity cases, which both offer similar feature strengths from the spectrally active species.

\section{Summary and conclusions}
\label{conclusions}

We used TESS, HARPS, PEST, and TRAPPIST-South data to confirm the planetary nature of TOI-5005~b and characterise its orbital and physical properties. Our analysis shows that TOI-5005~b is a super-Neptune with a radius of $R_{\rm p}$ = $6.25\pm 0.24$~$\rm R_{\rm \oplus}$ ($R_{\rm p}$ = $0.558\pm 0.021$ $\rm R_{\rm J}$) and a mass of $M_{\rm p}$~= $32.7\pm 5.9$ $\rm M_{\oplus}$ ($M_{\rm p}$ = $0.103\pm 0.018$ $\rm M_{\rm J}$). These properties, together with its orbital period ($P_{\rm orb}$ = $6.3085044^{+0.0000092}_{-0.0000088}$ days), place TOI-5005~b in the Neptunian savanna near the ridge, a recently identified over-density of Neptunes in the period range of  $\simeq$3-5 days. Having a density of $0.74 \pm 0.16$ $\rm g \, cm^{-3}$, TOI-5005~b becomes a new member of a population of low-density Neptunes ($<1$ $\rm g \, cm^{-3}$) in the savanna, in contrast to the planets in the ridge, which typically show higher densities ($>1$ $\rm g \, cm^{-3}$). This trend, together with the different occurrence rates, supports the existence of different processes populating both regimes.

We also identified a periodic modulation in the TESS data that matches the orbital period of TOI-5005~b. Given the sub-Alfvénic star-planet distance, this signal can be interpreted as a reflection of magnetic star-planet interactions (MSPIs). We studied the possibility that the modulation has an instrumental origin by performing the strongest possible CBV-based correction. We found that the amplitude decreased, but the periodicity holds, which supports a stellar origin. Additionally, we studied how the modulation correlates with the planetary orbit and found similar patterns to those observed in other systems with MSPIs. Overall, these results strongly support the MSPI explanation for the observed signal. However, we also consider that additional observations are needed to get a final confirmation. These can be performed in the optical with instrumentation subjected to different systematics \citep[e.g. CHEOPS;][]{2021ExA....51..109B}, or in radio wavelengths to observe the electron cyclotron maser emission \citep[e.g. GMRT;][]{1991ASPC...19..376S}.

Our internal structure modelling of TOI-5005~b constrains the overall planetary metal mass fraction to a value slightly lower than that of the Solar System ice giants ($Z_{\rm planet}$  = $0.76^{+0.04}_{-0.11}$). This value, together with the measured $Z_{\rm star}$, makes this system consistent with the well-known mass-metallicity trend, which suggests that TOI-5005~b was formed via core accretion. However, the core mass fraction (CMF = $0.74^{+0.05}_{-0.45}$) and envelope mass fraction ($Z_{\rm env}$ = $0.08^{+0.41}_{-0.06}$) are  degenerate. To break this degeneracy, a direct measurement of the atmospheric metallicity is needed. This shows the relevance of atmospheric observations not only in probing the atmosphere itself, but also the planetary interior. Interestingly, having a transmission spectroscopy metric of 84, TOI-5005~b is amenable to atmospheric studies. We simulated model spectra in both transmission and emission geometries and found the appearance of features such as H$_2$O, CO, CH$_4$, and CO$_2$, which would be detectable even with clouds at 1 mbar level. We also estimated the present-day atmospheric mass-loss of TOI-5005~b, and found inconsistent values depending on the choice of photo-evaporation model ($0.013\pm 0.008$~$\rm M_{\oplus}$~Gyr$^{-1}$ vs $0.17\pm 0.12$ ~$\rm M_{\oplus}$~Gyr$^{-1}$). Atmospheric escape mechanisms in super-Neptunes have received less attention compared to hot Jupiters and hot sub-Neptunes, as modelling efforts have been primarily focused on these latter categories. This knowledge gap has led to discrepancies that hinder a comprehensive understanding of the evolutionary processes in such bodies. Hence, the detection of planets similar to TOI-5005 b holds the potential to stimulate research into these intermediate-size planets.

TOI-5005~b orbits the brightest host of its surrounding neighbours in the period-radius space, which makes it a key target for atmospheric and orbital architecture observations. TOI-5005 is thus an excellent option to be observed by the large-scale survey of orbital architectures ATREIDES (Bourrier et al., in prep.), and by the Near-Infrared Gatherer of Helium Transits (NIGHT) spectrograph \citep{2024MNRAS.527.4467F}. Combining our current knowledge with additional constraints on the spin-axis angle and mass-loss rate will provide a clearer picture of the overall evolution of this unusual planetary system.


\newpage
\chapter{General discussion}
\label{ch:int_discussion}
\vspace{2.9cm}
\pagestyle{fancy}
\fancyhf{}
\lhead[\small{\textbf{\thepage}}]{\small{\textbf{\nouppercase{\leftmark}}}}
\rhead[\small{\textbf{\nouppercase{\rightmark}}}]{\small{\textbf{\thepage}}}

\bigskip

In chapters \ref{ch:k2_ojos}$-$\ref{ch:toi_5005}, we presented the analyses and specific discussions of this thesis, which covered three general topics: planet detection through the transit and radial velocity techniques, precise characterization of planetary systems, and population studies in different regions of the parameter space. Each chapter contains in-depth discussions about specific problems related to the corresponding planets and populations studied, such as the internal composition of small planets (e.g. chapters \ref{ch:k2_ojos} and \ref{ch:toi-244}), the formation and evolution of giant planets (e.g. chapters \ref{ch:nep_des} and~\ref{ch:toi_5005}), and the magnetic interactions between close-in planets and their stars (e.g. chapters \ref{ch:mspis} and \ref{ch:toi_5005}). The goal of this section is to put together the main results presented in the previous chapters and discuss them from a global and integrative perspective. This will allow us to put our findings in a broader context and potentially reach new conclusions thanks to the combination of insights acquired in the individual chapters. In Sect.~\ref{sec:systems_prop_overview}, we jointly discuss the main properties of all the systems detected and characterised in this thesis and contextualise them within the main studied populations and trends. This will possibly allow us to find new system-population connections, as well as to identify systems of interest for follow-up observations. In Sect.~\ref{sec:PLATO}, we discuss how our results can be integrated within the future plans of exoplanet exploration, focusing on the synergies with the upcoming large-scale space-based photometric survey PLATO.

\section{The system-population connection}
\label{sec:systems_prop_overview}

In this thesis, we have presented the discovery and statistical validation of four planets (K2-355~b, K2-356~b, K2-357~b, and K2-358~b), the RV confirmation and precise characterization of two other planets (TOI-244~b and TOI-5005~b), the detection of 14 new planet candidates, and the revision of the properties of 25 previously reported planets. Hence, a total of 45 planets and candidates (and their corresponding stellar hosts) have been studied in this dissertation. After the publication of our studies, different teams obtained additional constraints on some of these planets, which can help us to have a better view of their evolution. However, these constraints are either independent or have little influence on our system characterisations, so in the following sections, we exclusively discuss the results obtained in this thesis.

Regarding the published constraints, \citet{2023A&A...669A..63B} observed one spectroscopic transit of K2-105~b with HARPS-N and measured a tentative misaligned orbit. If confirmed, this could support
a disruptive dynamical past and the late arrival of the planet on its close-in orbit. Alternatively, the presence of additional planets (to be further explored) could support a primordial tilt of the star or protoplanetary disk, as this low-mass planet is likely far enough from its G-type star that its spin-orbit angle was not much influenced by tidal interactions. This target is being observed within the ATREIDES collaboration to unveil its spin-orbit misalignment definitively. K2-105~b has also been analysed within the helium survey with the GIANO-B instrument by \citet{2023A&A...676A.130G}, who were able to put a 3$\sigma$ upper limit to its mass-loss rate of 6.028 $\times$ $10^{10}$ $\rm g \, s^{-1}$. Another interesting example of follow-up observations of our studied targets corresponds to HD 118203, the eccentric system where we found signs of MSPIs. \citet{2024A&A...688A.172M} discovered an additional long-period companion in the system, 11 times more massive than Jupiter, with a moderately eccentric orbit of 14 years, constituting a hierarchical planetary system with the hot Jupiter. More recently, \citet{2024AJ....168..295Z} found that the orbit normal of the hot Jupiter is nearly aligned with the stellar spin axis, and constrained the line-of-sight mutual inclination between the hot Jupiter and the outer planet, finding low mutual inclinations between the outer giant planet, the inner hot Jupiter, and the host star. These results suggest that the system may have undergone coplanar high-eccentricity tidal migration.

In the following, we contextualise in different regions of the parameter space the main properties of the 45 planets and candidates studied in this work, and discuss the observed distributions, focusing on the most relevant populations studied here. 

\subsection{The period-radius space and the revisited exo-Neptunian landscape}

\begin{figure}
    \centering
    \includegraphics[width=\textwidth]{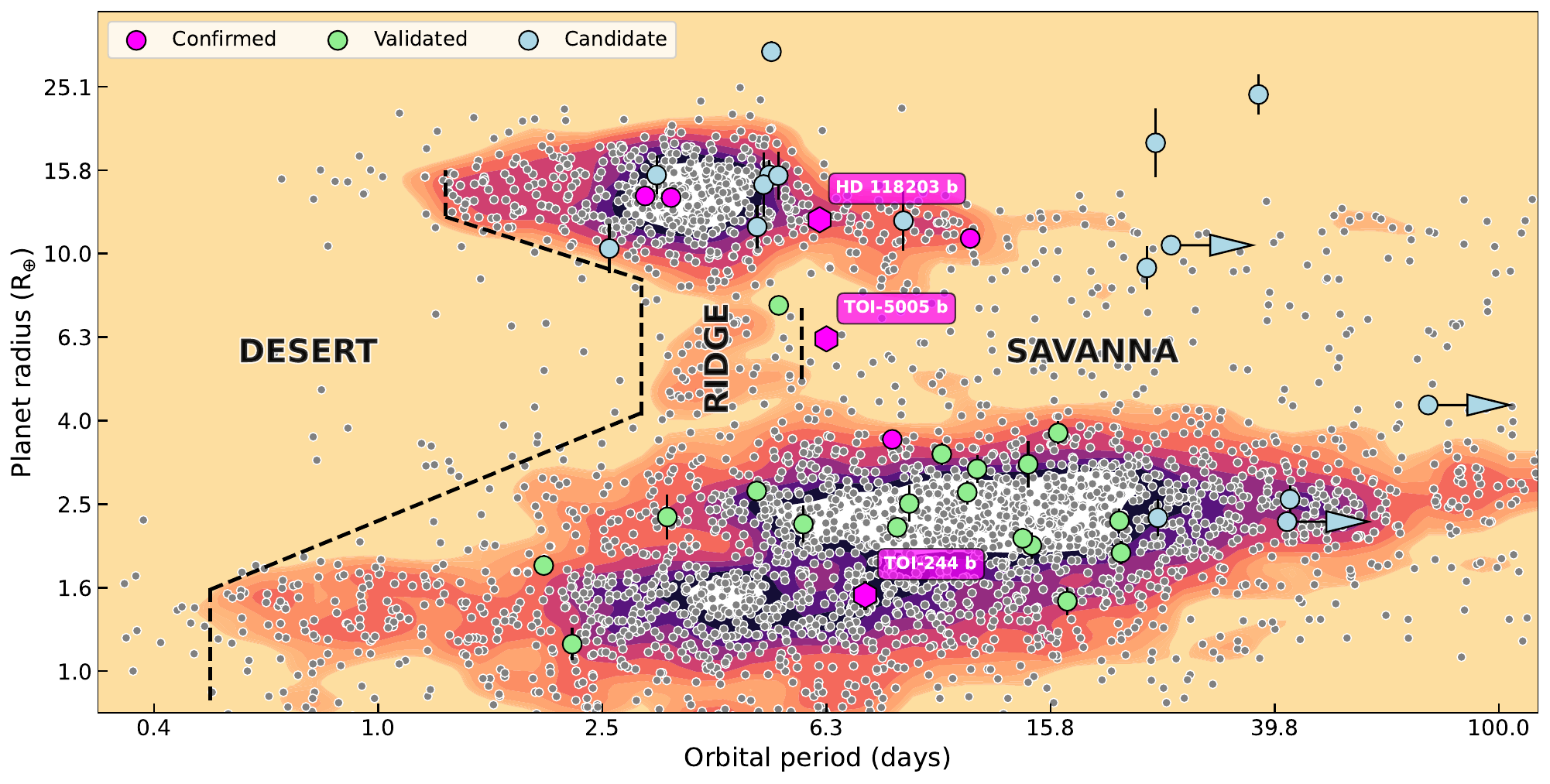}
    \caption[Period-radius diagram of close-in planets with the sample of confirmed, validated, and candidate planets studied in this thesis.]{Period-radius diagram of close-in planets, where we plot the sample of confirmed, validated, and candidate planets studied in this thesis. The arrows represent the minimum orbital periods of EPIC 211537087.03, EPIC 211590050.01, and EPIC 212008766.02. We highlight the location of the confirmed planets TOI-244~b, HD 118203~b, and TOI-5005~b, for which we conducted in-depth individualised studies (see chapters \ref{ch:toi-244}, \ref{ch:mspis}, and \ref{ch:toi_5005}, respectively). The boundaries for the desert, ridge, and savanna were derived in Chapter \ref{ch:nep_des}. The data were collected from the NASA Exoplanet Archive. This plot was generated through \texttt{nep-des} (\url{https://github.com/castro-gzlz/nep-des}). }
    \label{fig:period_radius_final}
\end{figure}

The 45 planets and candidates studied in this work were detected through the transit technique, mainly by the space-based photometers \textit{Kepler} (on its extended mission K2) and TESS. Therefore, we could measure the radii of the vast majority of the studied sample, which allows us to contextualise their properties in the period-radius diagram consistently. Most of these planets and candidates orbit their stars with short orbital periods, between 2 and 30 days.

Interestingly, in Chapter~\ref{ch:nep_des}, we estimated and analysed the occurrence rate of close-in planets ($P_{\rm orb}$ $<$ 30 days), which allowed us to identify new features and precisely delineate previously known characteristics of the underlying planet distribution in the period-radius space. In Fig.~\ref{fig:period_radius_final}, we plot the period-radius diagram of the planets and candidates studied in this thesis integrated within the landscape mapped in Chapter~\ref{ch:nep_des}, which includes the newly identified Neptunian ridge separating the desert from the savanna. We note that for four K2 planet candidates (EPIC 211319779.01, EPIC 211407755.01, EPIC 211480861.01, and EPIC 211791178.01), the stars hosting the transit signals could not be determined, and thus no planetary radii were reported (see Chapter~\ref{ch:k2_ojos}). Also, for the three mono-transit planet candidates EPIC 211537087.03, EPIC 211590050.01, and 212008766.02, we only constrained a lower limit for the orbital period (indicated with arrows in Fig.~\ref{fig:period_radius_final}). Overall, out of the 45 studied planets and candidates, we were able to determine precise periods and radii for 38. Out of these targets, we conducted in-depth individualised studies for TOI-244~b (Chapter~\ref{ch:toi-244}), HD 118203~b (Chapter~\ref{ch:mspis}),
and TOI-5005~b (Chapter~\ref{ch:toi_5005}), whose locations are highlighted in the diagram. 

As can be discerned in Fig.~\ref{fig:period_radius_final}, the distribution of our studied planets and candidates broadly follows the observed distribution of all known planets. Most planets and candidates cluster in the close-in sub-Neptune region ($R_{\rm p}$ $<$ 4 $\rm R_{\oplus}$ and $P_{\rm orb}$ $<$ 30 days) and the hot Jupiter pileup ($R_{\rm p}$ $>$ 10 $\rm R_{\oplus}$ and 3 days $<$ $P_{\rm orb}$ $<$ 6 days), and very few detections are found at intermediate radii. According to the original broad definition of the hot Neptune desert \citep{2016A&A...589A..75M}, three validated planets (K2-104 b, K2-275 b, and K2-121 b), one confirmed planet (TOI-5005~b), and one candidate planet (EPIC 211705502.01) would fall within its boundaries. In contrast, according to our revised population-based boundaries derived in Chapter \ref{ch:nep_des}, none of them (nor any other planet or candidate in our sample) are considered to be within this unpopulated region of the parameter space (although EPIC 211705502.01 is consistent within uncertainties at $\sim$1$\sigma$). According to our revised landscape, EPIC 211705502.01 belongs to the hot Jupiter pileup, and K2-104 b and K2-275 b belong to different regions of the hot sub-Neptune population, where planets are commonly detected. Interestingly, the statistically validated planet K2-121 b ($P_{\rm orb}$ = 5.2 days and $R_{\rm p}$ = 7.5 $\rm R_{\oplus}$) studied in Chapter \ref{ch:k2_ojos} lies within the Neptunian ridge identified in Chapter \ref{ch:nep_des}, which makes this target of special interest for follow-up observations to measure its mass, obliquity, and atmospheric escape and composition (see chapters \ref{ch:nep_des} and~\ref{ch:toi_5005}, for extended discussions). The host star K2-121 is relatively bright ($V$ = 13.2 mag), which, together with the expected sub-Jovian planetary mass ($M_{\rm p}$$\sim$70$-$90~$\rm M_{\oplus}$), makes K2-121~b amenable for mass measurements with spectrographs such as HARPS, CARMENES, and ESPRESSO. We note that measuring the mass of K2-121~b is of particular interest given its location near the Neptunian savanna, whose transition we estimated at $\sim$5.7 days (Chapter \ref{ch:nep_des}). 

In Chapter~\ref{ch:toi_5005}, we identified a density trend indicating a dichotomy between savanna planets and ridge planets, indicating that savanna planets tend to have low densities compared to the larger densities of the planets populating the ridge. Based on the current planet sample, this density transition is in good agreement with the planet occurrence transition. Identifying more precisely the location of these transitions and studying whether they correlate with other planetary or stellar properties will be of crucial importance to constrain planet formation and evolution theories, for which `frontier planets' such as K2-121~b (and TOI-5005~b on the other side of the ridge-savanna border) will likely play a relevant role. Regarding the study of atmospheric escape and atmospheric species, the detectability in K2-121~b needs to be specifically explored. The Rossiter–McLaughlin effect is detectable with ESPRESSO, which led to the scheduling of K2-121 b within the context of the ATREIDES collaboration, aimed to systematically measure the obliquities of Neptunes in the desert, ridge, and savanna (PI: Bourrier). We highlight that our ephemeris revision of K2-121 b based on C5 and C18 K2 data (Chapter~\ref{ch:k2_ojos}) allowed the scheduling of transit spectroscopy observations in 2025 within the context of ATREIDES, and will also allow its scheduling for the next decades, the era of extremely large telescopes. For example, in 2040, the propagated $T_{0}$ uncertainty will be 1.3~min, in contrast to the 8.4~h uncertainty propagated from the C5-based ephemeris reported in the discovery paper, which will make it possible to schedule future transit observations without needing additional ephemeris revisions.

In Fig.~\ref{fig:period_radius_final}, we can appreciate that the giant planet region ($R_{\rm p} > 10 \rm R_{\oplus}$) contains the larger fraction of planet candidates in our sample. This is due to our imposed radius threshold to avoid validating brown dwarfs as planets (see more details in Chapter ~\ref{ch:k2_ojos}). Hence, RV follow-up observations are necessary to confirm the planetary nature of these objects. In addition to the confirmation and mass measurements, RV observations would allow us to constrain their orbital eccentricities, which, as extensively discussed, are key tracers of the migration mechanisms that brought giant planets to their present-day locations. 

While the hot Jupiter pileup is predominantly dominated by planets with circular orbits, warm Jupiters are more commonly found with eccentric orbits. In that sense, the planet candidates EPIC 211335816.01, EPIC 211393988.01, EPIC 211418290.01, EPIC 211424769.01, and EPIC 211724246.01, being at the warm end of the pileup (i.e. with $P_{\rm orb}$ ranging from 4.8 days to 5.2 days), are of especial interest to gain further insight into how most hot Jupiters arrived onto the pileup. Their candidate-hosting stars are relatively bright (9.4 mag $<$ $V$ $<$ 13.9 mag), which makes this endeavour feasible through state-of-the-art spectrographs. Interestingly, HD 118203~b lies in this region of the period-radius space ($P_{\rm orb}$ = 6.1 days). In this eccentric hot Jupiter, we found signs of star-planet magnetic interactions (MSPIs) based on the detection of TESS photometric variability synchronised with the planetary orbit (Chapter~\ref{ch:mspis}). These kinds of planet-induced activity signals are unusual, and, based on the expected decoupled pseudo-synchronised planetary rotation in systems with non-circular orbits (which is expected to boost the signatures of MSPIs), we propose that systems with massive close-in planets in eccentric orbits might be the most favourable to search for these planet-induced signals (Chapter~\ref{ch:mspis}). We can thus now integrate this concept with the location of the closer-in orbits where giant planets start to be systematically found in eccentric orbits, which leads us to conclude that hot Jupiters at the warmer end of the pileup, being still very close to their stars (i.e. within the sub-Alfvénic radius), but frequently found in considerably eccentric orbits, could be the most amenable targets to search for signs of MSPIs. HD~118203~b resides in this exact region of the period-radius diagram. Interestingly, TOI-5005~b, for which we also found signs of MSPIs, also resides in this orbital region (although with a super-Neptune mass). In the case of TOI-5005~b, however, our RV data did not allow us to determine the orbital eccentricity, and, at a population level, super-Neptunes and sub-Jovians in this region are known to have more frequently circular orbits than hot Jupiters (see Sect.~\ref{sec:migration} and Chapter~\ref{ch:nep_des}). 

Overall, the possible connections between MSPIs, orbital eccentricity (and consequently planetary rotation rate), location within the period-radius plane, and orbital migration need to be further explored. To do so, more dedicated efforts to specifically search for these interactions are necessary, as well as maintained public databases that consistently gather the published relevant properties of MSPI detections. To date, based on a (non-exhaustive) search, we roughly estimate the existence of $\sim$20-30 works reporting signs of MSPIs on different targets (both spectroscopically and photometrically). This number will likely significantly increase with the launch of the PLATO mission and the start of operations of the ELT, opening the door to the first (time-resolved) population studies of MSPIs. 

\subsection{The mass-radius space and the emerging population of low-density super-Earths}

Among the 45 transit signals studied in this thesis, seven are confirmed as planets through RVs and thus have mass measurements: HAT-P-43 b, K2-114 b, K2-105 b, K2-34 b, TOI-244 b, HD 118203 b, and TOI-5005~b. We highlight their location in the period-radius space in Fig.~\ref{fig:period_radius_final} (i.e. magenta symbols). These planets span practically the entire planetary domain, from super-Earths to super-Jupiters, implying a wide range of compositions. Among them, the masses of two were measured for the first time in Chapter~\ref{ch:toi-244} (TOI-244~b) and Chapter~\ref{ch:toi_5005} (TOI-5005~b) of this thesis. As extensively discussed in this dissertation, measuring precise masses of transiting sub-Neptune planets is a hot topic in exoplanet research, since their densities are considered to be one of the key observables to solve the ambiguity on their internal structures (see Sect.~\ref{sec:small}, for an extended discussion). In our studied sample, a total of 22 planets and candidates lie within the sub-Neptune domain, of which only two have measured masses, K2-105 b and TOI-244 b (although we note that the mass detection of K2-105 b is marginal, at 1.6$\sigma$ level; $M_{\rm p}$ = 30 $\pm$ 19 $\rm M_{\oplus}$; \citealt{2017PASJ...69...29N}). Thus, the remaining 20 planets become targets of interest for follow-up RV observations. As of spring 2025, there are no published mass measurements for these planets. However, some of them have RV measurements in public archives or are being observed within different programs. We were granted with HARPS-N and CARMENES observing time to determine the mass of the sub-Neptune K2-274~b ($R_{\rm p}$ = $2.08^{+0.12}_{-0.07}$ $\rm R_{\oplus}$, $P_{\rm orb}$ = 14.1 days) and try to resolve the orbit of its outer warm Neptune-size candidate ($R_{\rm p}$ = $4.34^{+0.81}_{-0.66}$ $\rm R_{\oplus}$, $P_{\rm orb}$ $>$ 74.8 days) reported in Chapter ~\ref{ch:k2_ojos}. The RV data are under analysis. On its side, two different RV programs (i.e. the ESPRESSO GTO program and an open-time program led by V. Van Eylen) acquired a total of 46 ESPRESSO spectra to try to pin down the mass of the super-Earth K2-184 b  ($R_{\rm p}$ = $1.47^{+0.11}_{-0.05}$ $\rm R_{\oplus}$, $P_{\rm orb}$ $=$ 17.0 days). These data are public within the ESO Archive\footnote{\url{http://archive.eso.org/eso/eso_archive_main.html}}, and were analysed in the context of this thesis, but no significant RV signal was uncovered. We recall that in Chapter~\ref{ch:k2_ojos} we identified significant linear TTVs in this system (see Fig.~\ref{fig:ttv}), which, given the possible presence of an outer companion, makes this system of particular interest for follow-up spectroscopic and transit observations. Currently, K2-184~b is being observed by HARPS-N in the context of the THIRSTEE program \citep{2024A&A...692A.238L}. Hopefully, by joining the HARPS-N and ESPRESSO datasets, its planetary mass will be determined soon. Notably, two additional planets studied in Chapter \ref{ch:k2_ojos} are planned for RV measurements within THIRSTEE, so we foresee at least four targets to have their masses measured in the upcoming months. 

\begin{figure*}
    \centering
\includegraphics[width=\textwidth]{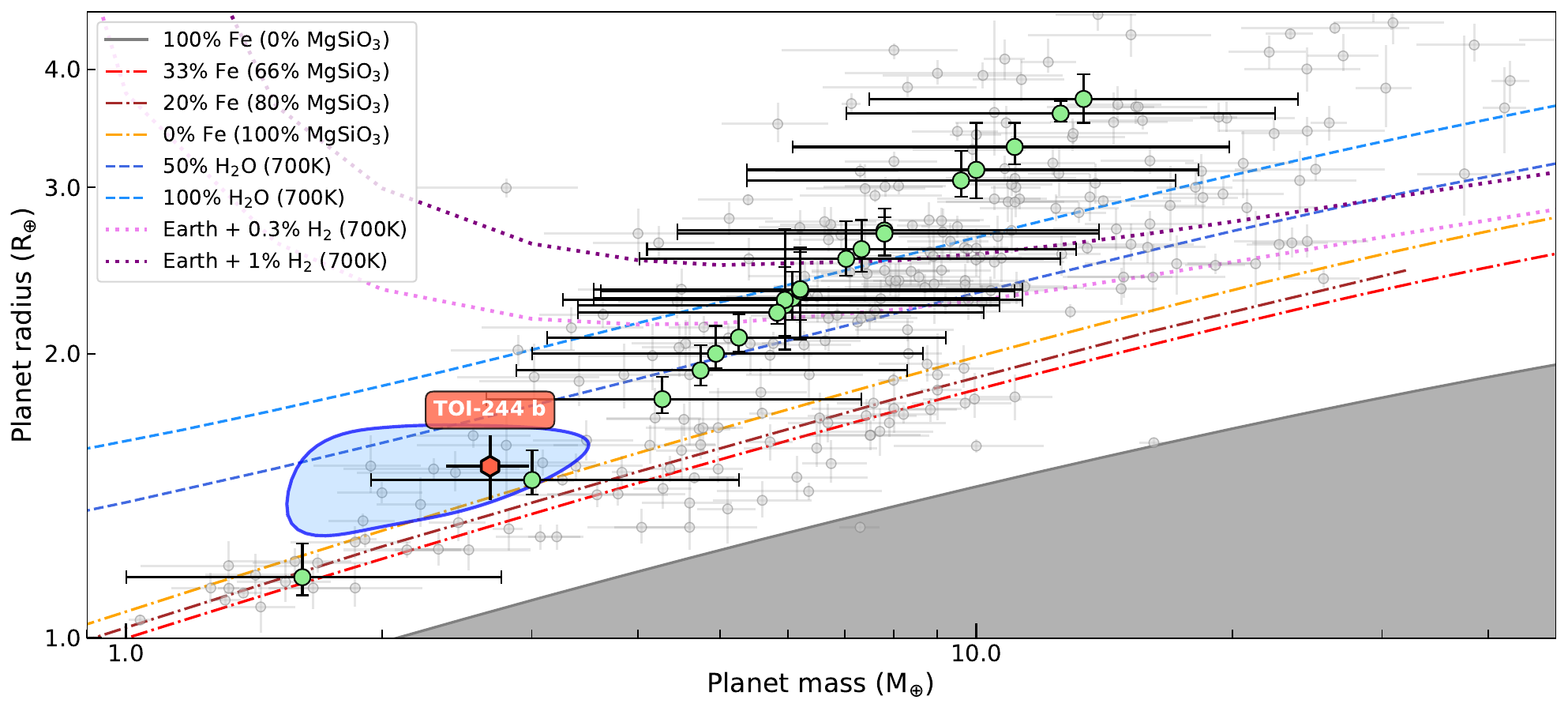}
    \caption[]{Mass-radius diagram of all known small planets with masses and radii measured with a precision better than 25$\%$. The green data points correspond to the small planets studied in this work, whose masses and uncertainties were estimated through a probabilistic algorithm \citep{2017ApJ...834...17C}. The dashed lines correspond to theoretical interior models that consider different mass percentages of Fe and $\rm MgSiO_{3}$ (red, brown, and orange), and condensed water in different percentages (light and dark blue). The dotted lines correspond to theoretical models for Earth-like planets with $\rm H_{2}$-dominated atmospheres (light pink and dark magenta). All the represented iso-composition lines were derived by \citet{2016ApJ...819..127Z,2019PNAS..116.9723Z}. Planet data were acquired from the NEA. This plot was prepared through \texttt{mr-plotter} (\url{https://github.com/castro-gzlz/mr-plotter}).}
    \label{fig:mr_final}
\end{figure*}

To visualise and contextualise the ability of our studied sub-Neptune sample to constrain the composition of this population, we estimated their expected masses through the probabilistic tool \texttt{forecaster}\footnote{Available at \url{https://github.com/chenjj2/forecaster}} \citep{2017ApJ...834...17C}. In Fig.~\ref{fig:mr_final}, we plot the mass-radius diagram of all known planets with masses and radii constrained to a precision better than 25$\%$. We over-plot the observed radii and predicted masses of our studied sub-Neptune sample, where we also highlight TOI-244~b, for which we measured a precise (12$\%$) mass in Chapter~\ref{ch:toi-244}. As expected, the 1$\sigma$ error bars of the predicted masses are large, given that for a certain sub-Neptune radius a wide range of masses is possible (e.g. Sect.~\ref{sec:small}). As we can appreciate from the iso-composition lines of Fig.~\ref{fig:mr_final}, the predicted masses are compatible with a wide range of bulk compositions. This situation is particularly critical for the four planets at around 2$\rm R_{\oplus}$, whose predicted bulk compositions can range from Earth-like to 100$\%$ condensed water. 

Interestingly, in this region of the mass-radius space, \citet{2021A&A...645A..41L} found that for planets hosted by M-dwarf stars there is a dichotomy between planets clustering around the 50$\%$$\rm H_{2}O$$-$50$\%$$\rm MgSiO_{3}$ iso-composition line and planets clustering around the rocky Earth-like model, which is interpreted to favour the water world hypothesis (see Sect.~\ref{sec:small}). In this regard, three out of four of the studied planets in this region (K2-103~b, K2-101~b, and K2-274~b) are hosted by GK stars, which makes them particularly interesting for testing whether the dichotomy found in M-dwarfs holds for earlier spectral types. 

\begin{figure*}
    \centering
    \includegraphics[width=0.49\textwidth]{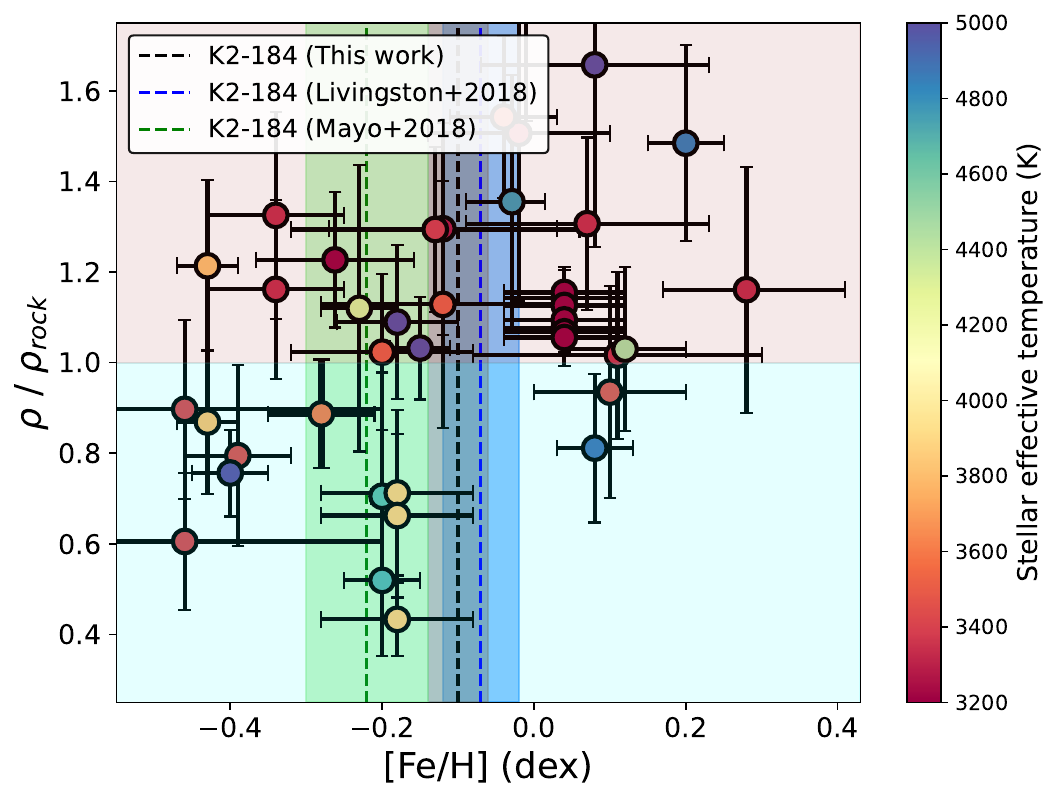}
    \includegraphics[width=0.49\textwidth]{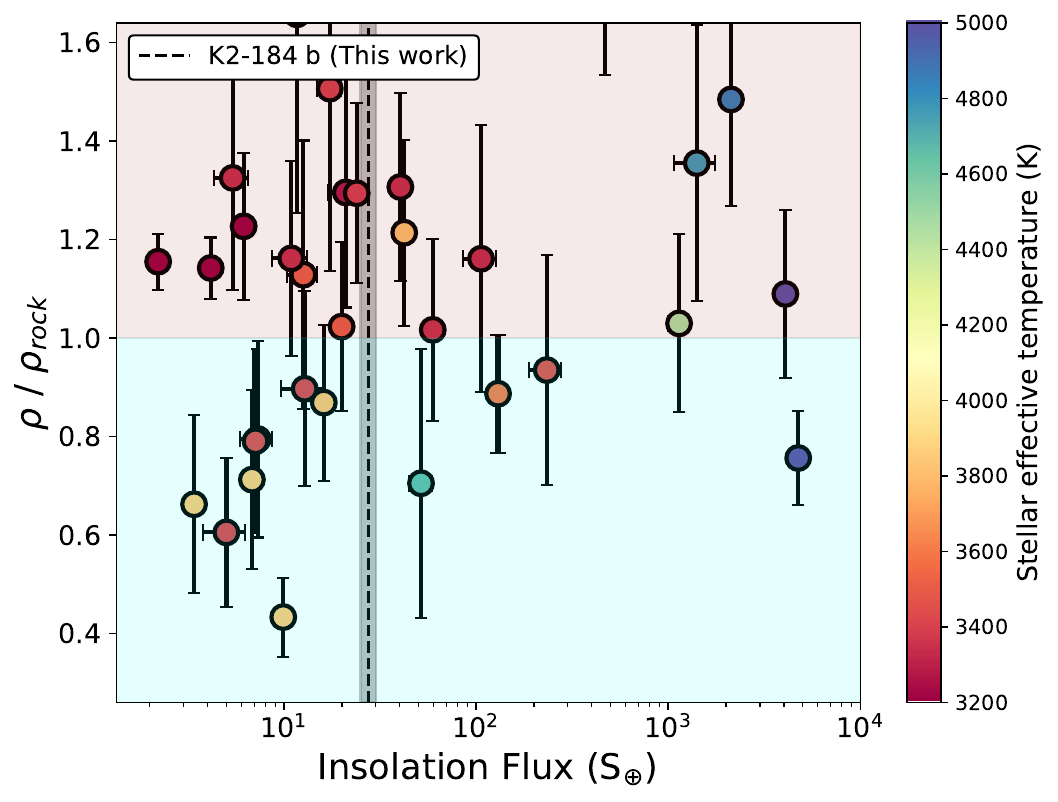}
    \caption[Planet density versus stellar metallicity and insolation flux for all confirmed planets with $R_{p}$ $<$ 2$\rm R_{\oplus}$, $M_{\rm p}$ $<$ 3.5$\rm M_{\oplus}$, and a mass precision better than 30$\%$.]{Normalized density to the density expected for a planet composed 100$\%$ of silicates \citep{2019PNAS..116.9723Z} versus the stellar metallicity of the stellar hosts (left) and the stellar insolation flux (right) for all the confirmed planets with $R_{p}$ $<$ 2$\rm R_{\oplus}$, $M_{\rm p}$ $<$ 3.5$\rm M_{\oplus}$, and a mass precision better than 30$\%$. The colour coding indicates the effective temperature of the host stars. The vertical dashed lines and corresponding vertical shades correspond to different estimates of the stellar metallicity of K2-184 and the insolation flux received by K2-184 b. Planet and stellar data were acquired from the NEA.}
    \label{fig:trends_tesis}
\end{figure*}

In this thesis, we have explored an emerging field of research contextualised within the general sub-Neptune composition problem: the internal structure of super-Earths and low-density super-Earths. Based on early mass measurements of transiting super-Earths (i.e small planets in the lower mode of the bi-modal radius distribution), it was widely assumed that these planets typically had canonical rocky Earth-like structures. However, over the last 3-5 years, thanks to systematic precise characterizations of small planets with state-of-the-art spectrographs such as ESPRESSO, it is becoming increasingly clear that a fraction of super-Earths show densities below the expected for a rocky composition, thus requiring the presence of a significant amount of volatile elements (see Sect.\ref{sec:small} and Chapter \ref{ch:toi-244}, for an extended discussion). In Fig.~\ref{fig:mr_final}, we show the low-density super-Earth region as identified in Chapter \ref{ch:toi-244}. As we can see, these planets typically occupy an intermediate region between the rocky models and the 50$\%$$\rm H_{2}O$$-$50$\%$$\rm MgSiO_{3}$ model, so that the dichotomy at larger radii found by \citet{2022Sci...377.1211L} appears to not hold in this region of the parameter space. 

Different hypotheses have been proposed to explain these observed densities. Currently, compositions based on canonical rocky structures surrounded by envelopes composed of high-mean molecular weight volatiles (i.e. such as water) are among the most promising possible scenarios (see Chapter \ref{ch:toi-244}). Interestingly, apart from TOI-244~b, in Fig.~\ref{fig:mr_final} we identify another planet studied in this thesis belonging to the super-Earth domain, K2-184 b. We recall that K2-184~b is the planet where we identified strong TTVs (Chapter \ref{ch:k2_ojos}) and whose mass measurement is underway through the THIRSTEE program. Being hosted by a bright star ($V$ = 10.35 mag), it has the potential to become a relevant target for future atmospheric studies. Its predicted mass ($M_{\rm pred}$ = $3.0^{+2.3}_{-1.1}$~$\rm M_{\oplus}$) is very uncertain, and is compatible with a wide range of bulk compositions. In Chapter~\ref{ch:toi-244}, we found that the observed sample of low-density super-Earths (LDSEs) tend to be hosted by metal-poor stars (although most of the observed hosts are M-dwarfs, so their metallicities have to be taken with care) and receive low insolation fluxes, typically below 10$\rm S_{\oplus}$. With this information, we can discuss whether K2-184 b has the potential to become a new member of the LDSE population. Based on the spectroscopic characterization from \citet{2018AJ....155...21P} and using the interpolation tool \texttt{isochrones}, in Chapter~\ref{ch:k2_ojos} we derived a metallicity of [Fe/H] = $-0.10 \pm 0.04$ dex. Independent spectroscopic observations from \citet{2018AJ....156..277L} and \citet{2018AJ....155..136M} resulted in metallicities of [Fe/H] = $-0.07 \pm 0.05$ dex and [Fe/H] = $-0.22 \pm 0.08$ dex, respectively. Two spectroscopic characterisations based on TRES spectra (uploaded by Prof. Allyson Bieryla to ExoFOP\footnote{ExoFOP site of K2-184: \url{https://exofop.ipac.caltech.edu/tess/target.php?id=21244210}}) indicate stellar metallicities of [Fe/H] = $-0.216 \pm 0.080$ dex and [Fe/H] = $-0.205 \pm 0.080$ dex. Therefore, independent metallicity derivations suggest that K2-184 has a slightly subsolar metallicity. The received insolation flux of K2-184 b can be obtained by combining its orbital distance with the \textit{Gaia} stellar luminosity: $S$ = 27.5 $\pm$ 0.3 $\rm S_{\oplus}$. In Fig.~\ref{fig:trends_tesis}, we show the metallicity and density trends with the measured values from the K2-184 system. Regarding stellar metallicity, K2-184 is compatible with the region where LDSEs are found. Regarding insolation flux, however, K2-184~b shows a higher value than that typically shown by the lowest dense LDSE. 

Overall, based on the trends identified in Chapter \ref{ch:toi-244}, K2-184 b will possibly not have a very low density. According to the preferred interpretation, it is plausible that K2-184 b has an enhancement of $\alpha$-elements in its composition, and that its original building blocks were water-rich. This abundance of water, which could have formed an extensive atmosphere of steam after the outgassing phase, however, could have been partially eroded through atmospheric escape, given its relatively high levels of irradiation. Ongoing spectroscopic observations will hopefully allow us to precisely estimate the iron-to-silicate mass fraction of the primordial protoplanetary disk, and most importantly, the planet's density, which will allow us to have a much more accurate view of the internal structure of this interesting planet.

\section{Entering the PLATO era: Synergies with the mission objectives}
\label{sec:PLATO}

Transiting planets are the `golden' targets for exoplanet exploration since their unique configurations facilitate the determination of different fundamental properties such as planetary masses, radii, atmospheric compositions, and spin-orbit angles. After \textit{Kepler} and TESS, the PLATO mission from the European Space Agency (ESA) will become the next large-scale transit survey, to be launched in 2026. During its primary mission, PLATO will observe a fixed region of the sky for two years. Based on the early results, it will be decided whether the field will be maintained or if it will be modified until completion of the primary mission, which will last four years. Overall, the PLATO mission concept and observing strategy are similar to that of \textit{Kepler} (see Sect.~\ref{sec:intro_kepler}), and the photometric precisions of both missions are expected to be comparable. However, there will be a fundamental difference between PLATO and \textit{Kepler}. PLATO will study brighter stars, with magnitudes between 8 mag and 11 mag, which, together with its long-baseline observing strategy and larger field of view, will allow us to systematically characterize long-period transiting planets subjected to low irradiation conditions\footnote{Today, ESA's CHEOPS mission \citep{2021ExA....51..109B} is following up several mono-transits and duo-transits in bright stars primarily detected by TESS with a similar objective \citep[e.g.][]{2023MNRAS.526..548O,2023A&A...674A..43U,2023MNRAS.523.3090T}.}.

The era of PLATO will coincide with the era of extremely large telescopes such as the European Extremely Large Telescope (ELT), which, with a 39-meter diameter, is expected to start operations in 2028. As it happened with the VLT, K2, and TESS, which, working together, revolutionised the exoplanet field through systematic characterisations of the smallest close-in planets in our Galaxy, the combination of PLATO and ELT will most likely initiate the `long-period revolution'. This powerful combination is expected to allow in-depth studies of small rocky worlds in the temperate regions of their stars, such as our own world, a major goal of exoplanet research that has been widely pursued since the first discoveries.

The PLATO Definition Study Report\footnote{Available at \url{https://sci.esa.int/documents/33240/36096/1567260308850-PLATO_Definition_Study_Report_1_2.pdf}}(DSR) details the following main objectives for the mission: 

\begin{enumerate}
  \item Determine the bulk properties (mass, radius, mean density) of planets in a wide range of systems,
including terrestrial planets in the habitable zone of solar-like stars
  \item Study how planets and planet systems evolve
  \item Study the typical architectures of planetary systems
  \item Analyse the correlation of planet properties and their frequencies with stellar parameters (e.g.
stellar metallicity, stellar type)
  \item Analyse the dependence of the frequency of terrestrial planets on the environment in which they
formed
\item Study the internal structure of stars and how it evolves with age
\item Identify good targets for spectroscopic follow-up measurements to investigate planet atmospheres
\end{enumerate}

This thesis is contextualized within the preparatory work for the PLATO mission, and our research primarily aligned with PLATO's objectives 1 (chapters~\ref{ch:k2_ojos}, ~\ref{ch:toi-244}, and \ref{ch:toi_5005}), 2 (chapters~\ref{ch:k2_ojos}$-$\ref{ch:toi_5005}), and 4 (Chapter~\ref{ch:toi-244}). In addition, the results presented in this thesis can indirectly contribute to achieving PLATO's objectives 3, 5, and 7. Thus, from the seven main PLATO objectives, this thesis has significant synergies with all of them except one (objective 6), which has a purely stellar nature. 

The PLATO scientific output will be of crucial importance to gain further insight into the two principal problems studied in this dissertation: the internal composition of super-Earths and the formation and evolution of giant planets. According to the DSR, PLATO aims at achieving radii precisions down to 3$\%$ for small planets, which is well below the typically reported uncertainties for most super-Earths known to date. Hence, through the detection and precise characterisation of new planets in this regime, PLATO will allow us to significantly improve the precision and accuracy of the distribution of super-Earths and low-density super-Earths (LDSEs). In Chapter \ref{ch:toi-244}, we found that LDSEs tend to receive lower insolation fluxes than canonical rocky super-Earths (S $<$ 10 $\rm S_{\oplus}$), which, interestingly, coincides with the parameter space that will be primarily targeted by PLATO. In addition, PLATO will focus more on GK stars, contrary to TESS, which focuses on M stars. Detecting super-Earths around GK stars will allow precise estimations of the chemical abundances of the stellar hosts, which will in turn allow us to confirm the metallicity trend for LDSEs identified in Chapter~\ref{ch:toi-244} based on the observed population, today dominated by M stars. Therefore, due to the increased number of precisely characterized super-Earths, which will facilitate the exploration of their distribution in the mass-radius space, and the focus on low-irradiated environments and early-type stars, which are two key regimes to confirm and expand the emerging trends identified in Chapter~\ref{ch:toi-244}, PLATO will likely play a relevant role in our understanding of the internal compositions of canonical and low-density super-Earths. 

PLATO will also significantly contribute to our understanding of the formation and evolution of giant planets. According to the DSR, PLATO will be able to determine the ages of a large sample of main-sequence stars down to a 10$\%$ precision, which will allow us to study compositional changes with time and the connection between age, inflation, and atmospheric loss rate of giant planets. In this thesis, we propose that the overabundance of planets detected in the Neptunian ridge (and the hot Jupiter pileup) could be related to a specific migration mechanism that preferentially brings planets to this specific region of the parameter space. In this regard, the different migration mechanisms have different predicted time scales, so precisely estimating the ages of systems across different features of the close-in period distribution can provide valuable insight into their origins. PLATO will also allow us to further probe the Neptunian ridge by focusing on specific stellar regimes. The \textit{Kepler} sample is too limited to study occurrences within the exo-Neptunian landscape across different stellar sub-samples. However, the increased homogeneously detected PLATO sample (possibly in combination with that of \textit{Kepler}) has the potential to allow the first partial characterization of the exo-Neptunian landscape restricted to certain spectral types, stellar metallicities, or even ages, which will provide critical insight into our understanding of the processes behind the inferred occurrence features. In addition, its focus on long-period orbits will facilitate the performance of more reliable occurrence studies at larger orbital distances, closer to the regions where these planets are thought to be formed. This will provide relevant insight into the runaway gas accretion process of giant planets and its likely interruption for super-Neptune and sub-Jovian planets. 

Another result presented in this thesis for which PLATO can play a relevant role is the density trend identified in Chapter \ref{ch:toi_5005}. Low-density savanna planets are known to dominate the density-period distribution of intermediate-size planets up to about 30 days, but it is not clear how this trend continues at larger orbits. Hence, measuring accurate densities of longer-period Neptunes will allow us to better characterise the extent of this trend and consequently the validity range of the inferences derived from it.

In summary, PLATO has great potential to keep exploring the close-in planet region, where TESS already is achieving outstanding results, and where this thesis is mostly focused. However, its true potential lies within the capability of exploring the still unknown warm regions of planetary systems. Regarding the small planet population, PLATO will be crucial to probe the irradiation conditions under which small rocky cores can retain different volatile elements. Regarding the giant planet population, which is known to be less sensitive to evaporation, PLATO will allow us to step into their birthplaces, securely providing invaluable insight into their formation as well as the evolution of their migrated counterparts. 

Beyond PLATO, the results presented in this thesis also have many synergies with different upcoming and currently operative space-based missions. ARIEL \citep{2022EPSC...16.1114T}, to be launched in 2029, will observe the transits of at least 1000 exoplanets to characterise their chemical compositions and thermal structures. For that purpose, having precise transit ephemerides is critical to allow efficient scheduling. In this regard, different targets whose ephemeris were improved in Chapter~\ref{ch:k2_ojos} belong to the current ARIEL target list \citep{2022AJ....164...15E}. From the scientific perspective, inferring the atmospheric compositions of transiting exoplanets is relevant to gaining insight into the problem of the formation and evolution of giant planets, since they could provide hints on the region of the protoplanetary disk where they formed. Most importantly, these observations will be crucial to disentangle the predominant atmospheric compositions of sub-Neptunes and super-Earths (see Sect.~\ref{sec:small}). We note that this endeavour is already being carried out by the JWST for a reduced group of targets with particularly favourable conditions. Interestingly, TOI-244 b is part of this group and has been recently chosen as one of the highest ranked targets for the Rocky Worlds JWST DDT program\footnote{\url{https://outerspace.stsci.edu/pages/viewpage.action?pageId=257035126}}. In addition, \textit{Gaia} is expected to publish its fourth data release in 2026. These data are expected to reveal the orbits of hundreds of giant planets in the outermost regions of their planetary systems. As extensively discussed here, coupling this upcoming catalogue with the currently known sample of systems with close-in giant planets will be crucial to gaining insights into the migration processes leading to the intriguing populations of hot Jupiters and Neptunes.


\newpage
\chapter{Conclusions}
\label{ch:conclusions}
\vspace{2cm}
\pagestyle{fancy}
\fancyhf{}
\lhead[\small{\textbf{\thepage}}]{\small{\textbf{\nouppercase{\leftmark}}}}
\rhead[\small{\textbf{\nouppercase{\rightmark}}}]{\small{\textbf{\thepage}}}


The results presented in this thesis allowed us to reach conclusions on different topics related to the formation and evolution of planetary systems,  which we reflect in chapters \ref{ch:k2_ojos}$-$\ref{ch:toi_5005}. In this final chapter, we aim to highlight the fundamental conclusions that we extracted from the whole investigation through a series of more general ideas and reflections that we made along the process.

The identification of the Neptunian ridge (both in the planet occurrence and density parameter spaces) separating the desert from the savanna is a central result of this thesis. In chapters \ref{ch:nep_des} and \ref{ch:toi_5005}, based on different observational constraints, we propose that the excess of planets in the ridge could be due to a specific migration mechanism named high-eccentricity tidal migration (HEM), bringing planets preferentially to this particular orbital region. However, as we also discuss in the aforementioned chapters, while promising, confirming this possibility requires additional observational evidence. Still, regardless of the possible (migration and/or atmospheric) mechanisms leading to the ridge, we can safely reach a more conservative yet suggestive conclusion: \textit{lying in orbits corresponding to periods of 3-5 days, in concordance with the hot Jupiter pileup, the Neptunian ridge suggests a common path in the evolution of the closest giant planets, from Neptune to Jupiter sizes}. This conclusion agrees with the fact that close-in Neptunes and Jupiters share several fundamental properties, such as similar (low) overall occurrence rates and a strong preference for stellar hosts with super-solar metallicities. Thus, from a purely observational perspective, the identification of the Neptunian ridge reinforces the idea that close-in giant planets across a wide range of masses and radii evolved similarly. However, we note that there might be second-order evolutionary differences between close-in Neptunes and Jupiters, as suggested by the different occurrence rates between the over-densities and the warmer regions, and the different eccentricity distributions. In the upcoming years, additional detections and characterisations of Neptunian and Jovian planets as well as systematic studies of their atmospheres and orbital architectures, will be crucial to unveil the evolutionary pathways leading to these differences. 

Another central result of this thesis is the identification and study of the emerging population of low-density super-Earths (LDSEs) and the characterisation of a new member of this population, TOI-244~b. Before this thesis, some authors used this terminology to refer to individual super-Earth detections with unusually low densities. However, these planets had not been studied as a group before. In this thesis, we propose that LDSEs should be considered as an independent population (or sub-population), motivated by different factors. Low-density super-Earths, such as canonical rocky super-Earths, lie within the lower mode of the bi-modal radius distribution. In this region, planets were thought to have purely rocky compositions and to have potentially evolved from the radius valley and upper mode of the distribution before their atmospheres were stripped. Thus, the fact of still finding some relatively large envelopes (or other structures integrating volatile elements) at these small planetary radii makes them worthy of special consideration (e.g. how did they avoid atmospheric erosion?). In addition, while in absolute terms the predicted envelopes (or amount of volatiles) of LDSEs are large if we compare them with the rocky planets in the Solar System, these are considerably small if we compare them with those of most sub-Neptunes in the radius valley and the upper mode of the distribution. Specifically, being located between the rocky and 50$\%$$\rm H_{2}O$$-$50$\%$$\rm MgSiO_{3}$ bulk composition models, LDSEs do not follow the density dichotomy found for sub-Neptunes around M-dwarfs, which further motivates a specific consideration for these planets. In Chapter~\ref{ch:toi-244}, based on two observed trends in the insolation and stellar metallicity spaces, we propose that LDSEs could be composed of rocky structures surrounded by steam water atmospheres. Again, while promising, this hypothesis has to be taken with care, since it involves a series of assumptions still not contrasted with observations, and the metallicity trend still needs to be tested with more reliable planet-hosting GK stars. However, we consider the insolation trend particularly relevant and suggestive by itself. Independently from any interpretation of the nature of the volatiles, this trend allows us to conclude that: \textit{the fact that the lowest dense super-Earths are observed to receive the lowest insolation fluxes suggests that the largest fraction of volatile elements in these planets is directly exposed to (or at least, influenced by) the received irradiation from their host stars.} This conclusion agrees with the most accepted view on the volatile storage of small planets: volatiles mostly lie within the planetary atmospheres. Two major questions to answer are `which volatiles are those', and `how these volatiles compare with those within the larger sub-Neptune population'. The answers will hopefully be unveiled soon through transit spectroscopy observations. 

This thesis also led us to conclude that, amidst the era of atmospheric exploration, where most resources are being invested, exoplanet research dedicated to detecting and characterising new planets is highly valuable. Thousands of planets have been identified today, spanning a wide range of properties. Still, many relevant regimes need further telescope investment to reach a more comprehensive understanding of their nature. Some of these regimes are the Earth/super-Earth and long-period planet domains, two observationally elusive regions where the upcoming PLATO mission will surely play a crucial role. From the detection side, we also reflected on the great importance of systematic planet detections and characterisations in all regions of the parameter space. Today, there is a sufficiently large number of planets that allow the identification of different regions and populations of particular interest from the formation and evolution perspective, where most detection surveys focus. Targeting regions of particular interest is highly valuable, but we should not forget to continue exploring regimes of apparently lower interest since this has the potential of revealing previously unnoticed features with much relevance to formation theories. An example of this situation corresponds to the growing population of LDSEs, a region devoid of planets five years ago, which led us to conclude on the dissimilarity between bare rock Earths and their higher super-Earth counterparts, where we are starting to see a larger variety of compositions. Thus, while `populations guide detections', we should not forget that `detections also build up populations'.

This thesis also led us to reflect on the relevance of archival data in exoplanet research. Building upon several detected and characterised planets and combining them with the known population of exoplanets, we found key (previously unexplored) aspects of the exoplanet diversity conundrum. An output that, on some occasions, was purely based on data from public databases. In that sense, we also highlight the potential of planet occurrence studies to explore specific regions of the parameter space. Planet occurrences across the hot Neptune desert were extensively computed soon after the first \textit{Kepler} data releases. However, many years later, there was not yet a dedicated occurrence study to try to unveil the underlying distribution of planets across the desert. Thus, the identification of the ridge and the computation for the first time of population-based desert boundaries highlight the great potential of these kinds of (still infrequent) studies, which can be applied to other relevant regions of the parameter space. 

Across this thesis, we also reflected on the great potential of star-planet interactions (both magnetic and tidal) in the exoplanet field, a topic that will most likely thrive during the upcoming decades. Today, the detection of these interactions is only possible for a very limited number of targets with particularly favourable detection advantages. Upcoming space-based and ground-based instrumentation, such as PLATO and the ELT, in combination with radio telescopes, will contribute to the increase in these detections, which will hopefully trigger more advanced theoretical frameworks that allow us to better understand these complex phenomena.

\clearpage\hbox{}\thispagestyle{empty}\newpage 

\setcounter{footnote}{0}

\thispagestyle{empty}
\vspace*{7cm}
\begin{flushright}
\textit{Antes de la iluminación, corto leña y acarreo agua. \\
Después de la iluminación, corto leña y acarreo agua. } \\
Proverbio Zen
\\

\textit{---}
\end{flushright}

\newpage
\thispagestyle{plain}
\vspace{9cm}
\lhead[\small{\textbf{\thepage}}]{\textbf{Bibliography}}
\rhead[\small{\textbf{Bibliography}}]{\small{\textbf{\thepage}}}
\phantomsection 
\addcontentsline{toc}{chapter}{Bibliography}
\bibliography{biblio}

\begin{center}
\aldine
\end{center}

\appendix

\renewcommand{\chaptername}{}
\titleformat{\chapter}[display]
  {\normalfont\Desmesurado\centering\color{black}}{\centering\chaptertitlename\\\ \MasDesmesurado{\thechapter}}{80pt}{\bfseries\Desmesurado\color{black}}
\titlespacing*{\chapter} 
  {0pt}{10pt}{40pt}


\renewcommand{\thetable}{A.\arabic{table}}   
\begin{center}
\chapter[Additional tables and figures from Chapter \ref{ch:toi-244}]{Additional tables and figures from Chapter \ref{ch:toi-244}}
\end{center}
\label{ch:Appendix_A}
\vspace{1cm}
\pagestyle{fancy}
\fancyhf{}
\lhead[\small{\textbf{\thepage}}]{\textbf{Additional tables from Chapter \ref{ch:toi-244}}}
\rhead[\small{\textbf{Appendix~\hyperref[ch:Appendix_A]{A}}}]{\small{\textbf{\thepage}}}

\begin{table*}
\caption[ESPRESSO radial velocities and activity indicators of GJ 1018.]{Complete ESPRESSO radial velocities and activity indicators of GJ 1018 acquired between 9 October 2021 and 2 October 2022 under the programs IDs 108.2254.002, 108.2254.005, and 108.2254.006. This table is available at the CDS (\url{https://cdsarc.cds.unistra.fr/viz-bin/cat/J/A+A/675/A52})}
\renewcommand{\arraystretch}{1.3}
\setlength{\tabcolsep}{4pt}
\fontsize{8.3pt}{8.3pt}\selectfont
	\begin{center}
		\begin{tabular}{cccccccc}
            \hline \hline
			{RJD (days)} & RV ($\rm m \, s^{-1}$) & FWHM ($\rm m \, s^{-1}$) & S-index (x1000)& Contrast ($\%$) & $\rm H_{\alpha}$ (x1000) & NaD (x1000) & BIS \\
			\hline
			59496.698 & $15139.30 \pm 0.73$ & $2747.3 \pm 1.5$ & $665.4 \pm 3.5$ & $34.337 \pm 0.018$ & $610.36 \pm 0.20$ & $70.79 \pm 0.12$ & $252.0 \pm 1.5$ \\
			59499.536 & $15139.54 \pm 0.79$ & $2748.4 \pm 1.6$ & $683.5\pm 4.1$ & $34.228 \pm 0.020$ & $618.31 \pm 0.20$ & $72.83 \pm 0.13$ & $324.6 \pm 1.6$ \\
			59514.774 & $15140.54 \pm 0.88$ & $2758.0 \pm 1.8$ & $820.7 \pm 5.1$ & $34.181 \pm 0.022$ & $601.96 \pm 0.23$ & $78.85 \pm 0.16$ & $135.3 \pm 1.8$ \\
			59517.632 & $15142.17 \pm 0.62$ & $2759.3 \pm 1.2$ & $812.0 \pm 2.6$ & $34.223 \pm 0.015$ & $598.70 \pm 0.15$ & $78.210 \pm 0.098$ & $225.1 \pm 1.2$ \\
			59520.697 & $15141.96 \pm 0.67$ & $2755.6 \pm 1.3$ & $745.0 \pm 3.0$ & $34.225 \pm 0.017$ & $588.32 \pm 0.17$ & $74.92 \pm 0.11$ & $252.0 \pm 1.3$ \\
			59523.770 & $15143.70 \pm 0.60$ & $2752.6 \pm 1.2$ & $691.3 \pm 2.6$ & $34.223 \pm 0.015$ & $570.90 \pm 0.14$ & $72.540 \pm 0.090$ & $185.1 \pm 1.2$ \\
			59529.697 & $15138.04 \pm 0.87$ & $2748.6 \pm 1.7$ & $662.5 \pm 4.8$ & $34.275 \pm 0.022$ & $567.23 \pm 0.23$ & $70.66 \pm 0.16$ & $270.6 \pm 1.7$ \\
			59532.530 & $15143.77 \pm 0.52$ & $2747.1 \pm 1.0$ & $633.8 \pm 1.9$ & $34.301 \pm 0.013$ & $580.01 \pm 0.13$ & $71.428 \pm 0.078$ & $238.2 \pm 1.0$ \\
			59535.617 & $15142.10 \pm 0.51$ & $2745.1 \pm 1.0$ & $613.5 \pm 1.8$ & $34.340 \pm 0.013$ & $572.37 \pm 0.12$ & $71.102 \pm 0.076$ & $229.7 \pm 1.0$ \\
			59538.593 & $15143.86 \pm 0.56$ & $2746.9 \pm 1.1$ & $627.4 \pm 2.1$ & $34.323 \pm 0.014$ & $589.11 \pm 0.14$ & $70.274 \pm 0.085$ & $190.7 \pm 1.1$ \\
			59541.659 & $15138.98 \pm 0.76$ & $2743.1 \pm 1.5$ & $630.8 \pm 3.5$ & $34.306 \pm 0.019$ & $586.31 \pm 0.20$ & $70.14 \pm 0.13$ & $329.2 \pm 1.5$ \\
			59544.621 & $15136.91 \pm 0.52$ & $ 	2744.6 \pm 1.0$ & $641.2 \pm 1.9$ & $34.310 \pm 0.013$ & $601.20 \pm 0.13$ & $71.598 \pm 0.078$ & $225.4 \pm 1.0$ \\
			59547.661 & $15137.17 \pm 0.74$ & $2745.2 \pm 1.5$ & $621.0 \pm 3.5$ & $34.299 \pm 0.018$ & $595.90 \pm 0.19$ & $70.72 \pm 0.12$ & $212.6 \pm 1.5$ \\
			59550.695 & $15138.33 \pm 0.53$ & $ 	2745.8 \pm 1.1$ & $674.3 \pm 2.0$ & $34.292 \pm 0.013$ & $592.22 \pm 0.12$ & $69.985 \pm 0.078$ & $214.5 \pm 1.1$ \\
			59553.568 & $15144.63 \pm 0.55$ & $2747.1 \pm 1.1$ & $713.2 \pm 2.1$ & $34.275 \pm 0.014$ & $605.78 \pm 0.13$ & $72.017 \pm 0.081$ & $273.4 \pm 1.1$ \\
			59556.631 & $15142.67 \pm 0.72$ & $2749.6 \pm 1.4$ & $868.8 \pm 3.5$ & $34.260 \pm 0.018$ & $661.87 \pm 0.19$ & $82.32 \pm 0.12$ & $250.5 \pm 1.4$ \\
			59559.584 & $15142.83 \pm 0.98$ & $2750.4 \pm 2.0$ & $758.5 \pm 5.6$ & $34.271 \pm 0.024$ & $604.98 \pm 0.27$ & $75.09 \pm 0.18$ & $332.5 \pm 2.0$ \\
			59562.558 & $15142.17 \pm 0.83$ & $2755.2 \pm 1.7$ & $787.5 \pm 4.5$ & $34.245 \pm 0.021$ & $611.19 \pm 0.22$ & $73.98 \pm 0.14$ & $253.2 \pm 1.7$ \\
			59580.587 & $15138.64 \pm 0.61$ & $2749.3 \pm 1.2$ & $791.4 \pm 2.6$ & $34.228 \pm 0.015$ & $593.87 \pm 0.15$ & $77.662 \pm 0.095$ & $230.3 \pm 1.2$ \\
			59583.554 & $15141.88 \pm 0.57$ & $2750.0 \pm 1.1$ & $752.6 \pm 2.3$ & $34.257 \pm 0.014$ & $584.92 \pm 0.14$ & $74.288 \pm 0.087$ & $230.0 \pm 1.1$ \\
			59744.857 & $15138.99 \pm 0.82$ & $2749.5 \pm 1.6$ & $647.6 \pm 4.3$ & $34.191 \pm 0.020$ & $588.62 \pm 0.21$ & $71.17 \pm 0.15$ & $198.2 \pm 1.6$ \\
			59760.864 & $15144.56 \pm 0.52$ & $2750.6 \pm 1.0$ & $680.0\pm 1.8$ & $34.172 \pm 0.013$ & $582.72 \pm 0.12$ & $73.414 \pm 0.079$ & $205.0 \pm 1.0$ \\
			59786.712 & $15144.9 \pm 1.1$ & $2750.0 \pm 2.1$ & $771.8 \pm 6.3$ & $34.195 \pm 0.027$ & $589.10 \pm 0.28$ & $73.44 \pm 0.21$ & $286.0 \pm 2.1$ \\
			59788.710 & $15144.87 \pm 0.69$ & $2752.1 \pm 1.4$ & $737.6 \pm 3.0$ & $34.199 \pm 0.017$ & $581.69 \pm 0.17$ & $73.32 \pm 0.11$ & $216.2 \pm 1.4$ \\
			59790.806 & $15144.61 \pm 0.63$ & $2751.2 \pm 1.3$ & $789.5 \pm 2.4$ & $34.213 \pm 0.016$ & $599.45 \pm 0.15$ & $76.86 \pm 0.10$ & $221.3 \pm 1.3$ \\
			59792.691 & $15140.70 \pm 0.92$ & $2751.7 \pm 1.8$ & $730.3 \pm 4.9$ & $34.236 \pm 0.023$ & $584.17 \pm 0.23$ & $73.20 \pm 0.17$ & $284.6 \pm 1.8$ \\
			59795.744 & $15138.93 \pm 0.82$ & $2754.0 \pm 1.6$ & $774.2 \pm 3.8$ & $34.229 \pm 0.020$ & $607.84 \pm 0.22$ & $77.12 \pm 0.15$ & $303.3 \pm 1.6$ \\
			59797.895 & $15139.5 \pm 1.1$ & $2755.8 \pm 2.3$ & $720.3 \pm 6.2$ & $34.190 \pm 0.028$ & $608.82 \pm 0.33$ & $76.21 \pm 0.24$ & $323.9 \pm 2.3$ \\
			59798.682 & $15141.05 \pm 0.75$ & $2751.9 \pm 1.5$ & $738.9 \pm 3.5$ & $34.240 \pm 0.019$ & $608.29 \pm 0.19$ & $76.36 \pm 0.13$ & $282.6 \pm 1.5$ \\
			59800.814 & $15140.88 \pm 0.53$ & $2754.3 \pm 1.1$ & $737.1 \pm 1.7$ & $34.203 \pm 0.013$ & $615.84 \pm 0.13$ & $75.087 \pm 0.082$ & $252.2 \pm 1.1$ \\
			59802.806 & $15142.63 \pm 0.54$ & $2751.0 \pm 1.1$ & $866.4 \pm 1.8$ & $34.181 \pm 0.013$ & $679.96 \pm 0.14$ & $83.944 \pm 0.085$ & $258.2 \pm 1.1$ \\
			59804.676 & $15142.86 \pm 0.71$ & $ 	2752.6 \pm 1.4$ & $669.5 \pm 2.9$ & $34.178 \pm 0.018$ & $603.54 \pm 0.18$ & $75.88 \pm 0.12$ & $217.0 \pm 1.4$ \\
			59805.877 & $15141.89 \pm 0.77$ & $2751.7 \pm 1.5$ & $752.9 \pm 3.3$ & $34.214 \pm 0.019$ & $611.17 \pm 0.21$ & $76.52 \pm 0.14$ & $273.8 \pm 1.5$ \\
			59807.711 & $15137.99 \pm 0.82$ & $2750.6 \pm 1.6$ & $687.6 \pm 3.8$ & $34.184 \pm 0.020$ & $611.05 \pm 0.22$ & $74.17 \pm 0.15$ & $318.6 \pm 1.6$ \\
			59811.828 & $15139.34 \pm 0.53$ & $ 	2750.1 \pm 1.1$ & $740.5 \pm 1.8$ & $34.225 \pm 0.013$ & $610.03 \pm 0.13$ & $75.519 \pm 0.083$ & $215.1 \pm 1.1$ \\
			59813.666 & $15139.80 \pm 0.96$ & $2751.8 \pm 1.9$ & $702.9 \pm 5.3$ & $34.222 \pm 0.024$ & $590.24 \pm 0.26$ & $82.35 \pm 0.18$ & $202.3 \pm 1.9$ \\
			59815.623 & $15137.99 \pm 0.89$ & $2752.5 \pm 1.8$ & $620.0 \pm 4.8$ & $34.206 \pm 0.022$ & $567.41 \pm 0.22$ & $87.95 \pm 0.16$ & $287.2 \pm 1.8$ \\
			59817.629 & $15140.49 \pm 0.61$ & $2749.0 \pm 1.2$ & $696.2 \pm 2.4$ & $34.256 \pm 0.015$ & $566.88 \pm 0.14$ & $73.399 \pm 0.096$ & $260.5 \pm 1.2$ \\
			59817.705 & $15141.42 \pm 0.60$ & $2749.0 \pm 1.2$ & $718.7 \pm 2.2$ & $34.284 \pm 0.015$ & $577.86 \pm 0.15$ & $74.411 \pm 0.099$ & $220.2 \pm 1.2$ \\
			59818.893 & $15142.96 \pm 0.66$ & $ 	2753.4 \pm 1.3$ & $669.7 \pm 2.5$ & $34.214 \pm 0.016$ & $573.71 \pm 0.17$ & $72.94 \pm 0.11$ & $346.4 \pm 1.3$ \\
			59823.850 & $15139.40 \pm 0.91$ & $2749.2 \pm 1.8$ & $669.6 \pm 4.4$ & $34.259 \pm 0.023$ & $566.65 \pm 0.25$ & $74.01 \pm 0.17$ & $239.7 \pm 1.8$ \\
			59827.827 & $15138.08 \pm 0.71$ & $2750.2 \pm 1.4$ & $688.1 \pm 2.9$ & $34.255 \pm 0.018$ & $580.15 \pm 0.19$ & $74.33 \pm 0.12$ & $183.0 \pm 1.4$ \\
			59828.825 & $15136.77 \pm 0.66$ & $2746.6 \pm 1.3$ & $645.8 \pm 2.5$ & $34.275 \pm 0.016$ & $564.14 \pm 0.17$ & $72.49 \pm 0.11$ & $189.7 \pm 1.3$ \\
			59830.775 & $15134.42 \pm 0.58$ & $2748.3 \pm 1.2$ & $827.7 \pm 2.0$ & $34.253 \pm 0.014$ & $630.62 \pm 0.15$ & $81.738 \pm 0.094$ & $253.0 \pm 1.2$ \\
			59832.849 & $15136.20 \pm 0.98$ & $ 	2745.3 \pm 2.0$ & $654.7 \pm 4.9$ & $34.141 \pm 0.024$ & $577.67 \pm 0.27$ & $72.18 \pm 0.19$ & $331.4 \pm 2.0$ \\
			59834.713 & $15142.24 \pm 0.65$ & $2748.2 \pm 1.3$ & $759.9 \pm 2.5$ & $34.251 \pm 0.016$ & $611.63 \pm 0.17$ & $77.76 \pm 0.11$ & $227.2 \pm 1.3$ \\
			59836.668 & $15144.5 \pm 1.0$ & $2753.3 \pm 2.0$ & $684.3 \pm 5.3$ & $34.220 \pm 0.025$ & $577.48 \pm 0.28$ & $74.86 \pm 0.20$ & $278.6 \pm 2.0$ \\
			59839.738 & $15144.60 \pm 0.85$ & $ 	2753.5 \pm 1.7$ & $734.0 \pm 4.0$ & $34.245 \pm 0.021$ & $577.22 \pm 0.23$ & $75.10 \pm 0.15$ & $215.0 \pm 1.7$ \\
			59841.810 & $15146.51 \pm 0.94$ & $2755.1 \pm 1.9$ & $754.1 \pm 4.7$ & $34.241 \pm 0.023$ & $575.28 \pm 0.26$ & $75.27 \pm 0.18$ & $372.5 \pm 1.9$ \\
			59842.586 & $15146.60 \pm 0.81$ & $2755.7 \pm 1.6$ & $698.5 \pm 3.8$ & $34.224 \pm 0.020$ & $582.39 \pm 0.21$ & $74.69 \pm 0.14$ & $231.0 \pm 1.6$ \\
			59844.569 & $15143.30 \pm 0.89$ & $2753.6 \pm 1.8$ & $784.1 \pm 4.7$ & $34.226 \pm 0.022$ & $581.37 \pm 0.23$ & $76.93 \pm 0.16$ & $260.2 \pm 1.8$ \\
			59846.556 & $15139.44 \pm 0.87$ & $2757.2 \pm 1.7$ & $876.1 \pm 4.5$ & $34.166 \pm 0.021$ & $624.29 \pm 0.23$ & $82.24 \pm 0.16$ & $281.8 \pm 1.7$ \\
			59848.563 & $15143.03 \pm 0.90$ & $ 	2758.3 \pm 1.8$ & $794.7 \pm 4.7$ & $34.178 \pm 0.022$ & $581.16 \pm 0.24$ & $76.12 \pm 0.16$ & $295.8 \pm 1.8$ \\
			59850.543 & $15141.32 \pm 0.96$ & $ 	2759.0 \pm 1.9$ & $776.4 \pm 5.4$ & $34.157 \pm 0.024$ & $572.63 \pm 0.25$ & $75.76 \pm 0.17$ & $165.4 \pm 1.9$ \\
			59852.544 & $15139.75 \pm 0.72$ & $ 	2754.6 \pm 1.4$ & $777.6 \pm 3.2$ & $34.215 \pm 0.018$ & $596.29 \pm 0.18$ & $78.35 \pm 0.12$ & $237.3 \pm 1.4$ \\
			59852.749 & $15139.74 \pm 0.66$ & $2758.4 \pm 1.3$ & $782.7 \pm 2.7$ & $34.199 \pm 0.016$ & $588.54 \pm 0.17$ & $77.55 \pm 0.11$ & $294.7 \pm 1.3$ \\
			59854.545 & $15140.52 \pm 0.85$ & $2754.3 \pm 1.7$ & $770.0 \pm 4.2$ & $34.199 \pm 0.021$ & $594.01 \pm 0.23$ & $76.82 \pm 0.15$ & $231.0 \pm 1.7$ \\
   \hline
		\end{tabular}
	\end{center}
 \label{tab:espresso_rvs}
\end{table*}

\begin{table*}
	\caption[TESS photometry of GJ 1018.]{TESS photometry of GJ 1018. The detrended light curve was obtained by dividing the PDCSAP by the trend obtained through the biweight method (see Section \ref{sec:obs_tess}). The complete table is accessible at the CDS (\url{https://cdsarc.cds.unistra.fr/viz-bin/cat/J/A+A/675/A52}).}
    \renewcommand{\arraystretch}{1.09}
	\begin{center}
		\begin{tabular}{cccc}
  \hline \hline
			{BJD (days)} & {PDCSAP ($\rm e^{-} s^{-1}$)} & Detrended PDCSAP & Sector \\
			\hline
			2458354.113 & $10687 \pm 14$ & $0.9997 \pm 0.0013$ & TESS02 \\
			2458354.114 & $10699 \pm 14$ & $1.0009 \pm 0.0013$ & TESS02 \\
			... & ... & ... & ... \\
			2459088.244 & $10517 \pm 13$ & $0.9990 \pm 0.0013$ & TESS29 \\
			2459088.246 & $10515 \pm 13$ & $0.9988 \pm 0.0013$ & TESS29 \\
			... & ... & ... & ... \\
   \hline 
		\end{tabular}
	\end{center}
 \label{tab:tess_phot}
\end{table*}
\begin{table}
\caption[HARPS radial velocities and FWHMs of GJ 1018.]{HARPS RVs and FWHMs of GJ 1018 acquired between 15 December 2018 and 8 January 2019 under the program ID 1102.C-0339. This table is accessible through CDS. The complete table is available at the CDS (\url{https://cdsarc.cds.unistra.fr/viz-bin/cat/J/A+A/675/A52}).}
\renewcommand{\arraystretch}{1.09}
\setlength{\tabcolsep}{12pt}
	\begin{center}
		\begin{tabular}{ccc}
  \hline \hline
			{RJD (days)} & {RV ($\rm m \, s^{-1}$)} & {FWHM ($\rm m \, s^{-1}$)} \\
			\hline
			58467.550 & $15127.3 \pm 1.8$ & $2960.4 \pm 1.8$ \\
			58467.635 & $15131.3 \pm 2.4$ & $2956.1 \pm 2.4$ \\
			58468.566 & $15126.5 \pm 1.9$ & $2959.4 \pm 1.9$ \\
			58469.627 & $15126.6 \pm 2.1$ & $2961.8 \pm 2.1$ \\
			58471.576 & $15131.1 \pm 2.0$ & $2968.6 \pm 2.0$ \\
			58474.620 & $15127.7 \pm 2.3$ & $2955.3 \pm 2.3$ \\
			58475.641 & $15134.9 \pm 3.0$ & $2954.2 \pm 3.0$ \\
			58477.600 & $15129.2 \pm 2.1$ & $2951.3 \pm 2.1$ \\
            ... & ... & ... \\
   \hline
		\end{tabular}
	\end{center}
 \label{tab:harps_rvs}
\end{table}
\begin{table}
\caption[ASAS-SN photometry of GJ 1018.]{ASAS-SN photometry \citep{2014ApJ...788...48S} of GJ 1018 computed by shifting the photometric aperture every 1 October in order to ensure the target centring. The complete table is accessible at the CDS (\url{https://cdsarc.cds.unistra.fr/viz-bin/cat/J/A+A/675/A52}).}
\renewcommand{\arraystretch}{1.09}
\setlength{\tabcolsep}{9pt}
	\begin{center}
		\begin{tabular}{cccc}
   \hline \hline
			{HJD (days)} & Flux (mJy) & Camera & Filter \\
			\hline
			2456788.907 & $32.464 \pm 0.080$ & bf & V \\
			2456789.906 & $32.402 \pm 0.075$ & bf & V \\
			2456794.907 & $32.41 \pm 0.13$ & bf & V \\
			2456805.887 & $31.331 \pm 0.066$ & bf & V \\
			2456808.883 & $31.868 \pm 0.049$ & bf & V \\
			2456809.884 & $30.971 \pm 0.072$ & bf & V \\
			2456810.872 & $32.599 \pm 0.085$ & bf & V \\
			... & ... & ... & ... \\
			2458014.683 & $16.229 \pm 0.029$ & bj & g \\
			2458015.680 & $16.303 \pm 0.028$ & bj & g \\
			2458016.678 & $16.282 \pm 0.049$ & bj & g \\
			... & ... & ... & ... \\
   \hline 
		\end{tabular}
	\end{center}
 \label{tab:asassn_phot}
\end{table}
\begin{table}
\fontsize{10.3pt}{10.3pt}\selectfont
\caption[Absolute fluxes of GJ 1018 in different filters.]{Absolute fluxes in different filters that make up the SED of GJ 1018. All the catalogs were accessed through the Spanish Virtual Observatory \citep[SVO;][]{bayo2008}. The complete table is accessible at the CDS (\url{https://cdsarc.cds.unistra.fr/viz-bin/cat/J/A+A/675/A52}).}
	\begin{center}
    \setlength{\tabcolsep}{12pt}
    \renewcommand{\arraystretch}{1.07}
		\begin{tabular}{cc}
  \hline \hline
			Filter ID & Abs. Flux ($\rm erg \, s^{-1} cm^{-2} \AA^{-1}$) \\
			\hline
			GALEX/GALEX.NUV & $\left(7.6 \pm 3.1\right) \times 10^{-17}$ \\
			Misc/APASS.B & $\left(6.48 \pm 0.15\right) \times 10^{-14}$ \\
			Generic/Johnson.B & $\left(6.503 \pm 0.057\right) \times 10^{-14}$ \\
			SLOAN/SDSS.g & $\left(9.85 \pm 0.18\right) \times 10^{-14}$ \\
			GAIA/GAIA3.Gbp & $\left(1.3338 \pm 0.0038\right) \times 10^{-13}$ \\
			Misc/APASS.V & $\left(1.527 \pm 0.039\right) \times 10^{-13}$ \\
			Generic/Johnson.V & $\left(1.5008 \pm 0.0054\right) \times 10^{-13}$ \\
			HST/ACS$\_$WFC.F606W & $\left(1.9170 \pm 0.0036\right) \times 10^{-13}$ \\
			GAIA/GAIA3.G & $\left(2.9298 \pm 0.0076\right) \times 10^{-13}$ \\
			SLOAN/SDSS.r & $\left(2.038 \pm 0.057\right) \times 10^{-13}$ \\
			Generic/Johnson.R & $\left(2.5561 \pm 0.0038\right) \times 10^{-13}$ \\
			SLOAN/SDSS.i & $\left(4.331 \pm 0.084\right) \times 10^{-13}$ \\
			GAIA/GAIA3.Grp & $\left(4.327 \pm 0.015\right) \times 10^{-13}$ \\
			HST/ACS$\_$WFC.F814W & $\left(4.8812 \pm 0.0055\right) \times 10^{-13}$ \\
			Generic/Johnson.I & $\left(4.9304 \pm 0.0062\right) \times 10^{-13}$ \\
			GAIA/GAIA3.Grvs & $\left(5.109 \pm 0.032\right) \times 10^{-13}$ \\
			SLOAN/SDSS.z & $\left(5.5047 \pm 0.0052\right) \times 10^{-13}$ \\
			PAN-STARRS/PS1.y & $\left(5.5937 \pm 0.0075\right) \times 10^{-13}$ \\
			2MASS/2MASS.J & $\left(4.492 \pm 0.095\right) \times 10^{-13}$ \\
			2MASS/2MASS.H & $\left(2.765 \pm 0.084\right) \times 10^{-13}$ \\
			2MASS/2MASS.Ks & $\left(1.354 \pm 0.036\right) \times 10^{-13}$ \\
			WISE/WISE.W1 & $\left(3.013 \pm 0.063\right) \times 10^{-14}$ \\
			WISE/WISE.W2 & $\left(9.75 \pm 0.16\right) \times 10^{-15}$ \\
			WISE/WISE.W3 & $\left(2.880 \pm 0.050\right) \times 10^{-16}$ \\
			WISE/WISE.W4 & $\left(2.46 \pm 0.30\right) \times 10^{-17}$ \\
			OAJ/JPAS.J0400 & $\left(3.494 \pm 0.091\right) \times 10^{-14}$ \\
			OAJ/JPLUS.J0410 & $\left(4.262 \pm 0.066\right) \times 10^{-14}$ \\
			OAJ/JPAS.J0410 & $\left(4.427 \pm 0.077\right) \times 10^{-14}$ \\
			OAJ/JPAS.J0420 & $\left(3.706 \pm 0.077\right) \times 10^{-14}$ \\
			OAJ/JPLUS.J0430 & $\left(4.107 \pm 0.067\right) \times 10^{-14}$ \\
			OAJ/JPAS.J0430 & $\left(4.154 \pm 0.082\right) \times 10^{-14}$ \\
			OAJ/JPAS.J0440 & $\left(5.994 \pm 0.094\right) \times 10^{-14}$ \\
			OAJ/JPAS.J0450 & $\left(8.53 \pm 0.11\right) \times 10^{-14}$ \\
			OAJ/JPAS.J0460 & $\left(1.017 \pm 0.012\right) \times 10^{-13}$ \\
			OAJ/JPAS.J0470 & $\left(9.77 \pm 0.13\right) \times 10^{-14}$ \\
			OAJ/JPAS.gSDSS & $\left(1.0074 \pm 0.0019\right) \times 10^{-13}$ \\
			OAJ/JPLUS.gSDSS & $\left(1.0286 \pm 0.0019\right) \times 10^{-13}$ \\
			OAJ/JPAS.J0480 & $\left(9.29 \pm 0.13\right) \times 10^{-14}$ \\
			OAJ/JPAS.J0490 & $\left(9.90 \pm 0.13\right) \times 10^{-14}$ \\
			OAJ/JPAS.J0500 & $\left(1.052 \pm 0.013\right) \times 10^{-13}$ \\
			OAJ/JPAS.J0510 & $\left(1.009 \pm 0.014\right) \times 10^{-13}$ \\
			OAJ/JPLUS.J0515 & $\left(1.079 \pm 0.0098\right) \times 10^{-13}$ \\
			OAJ/JPAS.J0520 & $\left(1.228 \pm 0.015\right) \times 10^{-13}$ \\
			OAJ/JPAS.J0530 & $\left(1.547 \pm 0.015\right) \times 10^{-13}$ \\
			OAJ/JPAS.J0540 & $\left(1.616 \pm 0.016\right) \times 10^{-13}$ \\
			OAJ/JPAS.J0550 & $\left(1.601 \pm 0.016\right) \times 10^{-13}$ \\
			OAJ/JPAS.J0560 & $\left(1.609 \pm 0.016\right) \times 10^{-13}$ \\
			OAJ/JPAS.J0570 & $\left(1.629 \pm 0.017\right) \times 10^{-13}$ \\
			OAJ/JPAS.J0580 & $\left(1.657 \pm 0.016\right) \times 10^{-13}$ \\
			OAJ/JPAS.J0590 & $\left(1.602 \pm 0.015\right) \times 10^{-13}$ \\
			OAJ/JPAS.J0600 & $\left(1.547 \pm 0.014\right) \times 10^{-13}$ \\
			OAJ/JPAS.J0610 & $\left(1.582 \pm 0.015\right) \times 10^{-13}$ \\
			OAJ/JPAS.rSDSS & $\left(2.1212 \pm 0.0024\right) \times 10^{-13}$ \\
			... & ... \\
   \hline
		\end{tabular}
	\end{center}
 \label{tab:svo_phot}
\end{table}

\begin{table*}[]
\centering
\fontsize{9.4pt}{9.4pt}\selectfont
\renewcommand{\arraystretch}{1.8}
\setlength{\tabcolsep}{9pt}
\caption[Inferred parameters of TOI-244 b.]{Inferred parameters of TOI-244 b obtained from the TESS light curve analysis, ESPRESSO and HARPS radial velocities analysis, and the final joint analysis based on the TESS, ESPRESSO and HARPS data. }
\begin{tabular}{lccc}
\hline \hline
Parameter                          & TESS photometry           & ESPRESSO and HARPS RVs & Joint fit                 \\ \hline
\multicolumn{4}{l}{Orbital parameters}                                                                                                                  \\ \hline
$P$ {[}days{]}                              & $7.397225^{+0.000026}_{-0.000024}$ & $7.39714^{+0.00074}_{-0.00075}$   & $7.397225^{+0.000026}_{-0.000023}$ \\
$T_{\rm 0}$ {[}JD{]}                       & 2458357.3627 $\pm$ 0.0020          & 2458357.3622 $\pm$ 0.017        & 2458357.3627 $\pm$ 0.0020          \\
$i$ {[}degrees{]}                           & 88.32 $\pm$ 0.16                   & --                              & 88.32 $\pm$ 0.16                   \\ \hline
\multicolumn{3}{l}{Planet parameters}                                                                              &                                    \\ \hline
$R_{\rm p}$/$R_{\rm \star}$                 & $0.0326^{+0.0017}_{-0.0018}$         & --                              & $0.0326^{+0.0017}_{-0.0018}$         \\
$K$ {[}$\rm m/s${]}                        & --                                 & 1.54 $\pm$ 0.16                 & 1.55 $\pm$ 0.16                    \\ \hline
\multicolumn{3}{l}{Stellar parameters}                                                                              &                                    \\ \hline
$M_{\rm \star}$ $[\rm M_{\odot}]$           & 0.428 $\pm$ 0.029                  & --                              & 0.428 $\pm$ 0.029                  \\
$R_{\rm \star}$ $[\rm R_{\odot}]$           & 0.426 $\pm$ 0.025                  & --                              & 0.426 $\pm$ 0.025                  \\
$u1$                                        & 0.2204 $\pm$ 0.0011                & --                              & 0.2204 $\pm$ 0.0011                \\
$u2$                                        & 0.4112 $\pm$ 0.0017                & --                              & 0.4112 $\pm$ 0.0017                \\ \hline
\multicolumn{3}{l}{Matérn-3/2 GP hyperparameters}                                                                  &                                    \\ \hline
$\eta_{\sigma_{S2}}$ {[}$\rm e^{-}/s${]}    & $0.88^{+0.42}_{-0.22}$             & --                              & $0.87^{+0.36}_{-0.21}$             \\
$\eta_{\rho_{S2}}$ {[}days{]}               & $1.13^{+0.95}_{-0.45}$             & --                              & $1.09^{+0.79}_{-0.45}$             \\
$\eta_{\sigma_{S29}}$ {[}$\rm e^{-}/s${]}   & $1.80^{+0.63}_{-0.34}$             & --                              & $1.79^{+0.63}_{-0.34}$             \\
$\eta_{\rho_{S29}}$ {[}days{]}              & $0.79^{+0.43}_{-0.25}$             & --                              & $1.09^{+0.44}_{-0.24}$             \\ \hline
\multicolumn{3}{l}{Quasiperiodic GP hyperparameters}                                                              &                                    \\ \hline
$\eta_{\rm 1,RV}$ {[}$\rm m/s${]}          & --                                 & $3.5^{+1.0}_{-0.7}$      & $3.5^{+1.0}_{-0.7}$       \\
$\eta_{\rm 2,RV}$ {[}days{]}                & --                                 & $86^{+26}_{-23}$                  & $85^{+26}_{-23}$                     \\
$\eta_{\rm 3,RV}$ {[}days{]}                & --                                 & $53.2^{+1.3}_{-1.1}$              & $53.3^{+1.2}_{-1.1}$                 \\
$\eta_{\rm 4,RV}$                           & --                                 & $0.62^{+0.11}_{-0.09}$            & $0.62^{+0.11}_{-0.09}$               \\
$\eta_{\rm 1,FWHM}$ {[}$\rm m/s${]}        & --                                 & $3.34^{+0.70}_{-0.53}$   & $3.35^{+0.70}_{-0.54}$      \\ \hline
\multicolumn{3}{l}{Instrument-dependent parameters (TESS)}                                                         &                                    \\ \hline
$F_{\rm 0,S2}$ {[}$\rm e^{-}/s${]}          & $0.04^{+0.31}_{-0.32}$             & --                              & $0.05^{+0.29}_{-0.30}$             \\
$F_{\rm 0,S29}$ {[}$\rm e^{-}/s${]}         & $0.17^{+0.56}_{-0.54}$             & --                              & $0.18^{+0.54}_{-0.55}$             \\
$\sigma_{\rm TESS,S2}$ {[}$\rm e^{-}/s${]}  & $0.21^{+0.23}_{-0.15}$             & --                              & $0.20^{+0.23}_{-0.15}$             \\
$\sigma_{\rm TESS,S29}$ {[}$\rm e^{-}/s${]} & $1.90^{+0.53}_{-0.76}$             & --                              & $1.92^{+0.51}_{-0.74}$             \\ \hline
\multicolumn{3}{l}{Instrument-dependent parameters (ESPRESSO and HARPS)}                                           &                                    \\ \hline
$\gamma_{\rm ESPR,RV}$ {[}$\rm m/s${]}     & --                                 & $15141.0^{+1.2}_{-1.3}$   & $15141.0^{+1.2}_{-1.3}$      \\
$\gamma_{\rm HAR,RV}$ {[}$\rm m/s${]}      & --                                 & $15128.5^{+2.5}_{-2.6}$   & $15128.5^{+2.4}_{-2.6}$      \\
$\gamma_{\rm ESPR,FWHM}$ {[}$\rm m/s${]}   & --                                 & 2750.9 $\pm$ 1.2             & 2750.9 $\pm$ 1.2                 \\
$\gamma_{\rm HAR,FWHM}$ {[}$\rm m/s${]}    & --                                 & 2959.9 $\pm$ 2.6              & 2959.9 $\pm$ 2.6                \\
$\sigma_{\rm ESPR,RV}$ {[}$\rm m/s${]}     & --                                 & $0.19^{+0.21}_{-0.13}$ & $0.19^{+0.20}_{-0.13}$    \\
$\sigma_{\rm HAR,RV}$ {[}$\rm m/s${]}      & --                                 & $2.9^{+1.3}_{-1.1}$    & $2.9^{+1.3}_{-1.1}$       \\
$\sigma_{\rm ESPR,FWHM}$ {[}$\rm m/s${]}   & --                                 & $0.55^{+0.39}_{-0.36}$ & $0.56^{+0.39}_{-0.37}$   \\
$\sigma_{\rm HAR,FWHM}$ {[}$\rm m/s${]}    & --                                 & $4.4^{+1.5}_{-1.2}$    & $4.4^{+1.5}_{-1.2}$    \\ \hline 
\end{tabular}
\label{tab:final_derived_params}
\end{table*}

\begin{figure}
    \centering
    \includegraphics[width=\textwidth]{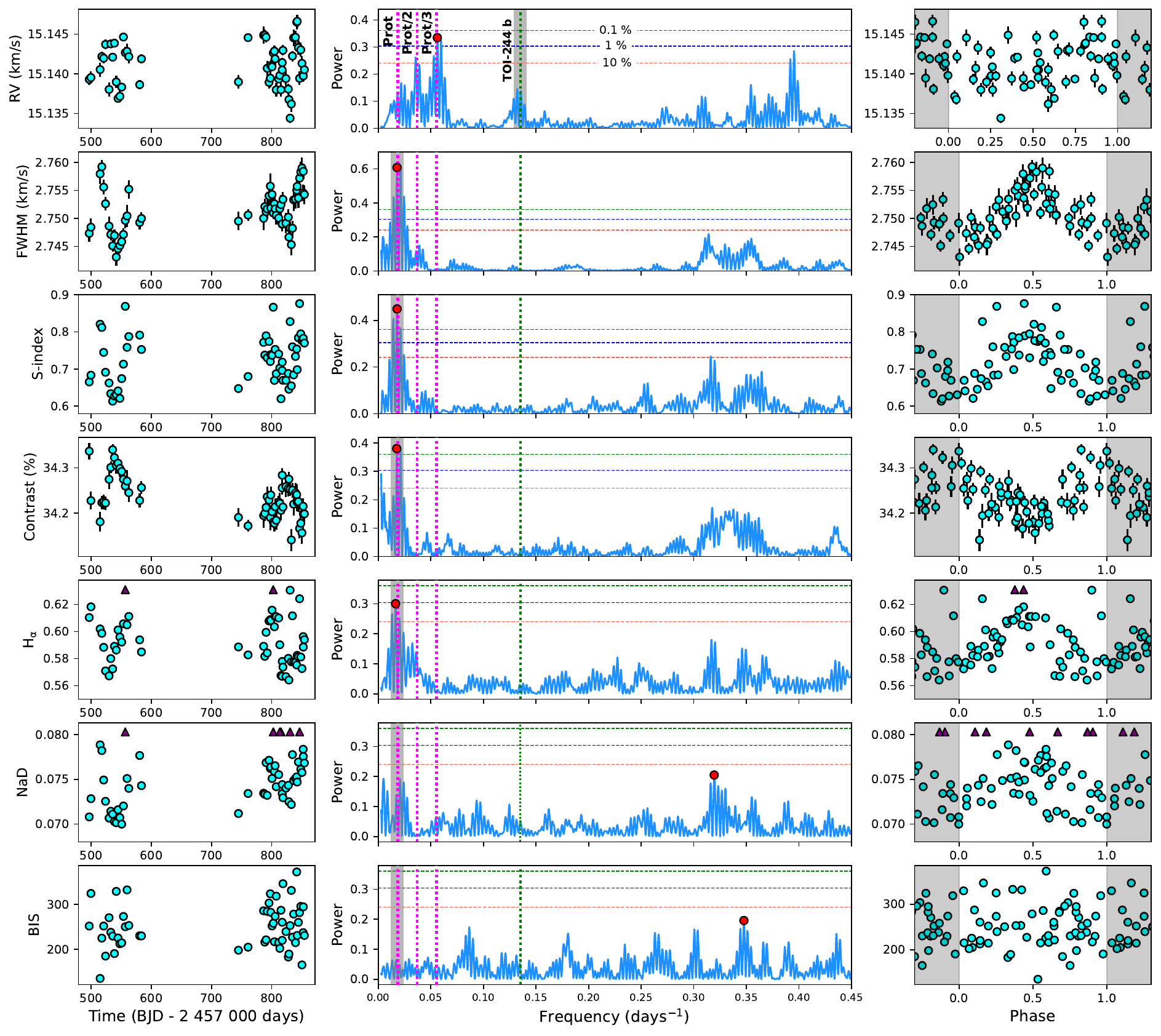}
    \caption[Time series and GLS periodograms of the ESPRESSO RVs and indicators of GJ 1018.]{\textit{Left panel:} Time series of the ESPRESSO RVs and activity indicators. \textit{Centre panel:} GLS periodograms of the time series. The red circles highlight the maximum power frequencies. The green dotted vertical lines indicate the location of the orbital period of TOI-244~b ($P_{\rm orb}$~=~7.4 days). The magenta dotted vertical lines indicate the rotation period of the star ($P_{\rm rot}$ $\sim$ 56 days) and its second and third harmonics. The grey vertical bands indicate the periods by which the time series in the right panel are folded. \textit{Right Panel:} ESPRESSO time series folded to the grey bands periods. The triangle markers within the left and right panels indicate the location of data points outside the boundaries of the plot.}
    \label{fig:gls_espresso}
\end{figure}

\begin{figure*}
    \centering
    \includegraphics[width=\textwidth]{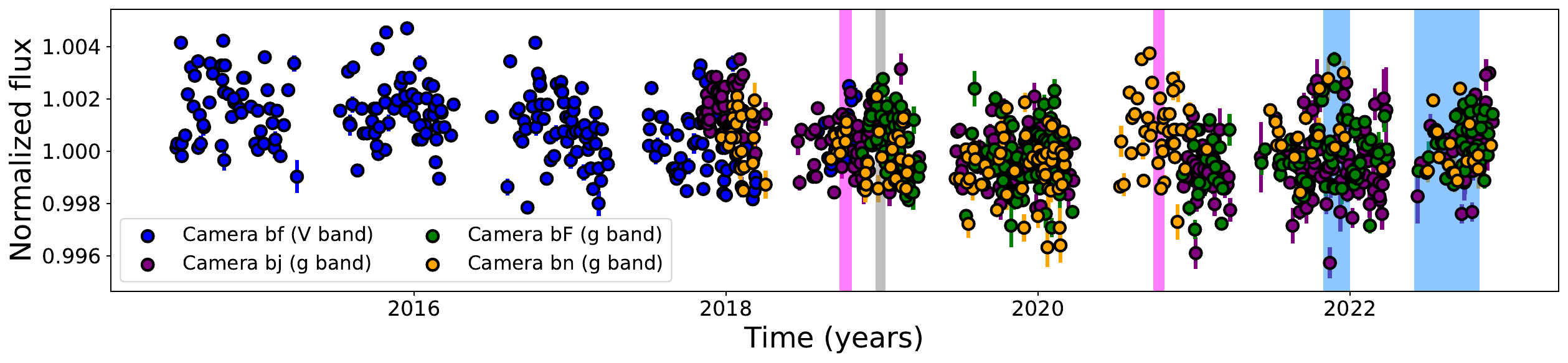}
    \includegraphics[width=\textwidth]{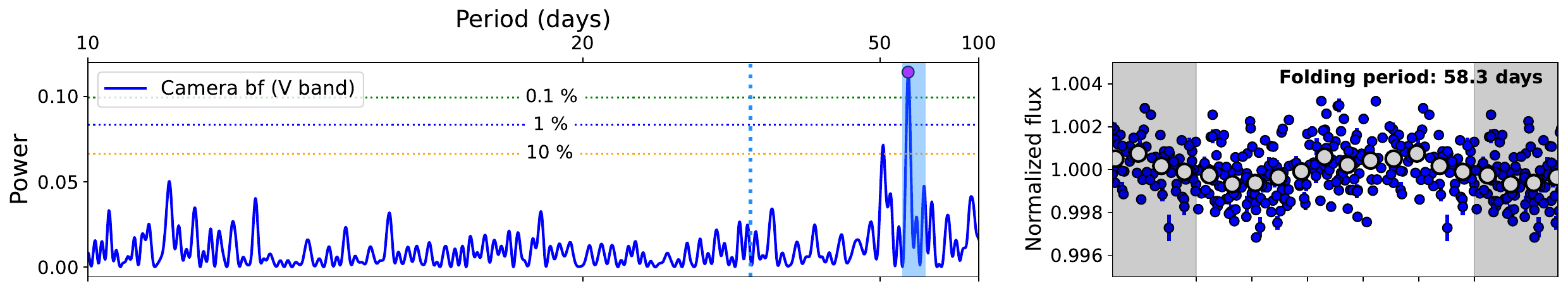}
    \includegraphics[width=\textwidth]{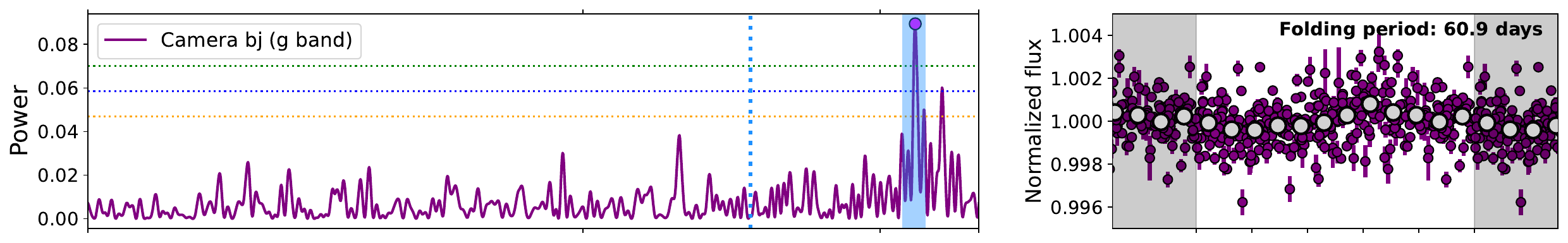}
    \includegraphics[width=\textwidth]{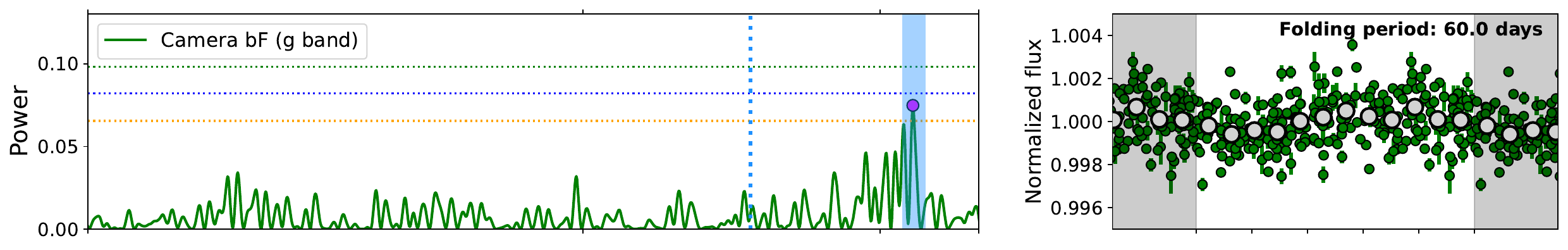}
    \includegraphics[width=\textwidth]{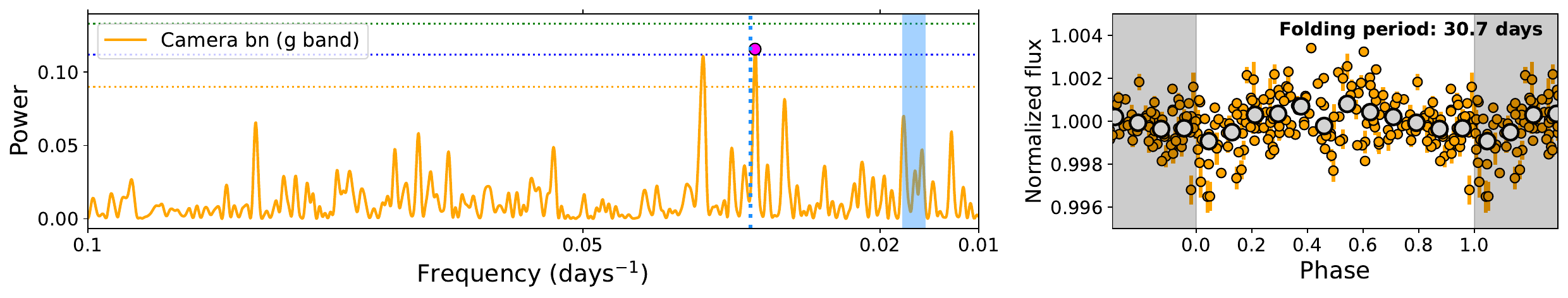}
    \caption[ASAS-SN photometric time series and GLS periodograms of GJ 1018]{\textit{Top panel:} ASAS-SN photometric time series of GJ 1018. The magenta, grey, and blue vertical lines correspond to the TESS, HARPS, and ESPRESSO observing windows, respectively. \textit{Left panels}: GLS periodograms of the ASAS-SN photometry corrected for the long-term trend. The vertical blue solid line indicates the median rotation modulation period obtained by the GLS, and the vertical blue dashed line indicates its second harmonic. \textit{Right panels:} Phase-folded light curves to the maximum power period obtained by the GLS periodogram.}
    \label{fig:asassn}
\end{figure*}


\renewcommand{\thetable}{B.\arabic{table}}   
\begin{center}
\chapter[Additional tables and figures from Chapter \ref{ch:mspis}]{Additional tables and figures from Chapter \ref{ch:mspis}}
\end{center}
\label{ch:Appendix_B}
\vspace{3cm}
\pagestyle{fancy}
\fancyhf{}
\lhead[\small{\textbf{\thepage}}]{\textbf{Additional tables and figures from Chapter \ref{ch:mspis}}}
\rhead[\small{\textbf{Appendix~\hyperref[ch:Appendix_B]{B}}}]{\small{\textbf{\thepage}}}


\begin{table*}[]
\centering
\fontsize{11pt}{11pt}\selectfont
\renewcommand{\arraystretch}{1.2}
\setlength{\tabcolsep}{10pt}
\caption[Parameters of HD~118203~b.]{Parameters of HD~118203~b from the joint analysis of TESS photometry and ELODIE RVs (Sect.~\ref{sec:joint_fit}).}
\label{tab:bestfit}
\begin{tabular}{lll}
\hline \hline
Parameter                                                       & Priors                          & Posteriors                      \\ \hline
\multicolumn{3}{l}{Orbital parameters}                                                                                              \\ \hline
Orbital period, $P_{\rm orb}$ (days)                            & $\mathcal{U}(6.0, 6.2)$         & $6.1349847 \pm 0.0000020$       \\
Time of mid-transit, $T_{0}$ (JD)                               & $\mathcal{U}(2459663, 2459664)$ & $2459663.58480 \pm 0.00025$     \\
Orbital inclination, $i$ (degrees)                              & $\mathcal{U}(50.0, 90.0)$       & $89.16^{+0.57}_{-0.65}$         \\
Ecc. parametrization, $cos(w) \sqrt{e}$                         & $\mathcal{U}(-1, 1)$            & $-0.499^{+0.014}_{-0.013}$      \\
Ecc. parametrization, $sin(w) \sqrt{e}$                         & $\mathcal{U}(-1, 1)$            & $0.259 \pm 0.029$               \\
Orbital eccentricity, $e$                                       & (derived)                       & $0.316 \pm 0.020$               \\
Argument of periastron, $w$ (degrees)                           & (derived)                       & $152.5 \pm 2.7$                 \\
Time of periastron, $T_{\rm p}$ (JD)                                & (derived)                       & $2453394.251 \pm 0.021$         \\
Transit duration, $T_{14}$ (hours)                              & (derived)                       & $5.631 \pm 0.015$               \\ \hline
Planet parameters                                               &                                 &                                 \\ \hline
RV semi-amplitude, $K$ ($\rm m \, s^{-1}$)                      & $\mathcal{U}(200, 250)$         & $218.0^{+4.0}_{-3.9}$           \\
Planet mass, $M_{p} \, (M_{\rm J})$                             & (derived)                       & $2.282 \pm 0.045$               \\
Scaled planet radius, $R_{p}/ R_{\star}$                        & $\mathcal{U}(0.0, 0.5)$         & $0.05546^{+0.00024}_{-0.00022}$ \\
Planet radius, $R_{p} \, (R_{\rm J})$                           & (derived)                       & $1.076 \pm 0.035$               \\
Planet density, $\rho_{p} \, (\rm g \, cm^{-3})$                & (derived)                       & $2.27 \pm 0.23$                 \\
Transit depth, $\Delta$ (ppt)                                   & (derived)                       & $3.075 \pm 0.025$               \\
Relative orbital separation, $a / R_{\star}$                    & (derived)                       & $9.32 \pm 0.31$                 \\
Orbit semimajor axis, $a$ (au)                                  & (derived)                       & $0.08635 \pm 0.00057$           \\
Planet surface gravity, $g \, (\rm m \, s^{-2})$                & (derived)                       & $48.9 \pm 3.4$                  \\
Impact parameter, $b$                                           & (derived)                       & $0.14 \pm 0.10$                 \\
Incident flux, $F_{\rm inc} \, (F_{\oplus})$                    & (derived)                       & $593.4 \pm 8.2$                 \\
Equilibrium temperature [A=0], $T_{\rm eq} \, (\rm K)$          & (derived)                       & $1360 \pm 23$                   \\ \hline
\multicolumn{3}{l}{Limb-darkening coefficients}                                                                                     \\ \hline
Limb-darkening coefficient, $u_{1}$                             & $\mathcal{U}(0, 1)$             & $0.347^{+0.041}_{-0.042}$       \\
Limb-darkening coefficient, $u_{2}$                             & $\mathcal{U}(0, 1)$             & $0.150^{+0.071}_{-0.069}$       \\ \hline
\multicolumn{3}{l}{GP hyperparameters}                                                                                              \\ \hline
$\eta_{\rm \sigma_{S15}}$ ($\rm e^{-}s^{-1}$)                   & $\mathcal{U}(0, 10^{3})$        & $76.7^{+5.6}_{-4.9}$            \\
$\eta_{\rm \rho_{S15}}$ (days)                                  & $\mathcal{U}(0, 10^{2})$        & $0.138^{+0.010}_{-0.009}$       \\
$\eta_{\rm \sigma_{S16}}$ ($\rm e^{-}s^{-1}$)                   & $\mathcal{U}(0, 10^{3})$        & $48.8^{+4.6}_{-3.9}$            \\
$\eta_{\rm \rho_{S16}}$ (days)                                  & $\mathcal{U}(0, 10^{2})$        & $0.229^{+0.025}_{-0.021}$       \\
$\eta_{\rm \sigma_{S22}}$ ($\rm e^{-}s^{-1}$)                   & $\mathcal{U}(0, 10^{3})$        & $37.1^{+5.7}_{-4.3}$            \\
$\eta_{\rm \rho_{S22}}$ (days)                                  & $\mathcal{U}(0, 10^{2})$        & $0.60^{+0.11}_{-0.09}$          \\
$\eta_{\rm \sigma_{S49}}$ ($\rm e^{-}s^{-1}$)                   & $\mathcal{U}(0, 10^{3})$        & $45.6^{+5.3}_{-4.3}$            \\
$\eta_{\rm \rho_{S49}}$ (days)                                  & $\mathcal{U}(0, 10^{2})$        & $0.310^{+0.048}_{-0.039}$       \\ \hline
\multicolumn{3}{l}{Instrument-dependent parameters}                                                                                 \\ \hline
TESS LC level S15, $F_{\rm 0,S15}$ ($\rm e^{-}s^{-1}$)          & $\mathcal{U}(-10^{3}, 10^{3})$  & $11.3^{+9.1}_{-9.3}$            \\
TESS LC level S16, $F_{\rm 0,S16}$ ($\rm e^{-}s^{-1}$)          & $\mathcal{U}(-10^{3}, 10^{3})$  & $8.1^{+7.6}_{-7.8}$             \\
TESS LC level S22, $F_{\rm 0,S22}$ ($\rm e^{-}s^{-1}$)          & $\mathcal{U}(-10^{3}, 10^{3})$  & $6.7^{+8.5}_{-8.3}$             \\
TESS LC level S49, $F_{\rm 0,S49}$ ($\rm e^{-}s^{-1}$)          & $\mathcal{U}(-10^{3}, 10^{3})$  & $5.4^{+8.6}_{-8.5}$             \\
TESS LC jitter S15, $\sigma_{\rm TESS,S15}$ ($\rm e^{-}s^{-1}$) & $\mathcal{U}(0, 10^{2})$        & $30.72 \pm 0.60$                \\
TESS LC jitter S16, $\sigma_{\rm TESS,S16}$ ($\rm e^{-}s^{-1}$) & $\mathcal{U}(0, 10^{2})$        & $31.79 \pm 0.63$                \\
TESS LC jitter S22, $\sigma_{\rm TESS,S22}$ ($\rm e^{-}s^{-1}$) & $\mathcal{U}(0, 10^{2})$        & $27.83 \pm 0.59$                \\
TESS LC jitter S49, $\sigma_{\rm TESS,S49}$ ($\rm e^{-}s^{-1}$) & $\mathcal{U}(0, 10^{2})$        & $36.81 \pm 0.57$                \\
ELODIE RV jitter, $\sigma_{\rm ELODIE}$ ($\rm m \, s^{-1}$)     & $\mathcal{U}(0, 10^{2})$        & $14.5^{+2.8}_{-2.5}$            \\ \hline
\multicolumn{3}{l}{RV linear trend}                                                                                                 \\ \hline
Systemic velocity, $v_{\rm sys}$ ($\rm m \, s^{-1}$)            & $\mathcal{U}(-30000, -29000)$   & $-29364.3 \pm 4.5$              \\
Slope, $\gamma$ ($\rm m \, s^{-1} \, day^{-1} $)                & $\mathcal{U}(-1, 1)$            & $0.139 \pm 0.011$               \\ \hline
\end{tabular}
\end{table*}

\begin{figure*}
    \includegraphics[width=0.99\textwidth]{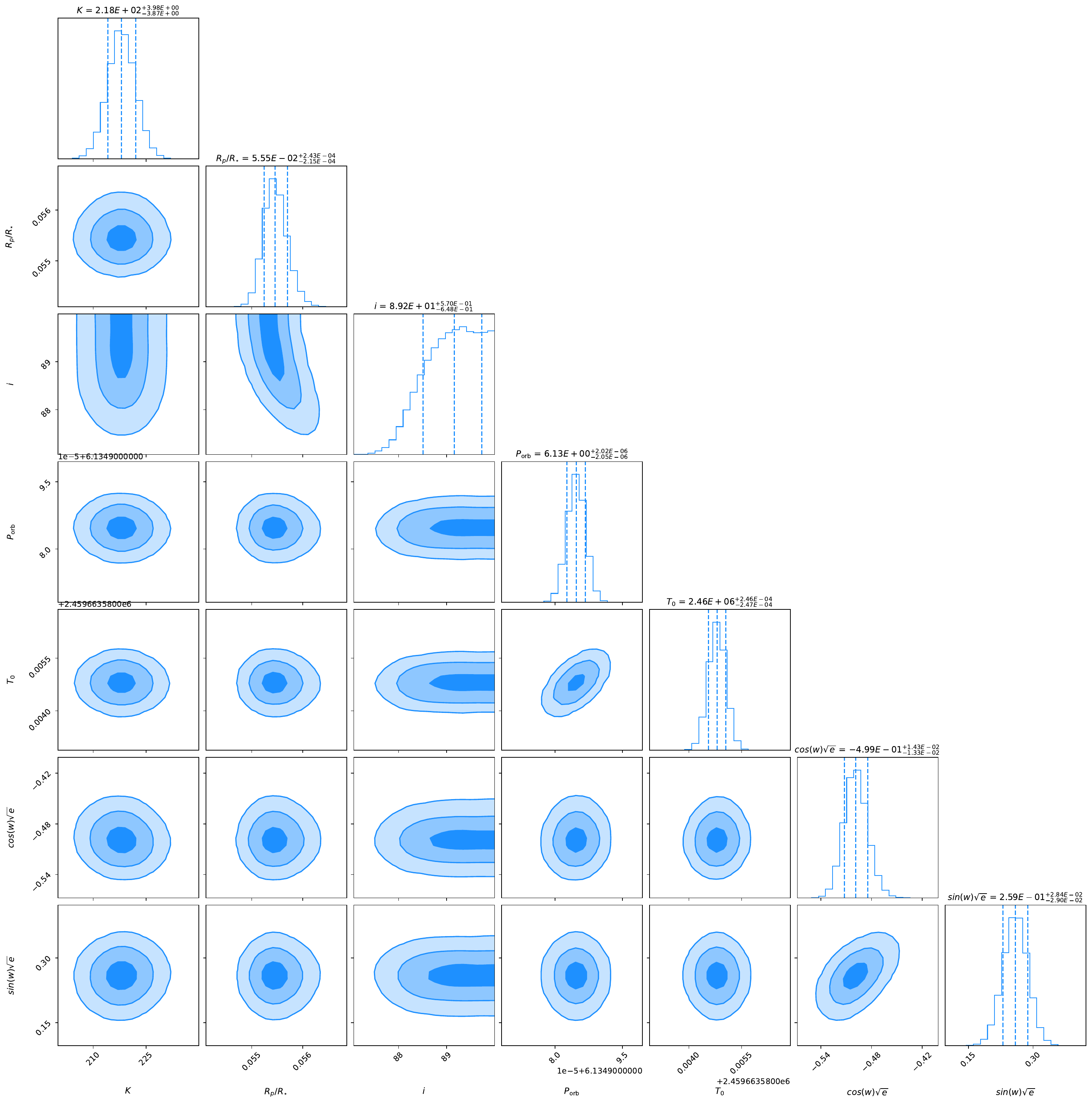}
    \caption[Corner plot of the main parameters describing the HD~118203 planetary system.]{Corner plot of the main parameters describing the HD~118203 planetary system obtained from the joint TESS light curve and ELODIE RV analysis.}
    \label{fig:corner_plot}
\end{figure*}

\begin{figure*}
    \includegraphics[width = 0.48\textwidth]{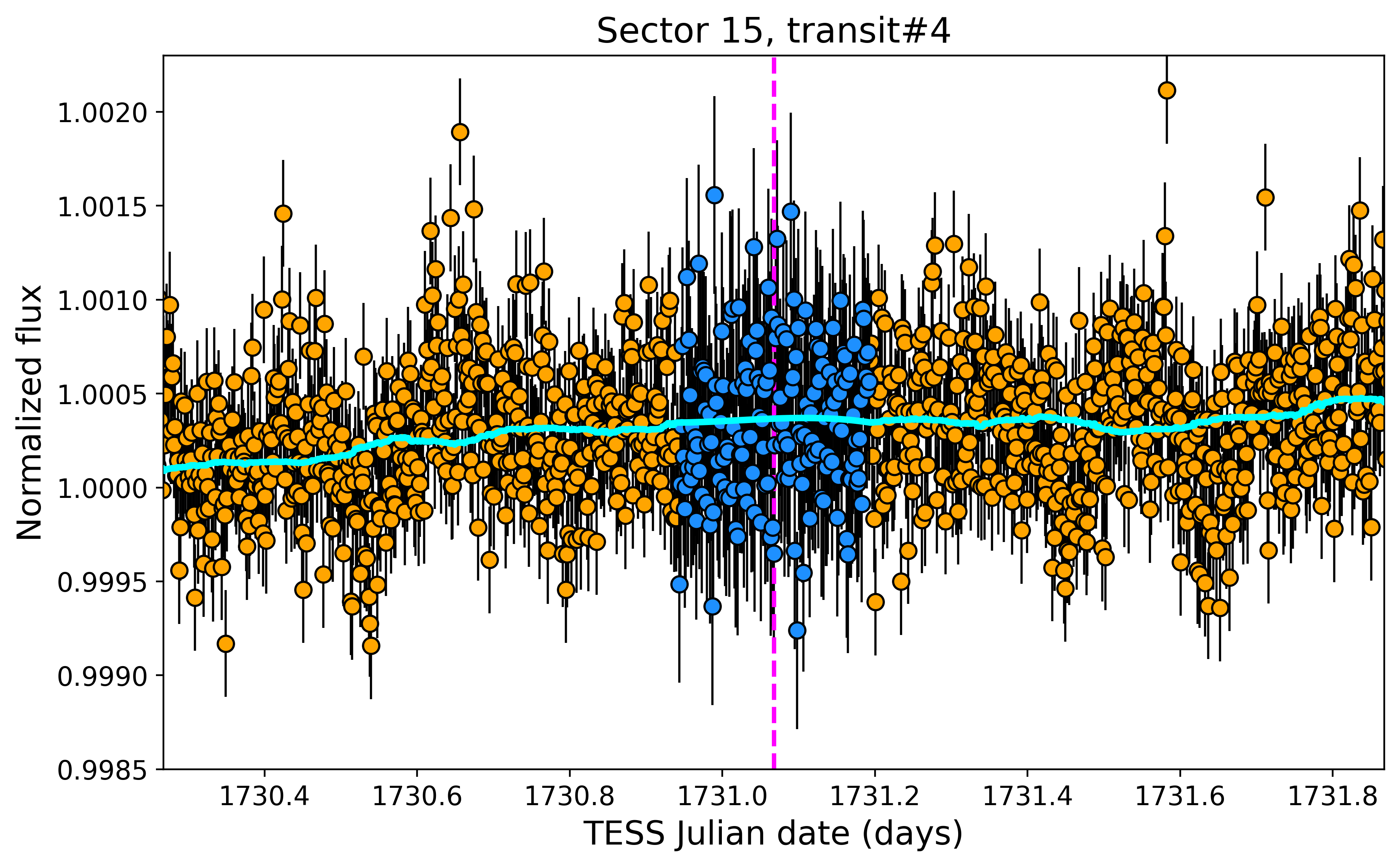}
    \includegraphics[width = 0.48\textwidth]{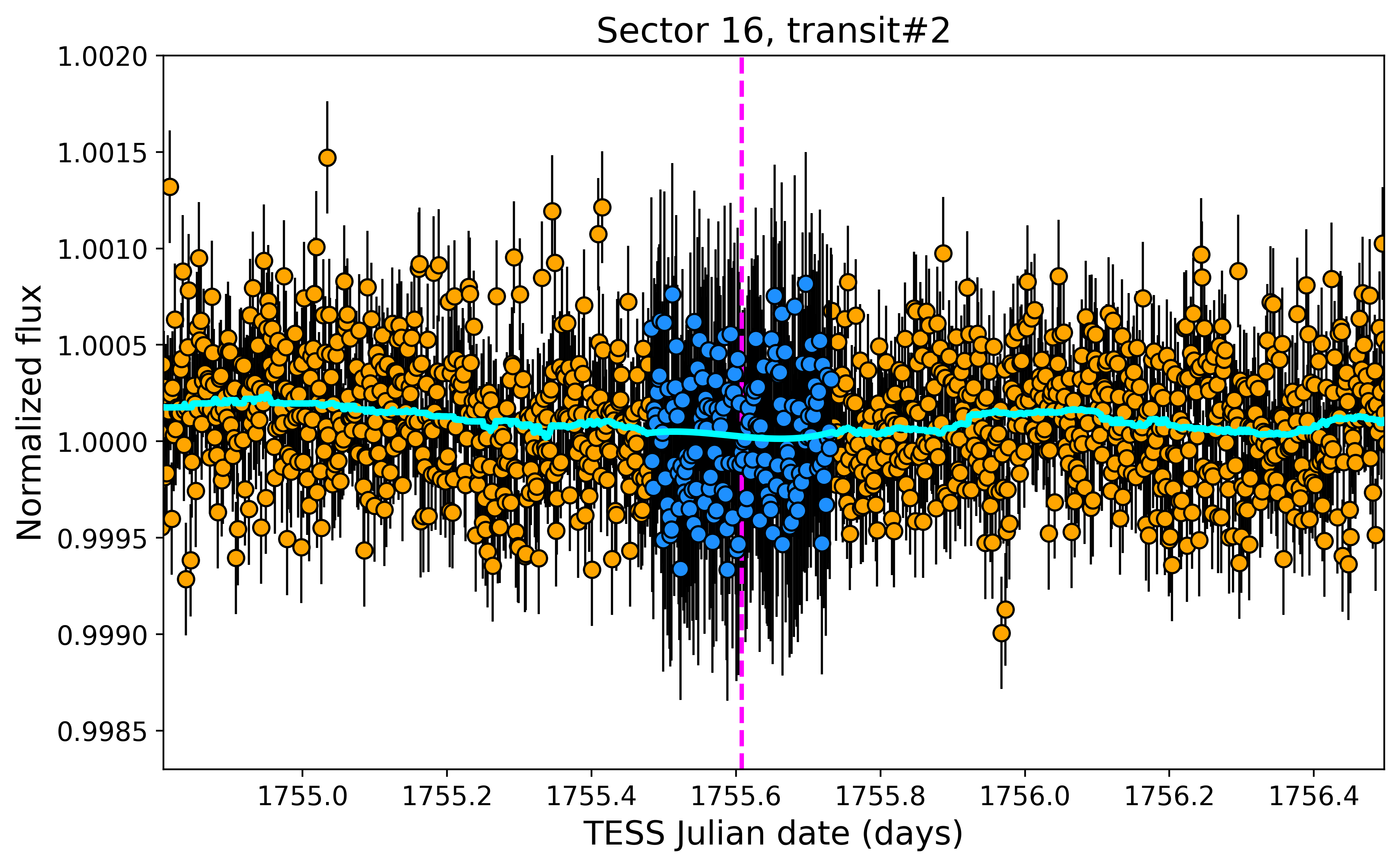}
    \caption[Examples of the mock dataset filling the TESS transit gaps of HD~118203.]{Examples of the mock dataset filling the TESS transit gaps. Orange data represent the TESS PDCSAP photometry. Blue data are the simulated 2-min cadence photometry filling the gaps of the masked transits. The cyan curve represents the cubic spline interpolation model. The vertical magenta dashed line indicates the mid-transit time of the masked transit. }
    \label{fig:mock_data_example}
\end{figure*}

\begin{figure}[t]
\centering
    \includegraphics[width = 0.8\textwidth]{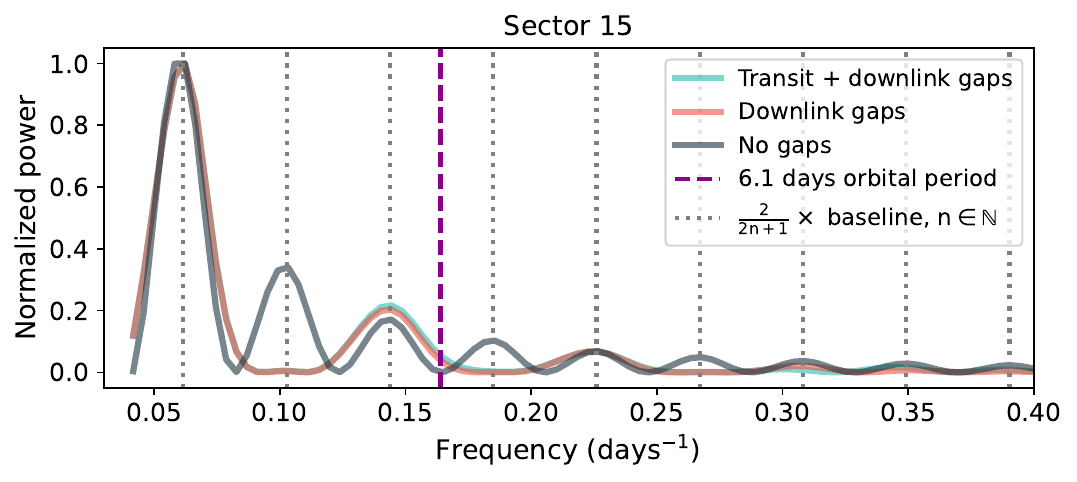}
    \includegraphics[width = 0.8\textwidth]{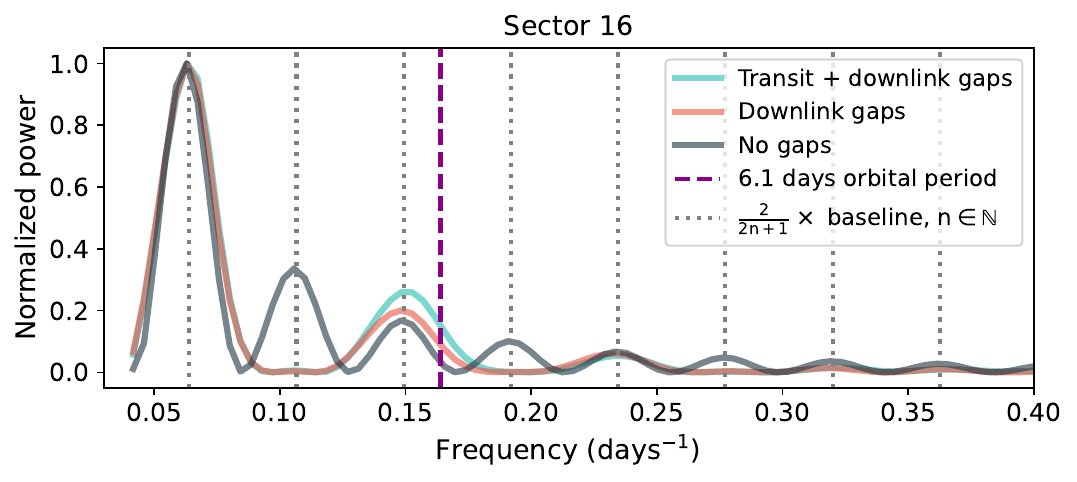}
    \includegraphics[width = 0.8\textwidth]{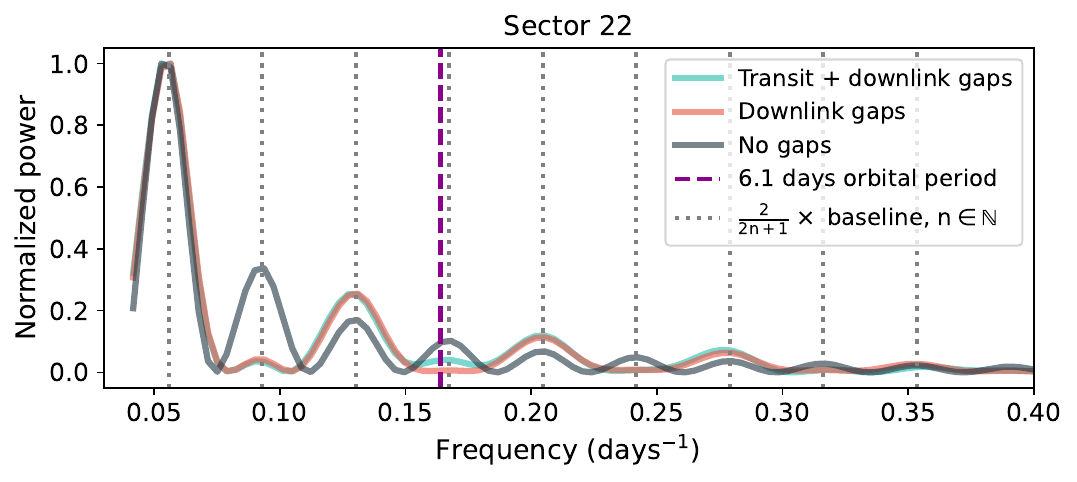}
    \includegraphics[width = 0.8\textwidth]{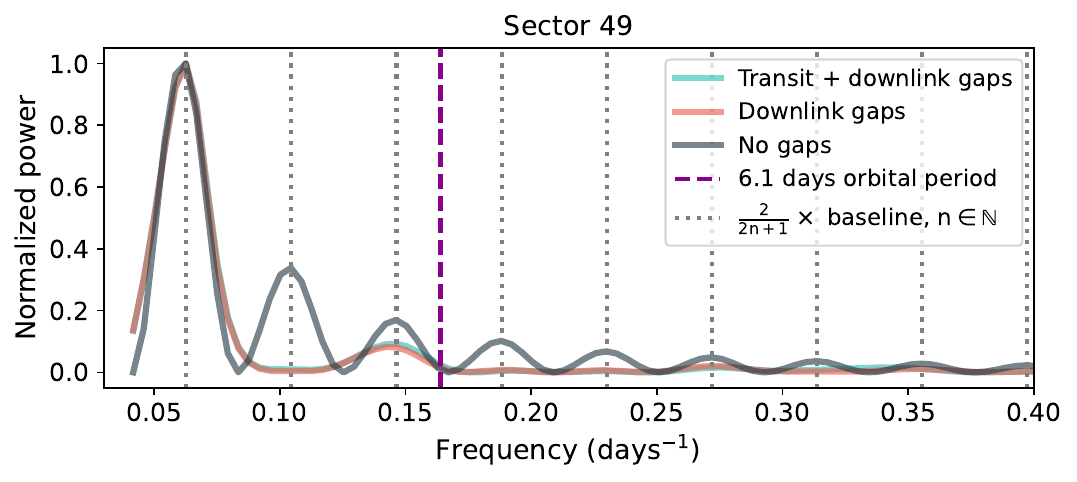}
    \caption[GLS periodograms of the TESS window functions with transit and downlink gaps, downlink gaps alone, and no gaps.]{GLS periodograms of the TESS window functions with transit and downlink gaps (blue), downlink gaps alone (red), and no gaps (grey). In all cases, the maximum power periods correspond to two-thirds of the total sector baselines, which indicates that those periodicities are not related to the gaps.}
    \label{fig:wf_comparison}
\end{figure}


\renewcommand{\thetable}{A.\arabic{table}}   
\begin{center}
\chapter[Additional figures from Chapter \ref{ch:nep_des}]{Additional figures from Chapter \ref{ch:nep_des}}
\end{center}
\label{ch:Appendix_C}
\vspace{3cm}
\pagestyle{fancy}
\fancyhf{}
\lhead[\small{\textbf{\thepage}}]{\textbf{Additional tables and figures from Chapter \ref{ch:mspis}}}
\rhead[\small{\textbf{Appendix~\hyperref[ch:Appendix_C]{C}}}]{\small{\textbf{\thepage}}}

\begin{figure*}
    \centering
    \includegraphics[width=0.4363\textwidth]{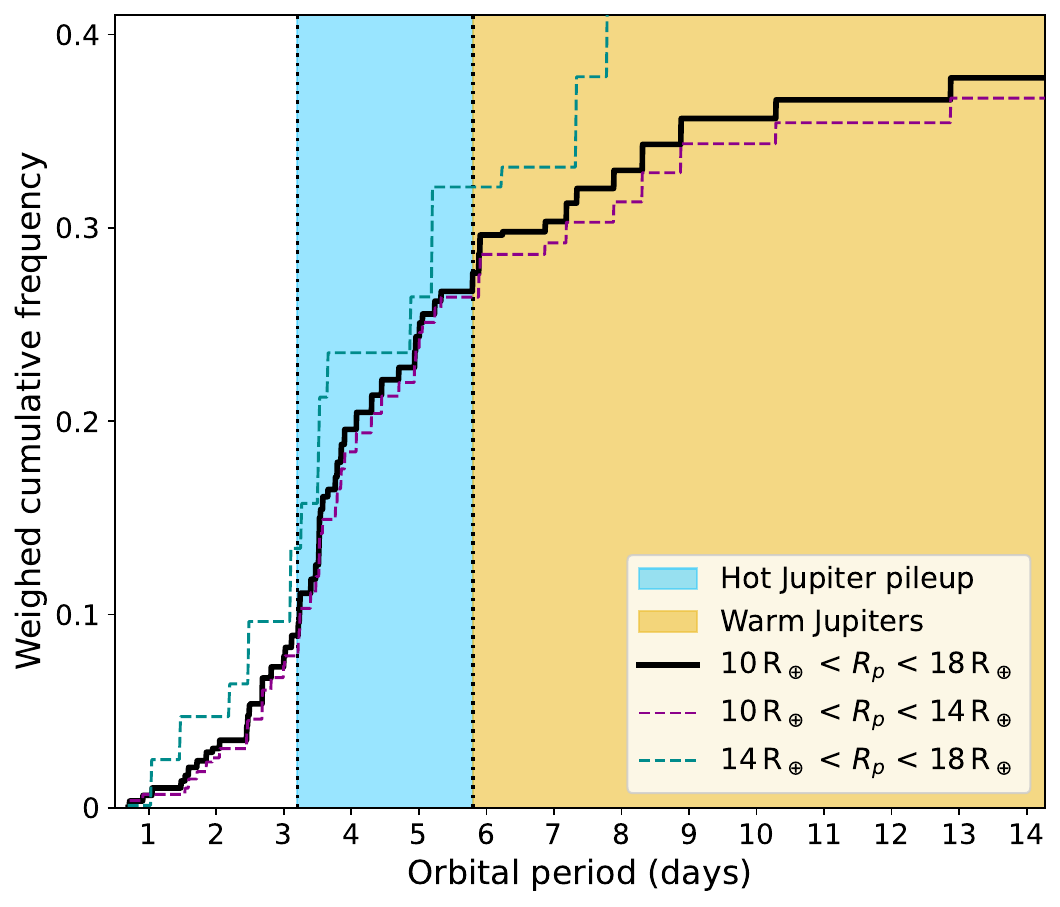}
    \includegraphics[width=0.44\textwidth]{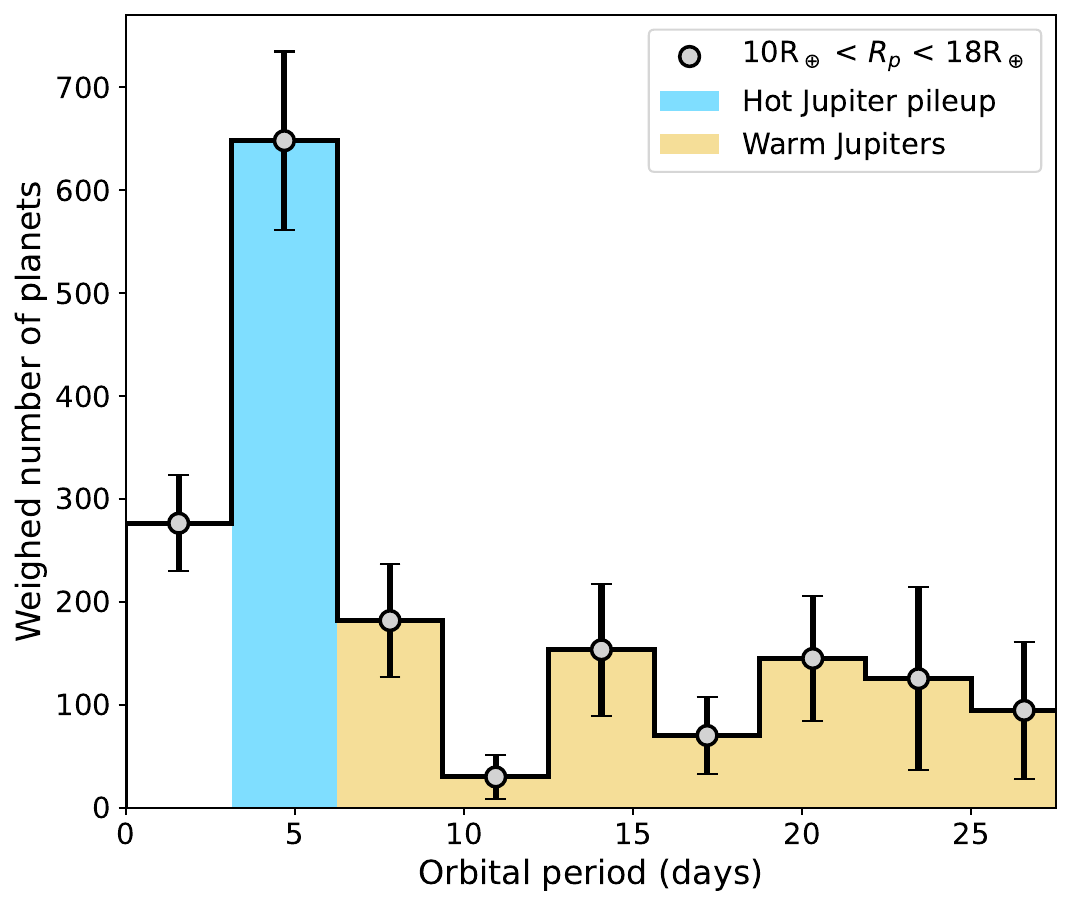}

     \vspace{0.7cm}

     \includegraphics[width=0.43 \textwidth]{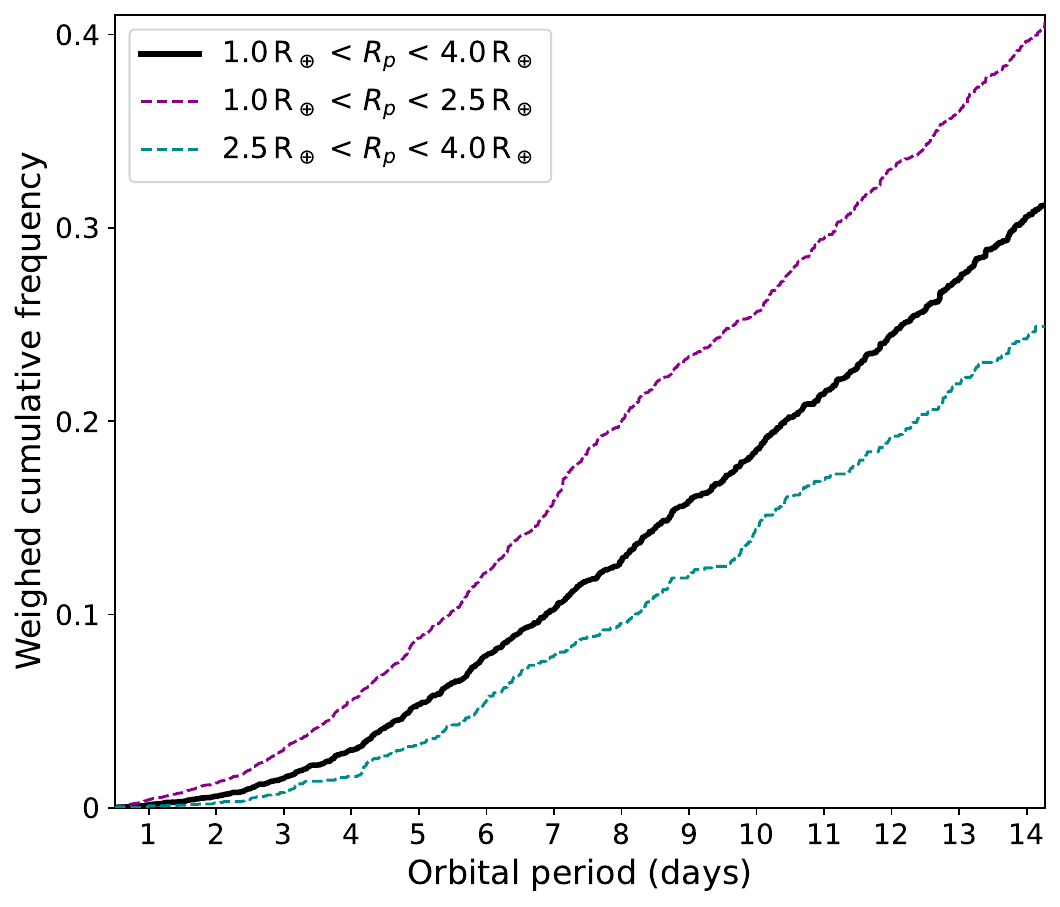}
    \includegraphics[width=0.457\textwidth]{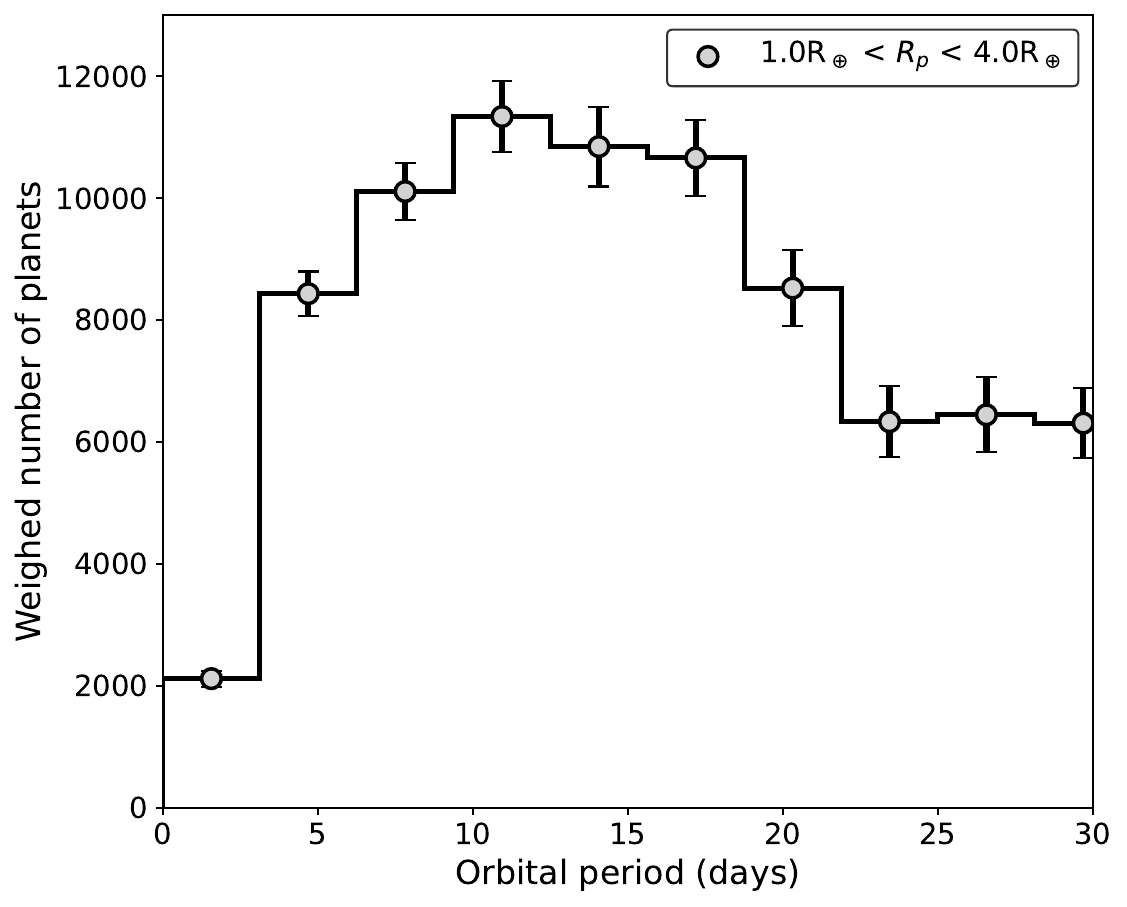}

    \caption[Occurrence of Jupiter-size planets across the orbital period space, and occurrence of sub-Neptune planets across the orbital period space.]{Top panels: Occurrence of Jupiter-size planets ($R_{\rm p}$ $>$ 10 $\rm R_{\oplus}$) across the orbital period space. Bottom panels: Occurrence of sub-Neptune planets ($R_{\rm p}$ $<$ 4 $\rm R_{\oplus}$) across the orbital period space. The histogram error bars were computed as the square root of the quadratic sum of the weights. }
    \label{fig:oc_jup}
\end{figure*}

\begin{figure}
    \centering
    \includegraphics[width=0.48\textwidth]{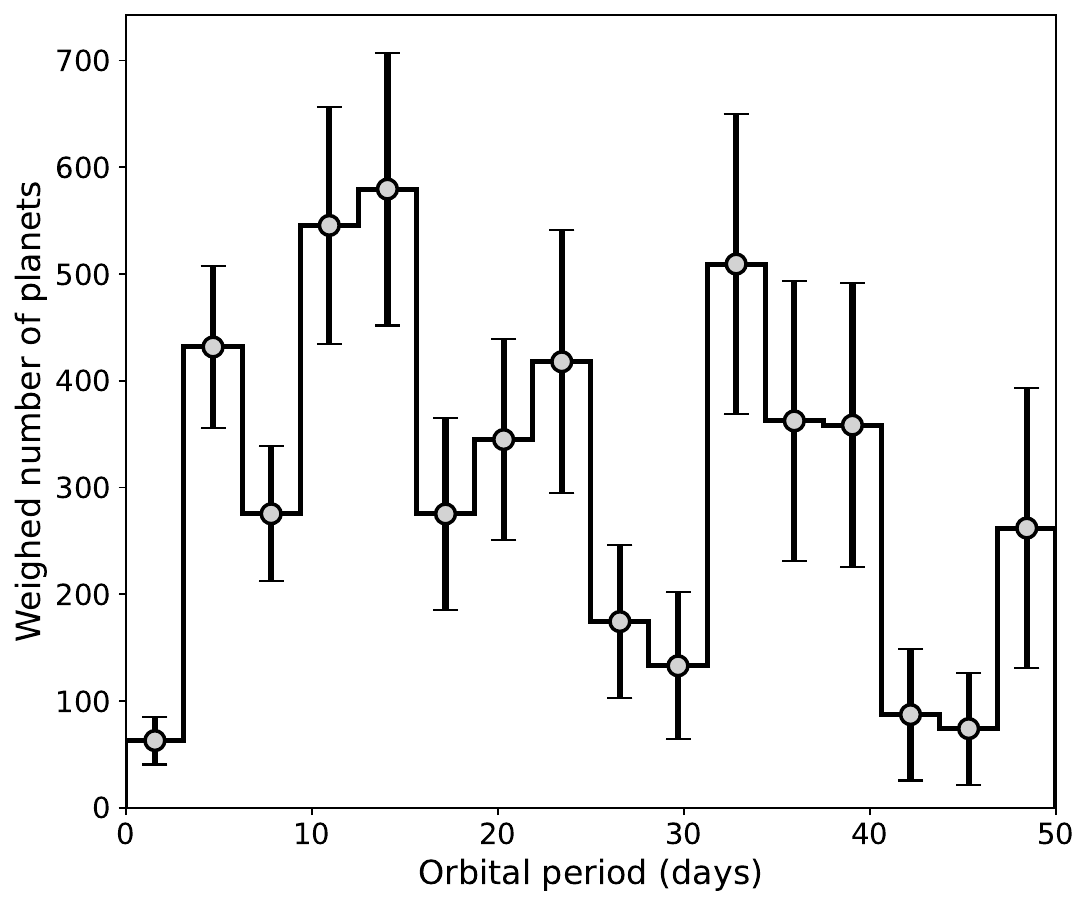}
    \includegraphics[width=0.48\textwidth]{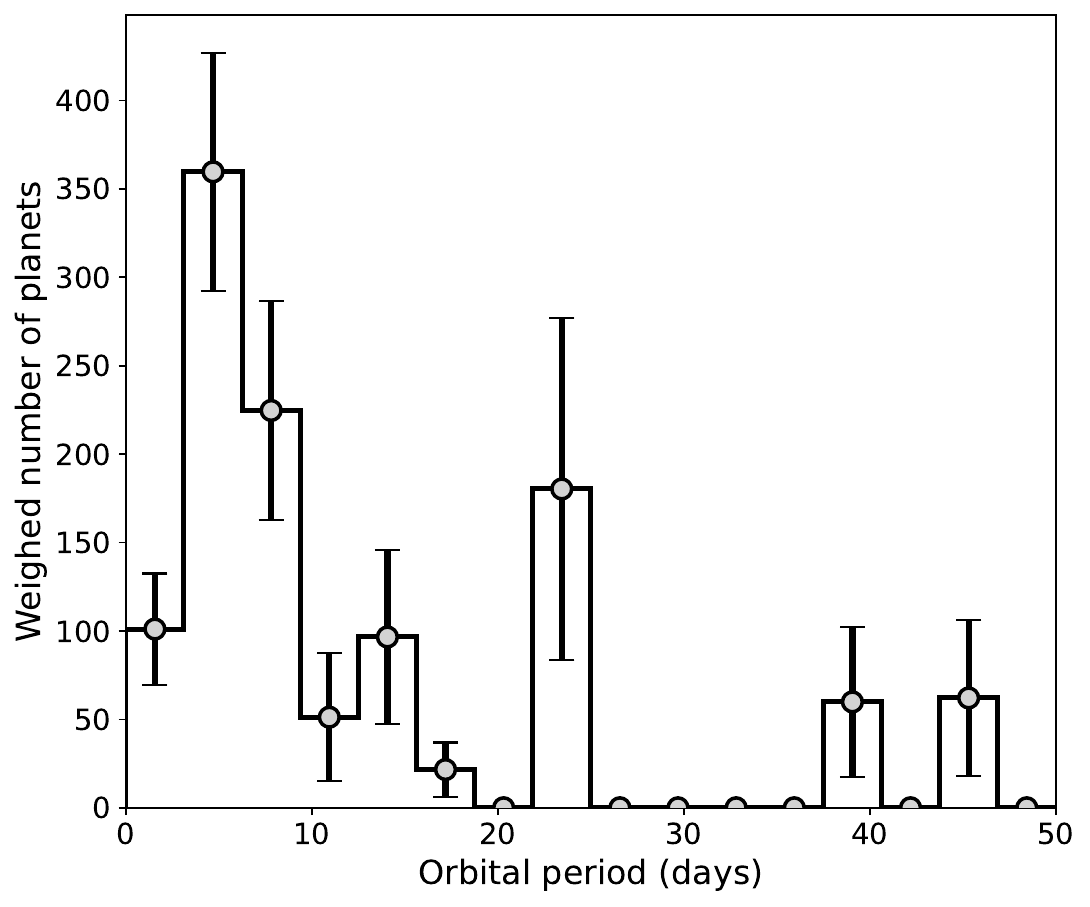}
    \caption[Occurrence of Neptunian planets in the frontier regimes.]{Occurrence of Neptunian planets in the frontier regimes (left, 4$\rm R_{\oplus}$ $<R_{\rm p}$ $<$ 5.5$\rm R_{\oplus}$; right, 8.5$\rm R_{\oplus}$ $<R_{\rm p}$ $<$ 10$\rm R_{\oplus}$). The histogram error bars were computed as the square root of the quadratic sum of the weights.}
    \label{fig:oc_transitional}
\end{figure}


\begin{center}
\chapter[Bayesian mathematical framework from Chapter \ref{ch:toi_5005}]{Bayesian mathematical framework from Chapter \ref{ch:toi_5005}}
\end{center}
\label{bayesian}
\vspace{3cm}
\pagestyle{fancy}
\fancyhf{}
\lhead[\small{\textbf{\thepage}}]{\textbf{Bayesian mathematical framework from Chapter \ref{ch:toi_5005}}}
\rhead[\small{\textbf{Appendix~\hyperref[ch:Appendix_D]{D}}}]{\small{\textbf{\thepage}}}

\renewcommand{\thetable}{B.\arabic{table}}   

\normalsize

We aim to estimate the probability that a physical model $M$ describes a data set $D$. This probability is the conditional probability that $M$ is true given that $D$ is true, and it can be computed according to Bayes' theorem \citep{1763RSPT...53..370B} as

\begin{equation}
    P\left(M|D\right) = \frac{P\left(D|M\right) P\left(M\right)}{P\left(D\right)}.
    \label{eq:bayes_theorem}
\end{equation}

$P\left(D|M\right)$ is the probability of observing $D$ if the model $M$ is true, and it is commonly called likelihood. $P\left(M\right)$ is the probability that $M$ is true before $D$ was observed, and it is commonly called prior probability. $P\left(D\right)$ is the probability of observing $D$ for all possible values of the model parameters, and it is commonly called marginal likelihood or model evidence.

For an observed data set $D$ with $N$ data points ($D_{1},...,D_{N}$), a synthetic set of $N$ values can be generated based on a physical parametric model $M$ ($M_{1},...,M_{N}$). The likelihood that a given data point $D_{i}$ is described by a synthetic point $M_{i}$ can be written as $P(D_{i}|M_{i})$, and the likelihood that the complete data set is described by the model is the product of each probability

\begin{equation}
    P\left(D|M\right) = \prod_{i=1}^{N} P\left(D_{i}|M_{i}\right).
\end{equation}

The form of the likelihood function depends on the measurement distribution and the noise properties. Assuming normally distributed data with uncorrelated noise, it can be written as

\begin{equation}
    P(D|M) = \prod_{i=1}^{N} \left[ \frac{1}{\sqrt{2\pi \delta_{i}^{2}}}  \right] e^{-\frac{1}{2}\chi^{2}},
\end{equation}

being 

\begin{equation}
    \chi^{2} = \sum_{i=1}^{N} \frac{\left(D_{i} - M_{i} \right)}{\delta_{i}^{2}},
\end{equation}

where $\delta_{i}$ is the uncertainty of the data point $D_{i}$. Given that formal uncertainties can be underestimated, we consider $\delta_{i}^{2}$ = $\sigma_{i}^{2}+\sigma_{jit}^{2}$, being $\sigma_{i}$ the estimated uncertainties and $\sigma_{jit}$ a jitter term introduced to account for additional uncorrelated noise. 

Equation \ref{eq:bayes_theorem} can lead to very large or small numbers that cause computational numerical overflows, so it is more appropriate to consider it in logarithmic form; that is, ${\rm ln}\, P\left(M|D\right) = {\rm \ln}\,P\left(D|M\right) + P(M) - P(D)$. Therefore, in practice, we implemented the log-likelihood function as

\begin{equation}
    {\rm ln}\, P\left(D|M\right) = -\frac{1}{2} \left[  \sum_{i}^{N} {\rm ln} \, 2\pi \left(\sigma_{i}^{2}+\sigma_{jit}^{2}\right) + \chi^{2} \right].
    \label{eq:log-like_uncorrelated}
\end{equation}

In order to deal with correlated noise, we used models composed of a mean function $f(x_{i})$ and an autocorrelation function $k(x_{i}, x_{j})$, with $i,j = 1,...,N$. These models are known as Gaussian Processes \citep[GPs;][]{2006gpml.book.....R}, and their log-likelihood functions can be written as

\begin{equation}
     {\rm ln}\, P\left(D|M\right) = -\frac{1}{2} r^{\rm T} K^{-1} r  - \frac{1}{2} {\rm ln} \, {\rm det} \, K - \frac{N}{2} {\rm ln}  (2\pi),
     \label{eq:log-like_correlated}
\end{equation}

where 

\begin{equation}
    r = \left[D_{1} - f(x_{1}), D_{2} - f(x_{2}),..., D_{N} - f(x_{N})  \right],
\end{equation}

and $K$ is the covariance matrix, whose elements are given by the autocorrelation function $\left[K\right]_{ij}$ = $k(x_{i}, x_{j})$.

We implemented Eq. \ref{eq:log-like_uncorrelated} and computed it directly since its computational cost depends only on the evaluation of the models. However, in Eq. \ref{eq:log-like_correlated}, besides the evaluation of the mean function, it is necessary to compute the inverse and determinant of a $N\times N$ matrix, which scales the computational cost as $N^{3}$. Therefore, when dealing with correlated noise, we used the newly developed fast methods from \citet{2015ITPAM..38..252A} and \citet{2017AJ....154..220F}, which are implemented in the \texttt{george}\footnote{Available at \url{https://github.com/dfm/george}.} and \texttt{celerite}\footnote{Available at \url{https://github.com/dfm/celerite}.} Python packages.

We considered three types of prior distributions $P(D)$, which we defined for each particular analysis according to our previous knowledge of the phenomena to analyse. Those types are uniform, Gaussian, and truncated Gaussian priors. Uniform priors are also called weakly informative or uninformative priors since they only constrain that a given parameter $\theta_{i}$ lies inside a range $\left[a,b\right]$ with no preferred values (i.e. with equal probabilities). Uniform priors can be expressed as 

\begin{equation}
\centering
\mathcal{U}(a,b) = \left\{
        \begin{array}{ll}
            \left(b-a\right)^{-1}& \quad  a < \theta_{i} < b \\
            \quad 0 & \quad {\rm otherwise}
        \end{array}
    \right.
    \label{eq:unifor_priors}
\end{equation}

Gaussian priors are also called informative priors since they contain previous information on model parameters (i.e. their mean values and 1-$\sigma$ uncertainties). This information is based on previous measurements of an independent data set not used within the analysis. We implemented Gaussian priors as 

\begin{equation}
    \mathcal{G}(a,b) = \frac{1}{\sqrt{2\pi b^{2}}} {\rm exp} \left[-\frac{\left(\theta_{i}-a\right)^{2}}{2b^{2}} \right],
    \label{eq:gaussian_priors}
\end{equation}

where $a$ and $b$ are the median and standard deviation of the measured model parameter, respectively. Even if they are highly constrained, Gaussians can take any value from -inf to +inf. To avoid non-physical values (e.g. negative stellar masses or radius, negative orbital periods of semimajor axes), we also used truncated Gaussian priors $\mathcal{TG}(a,b)$, which are Gaussian priors limited to a certain range of values. Given that non-physical values are typically negative values, we mainly consider zero-truncated Gaussians, which we refer to as $\mathcal{ZTG}$(a,b).

Finally, to obtain the posterior distribution of every parameter $\theta_{i}$, we integrated the non-normalized posterior distribution $P(D|M)P(M)$ over the remaining parameters $\theta_{j\neq i}$ (i.e. marginalization over $\theta_{i}$).


\begin{center}
\chapter[Uninformed CBV-based systematics correction from Chapter \ref{ch:toi_5005}]{Uninformed CBV-based systematics correction from Chapter \ref{ch:toi_5005}}
\end{center}
\label{sec:cbv_correction}
\vspace{3cm}
\pagestyle{fancy}
\fancyhf{}
\lhead[\small{\textbf{\thepage}}]{\textbf{Uninformed CBV-based systematics correction from Chapter \ref{ch:toi_5005}}}
\rhead[\small{\textbf{Appendix~\hyperref[ch:Appendix_E]{E}}}]{\small{\textbf{\thepage}}}

\renewcommand{\thetable}{E.\arabic{table}}   

\normalsize

The TESS PDC algorithm for systematics correction \citep{2012PASP..124.1000S,2012PASP..124..985S} consists of fitting SAP photometry to an ensemble of time series that represent the main instrumental trends in a given channel $\rm \left\lbrace Sector, Camera, CCD \right\rbrace$. Those time series are called co-trending basis vectors (CBVs) and are obtained through singular value decomposition (SVD) from the 50$\%$ most correlated SAP time series. CBVs contain information on a wide range of systematic effects detected within a given channel. However, not all those systematics necessarily affect the photometry of stars in different regions of the CCD in the same way. As detailed in \citet{2012PASP..124.1000S}, over-fitting is a major issue when dealing with CBVs. Coincidental correlations between instrumental effects and stellar variability can occur frequently, leading to stellar features being identified as systematics and consequently removed. To prevent the removal of astrophysical features, the PDC algorithm fits the CBVs to the observed data by using a Bayesian maximum a posteriori (MAP) approach. Bayesian inference allows for making conditioned fits that rely on a priori information on the expected phenomena to model. In the CBV correction context, the PDC algorithm assumes that the instrumental systematics affecting a given star are similar to the ones affecting its nearest stars. Hence, the algorithm imposes a priori constraints on the CBVs to be fitted to the raw data. This way, only photometric features present in stars in the neighbourhood are removed.

The SPOC PDC procedure has been proven to be very successful in preserving stellar signals, allowing the determination of many rotation periods \citep[e.g.][]{2020ApJS..250...20C}. However, it could lead to under-fitting if a given star is not affected by instrumental systematics in the same way as an average nearby star. Therefore, we aim to perform the strongest possible CBV correction, which consists of making no assumptions about the expected systematics behaviour. As discussed above, this approach considerably increases the risk of over-fitting and removing real astrophysical features. However, if the 6.3-day signal prevails after performing such an uninformed CBV correction, it would strongly favour a stellar origin, since no combination of the major TESS instrumental features would explain it.

For each channel $\rm \left\lbrace Sector, Camera, CCD \right\rbrace$ we built a CBV-based instrumental systematics model as

\begin{equation}
    M_{\rm CBV} (t) = \sum_{i = 1}^{N_{m_{1}}} b_{i} m_{1,i} (t) + \sum_{j = 1}^{N_{m_{2}}} c_{j} m_{2,j} (t) + \sum_{k = 1}^{N_{m_{3}}} d_{k} m_{3,k} (t) + \sum_{l = 1}^{N_{s}} e_{l} s_{l} (t), 
    \label{eq:systematics_model}
\end{equation}

where $m_{1,i}$, $m_{2,j}$, and $m_{3,k}$ are multi-scale CBVs containing systematic trends in specific wavelet-based band passes \citep{Stumpe2014}, $s_{l}$ are CBVs containing short spike systematics, and $b_{i}$, $c_{j}$, $d_{k}$, and $e_{l}$ are the linear combination coefficients. All the CBVs were obtained from MAST via TESS bulk downloads\footnote{ \url{https://archive.stsci.edu/tess/bulk_downloads/bulk_downloads_cbv.html}}.

\begin{figure*}
    \centering
    \includegraphics[width=\textwidth]{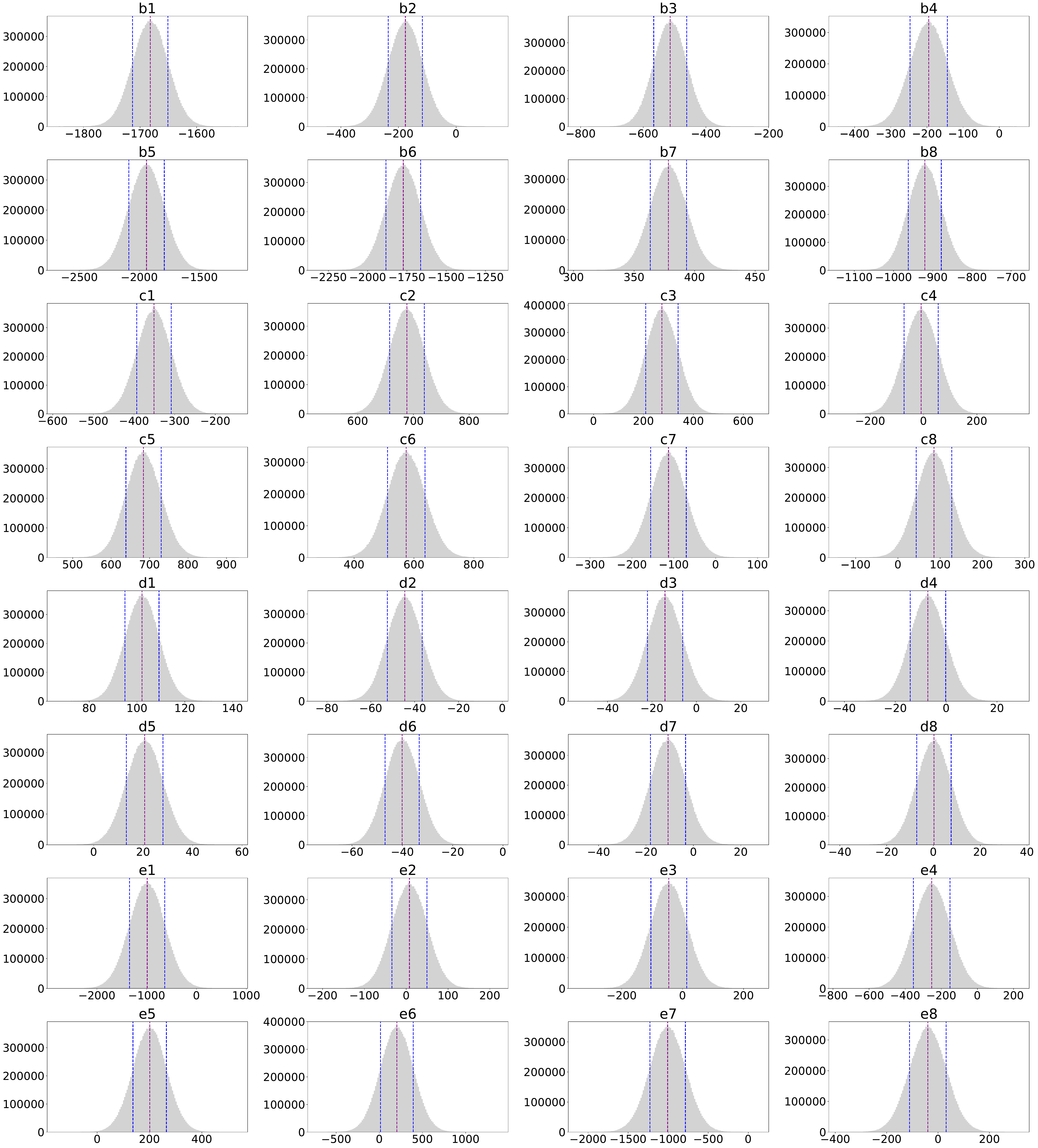}
    \caption[Posterior distribution of the coefficients of the TESS instrumental systematics model used to correct the photometry of TOI-5005 in sector 12 (Free-CBVs PDCSAP).]{Posterior distribution of the coefficients of the TESS instrumental systematics model (Eq.~\ref{eq:systematics_model}) used to correct the photometry of TOI-5005 in sector 12 (Free-CBVs PDCSAP). The magenta vertical lines indicate the median values of the distributions, and the blue vertical lines indicate the 68.3$\%$ (1$\sigma$) confidence intervals.}
    \label{fig:posterior_coeff_S12}
\end{figure*}

\begin{figure*}
    \centering
    \includegraphics[width=\textwidth]{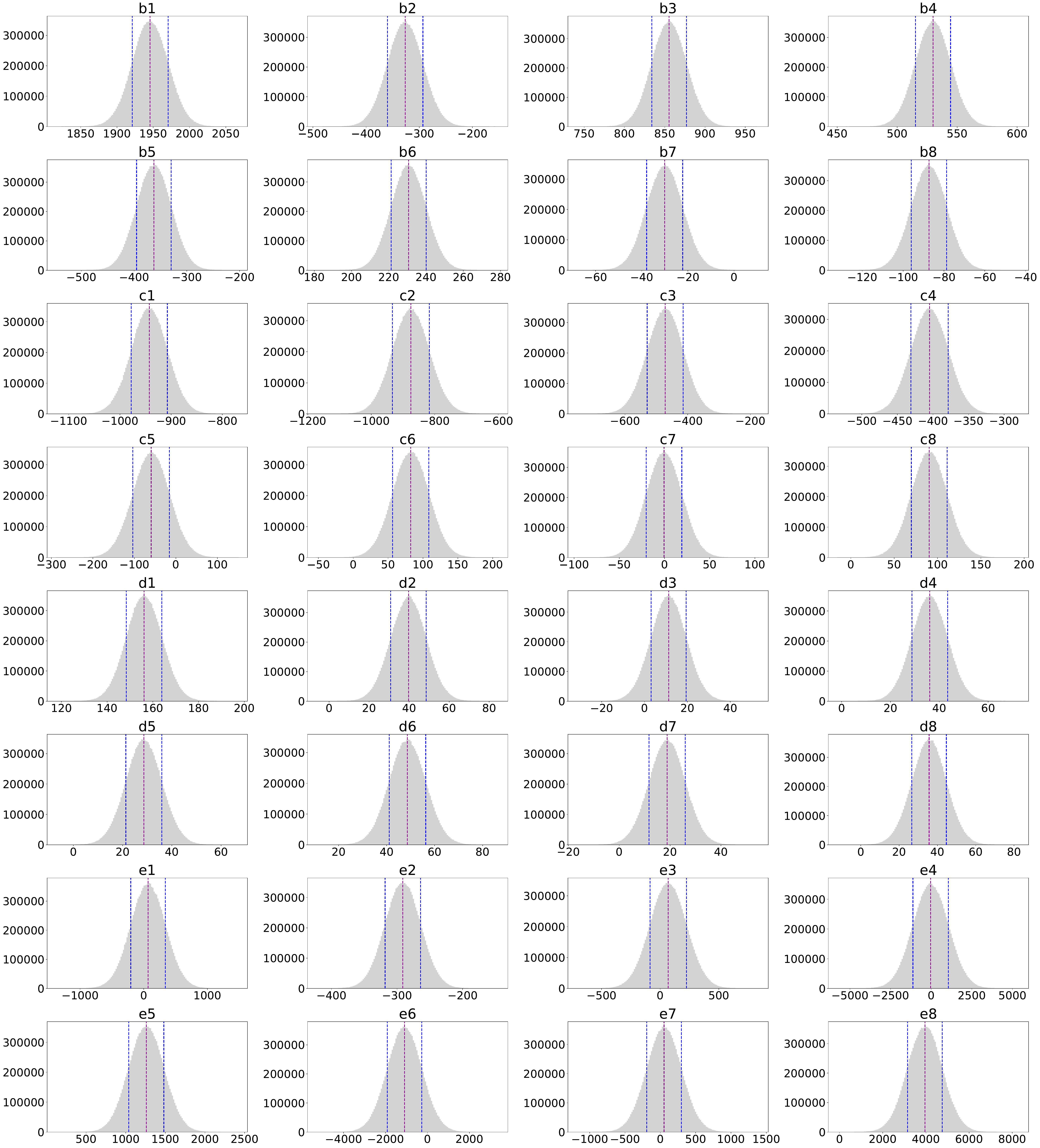}
    \caption[Posterior distribution of the coefficients of the TESS instrumental systematics model used to correct the photometry of TOI-5005 in sector 39 (Free-CBVs PDCSAP).]{Posterior distribution of the coefficients of the TESS instrumental systematics model (Eq.~\ref{eq:systematics_model}) used to correct the photometry of TOI-5005 in sector 39 (Free-CBVs PDCSAP). The magenta vertical lines indicate the median values of the distributions, and the blue vertical lines indicate the 68.3$\%$ (1$\sigma$) confidence intervals.}
    \label{fig:posterior_coeff_S39}
\end{figure*}

\begin{figure*}
    \centering
    \includegraphics[width=\textwidth]{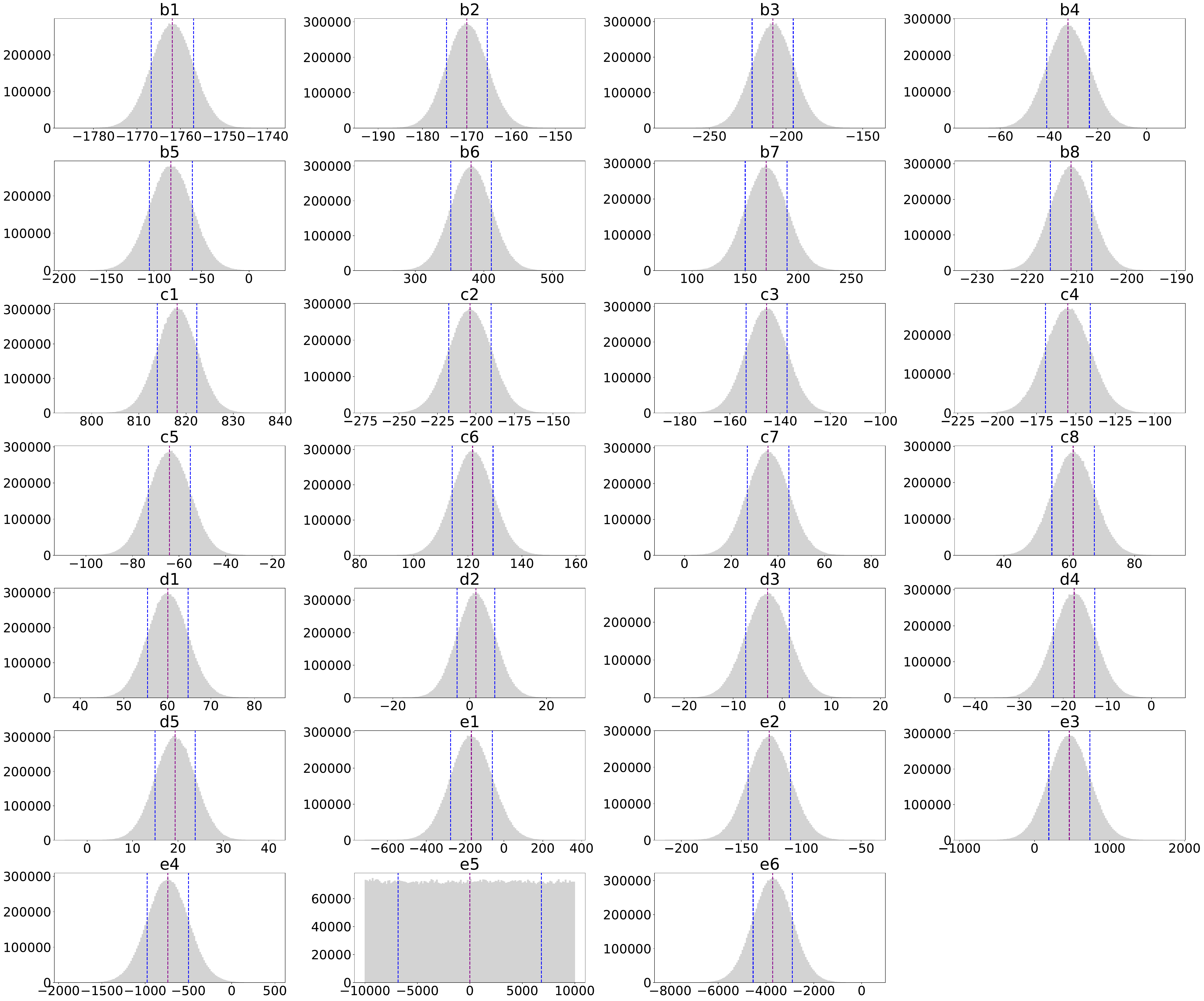}
    \caption[Posterior distribution of the coefficients of the TESS instrumental systematics model used to correct the photometry of TOI-5005 in sector 65 (Free-CBVs PDCSAP).]{Posterior distribution of the coefficients of the TESS instrumental systematics model (Eq.~\ref{eq:systematics_model}) used to correct the photometry of TOI-5005 in sector 65 (Free-CBVs PDCSAP). The magenta vertical lines indicate the median values of the distributions, and the blue vertical lines indicate the 68.3$\%$ (1$\sigma$) confidence intervals.}
   \label{fig:posterior_coeff_S65}
\end{figure*}

\begin{figure*}
    \centering
\includegraphics[width=\textwidth]{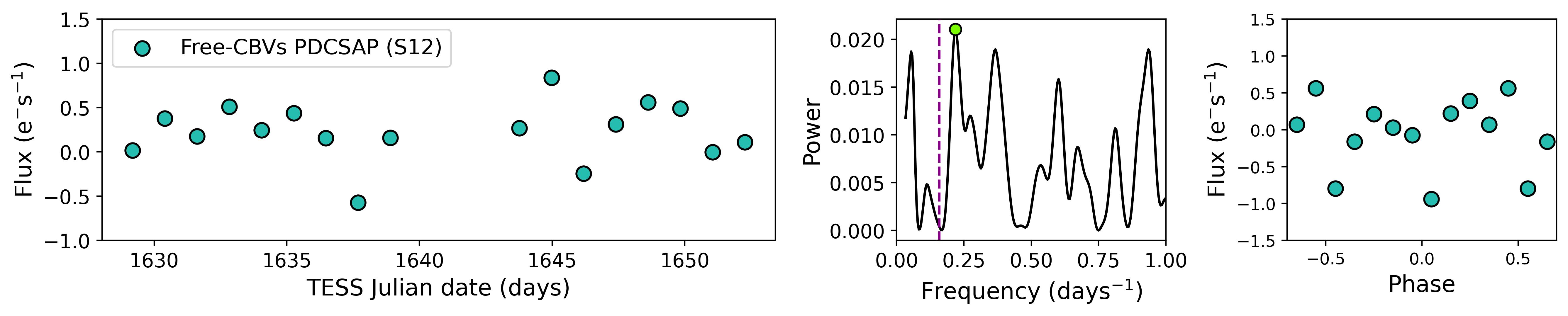}
\includegraphics[width=\textwidth]{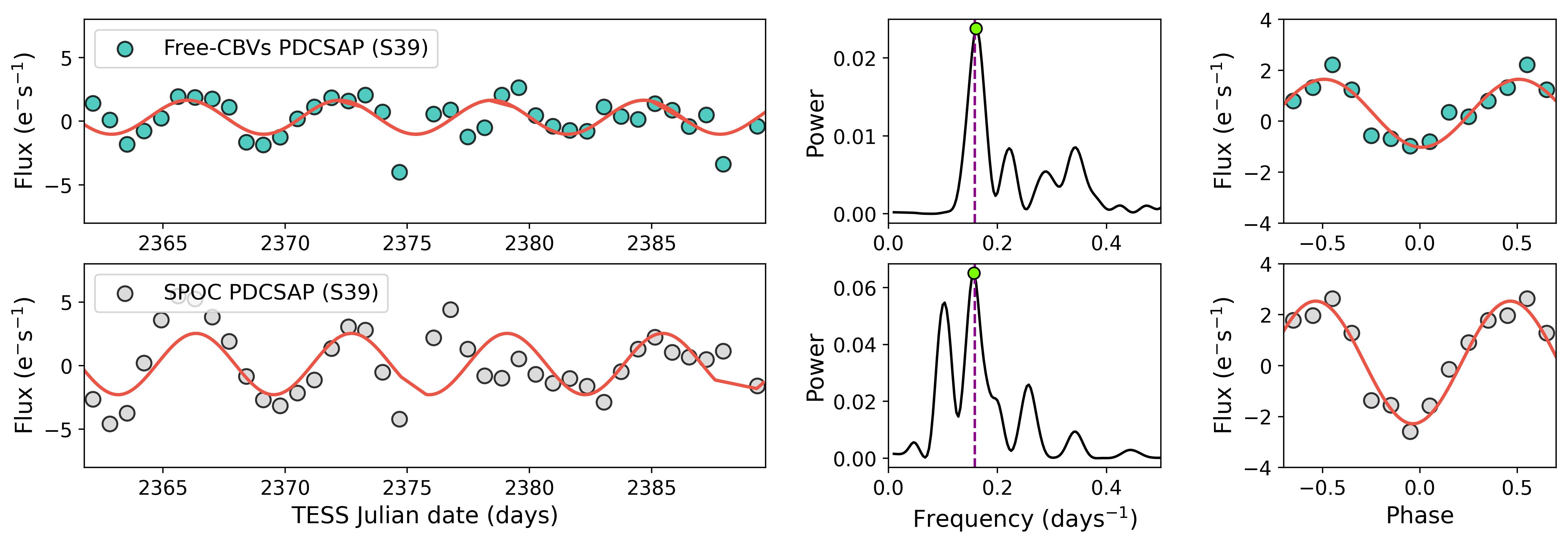}
\includegraphics[width=\textwidth]{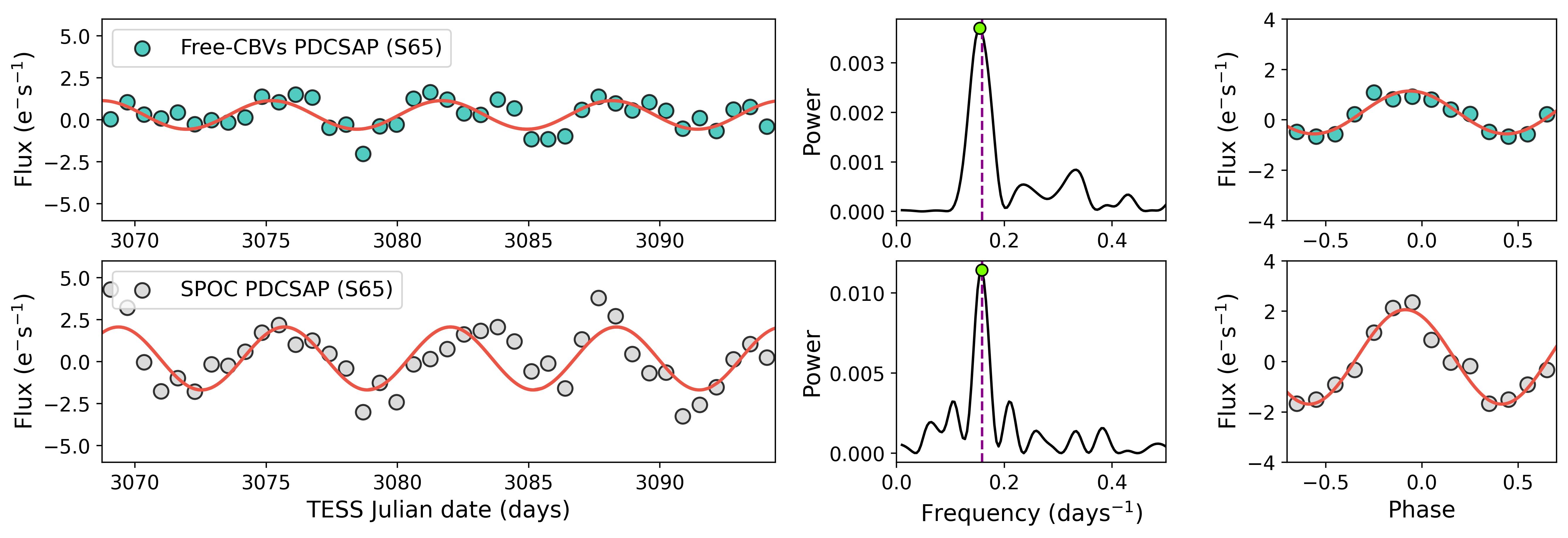}
    \caption[Sinusoidal signals within the TESS photometry of TOI-5005 corrected from systematics through SPOC and our uninformative CBV-based correction.]{Sinusoidal signals within the TESS photometry of TOI-5005 corrected from systematics through SPOC (PDCSAP photometry) and our uninformative CBV-based correction (Free-CBVs PDCSAP photometry). Left panels: TESS time series corresponding to PDCSAP and Free-CBVs PDCSAP photometry. The red lines are the sinusoidal models corresponding to the highest GLS powers. The time series are binned with 1.3-day bins for clarity. Centre panels: GLS periodograms of the unbinned time series. The vertical magenta dashed lines indicate the orbital period of TOI-5005~b. The green circles indicate the maximum power frequencies. Right panels: Photometric time series folded to the orbital period of TOI-5005~b. The phase refers to the time of inferior conjunction. The data is binned with bins of 10$\%$ of the orbital phase. }
    \label{fig:free_cbvs}
\end{figure*}

We modelled the SAP photometry of each sector through Bayesian inference as described in Sect.~\ref{sec:model_sel_param_det}. As discussed before, contrary to SPOC PDC, we did not impose any constraint on the linear combination coefficients, which we left to vary freely with uninformative priors $\mathcal{U} (-10^{4}, 10^{4})$. In Figs.~\ref{fig:posterior_coeff_S12}, \ref{fig:posterior_coeff_S39}, and \ref{fig:posterior_coeff_S65}, we show the posterior distributions obtained for each coefficient. In Fig.~\ref{fig:free_cbvs}, we show the  GLS periodograms of the SPOC PDCSAP photometry and our corrected photometry based on uninformative CBV coefficients (we refer to it as Free-CBVs PDCSAP). The S12 periodogram shows its highest peak (at $\simeq$4.6 days) surrounded by a forest of similar peaks, so no clear periodicity can be detected. The S39 and S65 periodograms, however, show maximum power peaks at 6.3 days, similar to the SPOC PDCSAP peaks. Although still present, the photometric modulation has decreased in amplitude, which is reflected in the smaller periodogram powers. Hence, the strongest possible CBV correction has absorbed part of the 6.3 days modulation detected within SPOC PDCSAP, but the absorption was not complete, still leaving clear evidence of a 6.3 days sinusoidal signal.


\begin{center}
\chapter[Additional tables and figures from Chapter \ref{ch:toi_5005}]{Additional tables and figures from Chapter \ref{ch:toi_5005}}
\end{center}
\label{sec:tables_figures_5005}
\vspace{3cm}
\pagestyle{fancy}
\fancyhf{}
\lhead[\small{\textbf{\thepage}}]{\textbf{Additional tables and figures from Chapter \ref{ch:toi_5005}}}
\rhead[\small{\textbf{Appendix~\hyperref[sec:cbv_correction]{F}}}]{\small{\textbf{\thepage}}}

\renewcommand{\thetable}{F.\arabic{table}}   
\normalsize

\newpage

\begin{table}[t]
\fontsize{10pt}{10pt}\selectfont
\renewcommand{\arraystretch}{1.6}
\setlength{\tabcolsep}{5.2pt}
\parbox{.47\linewidth}{
\caption[TESS Simple Aperture Photometry of TOI-5005.]{TESS Simple Aperture Photometry of TOI-5005. This table is available at the CDS (\url{https://cdsarc.cds.unistra.fr/viz-bin/cat/J/A+A/691/A233}).}
\begin{tabular}{cccc}
\hline \hline
BJD (days)  & SAP & QF  & Sector \\ \hline
2458624.995 & 1.0026 $\pm$ 0.0016     & 18432   & TESS12 \\
...         & ...                     & ... & ...    \\ \hline
\end{tabular}
\label{tab:TESS_SAP}
}
\hfill
\parbox{0.46\linewidth}{
\caption[TESS QLP and PDCSAP photometry of TOI-5005.]{TESS QLP and PDCSAP photometry of TOI-5005. This table is available at the CDS (\url{https://cdsarc.cds.unistra.fr/viz-bin/cat/J/A+A/691/A233}).}
\begin{tabular}{cccc}
\hline \hline
BJD (days)  & QLP/PDCSAP & QF  & Sector \\ \hline
2458629.891 & 1.0047 $\pm$ 0.0013     & 0   & TESS12 \\
...         & ...                     & ... & ...    \\ \hline
\end{tabular}
\label{tab:TESS_QLP_PDCSAP}
}
\end{table}

\begin{table*}[t]
\centering
\fontsize{10pt}{10pt}\selectfont
\caption[PEST photometry and de-trend parameters of TOI-5005.]{PEST photometry and de-trend parameters acquired on 5 May 2022. The de-trend parameters correspond to the position (\textit{x},\textit{y}) and distance (\textit{dist}) to the detector centre of the target star, the full width at half maximum of the target point spread function (\textit{fwhm}), airmass ($\chi$), and background flux (\textit{sky}). The complete table is available at the CDS (\url{https://cdsarc.cds.unistra.fr/viz-bin/cat/J/A+A/691/A233}).}
\renewcommand{\arraystretch}{1.6}
\setlength{\tabcolsep}{6.6pt}
\begin{tabular}{ccccccccc}
\hline \hline
BJD (days)  & Flux   & Flux error & \textit{x} & \textit{y} & \textit{dist} & \textit{fwhm} & $\chi$ & \textit{sky} \\ \hline
2459705.010 & 1.0023 & 0.0024     & 1797.5795  & 769.5013   & 438.5767      & 6.1665        & 1.9437 & 1009.1895    \\
...         & ...    & ...        & ...        & ...        & ...           & ...           & ...    & ...          \\ \hline
\end{tabular}
\label{tab:pest_data}
\end{table*}

\begin{table*}[]
\centering
\fontsize{10pt}{10pt}\selectfont
\caption[TRAPPIST-South photometry and de-trend parameters of TOI-5005.]{TRAPPIST-South photometry and de-trend parameters acquired on 22 March 2024. The de-trend parameters correspond to the displacement (\textit{dx},\textit{dy}) of the target star across the CCD, the full width at half maximum of the target point spread function (\textit{fwhm}), airmass ($\chi$), and background flux (\textit{sky}). The complete table is available at the CDS (\url{https://cdsarc.cds.unistra.fr/viz-bin/cat/J/A+A/691/A233}).}
\renewcommand{\arraystretch}{1.6}
\setlength{\tabcolsep}{10.2pt}
\begin{tabular}{cccccccc}
\hline \hline
BJD (days)   & Flux   & Flux error & \textit{dx} & \textit{dy} & \textit{fwhm} & $\chi$ & \textit{sky} \\ \hline
2460392.733 & 0.9982 & 0.0024     & -0.2685     & -0.5710     & 2.9760        & 1.2924 & 168.8333     \\
...          & ...    & ...        & ...         & ...         & ...           & ...    & ...          \\ \hline
\end{tabular}
\label{tab:trappist_south_data}
\end{table*}

\begin{table*}[]
\centering
\fontsize{9.2pt}{9.2pt}\selectfont
\caption[HARPS radial velocities and activity indicators of TOI-5005.]{HARPS radial velocities and activity indicators of TOI-5005 acquired between 15 March 2023 and 18 August 2023 under the programs 108.21YY.001 and 108.21YY.002. The complete table is available at the CDS (\url{https://cdsarc.cds.unistra.fr/viz-bin/cat/J/A+A/691/A233}).}
\renewcommand{\arraystretch}{1.6}
\setlength{\tabcolsep}{5.4pt}
\begin{tabular}{ccccccc}
\hline \hline
RJD (days) & RV ($\rm m \, s^{-1}$) & FWHM ($\rm m \, s^{-1}$) & S-index           & Ca                  & $\rm H_{\alpha}$    & Contrast (\%) \\ \hline
60018.834  & 17877.4 $\pm$ 3.5      & 7153.9 $\pm$ 8.7         & 0.208 $\pm$ 0.011 & 0.1742 $\pm$ 0.0044 & 0.2642 $\pm$ 0.0035 & 48.662        \\
...  & ...     & ...       & ... & ... & ... & ...        \\ \hline
\end{tabular}
\label{tab:harps_rvs}
\end{table*}




\begin{table}[]
\centering
\fontsize{9.2pt}{9.2pt}\selectfont
\renewcommand{\arraystretch}{1.14}
\setlength{\tabcolsep}{14pt}
\caption[Inferred parameters of TOI-5005~b.]{Inferred parameters of TOI-5005~b based on the joint analysis described in Sect.~\ref{subsec:joint_analysis}.}
\label{tab:parameters_joint}
\begin{tabular}{lll}
\hline \hline
Parameter                                                   & Priors                              & Posteriors                            \\ \hline
\multicolumn{3}{l}{Orbital and physical parameters}                                                                                       \\ \hline
Orbital period, $P_{\rm orb}$ (days)                        & $\mathcal{U}(5.0, 7.0)$             & $6.3085044^{+0.0000092}_{-0.0000088}$ \\
Time of mid-transit, $T_{0}$ (JD)                           & $\mathcal{U}(2460089.0, 2460091.0)$ & $2460090.0356 \pm 0.0010$             \\
Orbital inclination, $i$ (degrees)                          & $\mathcal{U}(50.0, 90.0)$           & $89.53^{+0.33}_{-0.44}$               \\
Scaled planet radius, $R_{p}/ R_{\star}$                    & $\mathcal{U}(0.0, 0.1)$             & $0.0616 \pm 0.0012$                   \\
RV semi-amplitude, $K \, (\rm m\,s^{-1})$                   & $\mathcal{U}(0.0, 10^{5})$          & $11.6 \pm 2.1$                        \\
Ecc. parametrisation, $\rm cos(\omega)\sqrt{e}$              & (fixed)                           & 0                                     \\
Ecc. parametrisation, $\rm sin(\omega)\sqrt{e}$               & (fixed)                           & 0                                     \\
Planet radius, $R_{p} \, (\rm R_{\rm \oplus})$              & (derived)                           & $6.25 \pm 0.24$                       \\
Planet radius, $R_{p} \, (\rm R_{\rm J})$                   & (derived)                           & $0.558 \pm 0.021$                     \\
Planet mass, $M_{p} \, (\rm M_{\rm \oplus})$                & (derived)                           & $32.7 \pm 5.9$                        \\
Planet mass, $M_{p} \, (M_{\rm J})$                         & (derived)                           & $0.103 \pm 0.018$                     \\
Planet density, $\rho_{p}$ ($\rm g \, cm^{-3}$)             & (derived)                           & $0.74 \pm 0.16$                       \\
Transit depth, $\Delta$ (ppt)                               & (derived)                           & $3.80 \pm 0.15$                       \\
Transit duration, $T_{\rm 14}$ (hours)                      & (derived)                           & $3.144 \pm 0.046$                     \\
Relative orbital separation, $a / R_{\star}$                & (derived)                           & $15.29 \pm 0.50$                      \\
Orbit semimajor axis, $a$ (au)                              & (derived)                           & $0.06614 \pm 0.00045$                 \\
Planet surface gravity, $g$ $(\rm m\,s^{-2})$               & (derived)                           & $8.2 \pm 1.6$                         \\
Impact parameter, $b$                                       & (derived)                           & $0.13 \pm 0.10$                       \\
Incident flux, $F_{\rm inc} \, (\rm F_{\oplus})$            & (derived)                           & $226 \pm 12$                          \\
Equilibrium temperature [A=0], $T_{\rm eq} \, (\rm K)$      & (derived)                           & $1040 \pm 20$                         \\ \hline
\multicolumn{3}{l}{Limb-darkening coefficients}                                                                                           \\ \hline
Limb-darkening coefficient, $q_{\rm 1, TESS}$               & $\mathcal{ZTG}(0.32, 0.32)$         & $0.41^{+0.20}_{-0.16}$                \\
Limb-darkening coefficient, $q_{\rm 2, TESS}$               & $\mathcal{ZTG}(0.36, 0.14)$         & $0.31^{+0.13}_{-0.12}$                \\
Limb-darkening coefficient, $q_{\rm 1, PEST}$               & $\mathcal{ZTG}(0.43, 0.37)$         & $0.66^{+0.20}_{-0.22}$                \\
Limb-darkening coefficient, $q_{\rm 2, PEST}$               & $\mathcal{ZTG}(0.38, 0.12)$         & $0.43^{+0.12}_{-0.12}$                \\
Limb-darkening coefficient, $q_{\rm 1, TRAPPIST-South}$     & $\mathcal{ZTG}(0.27, 0.29)$         & $0.37^{+0.20}_{-0.17}$                \\
Limb-darkening coefficient, $q_{\rm 2, TRAPPIST-South}$     & $\mathcal{ZTG}(0.34, 0.15)$         & $0.30^{+0.14}_{-0.13}$                \\ \hline
\multicolumn{3}{l}{GP hyper-parameters}                                                                                                    \\ \hline
$\eta_{\rm \sigma_{S12}}$                                   & $\mathcal{U}(0, 0.5)$               & $0.0060^{+0.0018}_{-0.0011}$          \\
$\eta_{\rm \sigma_{S39}}$                                   & $\mathcal{U}(0, 0.5)$               & $0.00423^{+0.00063}_{-0.00047}$       \\
$\eta_{\rm \sigma_{S65}}$                                   & $\mathcal{U}(0, 0.5)$               & $0.00353^{+0.00080}_{-0.00054}$       \\
$\eta_{\rm \rho_{S12}}$ (days)                              & $\mathcal{U}(0, 30)$                & $1.10^{+0.26}_{-0.18}$                \\
$\eta_{\rm \rho_{S39}}$ (days)                              & $\mathcal{U}(0, 30)$                & $0.633^{+0.125}_{-0.096}$             \\
$\eta_{\rm \rho_{S65}}$ (days)                              & $\mathcal{U}(0, 30)$                & $1.17^{+0.21}_{-0.16}$                \\ \hline
\multicolumn{3}{l}{Instrument-dependent parameters}                                                                                       \\ \hline
TESS LC level S12, $F_{\rm 0,S12}$                          & $\mathcal{U}(-0.1, 0.1)$            & $-0.00064^{+0.00175}_{-0.00195}$      \\
TESS LC level S39, $F_{\rm 0,S39}$                          & $\mathcal{U}(-0.1, 0.1)$            & $-0.00052^{+0.00098}_{-0.00096}$      \\
TESS LC level S65, $F_{\rm 0,S65}$                          & $\mathcal{U}(-0.1, 0.1)$            & $-0.00028^{+0.00112}_{-0.00111}$      \\
TESS LC jitter S12, $\sigma_{\rm TESS,S12}$                 & $\mathcal{U}(0, 0.005)$             & $0.000831 \pm 0.000030$               \\
TESS LC jitter S39, $\sigma_{\rm TESS,S39}$                 & $\mathcal{U}(0, 0.005)$             & $0.000750 \pm 0.000039$               \\
TESS LC jitter S65, $\sigma_{\rm TESS,S65}$                 & $\mathcal{U}(0, 0.005)$             & $0.000082^{+0.00084}_{-0.00057}$      \\
HARPS RV jitter, $\sigma_{jit,\rm HARPS}$ $(\rm m\,s^{-1})$ & $\mathcal{U}(0, 100)$               & $7.6^{+1.4}_{-1.2}$                   \\ \hline
\multicolumn{3}{l}{PEST and TRAPPIST-South de-trend parameters}                                                                            \\ \hline
$c_{\rm PEST, x}$                                           & $\mathcal{U}(-10^{4}, 10^{4})$      & $0.001 \pm 0.020$                     \\
$c_{\rm PEST, y}$                                           & $\mathcal{U}(-10^{4}, 10^{4})$      & $-0.011 \pm 0.016$                    \\
$c_{\rm PEST, dist}$                                        & $\mathcal{U}(-10^{4}, 10^{4})$      & $0.0151 \pm 0.0055$                   \\
$c_{\rm PEST, fwhm}$                                        & $\mathcal{U}(-10^{4}, 10^{4})$      & $-0.0047 \pm 0.0021$                  \\
$c_{\rm PEST, \chi}$                                        & $\mathcal{U}(-10^{4}, 10^{4})$      & $0.0008 \pm 0.0013$                   \\
$c_{\rm PEST, sky}$                                         & $\mathcal{U}(-10^{4}, 10^{4})$      & $-0.00033 \pm 0.00032$                \\
$c_{\rm TRAPPIST-South, dx}$                                & $\mathcal{U}(-10^{4}, 10^{4})$      & $0.0000078 \pm 0.0000013$             \\
$c_{\rm TRAPPIST-South, dy}$                                & $\mathcal{U}(-10^{4}, 10^{4})$      & $0.0000052 \pm 0.0000015$             \\
$c_{\rm TRAPPIST-South, fwhm}$                              & $\mathcal{U}(-10^{4}, 10^{4})$      & $-0.0041 \pm 0.0010$                  \\
$c_{\rm TRAPPIST-South, sky}$                               & $\mathcal{U}(-10^{4}, 10^{4})$      & $0.0027 \pm 0.0017$                   \\
$c_{\rm TRAPPIST-South, \chi}$                              & $\mathcal{U}(-10^{4}, 10^{4})$      & $0.0017 \pm 0.0013$                   \\ \hline
RV linear drift                                             &                                     &                                       \\ \hline
Systemic velocity, $v_{\rm HARPS}$ $(\rm m\,s^{-1})$        & $\mathcal{U}(17800, 17900)$         & $17858.6 \pm 2.3$                     \\
Slope, $\gamma_{\rm HARPS}$ $(\rm m\,s^{-1}\,day^{-1})$     & $\mathcal{U}(0, 1)$                 & $0.143 \pm 0.031$                     \\ \hline
\end{tabular}
\end{table}

\begin{figure*}
    \centering
    \includegraphics[width = \textwidth]{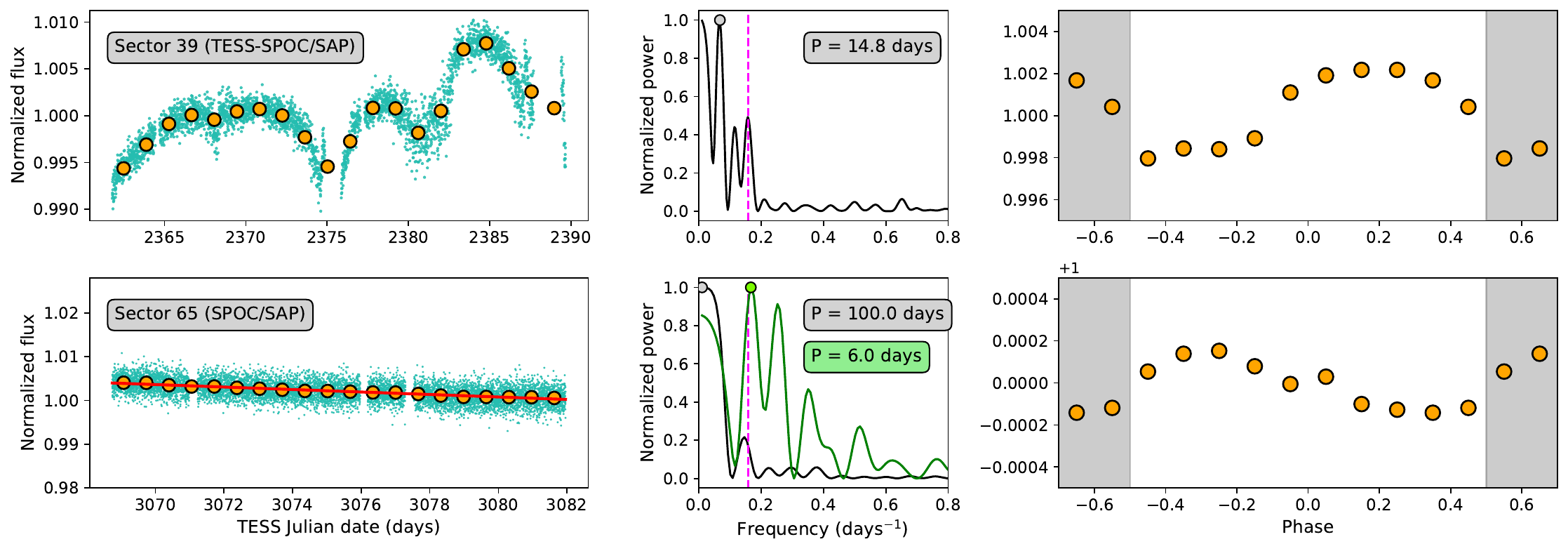}
    \caption[Sinusoidal signals within the S39 and S65 TESS SAP fluxes of TOI-5005.]{Sinusoidal signals within the S39 and S65 TESS SAP fluxes of TOI-5005. Left panels: SAP photometry with the TOI-5005.01 transit signatures masked. The orange circles correspond to 1.3-day binned data. Centre panels: Generalised Lomb-Scargle Periodograms of the time series. The vertical magenta dashed lines indicate the orbital period of TOI-5005.01. The grey circles and boxes indicate the maximum power frequencies. The green periodogram, circle, and box in the S65 panel illustrate the signal analysis over the detrended photometry. Right panel: SAP photometry folded to the orbital period of TOI-5005.01. The phase refers to the time of inferior conjunction. The orange circles correspond to binned data of 10$\%$ the orbital phase.}
    \label{fig:gls_to_TESS_sap}
\end{figure*}

\begin{figure*}
    \centering
    \includegraphics[width=0.48\textwidth]{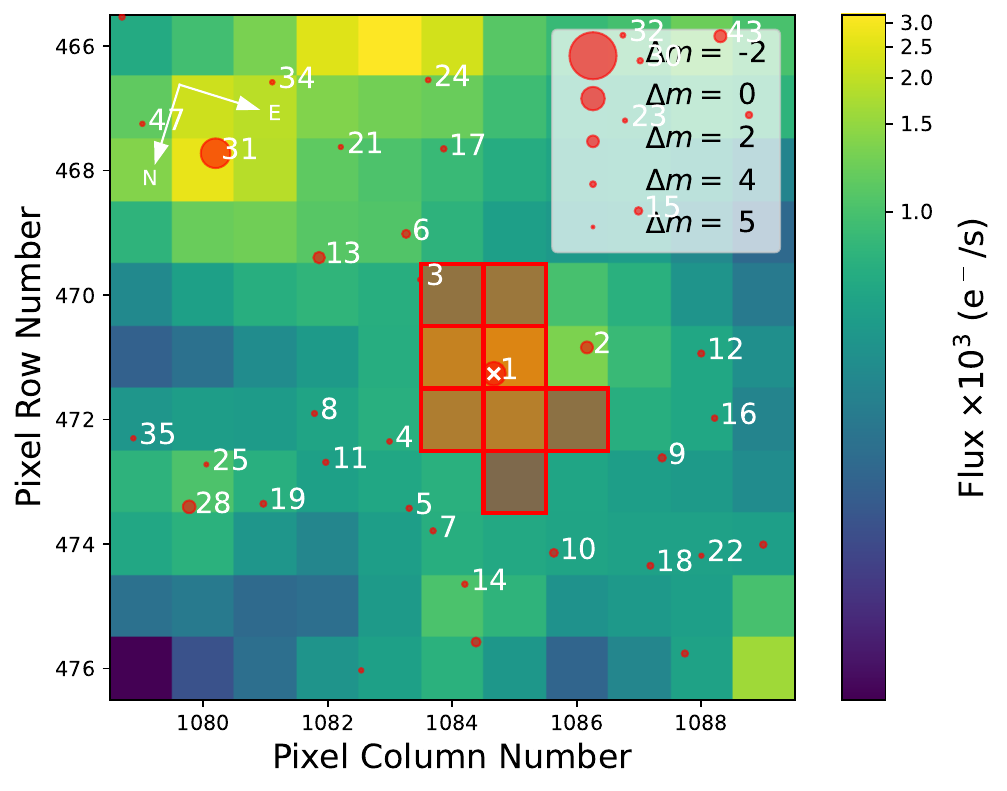}
    \includegraphics[width=0.48\textwidth]{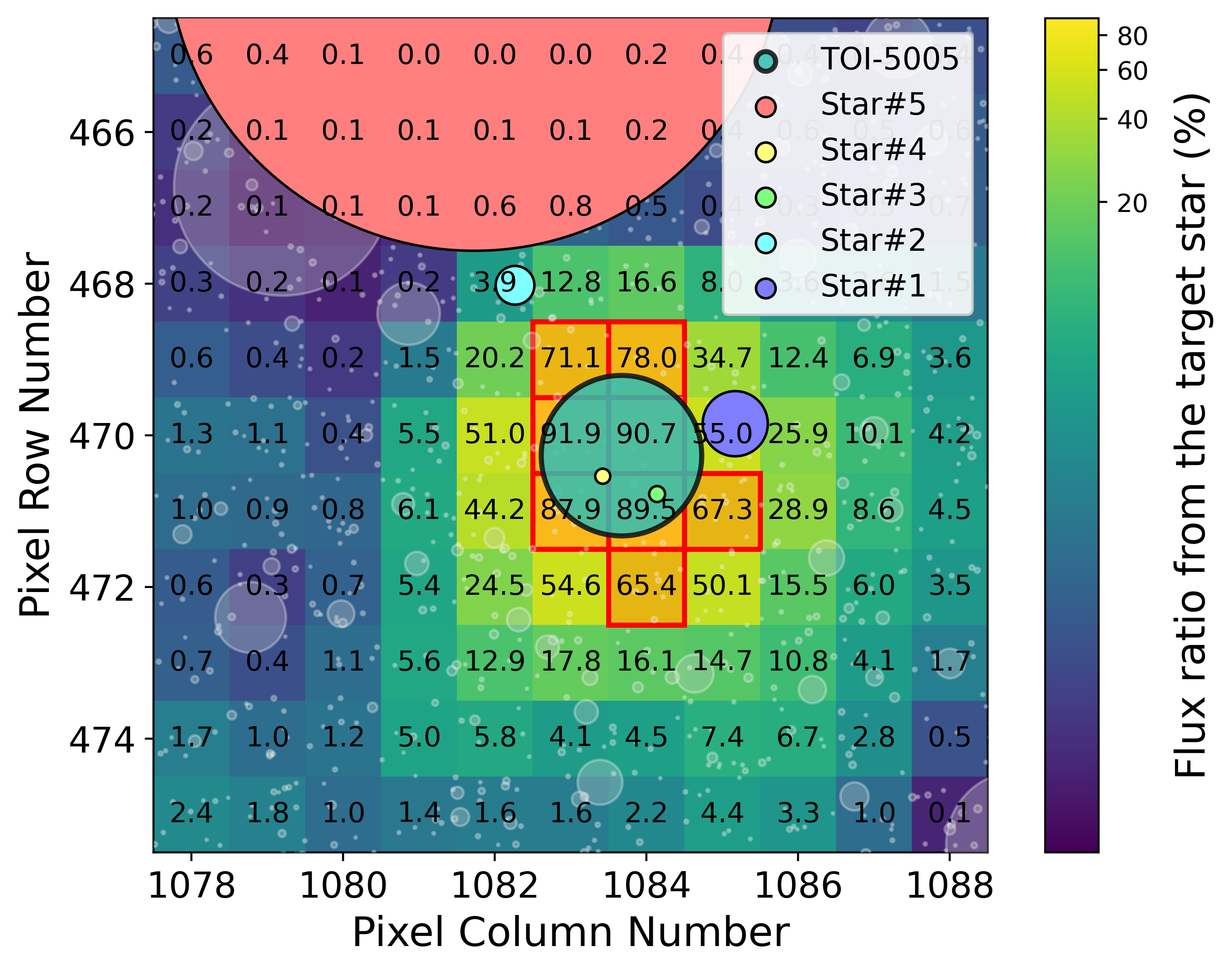}
    \caption[Pixel-based grid aperture used to extract SAP photometry in TESS S12.]{Pixel-based grid aperture used to extract SAP photometry in TESS S12. Left: 11 $\times$ 11 pixel FFI cutout showing the location of TOI-5005 and several \textit{Gaia} sources. Symbol sizes scale with their $G$ magnitudes. This plot has been made through \texttt{tpfplotter} \citep{2020A&A...635A.128A}. Right: 11 $\times$ 11 pixel heat-map representing the pixel-by-pixel target flux contribution. Symbol sizes for nearby \textit{Gaia} sources scale with their emitted fluxes. This plot has been prepared through our newly developed \texttt{TESS-cont} package. }
    \label{fig:S12_aperture}
\end{figure*}


\begin{figure*}
    \centering
    \includegraphics[width=0.24\textwidth]{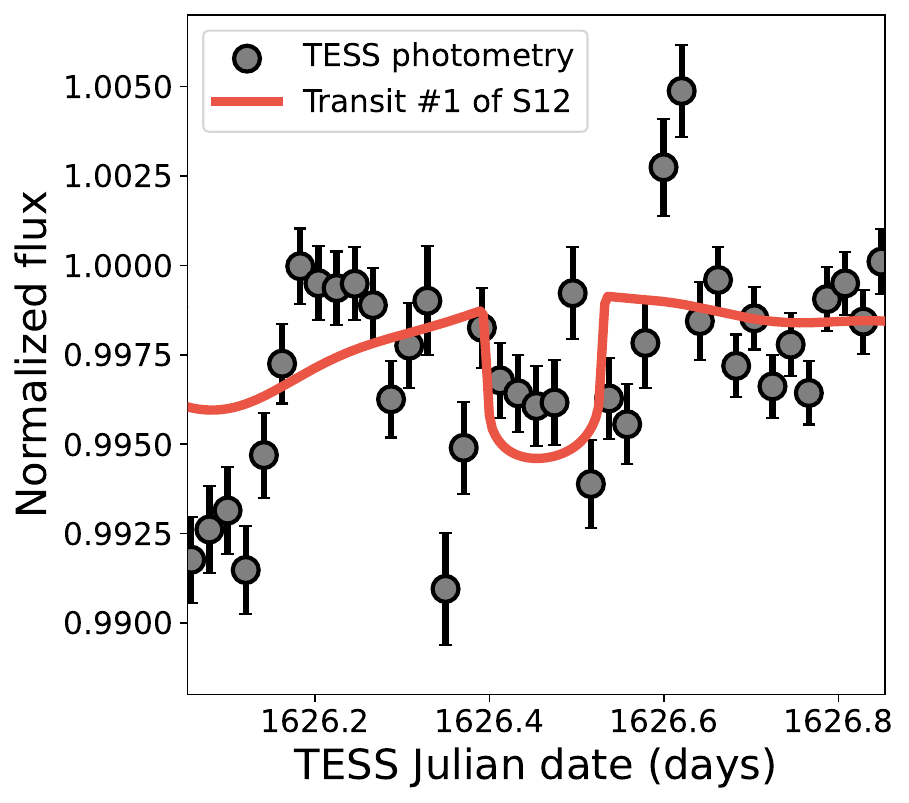}
    \includegraphics[width=0.24\textwidth]{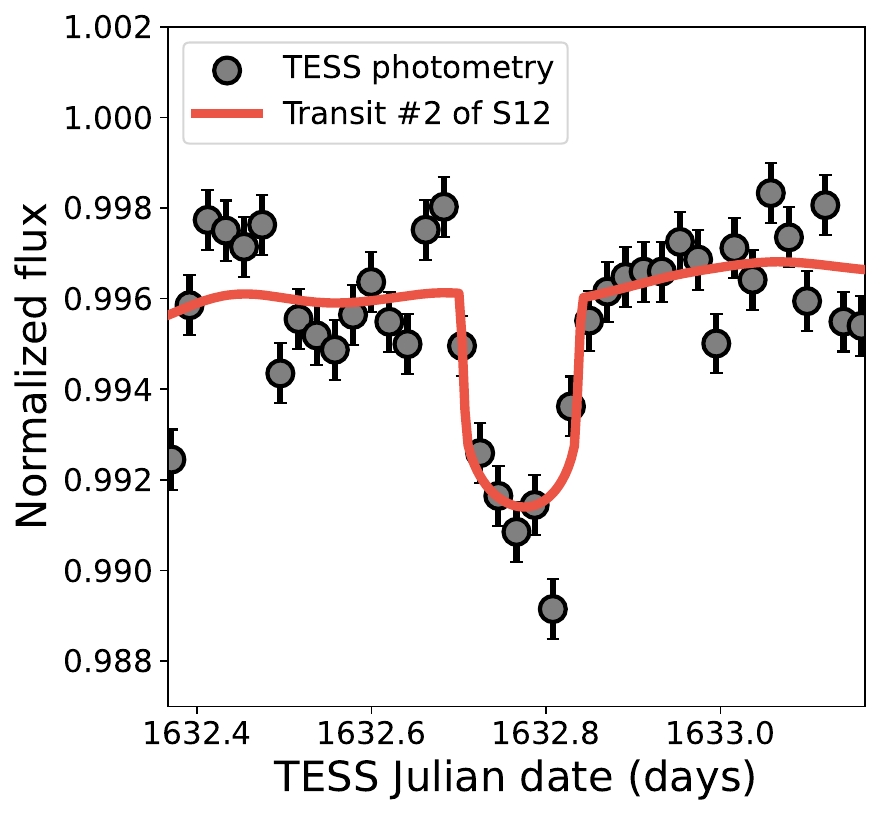}
    \includegraphics[width=0.24\textwidth]{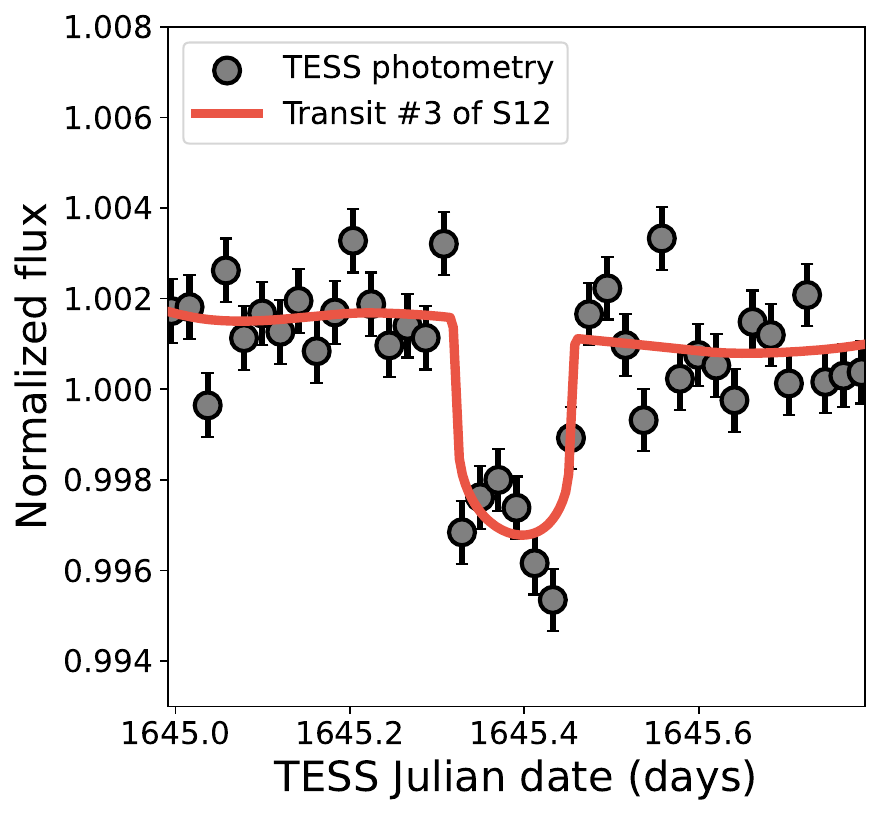}
    \includegraphics[width=0.24\textwidth]{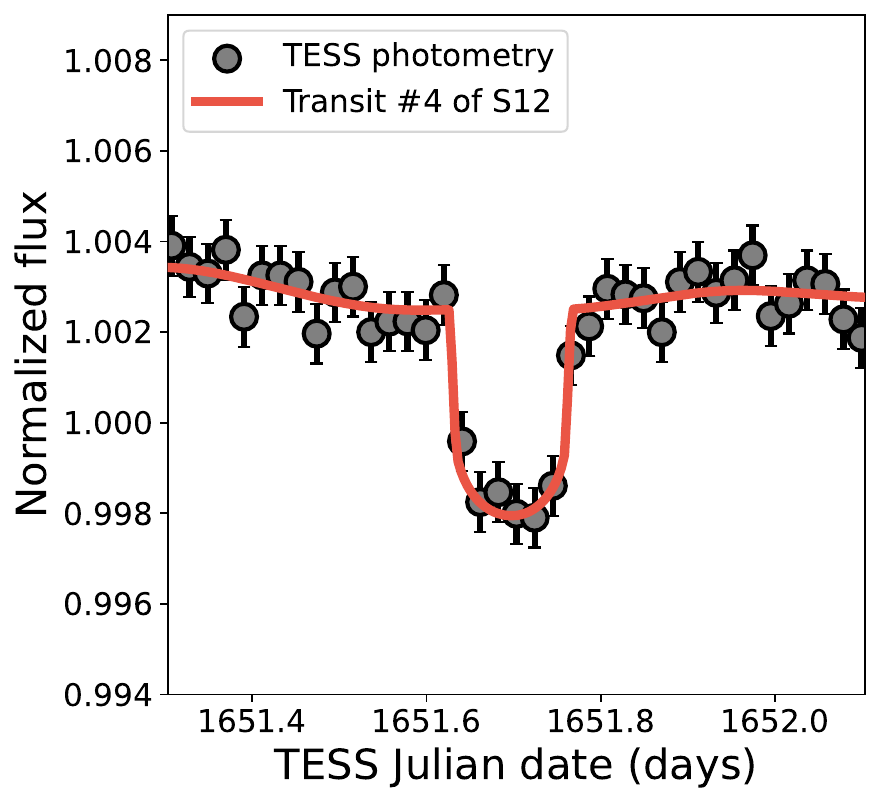}
    \includegraphics[width=0.24\textwidth]{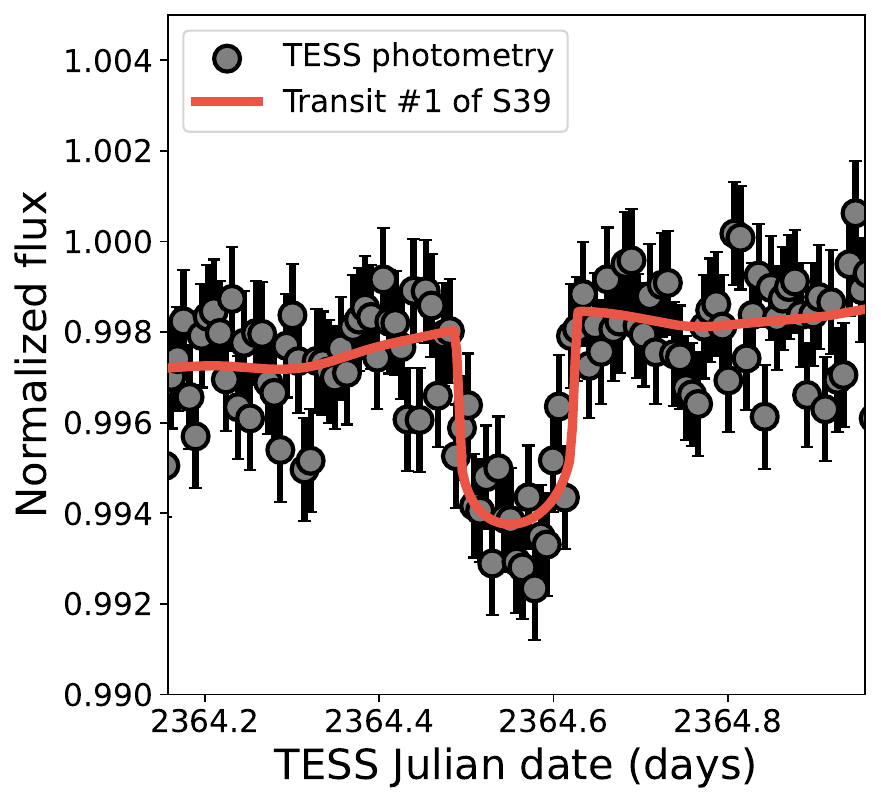}
    \includegraphics[width=0.24\textwidth]{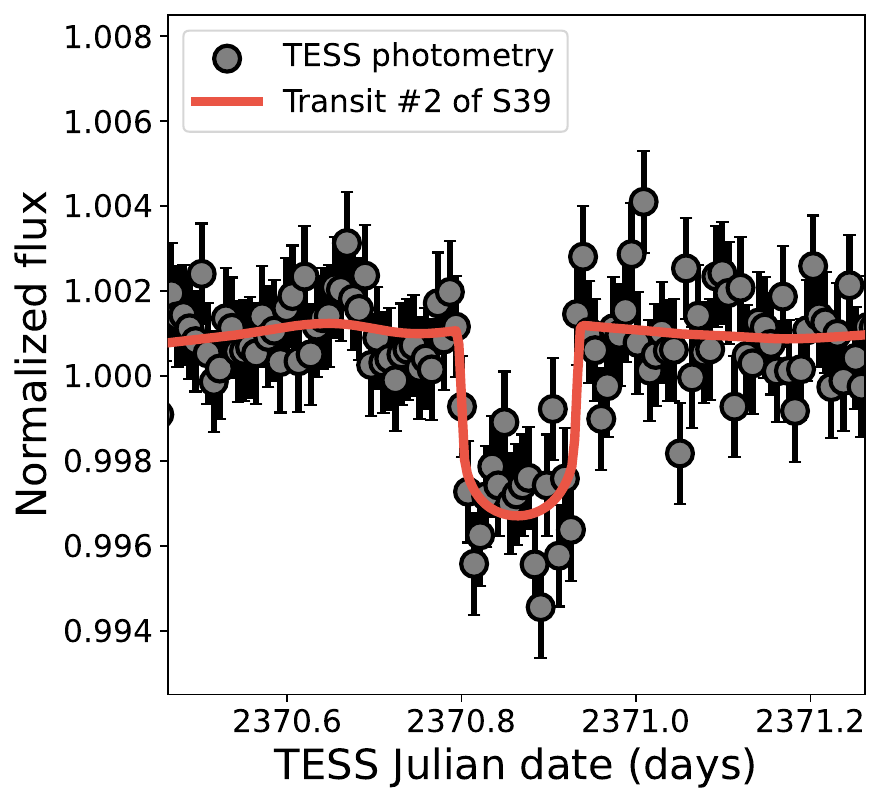}
    \includegraphics[width=0.24\textwidth]{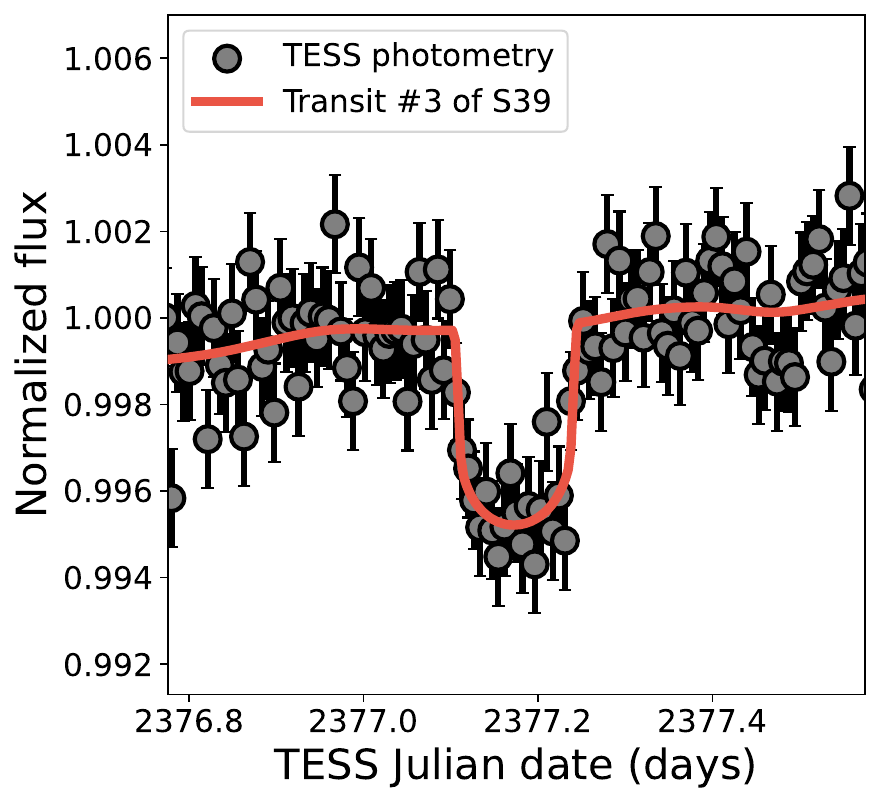}
    \includegraphics[width=0.24\textwidth]{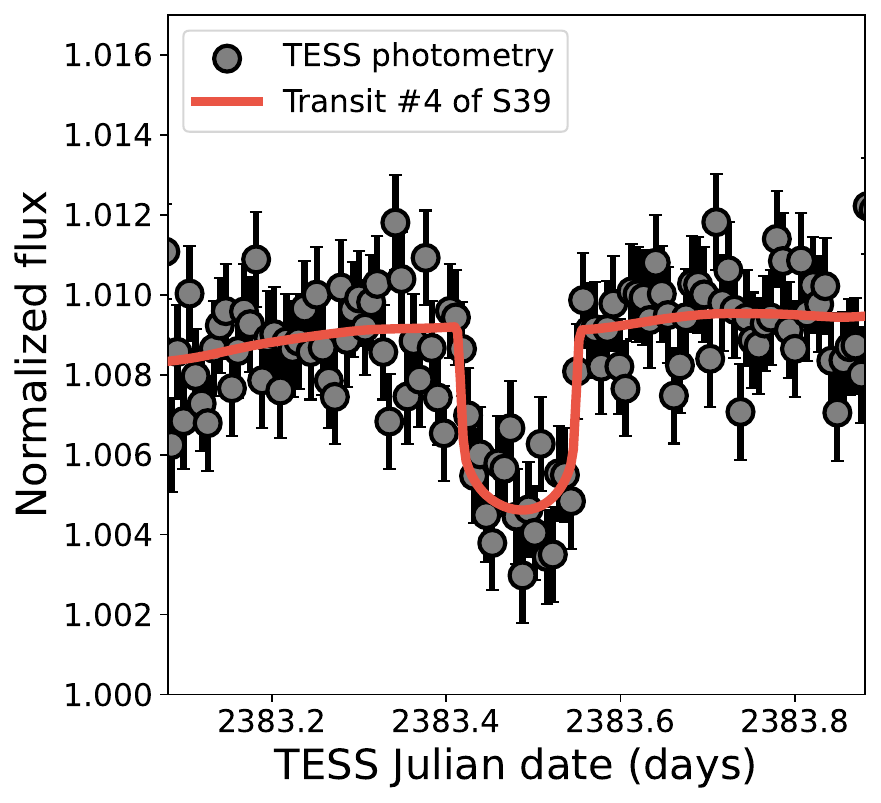}
    \includegraphics[width=0.24\textwidth]{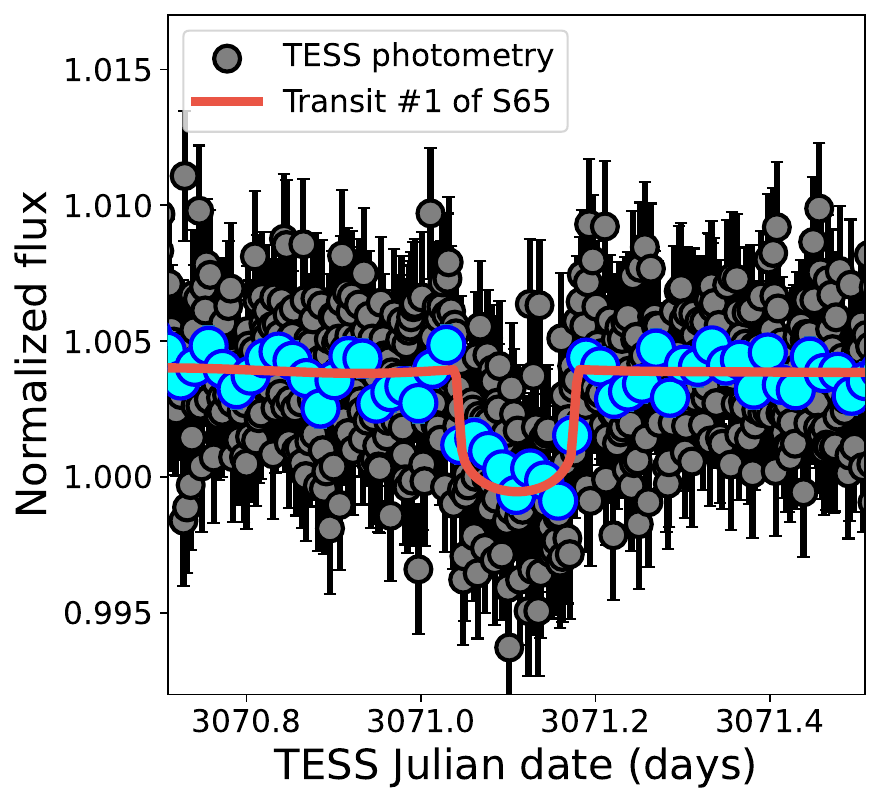}
    \includegraphics[width=0.24\textwidth]{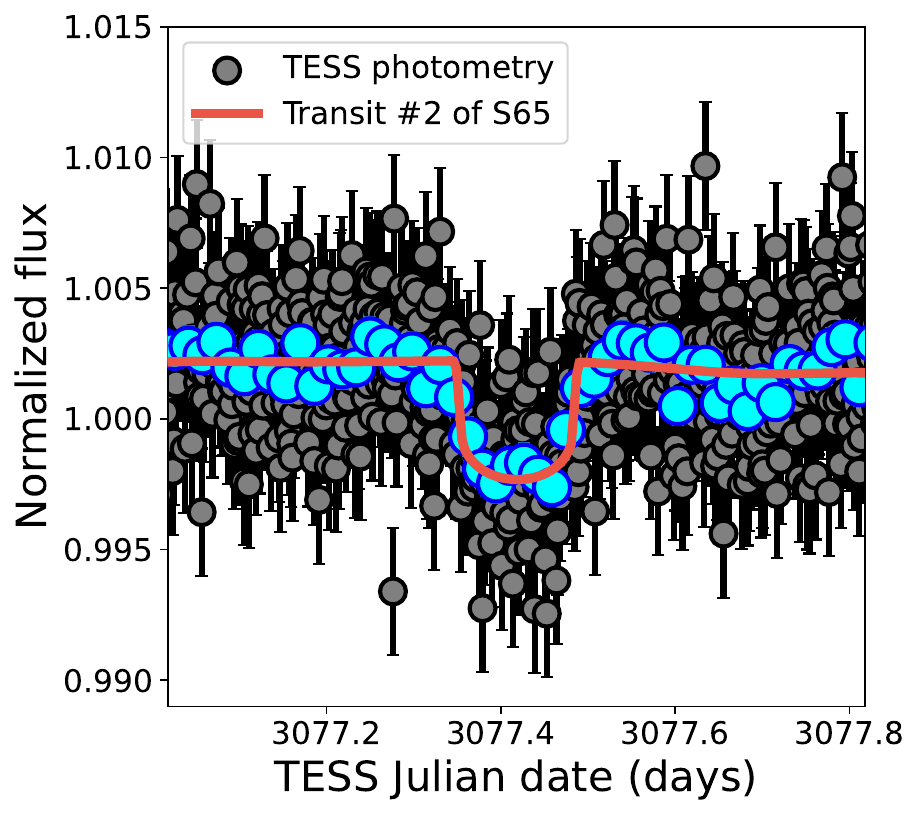}
    \includegraphics[width=0.24\textwidth]{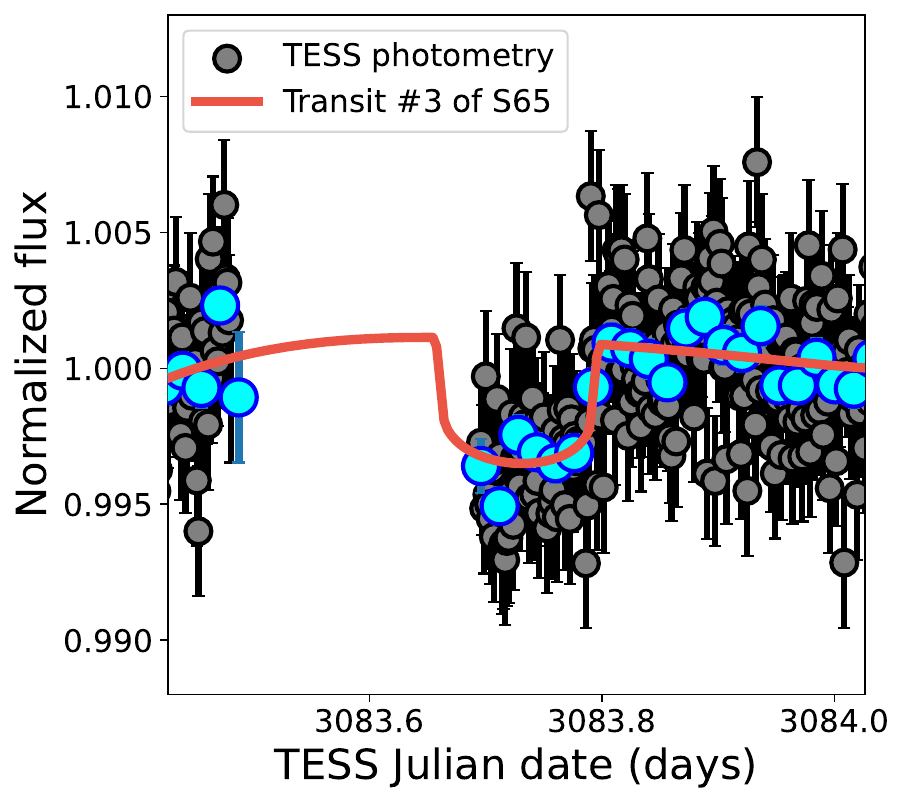}
    \includegraphics[width=0.24\textwidth]{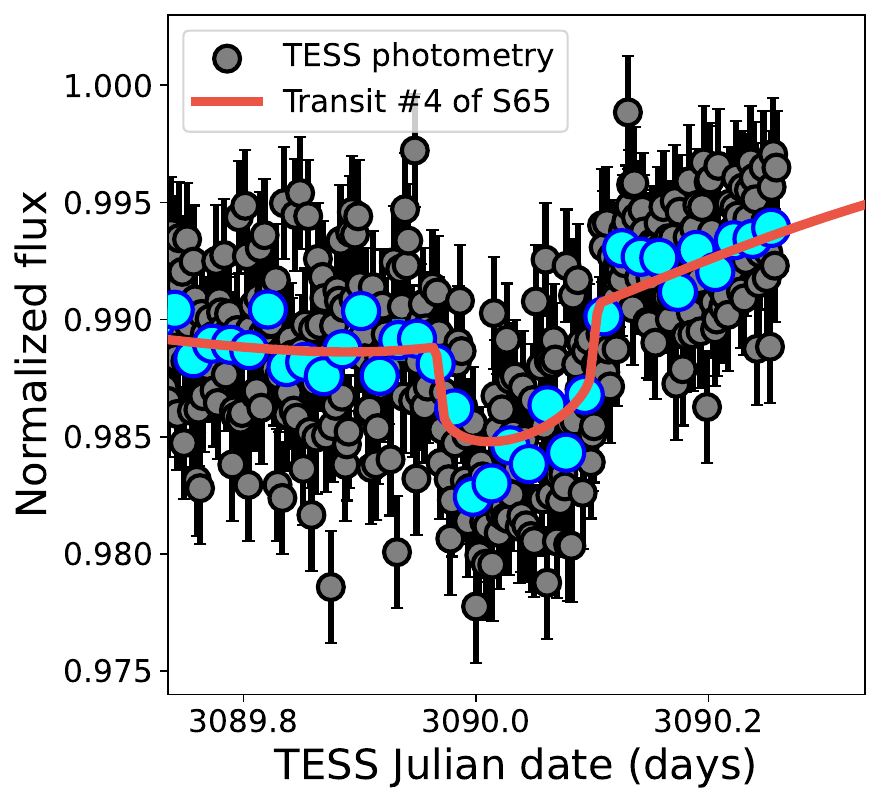}
    
    \caption[Individual transits of TOI-5005 b within the TESS SAP photometry together with the best-fit transit+GP model.]{Individual transits of TOI-5005 b within the TESS SAP photometry together with the best-fit transit+GP model. Blue data points in S65 data correspond to 25-min binned data. }
    \label{fig:individual_transits}
\end{figure*}


\begin{figure*}
    \centering
    \includegraphics[width=\textwidth]{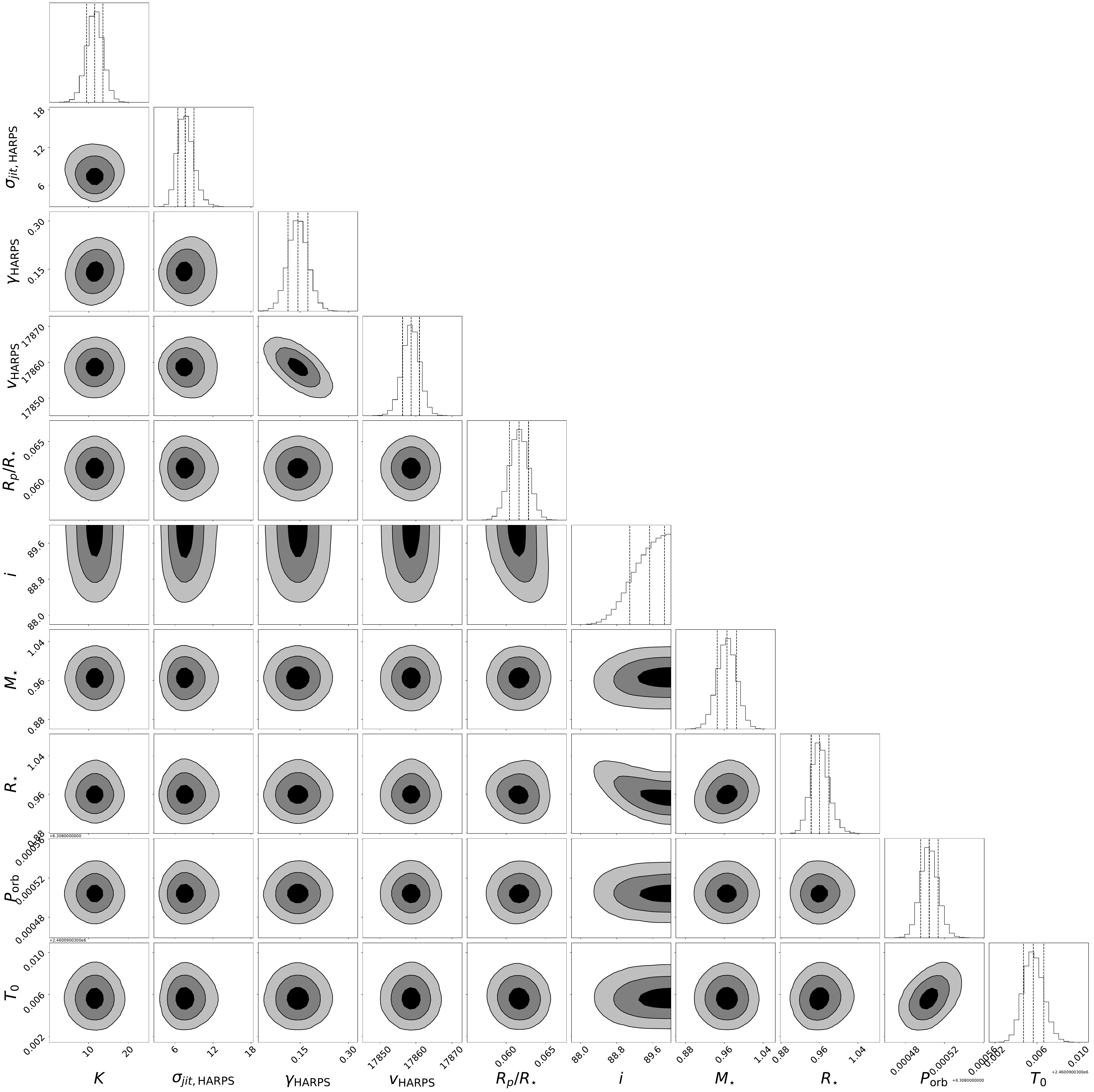}
    \caption[Corner plot showing the 1D and 2D posterior distributions of the main parameters of TOI-5005~b obtained in the joint analysis.]{Corner plot showing the 1D and 2D posterior distributions of the main parameters of TOI-5005~b obtained in the joint analysis (Sect.~\ref{subsec:joint_analysis}). The vertical dashed black lines indicate the median and 1$\sigma$ intervals.}
    \label{fig:corner_joint_analysis}
\end{figure*}


\begin{figure*}
    \centering
    \includegraphics[width=0.45\textwidth]{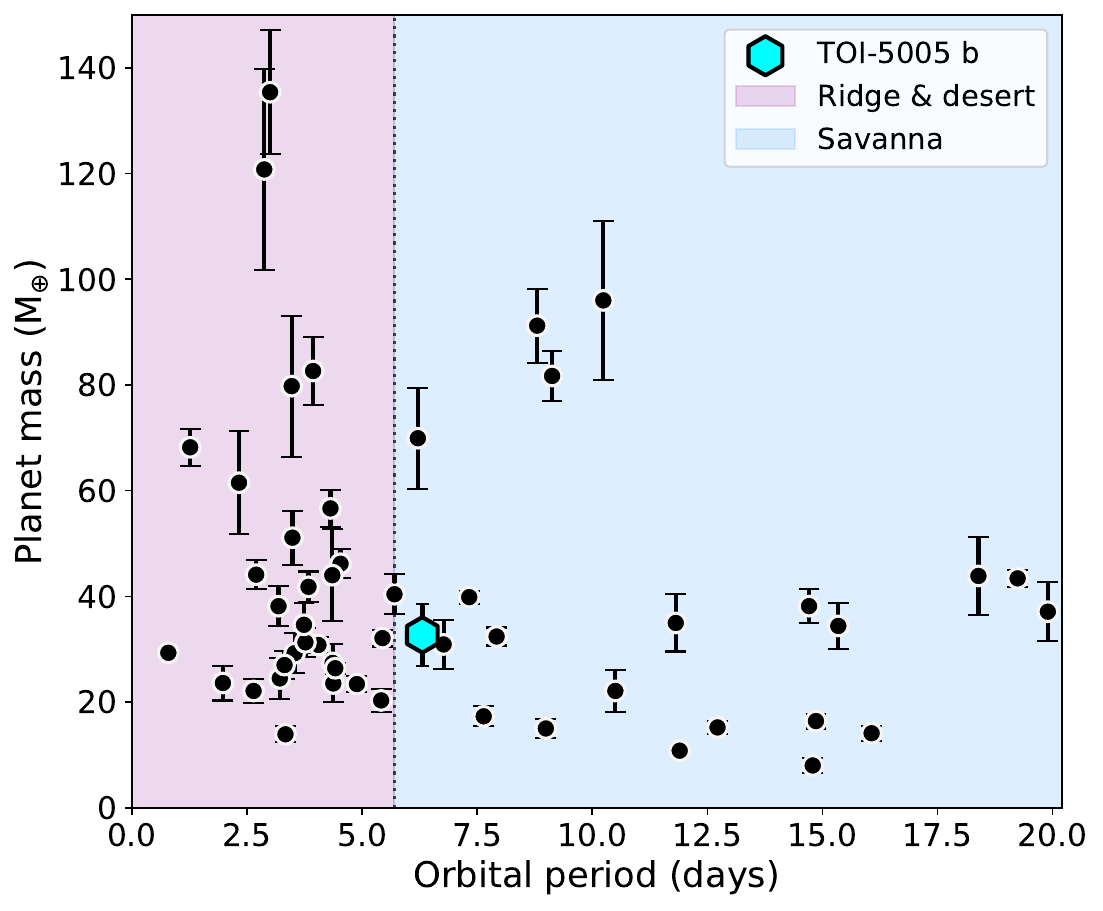}
    \includegraphics[width=0.44\textwidth]{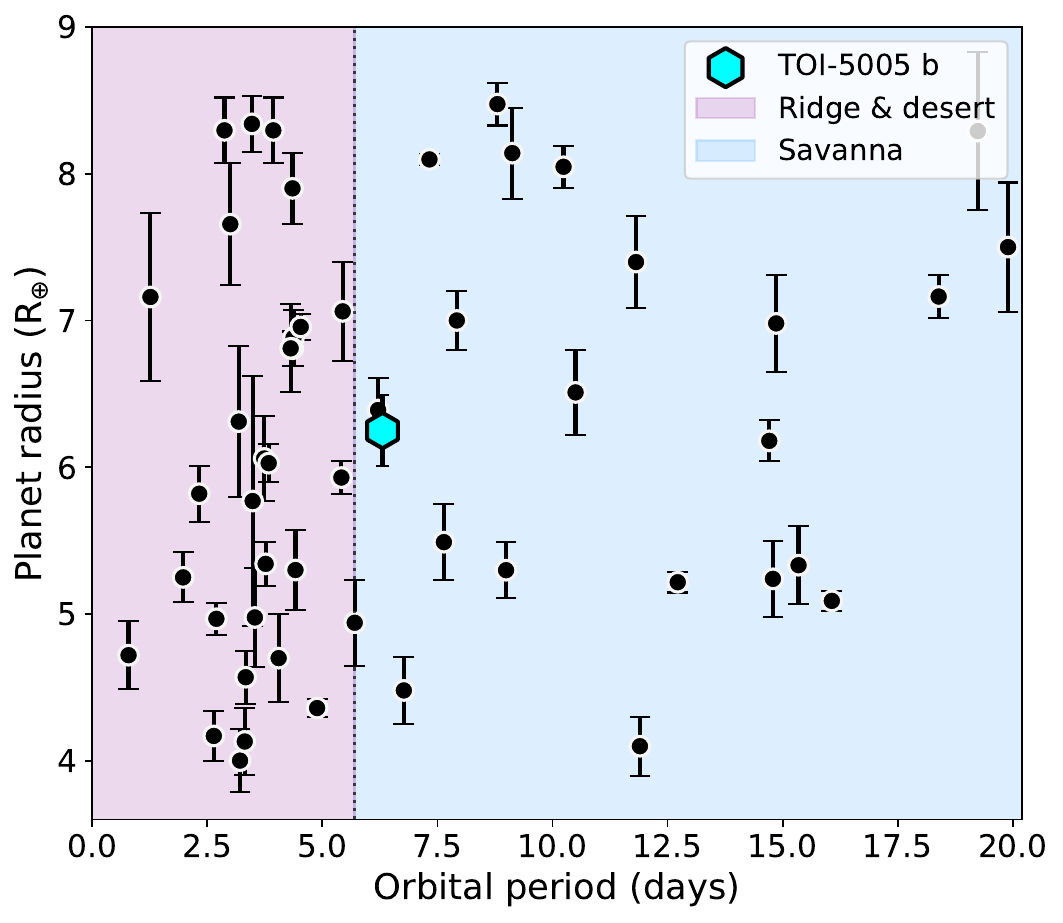}
    \caption[Mass-period and radius-period diagrams of all known Neptunian planets with masses and radii constrained to a precision better than 20$\%$.]{Mass-period and radius-period diagrams of all known Neptunian planets with masses and radii constrained to a precision better than 20$\%$. The KS statistical test (Sect.~\ref{subsec:LDSP}) indicates that planets in the ridge and desert tend to be more massive than those in the savanna ($p$-value = $0.037^{+0.030}_{-0.018}$), while the radius distribution is homogeneous in those regimes ($p$-value = $0.33^{+0.19}_{-0.13}$).}
    \label{fig:mass-period-radius-period}
\end{figure*}


\begin{figure*}
    \centering
\includegraphics[width=0.78\textwidth]{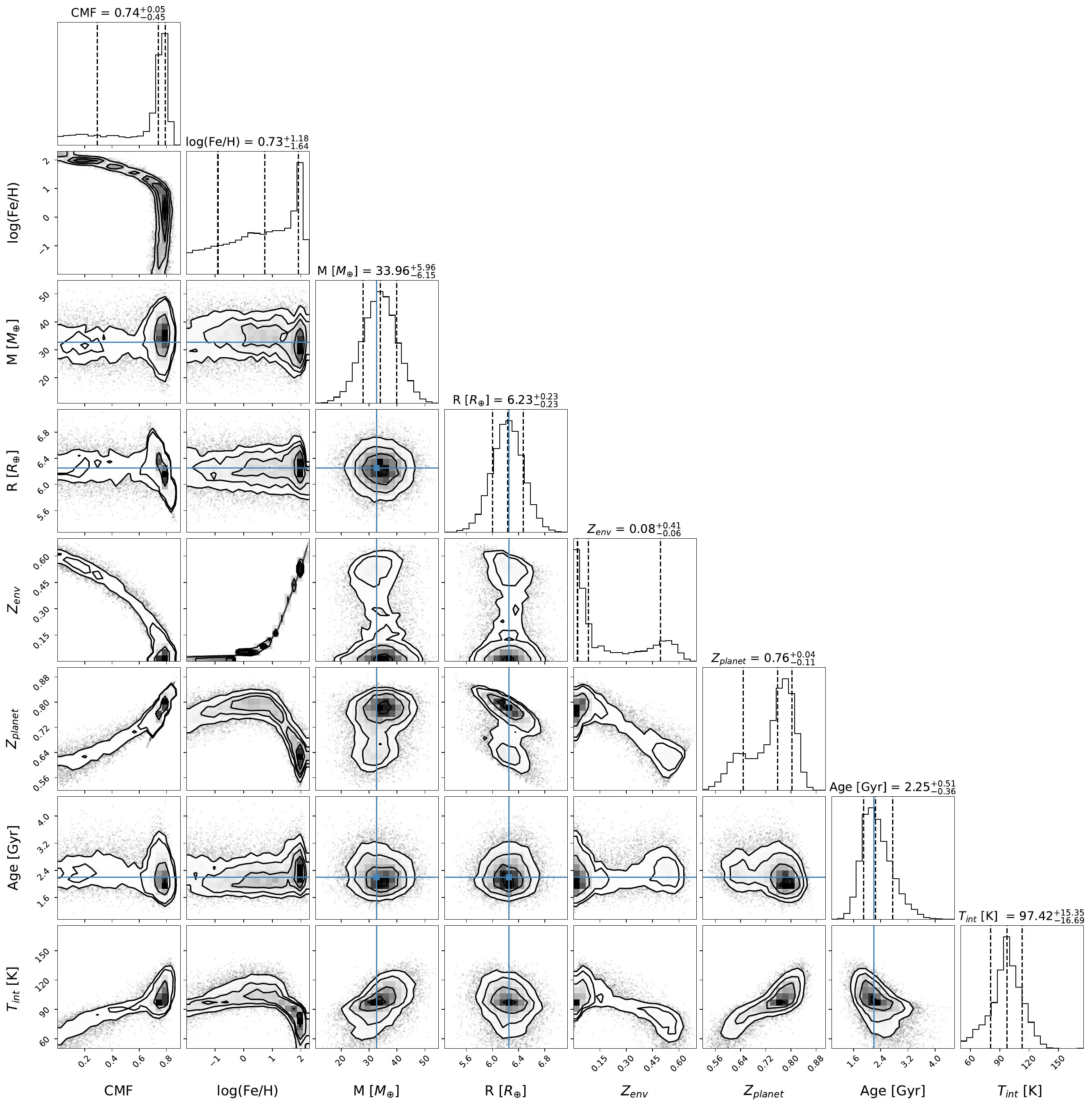}
    \caption[Corner plot showing the 1D and 2D posterior distributions obtained in the interior MCMC retrieval of TOI-5005~b.]{Corner plot showing the 1D and 2D posterior distributions obtained in the interior MCMC retrieval of TOI-5005~b. Light blue solid lines indicate the measured mean mass and radius, and the vertical dashed black lines indicate the median and 1$\sigma$ intervals.}
    \label{fig:interior_retrieval}
\end{figure*}



\begin{center}
\chapter[Codes and algorithms]{Codes and algorithms}
\end{center}
\label{sec:cbv_correction}
\vspace{3cm}
\pagestyle{fancy}
\fancyhf{}
\lhead[\small{\textbf{\thepage}}]{\textbf{Uninformed CBV-based systematics correction from Chapter \ref{ch:toi_5005}}}
\rhead[\small{\textbf{Appendix~\hyperref[ch:Appendix_E]{E}}}]{\small{\textbf{\thepage}}}

\renewcommand{\thetable}{E.\arabic{table}}   

\normalsize

In this Appendix, we include brief descriptions and links to the public repositories of the following codes developed for this thesis: \texttt{mr-plotter}, \texttt{nep-des}, and \texttt{TESS-cont}. 

\texttt{mr-plotter} \citep{2023A&A...675A..52C} is a Python tool developed to create paper-quality mass-radius diagrams based on a wide range of state-of-the-art models of planetary interiors and atmospheres. It can be used to contextualise any observed planet and infer its possible internal structure. It can also be used to search for correlations at a population level thanks to its colour-coding option based on any property collected in the NASA Exoplanet Archive, PlanetS, and Exoplanet.eu catalogues. The code and its corresponding documentation can be found at \url{https://github.com/castro-gzlz/mr-plotter}.

\texttt{nep-des} \citep{2024A&A...689A.250C} is a Python-based Jupyter Notebook to create period-radius diagrams with the population-based boundaries of the Neptunian desert, ridge, and savanna derived in Chapter~\ref{ch:nep_des}. The code and documentation can be found at \url{https://github.com/castro-gzlz/nep-des}.

\texttt{TESS-cont} \citep{2024A&A...691A.233C} is a Python tool to quantify the flux fraction coming from nearby stars in the TESS photometric aperture of any observed target. The package (1) identifies the main contaminant Gaia DR2/DR3 sources, (2) quantifies their individual and total flux contributions to the aperture, and (3) determines whether any of these stars could be the origin of the observed transit and variability signals. The TESS-cont algorithm is based on building the pixel response functions (PRFs) of nearby Gaia sources and computing their flux distributions across the TESS Target Pixel Files (TPFs) or Full Frame Images (FFIs). The full description of the algorithm can be found in Sect.~\ref{sec:tess-cont}. The code and its corresponding documentation can be found at \url{https://github.com/castro-gzlz/TESS-cont}.

\end{document}